\documentclass{raa_twocolumn}
\usepackage{natbib}
\usepackage{amssymb,amsmath}
\usepackage{url}
\usepackage{upgreek}
\bibpunct{(}{)}{;}{a}{}{,}
\usepackage{graphicx,times}
\usepackage{pdflscape}
\usepackage{longtable}
\usepackage{grffile}
\usepackage{gensymb}
\usepackage[figuresright]{rotating}
\usepackage{supertabular,multicol}
\usepackage{threeparttable} 
\usepackage{multirow}
\usepackage{makecell}
\usepackage[pagebackref=true]{hyperref}

\usepackage[final]{changes}

\hypersetup{pdftitle = The title of my PDF, pdfauthor = My name, pdfsubject= The subject, pdfkeywords = keyword1 keyword2 keyword3}
\hypersetup{colorlinks = true, linkcolor = blue, anchorcolor = blue, citecolor = blue, filecolor = blue, urlcolor = blue}
\footnotesize
\newdimen\digitwidth    %define ! a one digit width for tables
\setbox0=\hbox{\rm0}
\digitwidth=\wd0
\catcode`!=\active
\def!{\kern\digitwidth}
\normalsize

\begin{document}

\title{The FAST Galactic Plane Pulsar Snapshot Survey:
  \\ II. Discovery of 76 Galactic rotating radio transients and their enigma 
% \footnotetext{\small $*$ Supported by the National Natural Science Foundation of China.}
}
 \volnopage{ {\bf 2022} Vol.\ {\bf 21} No. {\bf 10}, A501}
   \setcounter{page}{1}
   \author{
     % -- Direct workers
     D. J. Zhou\inst{1,2},
     J. L. Han\inst{1,2},
     Jun Xu\inst{1}, 
     Chen Wang\inst{1},
     P. F. Wang\inst{1}, 
     Tao Wang \inst{1,2},
     Wei-Cong Jing\inst{1,2}, \\
     Xue Chen\inst{1,2},  
     Yi Yan\inst{1,2}, 
     Wei-Qi. Su\inst{1,2}, 
 %  Others in abc
     Heng-Qian Gan\inst{1,2},
     Peng Jiang\inst{1,2}, 
     Jing-Hai Sun\inst{1,2},
     Hong-Guang Wang\inst{3,4}, 
      Na Wang\inst{5,6}, 
      Shuang-Qiang Wang\inst{5},
     Ren-Xin Xu\inst{7},
and 
    Xiao-Peng You\inst{8}
   }

   \institute{
     National Astronomical Observatories, Chinese Academy of Sciences, 20A Datun
 Road, Chaoyang District, Beijing 100101, China;  {\it hjl@nao.cas.cn} \\ %1
     \and     
     School of Astronomy, University of Chinese Academy of Sciences, Beijing 100049, China; \\%3
%     \and
%     CAS Key Laboratory of FAST, NAOC, Chinese Academy of Sciences, Beijing 100101, China; \\%2
     % \and
     % Kavli Institute for Astronomy and Astrophysics, Peking University, Beijing 100871, China; \\%5
     \and
     % Department of Astronomy, GuangZhou University, GuangZhou 510006, China; \\%6
     Department of Astronomy, School of Physics and Materials Science, Guangzhou University, Guangzhou 510006, Guangdong Province, China %5
     \and 
     National Astronomical Data Center, Great Bay Area, Guangzhou 510006, Guangdong Province, China %6
     \and
     Xinjiang Astronomical Observatory, Chinese Academy of Sciences, 150 Science 1-street, Urumqi 830011, China; \\%7
     \and
     Key Laboratory of Radio Astronomy, Chinese Academy of Sciences, Nanjing 210008, China; \\   %8
     \and
     Department of Astronomy, Peking University, Beijing 100871, China; \\%4
     \and
     School of Physical Science and Technology, Southwest University, Chongqing 400715, China\\%9
    \vs \no
   {\small Received 2023 January 4; revised 2023 March 27; accepted 2023 March 30; published 2023 September 29}
 }

   \abstract{
%    Background
    We are carrying out the Galactic Plane Pulsar Snapshot (GPPS) survey by using the Five-hundred-meter Aperture Spherical radio Telescope (FAST), the most sensitive systematic pulsar survey in the Galactic plane.
%   Where we are doing
    In addition to about 500 pulsars already discovered through normal periodical search, we report here the discovery of 76 new transient radio sources with sporadic strong pulses, detected by using the newly developed module for a sensitive single pulse search. Their small DM values suggest that they all are the Galactic rotating radio transients (RRATs). They show different properties in the following-up observations. 
%    Results
%1) 
    %  Four of them have merely one pulse detected. These one-off
    %  sources cannot be redetected  In the following-up observations,
    %  Three one-off sources have a much larger dispersion measure (DM)
    %  than the predicated values from the Galactic electon density
    %  models, so that they are probably extragalactic and hence
    %  are fast radio bursts (FRBs), and one must be Galactic.
%
    More radio pulses have been detected from 26 transient radio sources but no periods can be found due to a limited small number of pulses from all FAST observations.
%2)
    The following-up observations show that 16 transient sources are newly identified as being the prototypes of RRATs with a period already determined from more detected sporadic pulses, and 10 sources are extremely nulling pulsars, and 24 sources are weak pulsars with sparse strong pulses.
%
%3) 
    On the other hand, 48 previously known RRATs have been detected by the FAST either during verification observations for the GPPS survey or through targeted observations of applied normal FAST projects. Except for 1 RRAT with four pulses detected in a session of five minute observation and 4 RRATs with only one pulse detected in a session, sensitive FAST observations reveal that 43 RRATs are just generally weak pulsars with sporadic strong pulses or simply very nulling pulsars, so that the previously known RRATs always have an extreme emission state together with a normal hardly detectable weak emission state. This is echoed by the two normal pulsars J1938+2213 and J1946+1449 with occasional brightening pulses.
   Though strong pulses of RRATs are very outstanding in the energy distribution, their polarization angle variations follow the polarization angle curve of the averaged normal pulse profile, suggesting that the predominant sparse pulses of RRATs are emitted in the same region with the same geometry as normal weak pulsars. % produced by the suddenly enhanced charge density rather than the reconnection of magnetic fields in the magnetosphere.
   }

\authorrunning{{\it D. J. Zhou et al.}: The FAST GPPS Survey: II. Discovery of 76 RRATs}            %author_head in even pages
\titlerunning{{\it D. J. Zhou et al.}: The FAST GPPS Survey: II. Discovery of 76 RRATs}  % title_head in odd pages

%   \authorrunning{D. J. Zhou et al. }            %author_head in even pages
%   \titlerunning{The GPPS survey: II. Bursts detection}  % title_head in odd pages
   \maketitle
%
%________________________________________________ sections below

\section{Introduction}
\label{sect1:intro}

% 1.  Introduction 
 
Time-domain astronomical observations reveal more and more the Galactic or extragalactic transients in many electromagnetic wave bands, rested on the new development of high sensitivity detectors which can outline the light curves of sources. Most successful were the detection of the Gamma-Ray bursts \citep[see review by][]{fm95}, Fast Radio Bursts \citep[FRB,][]{Lorimer2007} and extrasolar planets \citep[e.g.][]{mq95,bm96}. Radio pulsar detection, even in the first discovery \citep{hbp+68}, has been working in the time domain. It started with a high time resolution of seconds, later milliseconds \citep[e.g.][]{BurkeSpolaor2010}, and nowadays microseconds  or even to nanoseconds \citep[e.g.][]{hkwe03}. For a bright pulsar, each individual pulse from every pulsar period can be directly observed if a radio telescope is sensitive enough. For a weak pulsar, signals have to be integrated over many periods, so that a mean pulse profile is obtained. Because of the interstellar medium, the pulses must be dedispersed to diminish the delay between frequency channels in an observational radio band. %Pulsar pulses are highly polarized. 

Most known pulsars were discovered through searching in the Fourier domain by using the packages such as  PRESTO\footnote{\url{https://www.cv.nrao.edu/~sransom/presto}} \citep{Ransom2011} and SIGPROC\footnote{\url{http://sigproc.sourceforge.net}} \citep{Lorimer2011}, or using the fast-folding algorithm \citep[FFA, ][]{Staelin1969, Kondratiev2009, Cameron2017, Parent2018} in the time domain. 
The enhanced red noise from the observation system caused by the fluctuations of the receiver gain, system temperature and radio frequency interference (RFI) affects the detection of long-period pulsars in the Fourier domain. 
%of surveys, the emission energy variation, and/or the internal emission mechanism of long-period pulsars, like the lower spin-down luminosity, the flux over time of aggregated profiles of many pulsars is too weak to be discovered. %
%
\citet{Cordes2003} developed a new single pulse search technology, finding enhance points in dedispered data series. When the signal to noise ratio (S/N) is much larger than a given threshold, then a radio transients can be found as an isolated distribution in the image of dispersion measurement (DM) versus time. 
By using this technology, \citet{Mclaughlin2006} discovered the first rotating radio transients (RRATs), and \citet{Lorimer2007} discovered the first FRB. This technology is effective for the searching of bright radio transients, and has been used in many pulsar surveys \citep{Cordes2006,Hessels2008,Deneva2009,BurkeSpolaor2010,Keane2010MNRAS,Burke-Spolaor2011,Keane2011MNRAS,Coenen2014,Stovall2014,KarakoArgaman2015,Deneva2016,Keane2018}. 

As defined by \citet{BurkeSpolaor2010}, RRATs are often considered as a special group of pulsars that emit bright pulses sporadically, making them difficult to be found in a normal periodic search method. They are usually discovered by a single-pulse search method. Up to now, %In the decade since the first RRAT was discovered~\citep{Mclaughlin2006}, 
more than 160 RRATs have been discovered, see the list in the  \texttt{RRATalog}\footnote{\url{http://astro.phys.wvu.edu/rratalog/}} and also a recent summary by \citet{Abhishek2022arXiv220100295A}, plus more new RRATs recently discovered \citep{Patel2018,Tyul2018A&A,Tyul2018ARep,Good2021ApJ,Han2021RAA,bbc+22,Dong2022arXiv221009172D,tpk+22}. 
The number of known RRATs claimed in literature \citep{Abhishek2022arXiv220100295A} is about or more than five percent of the known pulsars. It is quite possible that many RRATs or long period pulsars are missed in some pulsar surveys, if the single-pulse search has not properly been carried out for surveys \citep[see][]{Keane2010MNRAS}. Discovery of more RRATs can help to understand the neutron star population in the Galaxy. 

To understand why the pulses radiate very occasionally, sensitive observations and detailed statistics of long term observations are desired \citep{Bhattacharyya2018MNRAS}. However, only a small fraction of RRATs have been well studied. The energy distribution or peak flux distribution of single pulses from a very sensitive observation could give the answer to how an RRAT behaves. For a generally very weak pulsar, it could appear as an RRAT if only a few bright pulses are detectable for a given sensitivity. The energy of normal pulses of a pulsar always follows a log-normal distribution, while abnormal bright pulses should show an extra component at the high energy end as if they are giant pulses. This has been verified by some RRATs, such as PSR J1846$-$0257  \citep{Mickaliger2018MNRAS}. 
For other RRATs, \citet{Cui2017ApJ} and \citet{Mickaliger2018MNRAS} did not find the tail beyond the log-normal distribution, indicating that the pulses are normal but just very nulling \citep{BurkeSpolaor2010}, so that they radiate pulses by the same emission mechanism as pulsars. Then, the concerns are turned to why RRATs are very nulling. Some RRATs, e.g. PSR J0941$-$39 \citep{BurkeSpolaor2010} and PSR J1752+2359 \citep{sywy21}, can have two emission states, one with occasional bright pulses and one with a very nulling state or normal pulsar state. \citet{Lu2019SCPMA} observed three RRATs by using the Five-hundred-meter Aperture Spherical radio Telescope \citep[FAST,][]{Nan2011IJMPD}, and they detected many bursts for PSRs J1538+2345 and J1854+0306, while PSR J1913+1330 radiates pulses continuously but clearly shows nulling nature.  Based on the FAST observation of PSR J0628+0909, \citet{Hsu2023MNRAS} concluded that this known RRAT has weak single pulses as a normal pulsar with transient-like strong pulses, similar to PSR J0659+1414~\citep[B0656+14,][]{Weltevrede2006}.

% + RRAT polarization, (single pulse polarization for J1819+1458 ?)
To understand the sporadic behavior of RRATs and the switching of different emission states, the polarization observations are desired not only for bright single pulses but also for weaker normal pulses. Previously available measurements were made for single pulses of PSR J1819$-$1458 by \citet{Karastergiou2009MNRAS}, of 17 RRATs by \citet{Caleb2019MNRAS}, of PSR J1752+2359 by \citet{sywy21} and of PSR J1905+0849g by \citet{Han2021RAA}. Mean polarization profiles of only a few RRATs have been measured \citep{Karastergiou2009MNRAS}, which in fact is the key to understand the radiation mechanism and radiation geometry.

Timing is fundamental for a rotating neutron star to get the period derivative measured, and hence the characteristic age and magnetic field derived. For RRATs, it is challenging to do timing, because only a small number of strong pulses can be detected in a limited observation time. The initial discovery position of detection is often poorly given. Single pulses probably emerge randomly in the averaged pulse emission window. A mean pulse profile can therefore rarely be obtained because of the sparseness of pulses, and the timing residual by using these individual pulses could be as large as hundred milliseconds \citep{Bhattacharyya2018MNRAS}. The available timing results for several RRATs in the P-Pdot diagram \citep{McLaughlin2009MNRAS,Keane2011MNRAS,Cui2017ApJ,Bhattacharyya2018MNRAS} show that long-period RRATs look like either magnetars if they have a large period derivative, or simply aged pulsars near the death line if they have small period derivatives. Statistical analysis of populations of known RRATs by  \citet{Abhishek2022arXiv220100295A} shows no correlation with nulling pulsars, nor tending to be older and closer to the death line.
Very surprising is a glitch detected from PSR J1819-1458\footnote{As we believe that RRATs are a sub-class of pulsars, therefore we name an RRAT in the general format of pulsars, such as PSR J1234+3456 or simply J1234+3456.} \citep{LyneAG2009MNRAS}, which indicates that some RRATs may be just young pulsars with merely bright single pulses ever detectable.

The FAST \citep{Nan2011IJMPD} has a larger collecting area with an aperture diameter of 300~m. Mounted with a 19-beam L-band receiver with a system temperature of 22~K \citep{Jiang2020}, it has a great sensitivity to study the universe in radio waves, such as the discovery of weak pulsars \citep{Han2021RAA,PanZC2021ApJ,WangP2021SCPMA}, and the discovery of new FRBs \citep{ZhuWW2020ApJ,NiuCH2021ApJ,NiuCH2021arXiv211007418N} or study the details of known pulsars \citep{whx+22} and FRBs \citep{lwm+20,XUHENG2021arXiv211111764X}.
At present, the FAST Galactic Plane Pulsar Snapshot (GPPS) survey\footnote{\url{http://zmtt.bao.ac.cn/GPPS/}} \citep{Han2021RAA} is hunting pulsars in the Galactic plane, and has discovered more than 500 pulsars already. In the first paper by \citet{Han2021RAA}, the GPPS survey strategy is introduced and the discoveries of 201 pulsars and 1 RRAT have been presented, together with parameter improvements for 64 previously known pulsars. %re than two hundred new pulsars, several of which were discovered based on single-pulse search method. For example, RRAT J1905+0849, later numbered gpps0275, only 12 bright pulses detected in 15 minutes. The PSR J1924+2037g (gpps0192), a weak and very nulling pulsar, was first discovered by the single-pulse module. PSR J1919+1527g (gpps0130), J1939+2352g (gpps0150) and PSR J1924+2037g (gpps0192) are also very nulling pulsars and detected a few individual pulses first by the single-pulse search method. The same is true for the long-period pulsar of 9.89012~s for PSR J1856+0211g (gpps0158).
In this second paper, we report the discoveries of 76 Galactic transient sources from our single-pulse search, and also present FAST observations of previously known RRATs. In Section~\ref{sect2:obs}, we briefly summarize the GPPS survey observations, and introduce our single pulse search module and the data processing. All the newly discovered results are presented in Section~\ref{sect3:newResults}. During the GPPS surveys,  the covers with known pulsars have been observed for system verification in every observation session. With these  data, together with targeted observations of some RRATs in
several normal applied FAST projects (see details below), 59 previously known RRATs have been observed by FAST as presented in Section~\ref{sect4:knownRRAT}, which can tell the RRAT enigma. A summary and discussions are given in  Section~\ref{sect5:Conclusions}.

\section{Observations and data processing}
\label{sect2:obs}

\subsection{The GPPS survey and other observations}

The FAST GPPS survey \citep[see details in][]{Han2021RAA} plans to hunt for pulsars in the Galactic plane within $\pm{10}^{\circ}$ visible in the zenith angle of 28.5$^{\circ}$, by using the L-band 19-beam receiver covering the frequency range of 1000 -- 1500 MHz. Up to now, about 10\% of the planned survey regions have been  observed\footnotemark[5]. The observations have been carried out with {\it the snapshot mode} specially designed for the survey, in which four pointings of 19 beams fully cover a hexagonal sky area of 0.1575 square degrees aligned with the Galactic plane. Each pointing tracks for 5 minutes, and a quick beam switching costs less than 20 seconds, so that {\it a cover} can be observed by $4\times19$ beams in 21 minutes. A cover is named with the Galactic coordinates, such as G31.91+0.42. The beams are then named in the format as being G31.91+0.42-MxxPn. Here, Mxx represents the beam number for one of 19 beams of the receiver, i.e. M01 to M19, and Pn is for one of 4 pointings, i.e. P1 to P4. 

During the survey, signals of the two polarization channels of XX and YY from each of the 19 beam receiver are channelized to 2048 channels for the total band of 500 MHz. The data are recorded in PSRFITS format with a sampling time of 49.152$\mu$s. In verification observations for a cover having a known pulsar located in the sky area, data of four polarization channels of XX, X$^*$Y, XY$^*$ and YY are recorded in the PSRFITS file. During FAST observations, data streams are continuously recorded in a series of PSRFITS files, each lasting for about 12.885~s (if for four polarization channels) or 25.770~s (for two polarization channels). All the GPPS survey data are searched for  radio transients.

\begin{figure*}[!th]
  \centering
  \includegraphics[width=0.9\textwidth]{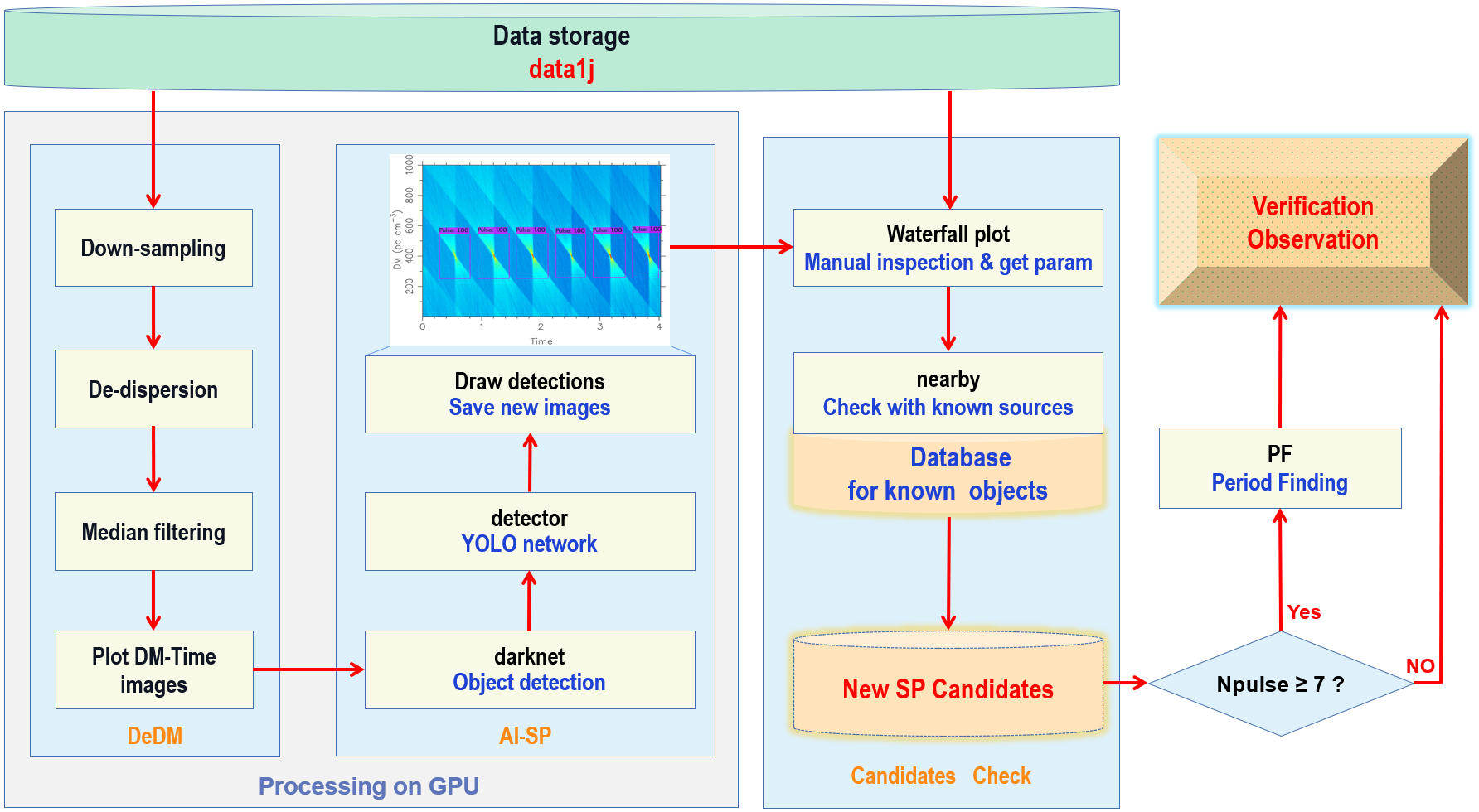}
  \caption{The flowchart for the single pulse search module for the GPPS survey.}
  \label{followchart}
\end{figure*}

Signals of newly discovered transients are often too weak to be detected by other telescopes, so we have to use the FAST for following-up observations. After a pulsar candidate or a transient source is found in a beam, a following-up observation is carried out with a {\it tracking mode} for generally 15 minutes, occasionally for a longer time, with the central beam of the L-band 19-beam receiver pointing to the position. Very often we record data for all 19 beams, all of which are used for slightly deeper pulsar searches.  We often name the beam with the source name and the beam number, such as J1849+0001-M09 is for data from the beam M09 of tracking observations for the source J1849+0001. The data of 4 polarization channels are recorded for the studies of polarization characteristics of a pulsar or a transient source. For some sources, we have also applied normal FAST project (PT2020\_0155 and PT2022\_0177,  PI: D.J. Zhou) to track even for one hour.

For known RRATs in the sky area of the GPPS survey, the verification observations take the data for 4 polarization channels on the cover with a high priority. For known RRATs outside the survey region, we take data from the tracking observations of other normal applied FAST projects, such as PT2021\_0051 (PI: Jun Xu) or PT2021\_0151 (PI: D.J. Zhou). For some objects with only a coarse position, we take data of $4\times19$ beams observed at the suggested position with a modified snapshot mode, called the {\it snapshotdec mode}, which covers a sky area with beams aligned along the equator plane.

\subsection{Data analysis}

In principle, single pulse search simply has three steps: 1) dedisperse data, 2) pick out pulses above a threshold of significance over the system noise, 3) sift candidates. 
For example, in the single pulse search pipeline on PRESTO \citep{Ransom2011}, the first step is to use the \texttt{prepsubband} to dedisperse data, and then the second step is to find the outstanding data points in each dedisperse data based on the  software~\texttt{single\_pulse\_search.py}. Finally, the classifier \texttt{RRATtrap}\footnote{\url{http://github.com/ckarako/rrattrap}} is used for candidates sifting, which is a single pulse sifting algorithm used in the Green Bank North Celestial Cap (GBNCC) survey and Low Frequency Array (LOFAR) survey, developed by \citet{KarakoArgaman2015} and updated by \citet{Patel2018}. 

After a huge number of outstanding data points are found in dedispersed data, a skillful job is the step for candidates sifting. Many software packages have been developed for `classifier'. 
In the package $\rm{Clusterrank}$\footnote{\url{http://github.com/juliadeneva/clusterrank}} used in Arecibo 327 MHz Drift Pulsar Survey (AO327), \citet{Devine2016MNRAS} developed a machine-learning classifier for identifying and classifying dedispered pulse groups in single pulse search output. 
\citet{Agarwal2020MNRAS} developed a new classifier to recognize the original achromatic waterfall image and the diagram of the occurrence DM-time of candidates. 
Other similar packages have been developed by \citet{Men2019MNRAS} and \citet{Zhang2021MNRAS} as well. 
\citet{Michilli2018MNRAS} introduce the machine-learning classifier  $\rm{L}{\text{-}}SPS$\footnote{\url{http://github.com/danielemichilli/LSPs}} \citep[based on $\rm SPS$\footnote{\url{http://github.com/danielemichilli/SPS}},][]{Michilli2018ascl} to discriminate astrophysical signals and RFI for the LOTAAS. 
All these classifiers are basically working on the bright points in the DM-time image, which is output by \texttt{single\_pulse\_search.py} or other similar codes. The parameters for such a single pulse search must be set differently for different projects by different telescopes, and they affect directly on candidate classification, otherwise narrow, weak or scattered pulses could be mis-eliminated. In general, these classifiers are hard to work in real time. 
  
With the developments of the Graphics Processing Unit (GPU) and Artificial Intelligence (AI), the single pulse search has adopted these technologies to improve the computation efficiency. For example, \citet{Petroff2014ApJ} searched for fast radio transients by using~$\rm{HEIMDALL}$\footnote{\url{http://sourceforge.net/projects/heimdall-astro}} based on GPU devices for dedispersion and candidates clustering, which can process data in real-time and directly produces a list of candidates for multi-beam surveys. \citet{Magro2011MNRAS} developed a GPU-based real time transient search machine.

\subsubsection{Single pulse search in the GPPS survey}
\label{sect2.1.1}

The single pulse search module has been developed (see Fig.~\ref{followchart}) and applied to the GPPS survey data. 

The original data of 2048 channels cover the frequency range of 1000 -- 1500 MHz. The channels within 31.25 MHz from each edge of the band are discarded due to their low gain. The original data of other channels for polarization channels XX and YY are scaled according to their root mean square (RMS) and added together for the total power, and then deposited into the {\it data1j} repository for pulsar searching. The details for data preparation are described in the first paper \citep{Han2021RAA}. The single pulse search module is taking these fits files from the {\it data1j} repository.

We have developed a source code to do quick dedispersion and generate DM-time images in GPU. The process includes three steps: loading data, dedispersion in the range from 3~pc\,cm$^{-3}$ to 1000~pc\,cm$^{-3}$ with a step of 1.0~pc\,cm$^{-3}$, and plotting DM-time images for every 4 seconds of data. This process consumes only about 0.25\,s of computation time in one node of the FAST computing cluster. In addition to recognize the pulse signal with a high DM, mostly for searching FRBs, images for three DM ranges are plotted and searched for radio transients independently: 900 -- 1900~pc\,cm$^{-3}$, 1800 -- 2800~pc\,cm$^{-3}$ and 2700 -- 3700~pc~cm$^{-3}$.
Because single pulses also rarely have a width less than one millisecond, after loading data, we down-sample the8 data from 49.152~$\mu$s to $4 \times 49.152 = 196.608$~$\mu$s and also combine every 4 frequency channels together. This not only improves the detection efficiency of single pulses, but also makes the subsequent data processing fast. 
After the data are dedispersed, in order to suppress the noise level and preserve the pulse features, the `fifth-order median filter' is applied to the dedispersed data, 
in which the median of in a sliding window of $5\times5$ data points in the two-dimensional Time-DM image is taken to replace the original point value. This filtering process can effectively smooth out strong RFI emerging in only one or two data bins and preserve the original features of the image.
Because some transient pulses have a steep spectrum or an inverse spectrum, we also make two more DM-time images for data in the upper half and lower half of the band, i.e. channels below and above 1250~MHz, and plot an image every 1.5~s. This operation ensures an effective detection of narrow and weak pulses with a steep spectrum, especially for the repeating FRBs \citep{ZhouDJ2022RAA}.

After the above-mentioned DM-time images are generated for a beam or even a cover, we used the \texttt{YOLO} object detection technology to recognize the outstanding points, which is developed in the  \texttt{Darknet}\footnote{\url{https://pjreddie.com/darknet/}} neural network framework for searching the distinctive features in the DM-time images. At the beginning of this single pulse search for the GPPS survey, the \texttt{yolov3.cfg} network file of \texttt{YOLOv3}\footnote{\url{https://pjreddie.com/darknet/yolo/}} \citep{yolov3} version was used. In 2021, we switched to  \texttt{YOLOv4} \citep{yolov4} and the new \texttt{yolov4{\_}csp.cfg}\footnote{\url{https://github.com/WongKinYiu/ScaledYOLOv4}} network file \citep{scaled-yolov4}. After the huge number of images sifted by this AI technology, the left images of single pulse candidates are manually examined, and the final burst candidates are listed with the DM and time of arrival (TOA, in MJD) for further verification observations and period-finding (see below).

Most of our single pulse searches of the GPPS survey have been carried out by using the GPU node of the computing cluster in the FAST data center. For data of 76 beams from one cover of the GPPS snapshot observation of 21 minutes, we organize the multitask parallel computing in the GeForce RTX 2080Ti GPU*4 with the program of GNU $\rm Parallel$~\citep{Parallel2018}. The processing costs about 9 minutes for single pulse search in the DM range from 3 to 1000~pc\,cm$^{-3}$.

\subsubsection{Sensitivity and pulse parameters }
\label{sect2.2.2}

For any pulse detected from the single pulse searches of the GPPS survey data, some parameters must be determined. 

The first is the DM value estimated by using the method given by~\citet{ZhuWW2020ApJ}, in which the most refined pulse structure is preserved when turning DM values, and the uncertainty is given as being the half-width of the Gaussian function fitted for the curve of $\sum~(\frac{d}{dt})^2$ over DMs. If a number of pulses are detected with the almost the same DM, they probably come from the same source. The final DM is determined by the brightest pulses.

Second is the time of arrival (TOA) of a pulse. We dedispered all the candidates and get the time of arrival (TOA, in MJD) at the pulse peak. This TOA must be converted to the time at the Solar barycentric center using the DE438 ephemeris.

For a single detected pulse, we have to get other necessary parameters, such as the observed pulse width ($W_{\rm obs}$) and the signal-to-noise ratio ($R=S/N$). Supposing the pulse in $n$ bins has an equivalent width of $W_{\rm obs}$ (or simply written as $W$ in units of ms in tables or figures), the $R=S/N$ of the pulse detection can be estimated from the total energy of $n$ on-pulse bins ($\sum{S_i}$) divided by the standard deviation ($\sigma$) of a nearby off-pulse range in the dedispersed data series, as %
\begin{equation}
 R = S/N = \frac{\sum{S_{i}}}{\sigma \cdot \sqrt{n}}. 
\end{equation}
%
%here the pulse width is naturally $W_{\rm pulse} = n \cdot t_{\rm bin}$. 

Note that the observed pulse width is a combination of the intrinsic pulse width with other factors. In addition to the sampling time   of the dedispersed data $t_{\rm bin} = 196.608~\mu$s, for a high DM pulse, the scattering broadening time $\tau_{s}$  has to be taken into account, which can be estimated as following \citep[e.g.][]{KarakoArgaman2015}:
\begin{equation}
    log(\tau_{s}) \approx a + b \cdot log(DM) + c \cdot log^2(DM) - \alpha \cdot log(\nu),
\end{equation}
here $a=-3.59$, $b=0.129$, $c=1.02$ and $\alpha=-4.4$. The dispersion time inside one frequency channel with a bandwidth $BW_{\rm chan} = 0.97656248$~MHz: 
\begin{equation}
    \Delta t_{\rm chan}=4.7{\mu}s\cdot\frac{BW_{\rm chan}}{\rm 24.414kHz}\cdot\left(\frac{\nu}{\rm 350MHz}\right)^{-3} \cdot \frac{DM}{\rm cm^{-3}pc},
\end{equation}
must be concerned. Therefore the observed pulse width $W_{\rm obs}$ could be estimated from the intrinsic pulse width $W_{\rm intrinsic}$ with these broadening, as being
\begin{equation}
    W_{\rm obs}  = \sqrt{ W_{\rm intrinsic}^2 + t_{\rm bin}^2 + \tau_{s}^2 + \Delta t_{\rm chan}^2}.
\end{equation}

The sensitivity of the single pulse detection at a given DM  
can be estimated from the pulse width  $W_{\rm obs}$ and $\sigma$ of the dedispersed data. The $\sigma$ of the data flow can be related to the system noise by 
\begin{equation}
\sigma = \frac{T_{\rm sys}} {G_0 \sqrt{n_p \cdot t_{\rm bin} \cdot BW }}, 
\label{sigma}
\end{equation}
here ${T_{\rm sys}}$ is the system noise temperature in units of K, $G_0 = 16.1$ K/Jy is the effective gain of the telescope \citep{Jiang2020}, $n_p=2$ is the number of polarization summed, $t_{\rm bin}$ is the time of a down-sampled bin, $BW$ = 437.5 MHz is frequency bandwidth (MHz) after removed the upper and lower sidebands of 31.25~MHz.  

%\subsubsection{Sensitivity of single pulse detection}
%\label{sect2.2.2}

\begin{figure}
\centering
  \includegraphics[width=0.9\columnwidth]{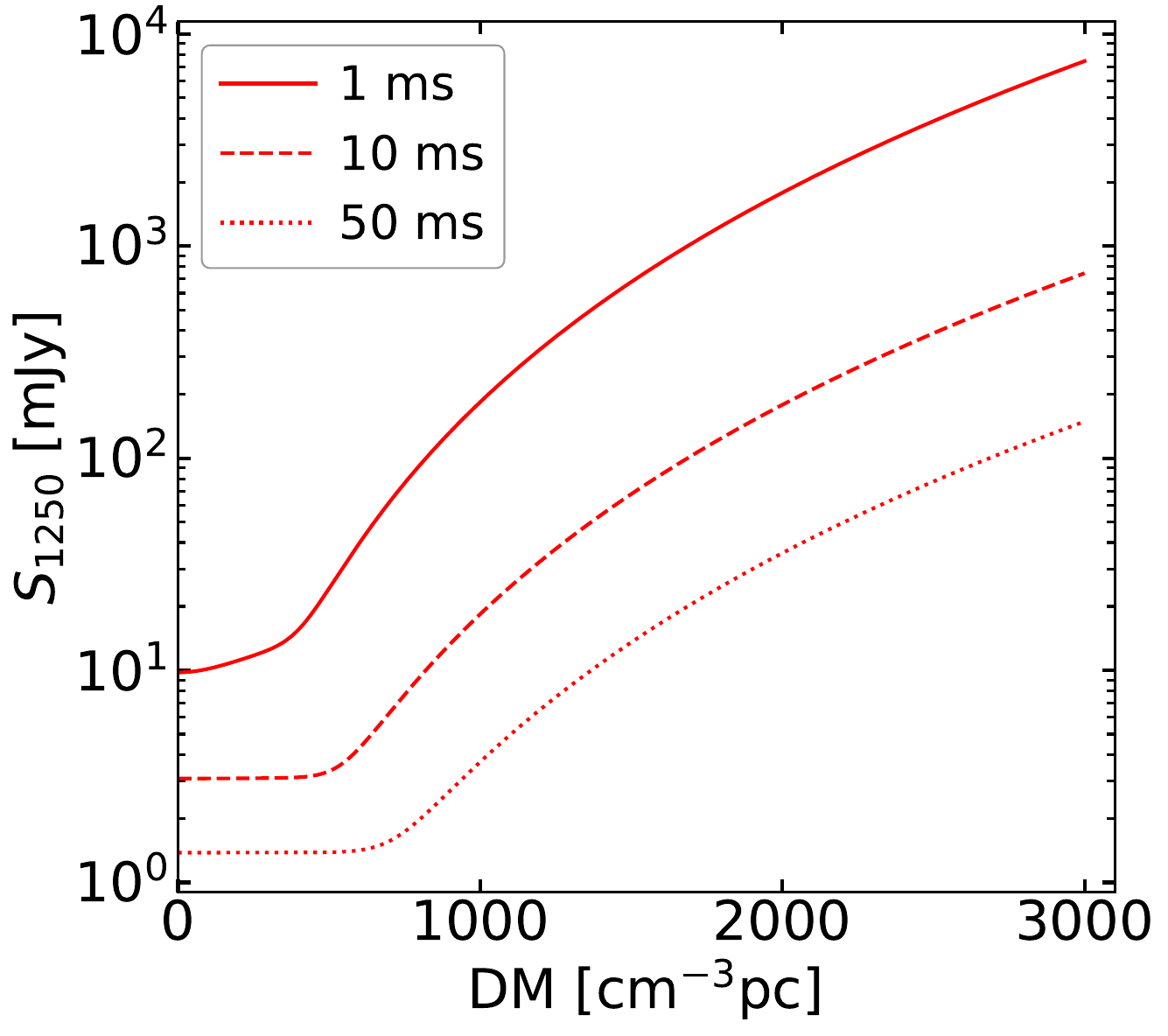}
\caption{FAST sensitivity for single pulse search with different intrinsic pulse widths.}
  \label{fig:sensitivity}
\end{figure}

The capability of the single pulse search must therefore consider the observed pulse width and the system noise of $\sigma$. For pulses with a high DM and an intrinsic pulse width $W_{\rm intrinsic}$, one can estimate the minimum flux density of detected pulse at a signal-to-noise ratio of $S/N > R_{\rm min}$ \citep{Cordes2003,Deneva2016} as being:
\begin{equation}
    % S_{1250} = R_{\rm min} \frac{W_{\rm obs}} {t_{\rm bin}} \cdot \sigma .
    S_{1250} = R_{\rm min} \frac{W_{\rm obs}} {W_{\rm intrinsic}} \frac{ T_{\rm sys}  }   {G_0 \cdot \sqrt {n_p \cdot BW \cdot W_{\rm obs} } }.
\end{equation}
Taking the minimum detection threshold of $R_{\rm min}$ = 7, one can get the
sensitivity curves for different DM values as shown in Fig.~\ref{fig:sensitivity}.

%Some sources which the average pulse profile are well for estimate the flux (S in units of $\mu$Jy at 1250 MHz) with the S/N by the follow equation
%\begin{equation}
%S = \frac{ S/N \cdot T_{sys} \cdot 10^{6}}{G_0 \cdot \sqrt{n_p \cdot %T_{obs} \cdot BW}}  \cdot \sqrt{\frac{W_{eq}}{P-W_{eq}}}
%\end{equation}
% where $T_{obs}$ is the observation duration, $W_{eq}$ and $P$, unit in s, is the equivalent pulse width and the rotation period, separately. The flux used in this paper is selected for the data with the highest $S/N = \sum{S_{i}/(\sigma \cdot \sqrt{n})$ of the integrated profiles if there are more observations of one source.
% Here the $n$ is the point number of the value greater than 3\sigma.
%The \sigma for the calculation of S/N for each single pulse and the average pulse profile is selected as the \sigma of the noise level in the off region of equal width adjacent to the on pulse region. If there is no pulse in this radiation window with a point greater than 3\sigma, the S/N is estimated with the value of the strongest point in the on region divided by the \sigma.
%

The fluence of each single pulse, $F$ in units of Jy$\cdot$s at 1250 MHz, is simply the integration of the pulse over the width ($t_{\rm bin}$, in units of s) as being 
% \begin{equation}
$F =\sum{S_{i} \cdot t_{\rm bin}}. $
% F = \frac{\rm \sum{S_{i}} \cdot T_{\rm sys} \cdot t_{bin} \cdot 10^{3}}{\sigma \cdot G_0 \cdot \sqrt{n_p \cdot t_{bin} \cdot BW}}
% \end{equation}
%Here $S_{i}$ should be scaled to mJy as being done for the sensitivity below.
The method for estimating the mean flux densities of pulsars with periods estimated is consistent with that described in Section 4 of Paper I~\citep{Han2021RAA}.

\subsubsection{Period finding}
\label{sect.pf}

A possible period for detected pulses with the almost same DM has always been searched. If a period is found, the bright pulses obviously come from a pulsar or an RRAT. 

For a sequence of TOAs of pulses with almost the same DM, one can get N(N-1)/2 differences for those TOAs. We test a number of periods in a range, and check the residual RMS for these TOAs and TOA differences for a given period of $P$. One would get complete noises for a series of residual RMS if there is no period and periodic derivative. Nevertheless, if there is a period, the very distinct minimum of residual RMS could be found, and the residual is then taken as the uncertainty of so-found period. The data are then fold with the best DM and the period $P$, and the average pulse profile is obtained. 
We wrote a small program ``PF'' for the purpose, and obtained the best periods from TOA sequences for many sources. After we got this job done, we found that the method is very coincident with that reported by \citet{Keane2010MNRAS}. 
%
%
%For those sources for which we got the rotation period, 35 of all the new sources, we need to make specific distinctions based on their radiation characteristics. 
% If the emission of each single pulse is separated, neither two single pulses are emit in two consecutive periods, we classify them as RRATs. For the other sources with all or several continuous emission of single pulses, we consider whether the average pulse profiles of all their single pulses with S/N~\textgreater~3 have a significant signal that they are classified as nulling pulsars or weak pulsars with few strong single pulses.The main differences and classification results are described in Section~\ref{sect3:newResults}.

For bright pulses and the mean profile of a pulsar or an RRAT, we get their polarization profile by using the package PSRCHIVE~\citep{Hotan2004PASA}, after the polarization data are calibrated and the Faraday rotation is corrected \citep{whx+22}.

% \onecolumn
% \begin{landscape}
\begin{table*}[!ht]
    \centering
    {\footnotesize
    \caption{76 new RRATs discovered via the single pulse module from the GPPS survey data.}
    \label{tab1}
    \setlength{\tabcolsep}{3pt}
    \begin{tabular}{lccrcclrrr}
    \hline\noalign{\smallskip}
  Name       &gpps&\multicolumn{1}{c}{P}  &\multicolumn{1}{c}{DM} & R.A.(J2000)    & Dec(J2000)  & \multicolumn{1}{c}{ObsDate/MJD: BeamName}   & T$_{\rm obs}$ &N$_d$/N$_p$  &$\langle S \rangle$  \\% &$S_{\rm 1.25GHz}$\\% & $D_{\rm NE2001}$ & $D_{\rm YMW16}$ &DM$^{\rm max}_{\rm NE2001}$& DM$^{\rm max}_{\rm YMW16}$ &RM\\
             &No. &\multicolumn{1}{c}{(s)}&\multicolumn{1}{c}{(cm$^{-3}$pc)}& (hh:mm:ss)   & ($\pm$dd:mm)   &                           & (min)       &    & ($\mu$Jy) \\
    (1)  &  (2) & (3) & (4) & (5)  & (6) & \multicolumn{1}{c}{(7)} & (8) & (9)  & (10)\\
\hline
\multicolumn{10}{c}{Transient sources with just few pulses, no period has been found yet} \\
\hline
J0637+0332g  &0528&  ...    &152(2) &06:37:41&+03:32  &20220909/59831: G208.21$-$1.36-M03P2    &5  &3/-&    \\
             &    &         &       &        &        &20230213/59988: J063740+033229-M01    &15 &0/-&    \\[0.5mm]%
J1828$-$0003g  &0501&  ...    &193(3) &18:28:53&$-$00:03  &20221114/59897: J182853$-$000340sp-M01  &15 &1/-&    \\%SP102
             &    &         &       &        &        &20220803/59794: G30.20+4.91-M19P2     &5  &1/-&    \\%
             &    &         &       &        &        &20220803/59794: G30.20+4.91-M19P4     &5  &1/-&    \\[0.5mm]%
J1847$-$0046g  &0282&  ...    &337(7) &18:47:47&$-$00:46  &20211113/59531: J184711$-$005148-M19    &15 &2/-&    \\%&86.4\\%    &  2.6&  2.6&1264.4&1773.8&        \\%SP074       18:47:46.7   -00:46:40
             &    &         &       &        &        &20211031/59518: G31.91+0.42-M06P4     &5  &1/-&    \\%     \\[0.5mm]%&    &     &      &      &        \\[0.5mm]%       18:47:46.6   -00:45:52
             &    &         &       &        &        &20220606/59735: J187446$-$004620-M01    &30 &3/-&    \\[0.5mm]%&
J1850$-$0004g  &0280&  ...    &154(1) &18:50:05&-00:04  &20220608/59737: J185005$-$000430-M01    &30 &4/-&    \\%
             &    &         &       &        &        &20200415/58953: J1849+0001-M09        &15 &2/-&    \\%&60.1\\%    & 4.4&  3.6&1270.7&1986.5&-316(9) \\%SP054       18:49:58.2   -00:03:56
             &    &         &       &        &        &20200902/59094: J1849+0001-M09        &15 &1/-&    \\%&    \\%    &     &     &      &      &        \\%            18:49:58.2   -00:03:56
             &    &         &       &        &        &20210624/59388: J184958$-$000356sp-M01  &15 &0/-&    \\%&    \\[0.5mm]%&    &     &      &      &        \\[0.5mm]%       18:49:58.2   -00:03:56
             &    &         &       &        &        &20221106/59889: J1850$-$0002-M01        &60 &1/-&    \\[0.5mm]%
J1853+0209g  &0502&  ...    &350(15)&18:53:07&+02:09  &20200812/59073: J1852+0159-M18        &15 &1/-&    \\%SP055
             &    &         &       &        &        &20210627/59392: J185306+020910sp-M01  &15 &0/-&    \\
             &    &         &       &        &        &20211014/59501: J185306+020910sp-M01  &60 &0/-&    \\
             &    &         &       &        &        &20220602/59731: J185306+020910-M01    &30 &0/-&    \\
             &    &         &       &        &        &20220824/59815: J1853+0209-M01        &60 &1/-&    \\
             &    &         &       &        &        &20221112/59895: J1853+0209-M01        &50 &0/-&    \\[0.5mm]
J1853+0353g  &0281&  ...    &379(2) &18:53:29&+03:53  &20210514/59347: J185404+035848-M13    &15 &2/-&    \\%&20.0\\%    &  7.6&  9.2& 844.9& 849.0&        \\%SP050       18:53:29.1    03:53:43
             &    &         &       &        &        &20210624/59388: J185329+035343sp-M01  &15 &1/-&    \\[0.5mm]%&    \\[0.5mm]%&    &     &      &      &        \\[0.5mm]%       18:53:29.1    03:53:43
J1855$-$0211g  &0526&  ...    &304(3) &18:55:32&$-$02:11  &20230305/60008: J185532$-$021119-M01    &15 &1/-&    \\
             &    &         &       &        &        &20221205/59918: J185556$-$022129-M16    &15 &3/-&    \\
             &    &         &       &        &        &20220810/59801: G31.61$-$2.12-M16P2     &5  &1/-&    \\[0.5mm]%
J1855$-$0154g  &0503&  ...    &417(1) &18:55:09&$-$01:54  &20210907/59464: G31.66$-$1.69-M01P1     &5  &3/-&    \\%&52.9\\%    &  8.3&  7.3& 824.7& 748.2&        \\%SP063       18:55:09.5   -01:54:06
             &    &         &       &        &        &20211004/59491: J185509$-$015406sp-M01  &15 &0/-&    \\%&    \\[0.5mm]%&    &     &      &      &        \\[0.5mm]%       18:55:09.5   -01:54:06
             &    &         &       &        &        &20221116/59899: J1855$-$0154-M01        &60 &0/-&    \\%
             &    &         &       &        &        &20230321/60024: J1855$-$0154-M01        &60 &0/-&    \\[0.5mm]%
J1855$-$0054g  &0504&  ...    &577(4) &18:55:21&$-$00:54  &20220602/59731: J185521$-$005410-M01    &30 &4/-&    \\%
             &    &         &       &        &        &20220210/59678: J185520$-$005434-M01    &15 &1/-&    \\%&47.3\\%    &  2.0&  1.4& 928.1& 893.7&        \\%SP079       18:55:20.0   -00:54:10
             &    &         &       &        &        &20211225/59573: G32.45$-$1.36-M19P1     &5  &1/-&    \\%&    \\%    &     &     &      &      &        \\%            18:55:20.6   -00:52:53
             &    &         &       &        &        &20211225/59573: G32.45$-$1.36-M19P2     &5  &1/-&    \\%&    \\%    &     &     &      &      &        \\%            18:55:15.3   -00:55:30
             &    &         &       &        &        &20211225/59573: G32.45$-$1.36-M19P4     &5  &1/-&    \\[0.5mm]%&    \\[0.5mm]%&    &     &      &      &        \\[0.5mm]%       18:55:27.0   -00:55:21
J1855+0033g  &0283&  ...    &554(1) &18:55:03&+00:33  &20210328/59300: J1855+0033-M01        &60 &3/-&    \\%&23.5\\%    &  8.3&  6.1&1146.2&1479.8&        \\%SP015       18:55:03.3    00:33:25
             &    &         &       &        &        &20200328/58935: G34.01$-$0.68-M16P3     &5  &1/-&    \\[0.5mm]%&    \\[0.5mm]%&    &     &      &      &        \\[0.5mm]%SP015  18:55:03.3    00:33:25
J1856+0528g  &0284&  ...    &307(2) &18:56:23&+05:28  &20210826/59452: J185623+052810-M01P1  &15 &3/-&    \\%&12.6\\%    &  6.6&  9.0& 812.8& 761.9&        \\%SP056       18:56:23.0   +05:28:10
             &    &         &       &        &        &20210608/59372: G38.36+1.61-M10P3     &5  &2/-&    \\%&    \\%    &     &     &      &      &        \\%SP056       18:56:25.2   +05:29:38
             &    &         &       &        &        &20210608/59372: G38.36+1.61-M11P4     &5  &1/-&    \\[0.5mm]%&    \\[0.5mm]%&    &     &      &      &        \\[0.5mm]%       18:56:19.9   +05:27:01 
J1859+0832g  &0505&  ...    &259(2) &18:59:27&+08:32  &20220526/59724: G41.40+2.12-M02P2     &5  &2/-&    \\%SP093
             &    &         &       &        &        &20221107/59890: J185927+083231sp-M01  &15 &1/-&    \\[0.5mm]%
J1900+0908g  &0527&  ...    &264(4) &19:00:19&+09:08  &20220522/59720: G42.13+2.03-M17P2     &5  &2/-&    \\
             &    &         &       &        &        &20221107/59890: J190019+090846sp-M01  &15 &0/-&    \\
             &    &         &       &        &        &20230305/60008: J190019+090846-M01    &15 &0/-&    \\[0.5mm]%
J1902+0557g  &0525&  ...    &414(2) &19:02:56&+05:57  &20221018/59870: G39.69+0.17-M05P2     &5  &2/-&    \\
             &    &         &       &        &        &20230212/59987: J190255+055718-M01    &15 &1/-&    \\[0.5mm]%
J1916+1142Ag &0287&  ...    &260(8) &19:16:59&+11:42  &20200302/58910: G46.34$-$0.17-M04P2     &5  &1/-&    \\%&10.1\\%    &  6.8&  6.1& 747.7& 813.5&        \\%SP021       19:16:58.9    11:42:12
             &    &         &       &        &        &20210108/59222: J1917+1142-M01        &60 &1/-&    \\[0.5mm]%&    \\[0.5mm]%&    &     &      &      &        \\[0.5mm]%       19:16:58.9    11:42:12
J1918+0342g  &0506&  ...    &174(5) &19:18:22&+03:42  &20211202/59550: G39.54$-$4.32-M15P3     &5  &1/-&    \\%&75.6& 6.2 &  4.8& 418.2& 576.0&211(4)  \\%SP075       19:18:21.9    +03:42:33
             &    &         &       &        &        &20220329/59667: J191821+034233-M01    &15 &0/-&    \\%&    &     &     &      &      &        \\[0.5mm]%     
             &    &         &       &        &        &20220602/57731: J191821+034233-M01    &30 &1/-&    \\[0.5mm]%
J1918+1514g  &0507&  ...    &134(2) &19:18:57&+15:14  &20200531/58999: G49.81+0.93-M15P3     &5  &2/-&    \\%                                              \\%SP088        19:18:57.3   15:14:22
             &    &         &       &        &        &20220602/57731: J191857+151422-M01    &15 &0/-&    \\
             &    &         &       &        &        &20230228/60003: J191857+151422-M01    &15 &0/-&    \\[0.5mm]%
J1921+1629g  &0288&  ...    &105(2) &19:21:47&+16:29  &20211004/59491: J192147+162934sp-M01  &15 &2/-&    \\%&34.1\\%    &  4.3&  3.1& 590.9& 773.7&369(3)  \\%SP067       19:21:47.0    +16:29:34
             &    &         &       &        &        &20210822/59448: G50.94+1.02-M02P4     &5  &1/-&    \\[0.5mm]%&    \\[0.5mm]%&    &     &      &      &        \\[0.5mm]%SP067  19:21:47.0    +16:29:34
J1924+1734g  &0289&  ...    &49(3)  &19:24:57&+17:34  &20211005/59492: J192457+173434sp-M01  &15 &3/-&    \\%&27.9\\%    &  2.8&  1.9& 589.8& 780.4&        \\%SP068       19:24:57.4    17:34:34
             &    &         &       &        &        &20210822/59448: G52.40+0.85-M01P3     &5  &2/-&    \\[0.5mm]%&    \\[0.5mm]%&    &     &      &      &        \\[0.5mm]%SP068  19:24:57.4    17:34:34
\hline
\end{tabular}}
	\begin{tablenotes}
	\item 
Notes: Column (1)-(2): source name and the GPPS survey discovery number; Column (3)-(4) is the period $P$ (in units of s, if obtained) and DM (in units of cm$^{-3}$pc) with uncertainty in brackets; Column (5)-(6): beam position of the detection in RA (2000) and Dec(2000); Column (7) is observation date together with MJD and also the beam name; Column (8): observation duration; Column (9): the number of pulses detected together with the number of periods (N$_d$/N$_p$)  if the period is available; Column (10): the averaged flux density $\langle S \rangle$ (in units of $\mu$Jy) for detected pulses. 
  \end{tablenotes}
\end{table*}  
\addtocounter{table}{-1}
\begin{table*}[!ht]
    \centering
    {\footnotesize
    \caption{{\it -- continued}.}
    \setlength{\tabcolsep}{3pt}
    \begin{tabular}{lccrcclrrr}
    \hline\noalign{\smallskip}
  Name       &gpps&\multicolumn{1}{c}{P}  &\multicolumn{1}{c}{DM} & R.A.(J2000)    & Dec(J2000)  & \multicolumn{1}{c}{ObsDate/MJD: BeamName}   & T$_{\rm obs}$ &N$_d$/N$_p$  & $\langle S \rangle$ \\%&$S_{\rm 1.25GHz}$\\%& $D_{\rm NE2001}$ & $D_{\rm YMW16}$ &DM$^{\rm max}_{\rm NE2001}$& DM$^{\rm max}_{\rm YMW16}$ &RM\\
             &No. &\multicolumn{1}{c}{(s)}&\multicolumn{1}{c}{(cm$^{-3}$pc)}& (hh:mm:ss)   & ($\pm$dd:mm)   &                           & (min)       &     &  ($\mu$Jy)             \\% & (kpc)            & (kpc)           & (cm$^{-3}$pc)             &(cm$^{-3}$pc)               &(rad~m$^{-2}$)     \\
    (1)  &  (2) & (3) & (4) & (5)  & (6) & \multicolumn{1}{c}{(7)} & (8) & (9)  & (10)\\
\hline
J1927+1940g  &0290&  ...    &347(2) &19:27:17&+19:40  &20190327/58569: G54.31+1.27-M19P4     &5  &1/-&    \\%&9.6 \\%    & 10.1&  8.9& 534.3& 717.6&        \\%SP024       19:27:17.2    19:40:06
             &    &         &       &        &        &20210624/59388: J192717+194006sp-M01  &15 &1/-&    \\%&    \\%    &     &     &      &      &        \\%            19:27:17.2    19:40:06
             &    &         &       &        &        &20201120/59173: J1927+1940-M01        &15 &0/-&    \\[0.5mm]%&    \\[0.5mm]%&    &     &      &      &        \\[0.5mm]%       19:27:17.2    19:40:06
J1932+2126g  &0508&  ...    &126(3) &19:32:51&+21:26  &20220323/59661: G56.51+1.02-M02P1     & 5 &1/-&    \\%                                               \\%SP090        19:32:51.4   21:26:09
             &    &         &       &        &        &20220608/59737: J193251+212609-M01    &15 &1/-&    \\%
             &    &         &       &        &        &20220720/59780: J193251+212609-M01    &15 &1/-&    \\[0.5mm]%
J1933+2401g  &0291&  ...    &185(3) &19:33:36&+24:01  &20210626/59390: J193335+240123sp-M01  &15 &1/-&    \\%&11.2\\%    &  6.8&  8.1& 445.1& 545.9&        \\%SP047       19:33:35.5    24:01:23
             &    &         &       &        &        &20210301/59274: G58.86+2.03-M19P2     &5  &1/-&    \\[0.5mm]%&    \\[0.5mm]%&    &     &      &      &        \\[0.5mm]%SP047  19:33:35.5    24:01:23
J1934+2341g  &0292&  ...    &252(2) &19:34:03&+23:41  &20210624/59388: J193402+234110sp-M01  &15 &5/-&    \\%&15.3\\%    &  8.5&  9.0& 462.3& 563.8&        \\%SP046       19:34:02.8    23:41:10
             &    &         &       &        &        &20210301/59274: G58.86+2.03-M12P2     &5  &3/-&    \\%&    \\[0.5mm]%&    &     &      &      &        \\[0.5mm]%SP046  19:34:02.8    23:41:10
             &    &         &       &        &        &20221106/59889: J1934+2341-M01        &15 &2/-&    \\[0.5mm]%
J2001+4209g  &0293&  ...    &153(2) &20:01:39&+42:09  &20211004/59491: J200139+420904sp-M01  &15 &2/-&    \\%&13.6\\%    &  7.2&  9.3& 208.7& 273.3&        \\%SP072       20:01:39.2   42:09:04
             &    &         &       &        &        &20210802/59427: G77.55+6.10-M19P3     &5  &1/-&    \\[0.5mm]%&    \\[0.5mm]%&    &     &      &      &        \\[0.5mm]%  
J2005+3154g  &0294&  ...    &225(1) &20:05:19&+31:54  &20211009/59496: J200519+315400sp-M01  &15 &1/-&    \\%&15.0\\%    &  7.3&  7.1& 458.4& 467.1&        \\%SP071       20:05:19.0   +31:53:60
             &    &         &       &        &        &20210804/59430: G69.43+0.17-M04P2     &5  &2/-&    \\%&    \\%    &     &     &      &      &        \\%            20:05:25.2   31:55:21
             &    &         &       &        &        &20210805/59431: G69.43+0.17-M05P4     &5  &1/-&    \\%&    \\%    &     &     &      &      &        \\%            20:05:11.4   31:55:28
             &    &         &       &        &        &20210805/59431: G69.43+0.17-M13P1     &5  &3/-&    \\%&    \\%    &     &     &      &      &        \\%            20:05:17.8   31:52:52
             &    &         &       &        &        &20211009/59496: J200530+315600sp-M01  &15 &2/-&    \\[0.5mm]%&    \\[0.5mm]%&    &     &      &      &        \\%            20:05:30.0    +31:56:00
J2030+3833g  &0295&  ...    &417(6) &20:30:31&+38:33  &20210317/59290: J203031+383329sp-M01  &15 &2/-&    \\%&13.9\\%    & 50.0& 15.1& 401.9& 431.7&        \\%SP044       20:30:31.6   +38:33:29
             &    &         &       &        &        &20210624/59388: J203031+383329sp-M01  &15 &2/-&    \\%&    \\%    &     &     &      &      &        \\%            20:30:31.6   +38:33:29
             &    &         &       &        &        &20210220/59265: G77.55$-$0.34-M09P1     &5  &2/-&    \\%&    \\[0.5mm]%&    &     &      &      &        \\[0.5mm]%SP044  20:30:31.6   +38:33:29
\hline
\multicolumn{10}{c}{ProtoRRATs with a good period identified}\\
\hline
J1857+0229g  &0296&0.584(3) &574(1) &18:57:19&+02:29  &20201108/59161: J1857+0229-M01        &60 &9/5749&    \\%&37.0\\%    &  8.5&  6.1&1146.4&1781.4&        \\%SP014       18:57:18.9    02:29:37
             &    &         &       &        &        &20200302/58910: G35.92$-$0.25-M06P2     &5  &1/513&    \\[0.5mm]%&    \\[0.5mm]%&    &     &      &      &        \\[0.5mm]%SP014  18:57:18.9    02:29:37
J1858+0453g  &0297&3.761(4) &429(1) &18:58:48&+04:53  &20210219/59264: J1859+0453-M01        &60 &7/923&    \\%&15.5\\%    &  7.5&  6.9&1011.4&1086.0&        \\%SP031       18:58:48.2    04:53:14
             &    &         &       &        &        &20200421/58959: G38.27+0.76-M12P2     &5  &1/79&    \\[0.5mm]%&    \\[0.5mm]%&    &     &      &      &        \\[0.5mm]%SP031  18:58:48.2    04:53:14
J1859+0251g  &0298&3.580(3) &286(3) &18:59:35&+02:51  &20210316/59288: J1900+0252-M01        &60 &14/969& \\%&18.3\\%    &  6.1&  4.9&1063.9&1305.5&        \\%SP025       18:59:35.7    02:51:40
             &    &         &       &        &        &20200221/58900: G36.26$-$0.51-M08P2     &5  &1/83&    \\[0.5mm]%&    \\[0.5mm]%&    &     &      &      &        \\[0.5mm]%SP025  18:59:35.7    02:51:40
J1904+0621g  &0299&1.232(3) &173(1) &19:04:55&+06:21  &20210318/59291: J190455+062136sp-M01  &15 &8/720&    \\%&34.7\\%    &  4.7&  4.2&1003.5&1234.6&-234(9) \\%SP042       19:04:55.0    06:21:36
             &    &         &       &        &        &20201225/59208: G40.03$-$0.08-M02P1     &5  &8/243&    \\%&    \\%    &     &     &      &      &        \\%SP042       19:04:55.0    06:21:36
             &    &         &       &        &        &20210213/59258: G40.03$-$0.08-M02P1     &5  &2/243&    \\[0.5mm]%&    \\[0.5mm]%&    &     &      &      &        \\[0.5mm]%       19:04:55.0    06:21:36
J1905+0156g  &0300&1.085(1) &137(1) &19:05:08&+01:56  &20210113/59227: J1905+0156-M01        &15 &20/818&  \\%&28.4\\%    &  2.8&  4.5& 786.6& 504.8&        \\%SP035       19:05:07.9    01:56:53
             &    &         &       &        &        &20201127/59180: J1904+0207-M10        &15 &7/818&    \\[0.5mm]%&    \\[0.5mm]%&    &     &      &      &        \\[0.5mm]%SP035  19:05:07.9    01:56:53
J1905+0558g  &0301&0.846(2) &472(1) &19:05:04&+05:58  &20210627/59391: J190500+055840sp-M01  &15 &11/1050& \\%&23.8\\%    &  8.1&  6.5&1001.7&1182.5&432(20) \\%SP019       19:05:04.2    05:58:40
             &    &         &       &        &        &20210306/59279: J190452+060345-M03    &15 &9/1063&    \\%&    \\%    &     &     &      &      &        \\%            19:05:04.2    05:58:40
             &    &         &       &        &        &20210109/59223: J1904+0558-M01        &30 &3/2127&    \\%&    \\%    &     &     &      &      &        \\%            19:04:53.4    05:58:41
             &    &         &       &        &        &20200219/58898: G39.64$-$0.25-M08P2     &5  &1/354&    \\%&    \\%    &     &     &      &      &        \\%SP019       19:04:53.4    05:58:41
             &    &         &       &        &        &20201123/59176: J1904+0558-M01        &15 &0/1063&    \\[0.5mm]%&    \\[0.5mm]%&    &     &      &      &        \\[0.5mm]%       19:04:53.4    05:58:41
J1908+0911g  &0510&5.1661(7)&132(4) &19:08:09&+09:11  &20221107/59890: J190808+091144sp-M01  &15 &8/172&    \\%SP018
             &    &         &       &        &        &20191226/58843: G43.06+0.42-M06P1     &5  &2/58&    \\[0.5mm]%
J1916+0937g  &0286&7.368(1) &186(2) &19:16:01&+09:37  &20221108/59891: J1916+0937-M01        &60 &10/472&   \\%&17.4\\%    &  5.4&  5.4& 734.1& 639.9&        \\%SP016       19:16:00.9    09:37:01
             &    &         &       &        &        &20200418/58956: G44.38$-$1.02-M05P1     &5  &2/40&    \\%&    \\%    &     &     &      &      &        \\%SP016       19:16:00.9    09:37:01
             &    &         &       &        &        &20201123/59176: J1916+0937-M01        &15 &1/122&    \\%&    \\[0.5mm]%&    &     &      &      &        \\[0.5mm]%       19:16:00.9    09:37:01
             &    &         &       &        &        &20210626/59390: J191600+093701sp-M01  &15 &4/122&    \\[0.5mm]
J1916+1142Bg &0303&1.188(3) &318(8) &19:16:59&+11:42  &20210108/59222: J1917+1142-M01        &60 &7/2929&    \\%&21.5\\%    &  7.8&  7.0& 747.7& 813.5&        \\%SP022       19:16:58.9    11:42:12
             &    &         &       &        &        &20201123/59176: G46.34$-$0.17-M04P2     &5  &3/252&    \\[0.5mm]%&    \\[0.5mm]%&    &     &      &      &        \\[0.5mm]%SP022  19:16:58.9    11:42:12
J1917+0834g  &0304&2.933(3) &101(3) &19:17:04&+08:34  &20210113/59227: J1917+0834-M01        &15 &6/303&    \\%&25.5\\%    &  3.8&  3.5& 648.1& 463.1&194(3)  \\%SP023       19:17:04.5    08:34:02
             &    &         &       &        &        &20201123/59176: J1917+0834-M01        &15 &5/303&     \\%&    \\%    &     &     &      &      &        \\%            19:17:04.5    08:34:02
             &    &         &       &        &        &20200419/58957: G43.55$-$1.78-M06P4     &5  &2/102&    \\[0.5mm]%&    \\[0.5mm]%&    &     &      &      &        \\[0.5mm]%SP023  19:17:04.5    08:34:02
J1921+1006g  &0511&3.345(9) &362(8) &19:21:44&+10:06  &20221107/59890: J192144+100630sp-M01  &15 &11/262& \\%SP107
             &    &         &       &        &        &20220821/59812: G45.36$-$2.20-M17P1     &5  &2/89&    \\%
             &    &         &       &        &        &20220821/59812: G45.36$-$2.20-M18P2     &5  &1/89&    \\[0.5mm]%
J1924+1446g  &0305&1.090(3) &336(3) &19:24:54&+14:46  &20201108/59139: J1924+1446-M01        &60 &9/3080&    \\%&34.4\\%    &  9.3&  7.0& 640.2& 716.4&        \\%SP-13       19:24:53.8    14:46:43
             &    &         &       &        &        &20200426/58964: G50.01$-$0.59-M15P1     &5  &1/275&    \\[0.5mm]%&    \\[0.5mm]%&    &     &      &      &        \\[0.5mm]%SP013  19:24:53.8    14:46:43
J1935+1841g  &0306&5.529(1) &290(3) &19:35:03&+18:41  &20211005/59492: J193502+184128sp-M01  &15 &7/161&    \\%&22.6\\%    &  8.8&  6.0& 569.0& 641.6&        \\%SP069       19:35:02.6    18:41:28
             &    &         &       &        &        &20210817/59442: G54.46$-$0.68-M03P2     &5  &4/54&     \\%&    \\[0.5mm]%&    &     &      &      &        \\[0.5mm]%SP069  19:35:02.6    18:41:28
             &    &         &       &        &        &20220925/59847: J1935+1841-M01        &15 &3/162&    \\
             &    &         &       &        &        &20221105/59888: J1935+1841-M01        &15 &4/162&    \\
             &    &         &       &        &        &20221128/59911: J1935+1841-M01        &15 &5/162&    \\[0.5mm]
J1948+2314g  &0512&1.471(4) &184(3) &19:48:40&+23:14  &20221107/59890: J194840+231423sp-M01  &15 &11/604& \\%SP100
             &    &         &       &        &        &20220418/59687: G59.99$-$1.10-M11P1     &5  &5/203&    \\[0.5mm]%
\hline
\end{tabular}}
\end{table*}  
\addtocounter{table}{-1}
\begin{table*}[!ht]
  \centering
      {\footnotesize
    \caption{{\it -- continued}.}
        \setlength{\tabcolsep}{3pt}
        \begin{tabular}{lccrcclrrr}
          \hline\noalign{\smallskip}
Name       &gpps&\multicolumn{1}{c}{P}  &\multicolumn{1}{c}{DM} & R.A.(J2000)    & Dec(J2000)  & \multicolumn{1}{c}{ObsDate/MJD: BeamName}   & T$_{\rm obs}$ &N$_d$/N$_p$ &  $\langle S \rangle$  \\%&$S_{\rm 1.25GHz}$\\%& $D_{\rm NE2001}$ & $D_{\rm YMW16}$ &DM$^{\rm max}_{\rm NE2001}$& DM$^{\rm max}_{\rm YMW16}$ &RM\\
             &No. &\multicolumn{1}{c}{(s)}&\multicolumn{1}{c}{(cm$^{-3}$pc)}& (hh:mm:ss)   & ($\pm$dd:mm)   &                           & (min)       &     & ($\mu$Jy) \\%&             \\%& (kpc)            & (kpc)           & (cm$^{-3}$pc)             &(cm$^{-3}$pc)               &(rad~m$^{-2}$)     \\
    (1)  &  (2) & (3) & (4) & (5)  & (6) & \multicolumn{1}{c}{(7)} & (8) & (9)  & (10)\\
\hline
J2005+3156g  &0307&2.146(1) &337(2) &20:05:30&+31:56  &20211009/59496: J200530+315600sp-M01  &15 &6/413&    \\%&82.2\\%    & 10.7& 10.7& 458.0& 466.3&-282(6) \\%SP070       20:05:30.0    +31:56:00
             &    &         &       &        &        &20211009/59496: J200519+315400sp-M01  &15 &4/413&    \\%&    \\%    &     &     &      &      &        \\%            20:05:19.0    +31:53:60
             &    &         &       &        &        &20210804/59430: G69.43+0.17-M04P1     &5  &2/139&    \\%&    \\%    &     &     &      &      &        \\%            20:05:32.6    31:57:50
             &    &         &       &        &        &20210804/59430: G69.43+0.17-M04P2     &5  &2/139&    \\%&    \\%    &     &     &      &      &        \\%            20:05:25.2    31:55:21
             &    &         &       &        &        &20210804/59430: G69.43+0.17-M04P4     &5  &2/139&    \\%&    \\%    &     &     &      &      &        \\%            20:05:39.0    31:55:14
             &    &         &       &        &        &20210805/59431: G69.43+0.17-M04P1     &5  &2/139&    \\%&    \\%    &     &     &      &      &        \\%            20:05:32.6    31:57:50
             &    &         &       &        &        &20210805/59431: G69.43+0.17-M04P2     &5  &2/139&    \\%&    \\%    &     &     &      &      &        \\%            20:05:25.2    31:55:21
             &    &         &       &        &        &20210822/59448: G69.43+0.17-M04P2     &5  &2/139&    \\%&    \\[0.5mm]%&    &     &      &      &        \\[0.5mm]%       20:05:25.2    31:55:21
J2014+3326g  &0524&0.9773(1)&333(2) &20:14:25&+33:26  &20230212/59987: J201424+332603-M01    &15 &8/910&    \\
             &    &         &       &        &        &20221201/59914: G71.58$-$0.51-M10P2     &5  &5/306&    \\[0.5mm]
\hline
\multicolumn{10}{c}{Extremely nulling pulsars} \\
\hline
J1828$-$0038g  &0509&2.426(3) &70(2)  &18:28:15&$-$00:38  &20221114/59897: J182815$-$003930sp-M01  &15 &21/370& \\%SP101
             &    &         &       &        &        &20220828/59819: J182828$-$004203-M06    &15 &14/365& \\
             &    &         &       &        &        &20220828/59819: J182828$-$004203-M05    &15 &1/371&    \\
             &    &         &       &        &        &20220828/59796: G29.80+4.74-M06P2     &15 &8/371&     \\
             &    &         &       &        &        &20220828/59796: G29.80+4.74-M16P4     &15 &5/371&     \\
             &    &         &       &        &        &20220828/59796: G29.80+4.74-M17P3     &15 &4/371&     \\
             &    &         &       &        &        &20220828/59796: G29.80+4.74-M16P1     &15 &2/371&     \\[0.5mm]
J1842+0114g  &0308&4.140(4) &307(8) &18:42:13&+01:14  &20210317/59289: J184213+011400sp-M01  &15 &23/215 &8.9 \\%&8.9 \\%    &  7.0&  7.8& 656.7& 649.8&120(5)  \\%SP040       18:42:13     +01:14:00
             &    &         &       &        &        &20210213/59258: G32.89+2.46-M18P1     &5  &5/72    &    \\%&    \\%    &     &     &      &      &        \\%            18:42:10.9   +01:14:37
             &    &         &       &        &        &20201225/59208: G32.89+2.46-M18P1     &5  &5/72    &    \\%&    \\%    &     &     &      &      &        \\%SP040       18:42:10.9   +01:14:37
             &    &         &       &        &        &20201225/59208: G32.89+2.46-M18P4     &5  &4/72    &    \\%&    \\%    &     &     &      &      &        \\%            18:42:17.2   +01:12:09
             &    &         &       &        &        &20210213/59258: G32.89+2.46-M18P4     &5  &2/72    &    \\[0.5mm]%&    \\[0.5mm]%&    &     &      &      &        \\[0.5mm]%       18:42:17.2   +01:12:09
J1845$-$0008g  &0309&1.268(3) &143(3) &18:45:08&$-$00:08  &20210626/59390: J184508$-$000800sp-M01  &15 &39/700  &9.6 \\%&9.6 \\%    &  4.0&  4.0& 918.6&1024.1&25(9)   \\%SP009       18:45:08     -00:08:00
             &    &         &       &        &        &20200422/58960: G32.01+1.27-M19P2     &5  &7/242    &    \\%&    \\%    &     &     &      &      &        \\%SP009 18:45:06.1   -00:07:09
             &    &         &       &        &        &20200422/58960: G32.01+1.27-M19P3     &5  &2/242    &    \\%&    \\%    &     &     &      &      &        \\%   18:45:12.4   -00:09:37
             &    &         &       &        &        &20201121/59174: J1845$-$0007-M01P1      &15 &0/710    &    \\%&    \\%    &     &     &      &      &        \\%            
             &    &         &       &        &        &20210707/59401: J184508$-$000800sp-M01  &15 &0/710    &    \\%&    \\%    &     &     &      &      &        \\%   18:45:08     -00:08:00  0253\\%
             &    &         &       &        &        &20210109/59223: J1845$-$0007-015P1      &30 &0/710    &    \\[0.5mm]%&    \\[0.5mm]%&    &     &      &      &        \\[0.5mm]%       
J1855+0240g  &0310&1.224(3) &397(3) &18:55:13&+02:40  &20210626/59390: J185513+024047sp-M01  &15 &5/725    &1.3 \\%&1.3 \\%    &  7.1&  5.6&1114.8&1486.5&95(5)   \\%SP004       18:55:13.1    02:40:47
             &    &         &       &        &        &20200321/58929: G35.63+0.42-M08P3     &5  &2/245    &    \\%&    \\%    &     &     &      &      &        \\%SP004       18:55:13.1    02:40:47
             &    &         &       &        &        &20201121/59174: J1855+0240-M01        &15 &2/725    &    \\%&    \\%    &     &     &      &      &        \\%            18:55:13.1    02:40:47
             &    &         &       &        &        &20210111/59225: J1855+0240-M01        &30 &0/1470   &    \\[0.5mm]%&    \\[0.5mm]%&    &     &      &      &        \\[0.5mm]%       18:55:13.1    02:40:47
J1858$-$0113g  &0513&1.532(1) &280(4) &18:58:52&$-$01:13  &20220403/59671: J185852$-$011306-M01P1  &15 &16/578   &7.8 \\%&7.8 \\%    &  6.6&  6.0& 704.2& 590.0&674(3)  \\%SP080       18:58:52.0   -01:13:06
             &    &         &       &        &        &20220127/59606: G32.74-2.03-M11P2     &5  &6/195    &    \\%&    \\%    &     &     &      &      &        \\%            
             &    &         &       &        &        &20220127/59606: G32.74-2.03-M11P4     &5  &7/195    &    \\[0.5mm]%&    \\[0.5mm]%&    &     &      &      &        \\%            
J1911+1017g  &0302&1.337(1) &162(2) &19:11:12&+10:17  &20201123/59176: J1911+1017b-M01       &15 &12/664& \\%&17.0\\%    &  4.3&  4.4& 844.2& 890.9&263(14) \\%SP008       19:11:13.8    10:17:21
             &    &         &       &        &        &20200321/58929: G44.53+0.25-M15P2     &5  &10/231&   \\[0.5mm]%&    \\[0.5mm]%&    &     &      &      &        \\[0.5mm]%SP008  19:11:13.8    10:17:21
J1921+0851g  &0234&0.957(5) &101(2) &19:21:11&+08:51  &20210429/59332: G44.23-2.29-M11P4     &5  &16/307   &72.9\\%&72.9\\%    &  4.2&  4.8& 543.7& 357.4&381.9(8)\\%SP049       19:21:11.3    08:51:33
             &    &         &       &        &        &20210624/59388: J192111+085133sp-M01  &15 &51/926   &    \\[0.5mm]%&    \\[0.5mm]%&    &     &      &      &        \\[0.5mm]%       19:21:11.3    08:51:33
J1921+1632g  &0311&0.493(1) &164(2) &19:21:37&+16:32  &20211004/59491: J192136+163201sp-M01  &15 &5/580    &    \\%&    \\%    &  5.6&  4.5& 586.7& 771.2&        \\%SP066       19:21:36.9   16:32:01
             &    &         &       &        &        &20210822/59448: G50.94+1.02-M02P1     &5  &1/195    &    \\[0.5mm]%&    \\[0.5mm]%&    &     &      &      &        \\[0.5mm]%       19:21:36.9   16:32:01
J1935+1901g  &0407&0.897(1) &365(2) &19:35:50&+19:01  &20220329/59667: J193550+190159-M01    &15 &11/990   &6.4 \\%&6.4 \\%    & 10.3&  9.0& 564.4& 628.2&82(7)   \\%SP076       19:35:50     +19:01:59
             &    &         &       &        &        &20211107/59525: G54.80$-$0.93-M18P2     &5  &9/334    &    \\[0.5mm]%&    \\[0.5mm]%&    &     &      &      &        \\[0.5mm]%       19:35:50.9   +19:01:59
J1940+2203g  &0312&11.906(1)&59(9)  &19:40:49&+22:03  &20210109/59223: J1940+2203-M01        &30 &9/150&    \\%&0.95\\%    &  3.2&  2.7& 548.3& 556.8&        \\%SP003       19:40:49.6    +22:03:13
             &    &         &       &        &        &20201123/59176: J1940+2203-M01        &15 &2/75 &    \\%&    \\%    &     &     &      &      &        \\%            19:40:49.6    +22:03:13
             &    &         &       &        &        &20200426/58964: G58.22$-$0.25-M05P4     &5  &1/25 &    \\%&    \\%    &     &     &      &      &        \\%            19:40:55.8    22:05:46 
             &    &         &       &        &        &20200426/58964: G58.22$-$0.25-M05P3     &5  &1/25 &    \\%&    \\[0.5mm]%&    &     &      &      &        \\[0.5mm]%SP003  19:40:49.6    22:03:13
             &    &         &       &        &        &20220925/59847: J1940+2203-M01        &15 &0/75 &    \\
             &    &         &       &        &        &20221107/59890: J1940+2203-M01        &15 &0/75 &    \\
             &    &         &       &        &        &20221128/59911: J1940+2203-M01        &15 &3/75 &    \\[0.5mm]
\hline
\multicolumn{10}{c}{Newly discovered weak pulsars with sparse strong pulses} \\
\hline
J1840$-$0245g  &0313&1.502(1) &277(2) &18:40:14&$-$02:45  &20211004/59491: J184014$-$024556sp-M01  &15 &4/591    & 6.7  \\%&6.7 \\%    &  5.6&  5.0&1026.9&1028.3&        \\%SP060       18:40:14.0   -02:45:56
             &    &         &       &        &        &20210816/59442: G29.41+1.36-M13P3     &5  &2/206    &      \\[0.5mm]%&    \\[0.5mm]%&    &     &      &      &        \\[0.5mm]%SP060  18:40:14.0   -02:45:56
J1843$-$0051g  &0314&0.580(1) &573(3) &18:43:32&$-$00:51  &20211009/59496: J184332$-$005100sp-M01  &15 &31/1528  & 6.0  \\%&6.0 \\%    & 10.1& 10.4& 927.8&1016.0&        \\%SP062       18:43:32.0   -00:51:00
             &    &         &       &        &        &20210811/59437: G31.27+1.36-M01P1     &5  &13/532   &      \\[0.5mm]%&    \\[0.5mm]%&    &     &      &      &        \\[0.5mm]%SP062    
J1845+0326g  &0521&0.968(1) &144(1) &18:45:42&+03:26  &20230218/59993: J184542+032600-M01    &15 &32/919   & 3.3  \\%
             &    &         &       &        &        &20221002/59854: G35.53+2.80-M15P3     &5  &2/309    &      \\%
             &    &         &       &        &        &20221002/59854: G35.53+2.80-M15P4     &5  &3/309    &      \\[0.5mm]%
J1845+0417g  &0515&1.697(6) &164(3) &18:45:33&+04:17  &20221111/59894: J184533+041730sp-M01  &15 &34/524   & 3.1  \\%SP092
             &    &         &       &        &        &20220524/59722: G35.97+3.39-M09P4     &5  &6/170    &     \\
             &    &         &       &        &        &20220524/59722: G35.97+3.39-M09P3     &5  &2/170    &      \\[0.5mm]
\hline
\end{tabular}}
\end{table*}
\addtocounter{table}{-1}
\begin{table*}[!ht]
  \centering
      {\footnotesize
    \caption{{\it -- continued and ended}.}
        \setlength{\tabcolsep}{3pt}
        \begin{tabular}{lccrcclrrr}
          \hline\noalign{\smallskip}
Name       &gpps&\multicolumn{1}{c}{P}  &\multicolumn{1}{c}{DM} & R.A.(J2000)    & Dec(J2000)  & \multicolumn{1}{c}{ObsDate/MJD: BeamName}   & T$_{\rm obs}$ &N$_d$/N$_p$ &  $\langle S \rangle$  \\%&$S_{\rm 1.25GHz}$\\%& $D_{\rm NE2001}$ & $D_{\rm YMW16}$ &DM$^{\rm max}_{\rm NE2001}$& DM$^{\rm max}_{\rm YMW16}$ &RM\\
             &No. &\multicolumn{1}{c}{(s)}&\multicolumn{1}{c}{(cm$^{-3}$pc)}& (hh:mm:ss)   & ($\pm$dd:mm)   &                           & (min)       &     & ($\mu$Jy) \\%&             \\%& (kpc)            & (kpc)           & (cm$^{-3}$pc)             &(cm$^{-3}$pc)               &(rad~m$^{-2}$)     \\
    (1)  &  (2) & (3) & (4) & (5)  & (6) & \multicolumn{1}{c}{(7)} & (8) & (9)  & (10)\\
\hline
J1849+0619g  &0522&2.011230(3)&110(1)&18:49:35&+06:19 &20230228/60003: J184935+061900-M01    &15 &4/442    &6.2   \\
             &    &         &       &        &        &20221001/59853: G38.56+3.30-M14P1     & 5 &1/149    &      \\
             &    &         &       &        &        &20221111/59894: J184950+062304-M04    &15 &1/442    &      \\[0.5mm]
J1851+0051g  &0315&4.027(2) &575(5) &18:51:40&+00:51  &20210207/59252: J1852+0051-M01        &60 &12/833   & 2.7  \\%&2.7 \\%    &  8.2&  6.1&1212.0&1775.6&        \\%SP018       18:51:39.9    00:51:14 
             &    &         &       &        &        &20200303/58911: G33.77+0.42-M04P1     &5  &1/74     &      \\%&    \\%    &     &     &      &      &        \\%SP018       18:51:39.9    00:51:14
             &    &         &       &        &        &20201228/59211: J1851+0056-M03        &20 &1/297    &      \\[0.5mm]%&    \\[0.5mm]%&    &     &      &      &        \\[0.5mm]%       18:51:38.7   +00:51:40
J1853$-$0130g  &0316&1.945(1) &344(1) &18:53:07&$-$01:30  &20211004/59491: J185307$-$013012sp-M01  &15 &15/454   & 6.4  \\%&6.4 \\%    &  6.6&  5.6&1026.1& 996.3&        \\%SP064       18:53:07.1   -01:30:12
             &    &         &       &        &        &20210901/59458: G31.76$-$0.85-M10P3     &5  &1/154    &      \\[0.5mm]%&    \\[0.5mm]%&    &     &      &      &        \\[0.5mm]%      
J1856+0029g  &0317&0.376(3) &234(3) &18:56:49&+00:29  &20210524/59357: J1856+0029-M01        &60 &23/8943  & 2.6  \\%&2.6 \\%    &  5.7&  4.7& 995.6& 969.9&        \\%SP011       18:56:49.8    00:29:39
             &    &         &       &        &        &20200422/58960: G33.96$-$1.10-M18P3     &5  &1/798    &      \\[0.5mm]%&    \\[0.5mm]%&    &     &      &      &        \\[0.5mm]%SP011  18:56:49.8    00:29:39
J1859+0239Bg &0529&0.848740(3)&624(4)&18:59:18&+02:39 &20220529/59728: J1859+0239-M01        &15 &1/1060    &      \\
             &    &         &       &        &        &20221227/59940: J1859+0239-M01        &5  &1/353     &      \\
             &    &         &       &        &        &20230125/59969: J1859+0239-M01        &15 &1/1060    &      \\
             &    &         &       &        &        &20221118/59901: J185918+023926sp-M01  &15 &0/1060    &      \\
             &    &         &       &        &        &20211226/59574: J1859+0239-M01        &15 &0/1060    &      \\
             &    &         &       &        &        &20230102/59946: J1859+0239-M01        &5  &0/353     &      \\
             &    &         &       &        &        &20230217/59992: J1859+0239-M01        &5  &0/353     &      \\
             &    &         &       &        &        &20230121/59965: J1859+0239-M01        &5  &0/353     &      \\
             &    &         &       &        &        &20221103/59886: J1859+0239-M01        &5  &0/353     &      \\
             &    &         &       &        &        &20221222/59935: J1859+0239-M01        &5  &0/353     &      \\[0.5mm]
J1900$-$0152g  &0516&1.384(6) &314(2) &19:00:56&$-$01:52  &20221109/59892: J190056$-$015215sp-M01  &15 &14/642    &14.5  \\%SP105
             &    &         &       &        &        &20220812/59803: G32.25$-$3.05-M19P2     &5  &7/216     &      \\[0.5mm]
J1900+0732g  &0318&1.709(1) &226(1) &19:00:15&+07:32  &20210626/59390: J190015+073203sp-M01  &15 &8/519     & 4.0   \\%&4.0 \\%    &  5.7&  8.2& 750.0& 697.0&373(17) \\%SP048       19:00:15.0    +07:32:03
             &    &         &       &        &        &20200512/58980: G40.57+1.36-M18P4     &5  &2/175     &       \\[0.5mm]%&    \\[0.5mm]%&    &     &      &      &        \\[0.5mm]%SP048  19:00:15.0    07:32:03
J1903+0319g  &0319&1.854(3) &307(3) &19:03:14&+03:19  &20210508/59341: J1903+0319-M01        &60 &11/1872   & 5.1   \\%&5.1 \\%    &  6.2&  5.6&1011.4&1086.0&        \\%SP032       19:03:13.9    03:19:45
             &    &         &       &        &        &20200521/58989: G37.29$-$1.27-M17P2     &5  &1/161     &       \\[0.5mm]%&    \\[0.5mm]%&    &     &      &      &        \\[0.5mm]%SP032  19:03:13.9    03:19:45
J1906+0335g  &0226&1.296(1) &213(1) &19:06:48&+03:35  &20210707/59401: J190649+033508-M01    &15 &34/685    & 8.0   \\%&8.0 \\%    &  5.3&  6.0& 740.1& 525.0&203(4)  \\%SP053       19:06:48.8    03:35:00
             &    &         &       &        &        &20210627/59391: J190648+033500sp-M01  &15 &25/685    &       \\%&    \\%    &     &     &      &      &        \\%            19:06:48.8    03:35:00  0226
             &    &         &       &        &        &20210606/59370: G37.97$-$1.78-M05P1     &5  &9/238     &       \\[0.5mm]%&    \\[0.5mm]%&    &     &      &      &        \\[0.5mm]%SP053  19:06:48.8    03:35:00
J1907+0555g  &0320&3.159(1) &150(5) &19:07:27&+05:55  &20201101/59154: J1907+0555-M01        &60 &31/1062   & 7.2   \\%&7.2 \\%    &  4.4&  4.0& 891.8& 794.2&        \\%SP010       19:07:27.6    05:55:22
             &    &         &       &        &        &20200514/58982: G39.93$-$0.93-M19P2     &5  &1/94      &       \\[0.5mm]%&    \\[0.5mm]%&    &     &      &      &        \\[0.5mm]%SP010  19:07:27.6    05:55:22
J1911+0310g  &0517&1.3326(5)&167.7(8)&19:11:18&+03:10 &20221111/59894: J191118+031005sp-M01&15 &11/667      & 7.7   \\%SP099
             &    &         &       &        &        &20220706/59765: G38.22$-$2.88-M13P2     &5  &7/225     &       \\[0.5mm]
J1912+1000g  &0321&3.053(4) &147(4) &19:12:46&+10:00  &20210113/59227: J1912+1000-M01        &15 &12/290    & 3.7   \\%&3.7 \\%    &  4.7&  5.6& 648.1& 463.1&        \\%SP038       19:12:46.3    10:00:30
             &    &         &       &        &        &20201123/59176: J1912+1000-M01        &15 &5/290     &       \\%&    \\%    &     &     &      &      &        \\%            19:12:46.3    10:00:30
             &    &         &       &        &        &20201206/59189: G43.55$-$1.78-M06P4     &5  &2/98      &       \\[0.5mm]%&    \\[0.5mm]%&    &     &      &      &        \\[0.5mm]%SP038  19:12:46.3    10:00:30
J1915+1045g  &0518&1.546(2) &123(3) &19:15:32&+10:45  &20221107/59890: J191532+104524sp-M01  &15 &15/575    &3.3    \\%SP109
             &    &         &       &        &        &20200219/58898: G45.21$-$0.25-M03P4     &5  &1/194     &       \\
             &    &         &       &        &        &20201205/59918: J191532+104524-M01    &15 &11/575    &       \\[0.5mm]%
J1919+1113g  &0322&0.766(5) &288(2) &19:19:27&+11:13  &20210113/59227: J1919+1113-M01        &15 &9/1159    & 4.8   \\%&4.8 \\%    &  7.6&  7.5& 681.3& 628.1&        \\%SP020       19:19:27.1    11:13:01
             &    &         &       &        &        &20201121/59174: J1919+1113-M01        &15 &6/1159    &       \\%&    \\%    &     &     &      &      &        \\%            19:19:27.1    11:13:01
             &    &         &       &        &        &20200509/58977: G46.24$-$1.02-M05P2     &5  &2/391     &       \\[0.5mm]%&    \\[0.5mm]%&    &     &      &      &        \\[0.5mm]%SP020  19:19:27.1    11:13:01
J1921+1227g  &0323&1.598(1) &259(2) &19:21:22&+12:27  &20211004/59491: J192122+122745sp-M01  &15 &4/556     & 0.3  \\%&0.28\\%    &  7.2&  6.8& 666.3& 660.8&        \\%SP065       19:21:22.4    12:27:45
             &    &         &       &        &        &20210805/59431: G47.37$-$0.93-M07P1     &5  &1/187     &       \\%&    \\[0.5mm]%&    &     &      &      &        \\[0.5mm]%    
             &    &         &       &        &        &20220629/59758: J192133+123801-M11    &15 &12/556    &       \\[0.5mm]%
J1927+1849g  &0389&0.3121(1)&200(3) &19:27:46&+18:49  &20220329/59667: J192746+184917-M01    &15 &13/2848   & 20.8  \\%SP081
             &    &         &       &        &        &20220210/59620: J1928+1852-M06P4      & 5 &2 /961    &       \\%
             &    &         &       &        &        &20220306/59644: J192733+185434-M03    &15 &11/2848   &       \\[0.5mm]%
J1938+1748g  &0523&7.1060(2)&56(1)  &19:38:30&+17:48  &20230218/59993: J193829+174846-M01    &15 &11/125    &1.8    \\
             &    &         &       &        &        &20221004/59856: G54.31$-$1.95-M15P3     & 5 & 2/42     &       \\[0.5mm]%
J1940+2231g  &0324&5.682(2) &198(7) &19:40:55&+22:31  &20201123/59176: J1940+2231-M01        &15 &8/156     & 2.9   \\%&7.0 \\%    &  6.8&  7.5& 547.4& 560.5&        \\%SP026       19:40:55.1    22:31:11
             &    &         &       &        &        &20200421/58960: G58.62$-$0.08-M05P2     &5  &4/52      &       \\%&    \\[0.5mm]%&    &     &      &      &        \\[0.5mm]%SP026  19:40:55.1    22:31:11
             &    &         &       &        &        &20220925/59847: J1940+2231-M01        &15 &5/156     &       \\
             &    &         &       &        &        &20221115/59898: J1940+2231-M01        &15 &3/156     &       \\
             &    &         &       &        &        &20221128/59911: J1940+2231-M01        &15 &9/156     &       \\[0.5mm]
J1948+2438g  &0519&1.9031(2)&450(4) &19:48:23&+24:38  &20221107/59890: J194823+243817sp-M01  &15 &1/467     & 5.0   \\%SP098
             &    &         &       &        &        &20220708/59767: G61.16$-$0.59-M07P2     &5  &1/157     &       \\[0.5mm]%
J1956+2911g  &0520&3.8163(2)&265(2) &19:56:36&+29:11  &20220329/59667: J195635+291113-M01    &15 &11/233    & 1.9   \\%&1.9 \\%    &  8.2&  8.0& 480.3& 482.7&        \\%SP077       19:56:35      +29:11:13
             &    &         &       &        &        &20210912/59469: G66.10+0.34-M04P2     &5  &3/78      &       \\%&    \\[0.5mm]%&    &     &      &      &        \\[0.5mm]%       19:56:35.6    29:11:13
\hline
\end{tabular}}
\end{table*}

\section{New Discoveries}
\label{sect3:newResults}
% \section{New Discoveries}
% \label{sect3:newResults}

% \input{Tex/tab_few_pulses}

\begin{figure*} %[!t]
\centering
\includegraphics[width=0.3\textwidth]{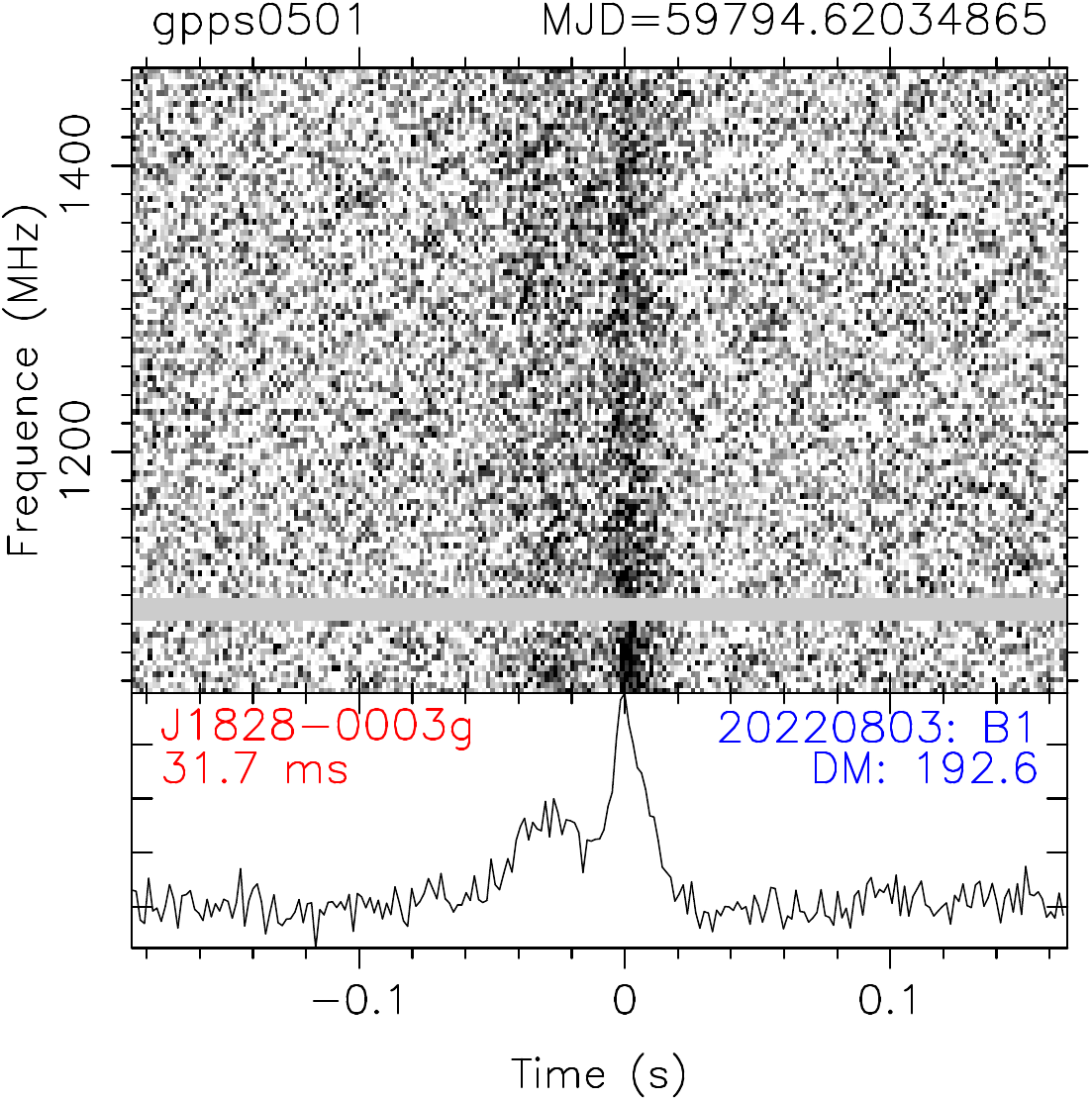}
\includegraphics[width=0.3\textwidth]{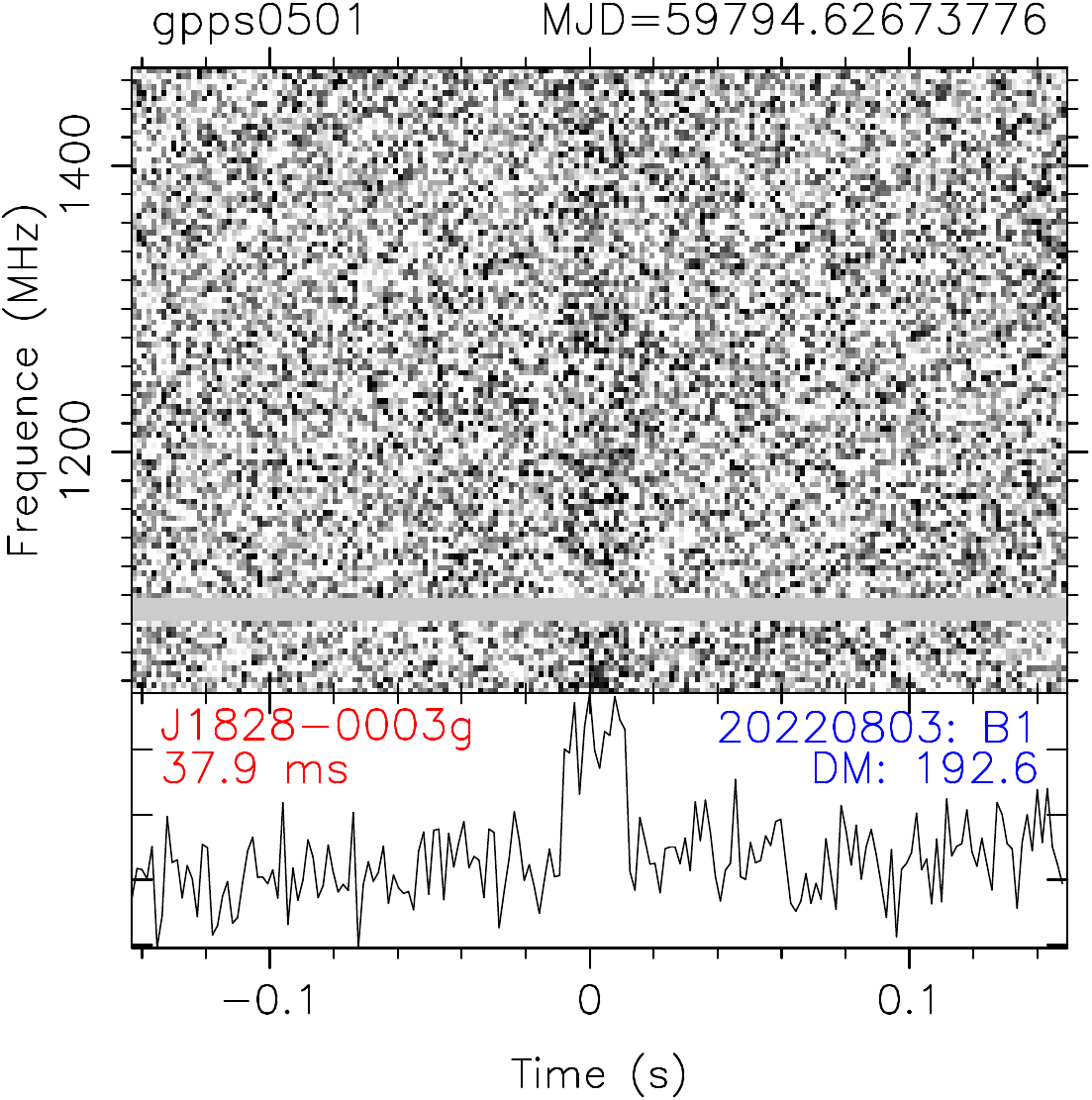}
\includegraphics[width=0.3\textwidth]{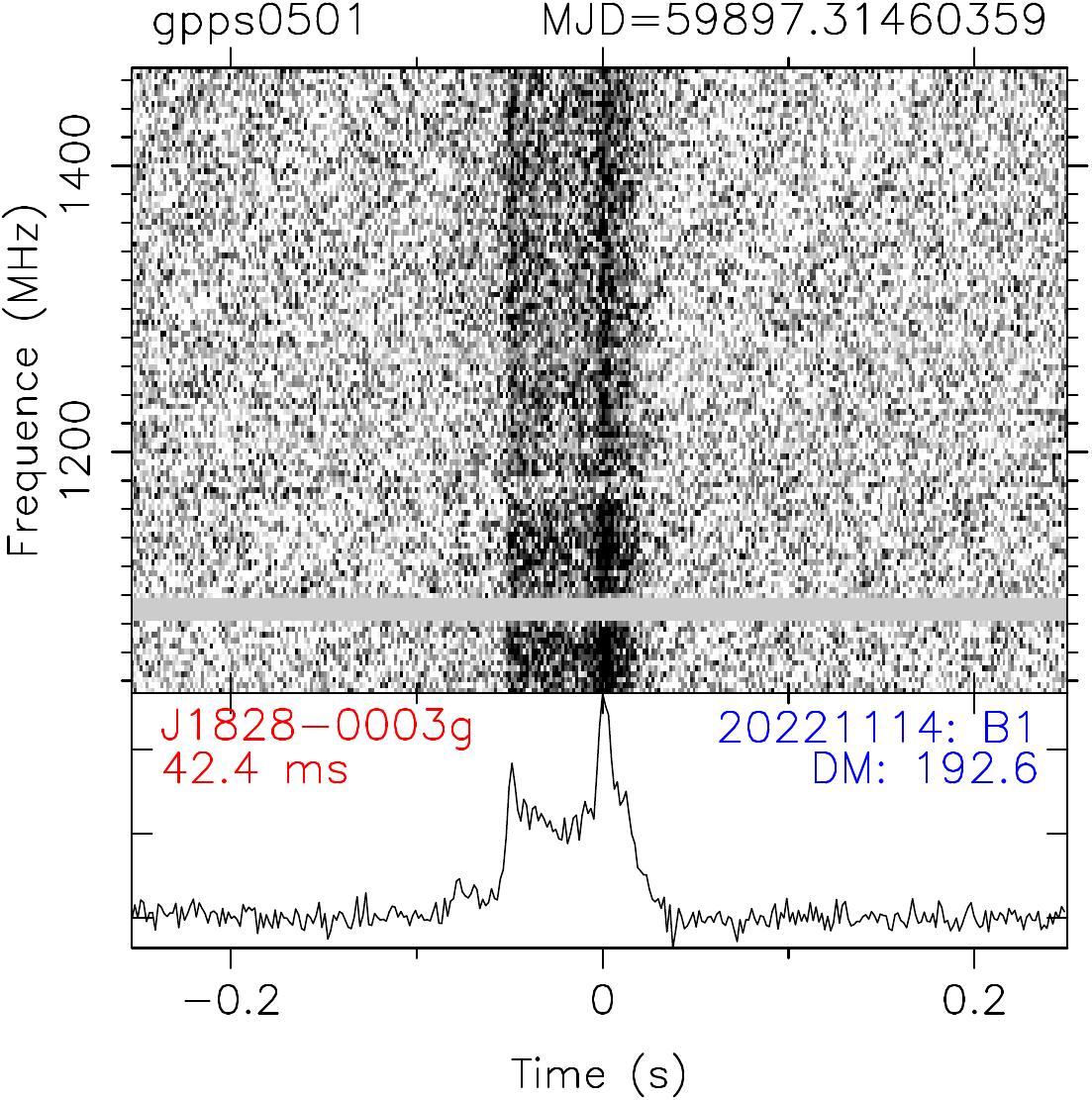}
\caption{Examples for the merely 3 pulses detected by the FAST at three epochs from a new FAST transient source, J1828-0003g. See Figure~\ref{fig:AppfewPulses} in the Appendix  for all 105  pulses detected by FAST from 26  radio transient sources. The two-dimensional dynamic spectrum over the frequency in the band and observation time is shown in the top-subpanel, and the de-dispersed averaged profile is shown in the bottom sub-panel. The GPPS object number is given on the top of the panel on the left, and epoch is marked on the right. The name of object and the pulse width (in ms) are marked in the top-left inside the lower sub-panel, observation date together with the pulse number (number after B) and the also DM value are marked in the top-right.  
}
\label{fig:fewPulses}
\end{figure*}

After processing the GPPS survey data obtained so far, we  discover 76 Galactic transient sources, which are all taken as RRATs since they are discovered in single pulses search rather than the periodicity search, see Table~\ref{tab1} for the list. 

Among them, we detect 105 pulses from 26 transient sources. Merely a few pulses from each source, as discussed in ection~\ref{sect2.1.2}, are not enough for the `PF' to find a period, though we believe they should be RRATs. We simply take these source as the first subclass of newly discovered RRAT with a period to be found. For the other 50 sources, based on  the discovered observation or follow-up observations, we classify them into three categories. The first is the prototypes of RRATs, in short `proto-RRATs', from which non-consecutive single pulses are detected in the available observations but a period can be identified from these sporadic bright pulses. In the other periods rather than these with sparse pulses, no significant pulsed emission is detected from any individual period, or only very weak emission is detected from the averaged for many periods. We discovered 16 such proto-RRATs. The second category is extremely nulling pulsars, which have no emission is detected in the most ($\gs90\%$) of observation duration, but some pulses are detected consecutively for some periods. We discover 10 such extremely nulling pulsars. The last category the weak pulsars with sparse strong pulses. They stay for most of them in a very weak emission state, but occasionally have strong single pulses. The two emission states are very clearly shown in the individual pulses. We discovered 24 weak pulsars with sparse strong pulses.

In the following, we discuss each kind of such newly discovered source first, and then present the polarization observation results for strong pulses or the averaged profiles of detected pulses.

%The last group is 15 normal weak pulsars with a few bright pulses detected by single pulse search method and their integral profiles for the single pulses of $\rm S/N \textless 3$ have faint signal. We fold the groups of RRATs, nulling pulsar and normal weak pulsars and obtain the average pulse profiles and that of the single pulses of $\rm S/N \textless or \textgreater 3$. All the groups and the special individual source are described particularly in the following sections.

\subsection{Galactic transient sources with merely few pulses}
\label{sect2.1.2}

Very few pulses have been detected by FAST for 26 transient sources, 
as listed in the first part of Table~\ref{tab1} and shown in  Figure~\ref{fig:fewPulses} for examples and Figure~\ref{fig:AppfewPulses} for pulses of all detected transient sources in the Appendix.  Mostly 1 or 2 pulses were detected in 5 or 15 minutes observations, and sometimes even non pulse detected in the following-up observations for 15 or even 60 minutes.
Note that the transient sources listed in Table~\ref{tab1} at least have two pulses, so they have `repeating radio bursts' with almost the same DM, and the DM in Table~\ref{tab1} is {\it obtained from the strongest pulse.}
%the weighted average of detected pulses. 
If only one pulse is detected in the first observation, we have to catch at least another one in the following observations. J1853+0209g, for example, one pulse was detected on 20200812, while no pulse is detected in many sessions such as the 15 minutes observation on 20210627, the 60 minutes observation on 20211014, one 30 minutes observation on 20220602, and one 50 minutes observation on 20221112. Fortunately, one pulse is detected in the 60 minutes observation on 20220824. Therefore, only two single pulses have been detected by FAST observations of in total 290 minutes. For J1855$-$0154g we get 3 pulses from the 5 minutes survey pointing  on 20210907, but none from the following-up observations for 15 minutes on 20211004 and for 60 minutes on 20221116.
For all these transient sources, the pulse detection rate is very very low. It is very hard to get a period from the TOAs of so few pulses from all observations of each source.

For these transients, we cannot do much but present the two-dimensional dynamic spectra of the dedispersed pulses, see Figure~\ref{fig:fewPulses} and Figure~\ref{fig:AppfewPulses} in the Appendix. % For J1918+0342g, we detected a burst with S/N=339 on 20211202 but not burst detected in the follow-up observation on 20220329.
The pulse morphology of these transient sources is rich. In addition to the single peak pulse, we do see that some pulses have two peaks, such as J1828$-$0003g at 20220803 and 20221114, J1853+0353 on 20210624, or one sharp peak plus a weak peak, e.g. J1847$-$0046g on 20200606. Pulses detected from one source in different epochs are often different in shape. Some pulses have a very wide profile, such as J1853+0209g, reaching hundreds of milliseconds.

In Table~\ref{tab:fewPulses} in the Appendix, we list parameters for each pulse, including the TOA of the pulse peak, signal-to-noise ratio of dedispersed pulse $R$, pulse width in millisecond, and the fluence discussed in Sect.\ref{sect2.2.2}. We will discuss the polarization of some strong pulses in Sect.\ref{sect3.4}.

The DM values of the pulses from these transient sources are all smaller than that the estimated upper limits given by the Galactic electron distribution models \citep{NE2001, YMW2016}, except for J2030+3833g who has a high DM of 417$\pm$6~pc\,cm$^{-3}$ which is near the DM upper limit of 402~pc\,cm$^{-3}$ and 432~pc\,cm$^{-3}$ given by NE2001 \citep{NE2001} and YMW2016 \citep{YMW2016}, respectively. Therefore we conclude that all these transient sources are inside our Milky Way.

\begin{figure}
  \centering
    \includegraphics[width=0.28\textwidth]{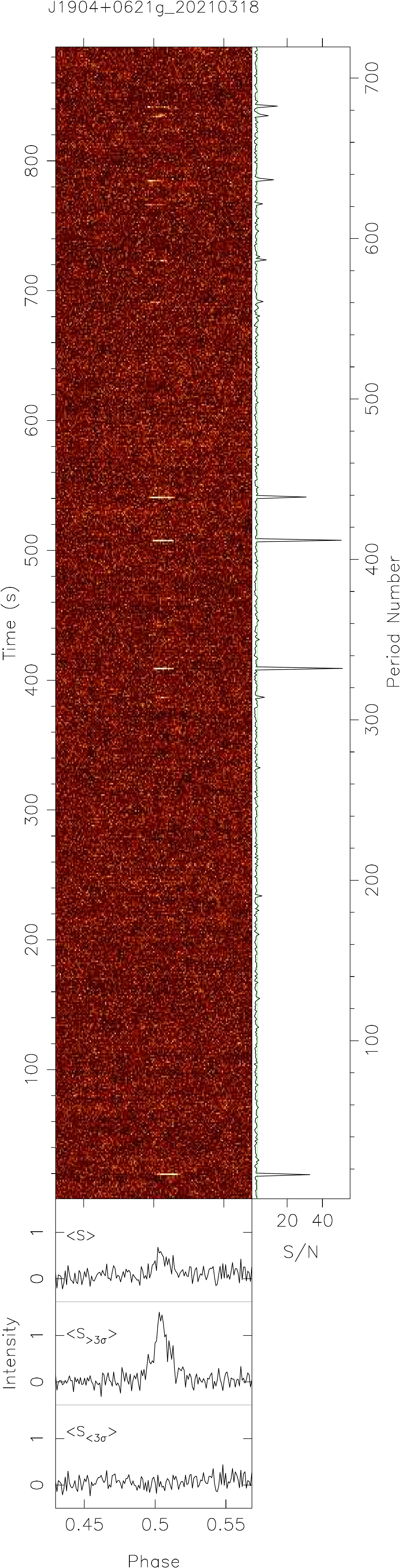} 
  \caption{An example of a newly discovered RRAT, PSR J1904+0621g. Plots for 16 newly discovered RRATs are presented in  Figure~\ref{fig:AppnewRRATs} in the Appendix. Pulse-stack is shown in the main left panel, with only a few pulses occasionally emit. The right panel shows the curve of S/N over pulse number (see Sect.\ref{sect2.2.2}), with the sigma calculated from a given width of off-pulse phase range. Three sub-panels below the main panel are the averaged profiles of all periods, of single pulses with the signal-to-noise ratio $>3\sigma$ and $<3\sigma$. }
\label{fig:newRRATs}
\end{figure}

\begin{figure}
    \centering
  \includegraphics[width=0.28\textwidth]{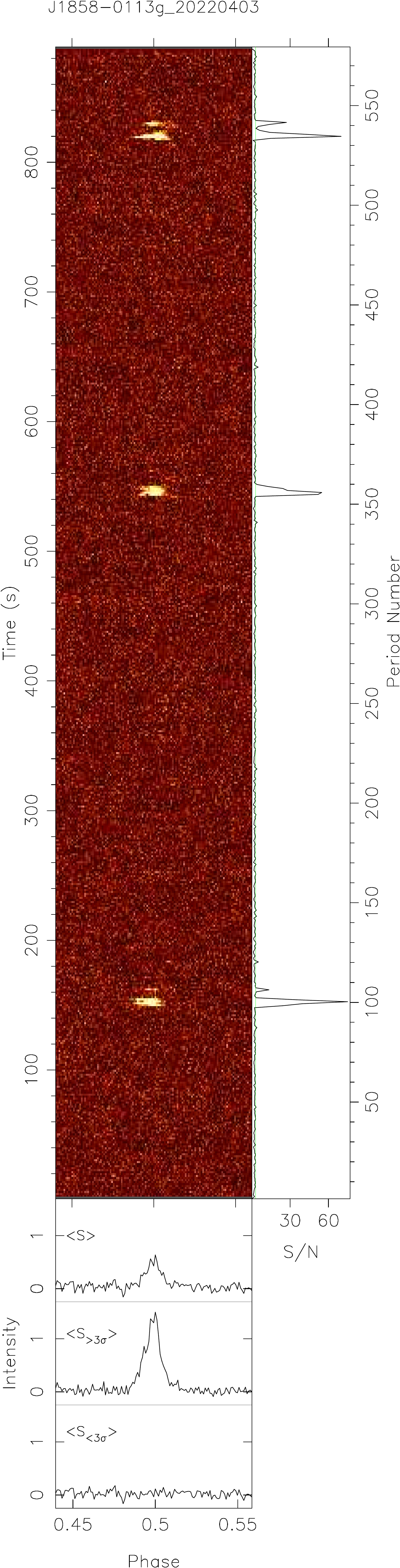} %SP080
    \caption{Same as Figure~\ref{fig:newRRATs} but for an example of extremely nulling pulsar, J1858$-$0113g, discovered by the GPPS survey. Plots for all observations of newly discovered nulling pulsars are presented in Figure~\ref{fig:Appnewnulling} in the Appendix.}
    \label{fig:newnullingpulsars}
\end{figure}

\subsection{Newly discovered proto-RRATs}

Based on the individual pulses detected by the single pulse module from the survey data and also from the longer following-up observations, we get a period from the TOA value of pulses for 50 transient sources (see Sect.~\ref{sect.pf}), see the second part of Table~\ref{tab1} for the list. We therefore confirm that these 16 sources are characterized as proto-RRATs with occasionally emitted pulses. With a newly found period, the detected pulses can be well aligned, see an example in Figure~\ref{fig:newRRATs}. Such plots for newly discovered RRATs in the GPPS survey are shown in Figure~\ref{fig:AppnewRRATs} in the Appendix. In general, we can get much better average profiles for these detected pulses over 3$\sigma$, but cannot detect significant emissions from other periods though they are all added together. Such proto-RRATs cannot be detected via general normal pulsar periodicity search, and their integrated profiles over all periods in an observation session do not have a good S/N.

%Their rotation periods are obtained by solving the TOAs of the detected single pulses, and their averaged profiles are obtained after folded and no significant signal that makes it difficult to be detected by period search methods. While the signal of the average pulse profile for single pulses of S/N~\textgreater~3 is obvious, but it is almost absent for that of S/N~\textless~3. 

%And the nulling pulsar J1940+2203g has a long-period of 11.906\,s, its fluence is just about 1.5\,$\mu$Jy\,s and pulse width is less than 0.5$\%$ of full pulse phase. 

For some proto-RRATs with several observation sessions, the pulse detect rate is somehow a constant, such as about 24 pulses per hour for J2005+3156g. 
For some RRATs, the pulse strengths and the detection rates vary a lot,  see N$_d$/N$_p$ and $\rm T_{obs}$ in Table~\ref{tab1} for J1905+0558g as examples. For some RRATs, however, a period is obtained fortunately from few detected pulses. For example, both J1857+0229g and J1924+1446g have only one pulse detected in the 5 minutes survey data, and their periods are obtained from only 9 pulses detected from the expensive one hour observations. No such fortune is available for many  transient sources discussed above, such as J1850$-$0004g, J1855+0033g and J1916+1142Ag.

\subsection{Newly discovered extremly nulling pulsars}

In addition to detecting sparse pulses from RRATs discussed above,  the single pulse search module can also pick up successive pulses for several periods  but merely emerge in a small fraction of observation time, so that they cannot be picked out via the periodicity pulsar search. No detection of the pulsed flux for other periods than bright pulses implies their nature of very nulling. We get 10 new extremely nulling pulsars from the single pulse searching, as listed in the third part of Table~\ref{tab1}. Figure~\ref{fig:newnullingpulsars} shows one example, while all newly discovered very nulling pulsars are shown in Figure~\ref{fig:Appnewnulling} in the Appendix.  

Such extremely nulling pulsars have an advantage for period determination. Catching the emission active episode sometimes is not easy for some objects, so none detection of any pulses (Np=0) in some observation sessions for e.g. PSR J1845$-$0008g and J1855+0240g is possible, see Table~\ref{tab1}. PSR J1845$-$0008g is a particularly interesting nulling pulsar. We detect pulses in three sessions, but not the other three. For PSR J1855+0240g, successive pulses are merely detected in a short episode during the 15-minute observation.  

%It is also difficult to see the signal in the integrated profile of the pulse with a S/N~\textgreater~3 and have no got the credible flux for J1921+1632g from it averaged profile. This may be related to the fact that the data from that observation are too noisy. 

%Almost all of these sources are characterized with subpulse drifting, although a few are not obvious due to the low signal-to-noise ratio and strong data noise. 

The mean flux density $\langle S \rangle$ of these pulsars  in Table~\ref{tab1}, calculated from the averaged flux over all periods in the observation session, should be taken as the upper limit. 

\begin{figure}
  \centering
  \includegraphics[width=0.33\textwidth]{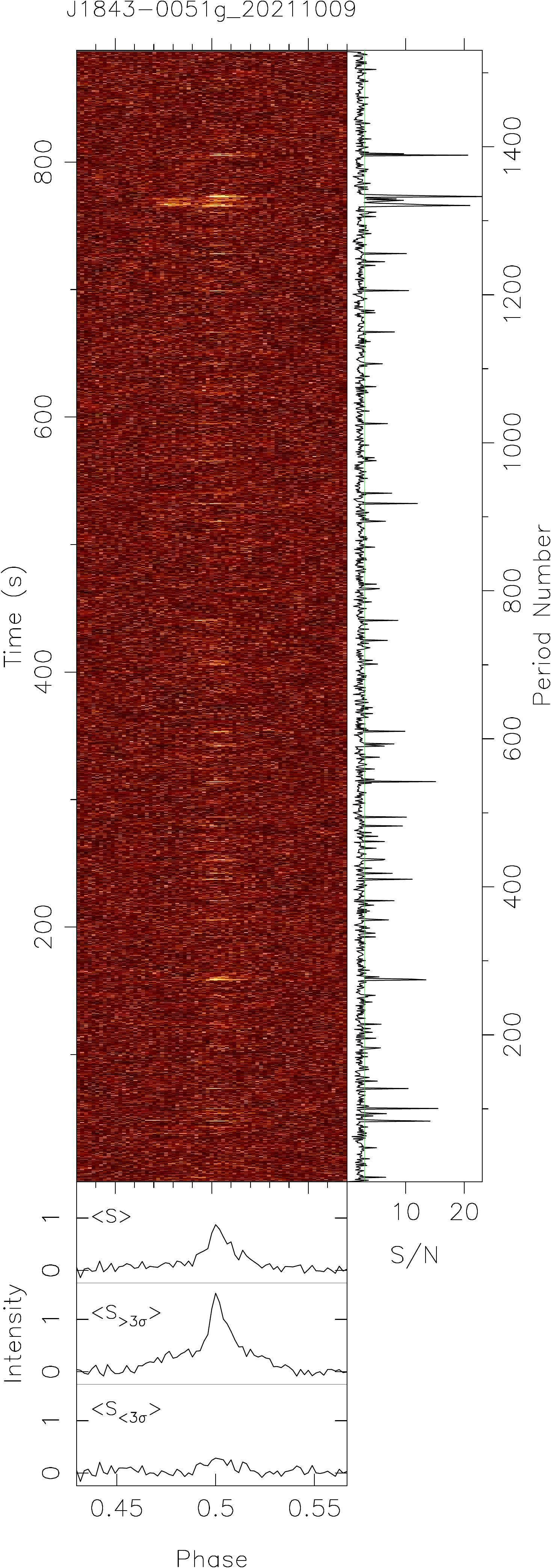} 
\caption{Same as Figure~\ref{fig:newRRATs} but for an example of a weak pulse with sparse bright pulses. Plots for many such new weak pulsars discovered by catching their bright pulses are presented in Figure~\ref{fig:AppnewweakPulsars} in the Appendix.}
\label{fig:newWeakPulsars}
\end{figure}

\begin{figure*}
\centering
\includegraphics[width=0.47\columnwidth]{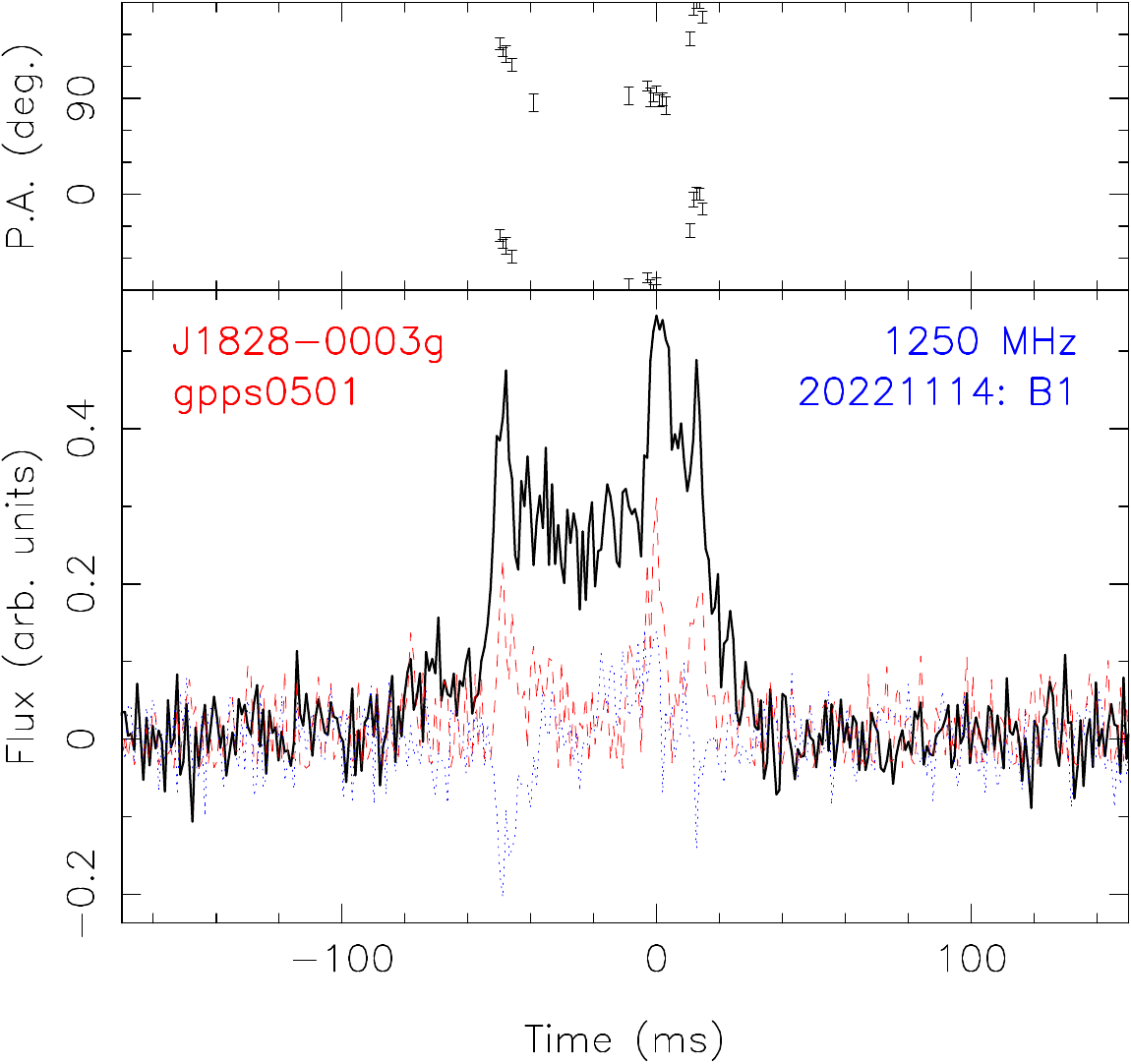}
\includegraphics[width=0.47\columnwidth]{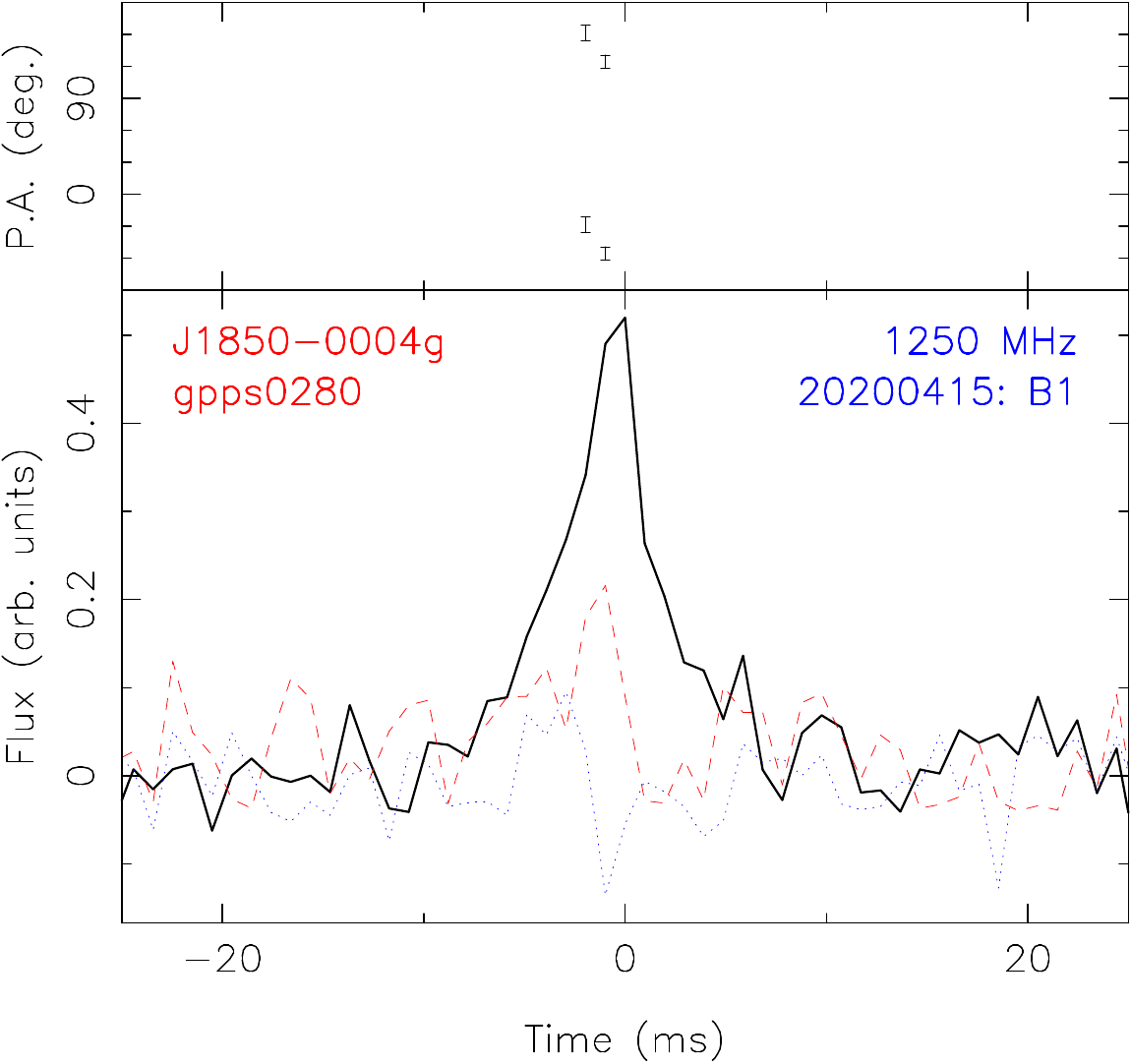}
\includegraphics[width=0.47\columnwidth]{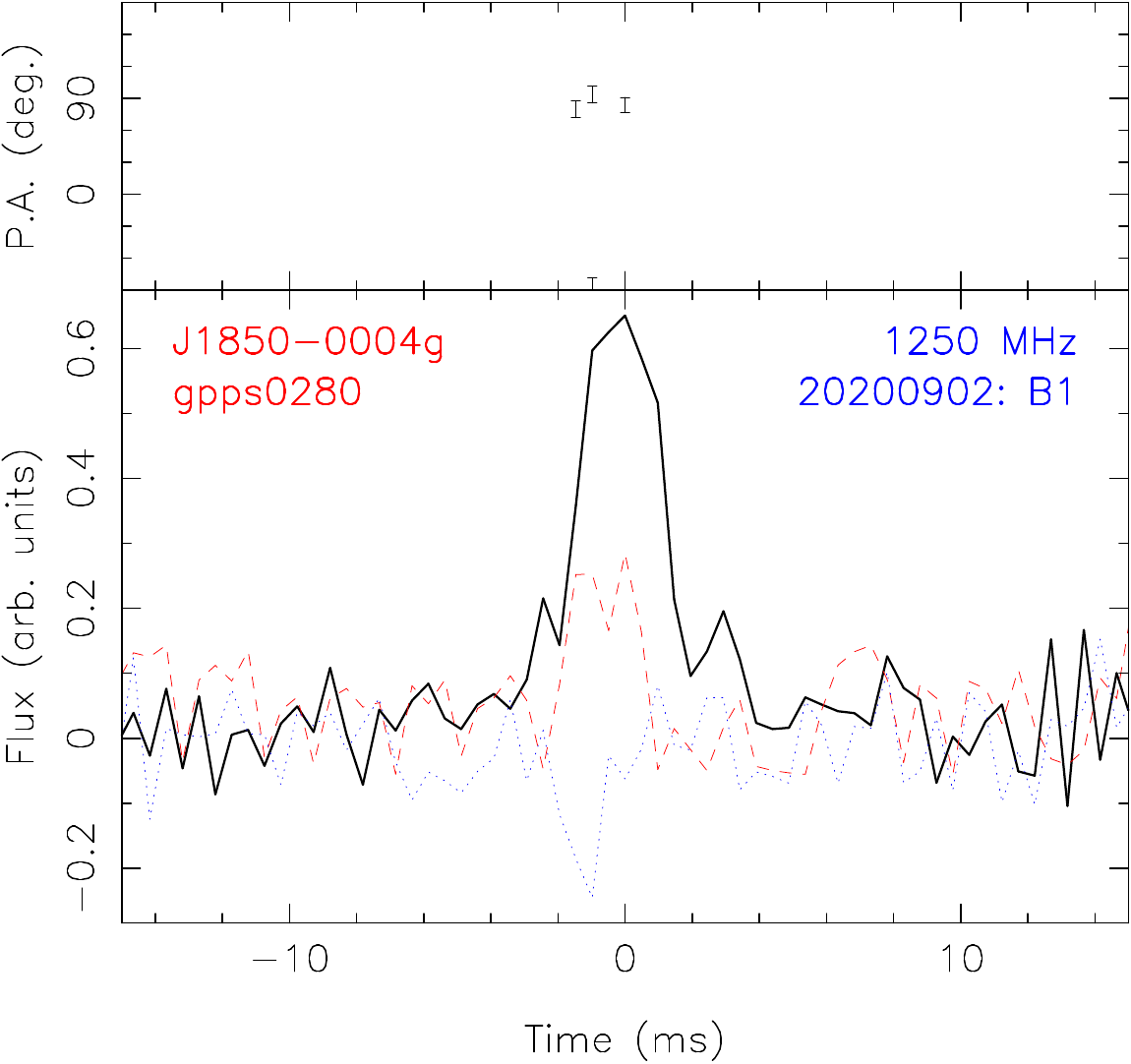}
\includegraphics[width=0.47\columnwidth]{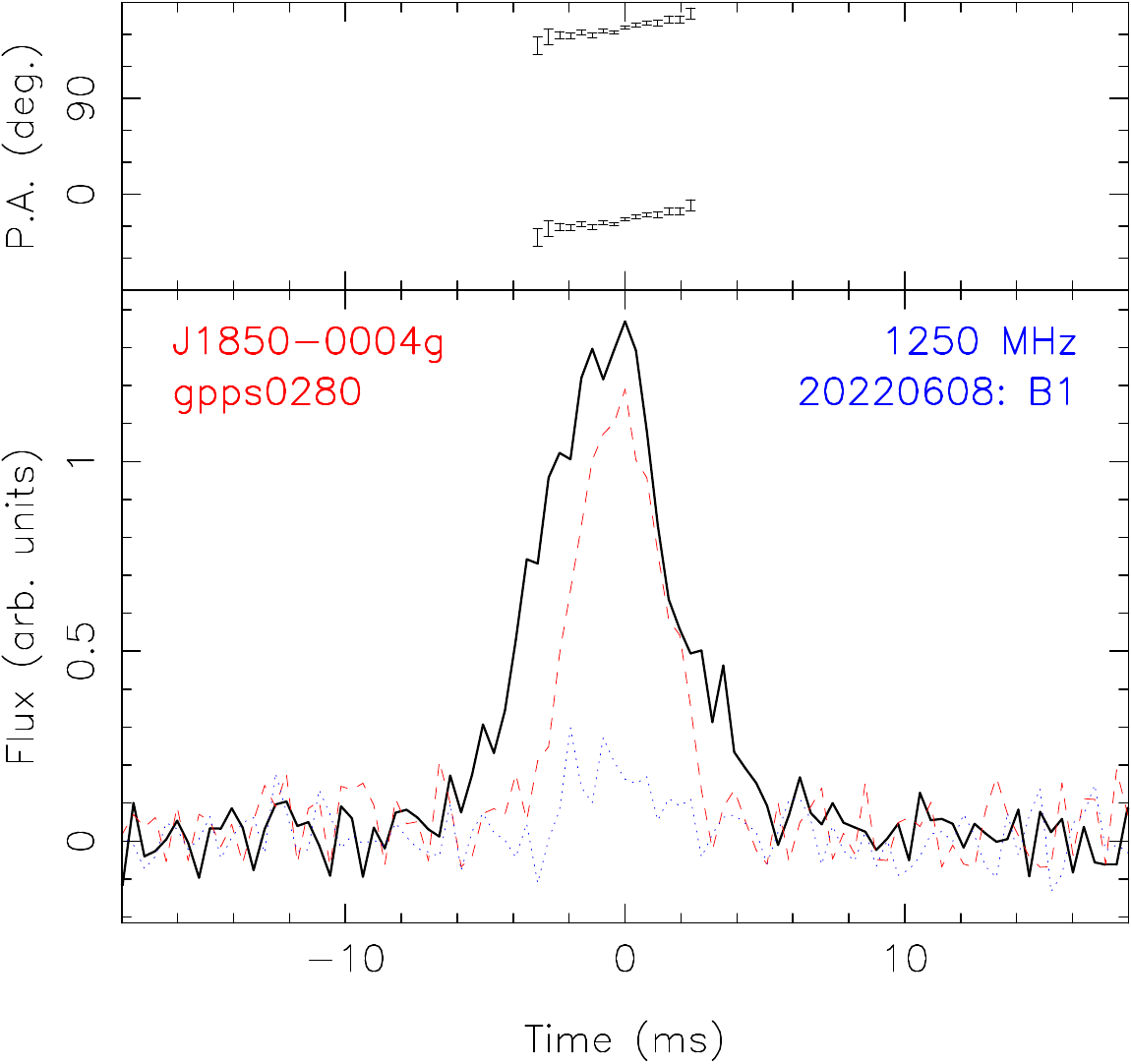}\\[1.0mm]
\includegraphics[width=0.47\columnwidth]{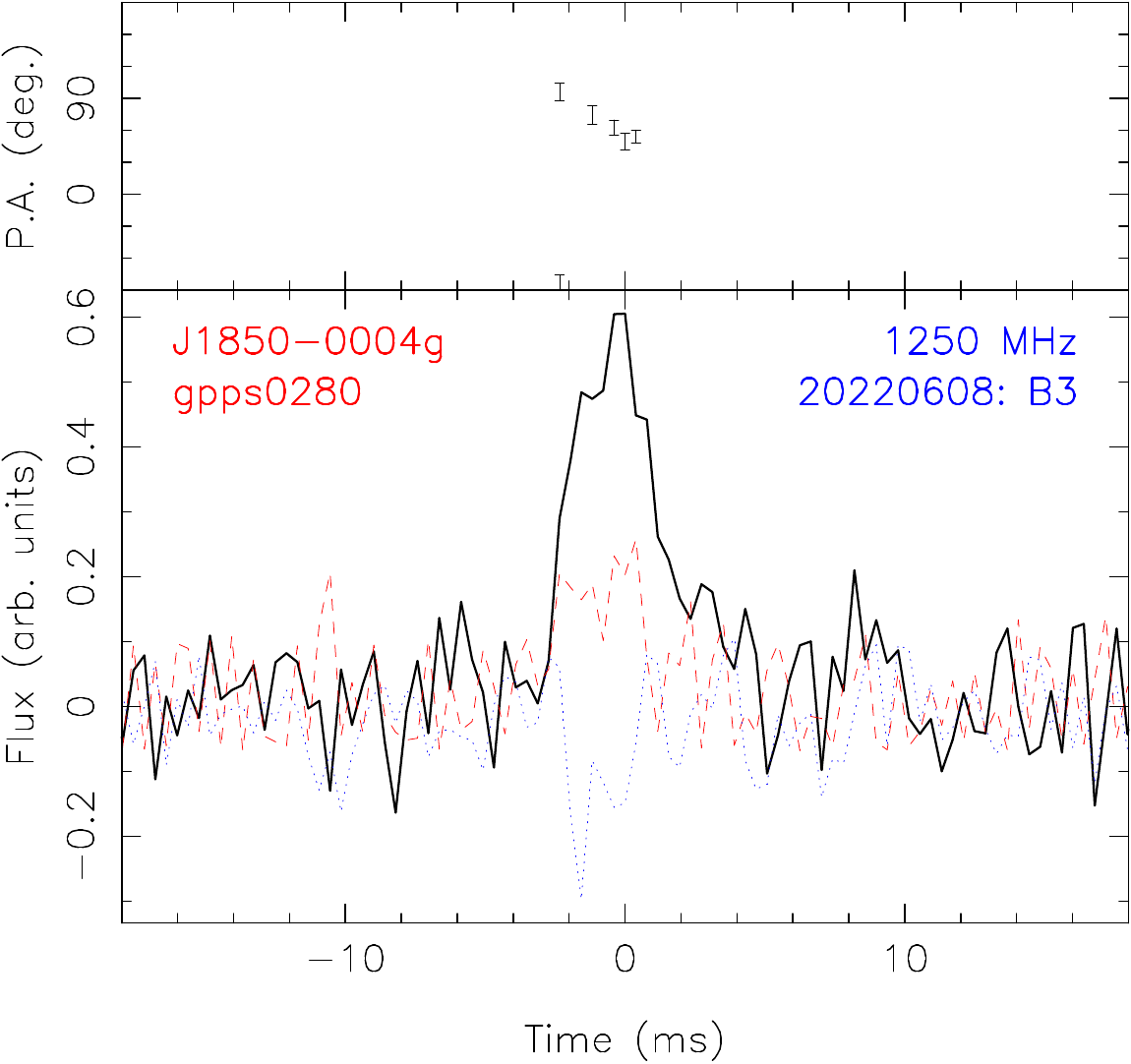}
\includegraphics[width=0.47\columnwidth]{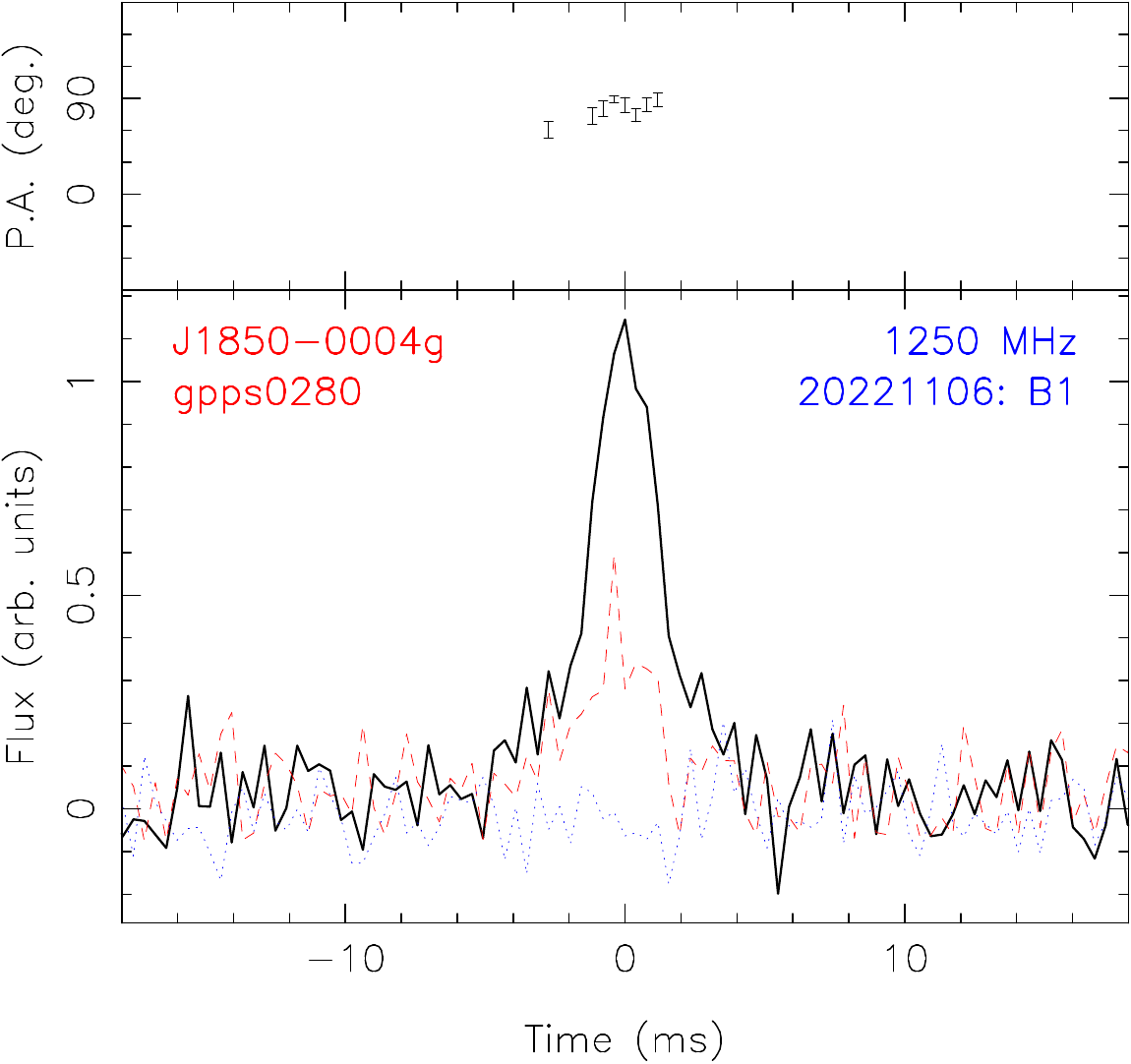}
\includegraphics[width=0.47\columnwidth]{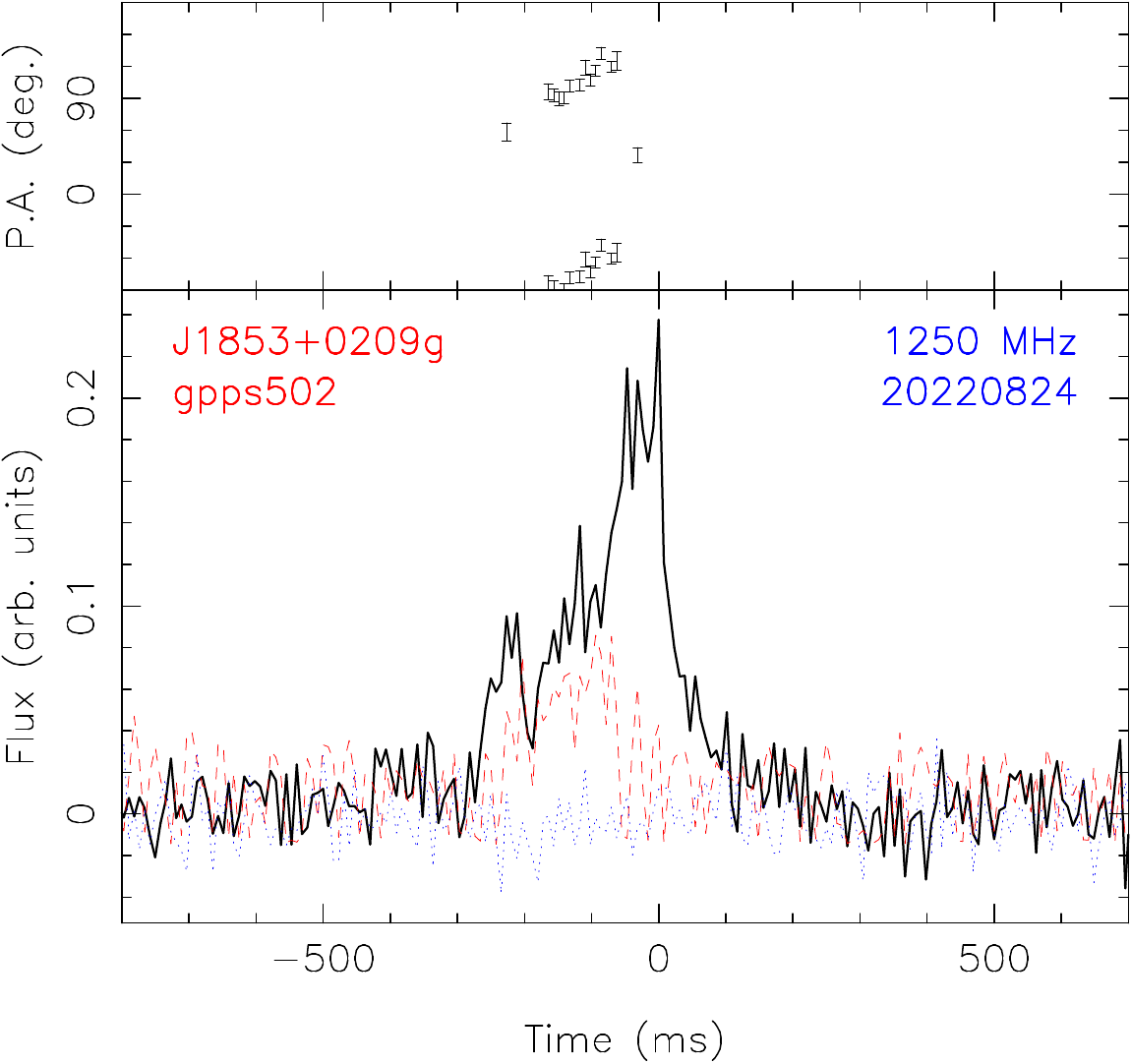}
\includegraphics[width=0.47\columnwidth]{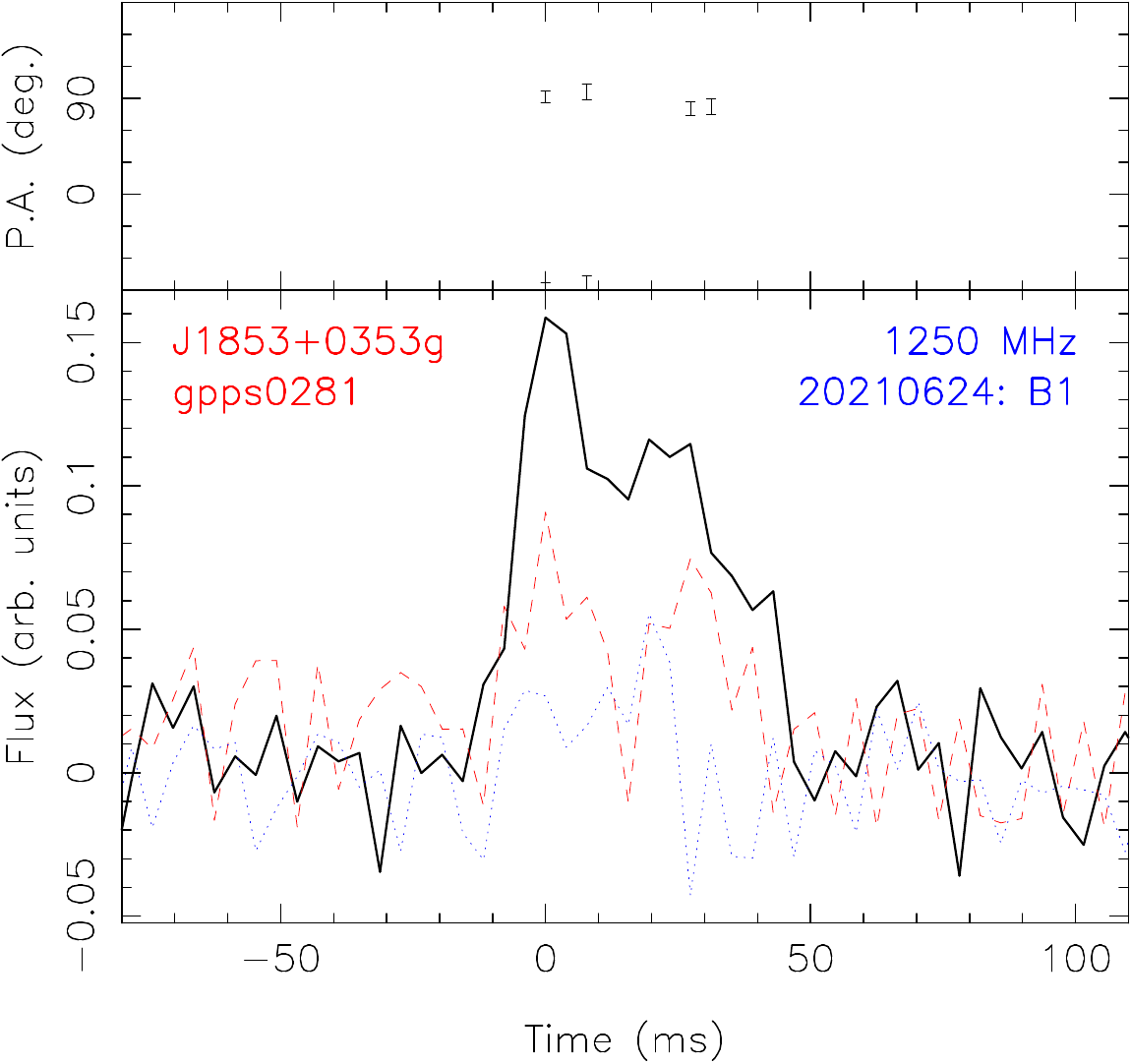} \\[1.0mm]
\includegraphics[width=0.47\columnwidth]{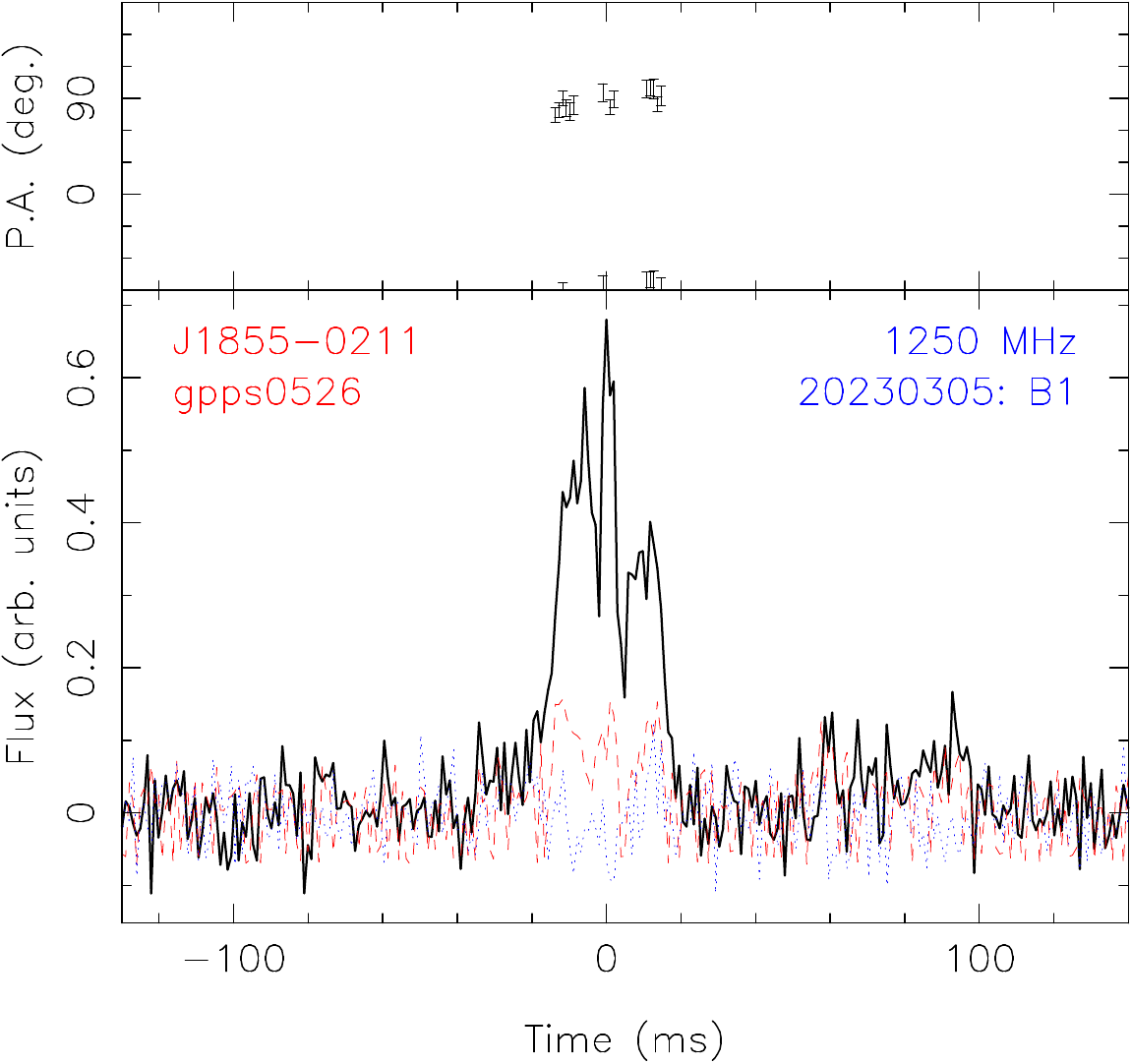}
\includegraphics[width=0.47\columnwidth]{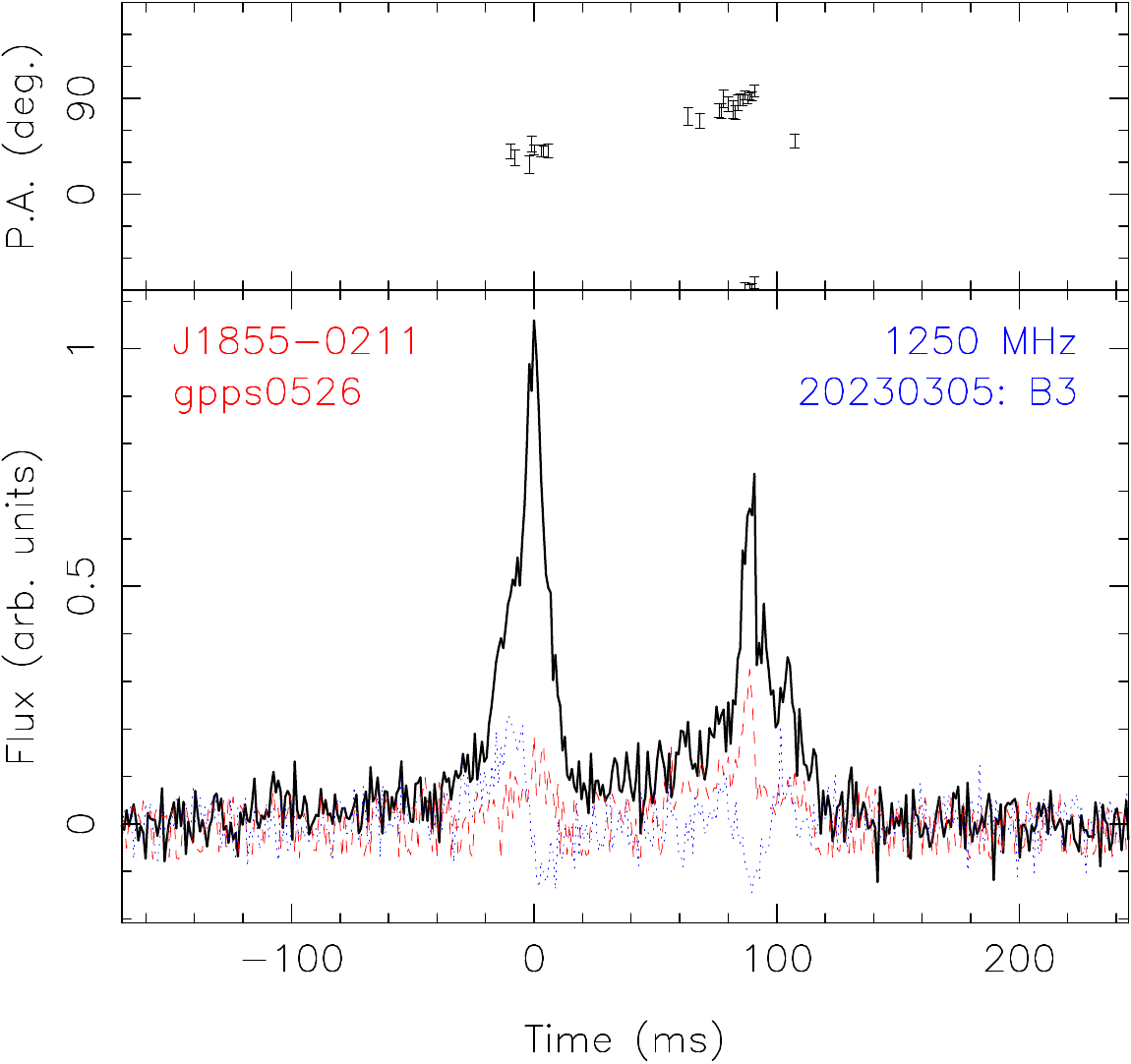} 
\includegraphics[width=0.47\columnwidth]{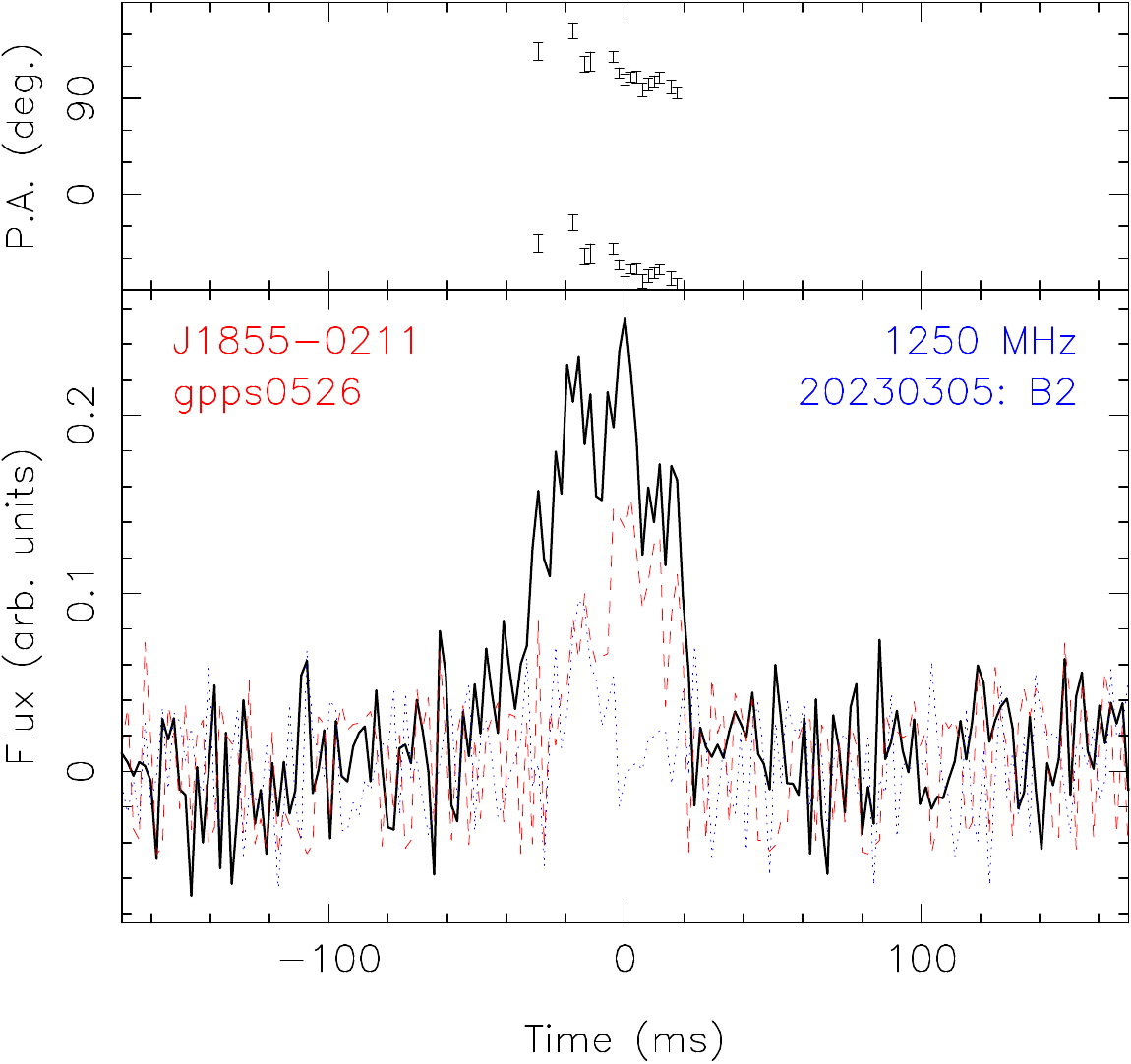}
\includegraphics[width=0.47\columnwidth]{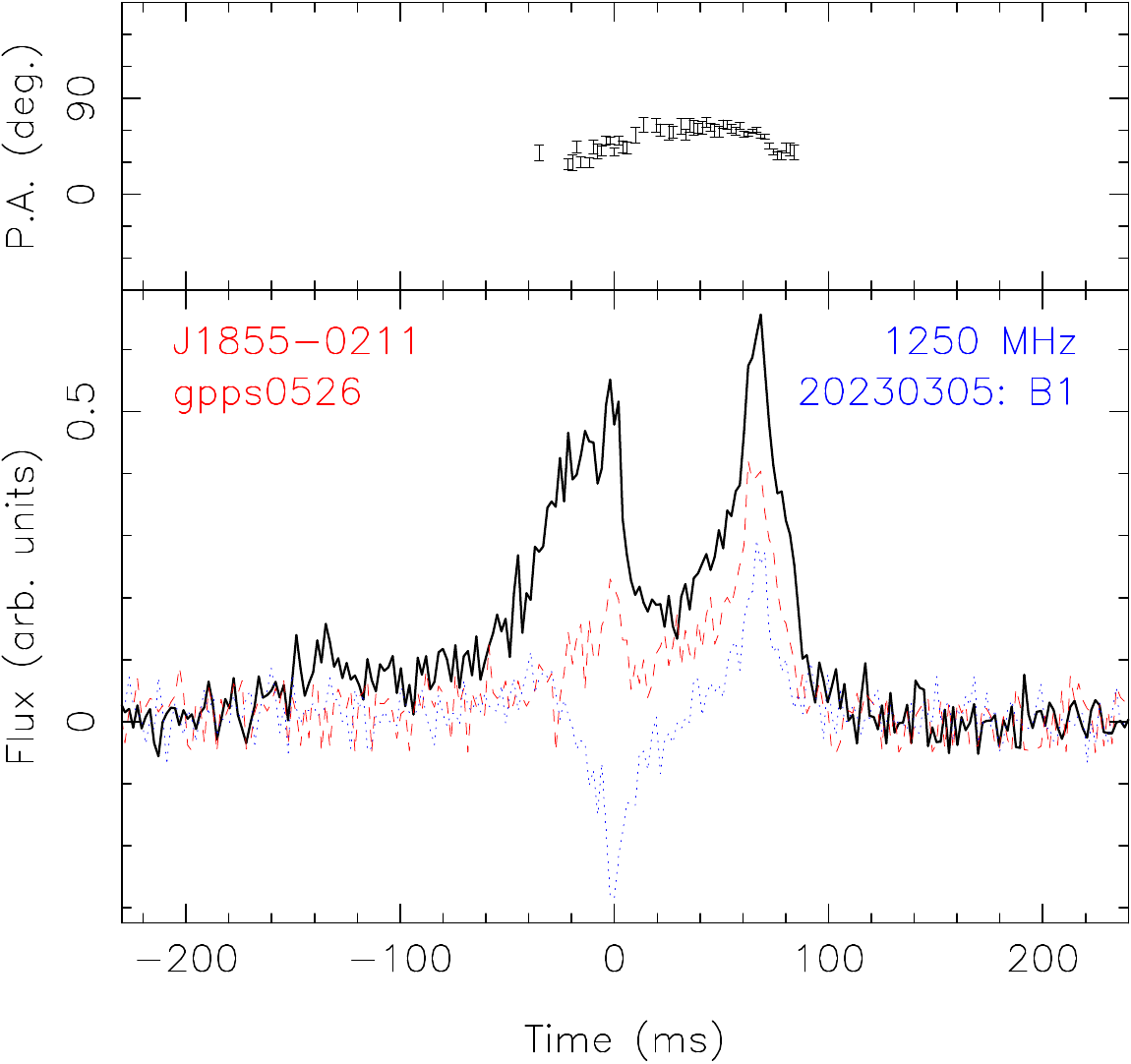} \\[1.0mm]
\includegraphics[width=0.47\columnwidth]{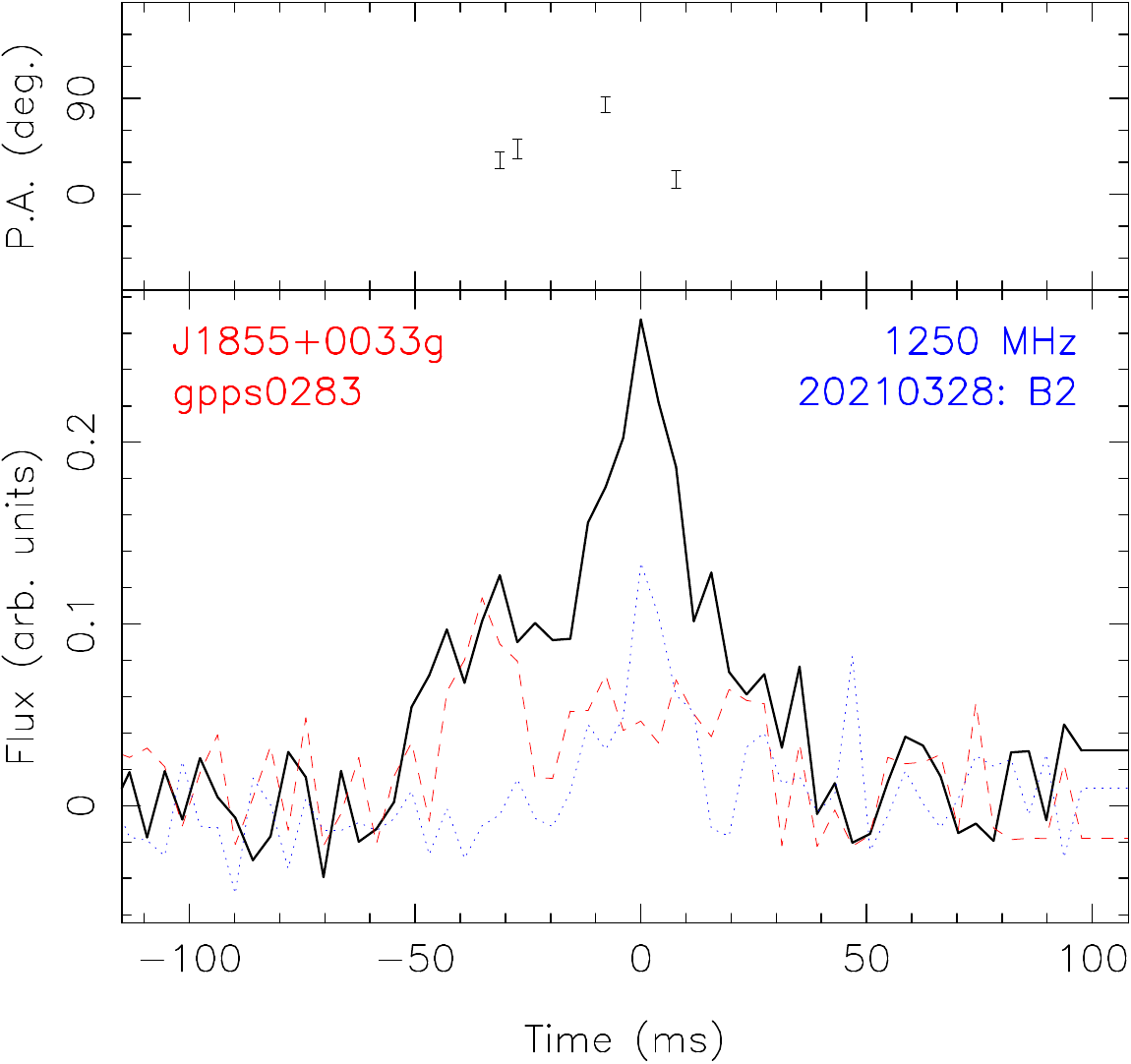}
\includegraphics[width=0.47\columnwidth]{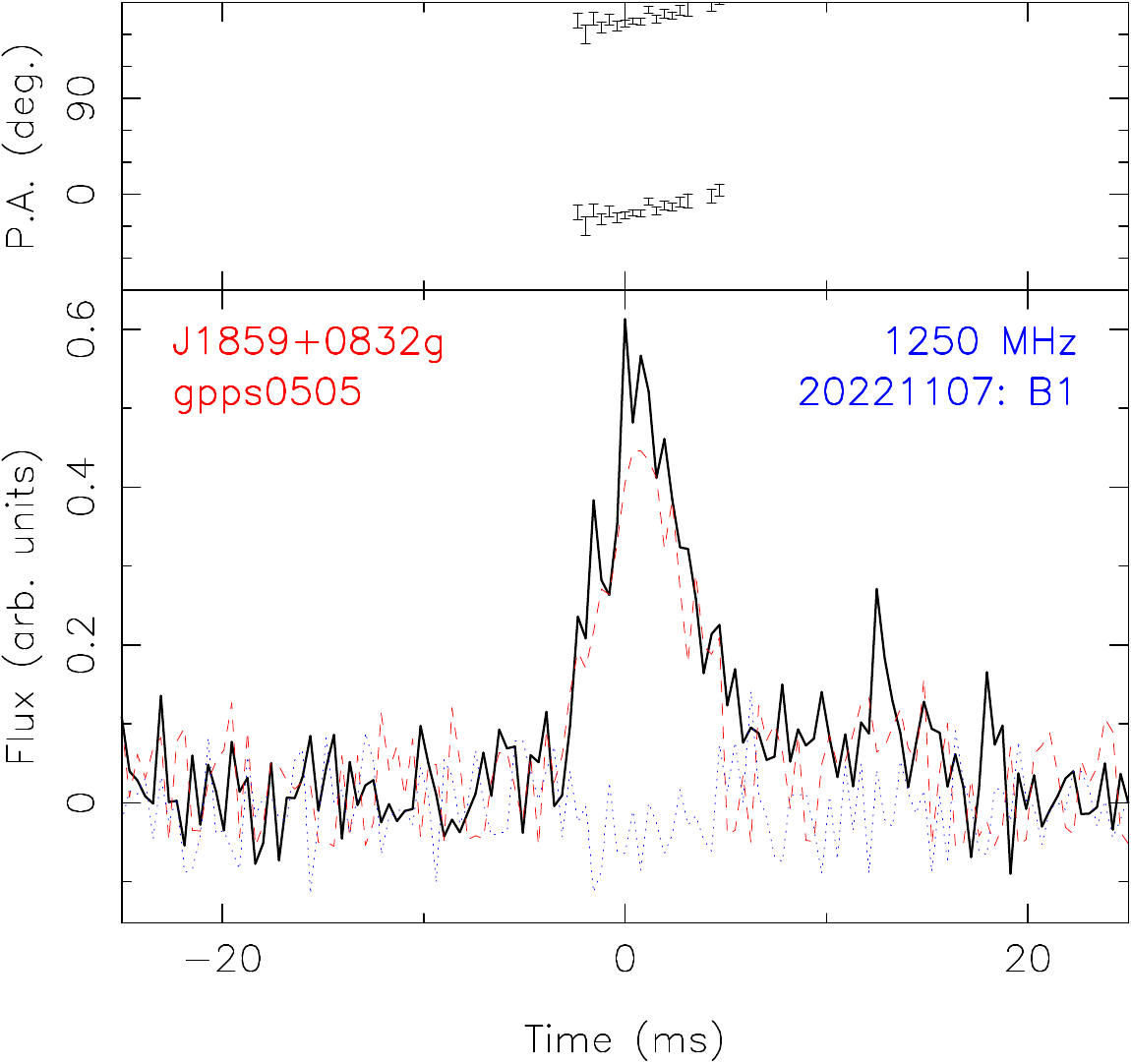}
\includegraphics[width=0.47\columnwidth]{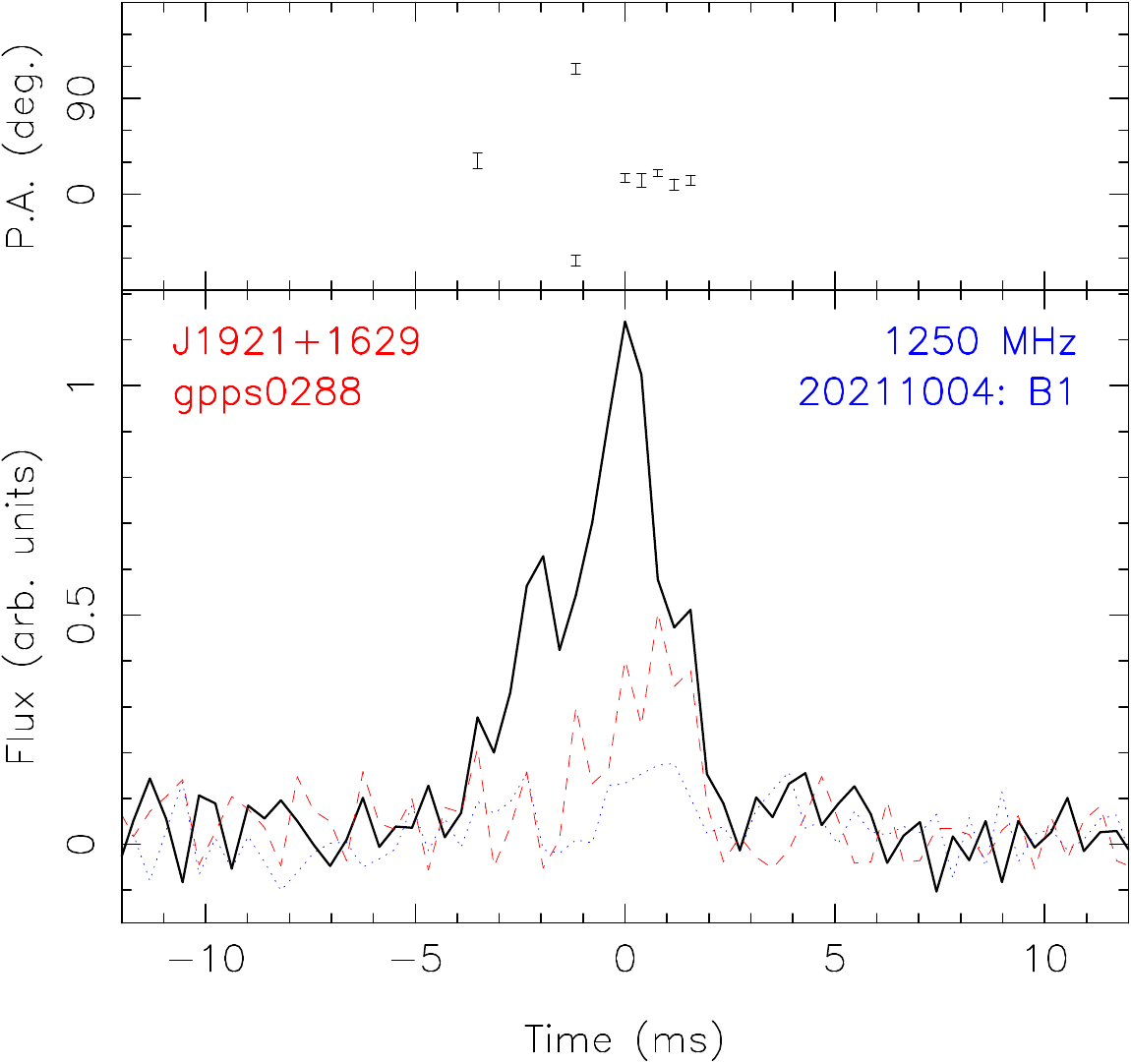} 
\includegraphics[width=0.47\columnwidth]{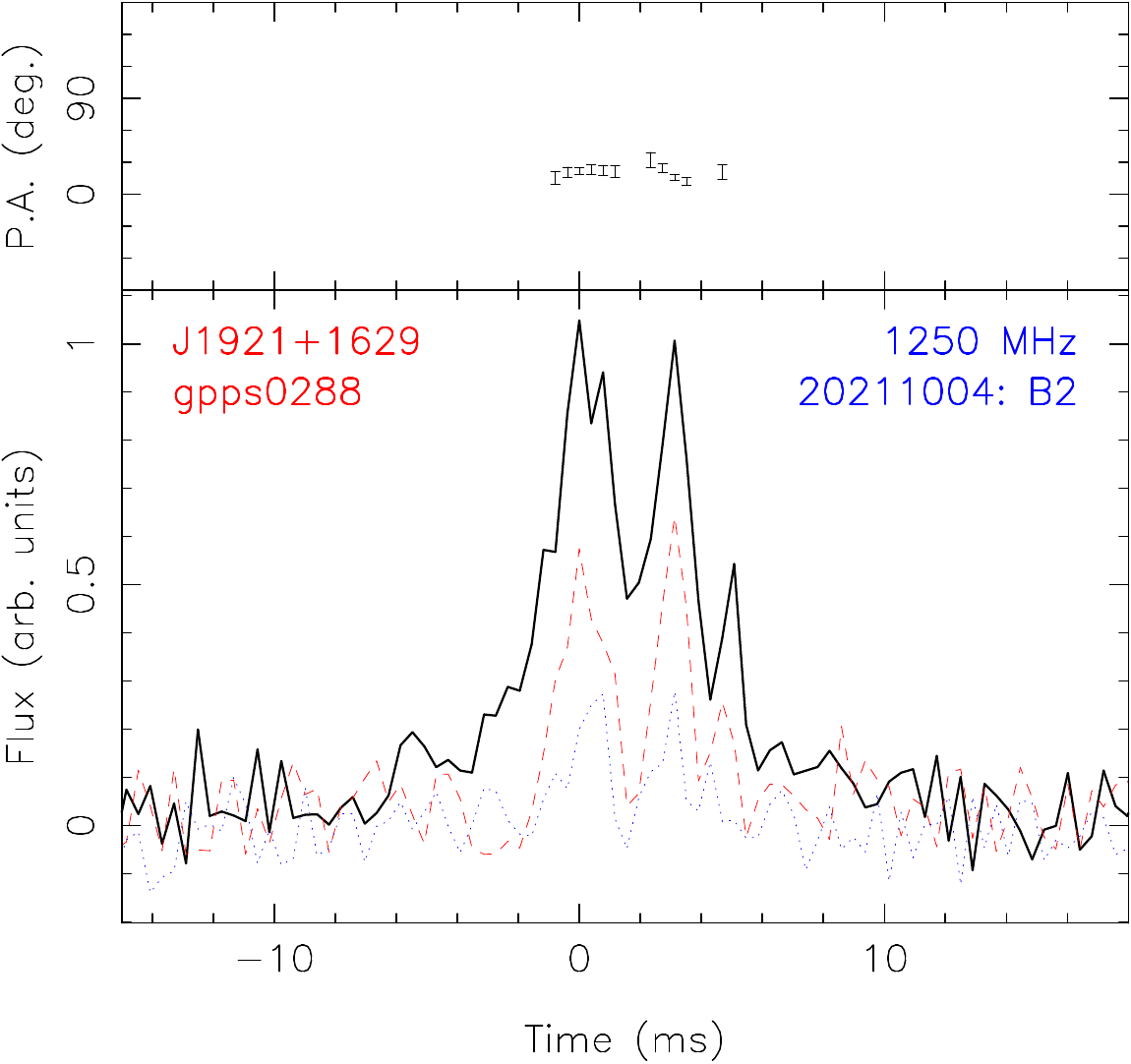} \\[1.0mm]
\includegraphics[width=0.47\columnwidth]{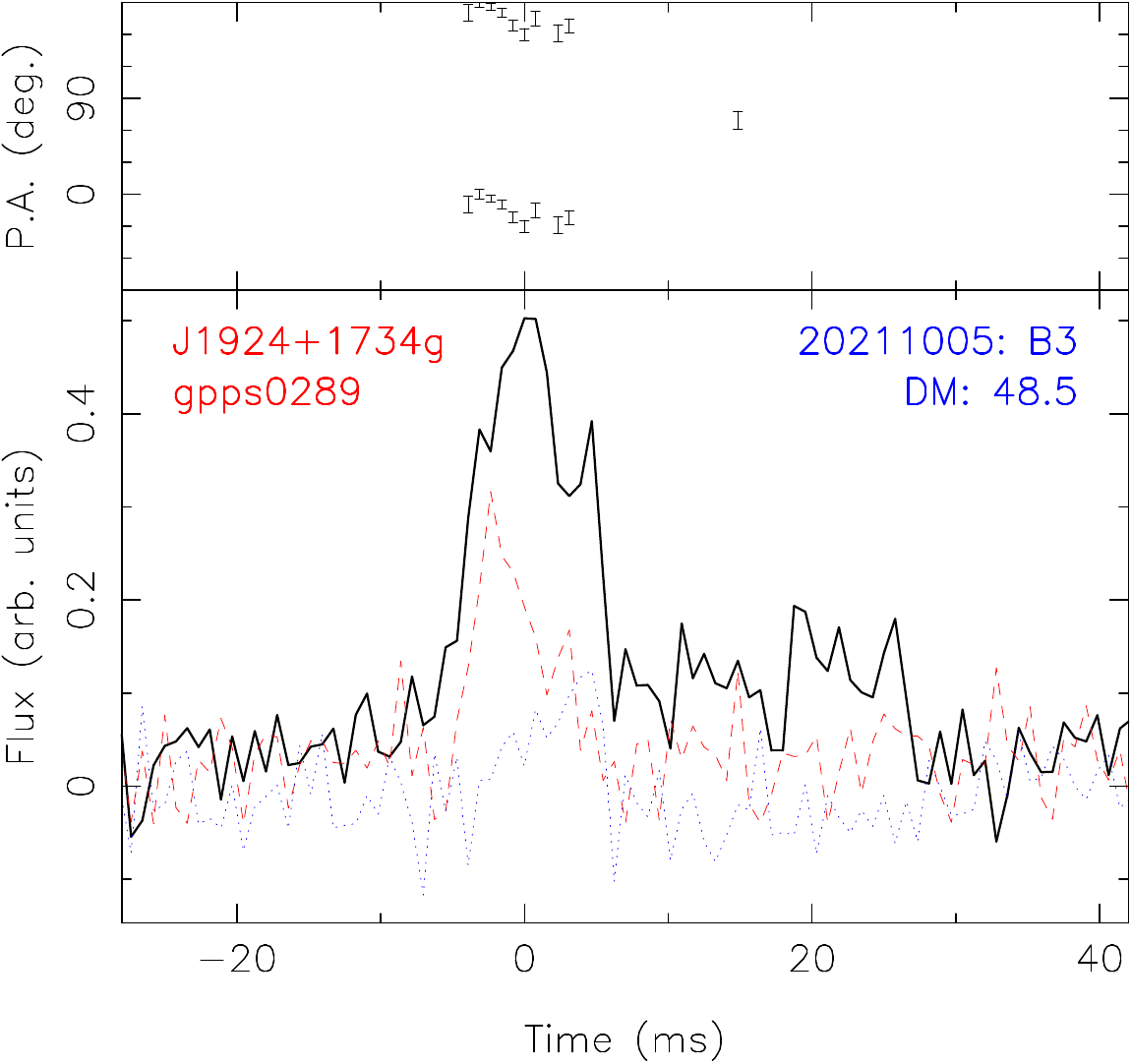}
\includegraphics[width=0.47\columnwidth]{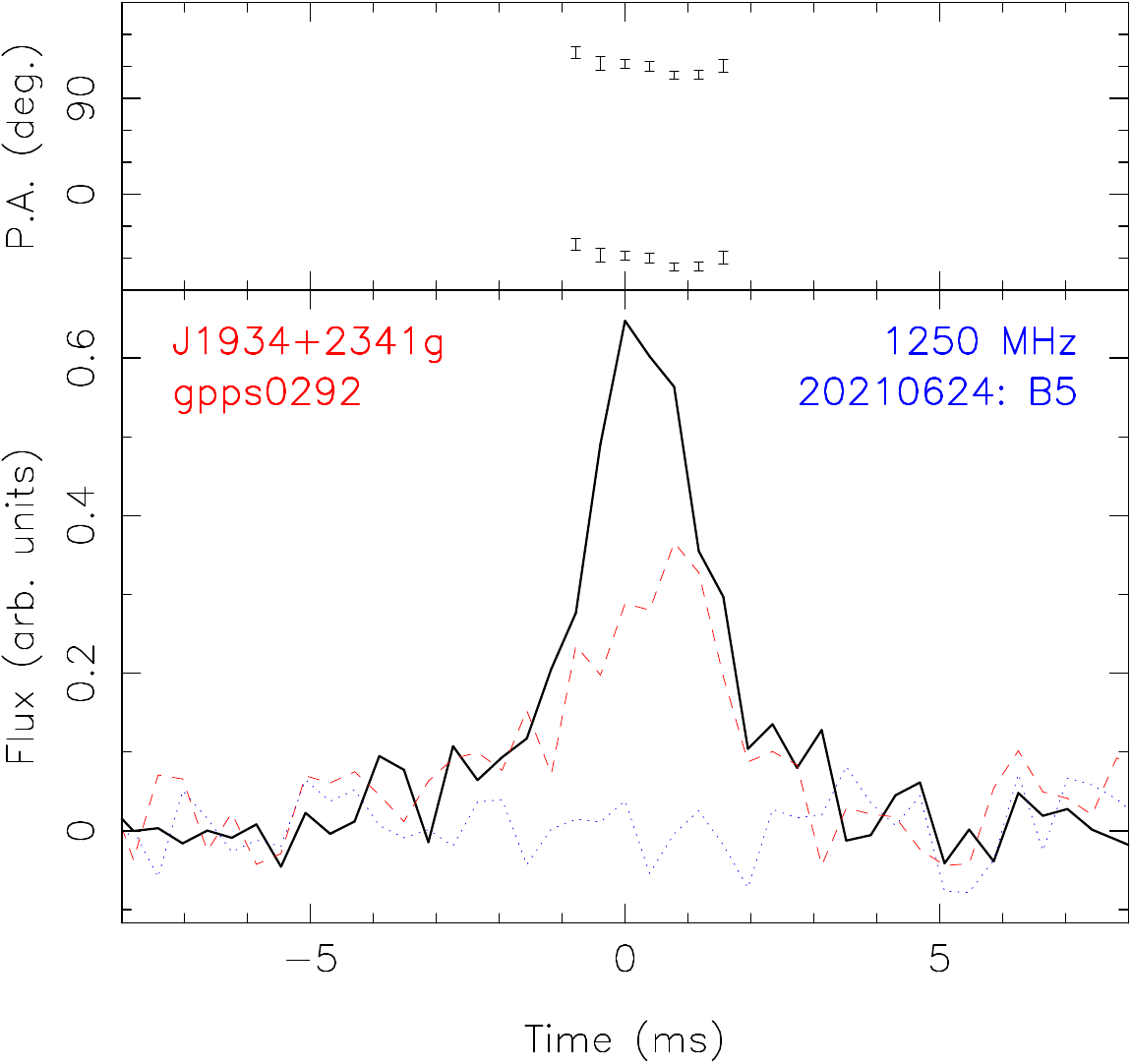}
\includegraphics[width=0.47\columnwidth]{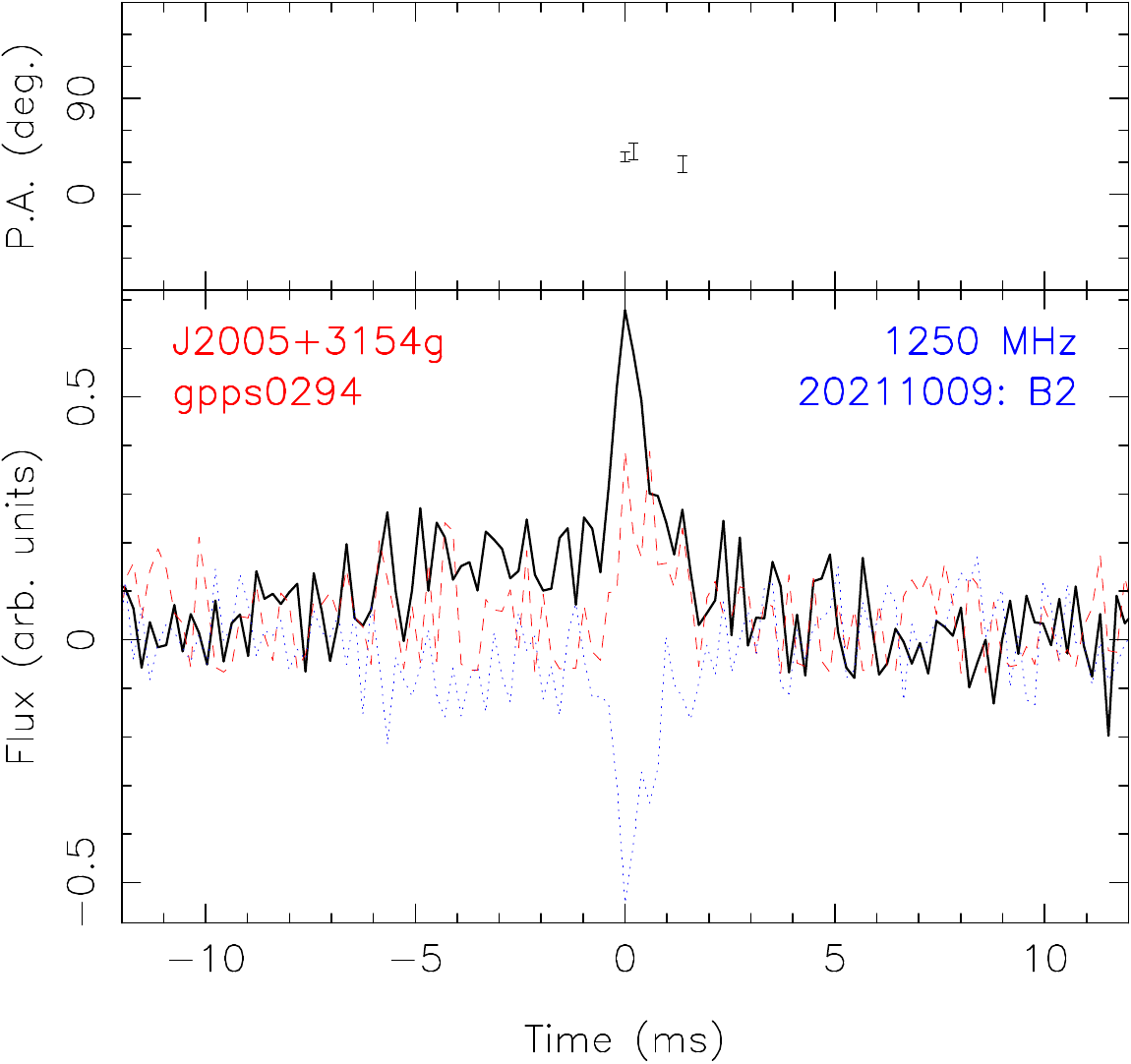}
\caption{Polarization profiles of strong pulses from newly discovered transient sources. The upper sub-panel is for polarization angle (PA), and  bottom sub-panel is for pulse profiles of total intensity $I$ (solid line), linear polarization angle $L$ (dashed line) and circular polarization $V$ (dotted line). Source name, the source number in GPPS survey, observe date, pulse number on that day are marked inside the bottom sub-panel.}
\label{fig:NewPol4fewPulse}
\end{figure*}
% \addtocounter{figure}{-1}

\begin{figure}[!ht]
    \centering
    \includegraphics[width=0.45\columnwidth]{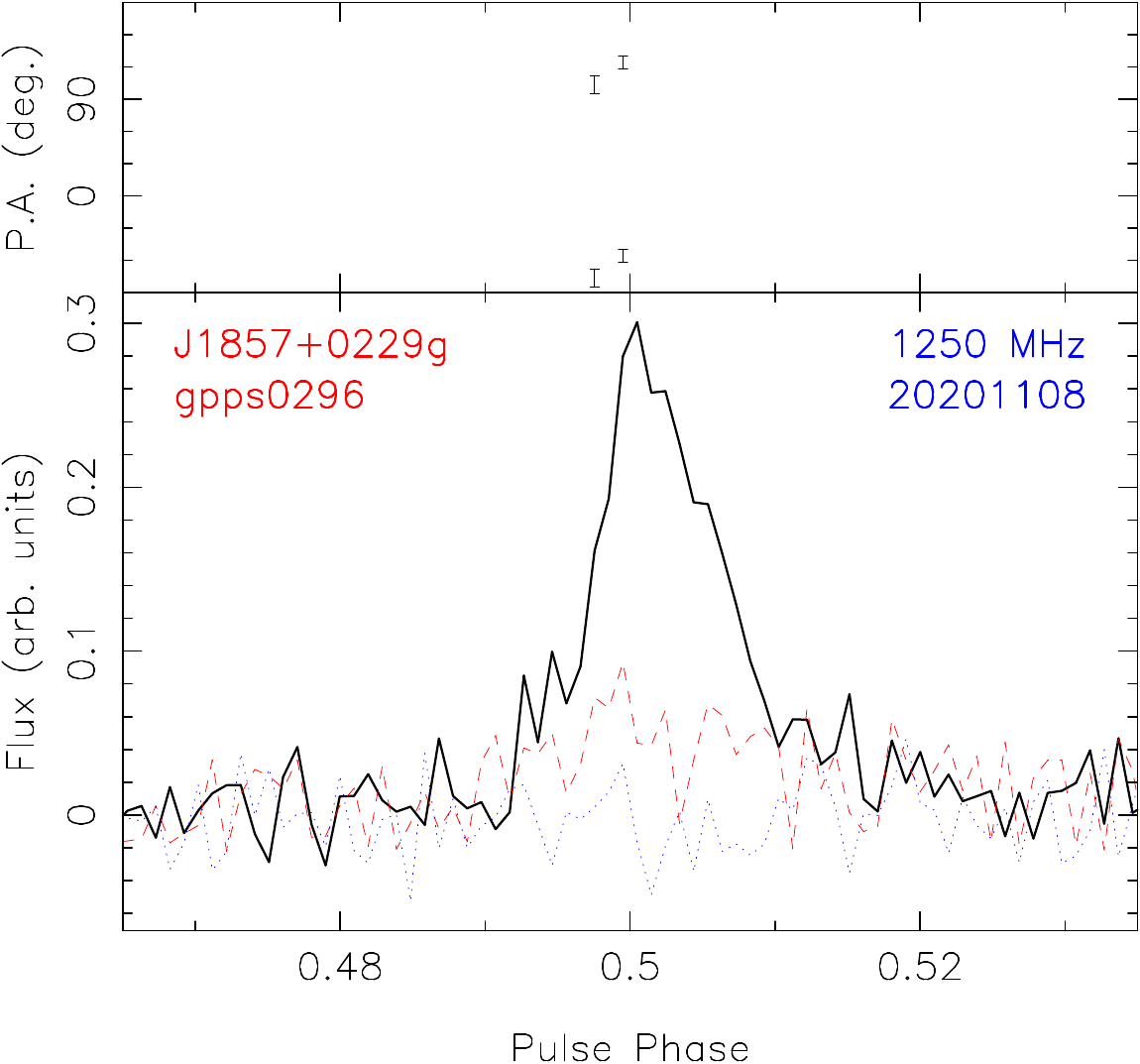} 
    \includegraphics[width=0.45\columnwidth]{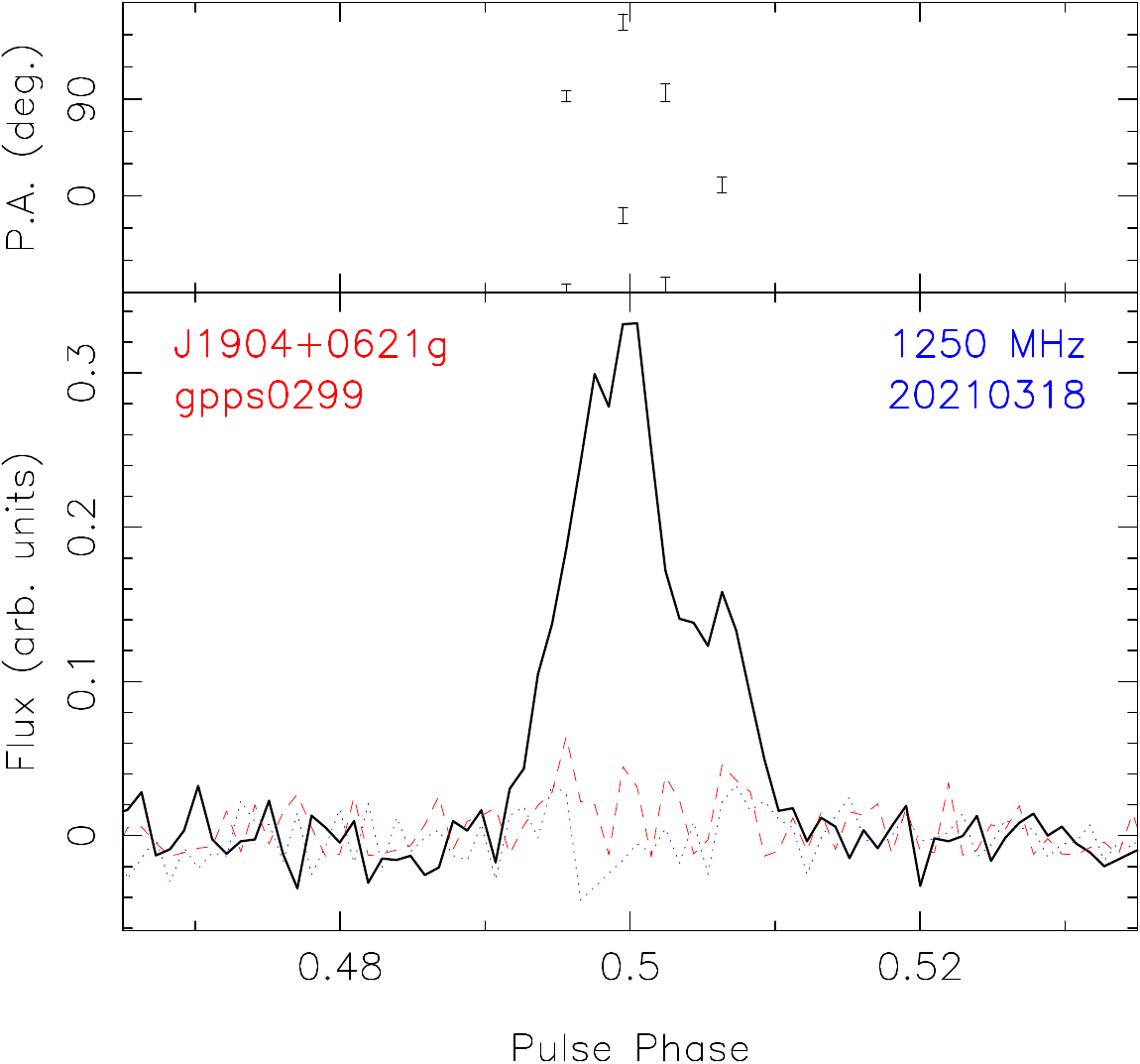}\\[1mm] 
    \includegraphics[width=0.45\columnwidth]{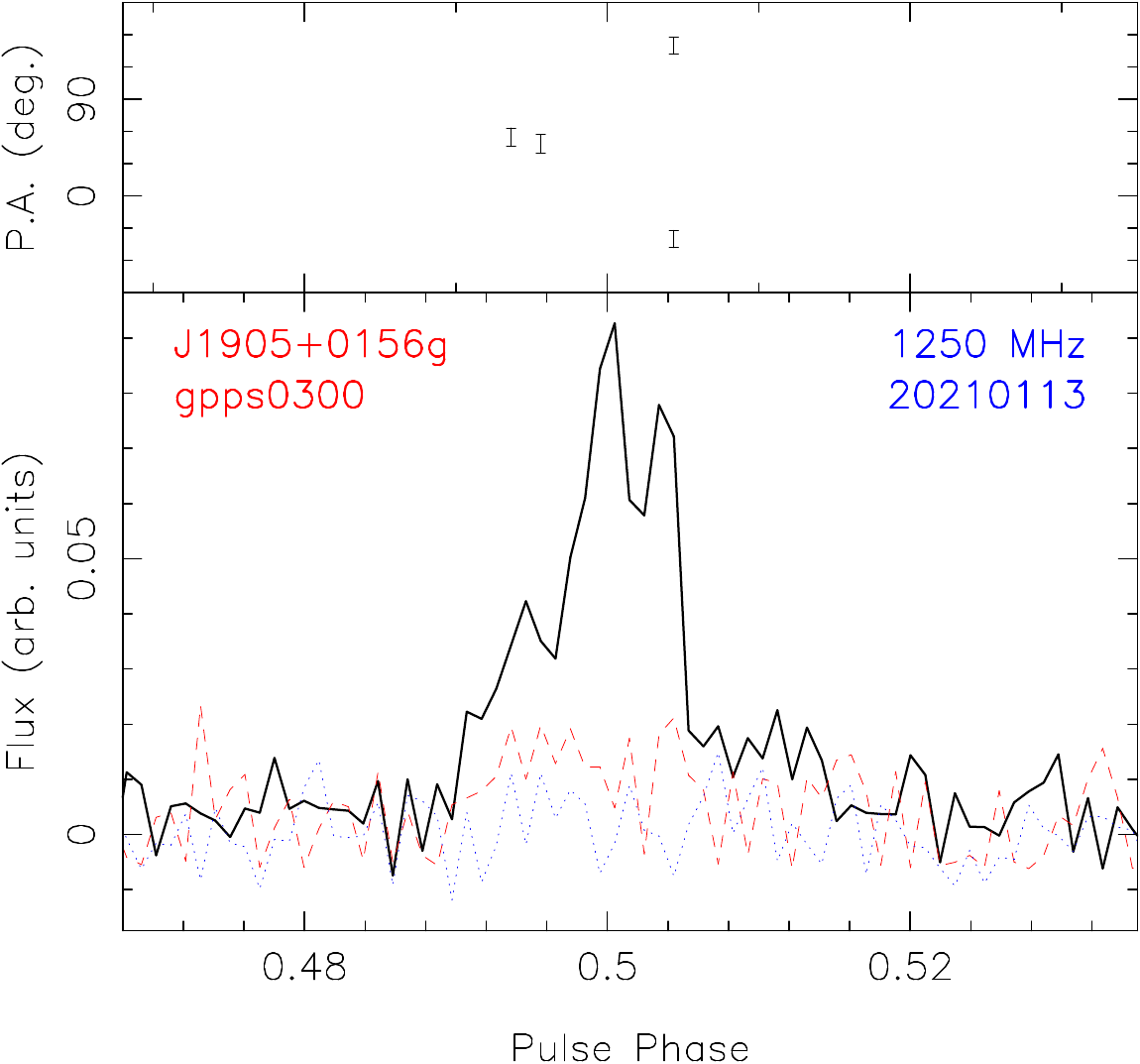} 
    \includegraphics[width=0.45\columnwidth]{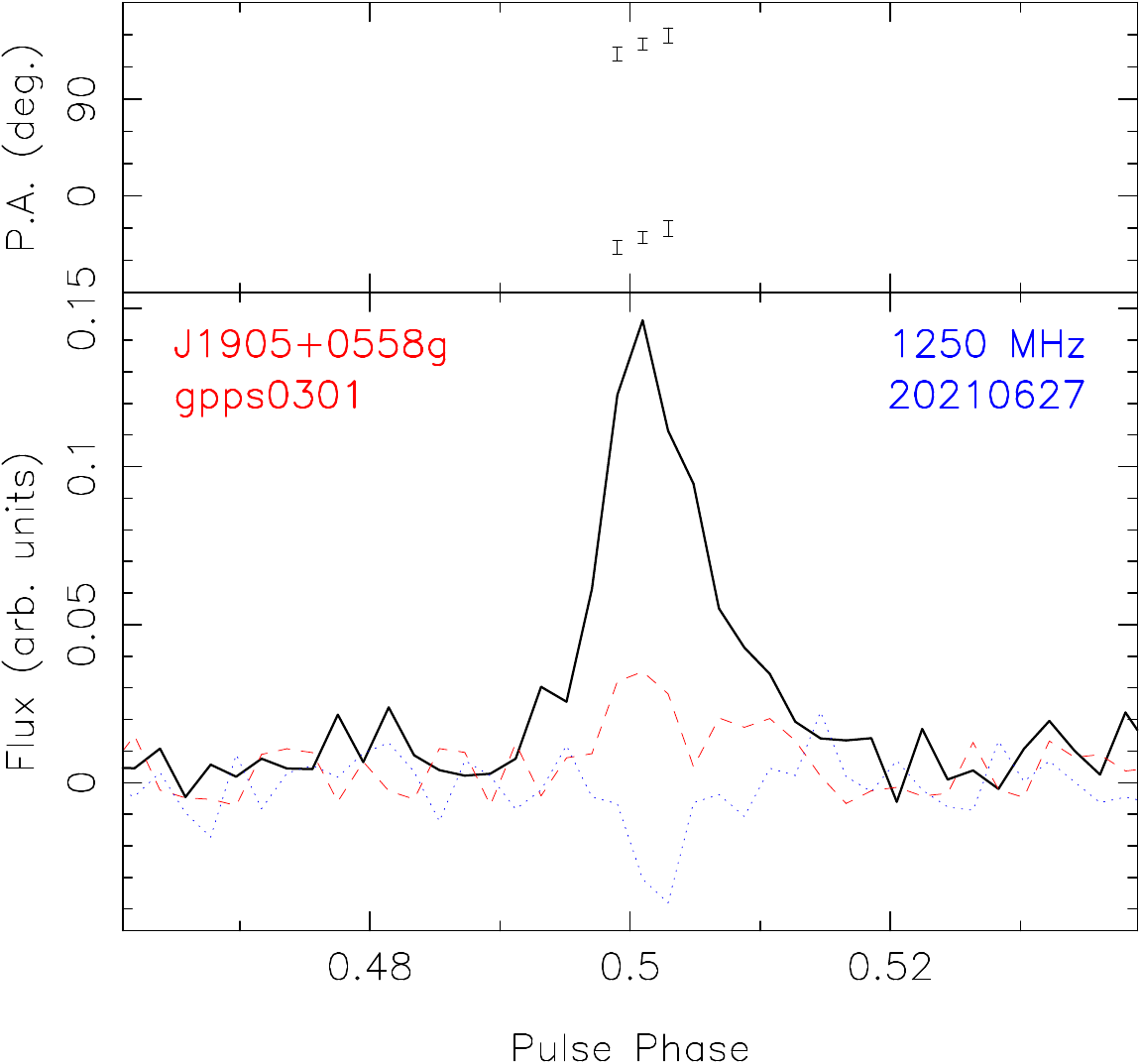} \\[1mm] 
    \includegraphics[width=0.45\columnwidth]{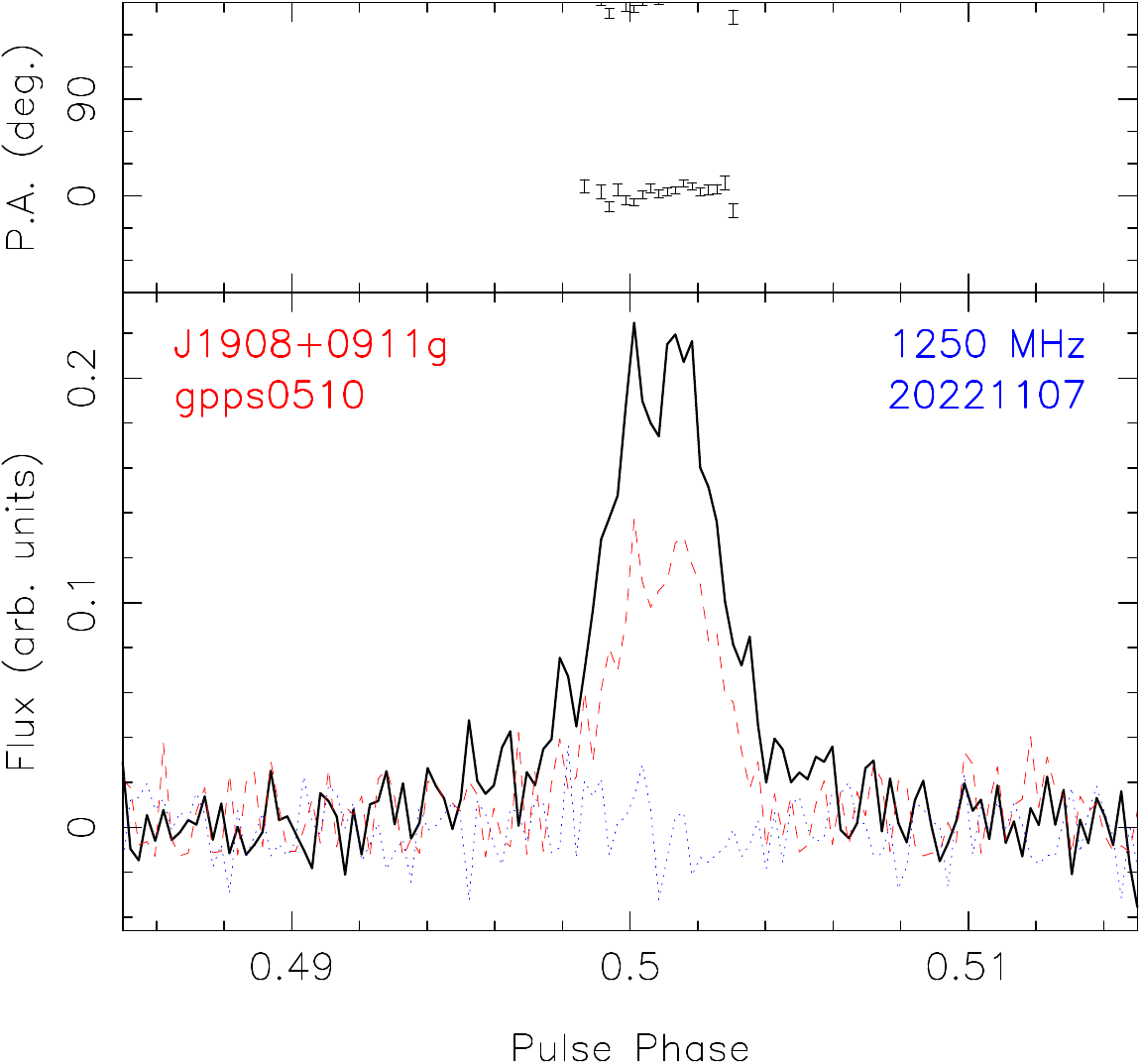} 
    \includegraphics[width=0.45\columnwidth]{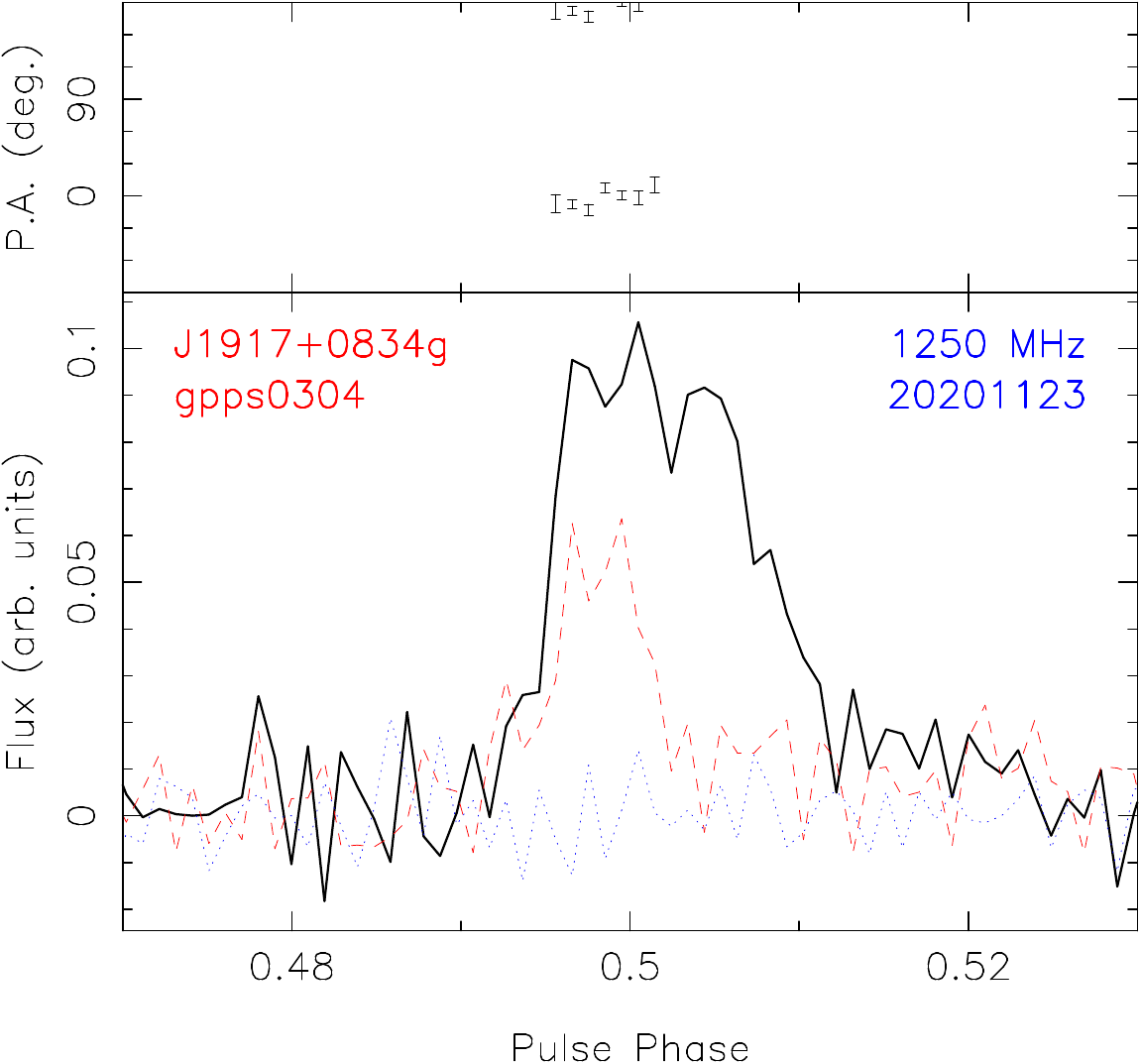} \\[1mm] 
    \includegraphics[width=0.45\columnwidth]{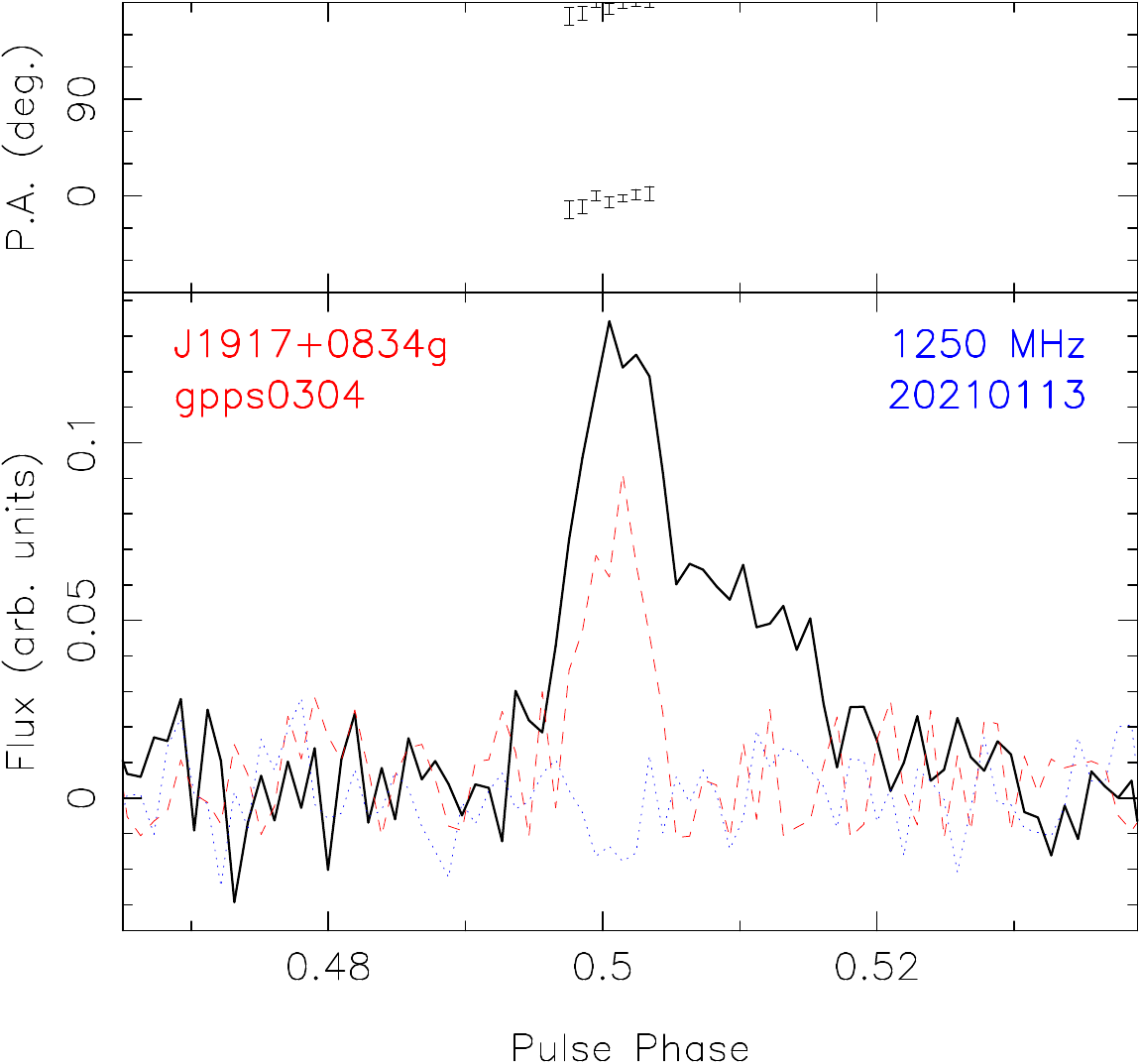} 
    \includegraphics[width=0.45\columnwidth]{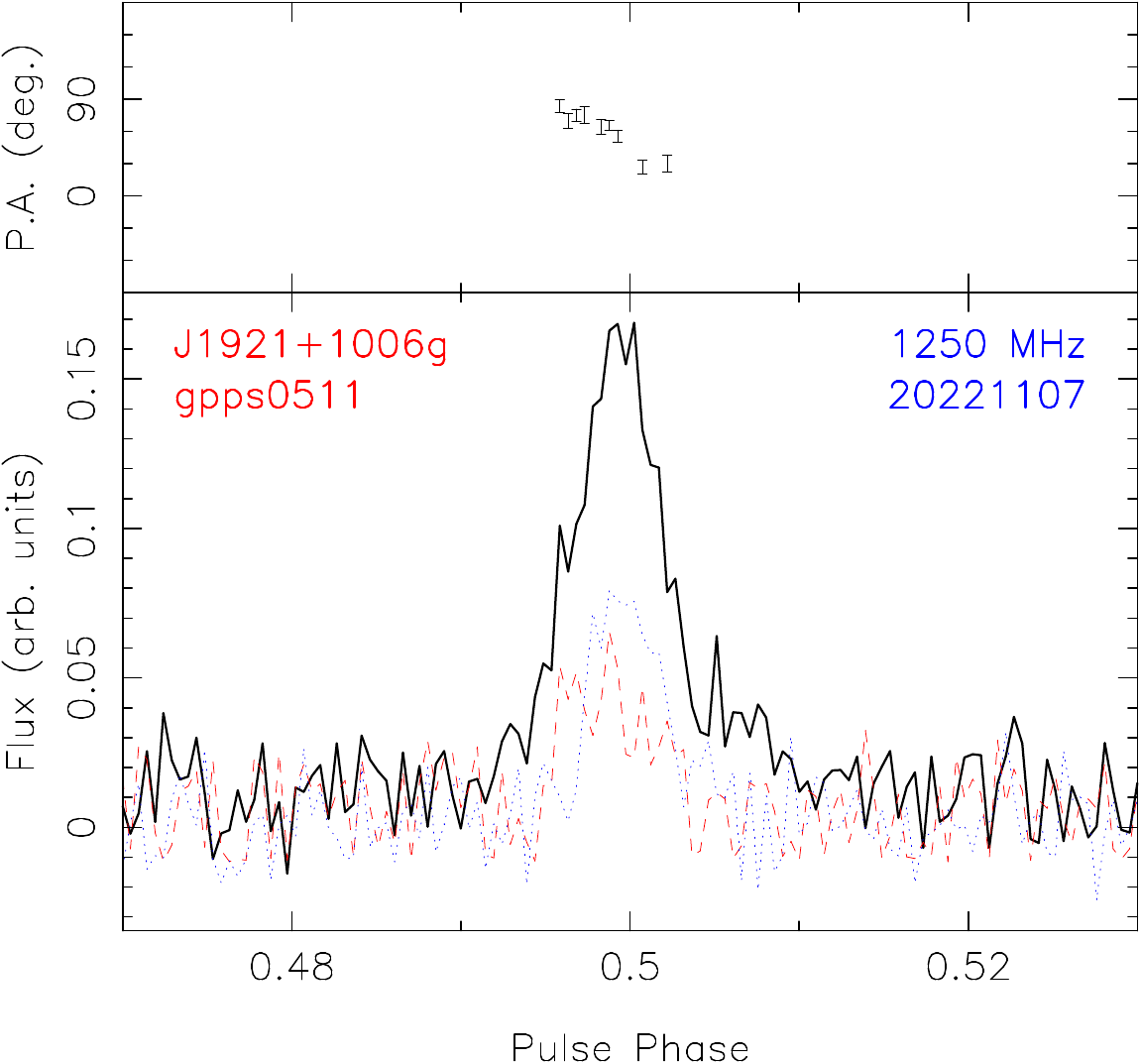} \\[1mm]
    \includegraphics[width=0.45\columnwidth]{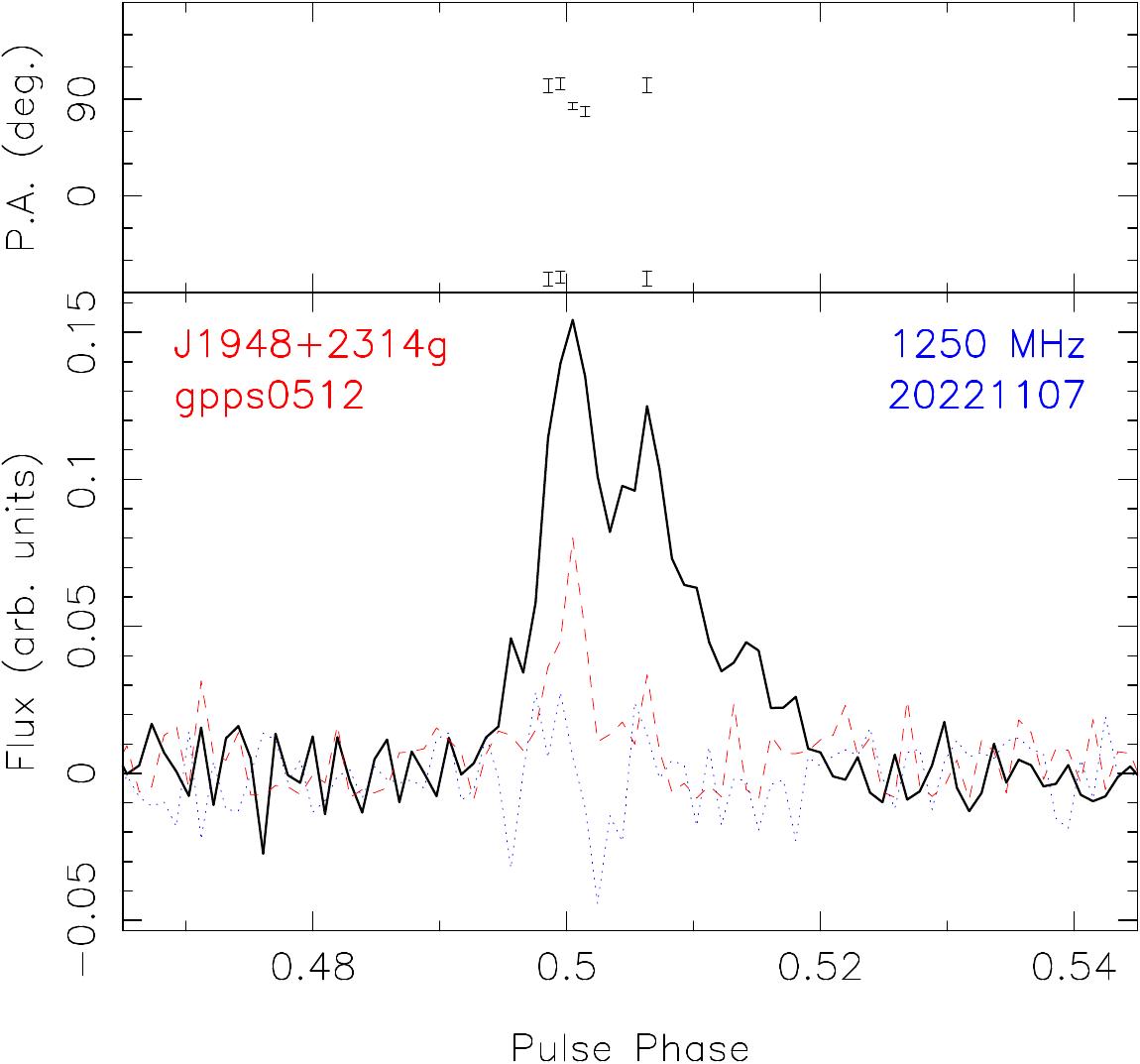} 
    \includegraphics[width=0.45\columnwidth]{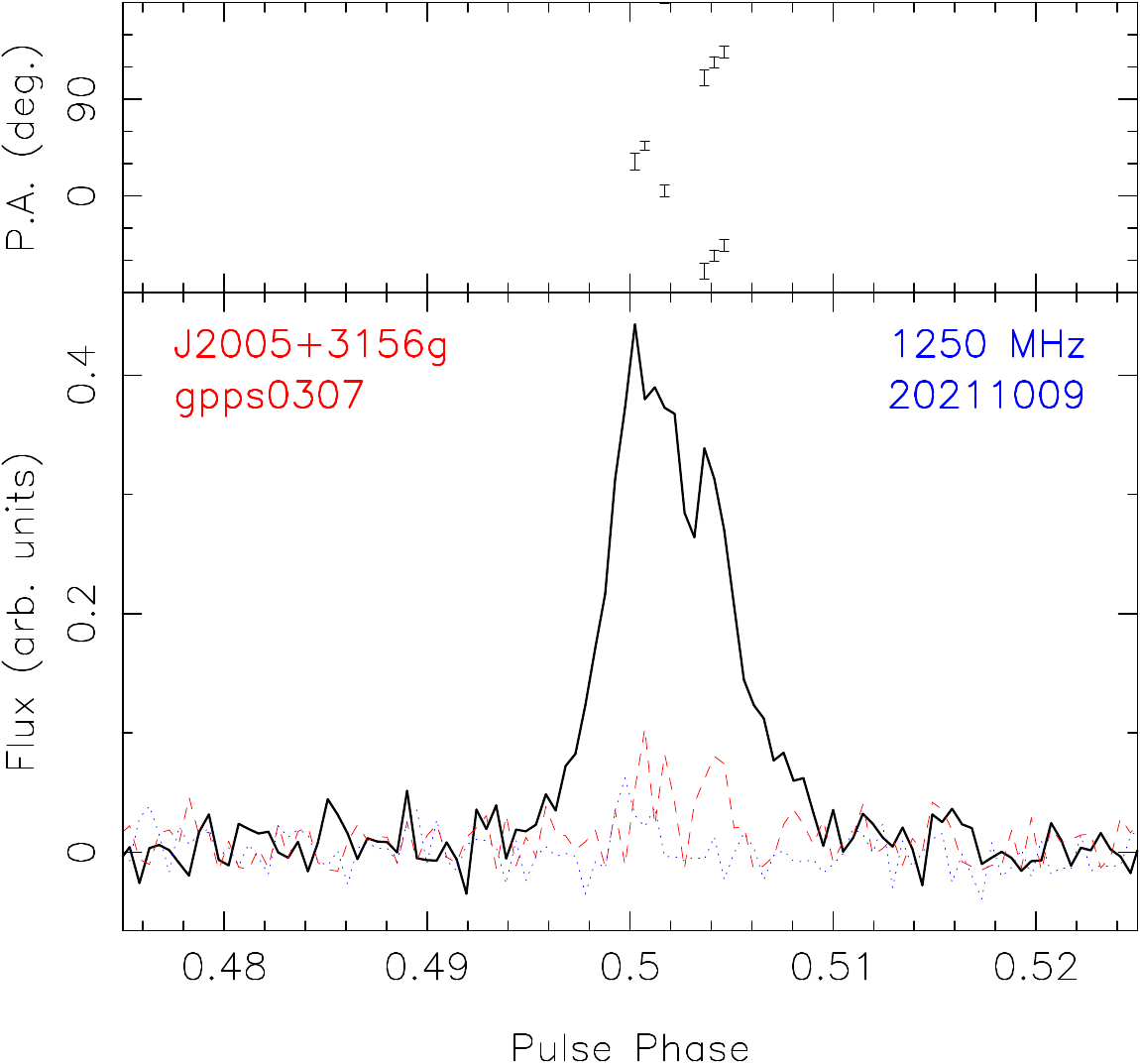}
    \caption{The same as Figure~\ref{fig:NewPol4fewPulse} but for the integrated polarization 
    profiles of the bright pulses of newly discovered RRATs in the GPPS survey.
    }
    \label{fig:NewPol4RRAT}
\end{figure}

\begin{figure}
    \centering
    \includegraphics[width=0.45\columnwidth]{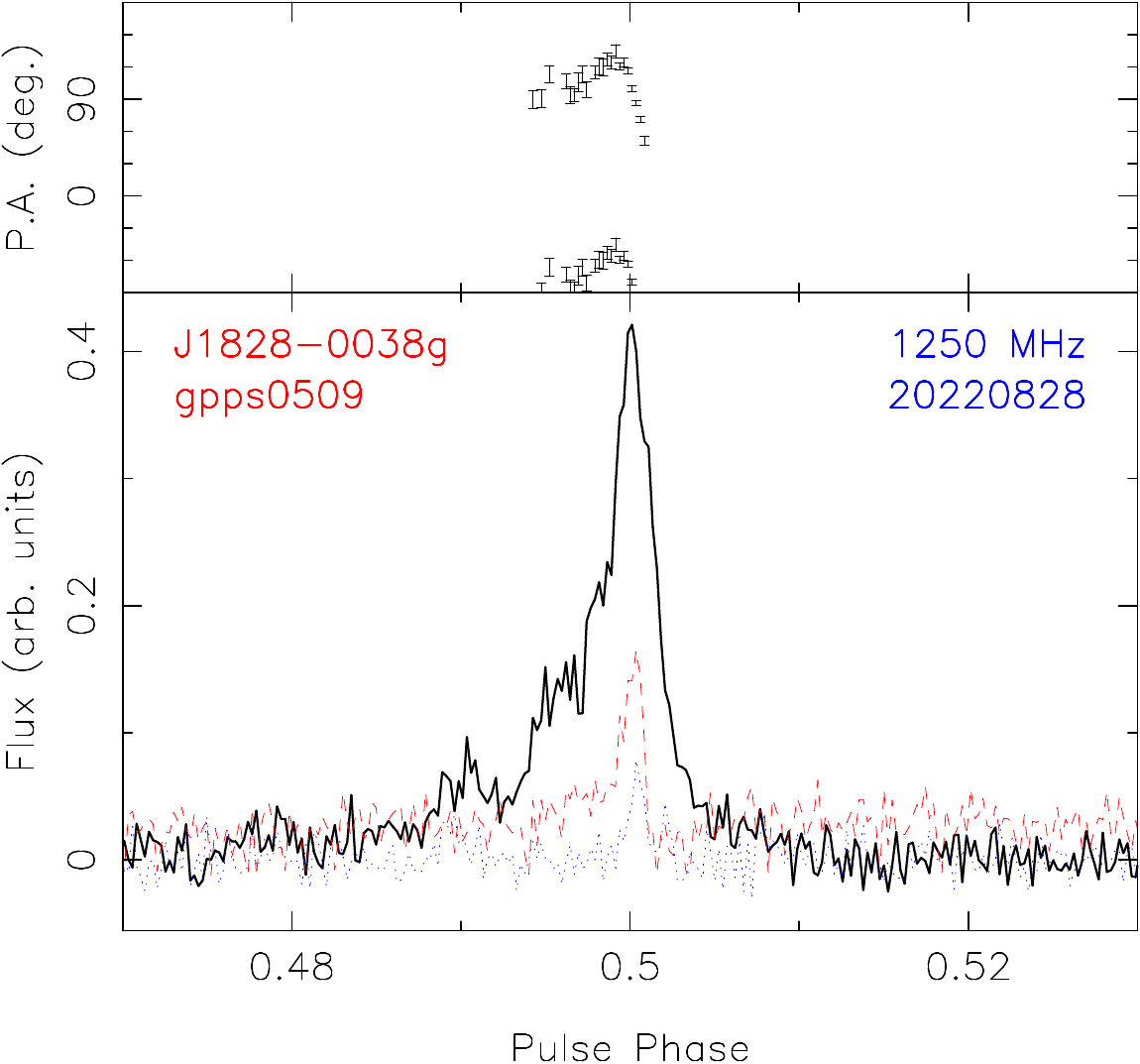} 
    \includegraphics[width=0.45\columnwidth]{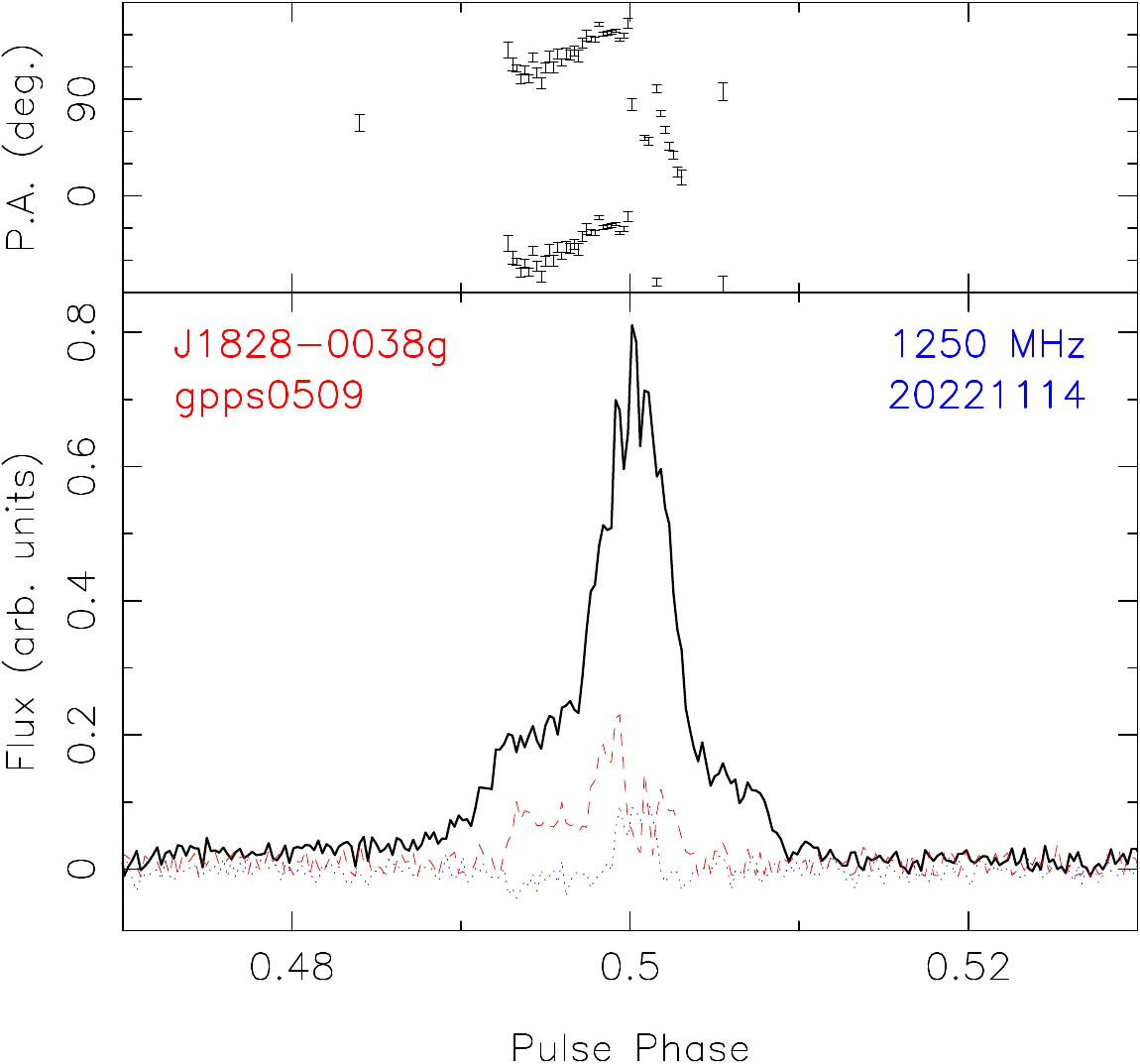} 
    \includegraphics[width=0.45\columnwidth]{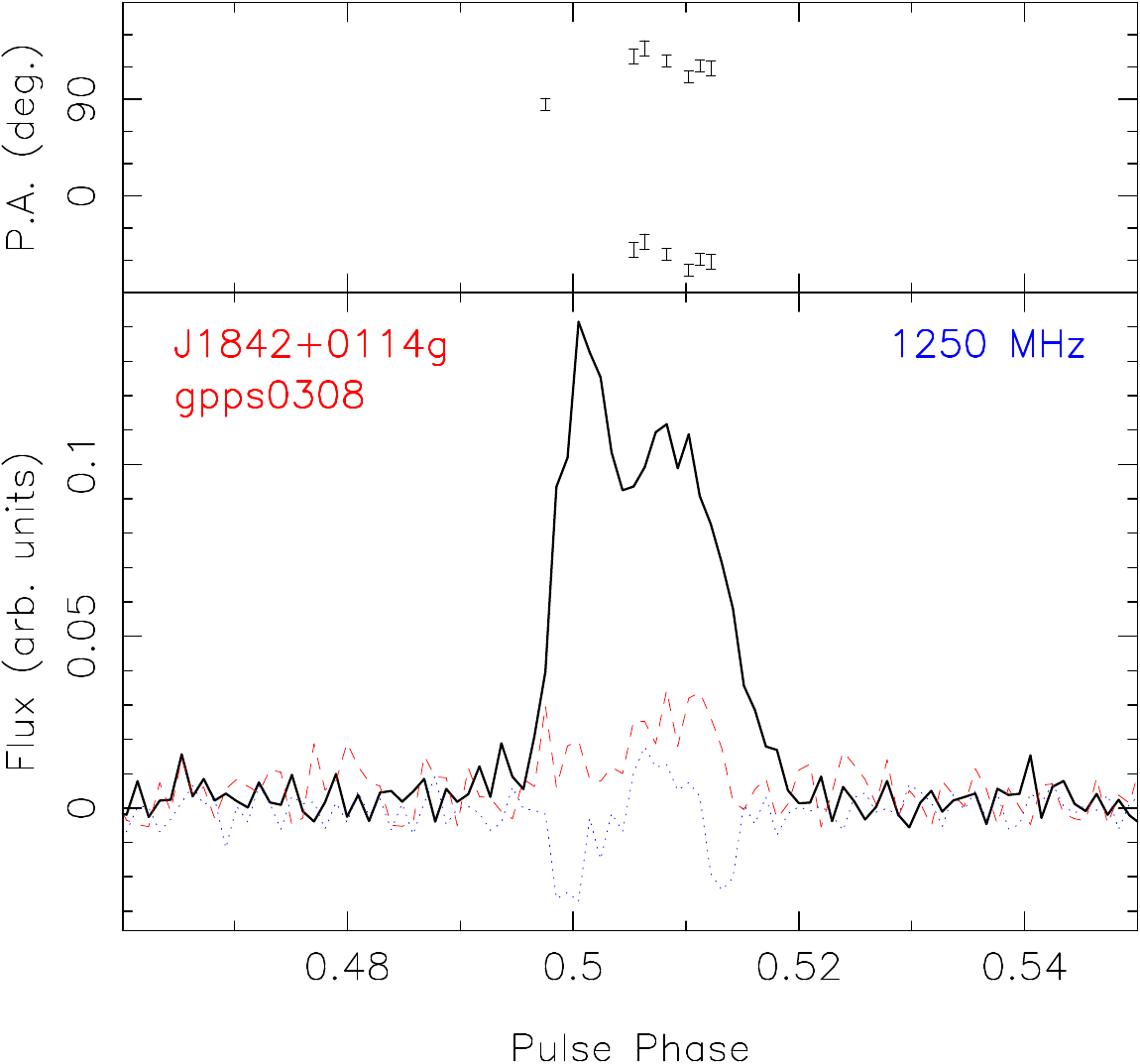} %SP040
    \includegraphics[width=0.45\columnwidth]{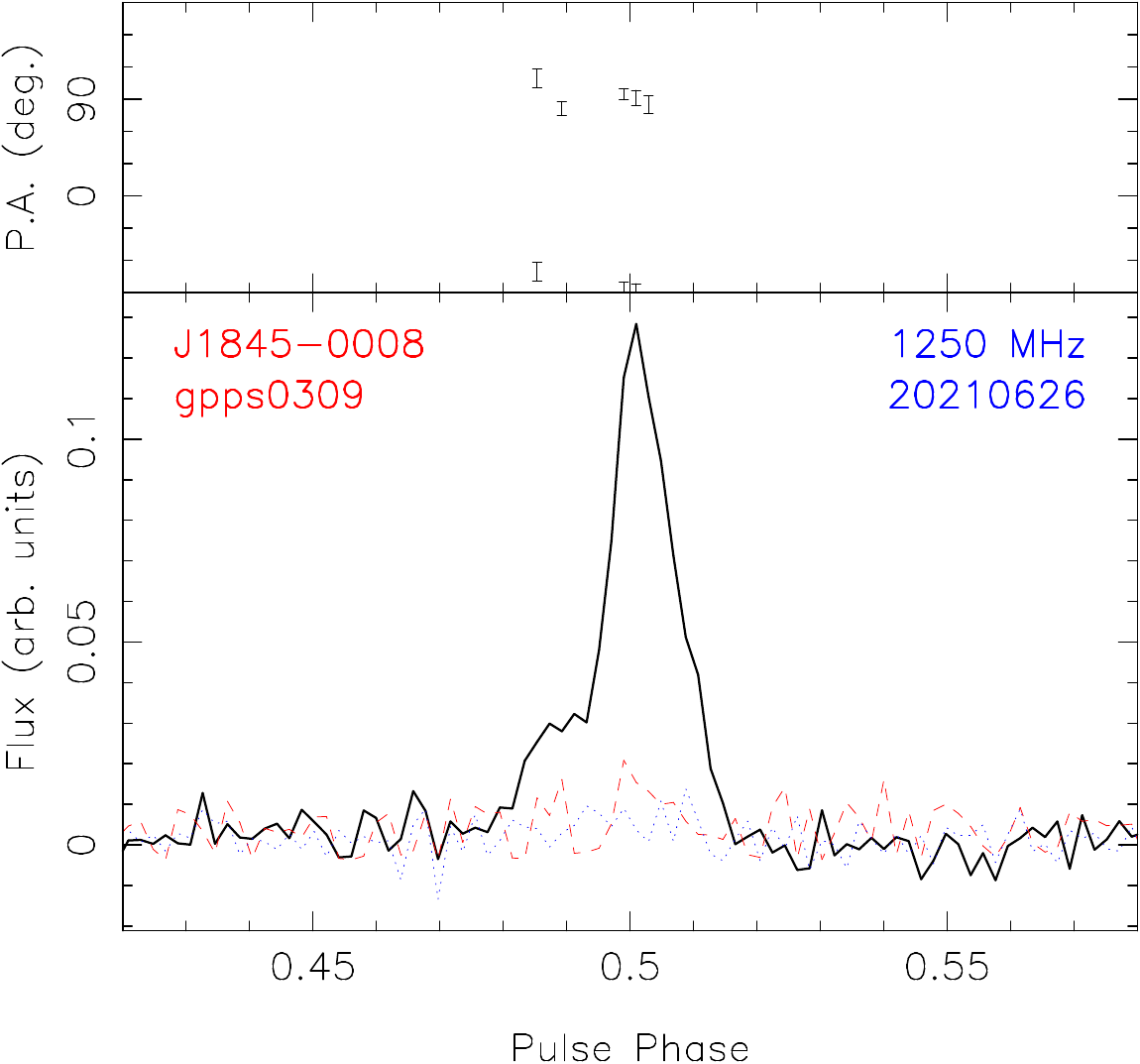}\\[1mm] %SP009
    \includegraphics[width=0.45\columnwidth]{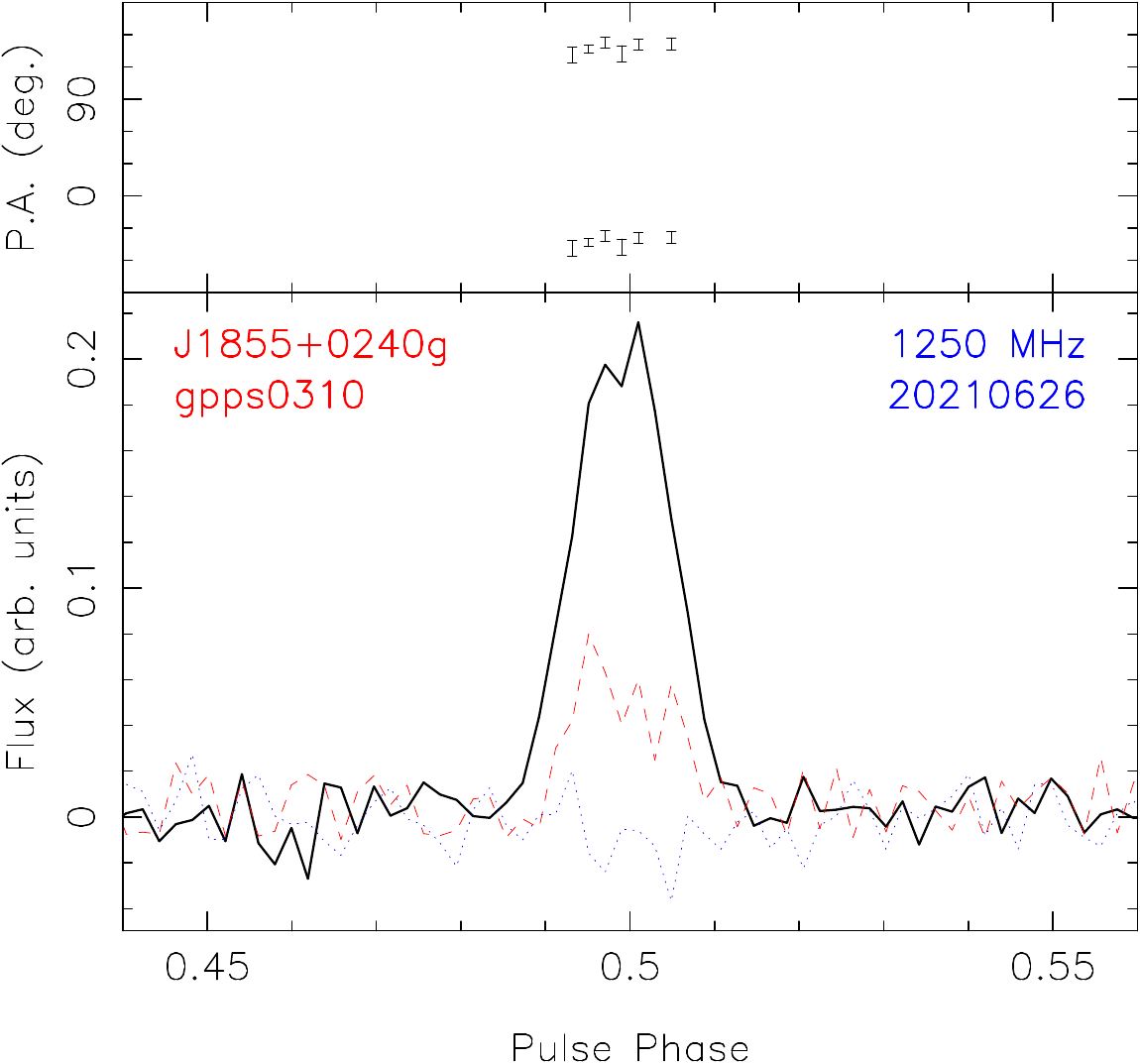} %SP004
    \includegraphics[width=0.45\columnwidth]{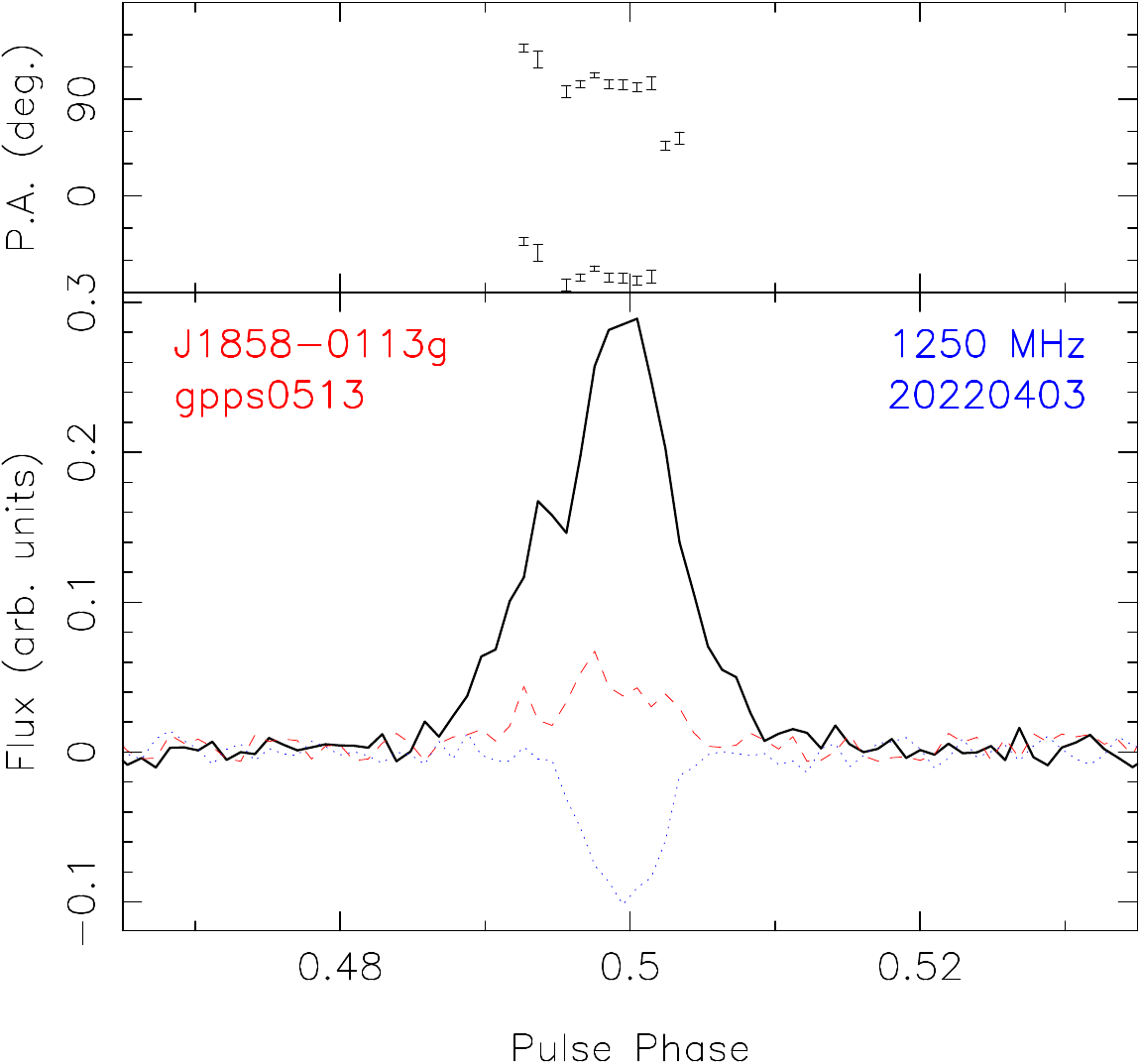} \\[1mm]%SP080
    \includegraphics[width=0.4\columnwidth]{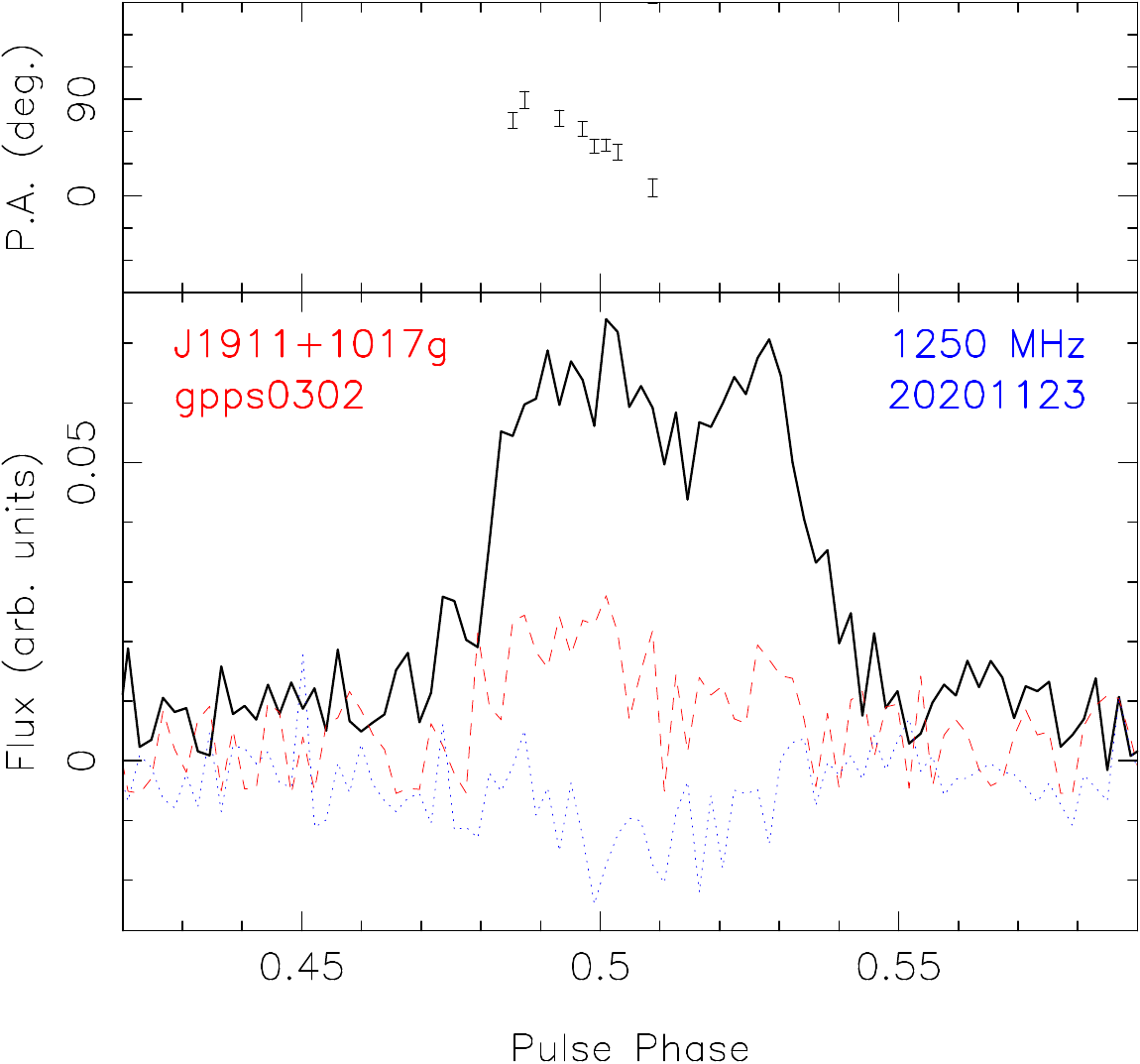} 
    \includegraphics[width=0.45\columnwidth]{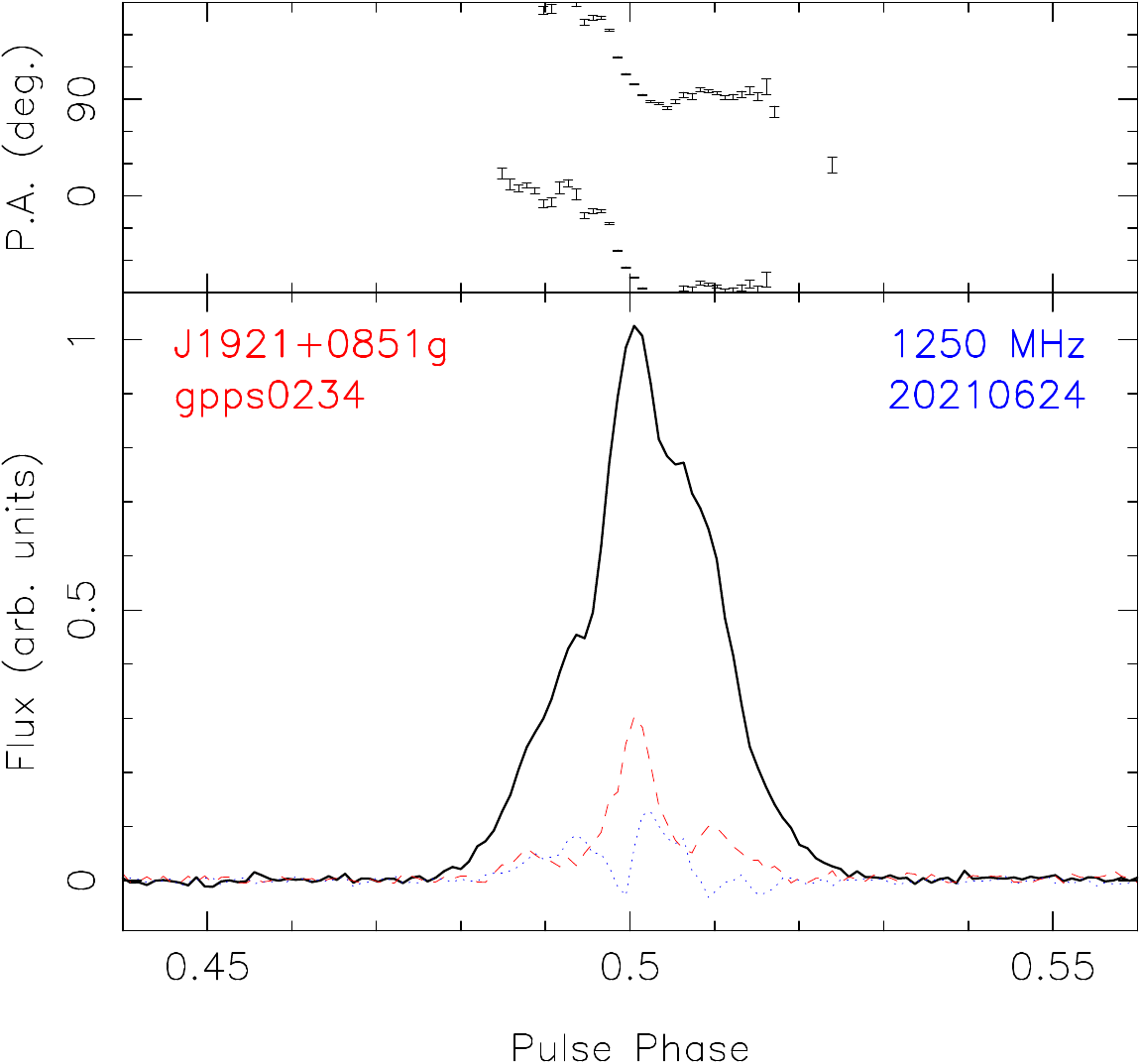} %SP049
    \includegraphics[width=0.45\columnwidth]{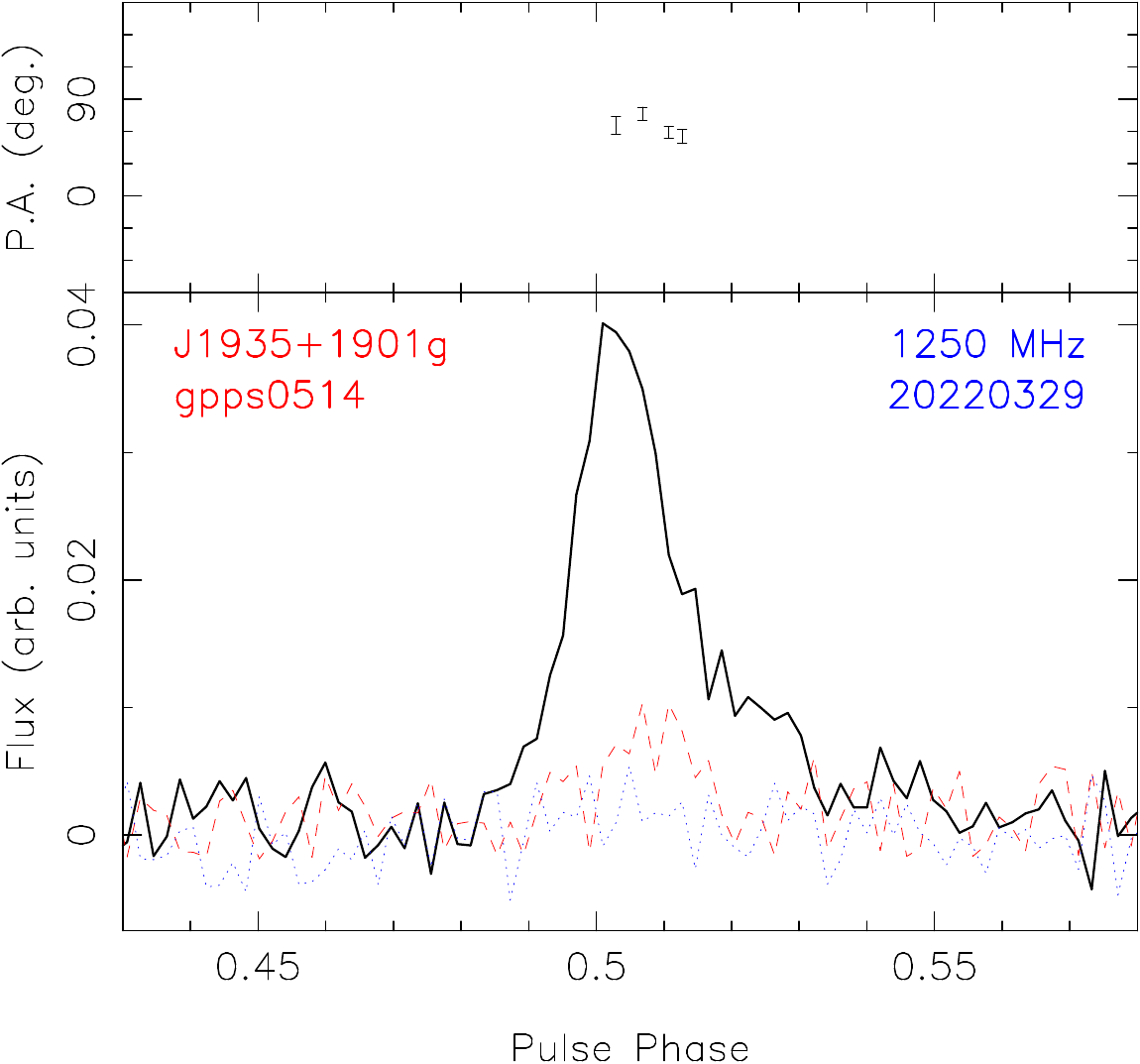}%SP076
    \caption{The same as Figure~\ref{fig:NewPol4fewPulse} but for the integrated polarization 
    profiles of the bright pulses of newly discovered extremely nulling pulsars  in the GPPS survey.
    }
    \label{fig:NewPol4NullingPulsars}
\end{figure}

\begin{figure}[!thp]
    \centering
    \includegraphics[width=0.45\columnwidth]{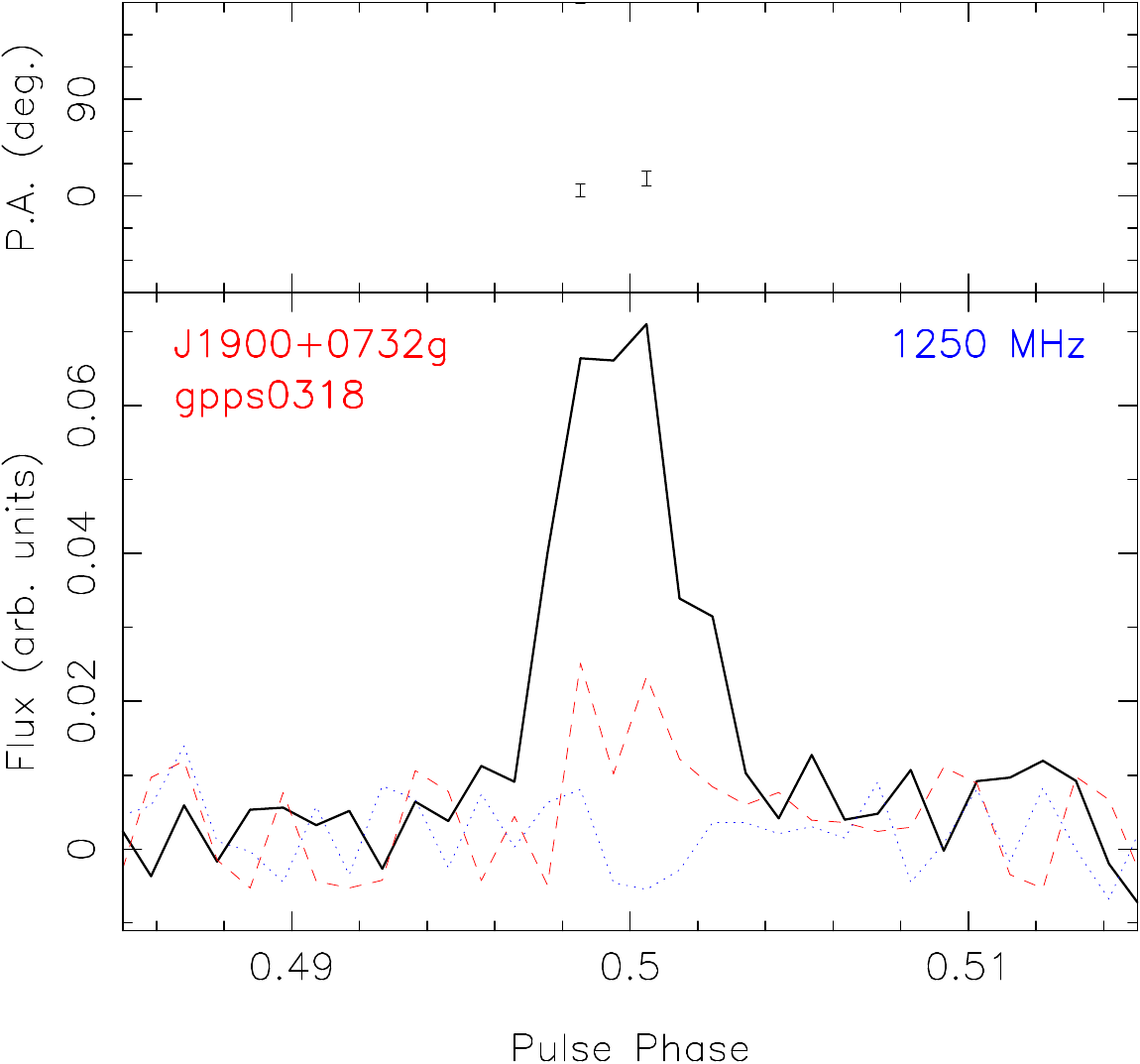} %SP048
    \includegraphics[width=0.45\columnwidth]{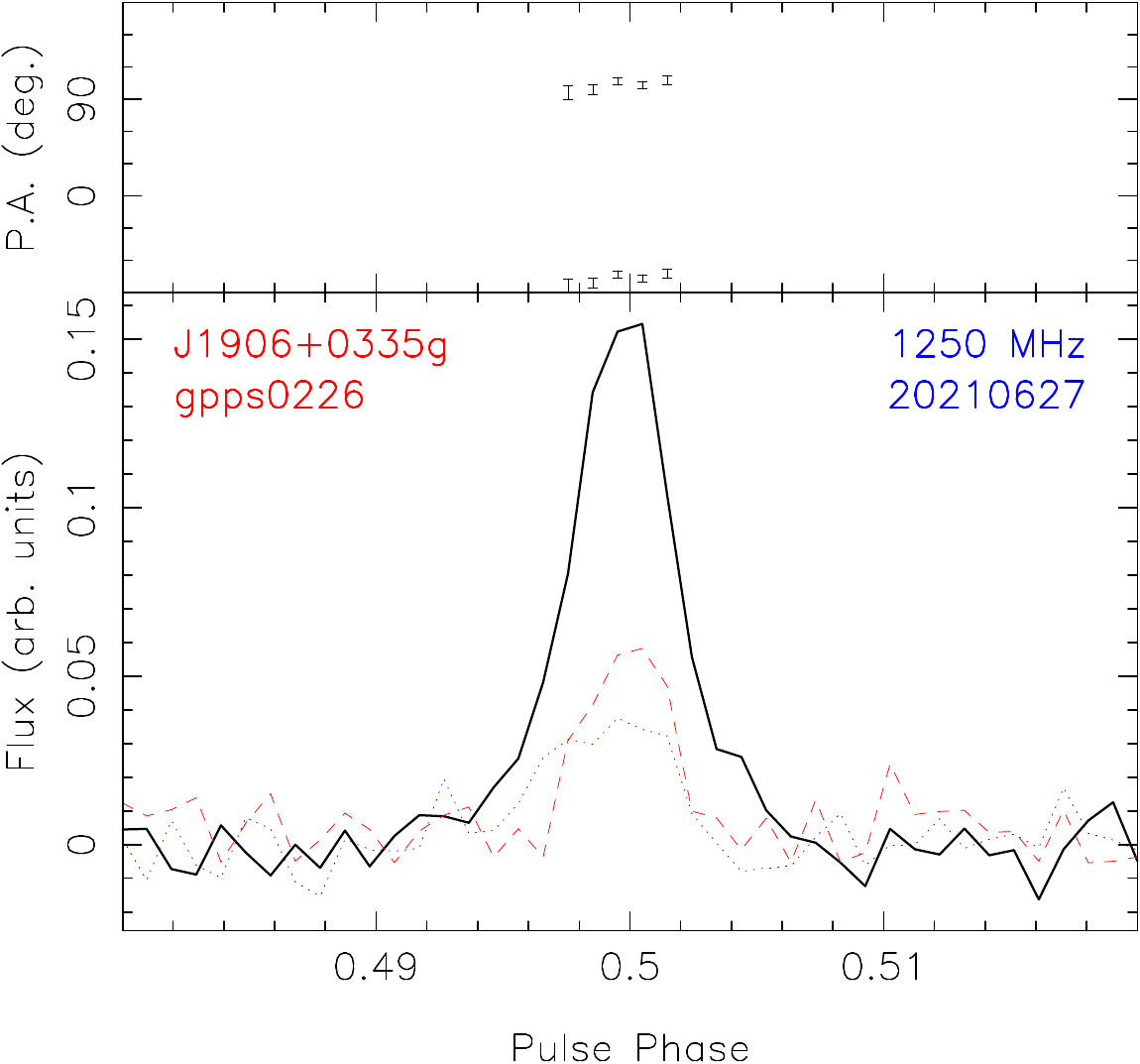} \\[1mm]%SP053
    \includegraphics[width=0.45\columnwidth]{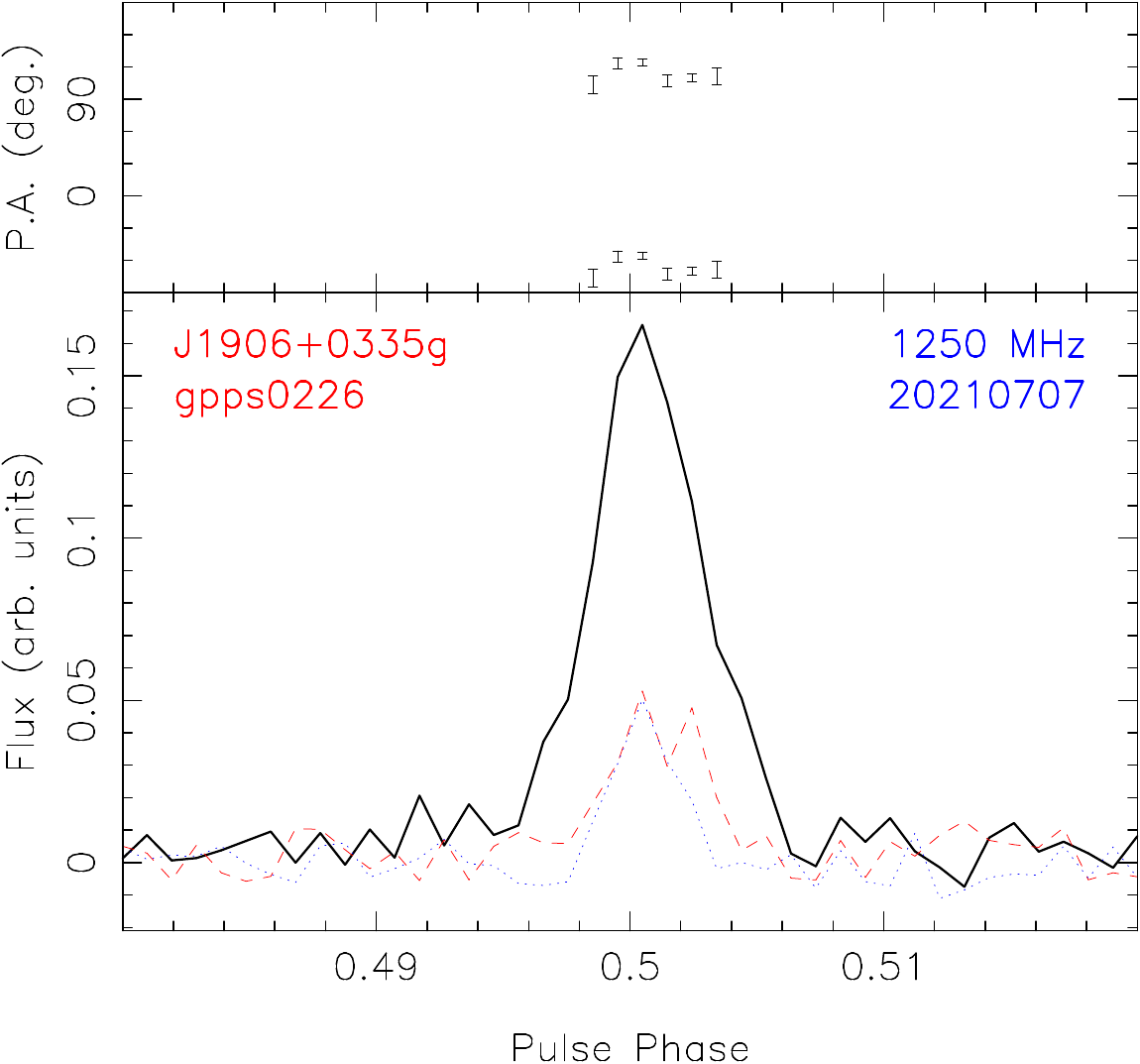} %SP053
    \caption{The same as Figure~\ref{fig:NewPol4fewPulse} but for the integrated polarization 
    profiles of the bright pulses of newly discovered weak pulsars  in the GPPS survey.
    }
    \label{fig:NewPol4WeakPulsars}
\end{figure}

\subsection{Weak pulsars with sparse strong pulses}

For 24 weak pulsars listed in the last part of Table~\ref{tab1}, the single pulse search module can also pick up sparse strong pulses from the 5 minutes snapshot observations. The normal pulsar searching cannot pick up these objects due to the low single-to-noise ratio. They may be detected as a pulsar with the same DM and period from the following-up 15 minutes tracking observations. Figure~\ref{fig:newWeakPulsars} shows one example, and plots for all the 24 weak pulsars are presented in Figure~\ref{fig:AppnewweakPulsars} in the Appendix.

Some newly discovered pulsars are remarkable in some aspects. PSR J1849+0619g (GPPS0522) is discovered via the single pulse search, one pulse in G38.56+3.30-M14P1 on 20221001 and another one pulse in J184950+062304-M04 on 20221111 with the same DM of 110$\pm$1~pc~cm$^{-3}$. Together with the 4 pulses discovered in a recent verification observation on 20230228, the period is found from this observation that is about 2.011230~s via the careful period search at this DM. 
PSR~J1859+0239Bg (GPPS0526) is discovered by only one pulse on 20220529 in the data for timing PSR~J1859+0239Ag \citep[GPPS0091, i.e. PSR~J1859+0239g reported in the paper I][]{Han2021RAA}. This new pulsar has only one pulse detected in each of the three sessions, 20220529, 20221227 and 20230125. Its period was found via the periodicity searching on 20230125 though with a very weak S/N  (see Figure~\ref{fig:AppnewweakPulsars} in Appendix), but not confirmed by the folded profiles with the same period on 20211226, 20230102 and 20230125. 
PSR~J1938+1748g has a period of 7.106~s, while the folded pulse-width is 8.7~ms which means an extremely narrow pulse with a duty cycle of only about 0.082\% or 0.29 degrees in the rotation phase longitude. 
PSR J1956+2911g is also a source with a narrow pulse width of only 3.7~ms, or 0.36 degrees in the rotation phase longitude.
PSR J1921+1227g has a low mean flux density of about 0.28~$\mu$Jy, which is probably one of the weakest pulsars. 
%this value would be lower if single pulse widths are considered that all the single pulse widths are smller than that of averaged profile. 
%
The pulse-stacks of PSR J1843$-$0051g and J1856+0029g show mode-changing with bright wider pulses detected only in some episodes and weaker narrow pulses in other periods. 
%
%Though There is a modulation of the emisison intensity for J1856+0029g. If we observe them with a more sensitive telescope, it may get a clearer picture of the mode change. 
%
They are similar to the case of the known pulsar PSR J1938+2213 with an occasionally bright mode, see Section~\ref{sect5:Conclusions}. 
PSR J1927+1849g was first picked up as 2 single pulses from the single pulse search in the 5 minutes observation, and the two tracking observations show a wide profile with these sparse narrow bright pulses.

%then monitoring it in long time and not only more single pulses detected but detected in the periodicity search module. It is interesting that the detected single pulses are the strong pulse of the pulsar and those single pulse width are less than 10~ms, but for the integrated pulse profile that the pulse width is larger than 40~ms. 
%It may exist two emission modes that one is it most emit weak and wide pulse, and these strong pulses could be another emission model for this pulsar.

\begin{table}
\centering
    {\footnotesize
    \caption{Polarization properties of individual pulses for the newly discovered transient sources.}
    \setlength{\tabcolsep}{4pt}
    \label{tab:NewPol4fewPulse}
    \begin{tabular}{lccccr}
    \hline\noalign{\smallskip}
    Name        & ObsDate:No.    & $L/I$ & $V/I$ &$|V|/I$&RM           \\
                &                & (\%)  & (\%)  & (\%)  &(rad~m$^{-2}$)\\
    \hline
J1828$-$0003g  & 20221114:B1 & 40.5  & -4.1  & 18.4  & 5.7(36) \\
J1850$-$0004g  & 20200415:B1 & 48.0  & -13.2 & 16.4  & -286(10)\\
             & 20200902:B1 & 49.1  & -30.7 & 29.7  & -317(9) \\% SP054   18:49:58.2   -00:03:56
             & 20220608:B1 & 73.4  &  12.3 & 13.1  & -298(1) \\
             & 20220608:B3 & 44.9  & -15.9 & 15.4  & -302(9) \\
             & 20221106:B1 & 38.2  & -3.9  &  0.0  & -286(7) \\
% \multicolumn{6}{r}{RM$_{mean}$ = -298(8)}\\[1mm]
% \cline{4-6}
J1853+0209g& 20220824:B1 & 57.4  & -1.6  & 2.7   & -27(10) \\
J1853+0353g& 20210624:B1 & 63.4  & 2.3   & 14.7  & 381(13) \\% SP050
J1855$-$0211g& 20221205:B1 & 31.5  & 1.8   & 12.0  & 715.9(46)\\
           & 20221205:B2 & 61.1  & 12.1  & 10.0  & 711.8(44)\\
           & 20221205:B3 & 31.3  &-2.2   & 13.5  & 712.1(37)\\
           & 20230305:B1 & 50.3  & 4.9   & 27.9  & 725.3(14)\\
J1855+0033g& 20210328:B2 & 53.5  & 17.5  & 15.3  & 381(9) \\% SP015
J1859+0832g& 20221107:B1 & 80.1  &-8.8   & 8.5   & 1004(2)\\
J1918+0342g& 20211202:B1 & 57.4  & -8.4  & 20.2  & 208(5) \\%SP075
J1921+1629g& 20211004:B1 & 52.7  & 18.5  & 18.1  & 362(4) \\
             & 20211004:B2 & 52.5  & 17.7  & 16.6  & 368(3) \\% SP067   19:21:47.0    +16:29:34
% \multicolumn{6}{r}{RM$_{mean}$ = 365(4)}\\[1mm]
% \cline{4-6}
J1924+1734g& 20211005:B3 & 51.4  & 7.5   & 10.9  &  108(6)\\%SP068
J1934+2341g& 20210624:B5 & 58.5  & 0.4   & 1.4   &  133(3)\\%SP046
J2005+3154g& 20211009:B2 & 54.5  & -70.4 & 70.2  & -132(9)\\%SP071
\hline
\end{tabular}}
\end{table}

\begin{table}
\centering
{\footnotesize
\caption{Polarization properties of new pulsars with bright pulses.}
\label{tab:NewPol4pulsar}
\setlength{\tabcolsep}{3pt}
\begin{tabular}{lcccccr}
\hline\noalign{\smallskip}
Name         & ObsDate  &$W_{50}$&$L/I$&$V/I$&$|V|/I$  & RM          \\
             &          & (ms) & (\%)  & (\%) &(\%)    & (rad~m$^{-2}$)\\
\hline
\multicolumn{7}{c}{RRATs with a good period identified}\\
\hline
J1857+0229g& 20201108 & 5.7  & 37.4  & 8.2  & 6.6    &  34(12)  \\%SP014
J1904+0621g& 20210318 & 9.6  & 22.8  & 4.4  & 7.2    & -238(25) \\%SP042
J1905+0156g& 20210113 &      & 42.7  & 10.4 & 19.8   & -84(5)   \\%SP035
J1905+0558g& 20210627 & 6.6  & 25.0  &-19.8 & 18.0   & 431(20)  \\%SP019
J1908+0911g& 20221107 & 18.9 & 57.0  &-0.7  &  5.0   & 73(2)    \\%SP108
J1917+0834g& 20201123 & 37.2 & 51.0  & 0.0  & 6.8    & 192(3)   \\%SP023
           & 20210113 & 22.9 & 53.3  & -6.8 & 8.5    & 180(4)   \\
J1921+1006g& 20221107 & 19.6 & 38.5  & 36.0 & 34.5   & 489(8)   \\
J1948+2314g& 20221107 & 14.4 & 35.6  & 6.0  & 7.5    & 70(4)    \\%SP100
J2005+3156g& 20211009 & 12.6 & 21.5  & 1.5  & 3.3    &-283(6)   \\%SP070
\hline
\multicolumn{7}{c}{Extremely nulling pulsars} \\
\hline
J1828$-$0038g& 20220828 & 8.3  & 31.6  & 6.4  & 6.6    &  33(2)   \\%SP101
           & 20221114 & 12.4 & 24.9  & 0.8  & 6.5    &  36.7(9) \\
J1842+0114g& 20210317 & 60.7 & 33.0  & 4.9  & 10.6   &119(5)    \\%SP040
J1845$-$0008g& 20210626 & 14.9 & 18.9  & 5.0  & 2.8    &24(9)     \\%SP009
J1855+0240g& 20210626 & 16.7 & 33.1  & -6.6 & 9.3    &99(6)    \\%SP004
J1858$-$0113g& 20220403 & 13.5 & 18.9  & -25.6& 25.5   &673(3)    \\%SP080
J1911+1017g& 20201123 & 70.5 & 38.0  & -19.1& 20.0   & 261(14)  \\%SP008
J1921+0851g& 20210624 & 14.0 & 16.7  & 6.3  & 8.2    &381.3(8)  \\%SP049
J1935+1901g& 20220329 & 14.0 & 31.1  & 5.4  & 2.1    &82(8)     \\%SP076
\hline
\multicolumn{7}{c}{Newly discovered weak pulsars with occasionally strong pulses} \\
\hline
J1900+0732g& 20210626 & 6.7  & 35.3  & 1.9  & 5.3    &372(17)   \\%SP048
J1906+0335g& 20210627 & 6.3  & 37.5  & 26.4 & 26.4   & 203(4)   \\%SP053
           & 20210707 & 6.3  & 27.4  & 19.5 & 19.8   & 205(4)   \\
% J1929+1731g& 20201120 & 50.7 & 62.5  & -9.2 & 8.5    & 177(3)   \\%SP006
\hline
\end{tabular}}
\end{table}

\subsection{Polarization of new sources}
\label{sect3.4}

Fortunately, polarization data are recorded for all tracking observations for verification of the detected single pulses.
Following \citet{Han2021RAA} and \citet{whx+22}, we calibrate the polarization data, and obtain the polarization profiles for some
strong single pulses of the transient sources (see Figure~\ref{fig:NewPol4fewPulse}) and the mean polarization profiles of bright pulses of the newly discovered proto-RRATs, extremely nulling pulsars and weak pulsars (see Figures~\ref{fig:NewPol4RRAT}, \ref{fig:NewPol4NullingPulsars} and \ref{fig:NewPol4WeakPulsars}). 
We derive the polarization properties and also the Faraday rotation measures (RMs) of the bright pulses for the transient sources, as listed in Table~\ref{tab:NewPol4fewPulse}, and of the mean for bright pulses of RRATs, nulling pulsars and weak pulsars as listed in Table~\ref{tab:NewPol4pulsar}. The new measurements of RMs can be used for revealing the interstellar magnetic fields \citep{Han2017ARAA,Han2018ApJS,xhw+22a}.

Some features of these polarization profiles of transient sources are remarkable.
%
%Some sources have few strong pulses detected that they also have the polarization results and their parametersand polarization profiles shown in Figure~\ref{fig:NewPol4fewPulse}.
Pulse No.1 of J1850$-$0004g observed on 20220608 is highly linearly polarized, with the polarization angles (PAs) clearly sweep up. So do the pulse on 20220824 of J1853+0209g and pulse No.1 on 20221107 of J1859+0832g. The PAs in part of the profiles sweep down are seen in the pulse No.1 on 20221114 of J1828$-$0003g, pulse No.3 on 20211005 of J1924+1734g and pulse No.5 on 20210824 of J1934+2341g. Such PA sweeping  is a typical feature of pulsar signals produced in pulsar magnetosphere. Therefore we trust that the pulses of these transient sources are produced by neutron stars, and they are probably just bright pulses of undetectable weaker pulsars, see more discussion of FAST observations of previously known RRATs below.

The polarized profiles of the pulse No.2 on 20221009 of J2005+3154g have a highly circular polarization of $V/I = -0.702$.

As seen in Table~\ref{tab:NewPol4fewPulse} and \ref{tab:NewPol4pulsar}, for pulses detected from one source or observed at different days, their RMs are consistent with each other within uncertainties. The largest RM value is detected for the pulse No.1 on 20221107 of J1859+0832g, which is 1004$\pm$2\,rad~m$^{-2}$ for the pulse with a DM of 259$\pm$2\,pc~cm$^{-3}$. The result is understandable that this object is located near the Galactic plane in the spiral arm tangent where the interstellar magnetic fields could be very strong \citep{Han2017ARAA,Han2018ApJS,xhw+22a}.

No question that the polarization profiles of
the mean bright pulses of RRATs, nulling pulsars and weak pulsars are more or less similar to polarization of normal pulsars \citep{whx+22}. 
\begin{sidewaystable*} !htp
\renewcommand\arraystretch{0.5}
\vspace{16.6cm} 
% \onecolumn
% \begin{landscape*}
% \begin{table*}
\centering
\caption{FAST observations of Previously Known RRATs: Parameters}
%\caption{FAST observations of previously known RRATs and RRAT-like pulsars: parameters}
\label{tab:RRATcat}
\setlength{\tabcolsep}{3.0pt}
\footnotesize
\begin{threeparttable}
\begin{tabular}{lcllrrrlcccrrrrr}
% \begin{tabularhtx}{cclcrrlcccrrrrrr}
\hline%\noalign{\smallskip}
Name     &Ref.  & \multicolumn{1}{c}{Old name} & \multicolumn{1}{c}{FAST ObsDate/MJD: BeamName}  & T$_{\rm obs}$ &\multicolumn{1}{c}{P} & DM       & R.A.(J2000)   & Decl(J2000) & N$_d$/N$_p$ & $\langle S \rangle$ & $W_{50}$   & $L/I$ & $V/I$ &$|V|/I$& RM           \\
   &      &   &               &(min)       &\multicolumn{1}{c}{(ms)}&(cm$^{-3}$pc)&(hh:mm:ss)& (dd:mm)   &            & ($\mu$Jy)        & (ms)      & (\%)  & (\%)  &(\%)   &(rad m$^{-2}$)     \\
(1)  & (2) & \multicolumn{1}{c}{(3)}&  \multicolumn{1}{c}{(4)}& \multicolumn{1}{c}{(5)} & \multicolumn{1}{c}{(6)} & (7) & \multicolumn{1}{c}{(8)}& (9) & (10) & (11) & (12) & (13) & (14) & (15) & (16) \\
\hline
\multicolumn{16}{c}{Sources with no period searched by FAST or just a RRAT:}\\
%\hline
J0534+3425  & [1]      &  J0534+3407      &20221130/59913: G174.26+0.85-M15P1  & 5  &          &24.7!  & 05:34:38*&+34:25*&4/-  &  &  &  &  &  &  \\
J0550+0948  & [2]      &  J0550+09        &20211215/59562: J0550+0900-M02P3    & 5  & 1745!    &86.6!  & 05:50:10*&+09:48!&1/-  &  &  &  &  &  &  \\
J0625+1254  & [3]      &  J0625+12        &20221201/59913: G198.72+0.17-M15P3  & 5  &          &102.8! & 06:25:26*&+12:54*&1/-  &  &  &  &  &  &  \\
J0640+0744  & [1]      &                  &20200102/58850: G204.79+1.19-M11P1  & 5  &          &55.6*  & 06:40:08*&+07:42*&1/-  &  &  &  &  &  &  \\
J1859+0759  & [2]      &  J1859+07        &20221016/59868: G41.01+1.95-M11P1   & 5  &          &303.0! & 18:59:52!&+07:59!&1/-  &  &  &  &  &  &  \\
\hline
\multicolumn{16}{c}{Just a pulsar with nulling features in FAST observations:}\\
%\hline
J0156+0358 & [2]      & J0156+04 &20211210/59558: J0156+0400-M04P3  &  5 & 1359.077*& 27.5! & 01:55:22*&+03:58*& 68/226   &80   & 15.9  &       &       &       &            \\%First P0 
J0302+2252 & [1,4]    & J0301+20 &20210921/59477: J0302+2252-M01P1  & 15 & 1207.073!& 19.0! & 03:02:32!&+22:52!& 732/739  &2142 & 20.2  & 24.2  & -0.3  & 6.4   & -7.20(3)   \\% redo, first discovered and 
J0608+1635 & [1]      &J0609+1635&20220308/59646: G193.34-1.36-M10P3& 5  &  945.853*& 86.9* & 06:08:52*&+16:35!& 190/340  &89   & 18.9  & 14.7  & 10.2  &  9.9  & -106(2)    \\
J0625+1714 & [1]      &J0625+1730&20220308/59646: G194.86+2.12-M15P2& 5  & 2518.400*& 57.9! & 06:25:19!&+17:14*& 37/127   &27   & 27.1  & 42.7  & -1.1  & 7.1   & 107(2)     \\
J0630+1933 & [2,5]    & J0630+19 &20190418/58591: G193.24+4.24-M06P3& 5  & 1248.661!& 47.2* & 06:30:04!&+19:33*& 92/247   &30   & 8.5   &       &       &       &            \\% 9:paper: PS
J1354+2452 & [6,7]    & J1354+24 &20211209/59557: J1354+2400-M01P4  & 5  &  851.001*& 19.8! & 13:54:10*&+24:52*& 37/408   &25   & 5.0   &       &       &       &            \\
J1538+2345 & [6]      &          &20211122/59540: J1538+2345-M01    & 15 & 3449.385!& 14.9! & 15:38:06!&+23:45!& 106/253  &1305 & 65.7  & 47.7  & -5.9  & 6.9   & 9.2(2)     \\% J1538+2345
J1838+0414 & [8]      & J1838+04 &20221213/59926: G35.38+4.74-M15P1 & 5  & 1330.681*& 154.2*& 18:38:24!&+04:14!& 17/213   &19   & 7.8   &       &       &       &            \\
J1843+0118 & [8,9]    & J1843+01 &20201218/59201: G33.23+2.20-M17P2 & 5  & 1267.056*& 248.0!& 18:43:27!&+01:18!& 62/243   &48   & 27.2  & 44.5  & -8.8  & 9.0   & 49(5)      \\
J1843+0527 & [8]      & J1843+05 &20221213/59926: G36.80+4.15-M19P4 & 5  & 2034.918*& 261.1!& 18:43:44*&+05:27*& 33/147   &26   & 24.8  & 22.8  & 0.4   & 10.5  & 221(5)     \\
J1849+0106 & [8,9,10] &          &20211014/59501: J1849+0106sp-M01  & 15 & 1832.183!& 217.2!& 18:49:51 &+01:06!& 256/483  &150  & 7.2   & 31.3  & -5.8  & 6.0   & 107.4(6)   \\
J1850+1532 & [0,11]   & J1850+15 &20210806/59432: G46.83+7.29-M05P4 & 5  & 1383.978*& 22.35*& 18:50:22*&+15:32!& 30/212   &43   & 31.1  & 49.0  & -17.7 & 22.4  & 67(7)      \\
J1853+0427 & [0,8,9]  & J1853+04 &20211004/59491: J1853+0427sp-M01  & 15 & 1320.595*& 549.3!& 18:53.46!&+04:27!& 426/670  &123  & 9.0   & 33.2  & 15.3  & 15.9  & 379.9(5)   \\% 1 : say is pulsar, not RRAT
J1856+0912 & [0,8]    & J1856+09 &20211009/59496: J1856+0912sp-M01  & 15 & 2170.894*& 193.4!& 18:56:35*&+09:12*& 357/409  &155  & 14.8  & 40.1  & 0.9   & 8.7   & 618.3(7)   \\
J1857+0719 & [0,8]    & J1859+07 &20220614/59743: G40.27+2.03-M05P2 & 5  & 1070.639*& 308.1*& 18:57:20*&+07:19*& 3/279    &16   & 23.0  &       &       &       &            \\
J1905+0902 & [0,12,13]&          &20200418/58956: G42.77+1.10-M14P4 & 5  &  218.253!& 433.4!& 19:05:19!&+09:02!& 249/1347 &78   & 3.8   & 20.7  & 1.7   & 2.2   & 520(3)     \\%
J1908+1351 & [8,10]   &          &20211004/59491: J1908+1351sp-M01  & 15 & 3175.110!& 180.4!& 19:08:36!&+13:51!& 104/278  &60   & 31.0  & 27.6  & -8.3  & 10.2  & 634(2)     \\%
J1909+0641 & [0,13,14]&          &20211122/59540: J1909+0641-M01    & 15 &  741.762!&  36.7!& 19:09:29!&+06:41!& 743/1197 &164  & 7.2   & 15.8  & -4.1  & 8.7   & -18(2)     \\% J1909+0641 
J1915+0639 & [8,10]   &          &20211009/59496: J1915+0639sp-M01  & 15 &  644.140!& 212.3!& 19:15:55!&+06:39!& 114/1398 &13   & 5.0   & 20.5  & -10.2 & 10.9  & 201(6)     \\%
J1919+1745 & [0,13,14]&          &20210709/59404: G51.82+2.20-M10P4 & 5  & 2081.343!& 142.3!& 19:19:43!&+17:45!& 100/149  &520  & 14.2  & 41.3  & 2.4   & 3.9   & 520.0(5)   \\%
J1952+3021 & [0,8,10] &          &20211009/59496: J1952+3021sp-M01  & 15 & 1665.763!& 189.8!& 19:52:20!&+30:21!& 324/533  &114  & 24.4  & 10.8  & -3.0  & 2.7   & -8(2)      \\
J1958+3033 & [0,8,10] &          &20211009/59496: J1958+3033sp-M01  & 15 & 1098.646!& 200.3!& 19:58:07!&+30:33!& 242/808  &40   & 5.4   & 16.5  & 1.5   & 7.5   & -15(3)     \\
J2000+2920 & [0,8,10] &          &20211009/59496: J2000+2920sp-M01  & 15 & 3073.983!& 132.5!& 20:00:12!&+29:20!& 233/289  &265  & 15.0  & 22.5  & -4.0  & 5.1   & 63.3(3)    \\ 
J2008+3758 & [25,26]  & J2008+37 &20211225/59573: G74.56+2.80-M19P1 & 5  & 4352.104*& 142.5!& 20:08:03!&+37:57!& 38/70    &95   & 31.9  & 31.8  & 0.5   & 3.4  & 277(2)     \\% J2008+3757, just RRAT-like 
J2033+0042 & [0,11,23,24] &      &20211122/59540: J2033+0042-M01    & 15 & 5013.850!&  37.8!& 20:33:31!&+00:42!& 66/177   &1040 & 75.9  & 15.0  & -2.2  & 4.1  & -71.2(3)   \\
\hline
\multicolumn{16}{c}{Extremely nulling pulsars in FAST observations:} \\
J0103+5354 & [6]       & J0103+54 &20210927/59483: J0103+54-M04P4      & 5  &  354.304!&  55.6!& 01:03:06*&+53:54*& 24/866  &8    & 6.2   & 26.1  & 6.5   & 6.4  & -56(4)     \\
J1717+0305 & [2,5]     & J1717+03 &20211028/59515: J1717+03-M04P1      & 5  & 3901.603*& 25.26!& 17:17:44*&+03:05*& 10/79   &6    & 13.3  & 39.3  & -3.1  & 6.8  & 22(4)      \\% PT2021\_0051
J1720+0040 & [2,5]     & J1720+00 &20211028/59515: J1720+00-M05P1      & 5  & 3356.875*& 46.0! & 17:20:32*&+00:40!& 7/91    &9    & 8.2   & 17.9  & -1.0  & 9.5  & 1(5)       \\% PT2021\_0051
J1839-0141 & [15,16,17]&          &20211122/59540: J1839-0141-M01      & 15 &  933.265!& 293.2!& 18:39:07!&-01:41!& 29/951  &42   & 9.1   & 27.6  & -16.1 & 17.3 & 355.8(3)   \\% J1839-0141
J1928+1725 & [8,10]    &          &20210514/59347: J192904+173105-M04  & 15 &  289.807!& 136.0!& 19:28:52!&+17:25!& 37/3064 &2    & 1.8   & 95.9  & -1.8  & 2.4  & 215.6(5)   \\ 
\hline
\multicolumn{16}{c}{Very weak pulsars with sparse strong pulses in FAST observations:}\\
J0623+1536 & [3]       & J0623+15 &20230213/59988: J062319+153611-M01  & 15 & 2.638545*& 92.7* & 06:23:19*&+15:36*& 18/336  &19   & 16.7  & 46.5  & -7.2  & 11.1 & 28.1(3)    \\%
J0627+1612 & [0,14]    & J0627+16 &20211004/59490: J0627+16sp-M01      & 15 & 2180.066*& 113.0!& 06:27:13!&+16:12!& 43/406  &3    & 2.1   & 78.9  & -4.2  & 0.0  & 166(8)     \\
J0628+0909 & [0,13,14] &          &20210922/59478: J0628+0909-M01P1    & 15 & 1241.425!& 88.3! & 06:28:36!&+09:09!& 42/258  &84   & 6.1   & 31.4  & 15.9  & 16.7 & 129.4(3)   \\ %% rratalog: RM: 124
J1841+0328 & [8]       & J1841+03 &20221213/59926: G34.89+3.73-M17P4   & 5  &  444.629*&153.1* & 18:41:13!&+03:28!& 15/625  &1.6  & 4.3   &       &       &      &            \\
J1846-0257 & [15,18]   &          &20211122/59540: J1846-0257-M01      & 15 & 4477.094!& 237.0!& 18:46:15!&-02:57!& 15/198  &10   & 21.9  &       &       &      &            \\% J1846-0257
J1848+1516 & [1,19,20] &          &20211122/59540: J1848+1516-M01      & 15 & 2233.770!& 77.4! & 18:48:56!&+15:16!& 100/397 &420  & 93.8  & 40.8  & -2.1  & 15.2 & 240.4(6)   \\% J1848+1516
J1854+0306 & [14,21,22]&          &20211122/59540: J1854+0306-M01      & 15 & 4557.820!& 192.4!& 18:54:03!&+03:06!& 54/192  &92   & 20.0  & 52.7  & -2.4  & 3.4  & -43.3(5)   \\%J1854+0306
J1905+0414 & [8]       &          &20210624/59388: J190511+041400-M01  & 15 &  894.124*& 383.0!& 19:05:11!&+04:14!& 41/993  &2    & 41.9  & 63.5  & -4.0  & 11.6 & 1089(3)    \\%
J1913+1330 & [15,18]   &          &20201205/59188: G47.61+1.19-M16P2   & 5  &  923.441!& 175.6!& 19:13:18!&+13:30!& 44/330  &18   & 7.2   &       &       &      &            \\
J1924+1006 & [8]       & J1924+10 &20211202/59550: G45.70-2.46-M11P1   & 5  & 4619.757*& 178.1!& 19:24:29!&+10:06!& 12/65   &9    & 20.3  & 50.1  & -3.2  & 2.7  & 425(4)     \\% J1924+10, P0: 5.281s 
J1929+1155 & [8]       & J1929+11 &20211202/59550: G47.90-2.71-M04P1   & 5  & 3216.892*&  81.2*& 19:29:15*&+11:55*& 7/96    &11   & 20.4  &       &       &      &            \\
J1945+2357 & [14]      & J1946+24 &20220628/59757: J194522+240756-M10  & 23 & 4717.624*&  87.5*& 19:45:48*&+23:57*& 24/292  &7    & 27.6  & 38.0  & 13.3  & 20.3 & 20(8)      \\ 
J2215+4524 & [26]      &          &20211212/59560: G96.58-9.40-M16P3   & 5  & 2723.222!&  18.5!& 22:15:46!&+45:24!& 15/113  &13   & 13.3  &       &       &      &            \\
\hline
%\multicolumn{16}{c}{Normal pulsars with single-pulses similar to known RRATs in FAST observations:}\\
%J1938+2213 & 27,28,29 &          &20200404/58943: G57.93+0.42-M03P3   & 5  &  166.123!& 92.7! & 19:38:14!&+22:13!& 87.1  &47   & 5.8   & 55.8  & 5.1   & 5.2  & 143.4(4)  \\% J1938+2213, nomal pulsar
%J1946+1449 & 1,5      & J1946+14 &20210803/59428: G52.45-5.17-M07P2   & 5  & 2282.332!&  50.6!& 19:46:39*&+14:49*& 25.2  &50   & 14.5  & 16.9  & 3.7   & 7.5  & -54(1)    \\% J1946+1449, not RRAT, just  
\hline
\end{tabular}
\begin{tablenotes}
% \vspace{0.3mm}
\footnotesize
\renewcommand\arraystretch{0.3}
%    \multicolumn{16}{l}{
\item 
    Reference in Column (2):
    [0]  = \url{http://astro.phys.wvu.edu/rratalog/};
    [1]  = \citet{Tyul2018A&A};
    [2]  = \citet{Deneva2016};
    [3]  = \citet{Patel2018};
    [4]  = \citet{Tyul2016ARep,Sanidas2019};
    [5]  = \url{http://www.naic.edu/~deneva/drift-search/};
    [6]  = \citet{KarakoArgaman2015};
    [7]  = \url{http://www.physics.mcgill.ca/~chawlap/GBNCC_RRATs};
    [8]  = \url{http://www.naic.edu/~palfa/newpulsars/};
    [9]  = \citet{Han2021RAA};
    [10]  = \citet{Parent2022ApJ};
    [11] = \citet{BurkeSpolaor2010};
    [12] = \citet{Cordes2006};
    [13] = \citet{Nice2013};
    [14] = \citet{Deneva2009};
    [15] = \citet{Mclaughlin2006};
    [16] = \citet{Cui2017ApJ};
    [17] = \citet{Jiang2017ApJ};
    [18]  = \citet{McLaughlin2009MNRAS};
    [19]  = \citet{Michilli2018MNRAS};
    [20]  = \citet{Michilli2020MNRAS};
    [21]  = \citet{Keane2010MNRAS};
    [22]  = \citet{Keane2011MNRAS};
    [23]  = \citet{Lynch2013ApJ};
    [24]  = \citet{Lower2020MNRAS};
    [25]  = \citet{Dong2021}
    [26]  = \citet{Dong2022arXiv221009172D};
    [27]  = \citet{Chandler2003PhDT};
    [28]  = \citet{Lorimer2013MNRAS};
    [29]  = \citet{Serylak2021MNRAS}. 
    In column (4), observations with a beam name starting with `G' were made by the GPPS survey, and others with "J" by applied FAST PI projects.\\
    $^a$ Newly determined period, DM, and RA. and Decl. by FAST in this paper. All polarization properties and RMs are obtained from the averaged of single-pulses above $3\sigma$. %    \item 
\end{tablenotes}
%\end{table*}
\end{threeparttable}
\end{sidewaystable*}

% J0544+20, J1059-01, J1433+00, J1554+18, J1603+18, J1611-01, J1911+00(McLaughlin2006) not detected
% J0550+0900_20211215_snapshot-M02-P3: a weak pulse

% \newpage
\begin{sidewaystable*} !htp
\renewcommand\arraystretch{0.5}
\vspace{16.6cm} 
% \onecolumn
% \begin{landscape*}
% \begin{table*}
\centering
\caption{FAST observations of Previously Known RRATs: Parameters}
\label{tab:RRATcat}
\setlength{\tabcolsep}{8.0pt}
\footnotesize
\begin{threeparttable}
\begin{tabular}{lccccccccccc}
% \begin{tabularhtx}{cclcrrlcccrrrrrr}
\hline%\noalign{\smallskip}
Name     &Ref.  & \multicolumn{1}{c}{Old name} & \multicolumn{1}{c}{FAST ObsDate/MJD: BeamName}  & T$_{\rm obs}$ &\multicolumn{1}{c}{P} & DM       & R.A.(J2000)   & Decl(J2000) & N$_d$/N$_p$ & $\langle S \rangle$ & $W_{50}$   \\
   &      &   &               &(min)       &\multicolumn{1}{c}{(ms)}&(cm$^{-3}$pc)&(hh:mm:ss)& (dd:mm)   &            & ($\mu$Jy)        & (ms)      \\
(1)  & (2) & \multicolumn{1}{c}{(3)}&  \multicolumn{1}{c}{(4)}& \multicolumn{1}{c}{(5)} & \multicolumn{1}{c}{(6)} & (7) & \multicolumn{1}{c}{(8)}& (9) & (10) & (11) & (12) \\
\hline
\multicolumn{12}{c}{Sources with no period searched by FAST or just a RRAT:}\\
%\hline
J0534+3425  & [1]      &  J0534+3407      &20221130/59913: G174.26+0.85-M15P1  & 5  &          &24.7    & 05:34:38$^a$&+34:25$^a$&4/-  &  &  \\
J0550+0948  & [2]      &  J0550+09        &20211215/59562: J0550+0900-M02P3    & 5  & 1745     &86.6    & 05:50:10$^a$&+09:48    &1/-  &  &  \\
J0625+1254  & [3]      &  J0625+12        &20221201/59913: G198.72+0.17-M15P3  & 5  &          &102.8   & 06:25:26$^a$&+12:54$^a$&1/-  &  &  \\
J0640+0744  & [1]      &                  &20200102/58850: G204.79+1.19-M11P1  & 5  &          &55.6$^a$& 06:40:08$^a$&+07:42$^a$&1/-  &  &  \\
J1859+0759  & [2]      &  J1859+07        &20221016/59868: G41.01+1.95-M11P1   & 5  &          &303.0   & 18:59:52    &+07:59    &1/-  &  &   \\
\hline
\multicolumn{12}{c}{Just a pulsar with nulling features in FAST observations:}\\
%\hline
J0156+0358 & [2]      & J0156+04 &20211210/59558: J0156+0400-M04P3  &  5 & 1359.077$^a$& 27.5     & 01:55:22$^a$&+03:58$^a$& 68/226   &80   & 15.9  \\%First P0 
J0302+2252 & [1,4]    & J0301+20 &20210921/59477: J0302+2252-M01P1  & 15 & 1207.073    & 19.0     & 03:02:32    &+22:52    & 732/739  &2142 & 20.2  \\% redo, first discovered and 
J0608+1635 & [1]      &J0609+1635&20220308/59646: G193.34$-$1.36-M10P3& 5  &  945.853$^a$& 86.9$^a$ & 06:08:52$^a$&+16:35    & 190/340  &89   & 18.9  \\
J0625+1714 & [1]      &J0625+1730&20220308/59646: G194.86+2.12-M15P2& 5  & 2518.400$^a$& 57.9     & 06:25:19    &+17:14$^a$& 37/127   &27   & 27.1  \\
J0630+1933 & [2,5]    & J0630+19 &20190418/58591: G193.24+4.24-M06P3& 5  & 1248.661    & 47.2$^a$ & 06:30:04    &+19:33$^a$& 92/247   &30   & 8.5   \\% 9:paper: PS
J1354+2452 & [6,7]    & J1354+24 &20211209/59557: J1354+2400-M01P4  & 5  &  851.001$^a$& 19.8     & 13:54:10$^a$&+24:52$^a$& 37/408   &25   & 5.0   \\
J1538+2345 & [6]      &          &20211122/59540: J1538+2345-M01    & 15 & 3449.385    & 14.9     & 15:38:06    &+23:45    & 106/253  &1305 & 65.7  \\% J1538+2345
J1838+0414 & [8]      & J1838+04 &20221213/59926: G35.38+4.74-M15P1 & 5  & 1330.681$^a$& 154.2$^a$& 18:38:24    &+04:14    & 17/213   &19   & 7.8   \\
J1843+0118 & [8,9]    & J1843+01 &20201218/59201: G33.23+2.20-M17P2 & 5  & 1267.056$^a$& 248.0    & 18:43:27    &+01:18    & 62/243   &48   & 27.2  \\
J1843+0527 & [8]      & J1843+05 &20221213/59926: G36.80+4.15-M19P4 & 5  & 2034.918$^a$& 261.1    & 18:43:44$^a$&+05:27$^a$& 33/147   &26   & 24.8  \\
J1849+0106 & [8,9,10] &          &20211014/59501: J1849+0106sp-M01  & 15 & 1832.183    & 217.2    & 18:49:51    &+01:06    & 256/483  &150  & 7.2   \\
J1850+1532 & [0,11]   & J1850+15 &20210806/59432: G46.83+7.29-M05P4 & 5  & 1383.978$^a$& 22.35$^a$& 18:50:22$^a$&+15:32    & 30/212   &43   & 31.1  \\
J1853+0427 & [0,8,9]  & J1853+04 &20211004/59491: J1853+0427sp-M01  & 15 & 1320.595$^a$& 549.3    & 18:53.46    &+04:27    & 426/670  &123  & 9.0   \\% 1 : say is pulsar, not RRAT
J1856+0912 & [0,8]    & J1856+09 &20211009/59496: J1856+0912sp-M01  & 15 & 2170.894$^a$& 193.4    & 18:56:35$^a$&+09:12$^a$& 357/409  &155  & 14.8  \\
J1857+0719 & [0,8]    & J1859+07 &20220614/59743: G40.27+2.03-M05P2 & 5  & 1070.639$^a$& 308.1$^a$& 18:57:20$^a$&+07:19$^a$& 3/279    &16   & 23.0  \\
J1905+0902 & [0,12,13]&          &20200418/58956: G42.77+1.10-M14P4 & 5  &  218.253    & 433.4    & 19:05:19    &+09:02    & 249/1347 &78   & 3.8   \\%
J1908+1351 & [8,10]   &          &20211004/59491: J1908+1351sp-M01  & 15 & 3175.110    & 180.4    & 19:08:36    &+13:51    & 104/278  &60   & 31.0  \\%
J1909+0641 & [0,13,14]&          &20211122/59540: J1909+0641-M01    & 15 &  741.762    &  36.7    & 19:09:29    &+06:41    & 743/1197 &164  & 7.2   \\% J1909+0641 
J1915+0639 & [8,10]   &          &20211009/59496: J1915+0639sp-M01  & 15 &  644.140    & 212.3    & 19:15:55    &+06:39    & 114/1398 &13   & 5.0   \\%
J1919+1745 & [0,13,14]&          &20210709/59404: G51.82+2.20-M10P4 & 5  & 2081.343    & 142.3    & 19:19:43    &+17:45    & 100/149  &520  & 14.2  \\%
J1952+3021 & [0,8,10] &          &20211009/59496: J1952+3021sp-M01  & 15 & 1665.763    & 189.8    & 19:52:20    &+30:21    & 324/533  &114  & 24.4  \\
J1958+3033 & [0,8,10] &          &20211009/59496: J1958+3033sp-M01  & 15 & 1098.646    & 200.3    & 19:58:07    &+30:33    & 242/808  &40   & 5.4   \\
J2000+2920 & [0,8,10] &          &20211009/59496: J2000+2920sp-M01  & 15 & 3073.983    & 132.5    & 20:00:12    &+29:20    & 233/289  &265  & 15.0  \\ 
J2008+3758 & [25,26]  & J2008+37 &20211225/59573: G74.56+2.80-M19P1 & 5  & 4352.104$^a$& 142.5    & 20:08:03    &+37:57    & 38/70    &95   & 31.9  \\% J2008+3757, just RRAT-like 
J2033+0042 & [0,11,23,24] &      &20211122/59540: J2033+0042-M01    & 15 & 5013.850    &  37.8    & 20:33:31    &+00:42    & 66/177   &1040 & 75.9  \\
\hline
\multicolumn{12}{c}{Extremely nulling pulsars in FAST observations:} \\
J0103+5354 & [6]       & J0103+54 &20210927/59483: J0103+54-M04P4      & 5  &  354.304    &  55.6 & 01:03:06$^a$&+53:54$^a$& 24/866  &8    & 6.2   \\
J1717+0305 & [2,5]     & J1717+03 &20211028/59515: J1717+03-M04P1      & 5  & 3901.603$^a$& 25.26 & 17:17:44$^a$&+03:05$^a$& 10/79   &6    & 13.3  \\% PT2021\_0051
J1720+0040 & [2,5]     & J1720+00 &20211028/59515: J1720+00-M05P1      & 5  & 3356.875$^a$& 46.0  & 17:20:32$^a$&+00:40    & 7/91    &9    & 8.2   \\% PT2021\_0051
J1839$-$0141 & [15,16,17]&          &20211122/59540: J1839$-$0141-M01    & 15 &  933.265    & 293.2 & 18:39:07    &$-$01:41    & 29/951  &42   & 9.1   \\% J1839-0141
J1928+1725 & [8,10]    &          &20210514/59347: J192904+173105-M04  & 15 &  289.807    & 136.0 & 19:28:52    &+17:25    & 37/3064 &2    & 1.8   \\ 
\hline
\multicolumn{12}{c}{Very weak pulsars with sparse strong pulses in FAST observations:}\\
J0623+1536 & [3]       & J0623+15 &20230213/59988: J062319+153611-M01  & 15 & 2.638545$^a$& 92.7$^a$ & 06:23:19$^a$&+15:36$^a$& 18/336  &19   & 16.7  \\%
J0627+1612 & [0,14]    & J0627+16 &20211004/59490: J0627+16sp-M01      & 15 & 2180.066$^a$& 113.0    & 06:27:13    &+16:12    & 43/406  &3    & 2.1   \\
J0628+0909 & [0,13,14] &          &20210922/59478: J0628+0909-M01P1    & 15 & 1241.425    & 88.3     & 06:28:36    &+09:09    & 42/258  &84   & 6.1   \\ %% rratalog: RM: 124
J1841+0328 & [8]       & J1841+03 &20221213/59926: G34.89+3.73-M17P4   & 5  &  444.629$^a$&153.1$^a$ & 18:41:13    &+03:28    & 15/625  &1.6  & 4.3   \\
J1846$-$0257 & [15,18]   &          &20211122/59540: J1846$-$0257-M01    & 15 & 4477.094    & 237.0    & 18:46:15    &$-$02:57    & 15/198  &10   & 21.9  \\% J1846-0257
J1848+1516 & [1,19,20] &          &20211122/59540: J1848+1516-M01      & 15 & 2233.770    & 77.4     & 18:48:56    &+15:16    & 100/397 &420  & 93.8  \\% J1848+1516
J1854+0306 & [14,21,22]&          &20211122/59540: J1854+0306-M01      & 15 & 4557.820    & 192.4    & 18:54:03    &+03:06    & 54/192  &92   & 20.0  \\%J1854+0306
J1905+0414 & [8]       &          &20210624/59388: J190511+041400-M01  & 15 &  894.124$^a$& 383.0    & 19:05:11    &+04:14    & 41/993  &2    & 41.9  \\%
J1913+1330 & [15,18]   &          &20201205/59188: G47.61+1.19-M16P2   & 5  &  923.441    & 175.6    & 19:13:18    &+13:30    & 44/330  &18   & 7.2   \\
J1924+1006 & [8]       & J1924+10 &20211202/59550: G45.70$-$2.46-M11P1 & 5  & 4619.757$^a$& 178.1    & 19:24:29    &+10:06    & 12/65   &9    & 20.3  \\% J1924+10, P0: 5.281s 
J1929+1155 & [8]       & J1929+11 &20211202/59550: G47.90$-$2.71-M04P1 & 5  & 3216.892$^a$&  81.2$^a$& 19:29:15$^a$&+11:55$^a$& 7/96    &11   & 20.4  \\
J1945+2357 & [14]      & J1946+24 &20220628/59757: J194522+240756-M10  & 23 & 4717.624$^a$&  87.5$^a$& 19:45:48$^a$&+23:57$^a$& 24/292  &7    & 27.6  \\ 
J2215+4524 & [26]      &          &20211212/59560: G96.58$-$9.40-M16P3 & 5  & 2723.222    &  18.5    & 22:15:46    &+45:24    & 15/113  &13   & 13.3  \\
\hline
\hline
\end{tabular}
\begin{tablenotes}
% \vspace{0.3mm}
\footnotesize
\renewcommand\arraystretch{0.3}
%    \multicolumn{16}{l}{
\item 
    Reference in Column (2):
    [0]  = \url{http://astro.phys.wvu.edu/rratalog/};
    [1]  = \citet{Tyul2018A&A};
    [2]  = \citet{Deneva2016};
    [3]  = \citet{Patel2018};
    [4]  = \citet{Tyul2016ARep,Sanidas2019};
    [5]  = \url{http://www.naic.edu/~deneva/drift-search/};
    [6]  = \citet{KarakoArgaman2015};
    [7]  = \url{http://www.physics.mcgill.ca/~chawlap/GBNCC_RRATs};
    [8]  = \url{http://www.naic.edu/~palfa/newpulsars/};
    [9]  = \citet{Han2021RAA};
    [10]  = \citet{Parent2022ApJ};
    [11] = \citet{BurkeSpolaor2010};
    [12] = \citet{Cordes2006};
    [13] = \citet{Nice2013};
    [14] = \citet{Deneva2009};
    [15] = \citet{Mclaughlin2006};
    [16] = \citet{Cui2017ApJ};
    [17] = \citet{Jiang2017ApJ};
    [18]  = \citet{McLaughlin2009MNRAS};
    [19]  = \citet{Michilli2018MNRAS};
    [20]  = \citet{Michilli2020MNRAS};
    [21]  = \citet{Keane2010MNRAS};
    [22]  = \citet{Keane2011MNRAS};
    [23]  = \citet{Lynch2013ApJ};
    [24]  = \citet{Lower2020MNRAS};
    [25]  = \citet{Dong2021}
    [26]  = \citet{Dong2022arXiv221009172D};
    [27]  = \citet{Chandler2003PhDT};
    [28]  = \citet{Lorimer2013MNRAS};
    [29]  = \citet{Serylak2021MNRAS}. 
    In column (4), observations with a beam name starting with `G' were made by the GPPS survey, and others with "J" by applied FAST PI projects.\\
    $^a$ Newly determined period, DM, and RA. and Decl. by FAST in this paper. All polarization properties and RMs are obtained from the averaged of single-pulses above $3\sigma$. %    \item 
\end{tablenotes}
%\end{table*}
\end{threeparttable}
\end{sidewaystable*}

% \newpage
% \begin{sidewaystable*} !htp
% \renewcommand\arraystretch{0.5}
% \vspace{16.6cm} 
% \onecolumn
% \begin{landscape*}
\begin{table}
\centering
\caption{FAST observations of Previously Known RRATs: Polarization Parameters}
%\caption{FAST observations of previously known RRATs and RRAT-like pulsars: parameters}
\label{tab:RRATcat}
\setlength{\tabcolsep}{5.0pt}
\footnotesize
% \begin{threeparttable}
\begin{tabular}{lcccc}
% \begin{tabularhtx}{cclcrrlcccrrrrrr}
\hline%\noalign{\smallskip}
Name     & $L/I$ & $V/I$ &$|V|/I$& RM           \\
   & (\%)  & (\%)  &(\%)   &(rad m$^{-2}$)     \\
\hline
\multicolumn{5}{c}{Just a pulsar with nulling features in FAST observations:}\\
%\hline
J0302+2252 & 24.2  & -0.3  & 6.4   & -7.20(3)   \\% redo, first discovered and 
J0608+1635 & 14.7  & 10.2  &  9.9  & -106(2)    \\
J0625+1714 & 42.7  & -1.1  & 7.1   & 107(2)     \\
J1538+2345 & 47.7  & -5.9  & 6.9   & 9.2(2)     \\% J1538+2345
J1843+0118 & 44.5  & -8.8  & 9.0   & 49(5)      \\
J1843+0527 & 22.8  & 0.4   & 10.5  & 221(5)     \\
J1849+0106 & 31.3  & -5.8  & 6.0   & 107.4(6)   \\
J1850+1532 & 49.0  & -17.7 & 22.4  & 67(7)      \\
J1853+0427 & 33.2  & 15.3  & 15.9  & 379.9(5)   \\% 1 : say is pulsar, not RRAT
J1856+0912 & 40.1  & 0.9   & 8.7   & 618.3(7)   \\
J1905+0902 & 20.7  & 1.7   & 2.2   & 520(3)     \\%
J1908+1351 & 27.6  & -8.3  & 10.2  & 634(2)     \\%
J1909+0641 & 15.8  & -4.1  & 8.7   & -18(2)     \\% J1909+0641 
J1915+0639 & 20.5  & -10.2 & 10.9  & 201(6)     \\%
J1919+1745 & 41.3  & 2.4   & 3.9   & 520.0(5)   \\%
J1952+3021 & 10.8  & -3.0  & 2.7   & -8(2)      \\
J1958+3033 & 16.5  & 1.5   & 7.5   & -15(3)     \\
J2000+2920 & 22.5  & -4.0  & 5.1   & 63.3(3)    \\ 
J2008+3758 & 31.8  & 0.5   & 3.4   & 277(2)     \\% J2008+3757, just RRAT-like 
J2033+0042 & 15.0  & -2.2  & 4.1   & -71.2(3)   \\
\hline
\multicolumn{5}{c}{Extremely nulling pulsars in FAST observations:} \\
J0103+5354 & 26.1  & 6.5   & 6.4  & -56(4)     \\
J1717+0305 & 39.3  & -3.1  & 6.8  & 22(4)      \\% PT2021\_0051
J1720+0040 & 17.9  & -1.0  & 9.5  & 1(5)       \\% PT2021\_0051
J1839$-$0141 & 27.6  & -16.1 & 17.3 & 355.8(3)   \\% J1839-0141
J1928+1725 & 95.9  & -1.8  & 2.4  & 215.6(5)   \\ 
\hline
\multicolumn{5}{c}{Very weak pulsars with sparse strong pulses in FAST observations:}\\
J0623+1536 & 46.5  & -7.2  & 11.1 & 28.1(3)    \\%
J0627+1612 & 78.9  & -4.2  & 0.0  & 166(8)     \\
J0628+0909 & 31.4  & 15.9  & 16.7 & 129.4(3)   \\ %% rratalog: RM: 124
J1848+1516 & 40.8  & -2.1  & 15.2 & 240.4(6)   \\% J1848+1516
J1854+0306 & 52.7  & -2.4  & 3.4  & -43.3(5)   \\%J1854+0306
J1905+0414 & 63.5  & -4.0  & 11.6 & 1089(3)    \\%
J1924+1006 & 50.1  & -3.2  & 2.7  & 425(4)     \\% J1924+10, P0: 5.281s 
J1945+2357 & 38.0  & 13.3  & 20.3 & 20(8) \\     
\hline
\end{tabular}
\end{table}
% \end{threeparttable}
% \end{sidewaystable*}

% J0544+20, J1059-01, J1433+00, J1554+18, J1603+18, J1611-01, J1911+00(McLaughlin2006) not detected
% J0550+0900_20211215_snapshot-M02-P3: a weak pulse

\section{Observations of known RRATs}
\label{sect4:knownRRAT}

%\section{Observations of known RRATs}
% \label{sect4:knownRRAT}
%  Sect.4 Known RRAT

% \input{Tex/tab_knownRRATs.tex}

\begin{table*}[!thp]
\centering
{ 
\caption{Previously known RRATs observed but not detected by FAST observations}
\label{tab:UnconfirmedRRAT}
\footnotesize
\setlength{\tabcolsep}{4.0pt}
\begin{threeparttable}
\begin{tabular}{llcccccrc}
\hline\noalign{\smallskip}
Name      &Ref. & R.A.(J2000)  & Dec(J2000)& P      & DM          & FAST     & FAST beam Name /ObsDate    & T$_{\rm obs}$ \\%& Rate\\
          &     & (hh:mm:ss)& (dd:mm)  & (s)    & (cm$^{-3}$pc) &     Project ID          &                       & (min)         \\%& (hr$^{-1}$)\\
(1)  & (2) & \multicolumn{1}{c}{(3)}&  \multicolumn{1}{c}{(4)}& \multicolumn{1}{c}{(5)} & \multicolumn{1}{c}{(6)} & (7) & \multicolumn{1}{c}{(8)}& (9)  \\

\hline
J0544+20    & [1]    & 05:44:12   & +20:50 &         &56.9       &PT2021\_0151    & J0544+2000-M01 /20211210   & 5    \\%& 4     \\% no P0, 4/hr,1, rratcatlog: J0544-20, 7.5`
J1059$-$01  & [2]    & 10:59!!!   & $-$01:02 &         &18.7       &PT2021\_0151    & J1059$-$0100-M01 /20211209   & 5    \\%&       \\% no P0, , 2/7
J1433+00    & [1]    & 14:33:30   & +00:28 &         &23.5       &PT2021\_0151    & J1433+0000-M01 /20211210   & 5    \\%& 2     \\% no P0, 2, 7.5`
J1554+18    & [1]    & 15:54:17   & +18:04 &         &24.0       &PT2021\_0151    & J1554+1800-M01 /20211210   & 5    \\%& 11    \\% no P0, 11/hr, 1, 7.5
J1603+18    & [1]    & 16:03:34   & +18:51 &0.503    &29.7       &PT2021\_0151    & J1603+1800-M01 /20211210   & 5    \\%& 4     \\% 0.503s, 4/hr, 1, 7.5
J1611$-$01  & [3]    & 16:11!!!   & $-$01:28 &1.297    &27.2       &PT2021\_0151    & J1611$-$0100-M01 /20211210   & 5    \\%& 51    \\% 1.297s, 51/hr, , 36
J1911+00    & [4]    & 19:11:48   & +00:37 &6.94!    &100        &PT2021\_0151    & J1911+0000-M01 /20220126   & 3    \\%& 0.31  \\% 6.94s, 0.31/hr, 4, 1/7
J1912+08    & [5]    & 19:12:57   & +08:29 &         & 96        &GPPS            & G42.91$-$0.85-M02P3 /20200514 & 5    \\
J1917+11    & [6]    & 19:17:01   & +11:42 &         & 38.0      &PT2020\_0155    & J1917+1142-M01 /20210108   & 60    \\
J1928+15    & [5]    & 19:28:20   & +15:13 &         & 242.0     &GPPS            & G50.69$-$1.10-M06P1 /20200528 & 5    \\
J2007+20    & [3]    & 20:07!!!   & +20:21 & 4.634   & 67.0      &GPPS            & G59.74$-$6.44-M07-P3 /20210710 & 5    \\%36` error
\hline
\end{tabular}
\begin{tablenotes}
% \vspace{0.3mm}
%\footnotesize
    \item References in Column (2): % in Column (2):
    [1] = \citet{Deneva2016};
    [2] = \url{http://astro.phys.wvu.edu/rratalog/};
    [3] = \citet{KarakoArgaman2015};
    [4] = \citet{Mclaughlin2006}
    [5] = \url{http://www.naic.edu/~palfa/newpulsars/}
    [6] = \citet{Tyul2018A&A}. In column (8), observations with a beam name starting with `G' were made by the GPPS survey, and others with `J' by applied FAST PI projects.
\end{tablenotes}
\end{threeparttable}
}
\end{table*}

\begin{figure*}
  \centering
  \includegraphics[width=0.24\textwidth]{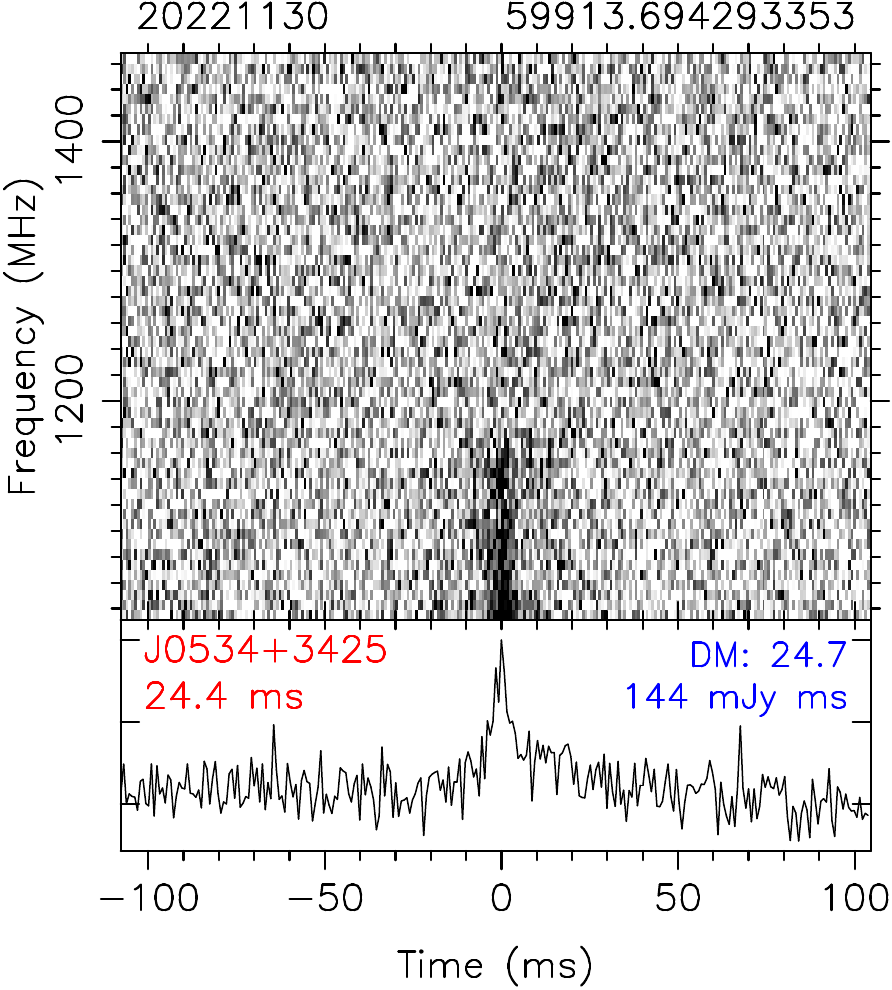}
  \includegraphics[width=0.24\textwidth]{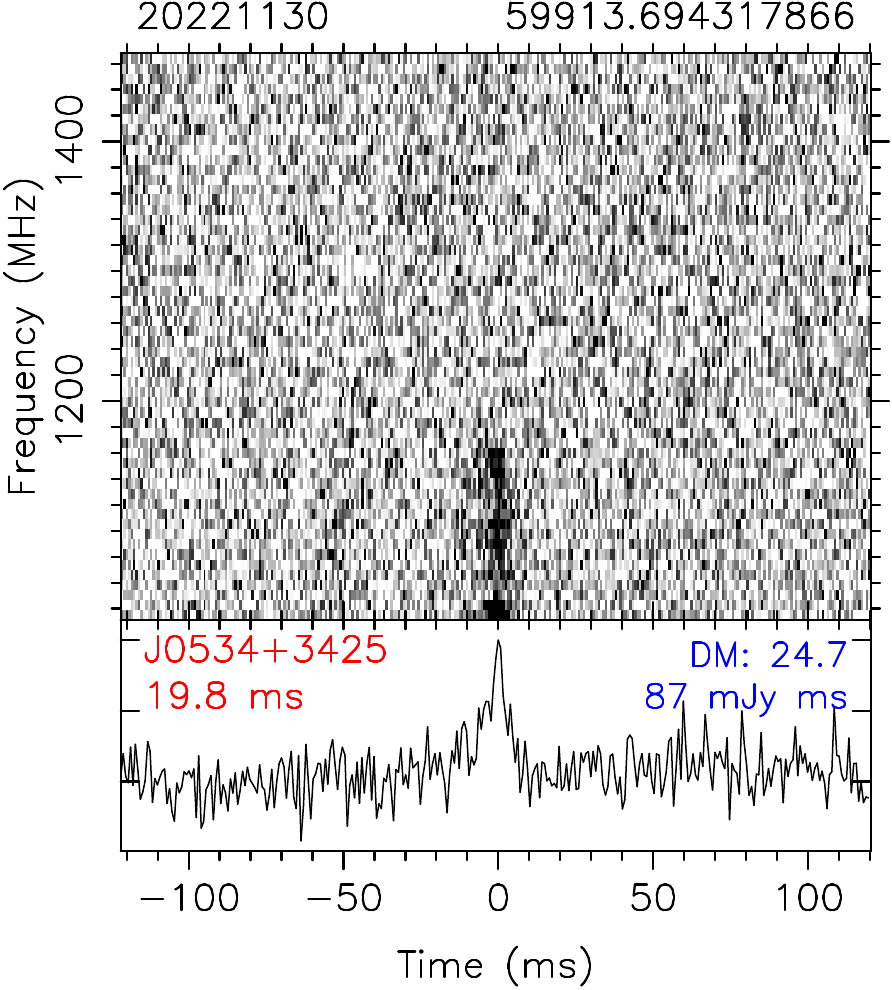}
  \includegraphics[width=0.24\textwidth]{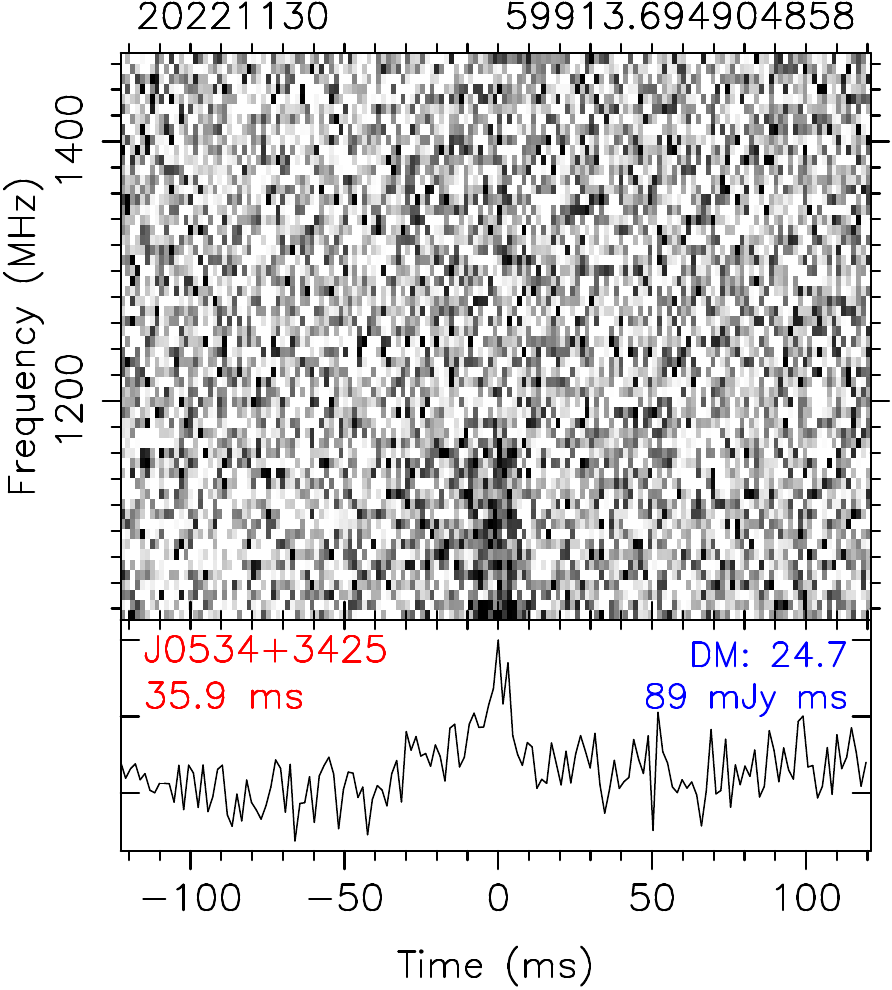}
  \includegraphics[width=0.24\textwidth]{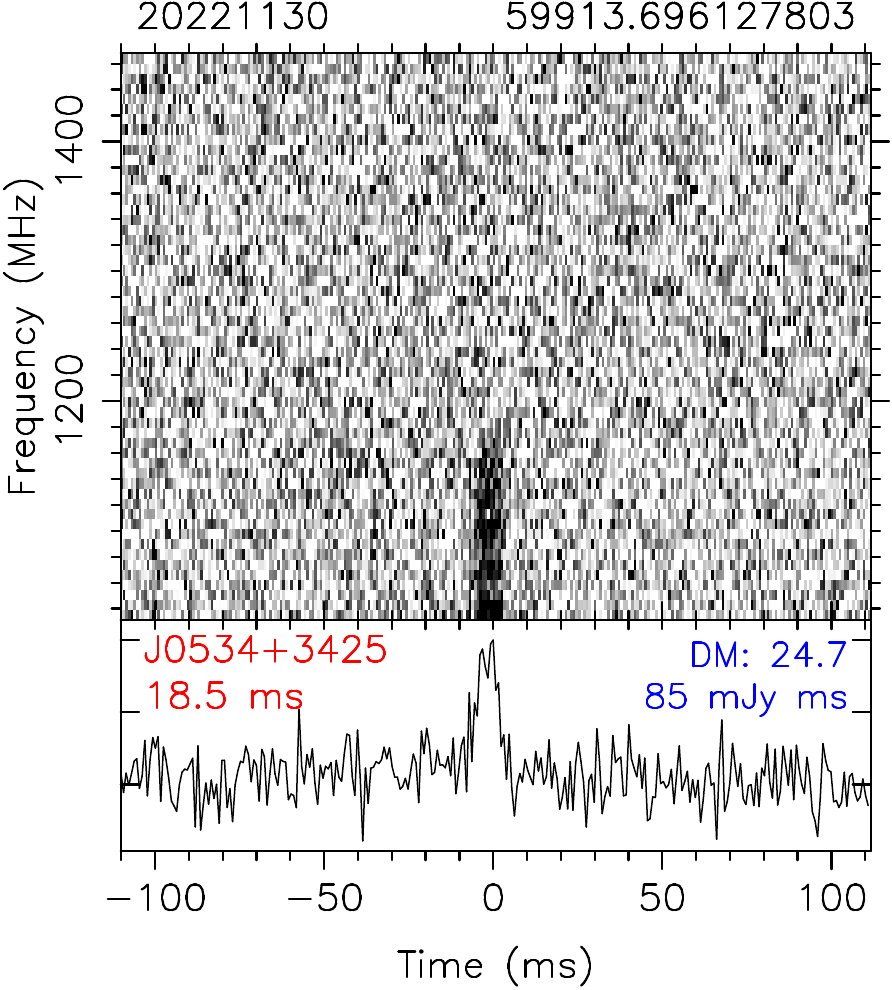} \\[1mm]
  \includegraphics[width=0.24\textwidth]{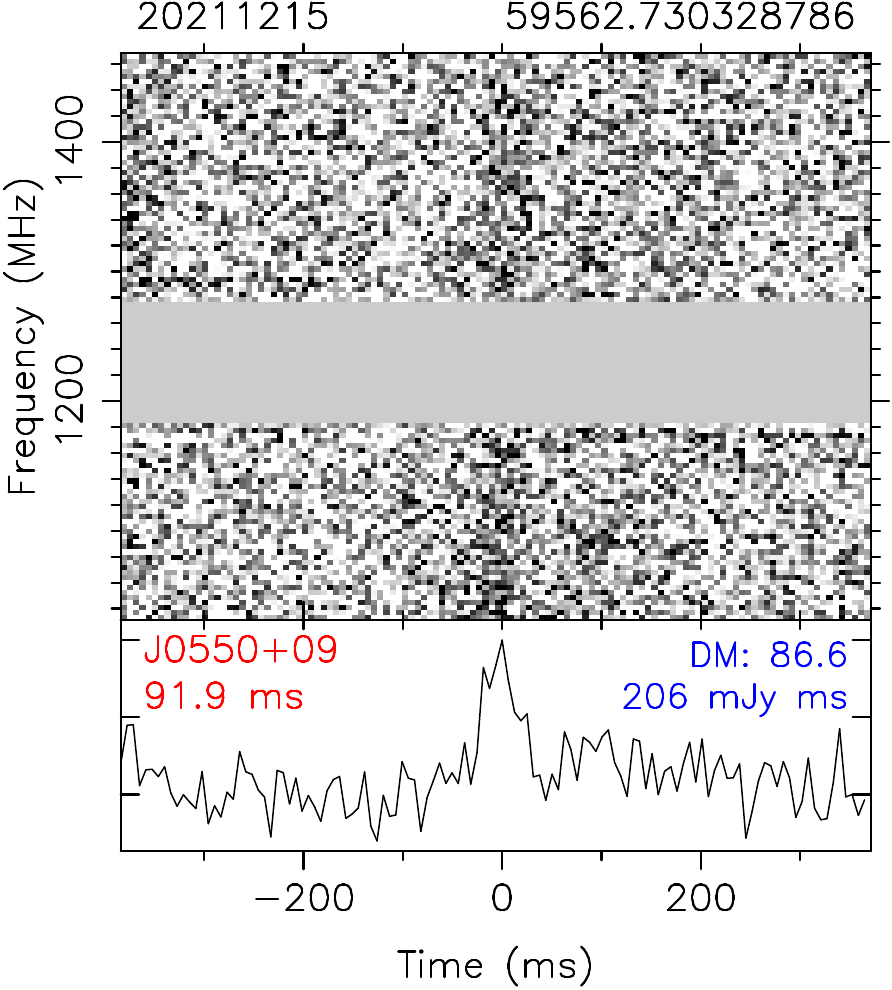}
  \includegraphics[width=0.24\textwidth]{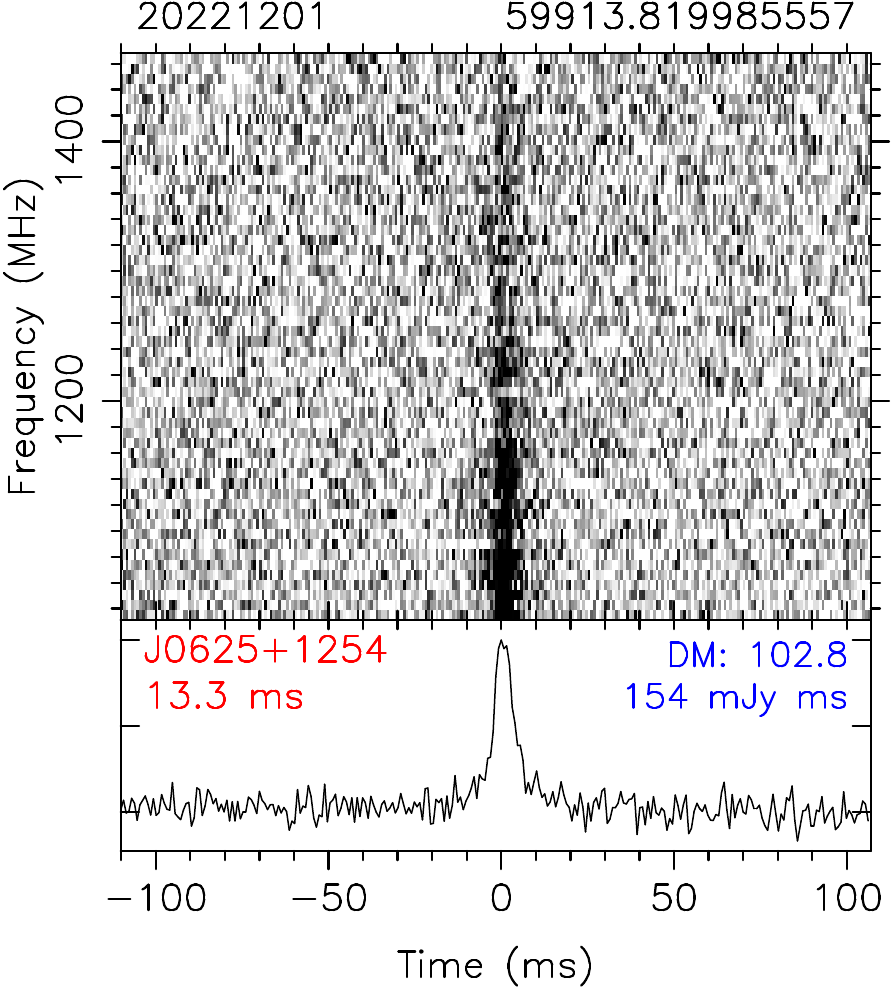}
  \includegraphics[width=0.24\textwidth]{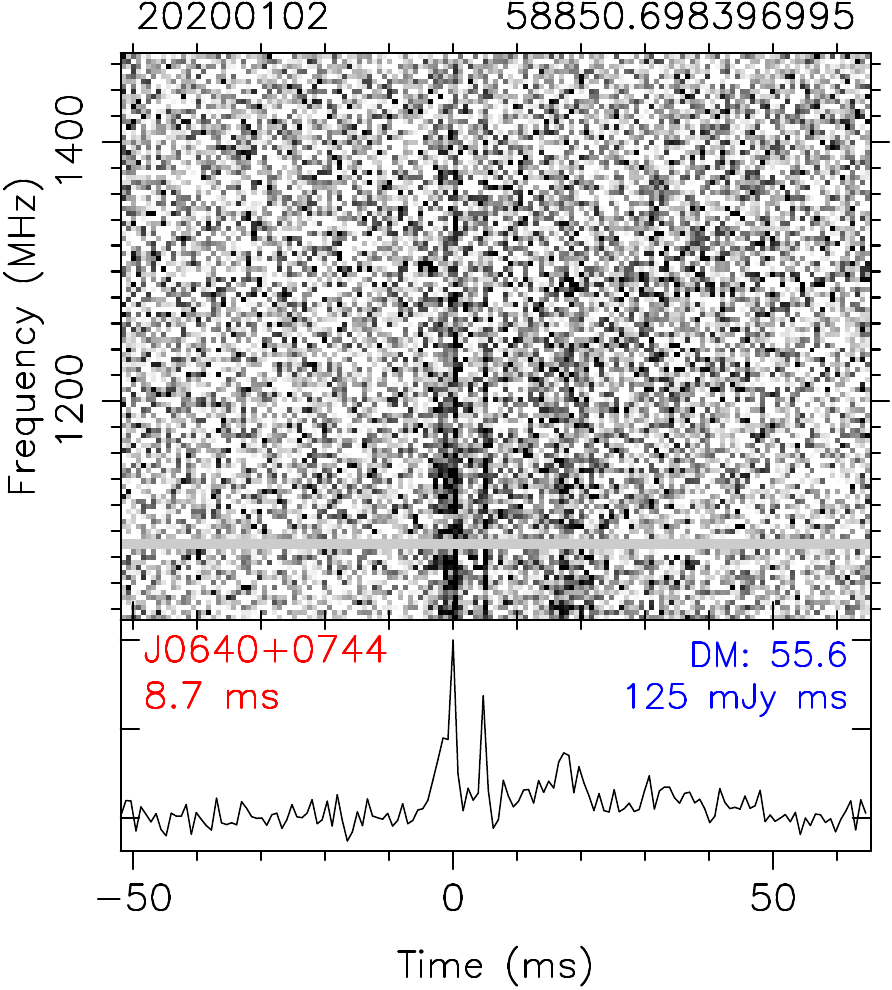}
  \includegraphics[width=0.24\textwidth]{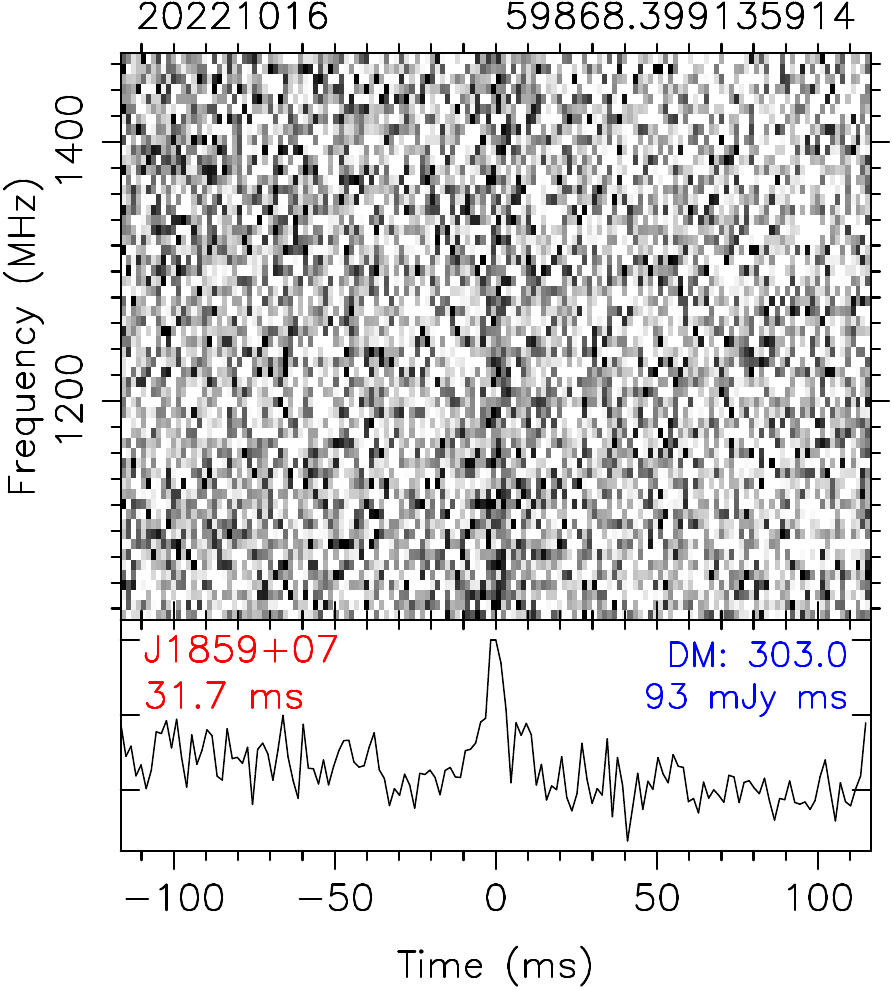}
  \caption{
  The same as Figure~\ref{fig:fewPulses} but for no period is found for previously known RRATs J0534+3425, J0550+09, J0625+1254, J0640+0744 and J1859+07. Some channels in dynamic spectra are removed due to RFI.
  }
  \label{fig:J055009}
\end{figure*}

Sensitive observations of previously known RRATs can help to understand the enigma of RRATs. 

In the visible sky area of FAST, there are more than 60 previously known RRATs.  Since these RRATs were discovered via detection of sparse strong pulses by other telescopes \citep{Mclaughlin2006,Hessels2008,Deneva2009,Keane2010MNRAS,BurkeSpolaor2010,Burke-Spolaor2011,Keane2011MNRAS,Coenen2014,Stovall2014,KarakoArgaman2015,Deneva2016,Keane2018,Patel2018,Tyul2018A&A,Tyul2018ARep,Good2021ApJ,Dong2022arXiv221009172D}, very sensitive observations by FAST may reveal possible weaker pulses from these rotating neutron stars. Some known RRATs have basic parameters (period, DM and position) with a poor accuracy, and our new observations with the FAST snapshot observations can improve these parameters (see Table~\ref{tab:RRATcat}), as done for other pulsars in the first paper \citep[see ][]{Han2021RAA}.

With the observations of the GPPS survey and also the project for the known pulsar positioning  (PT2021\_0051, PI: J. Xu) and the projects for known RRATs studies (PT2021\_0151, PI: D.J. Zhou), we have got 59 RRATs observed by chance or by targeting observations, and detected 48 RRATS (see the list in Table~\ref{tab:RRATcat}). 11 RRATs have been observed but not detected (see Table~\ref{tab:UnconfirmedRRAT}), 
%There are some known RRATs %~\footnote{The RRAT J0544+20 been incorrectly named as J0544-20 in RRATalog at http://astro.phys.wvu.edu/rratalog/ } 
%that the locations provided by the reports and nearing the covers of our observations but neither have pulse detected by using single-pulse search nor have results searched by period search, and the observation parameters of them are listed in Tabel~\ref{tab:UnconfirmedRRAT}. 
among which 8 RRATs could not be detected in targeted observations and the other 3 could not be detected in the snapshot observations for the GPPS survey. It is 
possible that these RRATs were not active when they were observed by FAST, e.g. for J1911+00 only 0.31 pulses~hr$^{-1}$ in the original discovery paper. Some RRATs were discovered at a very low-frequency band, and do not have enough flux at the L-band for the FAST to detect, e.g. J0534+3407. Otherwise, non-detection could be caused by the very large uncertain position.
%Most of the known RRATs have no spin period report previously and only a few pulses of each sources been detected. The position uncertainty of all the sources are more than 7 arcminutes, larger than the coverage radius of FAST-19 beam receiver. Considering the coverage area of about 0.1575 degree$^2$ for the FAST L-band 19-beam receiver and our observation time duration, the low burst rate and the poor positioning, these sources are indeed difficult to be detected in our observations.
% Considering the 7.5$\prime$ uncertainty in coordinates and the detected beam number, it may be located in the periphery of the beam near M02P3.

For 48 known RRATs we detect, their parameters are presented in Table~\ref{tab:RRATcat}, and % For the well detected known RRATs, FAST observations, with its highly sensitive 19-beam receiver, could improve 
%thbasic parameters for most of them, 
period, DM or position accuracy has been improved for 28 RRATs. 
The period of an RRAT is obtained either from the TOAs of single pulses or even via a normal periodic search for the survey data (see  Table~\ref{tab:RRATcat}). The spin period of J0156+0358, J1857+0719 and J1905+0414 are obtained for the first time here by our FAST observations.
The periods of J1838+0414 and J1924+1006 were 3.640~s and 5.281~s, respectively, in the discovery paper for the PALFA\footnote{\url{http://www.naic.edu/~palfa/newpulsars/}}, however, we detect the periods of 1.330681~s and 4.619757~s by analyzing the TOA of the detected single pulses in our FAST data with the period analysis program {\it PF}. 
The DM values of these RRATs were estimated by using some bright single pulses, and we get DMs of 10 RRATs improved. J1945+2357 (J1946+24) was reported to have a DM of 96\,$\rm pc~cm^{-3}$~\citep{Deneva2009}, but our FAST observations give a much better determined DM as being 87.53$\pm$0.18\,$\rm pc~cm^{-3}$.
For some RRATs, their position can be constrained by the FAST observations to a half beam size of FAST \citep[see discussion in Sect.4 and Sect.5 of][]{Han2021RAA}, and we get positions of 19 RRATs, significantly improved from these in references. Their new names accordingly to the latest coordinates are given in the first column in Table~\ref{tab:RRATcat}. 

We emphasize that most of the previously known RRATs are very bright for the FAST, as indicated by the high detection percentage of individual pulses over the $3\sigma$ criteria as listed in column (10). Nevertheless, for 1 known RRAT, we detect only 4 pulses, and for other 4 known RRATs, we detect only one pulse each.
PSR J0534+3425 was first discovered at 111~MHz~\citep{Tyul2018A&A}, and we detect 4 pulses in a 5-minute tracking observation of the GPPS survey, but no pulse was detected in the following 15 monitoring observation. For these four pulses, only signals below 1180 MHz are detected in our observations. Considering the low DM of about 24.7~pc~cm$^{-3}$, the detection in a partial band is likely to be caused by interstellar scintillation. No period can be found from our FAST observation yet.
PSR J0550+0948 was reported by \citet{Deneva2016} with  $P= 1.745$~s and $DM=86.6$~pc~cm$^{-3}$. Only one weak pulse is detected by FAST with $\rm S/N = 17.2$ during the snapshot observation on 20211215 around a candidate position in the beam of J0550+0900-M02P3. 
PSRs J0550+0948 \citet{Deneva2016}, J0625+1254~\citep{Patel2018}, J0640+0744 \citep{Tyul2018A&A} and J1859+0759 \citep{Patel2018} have no previously known period. In our FAST observation, only one pulse is detected in a 5~minutes observation for each source with S/N=17.2, S/N=32.8, S/N=25.0 and S/N=15.0, respectively. For these sources, we cannot get a period from our data even after folding for J0550+0948. The dynamic spectra for these eight pulses are shown in Figure~\ref{fig:J055009}.  

\begin{figure}% [!htp]
  \centering
\includegraphics[width=0.23\textwidth]{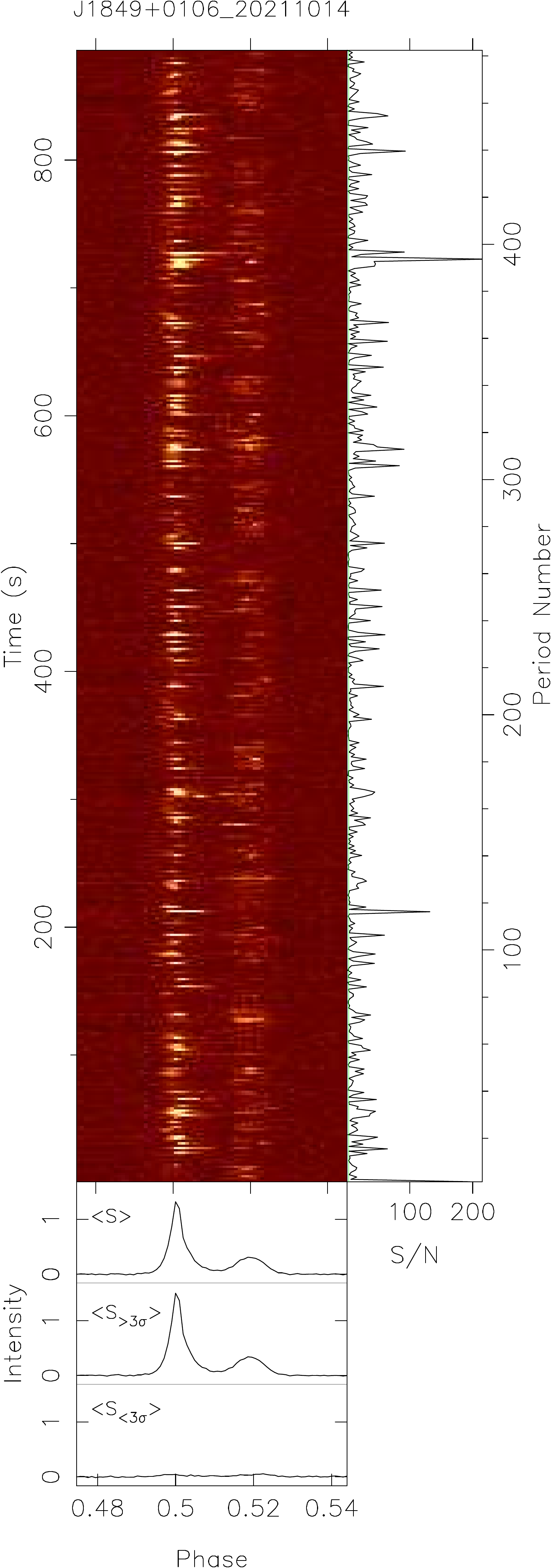}
\includegraphics[width=0.23\textwidth]{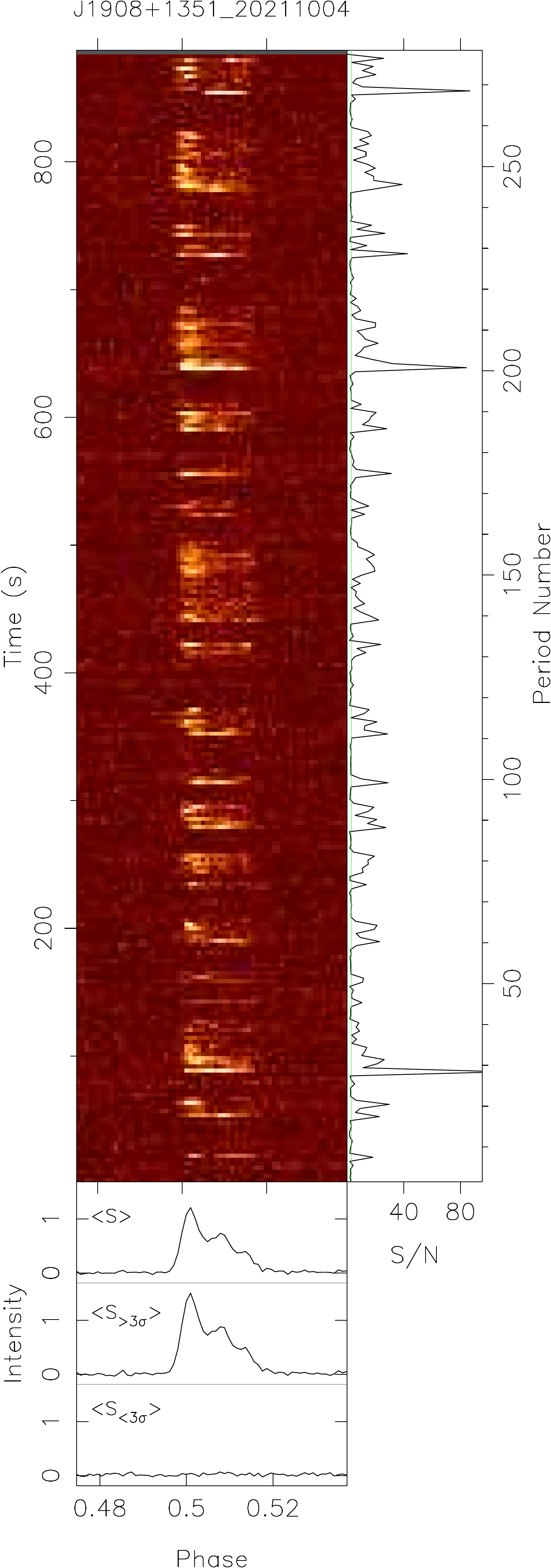} 
  \caption{Two examples (J1849+0106 and J1908+1351) for the known RRATs shown as normal pulsars in the FAST observations though some bright pulses occasionally emerge. All such cases are presented in Figure~\ref{fig:APPknownRRAT1} in Appendix.}
  \label{knownRRAT1}
\end{figure}

\begin{figure}%[!htp]
  \centering
\includegraphics[width=0.23\textwidth]{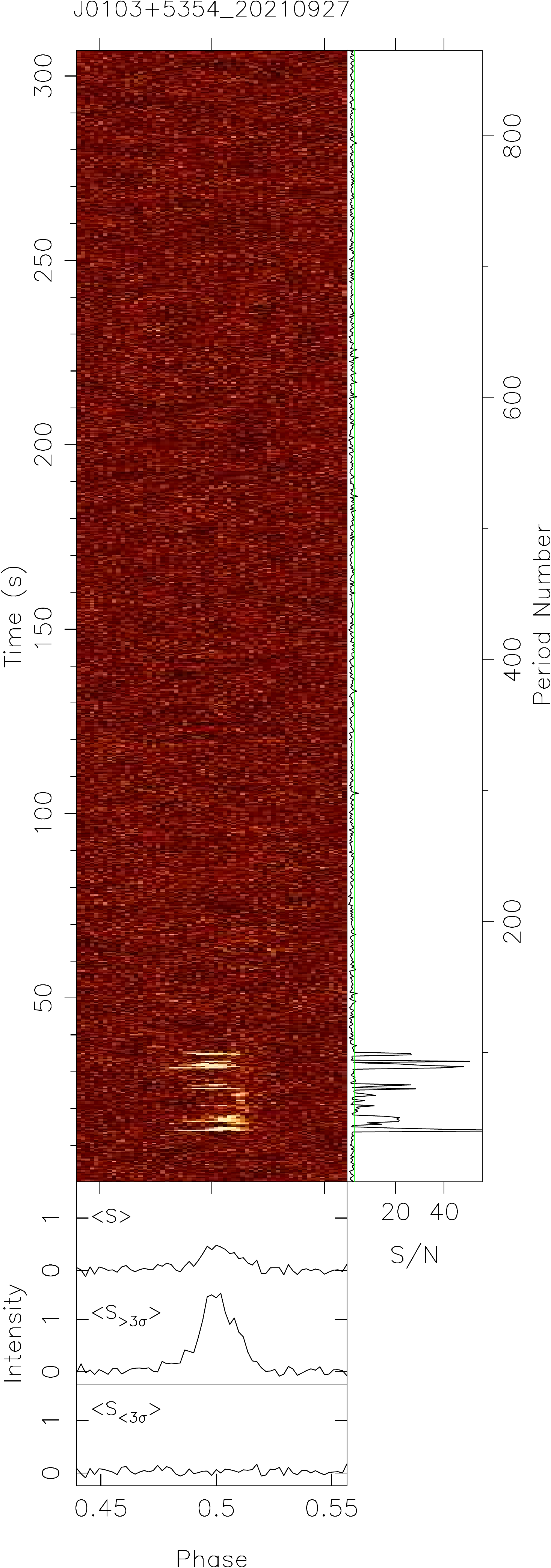}
  \includegraphics[width=0.23\textwidth]{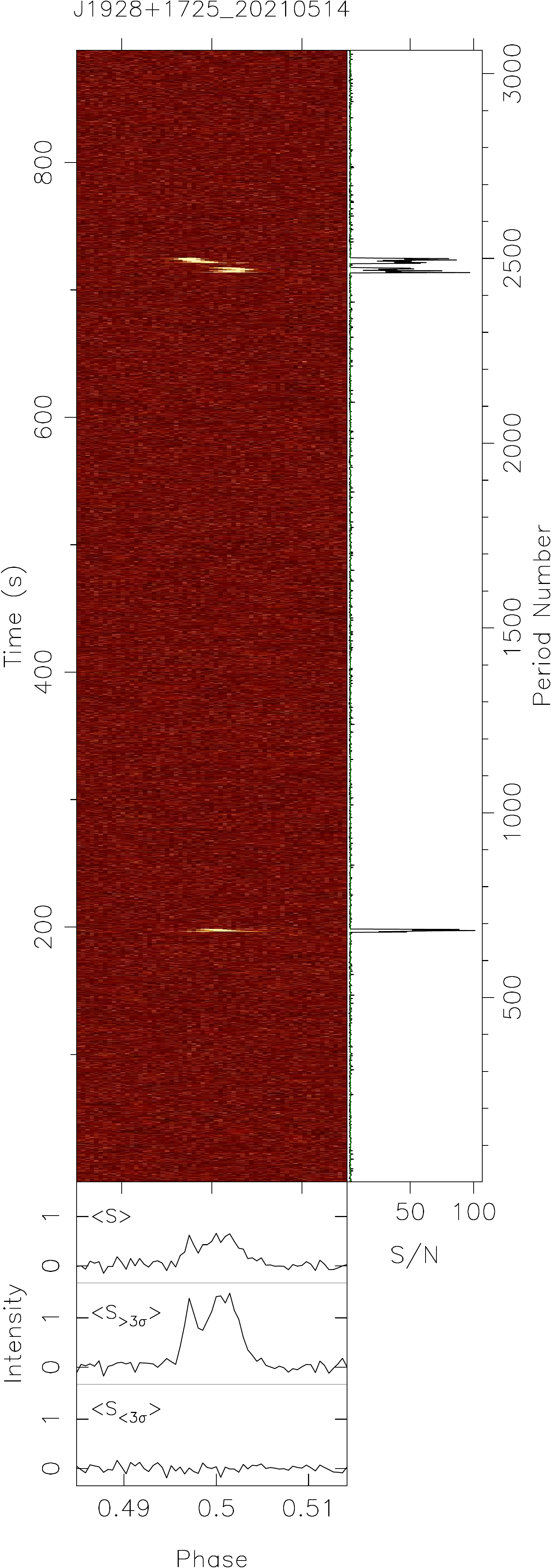} %J1928+1725
  \caption{The same as Figure~\ref{knownRRAT1} but for two examples of the known RRATs shown as extremely nulling pulsars. All such cases are presented in Figure~\ref{fig:APPknownRRAT2} in Appendix.}
    \label{knownRRAT2}
\end{figure}

\begin{figure}%[!htp]
  \centering
  \includegraphics[width=0.23\textwidth]{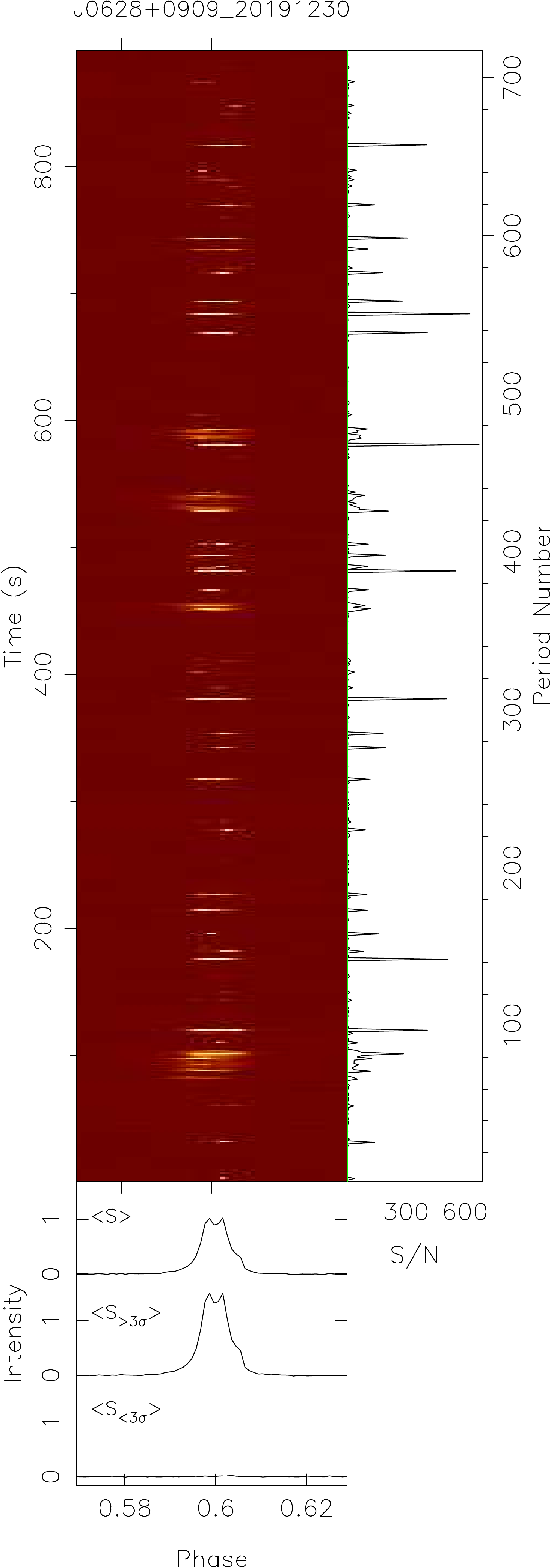} 
  \includegraphics[width=0.23\textwidth]{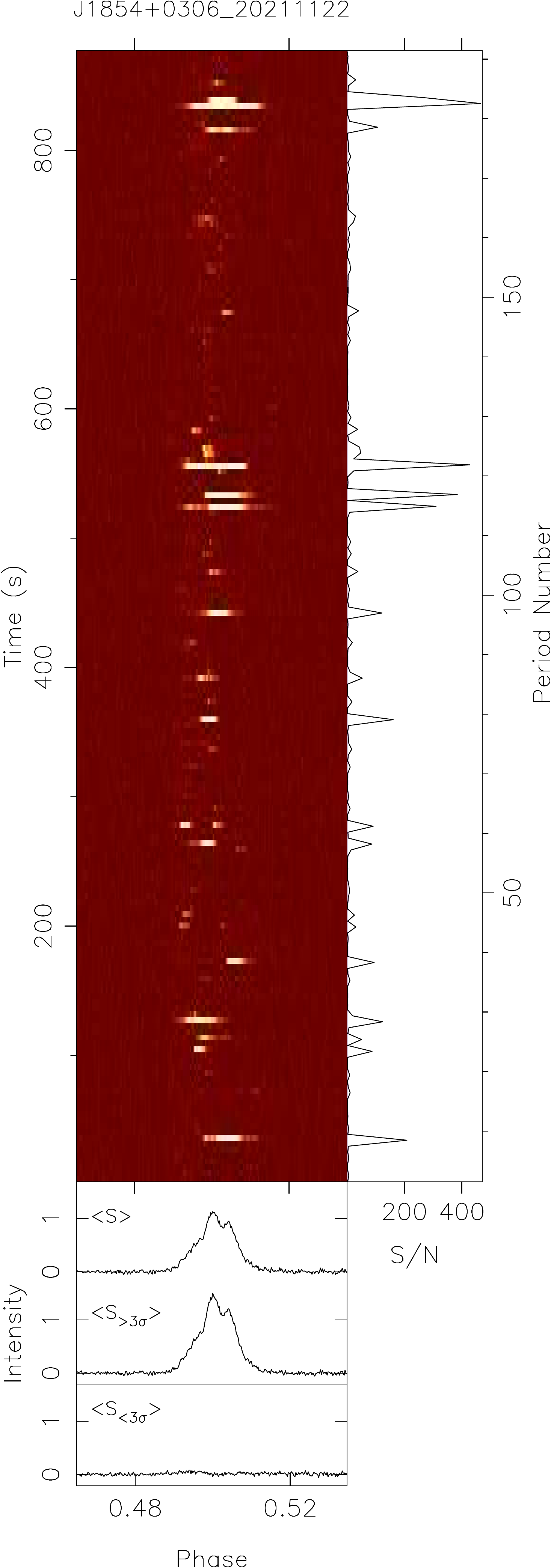} %J1854+0306
  \caption{Two examples for the known RRATs shown as generally very weak pulsars with sparse strong pulses. See all such cases Figure~\ref{fig:APPknownRRAT3} in Appendix.}
    \label{knownRRAT3}
\end{figure}

For the other 43 known RRATs detected by FAST, the folded data according to the period and DM 
%with periodic results in a single pulse. The results show that there is no source with the characteristics of emit single pulses occasionally. 
show that 25 of them are just normal pulsars (see Figure~\ref{knownRRAT1}), 5 of them are very nulling pulsars  (see  Figure~\ref{knownRRAT2}), and others are pulsars with sparse strong pulses  (see  Figure~\ref{knownRRAT3}). %, or have a mixture of multiple pulse emission modes. %For some of other 39 previously known RRATs, the pulsar-stacks look like normal pulsars with few bright single pulses. Some other RRATs are too faint to be detected by period search. Few bright single pulses are detected during our single pulse module . The integrated profiles of these three sources show two-component structure, and the each bright pulse appears to have one main peak. 
% These extremely nulling pulsars and other RRATs are almost only could be timing by their single pulse, such as RRAT J1928+1725\citep{Parent2022ApJ}, and it is difficult to get a better result. The new results we provide here can provide the basis for subsequent analysis.
In the following, we discuss such these cases in detail. 
%We classified them as two groups. The first group is that they are obvious with a large number of fractions with radiation interruptions in pulse-stacks, called extremely nulling pulsars. The other groups is the ordinary pulsars, which are almost continuously radiating, or with occasional short nulling fractions. The following section will discuss the sources in these two groups in detail. 

\begin{figure} %[!htp]
  \centering
  \includegraphics[width=0.7\columnwidth]{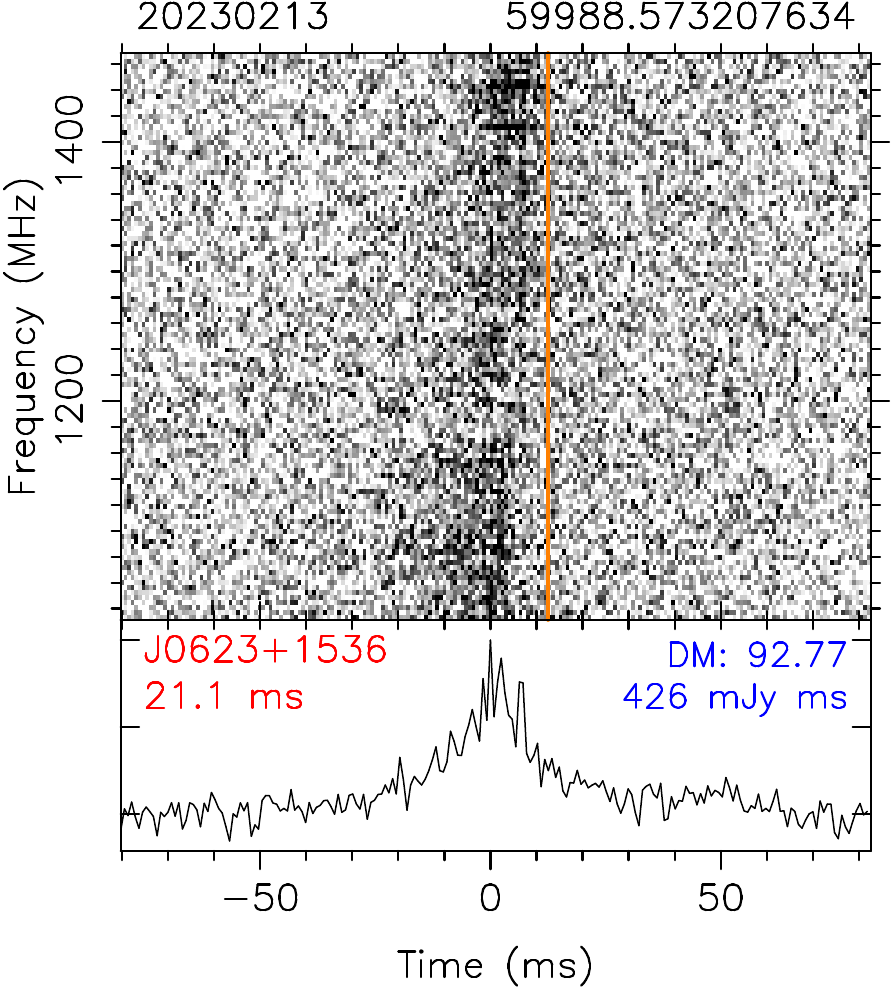}
  \includegraphics[width=0.7\columnwidth]{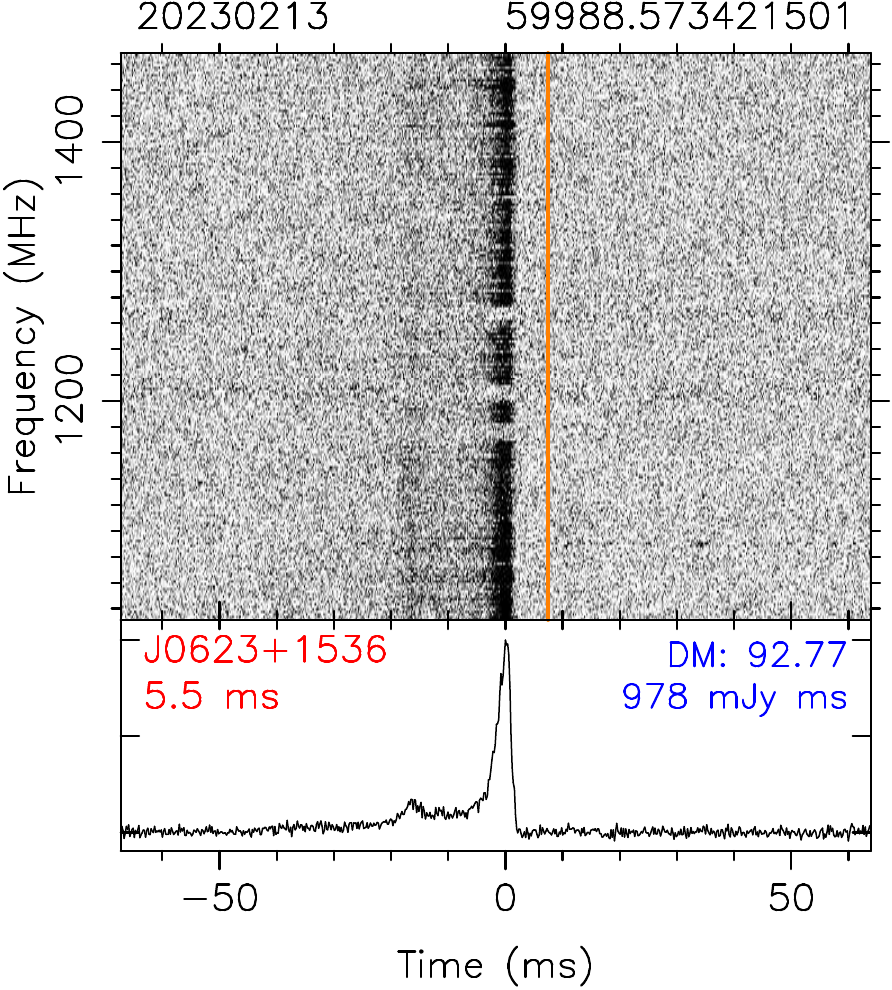}
  \caption{The same as Figure~\ref{fig:fewPulses} but for two single pulses of PSR J0623+1536, dedispersed by using the DM of $\rm~92.77~cm^{-3}~pc$. A dark yellow vertical line is used to indicate the proper dedispersion for the two pulses. The dynamic spectrum of the upper one for the period No.330 shows that emission first appears at the lower part of frequency band but later the emission appears at the higher part of the band. The other pulse in the lower panel for the period No.337 does not show such a frequency drifting.}
  \label{J0623+1536-2sp}
\end{figure}

\begin{figure} %[!htp]
  \centering
  \includegraphics[width=38mm]{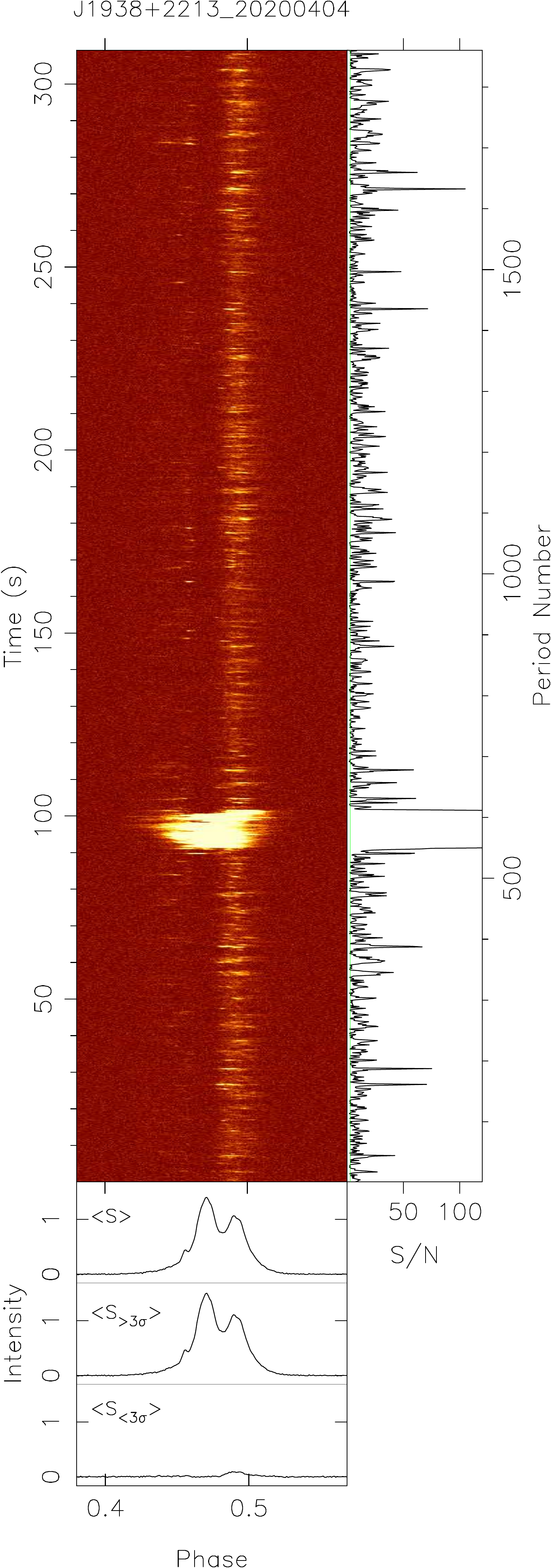} %J1938+2213, nomal pulsar with GP
  \includegraphics[width=38mm]{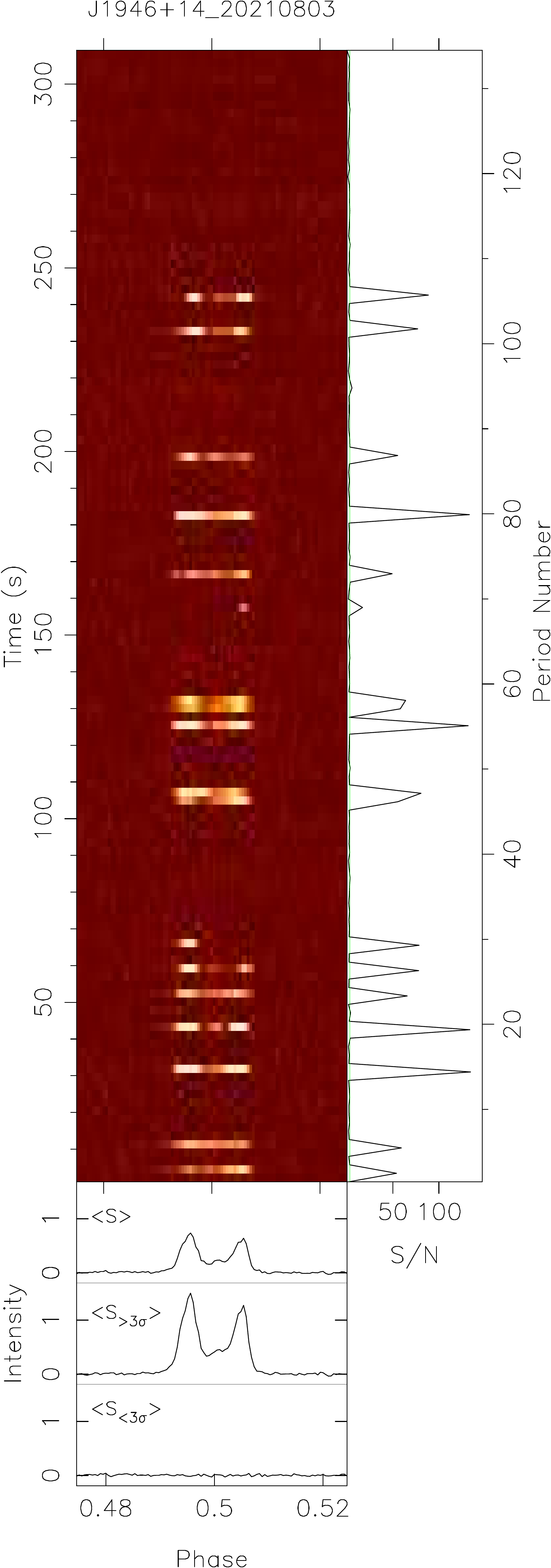}%J1946+1449, not RRAT, just RRAT-like nulling pulsar
  \caption{Same as Figure~\ref{knownRRAT1}, but for special RRAT-like pulsars. The left plot for J1938+2213 has zoomed in for the maximum of S/N to 120.}
  \label{knownpulsarRRATlike}
\end{figure}

% polarization
\begin{figure*}[!htp]
    \centering
        \includegraphics[width=0.45\columnwidth]{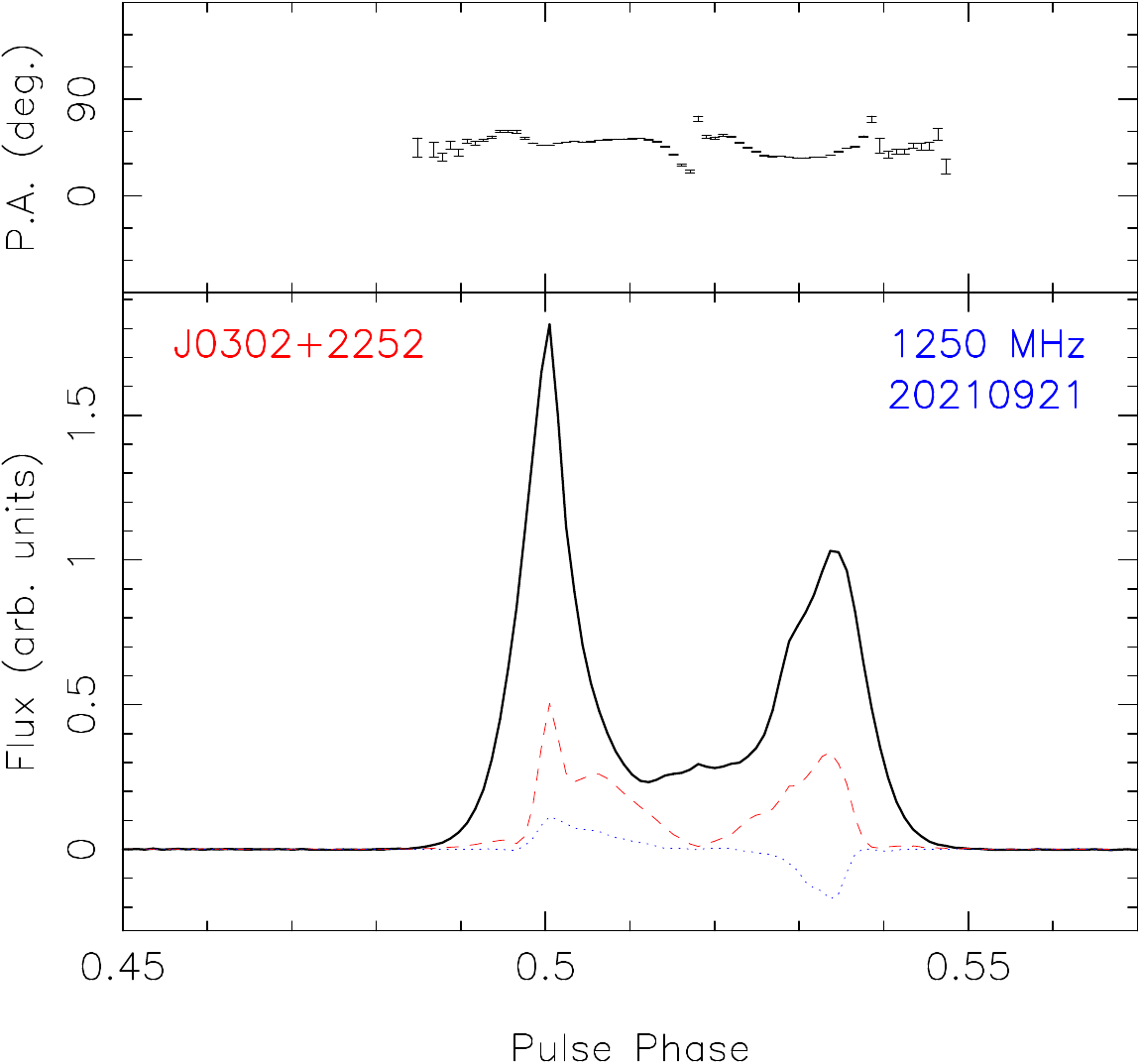} 
        \includegraphics[width=0.45\columnwidth]{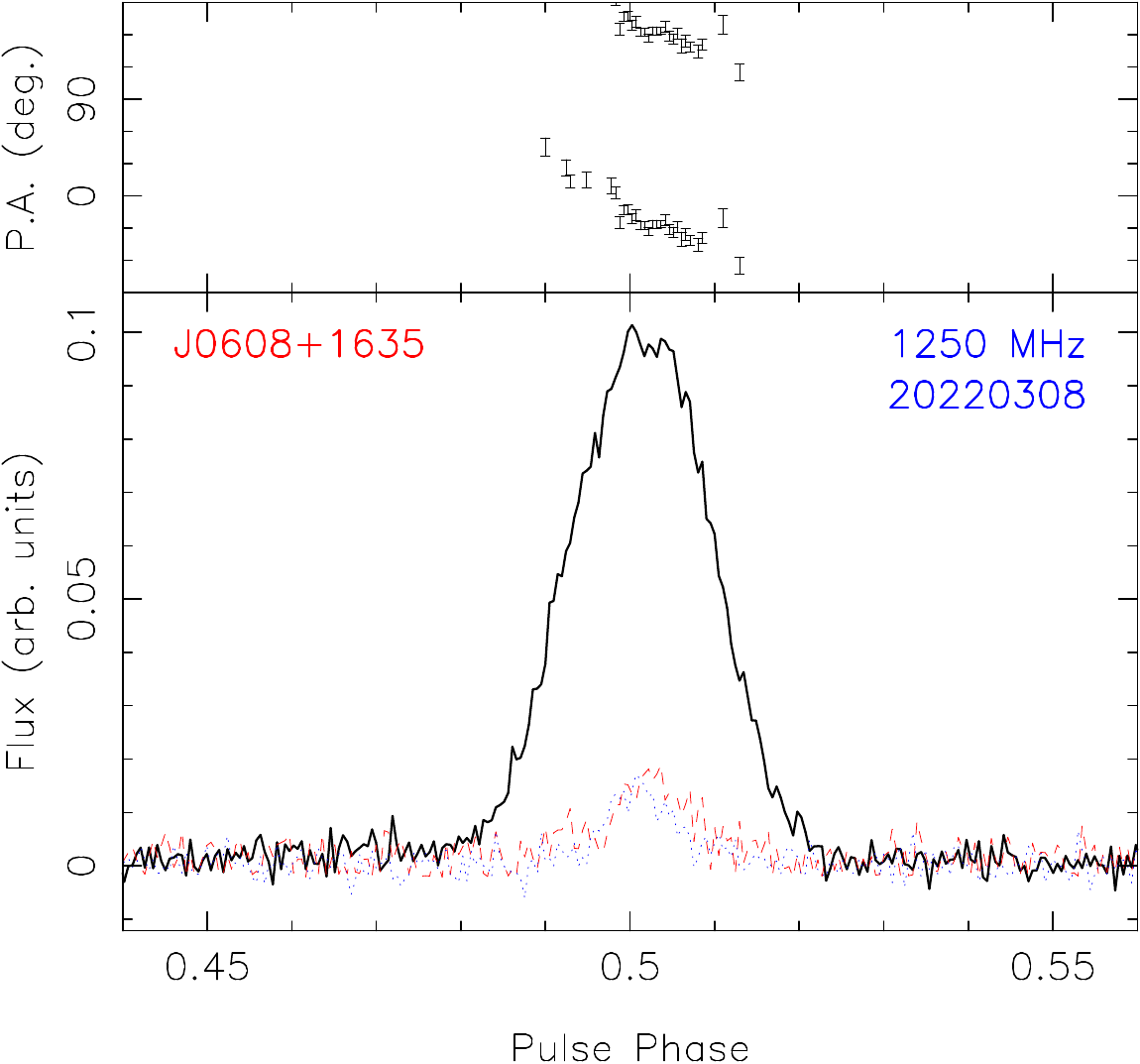} 
        \includegraphics[width=0.45\columnwidth]{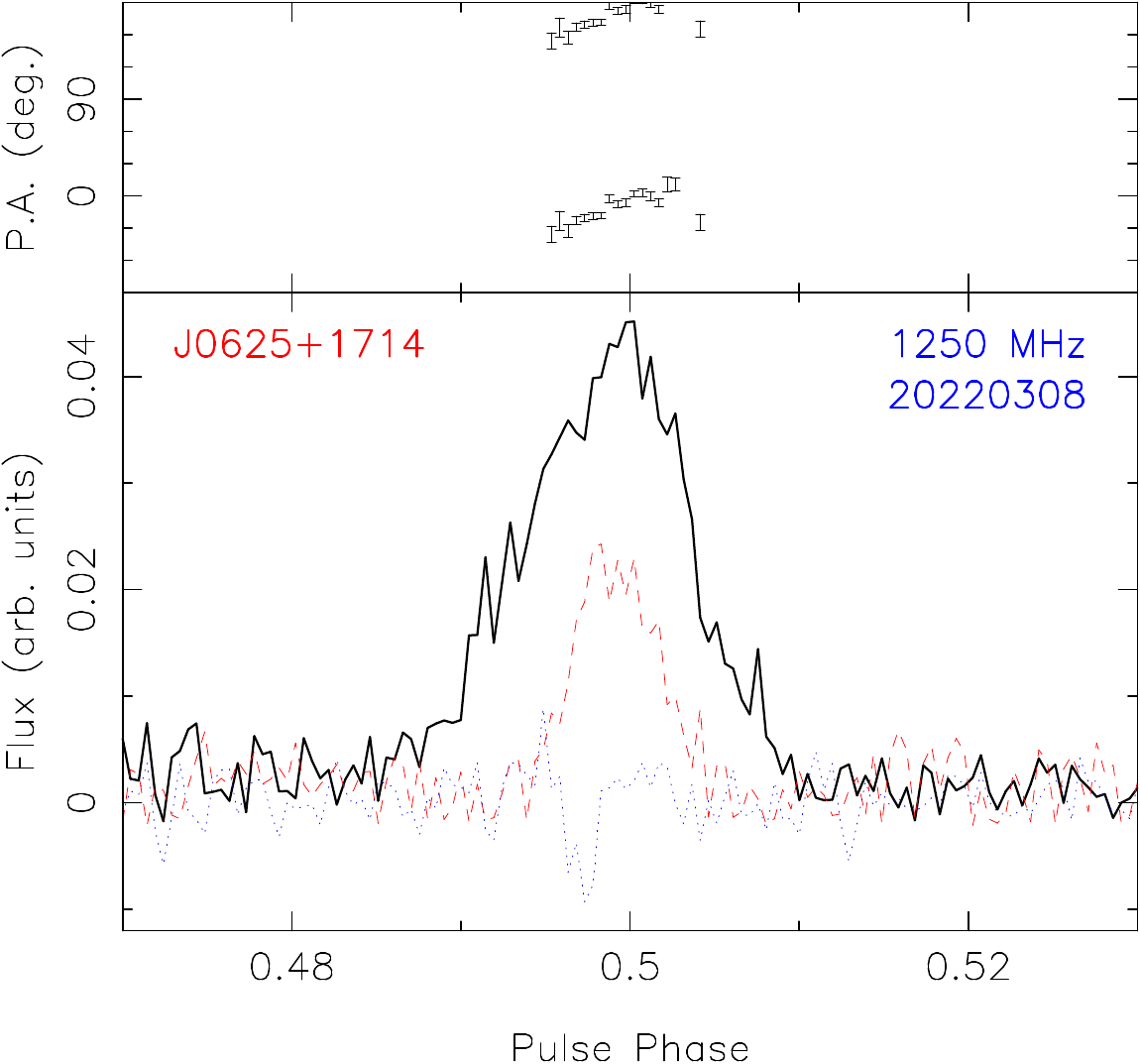} 
        \includegraphics[width=0.45\columnwidth]{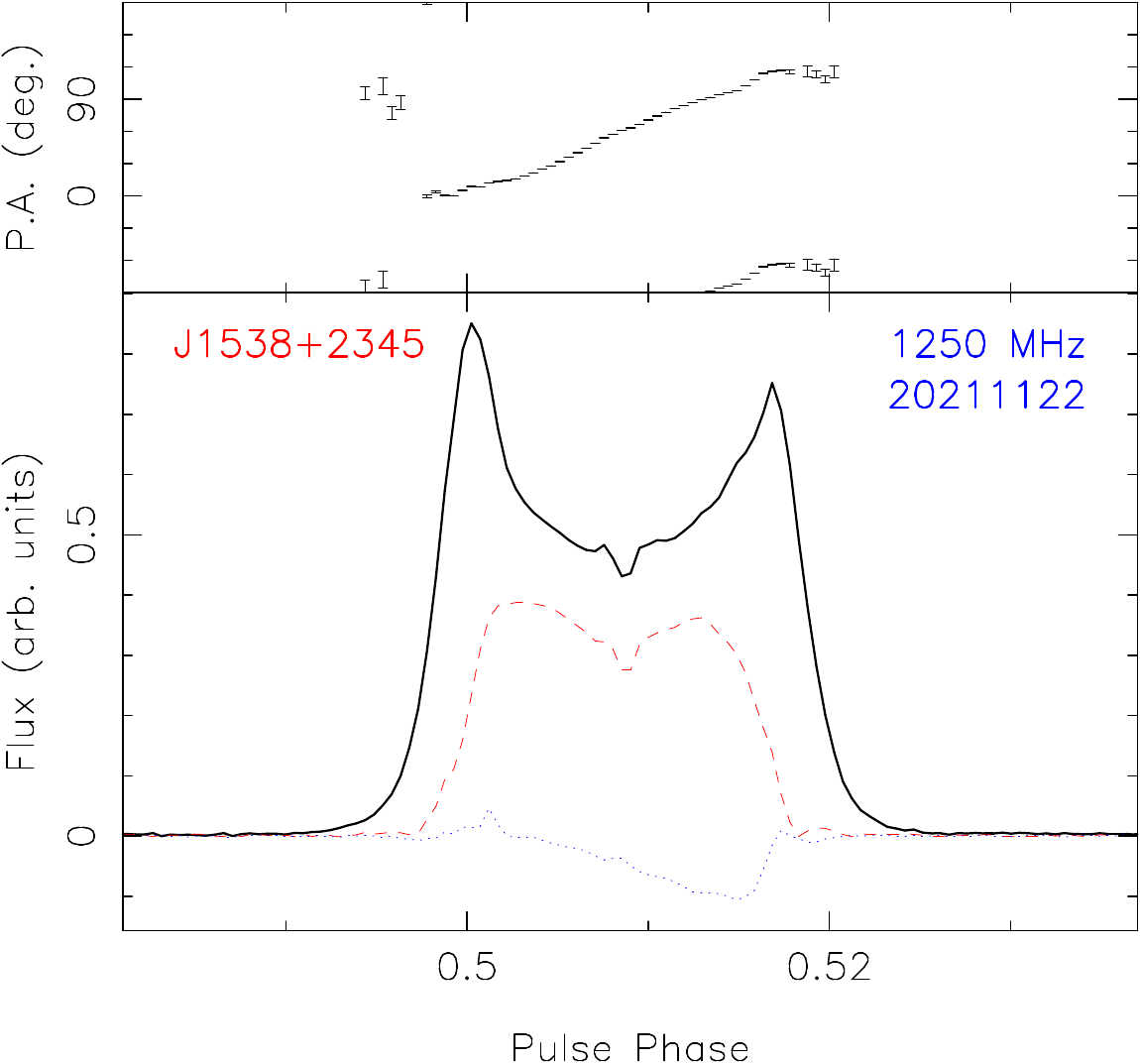} 
        \includegraphics[width=0.45\columnwidth]{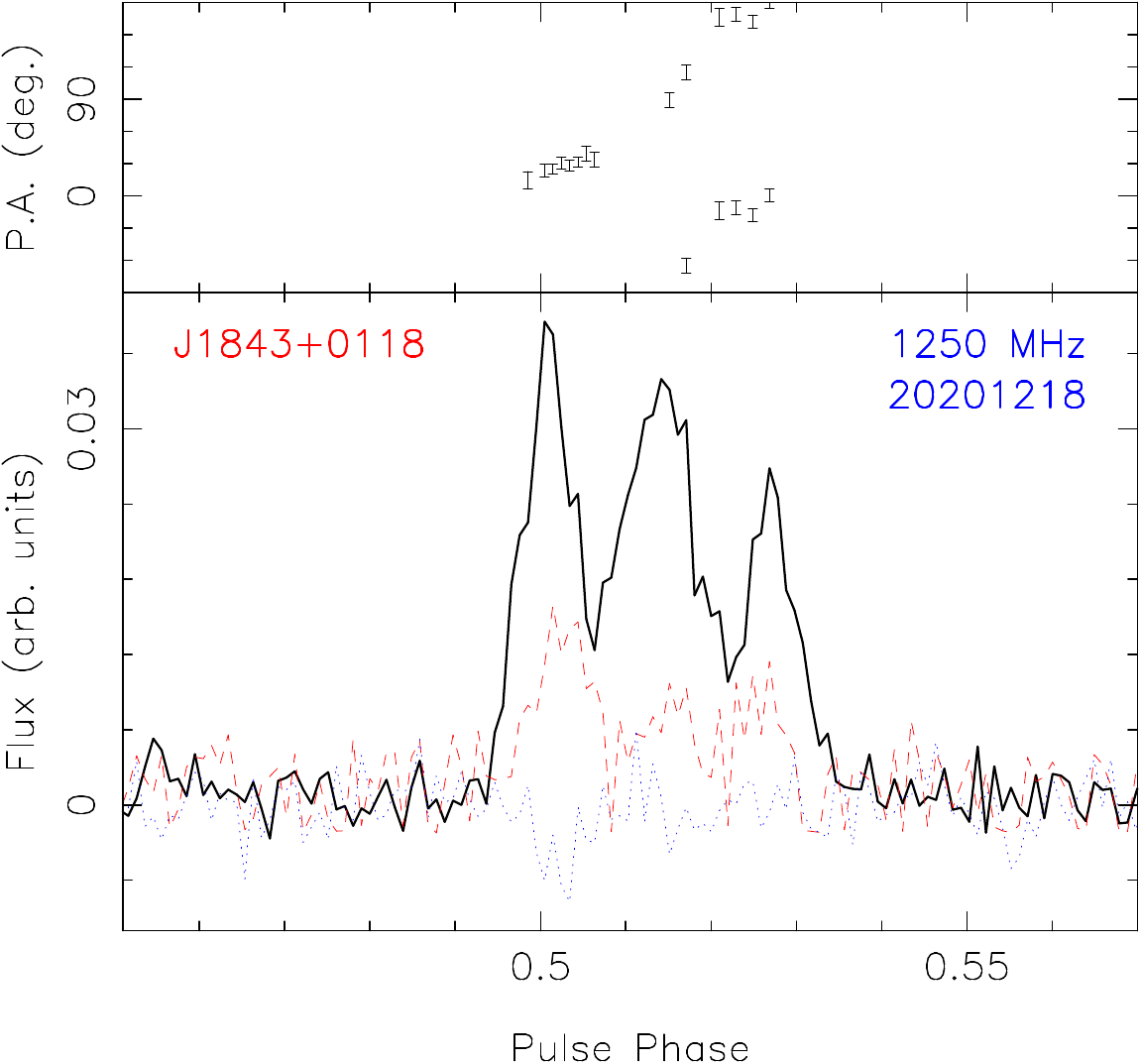} 
        \includegraphics[width=0.45\columnwidth]{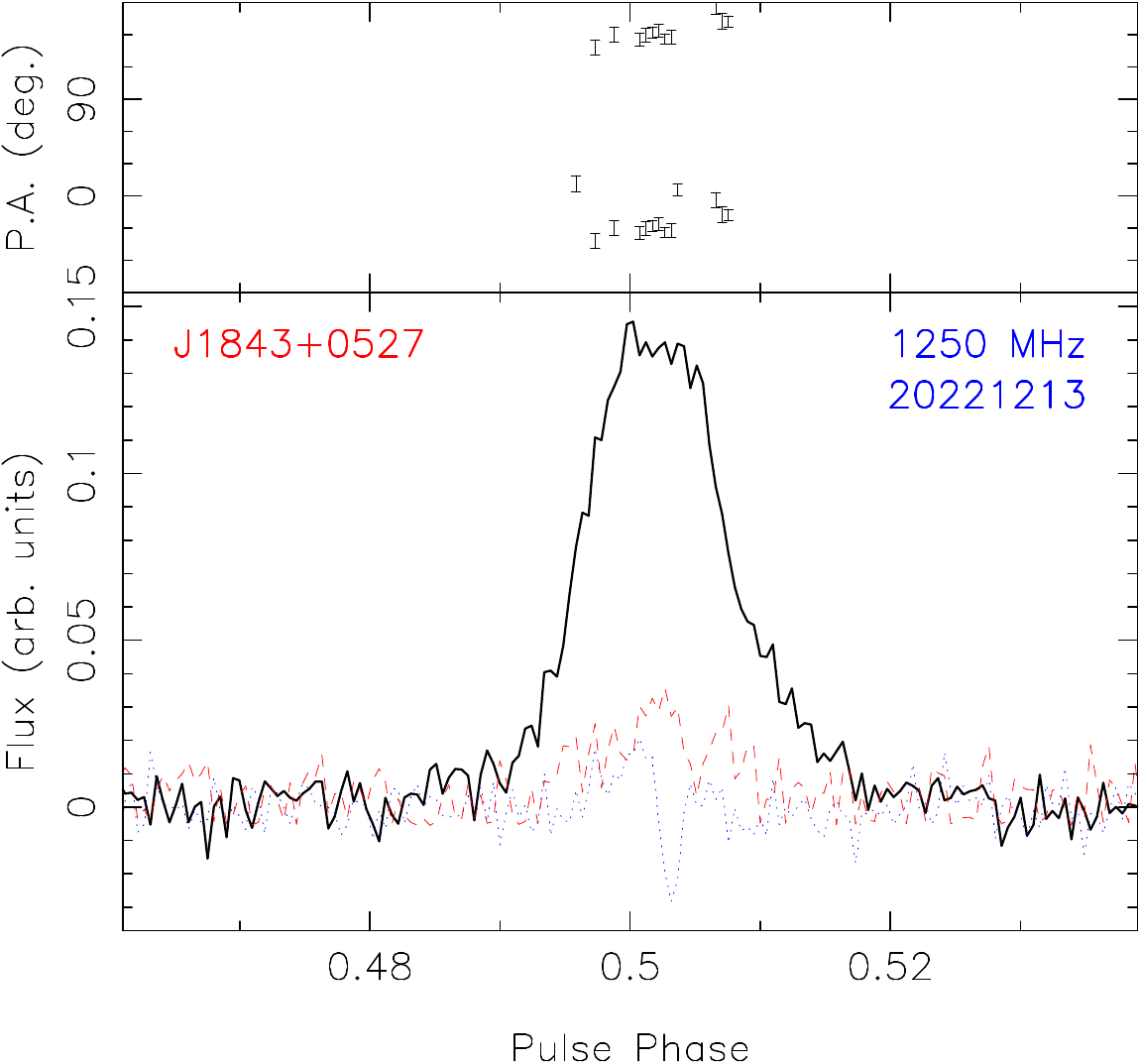}
        \includegraphics[width=0.45\columnwidth]{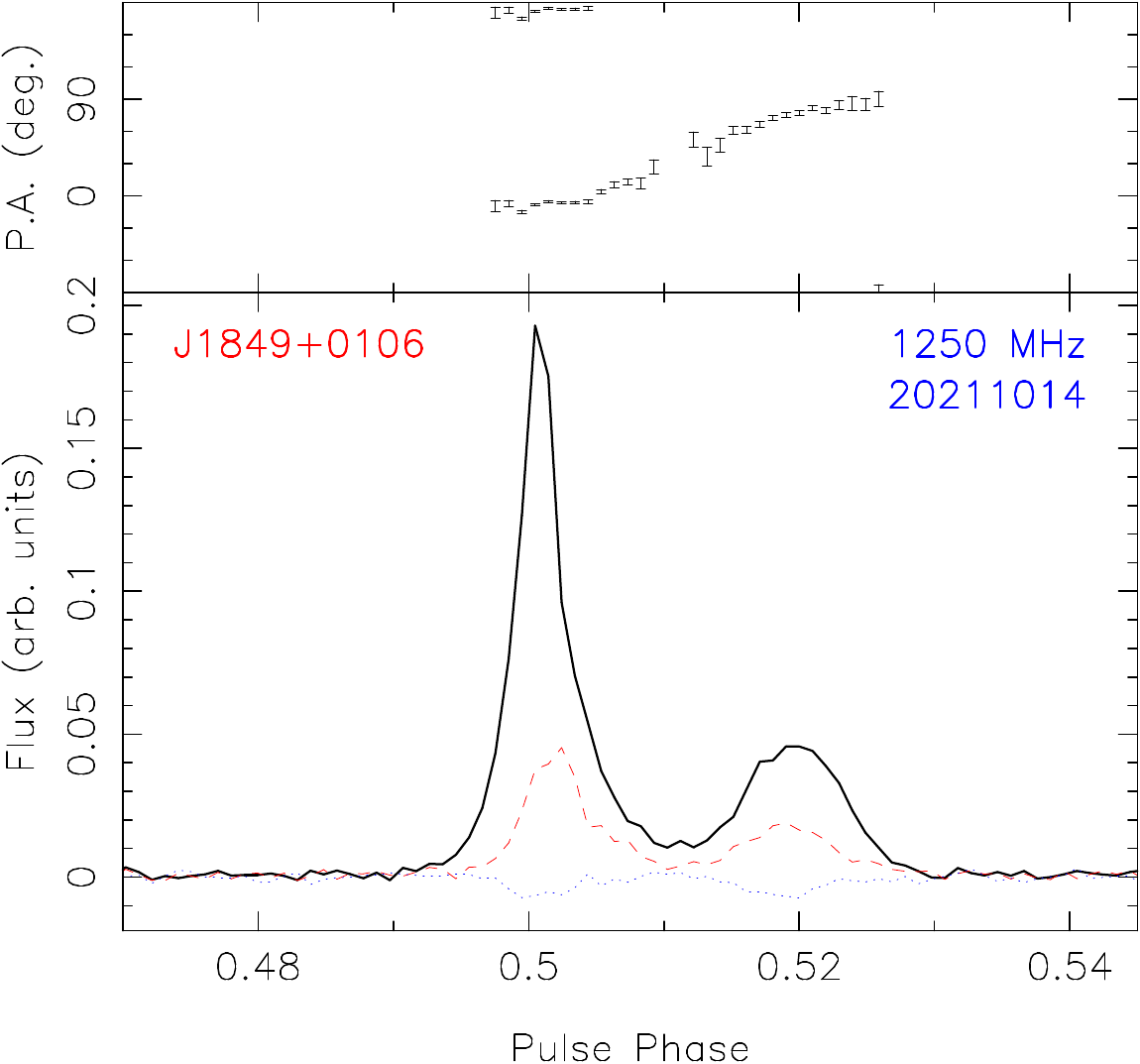} 
        \includegraphics[width=0.45\columnwidth]{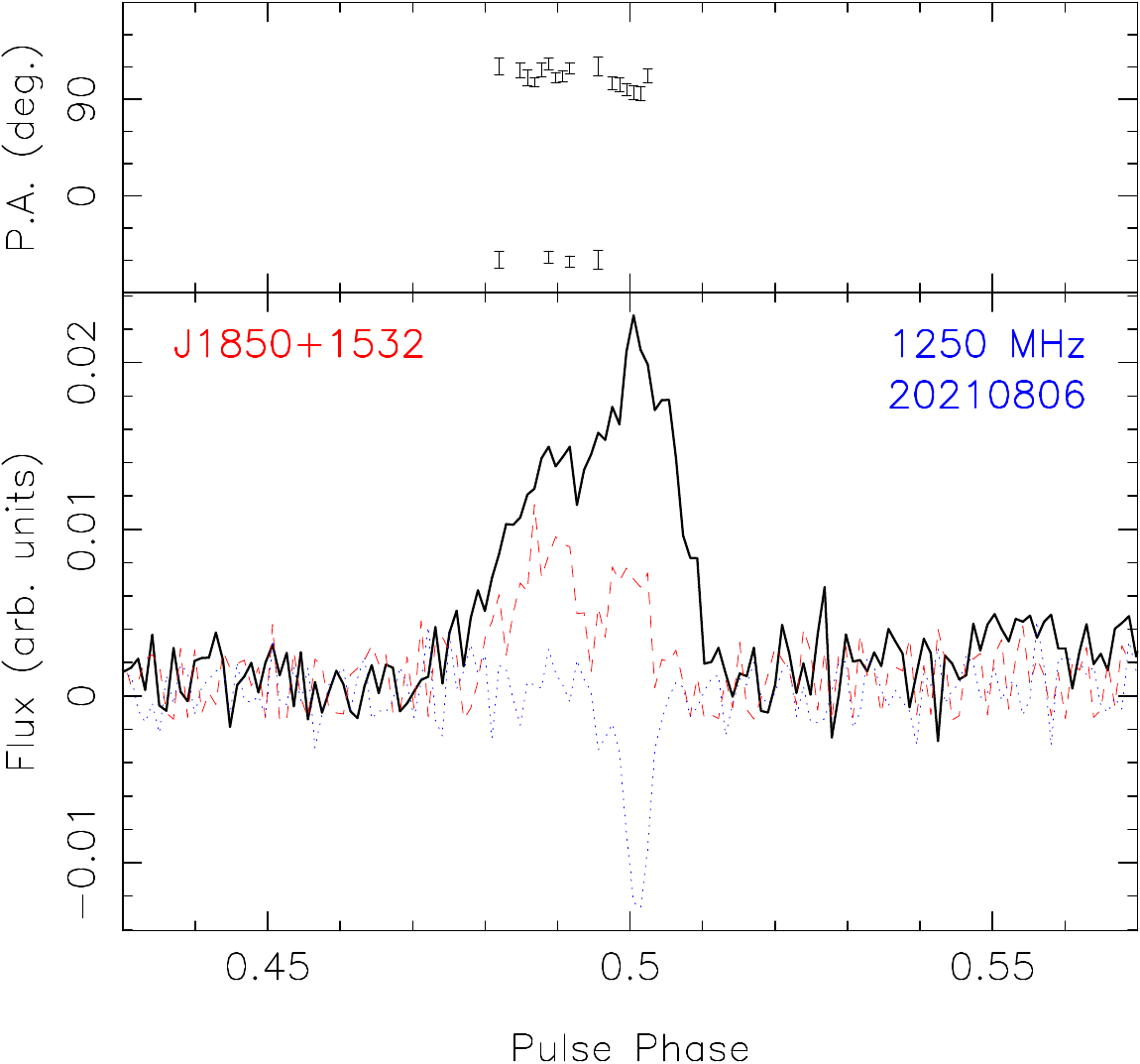} 
        \includegraphics[width=0.45\columnwidth]{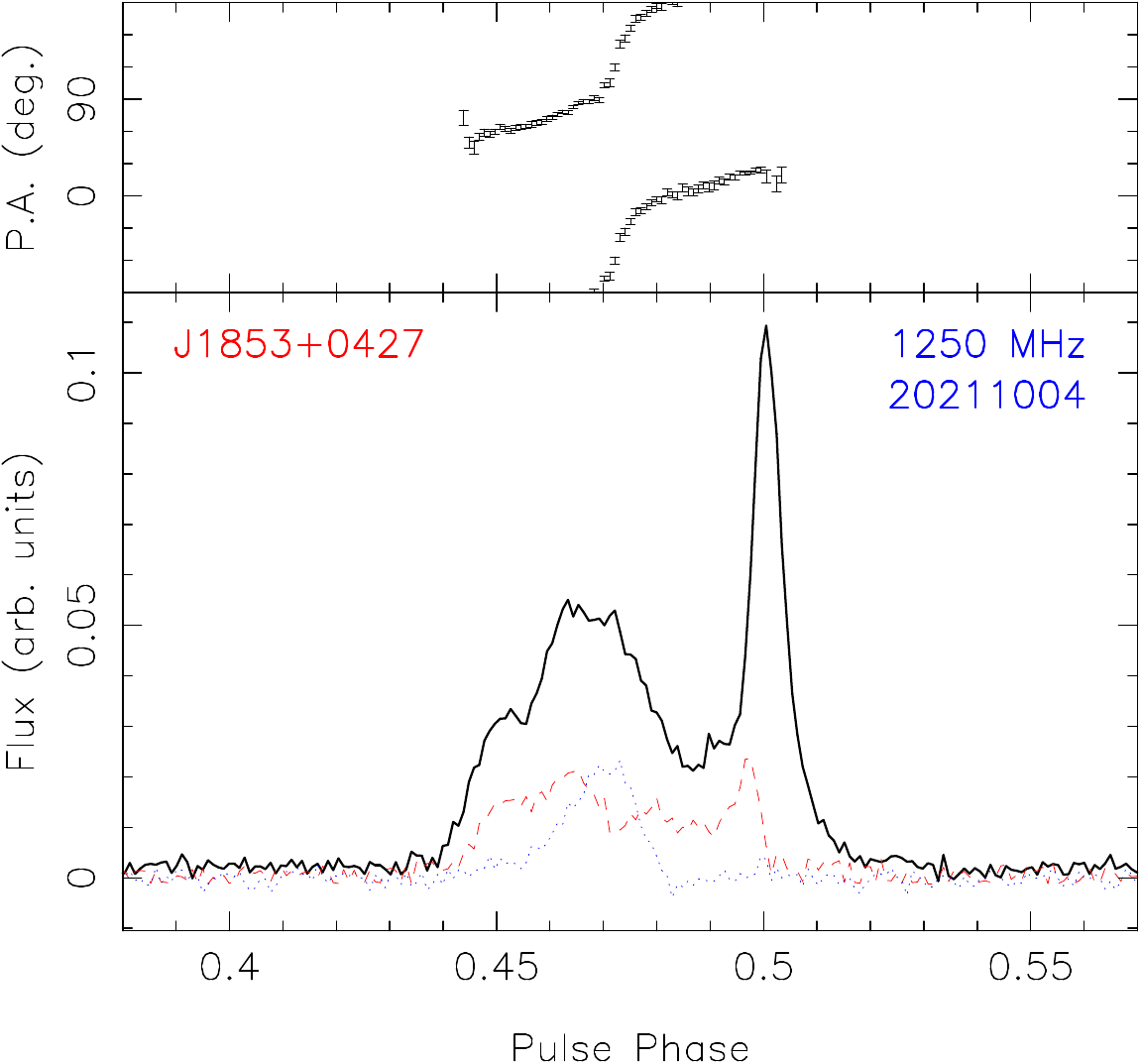} 
        \includegraphics[width=0.45\columnwidth]{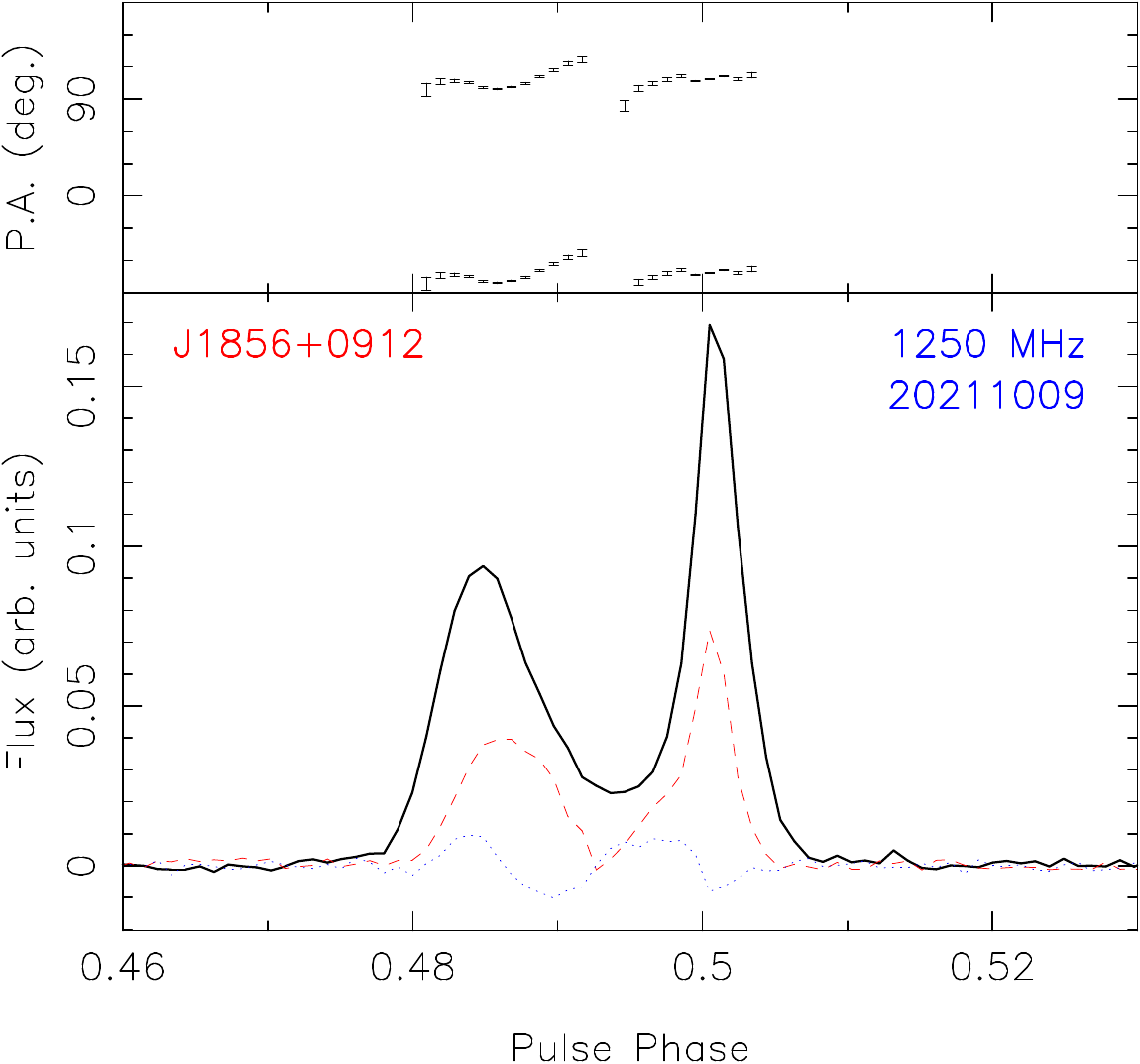} 
        \includegraphics[width=0.45\columnwidth]{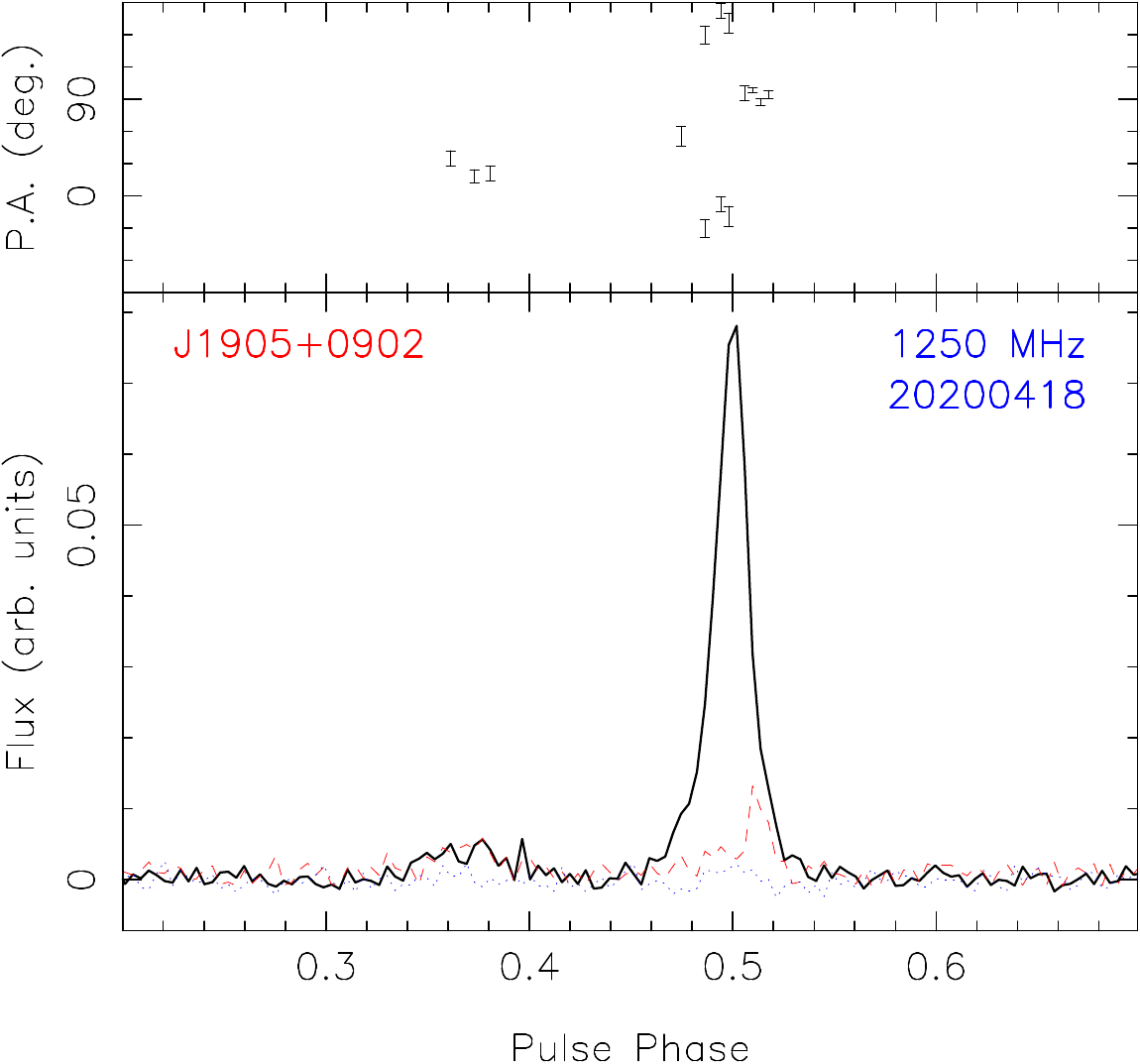} 
        \includegraphics[width=0.45\columnwidth]{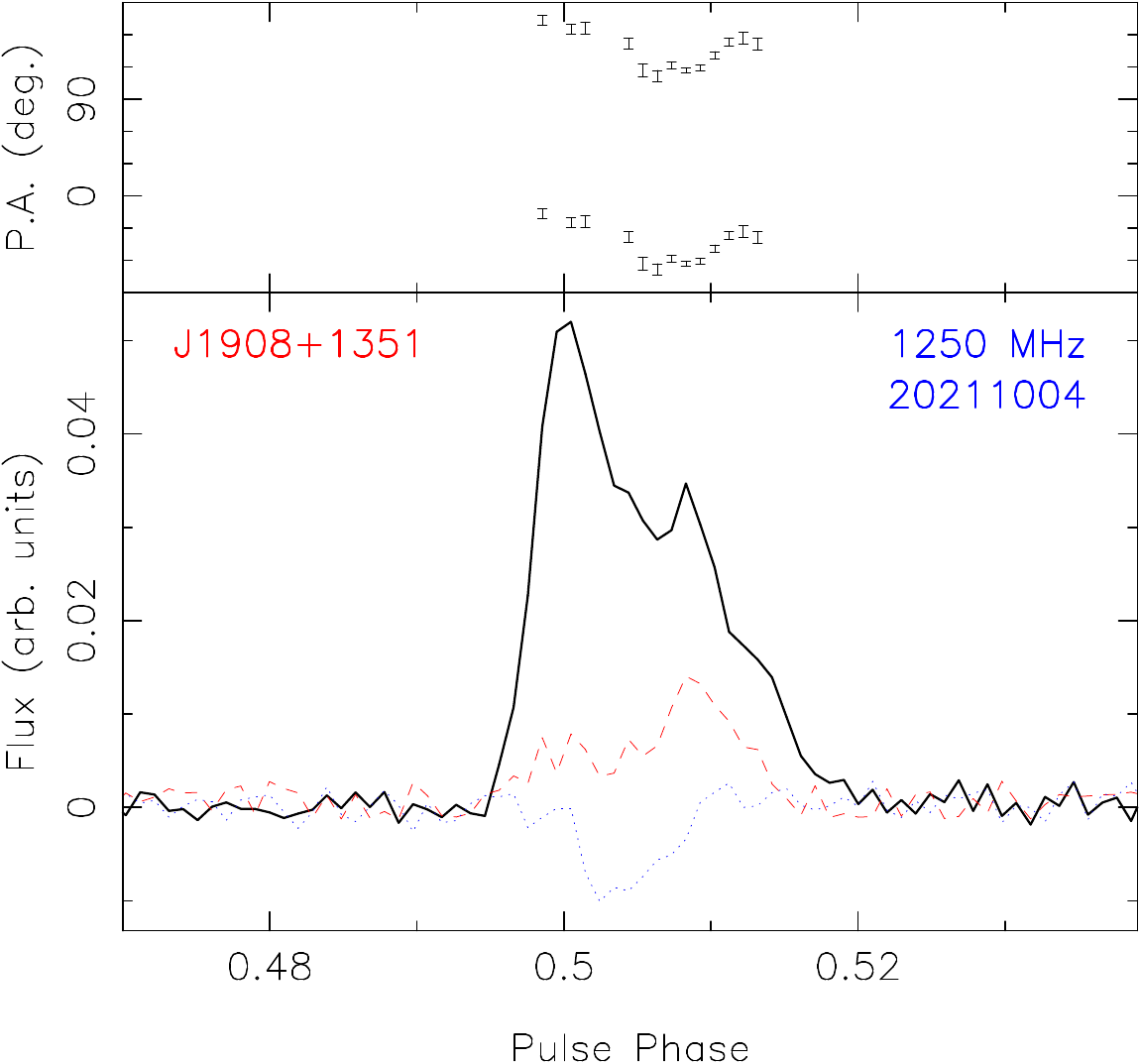} 
        \includegraphics[width=0.45\columnwidth]{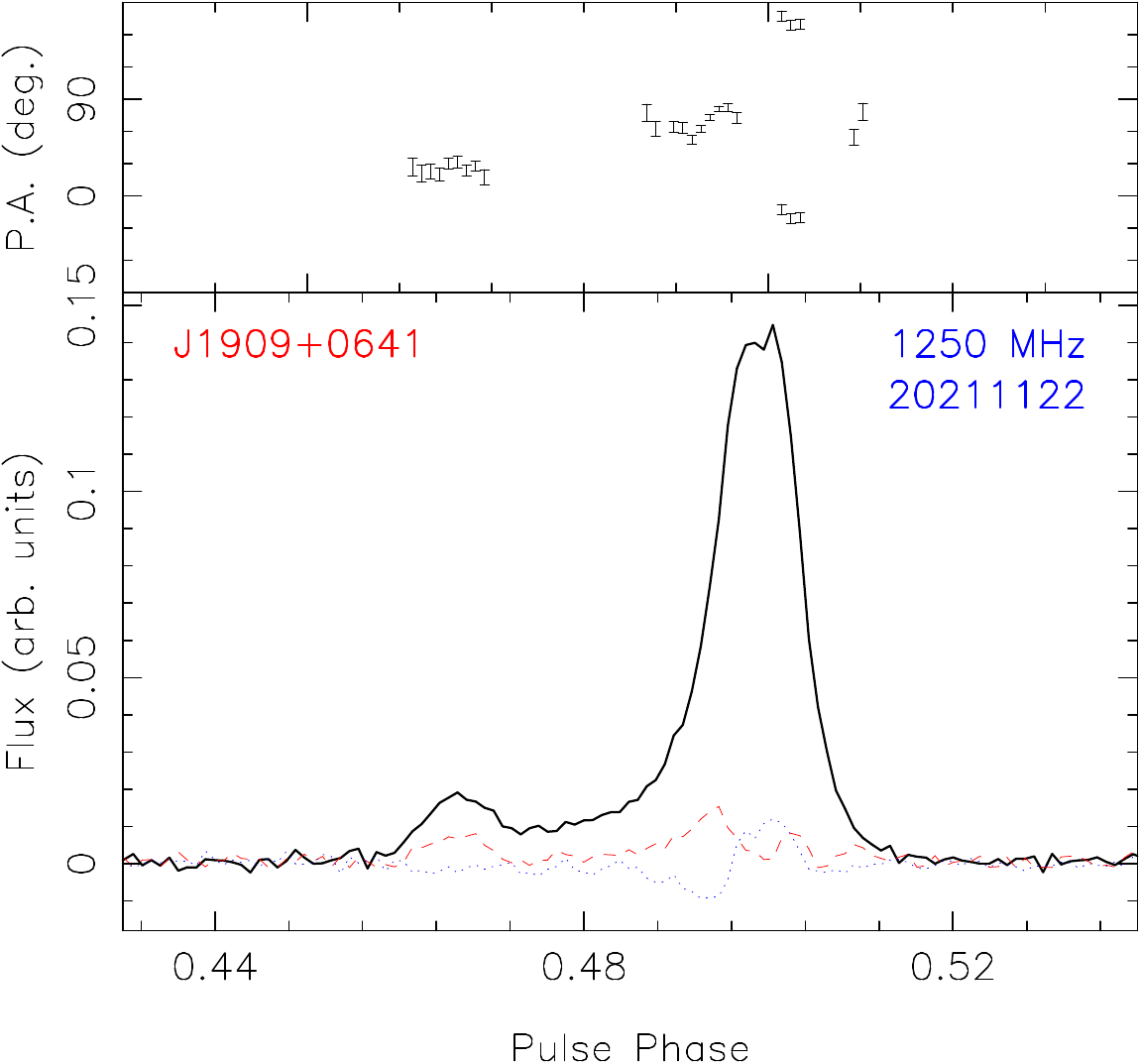} %J1909+0641
        \includegraphics[width=0.45\columnwidth]{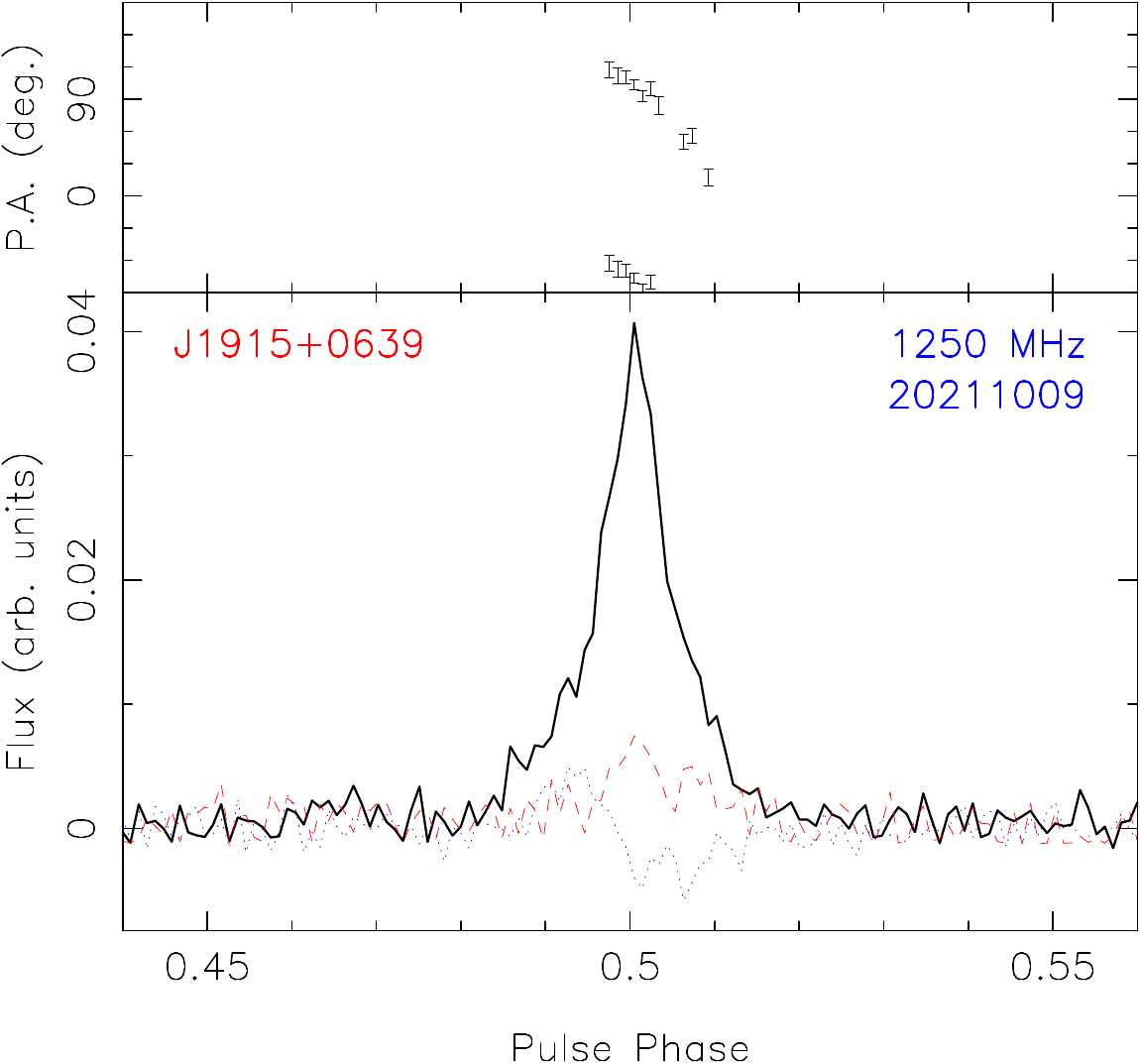} 
        \includegraphics[width=0.45\columnwidth]{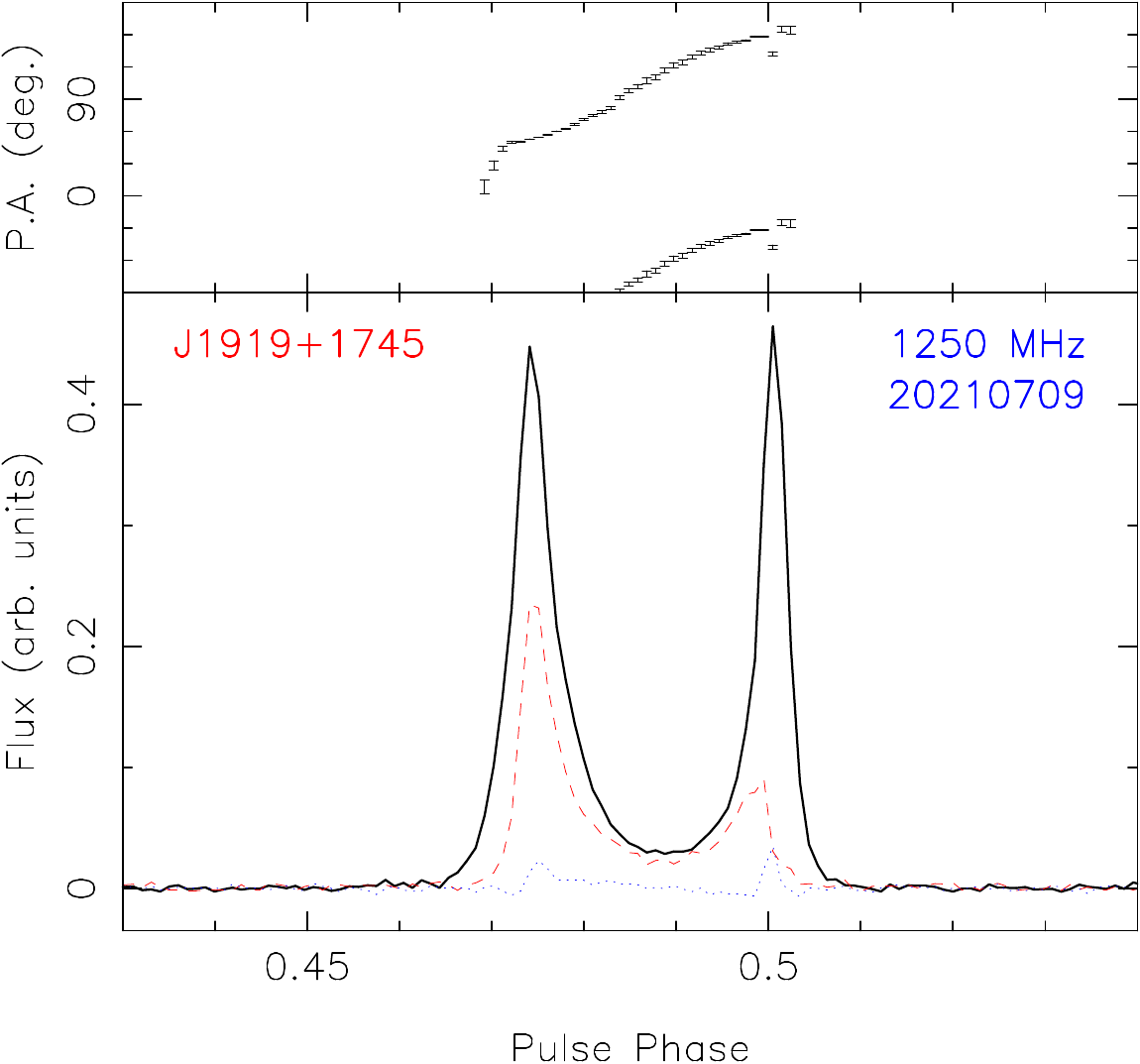} 
        \includegraphics[width=0.45\columnwidth]{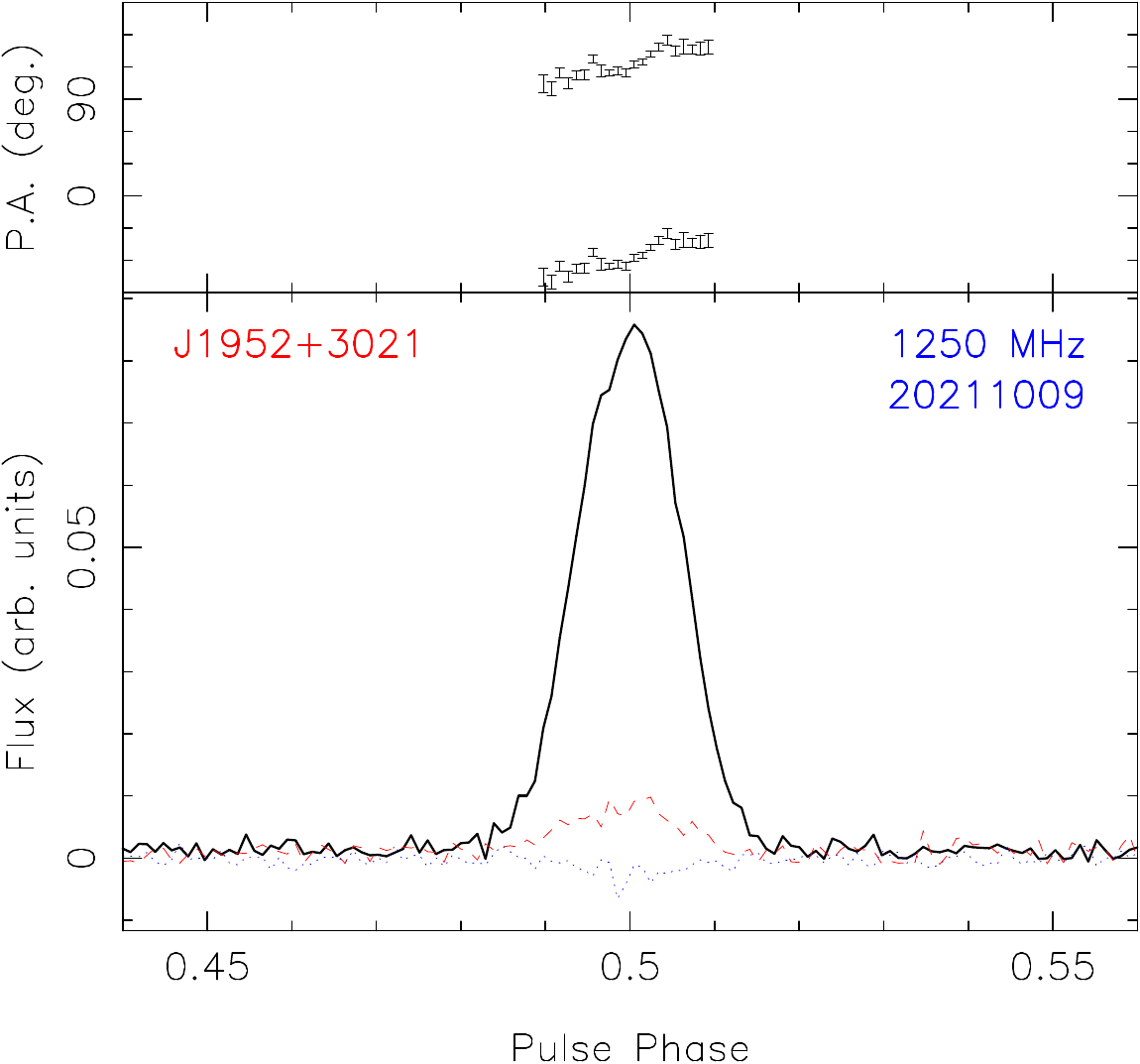} 
        \includegraphics[width=0.45\columnwidth]{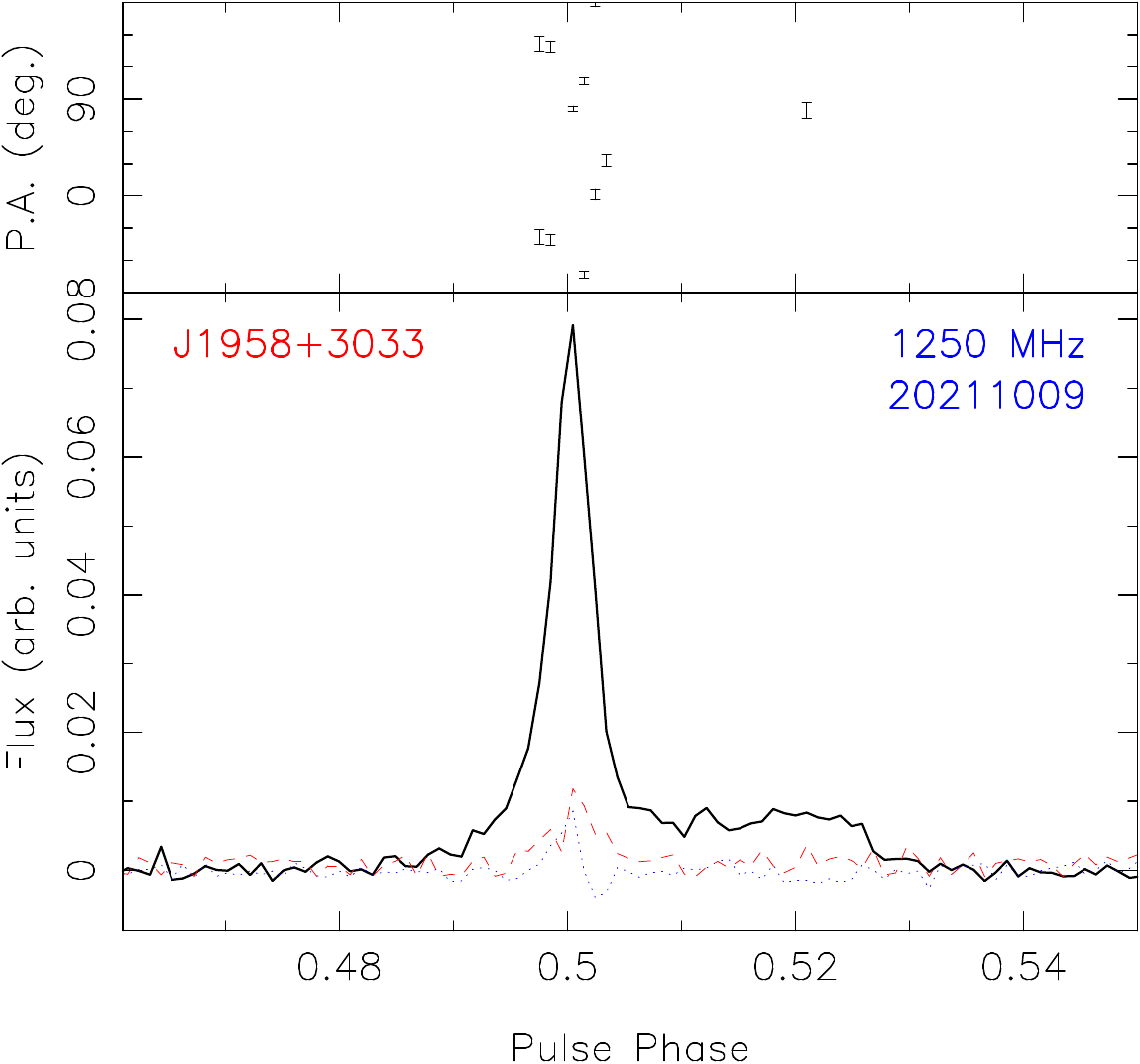} 
        \includegraphics[width=0.45\columnwidth]{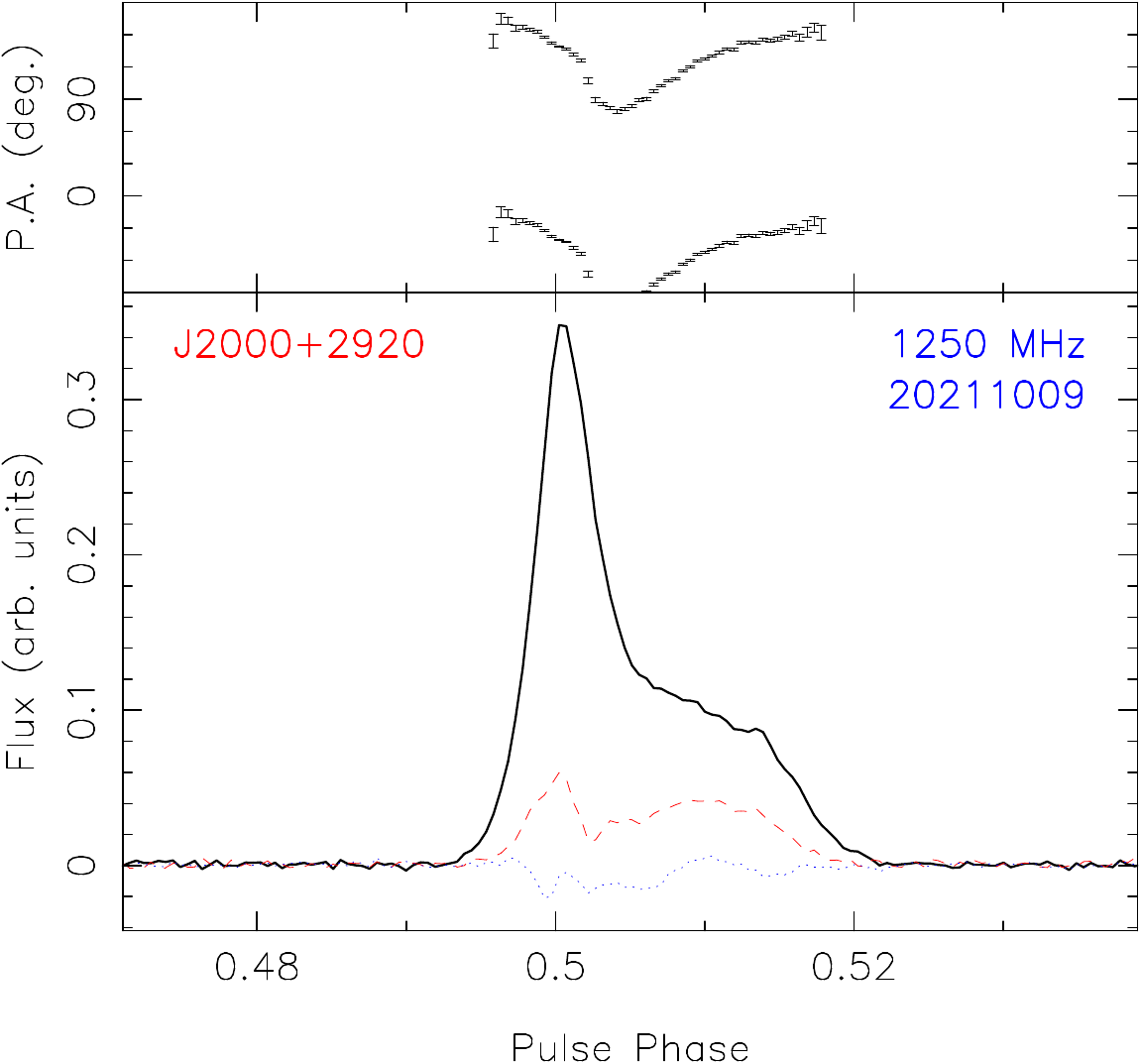} 
        \includegraphics[width=0.45\columnwidth]{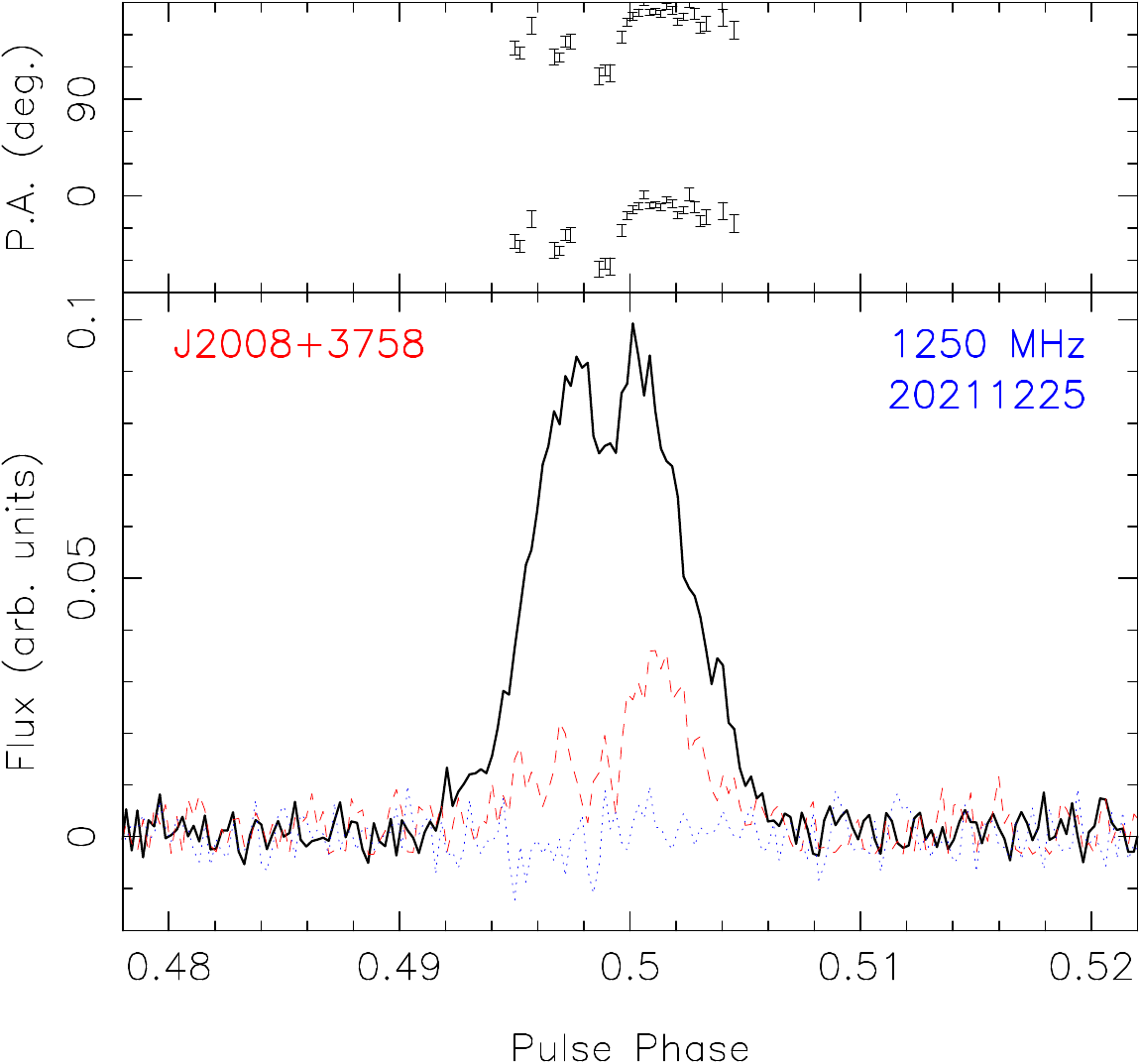} 
        \includegraphics[width=0.45\columnwidth]{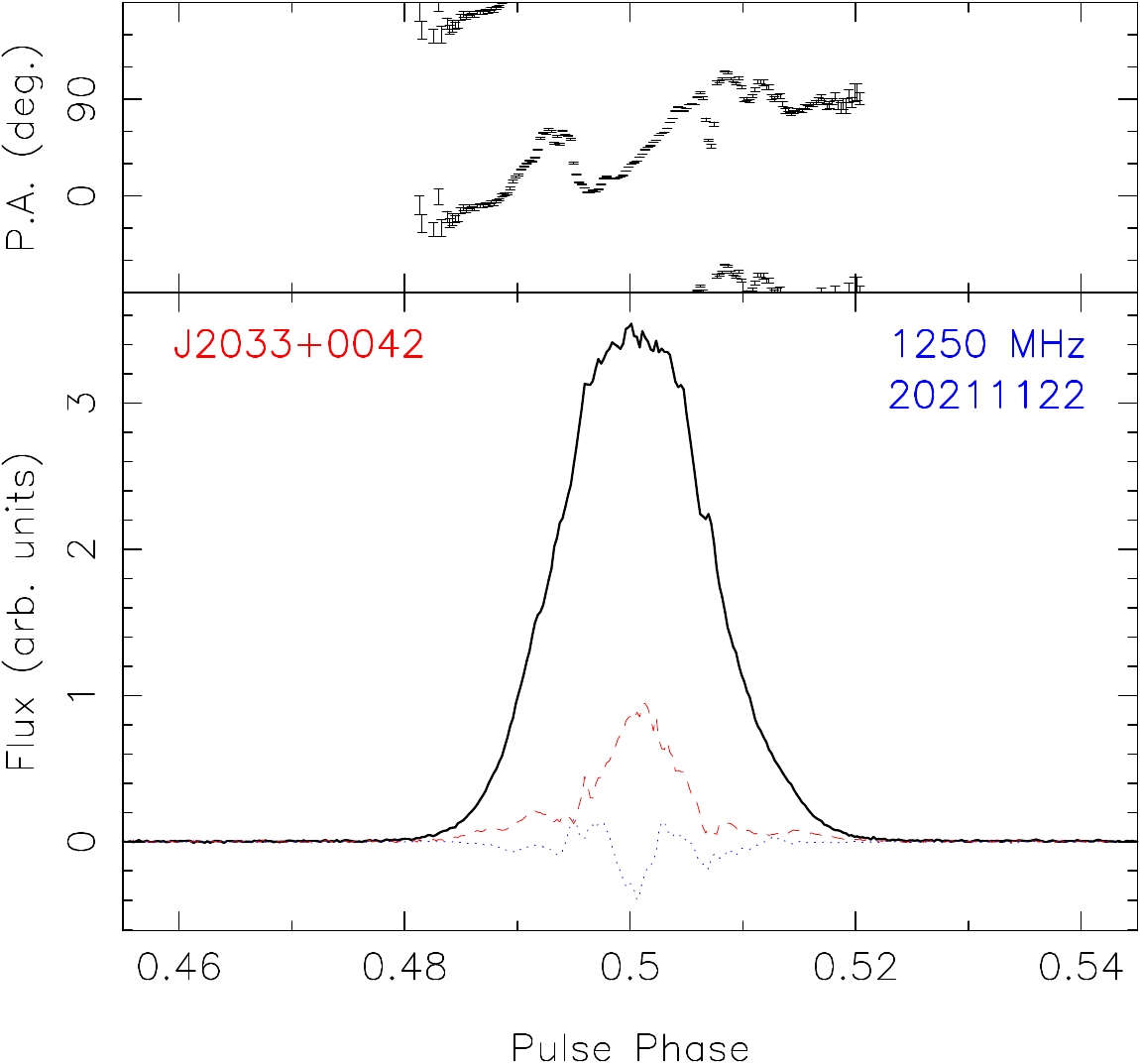} 
    \caption{The polarization angle and integrated polarization profiles of single pulses with $\rm S/N>3.0$ of known RRATs as normal pulsars.}
    \label{fig:knownRRAT1pol}
\end{figure*}

\begin{figure}[!htp]
\centering
\includegraphics[width=0.45\columnwidth]{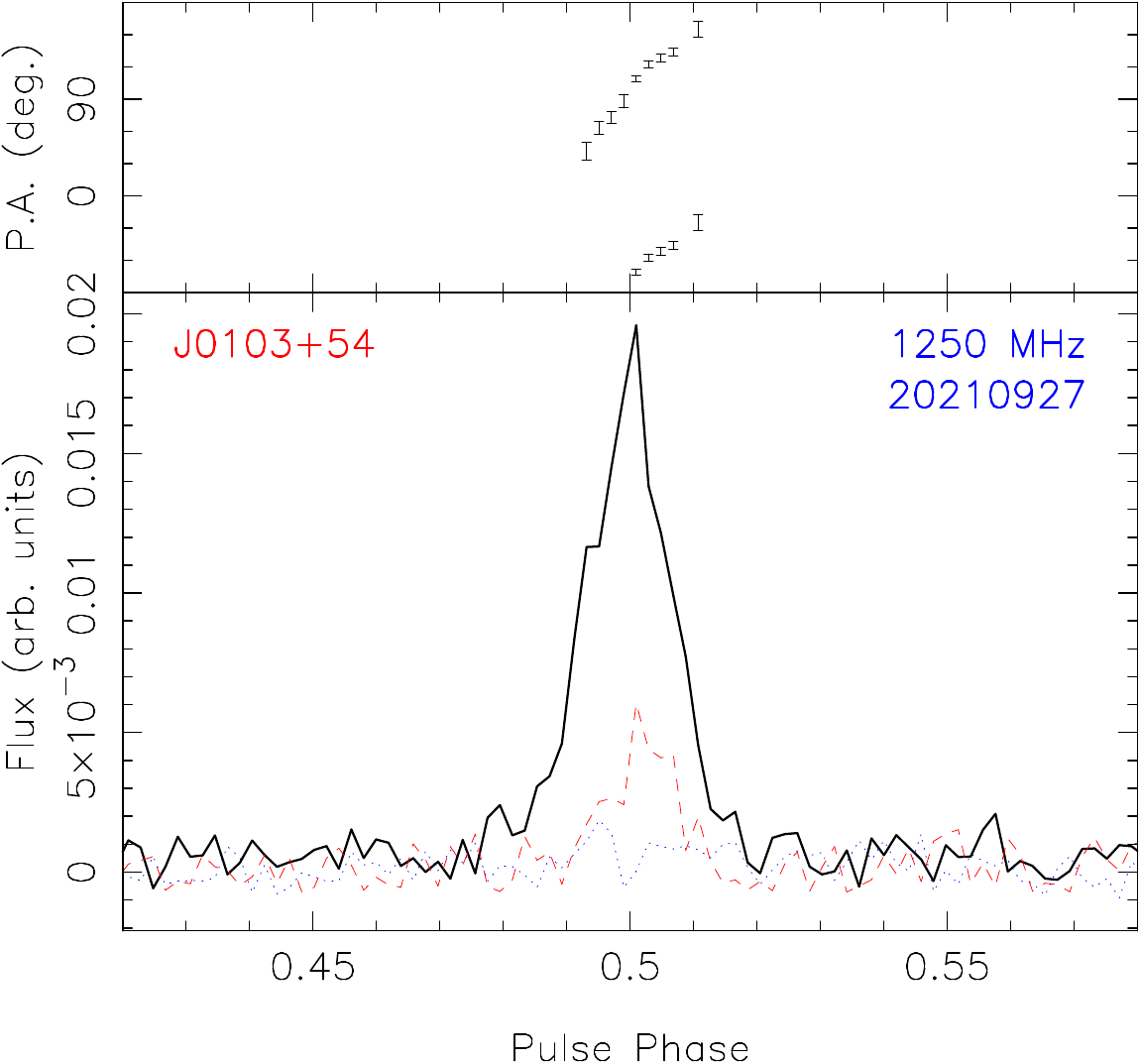} 
\includegraphics[width=0.45\columnwidth]{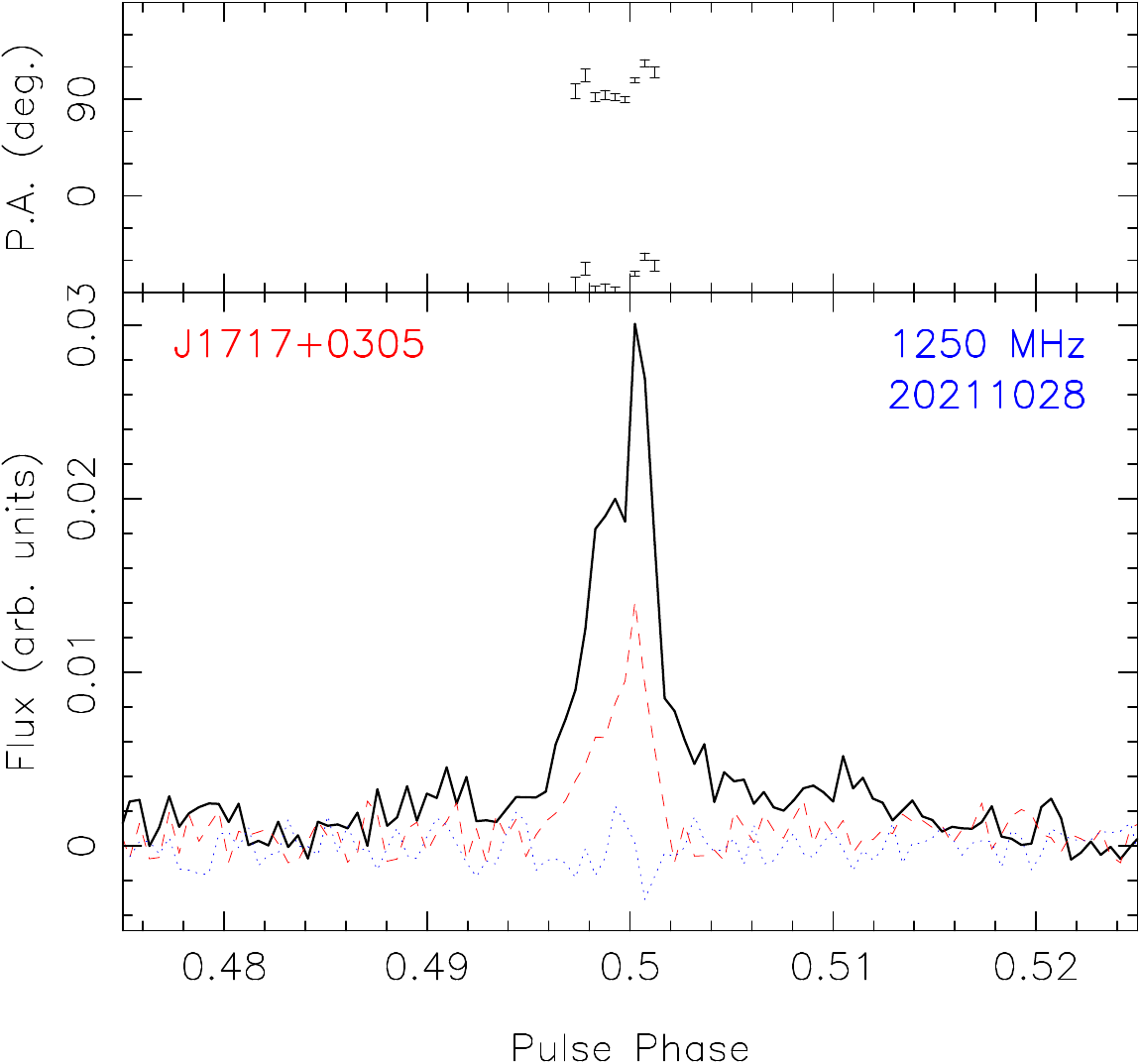} 
\includegraphics[width=0.45\columnwidth]{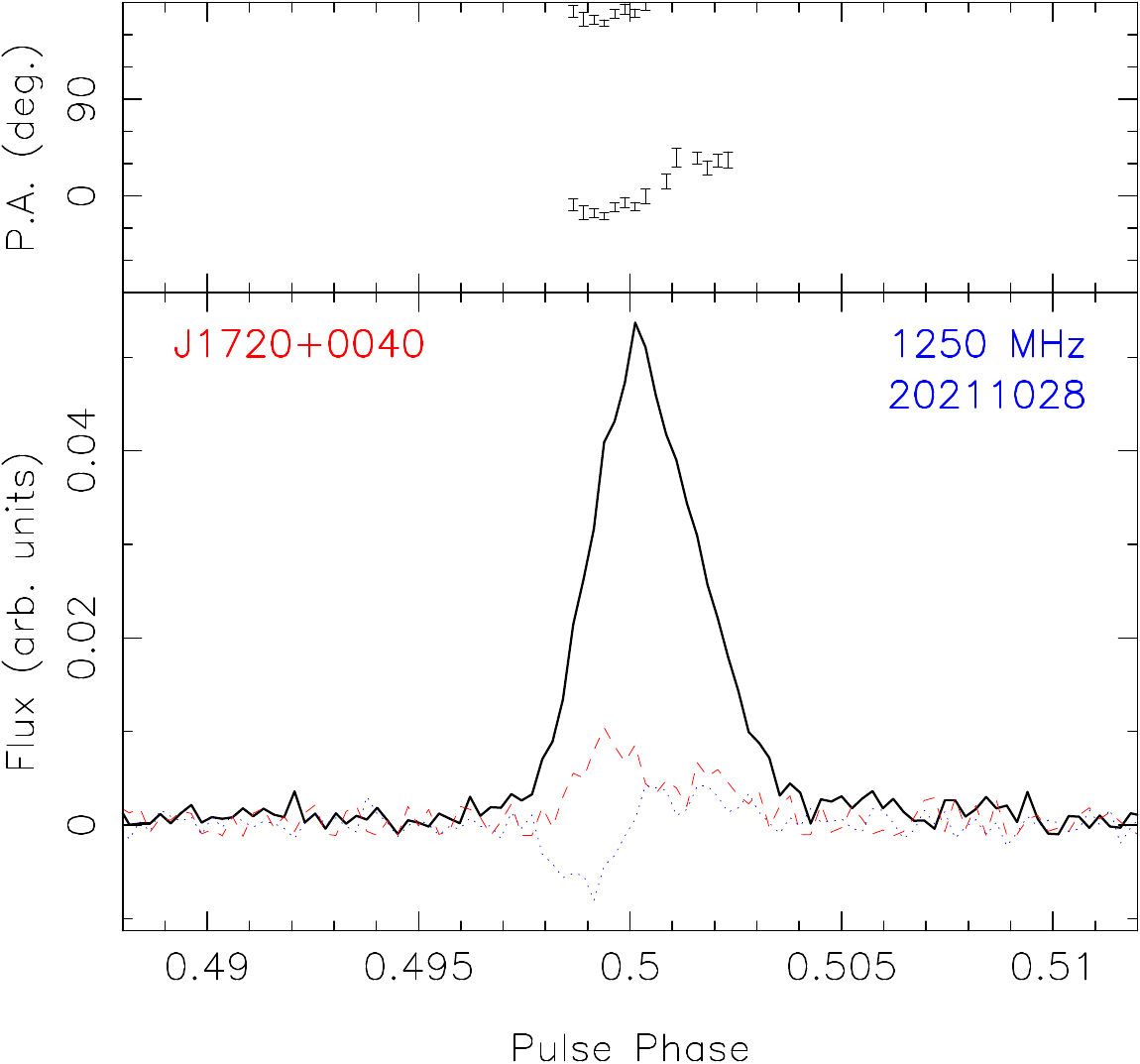} 
\includegraphics[width=0.45\columnwidth]{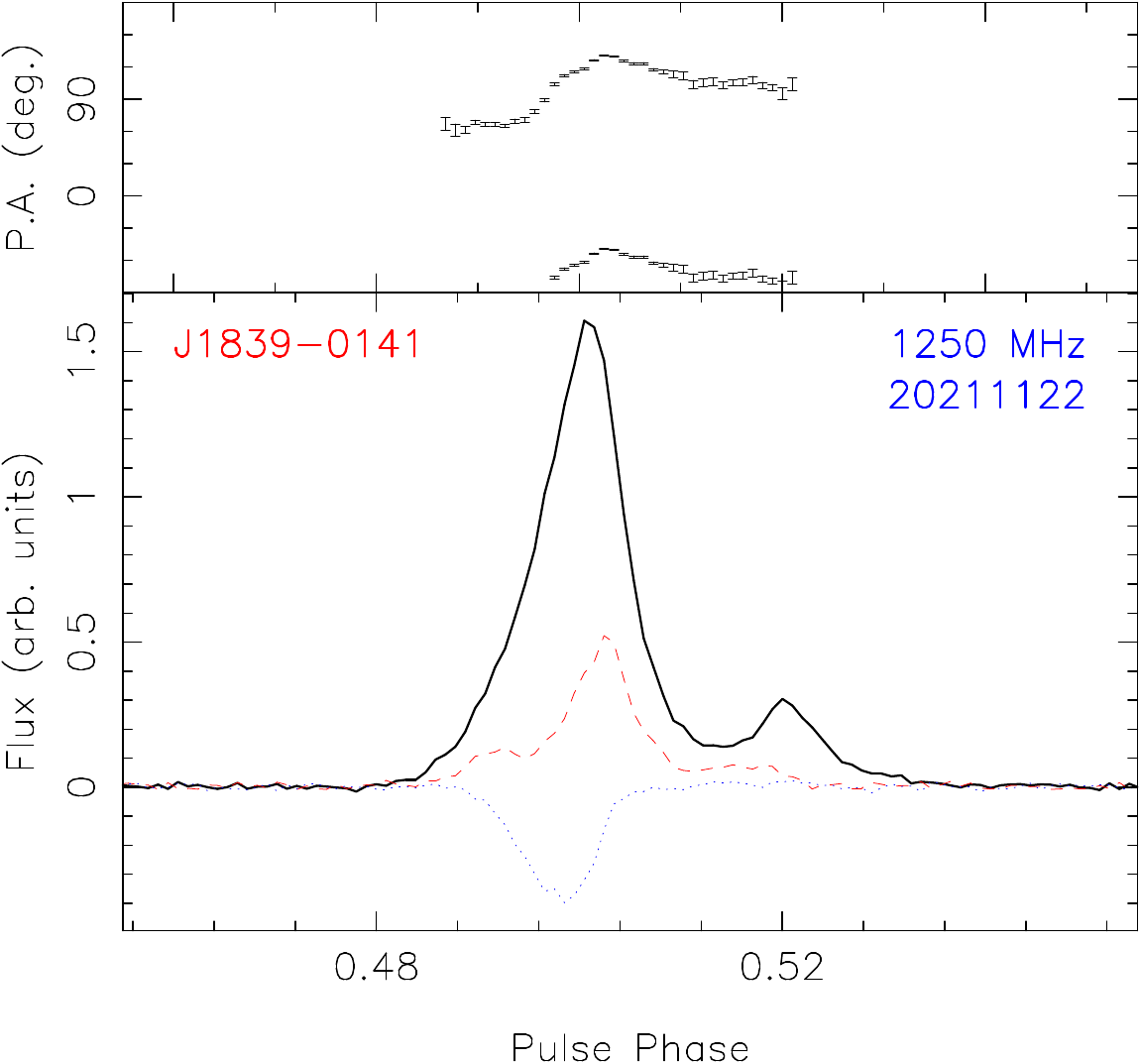} 
\includegraphics[width=0.45\columnwidth]{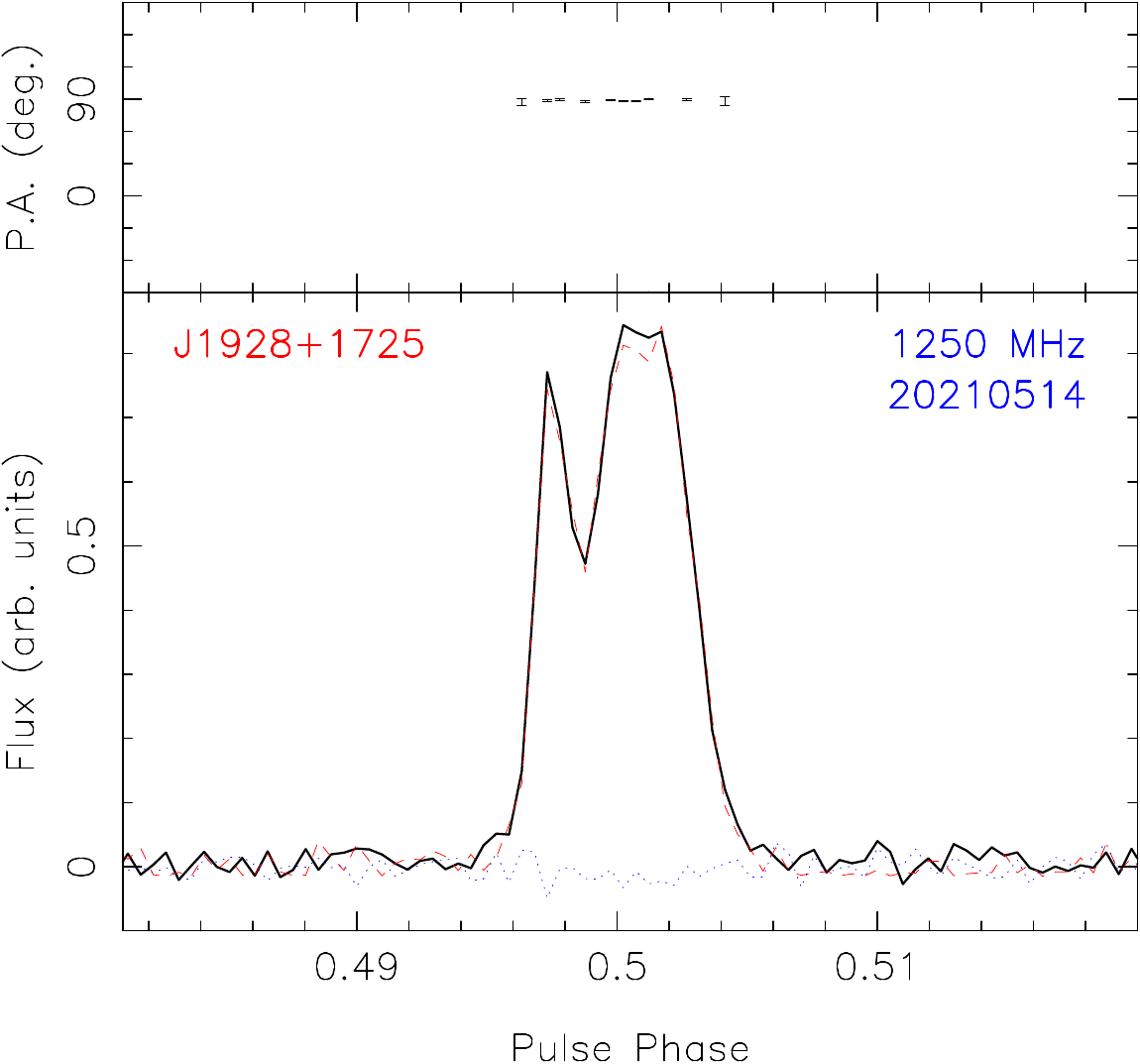} 
\caption{Same as Figure~\ref{fig:knownRRAT1pol} but for known RRATs as just pulsars with extremely nulling features. 
% The solid line is the total intensity line, the dashed line is the linearly polarized intensity line, and the dotted line is the circular polarized line. The source name and observation central frequency and date are show on the bottom panel of each one.
}
\label{fig:knownRRAT2pol}
\end{figure}

\begin{figure}[!htp]
\centering
\includegraphics[width=0.45\columnwidth]{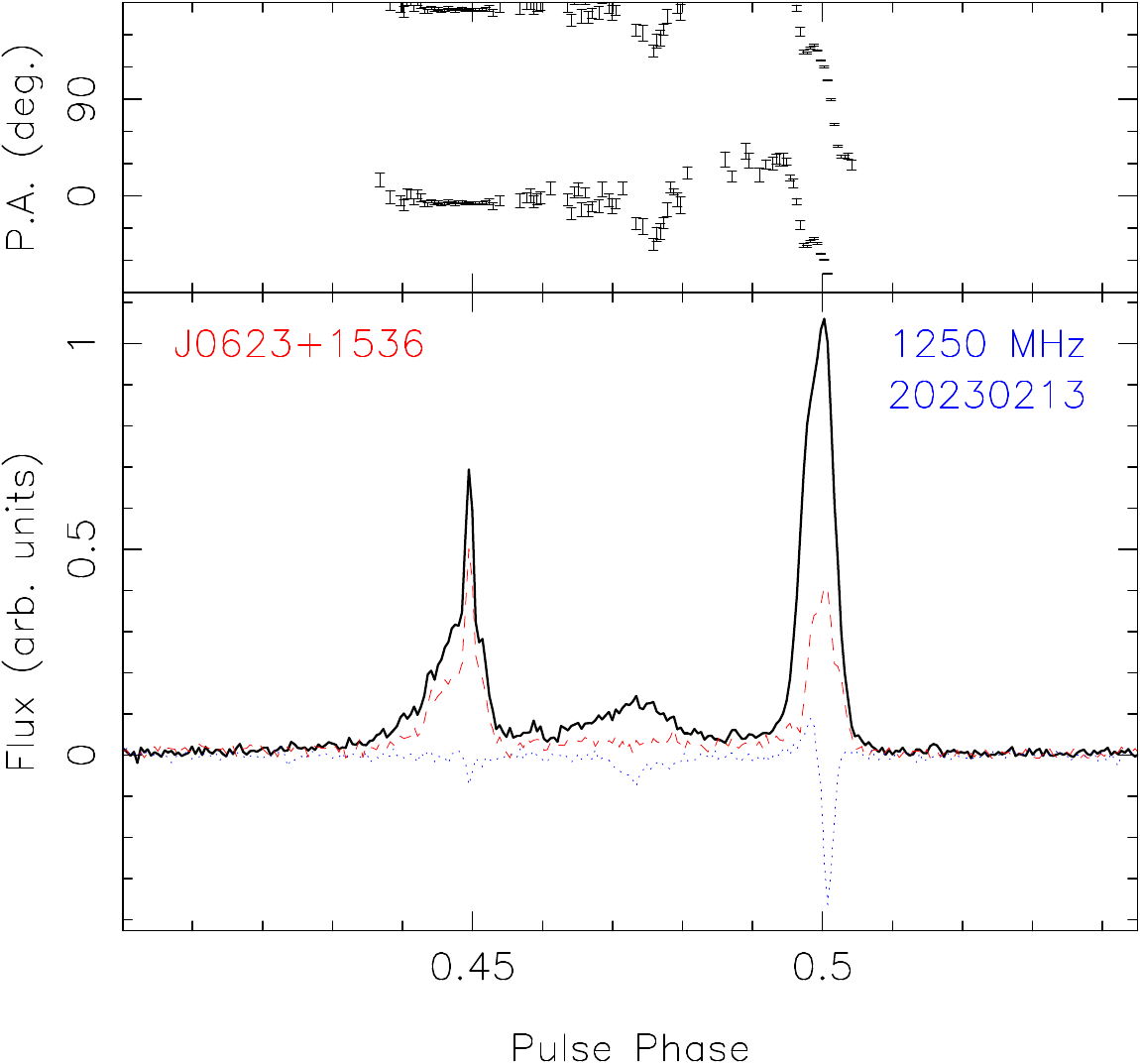}
\includegraphics[width=0.45\columnwidth]{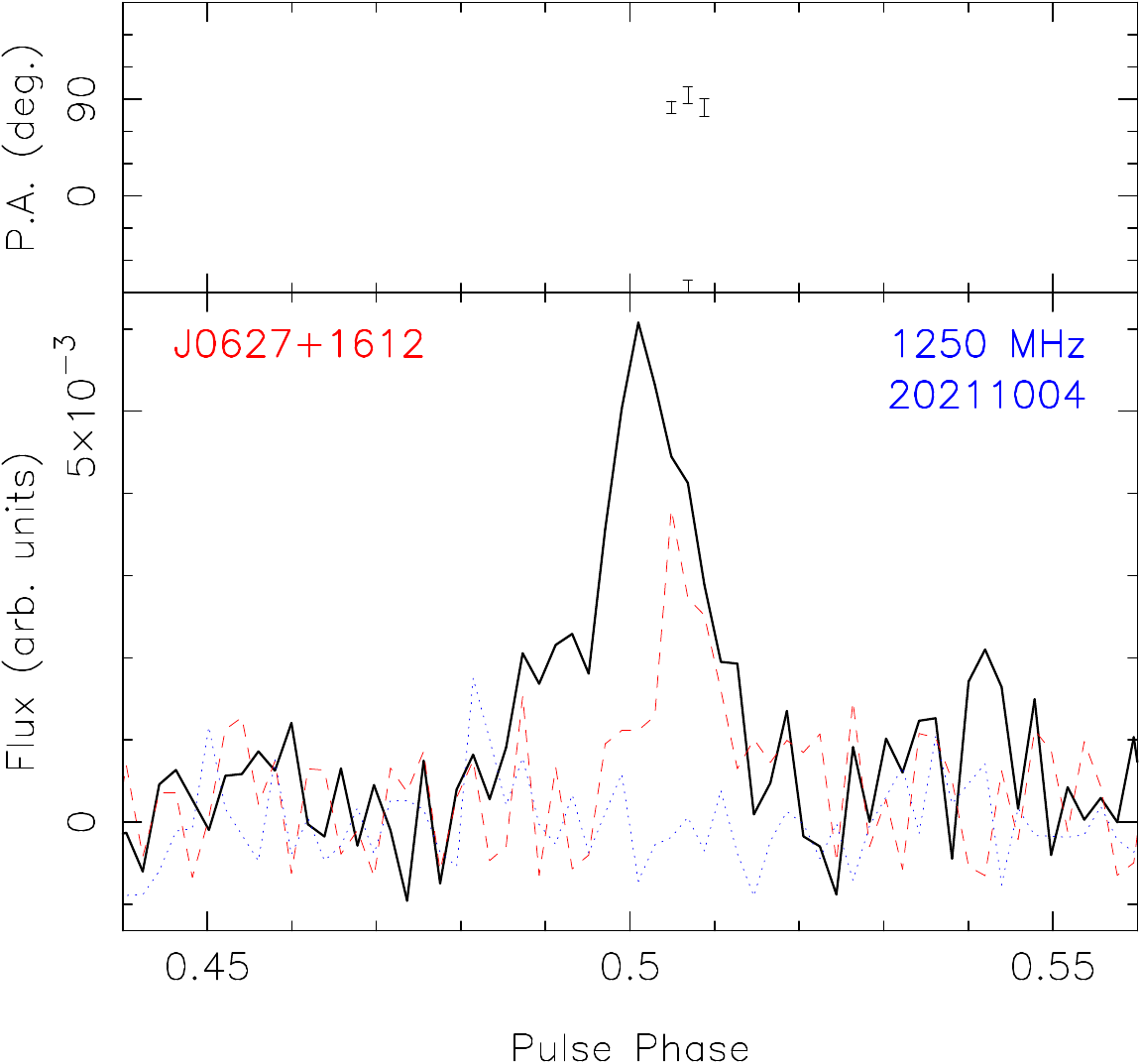} 
\includegraphics[width=0.45\columnwidth]{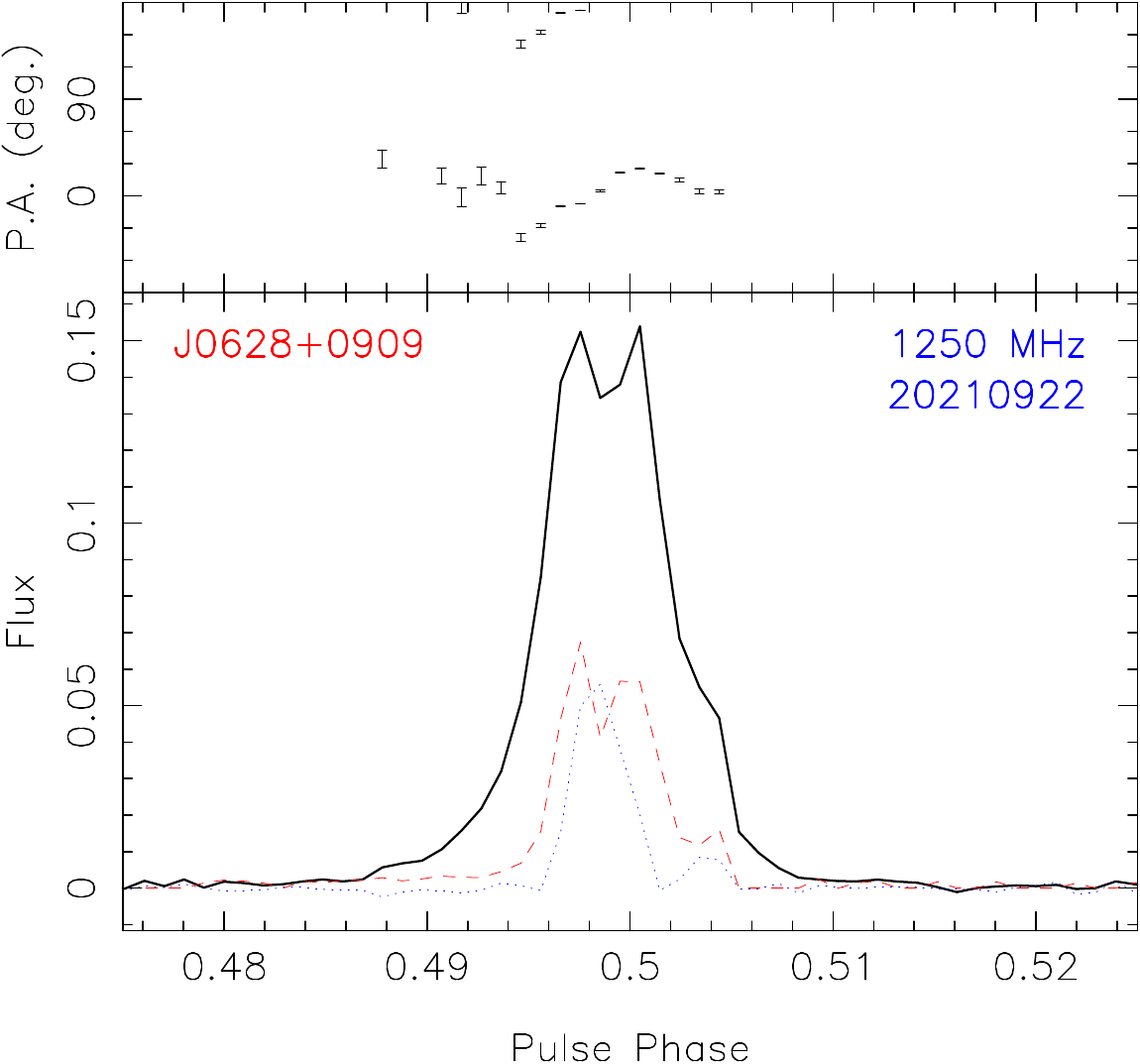} 
\includegraphics[width=0.45\columnwidth]{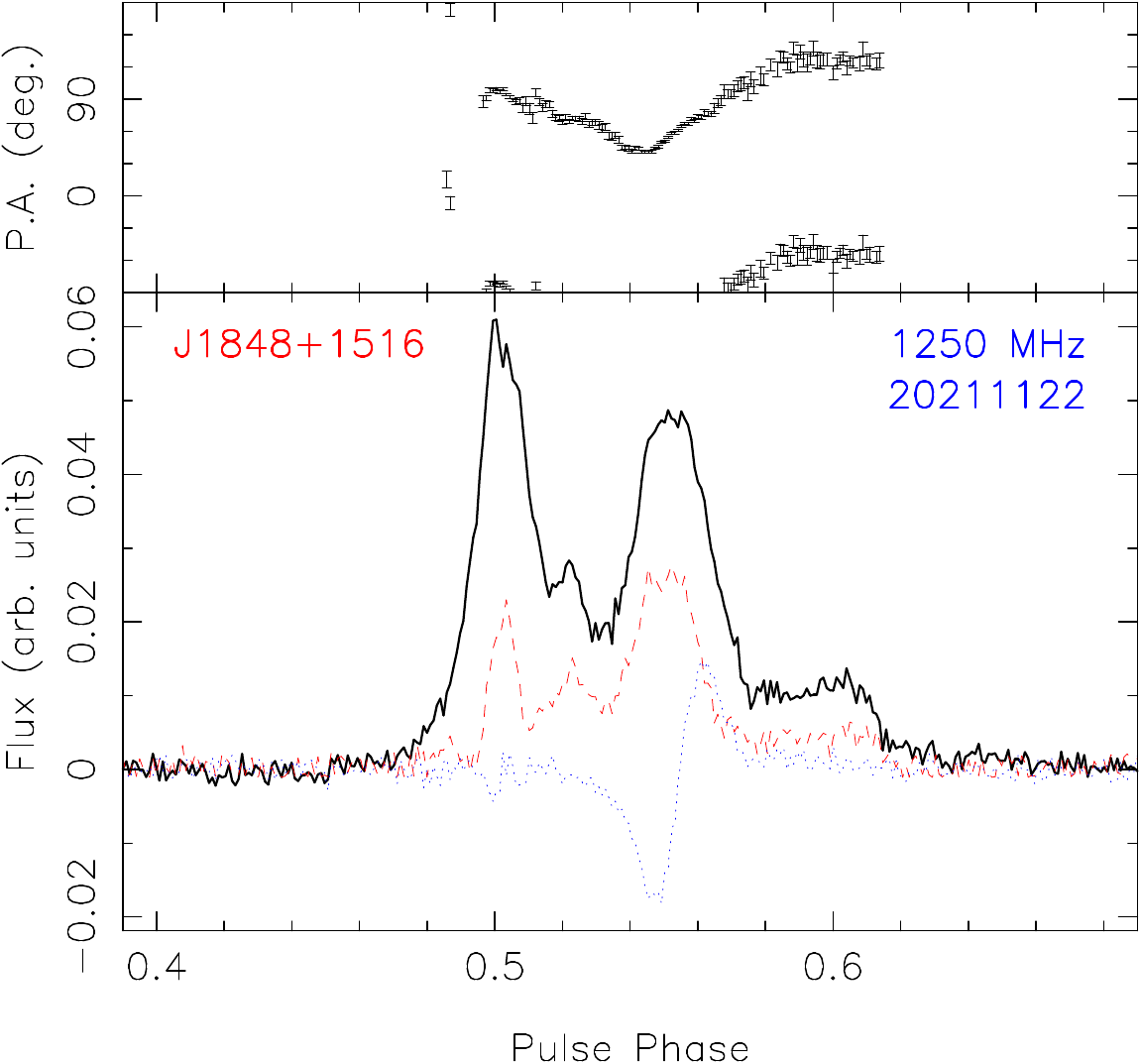} 
\includegraphics[width=0.45\columnwidth]{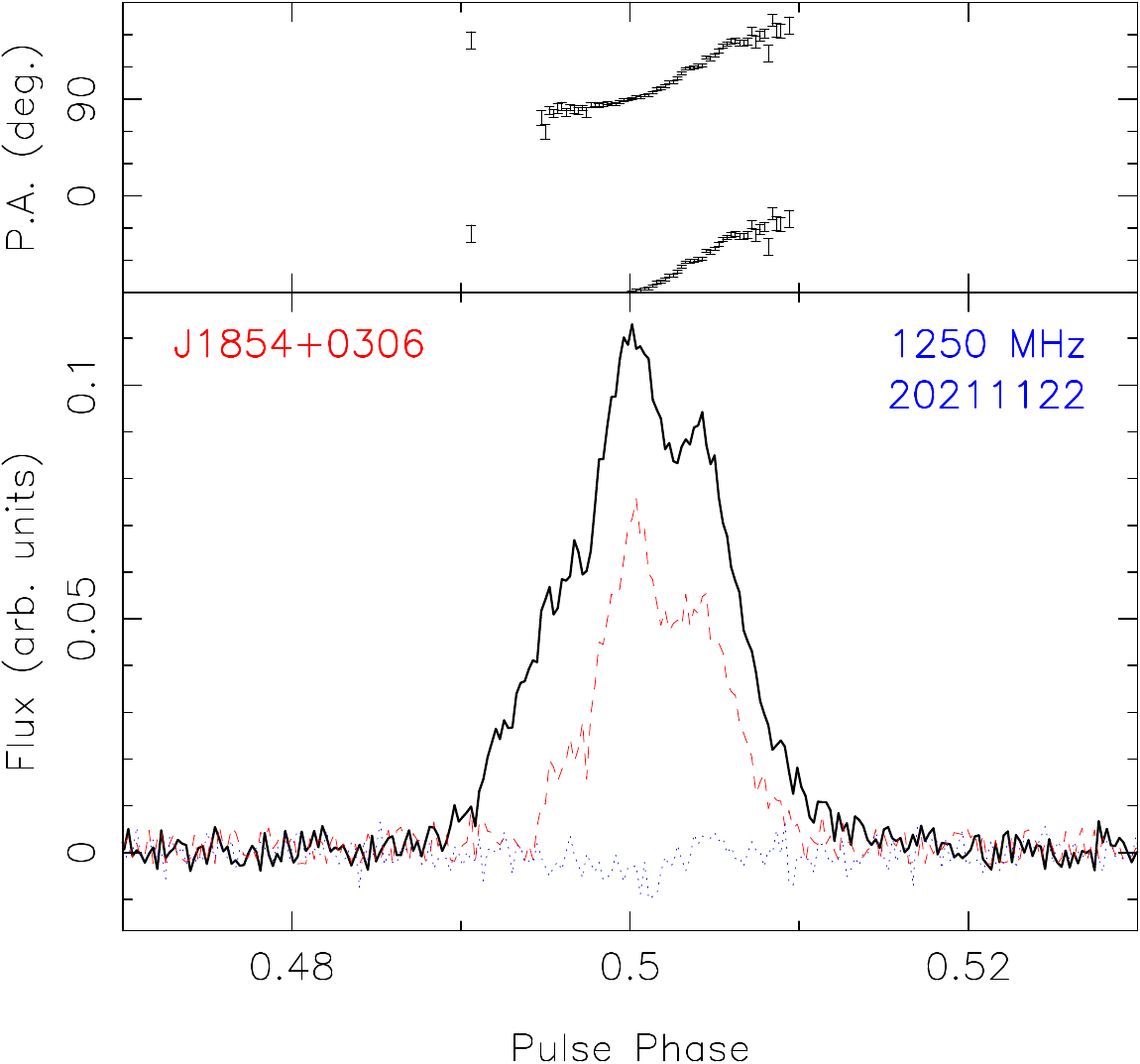} 
\includegraphics[width=0.45\columnwidth]{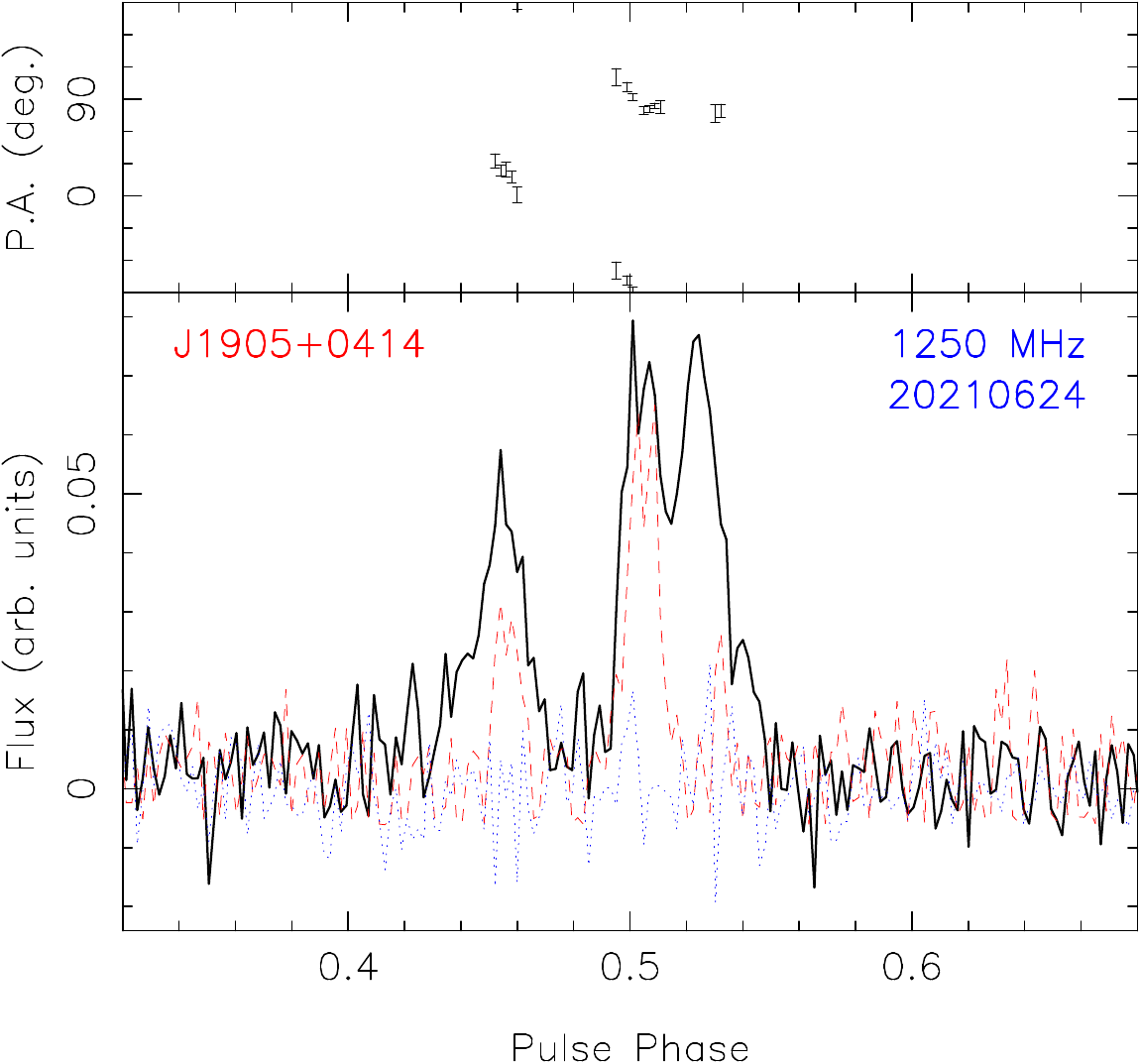} 
\includegraphics[width=0.45\columnwidth]{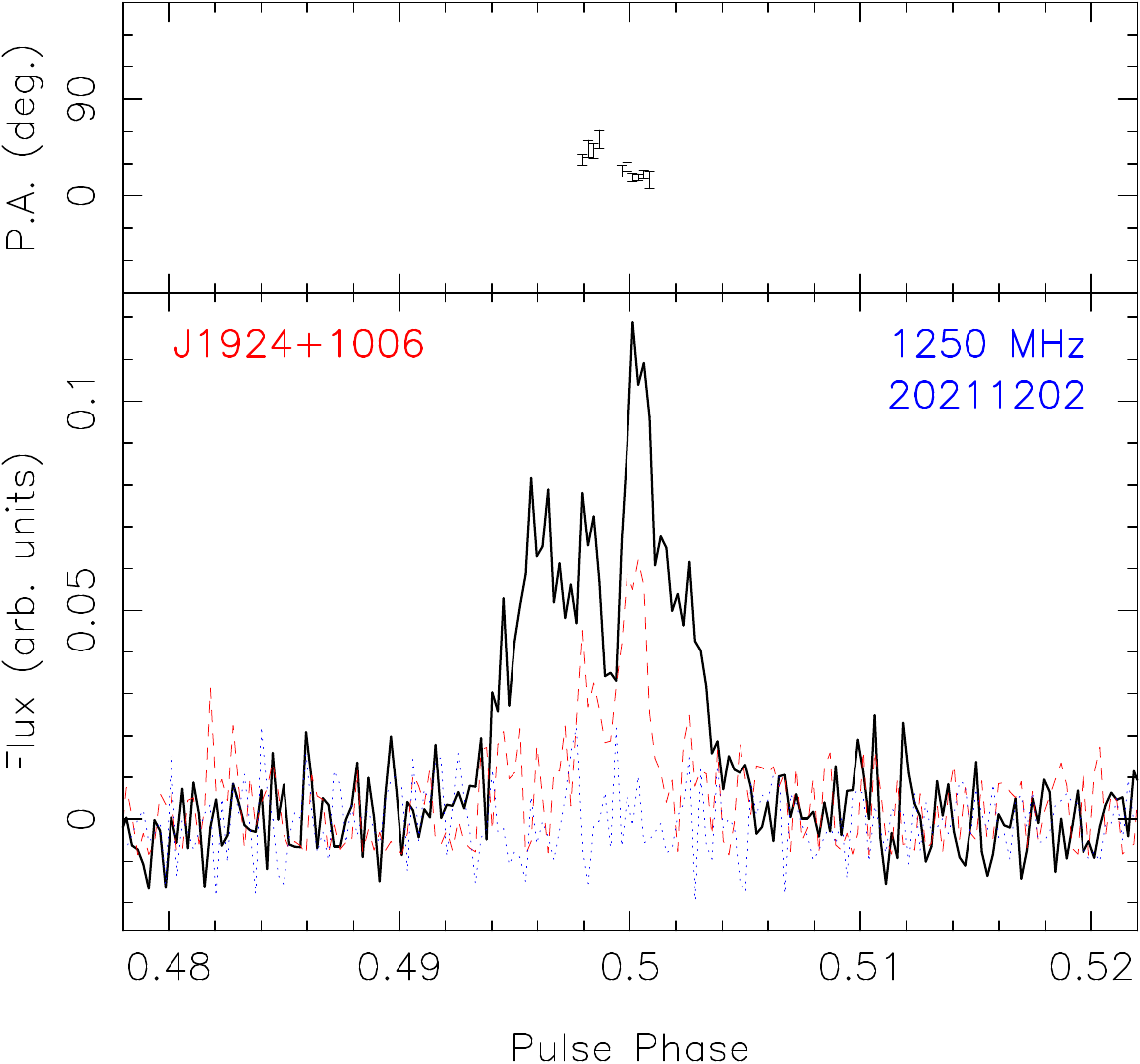} 
\includegraphics[width=0.45\columnwidth]{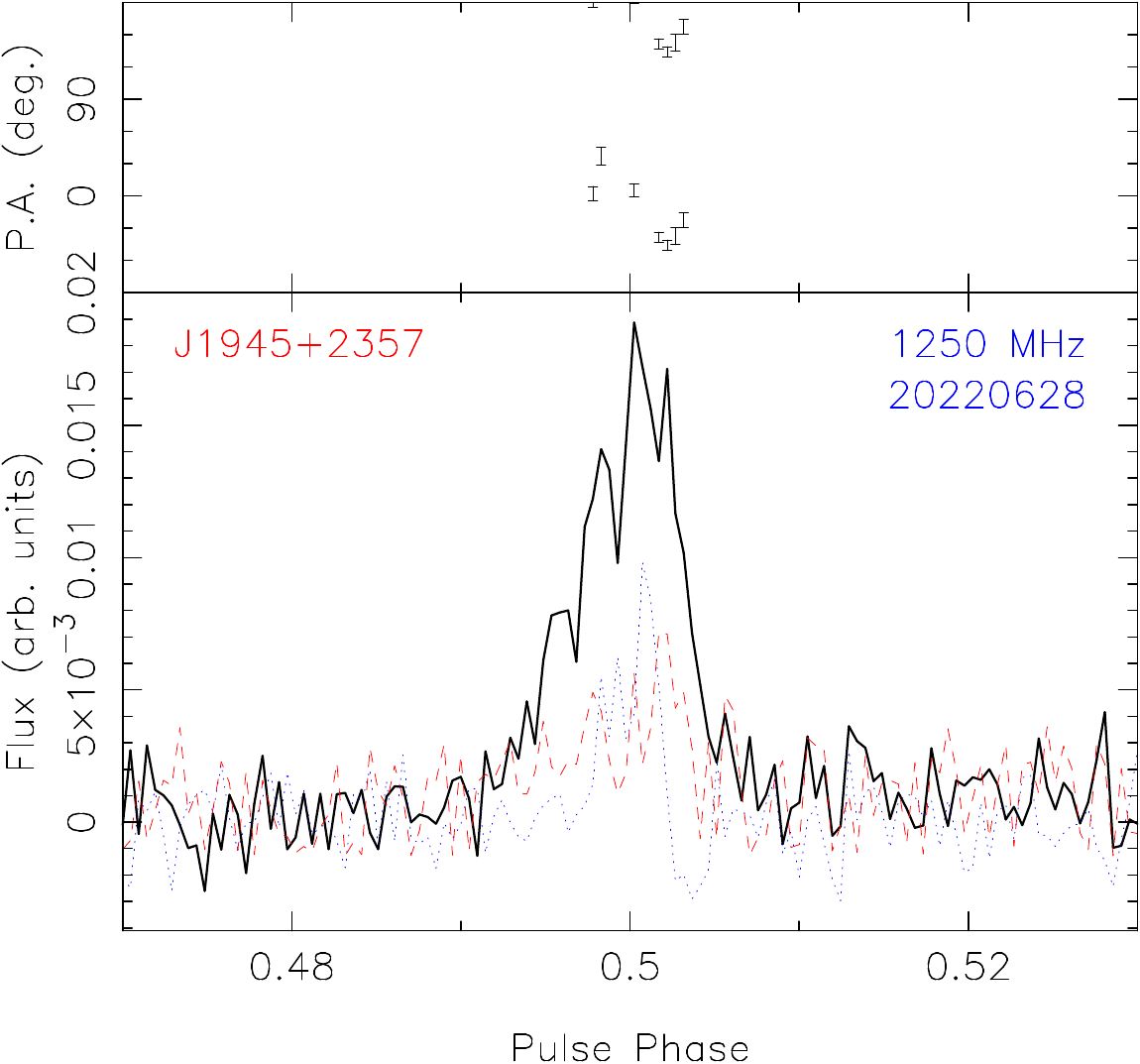}
\includegraphics[width=0.45\columnwidth]{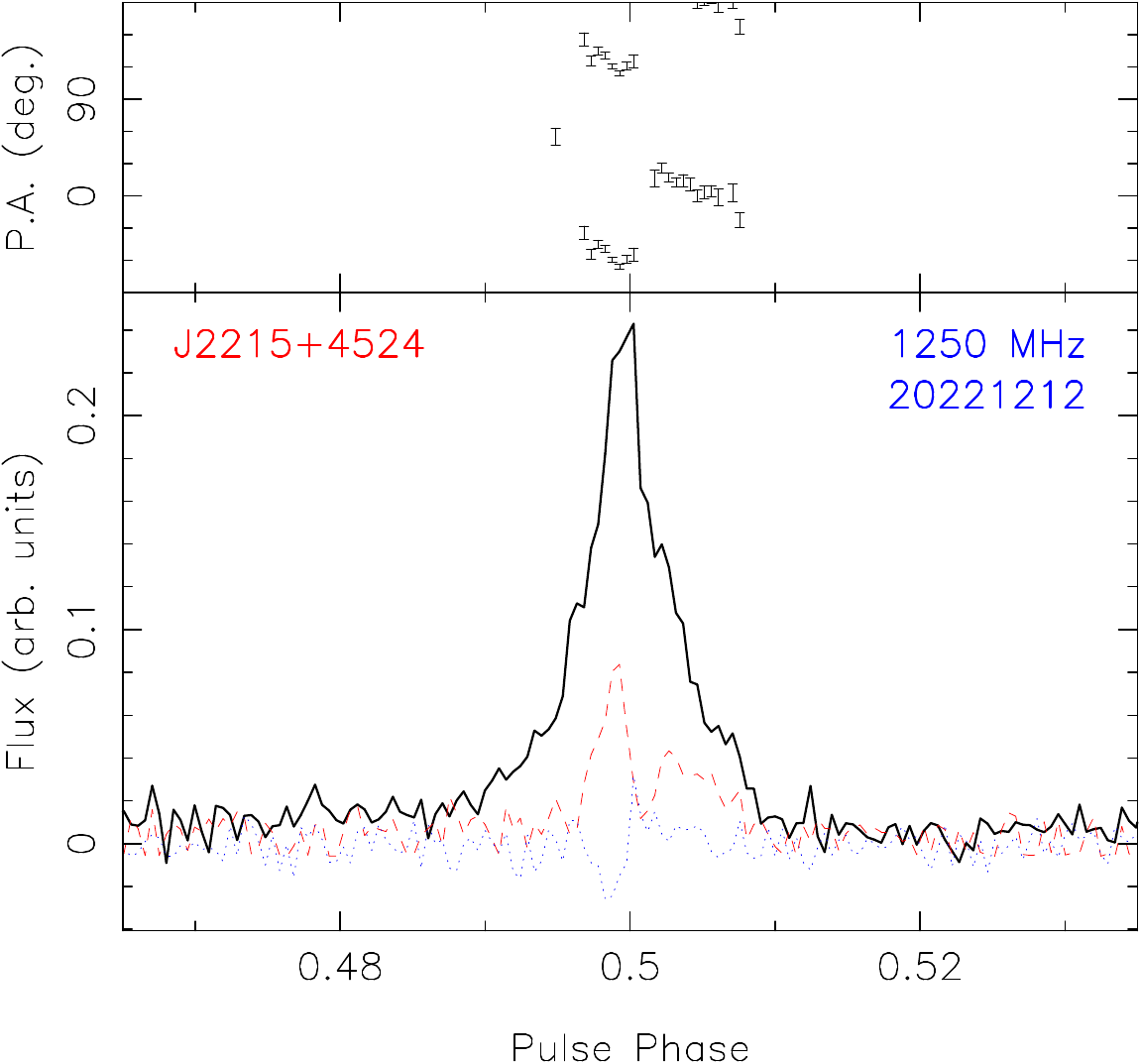}
\caption{Same as Figure~\ref{fig:knownRRAT1pol} but for known RRATs with sparse strong pulses. 
}
\label{fig:knownRRAT3pol}
\end{figure}

%%%%%%%%%%%%%%%%%%%%%%%%%%%%%%%%%% RRAT as normal Pulsars
\subsection{Known RRATs as normal pulsars}

Because of extremely high sensitivity of the FAST, individual pulses of 25 previously known RRATs are detected over a large fraction of observation time, and they appear as just normal pulsars as shown in Figure~\ref{knownRRAT1} though bright pulses or nulling features occasionally emerge. It is understandable that when the detection threshold is tens times or hundreds times worse than the current 3$\sigma$ threshold of FAST observations, only a few strong pulses would be detected, and they would definitely appear as RRATs as shown in original discovery papers \citep{Mclaughlin2006,Cordes2006,Deneva2009,Patel2018}, and a period of an RRAT may be hard to find, for example, J1849+0106 and J1908+1351 by Arecibo, see the web-page\footnote{see \url{http://www.naic.edu/~palfa/newpulsars/}}. 
%
%as ordinary pulsars with occasional small amounts of strong pulses. The average pulse profiles of the single pulses with S/N~\textless~3 in most sources of the group show that there are still significant residual pulse structure, implying the presence of weak emission for these known RRATs in these cycles and they are difficult to be detected by the single pulse search modules.
%
%J0608+1635 show two components in its averaged profile that the left component shows the continuous emission of ordinary pulsar, while the right component has intermittent emission like RRAT. 
%By FAST, they are detected as a normal pulsar, with occasionally strong pulses. 

Some interesting features are revealed in our FAST observations. For example, nulling for a small fraction of periods in the single pulse series is observed for J1919+1745, and even one weak dwarf pulse has been detected (see Fig.~\ref{fig:APPknownRRAT1}). Some pulsars are very normal so that we can see a quasi-periodic intensity modulation in their single pulse sequence, such as J1915+0639 and J1952+3021, or subpulse drifting, such as J1843+0527 and J1952+3021. Analyses of the single pulse behaviors for a large number of pulsars will be reported by Y. Yan et al.(2023, in preparation).

%%%%%%%%%%%%%%%%%%%%%%%%%%%%%%%%%%%%%%%%%%%%%%%% RRAT as very nulling Pulsars
\subsection{Known RRATs as extremely nulling pulsars and  pulsars with sparse strong pulses}

Five previously known RRATs appear as extremely nulling pulsars in the FAST observation as shown in Figure~\ref{knownRRAT2}. For example, the PSR J0103+5354 and J1839$-$0141 have only one, J1717+0305 and J1928+1725 have two brief continuous radiation episodes, and J1720+0040 have several interspersed emission fractions in our observations. Outside the active episodes, no detectable pulse is found in FAST data.

%\subsection{Known RRATs as generally very weak pulsars with sparse strong pulses}

We have got the other 13 known RRATs observed by the FAST, and they generally appear as many sparse strong pulses as shown in Figure~\ref{knownRRAT3} and also shown Figure~\ref{fig:APPknownRRAT3}, much more pulses than in their original discovery papers. They `must' be RRATs in the eyes of other telescopes. In our FAST observations, we can get them easily detected not only by single pulse search module, and most of them can be picked out through normal pulsar searches. 

Some objects are worth to give a note.
PSR J1720+0040 is an interesting RRAT. We detect 7 pulses during a 305~s observation (see Figure~\ref{knownRRAT3}), and we fortunately can get its period as being 3.356875~s. We can determine the position according to the position of the snapshot beam with significant pulse signal detection.
The occasional brightened state, for example, has been recently detected for PSR J1938+2213 \citep{Chandler2003PhDT, Sun2022ApJ}, which is shown in Figure~\ref{knownpulsarRRATlike}. The single pulses in the bright state are suddenly 100 times brighter. If it is observed by a small telescope, probably only the brightened part would be observable, it could be shown as an extremely nulling pulsar, like PSR J1839$-$0141 \citep{Lu2019SCPMA} and PSR B0823+26 \citep{Sobey2015MNRAS} which have a `bright' mode that regularly emits bright pulses and a `quiet' mode that occasionally emits weak pulses. PSR J1946+1449 (J1946+14) \citep{Deneva2016} is also shown RRAT-like pattern that emission is intermittent, and we detect few bright pulses. If they are observed by a not-so-sensitive telescope, they must be classified as RRATs as being pulsars with sparse strong pulses.

J0623+1536 is a known RRAT J0623+15 discovered in PALFA survey \citep{Patel2018} with the previopusly given DM of $92.5\pm1.6~cm^{-3}~pc$. In the 15-minute observation on 20230213, ten single pulses have been detected. We analyze and get a DM of $\rm~92.77~\pm~0.24~cm^{-3}~pc$ and a period of about 2.638545(18)~s. Interestingly, the single pulse at the period No.330 shows the emission frequency drifting in the dynamic spectrum, while other pulses are very normal without any drifting, see Figure~\ref{J0623+1536-2sp}. The dynamic spectrum of this anomalous pulse shows well-dedispersed structures with the DM of $\rm~92.77~cm^{-3}~pc$. Such a phenomenon has been identified for some bursts of FRBs \citep{Hessels2019ApJ,ZhouDJ2022RAA} but rarely seen in pulsars. The previously known case is the emission drifting structure of PSR J0953+0755 at low frequency~\citep{Bilous2022AA}.

\begin{figure*}
    \centering
    \includegraphics[width=0.24\textwidth]{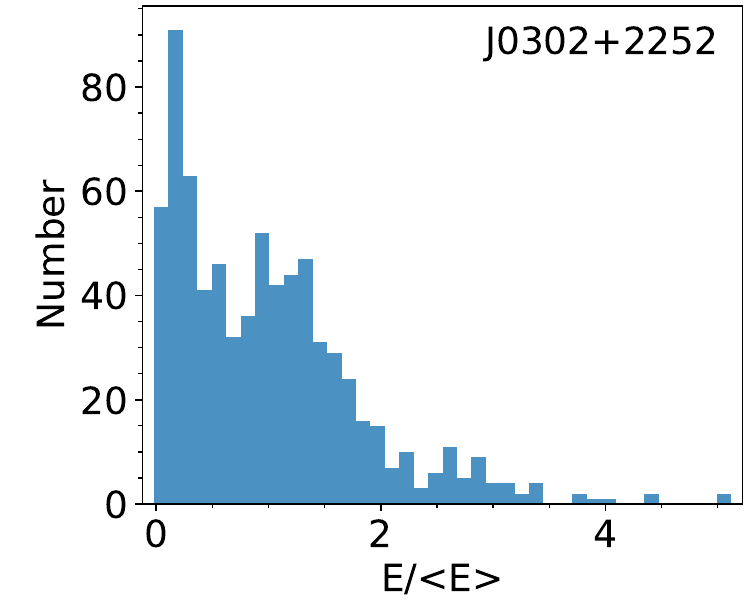} 
    \includegraphics[width=0.24\textwidth]{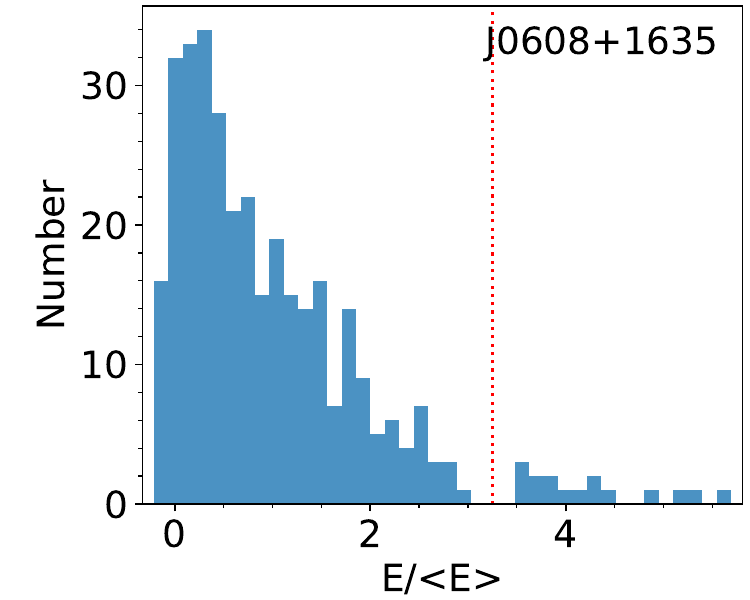} 
    \includegraphics[width=0.24\textwidth]{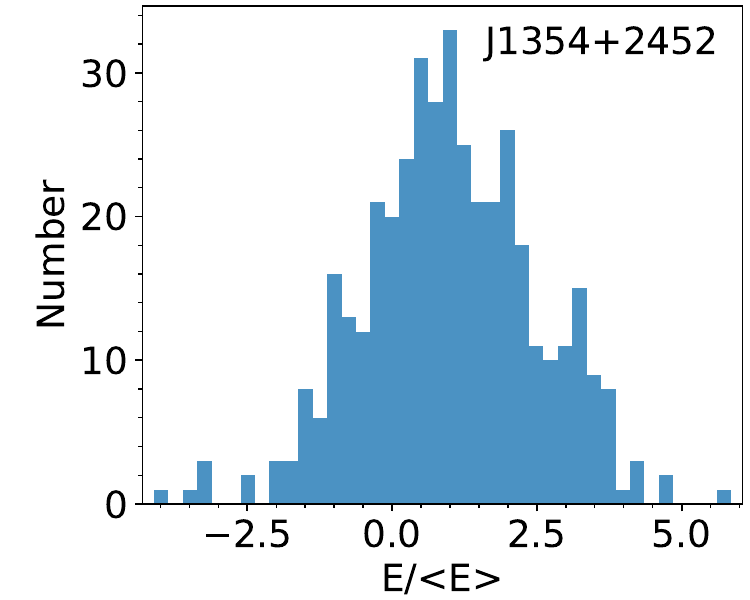} %J0630+1933
    \includegraphics[width=0.24\textwidth]{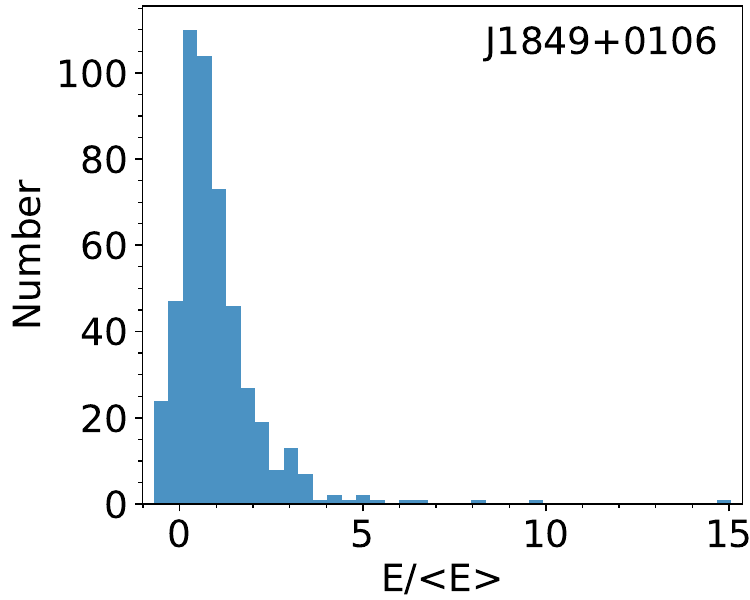}
    \includegraphics[width=0.24\textwidth]{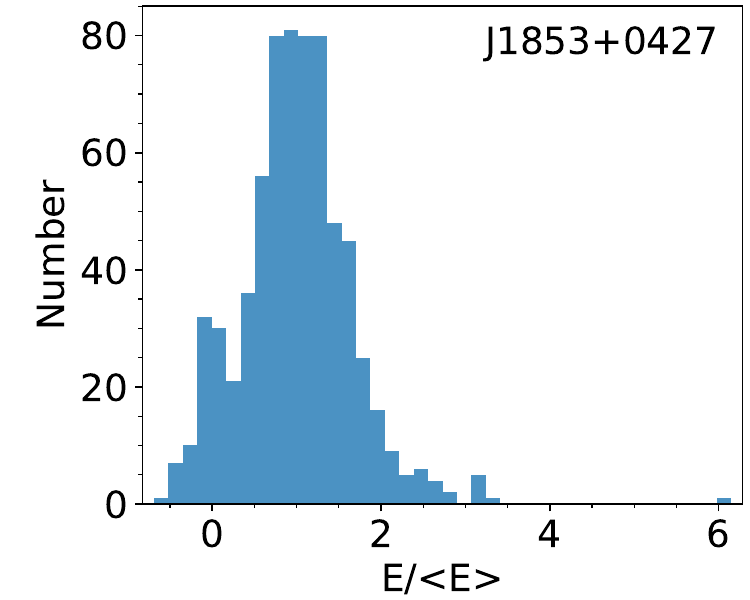} 
    \includegraphics[width=0.24\textwidth]{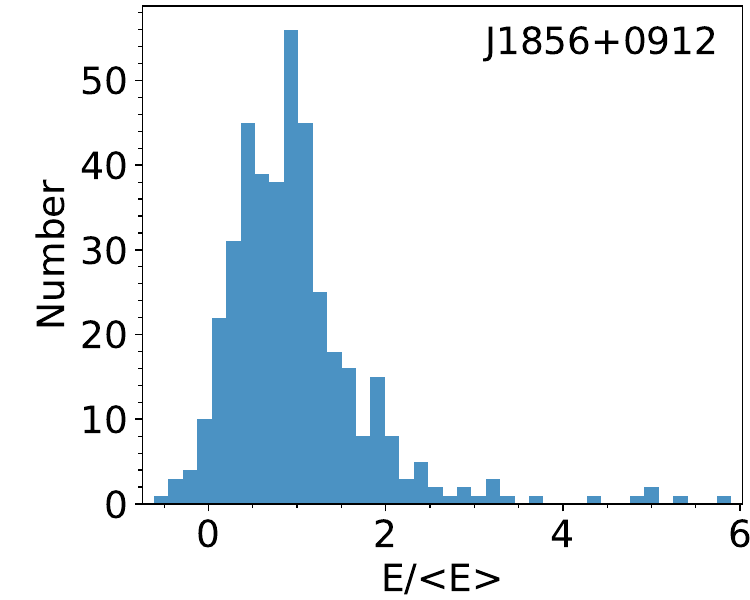} 
    \includegraphics[width=0.24\textwidth]{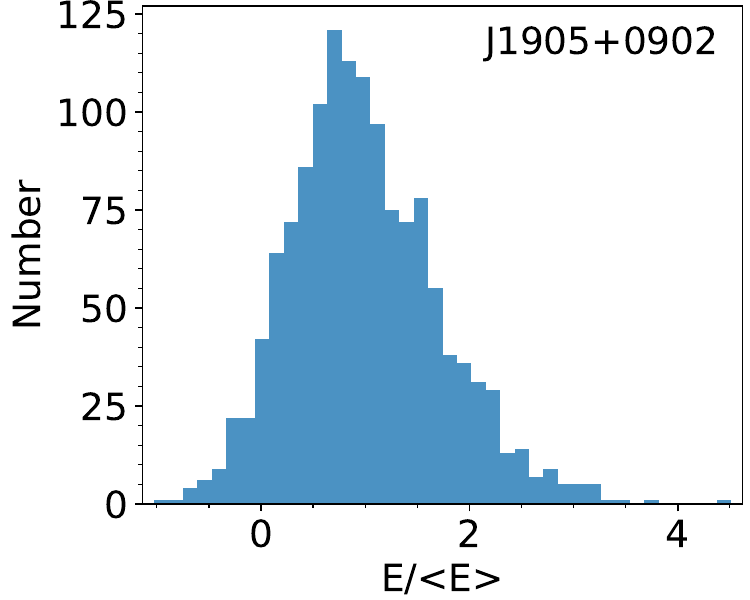} 
    \includegraphics[width=0.24\textwidth]{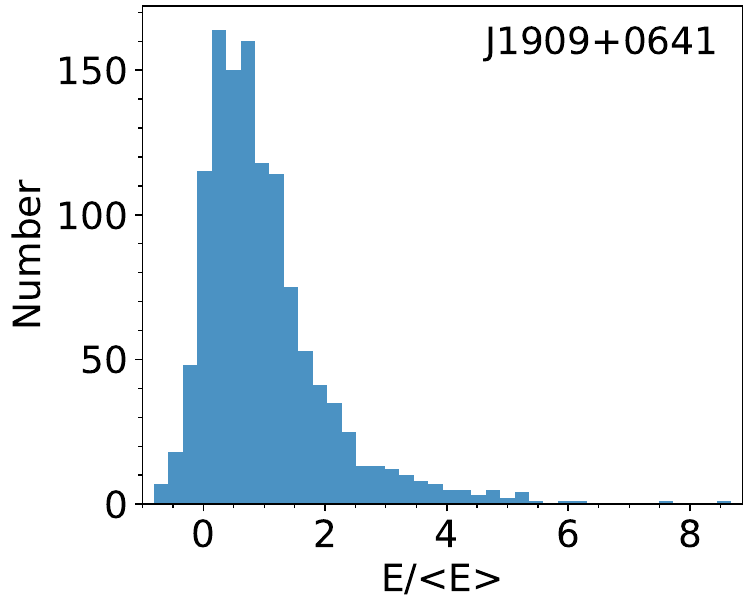} 
    \includegraphics[width=0.24\textwidth]{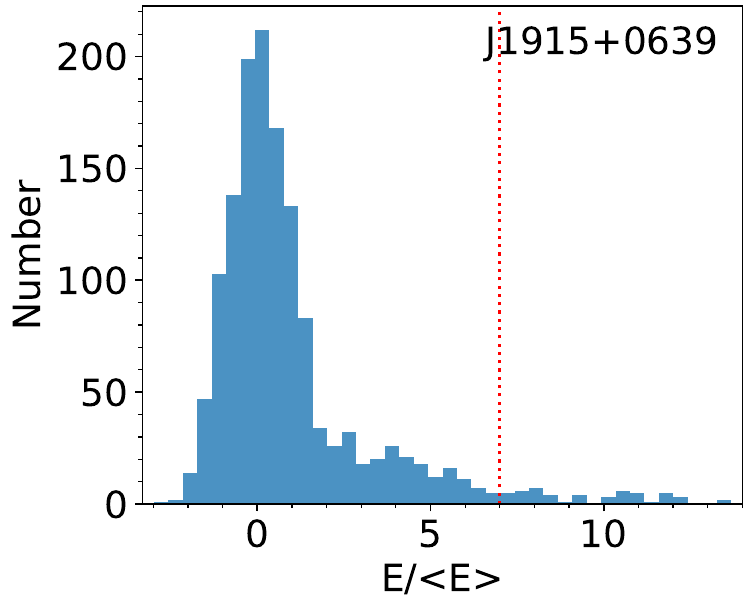} 
    \includegraphics[width=0.24\textwidth]{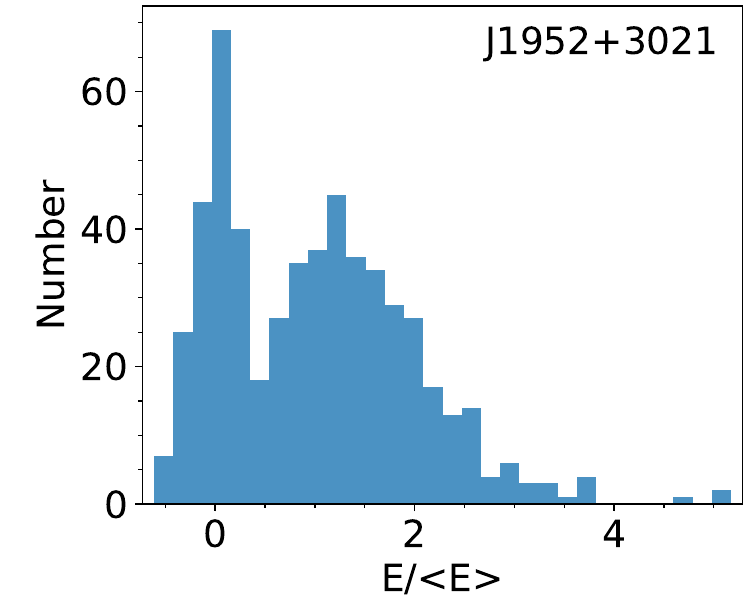} 
    \includegraphics[width=0.24\textwidth]{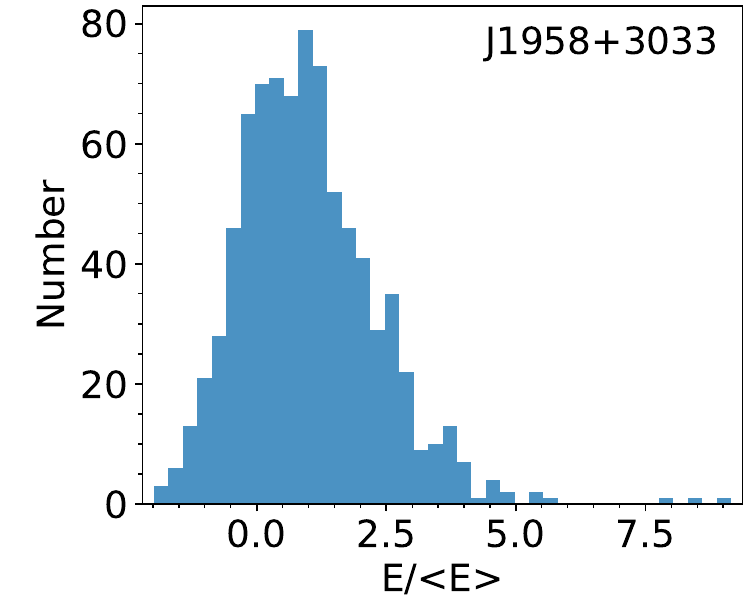} 
    \includegraphics[width=0.24\textwidth]{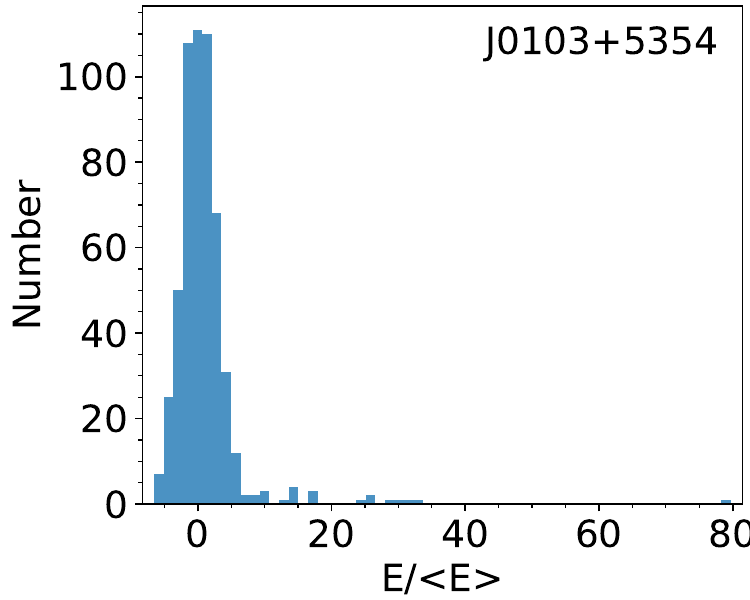} 
    \includegraphics[width=0.24\textwidth]{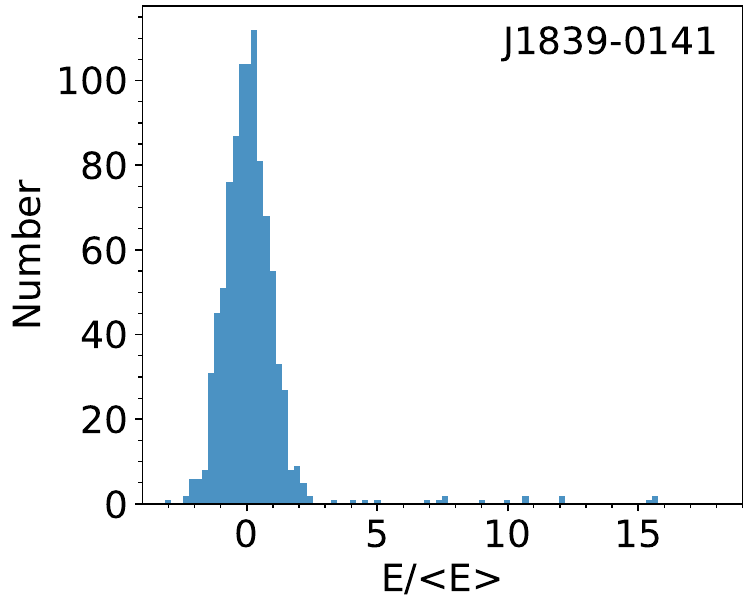} 
    \includegraphics[width=0.24\textwidth]{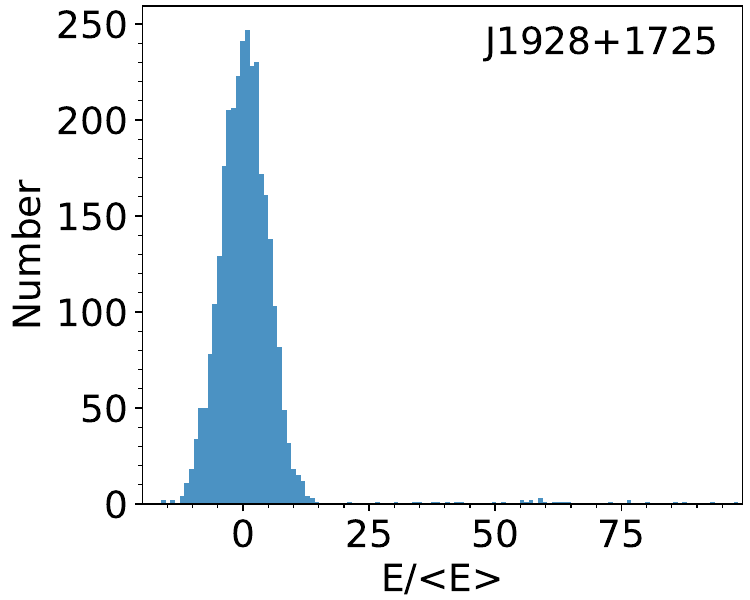} 
    \includegraphics[width=0.24\textwidth]{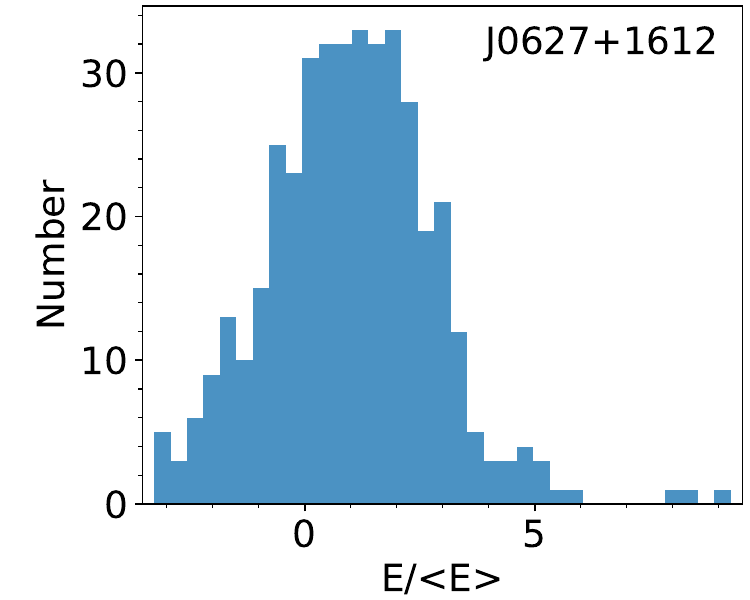} 
    \includegraphics[width=0.24\textwidth]{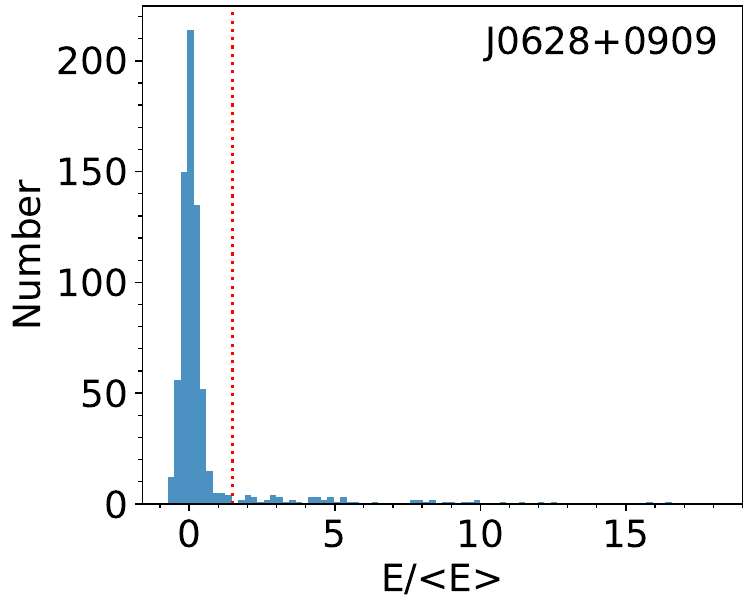} 
    \includegraphics[width=0.24\textwidth]{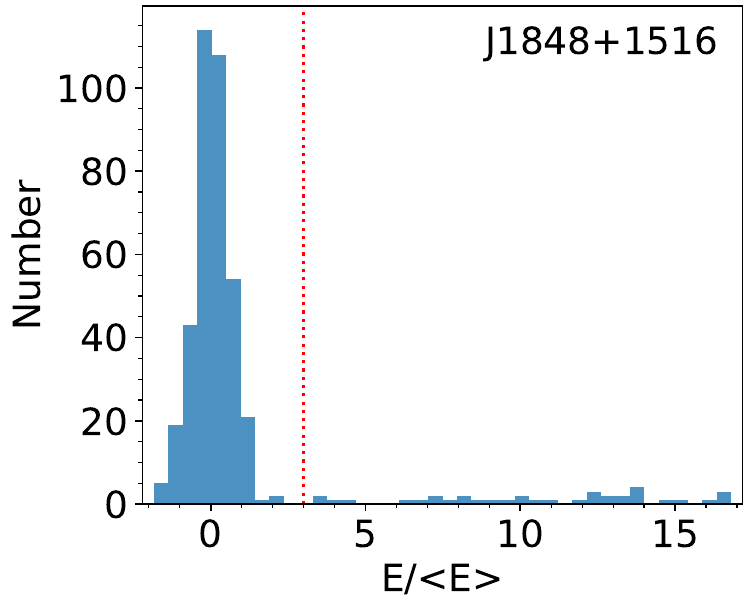} 
    \includegraphics[width=0.24\textwidth]{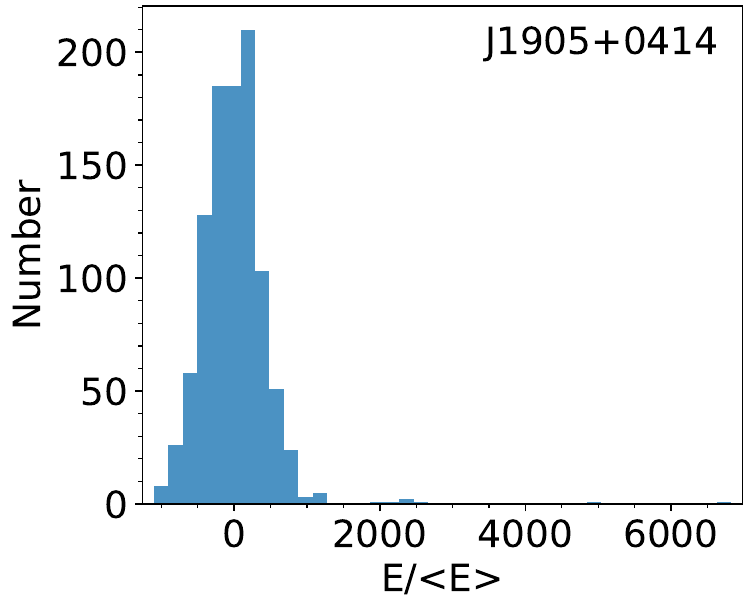}
    \includegraphics[width=0.24\textwidth]{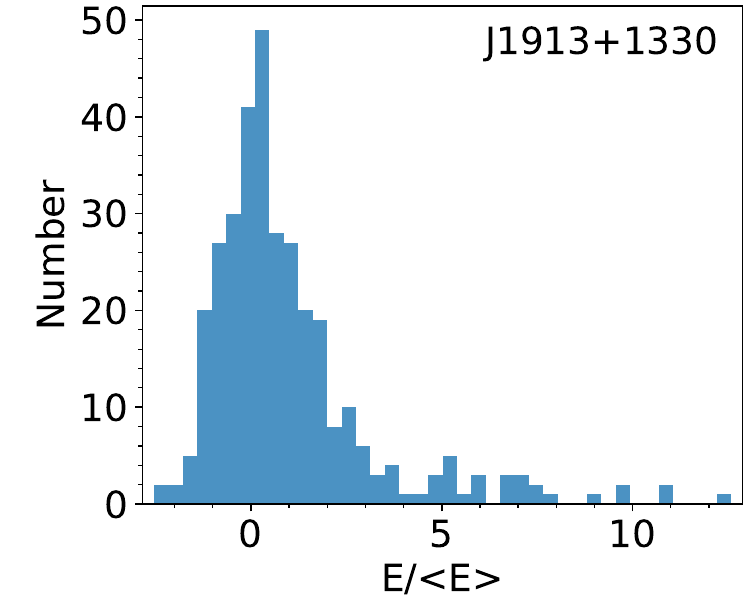} 
\caption{The energy distributions of individual pulses for known RRATs with FAST observations for more than 300 pulses, normalized by the mean energy of all observed periods. The vertical line indicates an easy criterion to separate normal pulses and sparsely strong pulses, for them the polarization profiles are compared in Figure~\ref{fig:know2pol}.
}
\label{fig:knowEnergy3}
\end{figure*}

\subsection{Polarization of known RRATs}

When these known RRATs were observed by FAST, the polarization signals are recorded. We use the standard polarization calibration procedure \citep{whx+22} and process all data. We obtain the polarization profiles of these known RRATs as shown in Figure~\ref{fig:knownRRAT1pol}, Figure~\ref{fig:knownRRAT2pol} and Figure~\ref{fig:knownRRAT3pol}. To maximize the signal to noise ratio,
we have to ignore these periods with undetectable signal, i.e. only work on the periods with a signal of $S/N >3$. We extract polarization information from the Faraday RMs as listed in Table~\ref{tab:RRATcat}. 

As one can see from these polarization profiles, in general, they are not unusual compared to normal pulsar profiles. For these RRATs which appear as normal pulsars in the FAST observations, the PA curves of J1849+0106, J1853+0427 and J1919+1745 show a smooth variation with an  `S' shape.  Some pulsars show a steep polarization angle sweep at the profile center with a reversal of circular polarization, such as J0302+2252 and J1856+0912. The J1850+1532 has a highly circular polarized component. Orthogonal polarized modes have been detected in the PA curve of J1538+2345.

For five RRATs as extremely nulling pulsars very interesting is the extremely highly linear polarization of J1928+1725 with a flat PA curve (see Figure~\ref{fig:knownRRAT2pol}). For pulsars with sparse strong pulses, because of limited periods accumulated, it is hard to say the profiles in Figure~\ref{fig:knownRRAT3pol} are stabilized. 
Interestingly the RM value for J1905+0414, which has a large RM value of 1090$\pm$3 rad\,m$^{-2}$. Combined with the DM of this RRAT $383.0\,\rm pc~cm^{-3}$, the averaged magnetic fields in the line of sight to this distant RRAT is $1.232 \times 1090/383 = 3.5~\mu$G. This seems to be fine \citep{Han2017ARAA} for a pulsar at a spiral arm tangent at $(l,b)$ =  ($38.246^{\circ}$, $-1.079^{\circ}$). J1720+0040 and J1848+1516 also have reversal of circular polarization.

\begin{figure}[!htp]
  \centering
  \includegraphics[width=0.45\columnwidth]{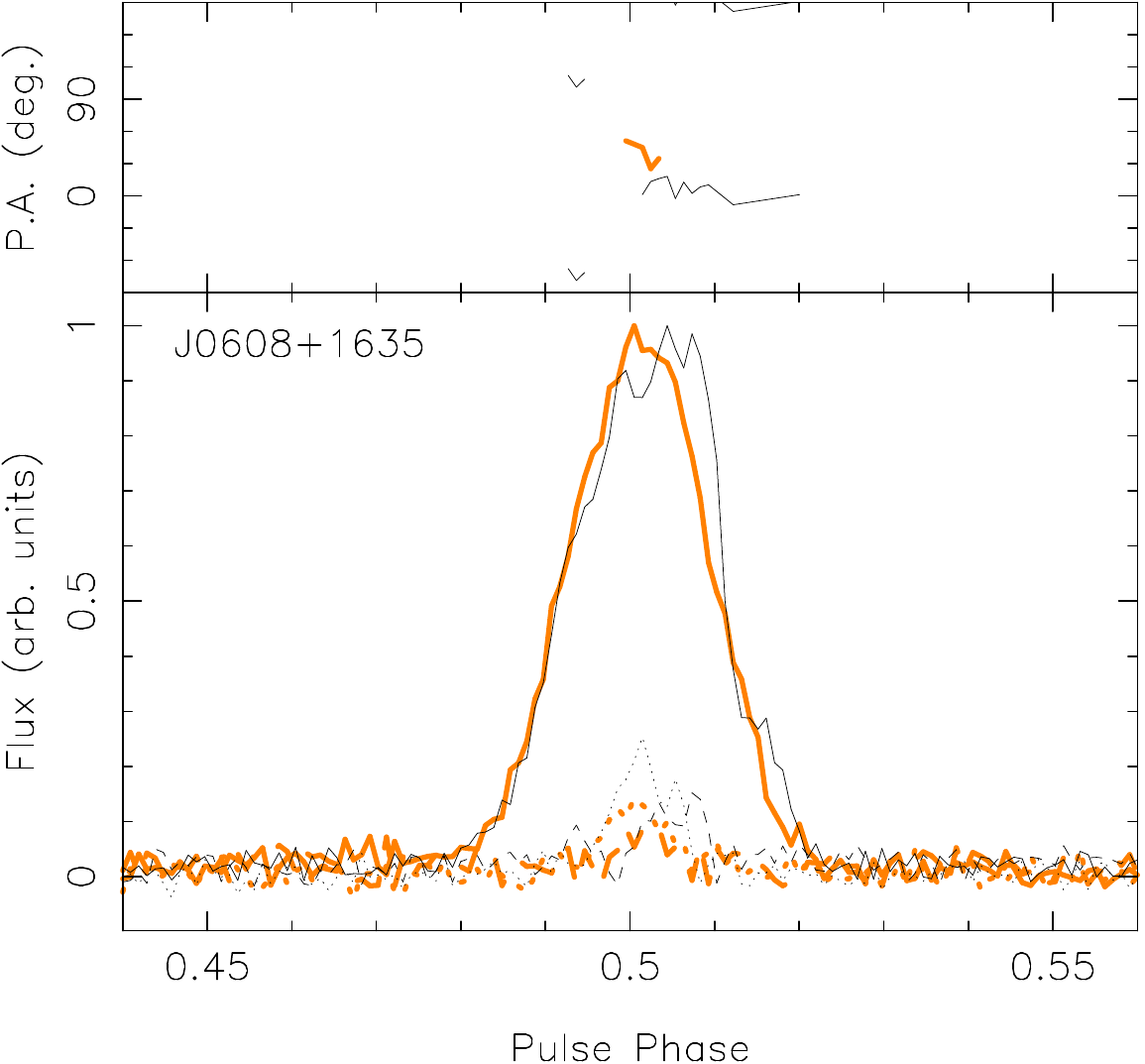}
  \includegraphics[width=0.45\columnwidth]{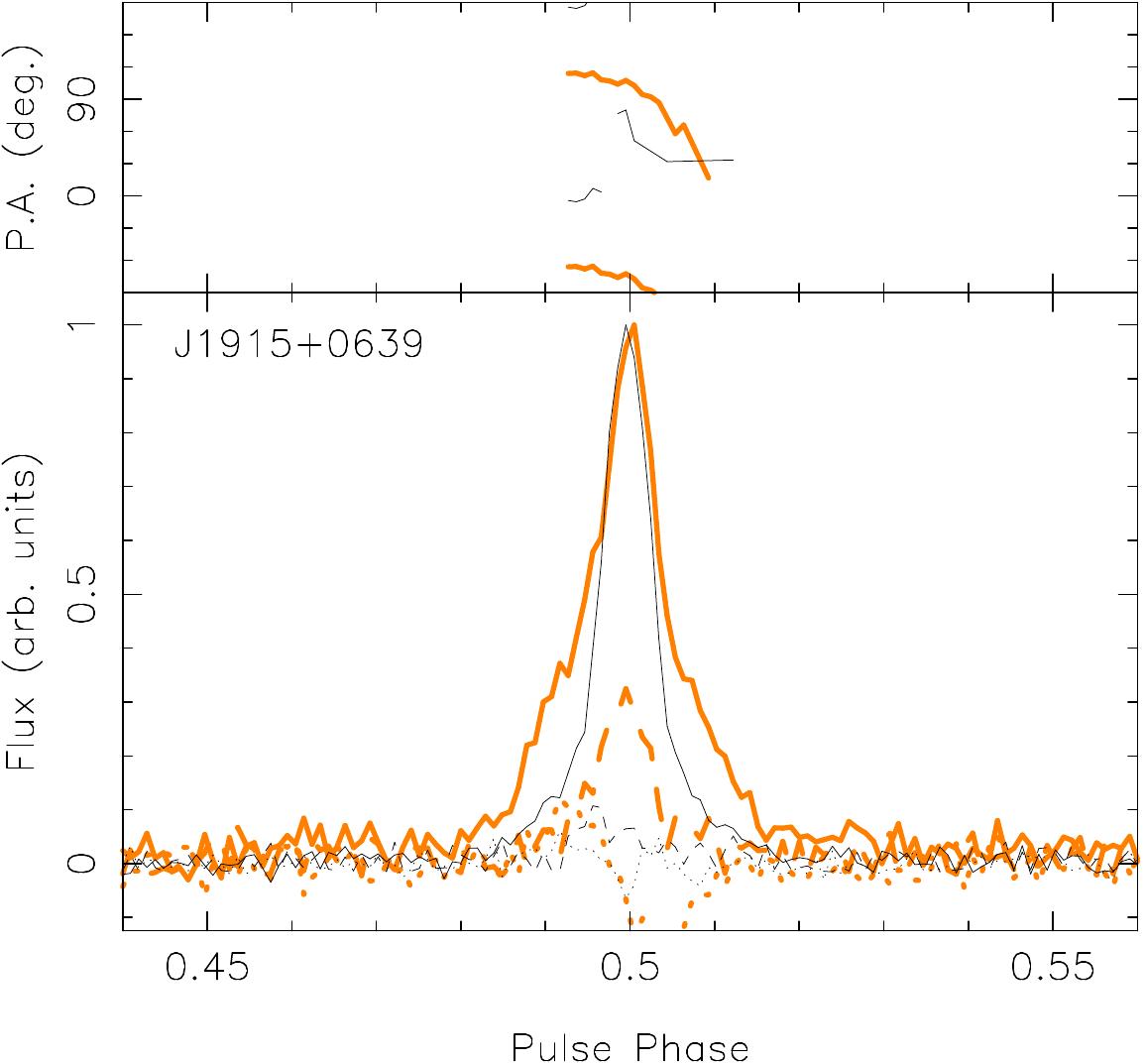}
  \includegraphics[width=0.45\columnwidth]{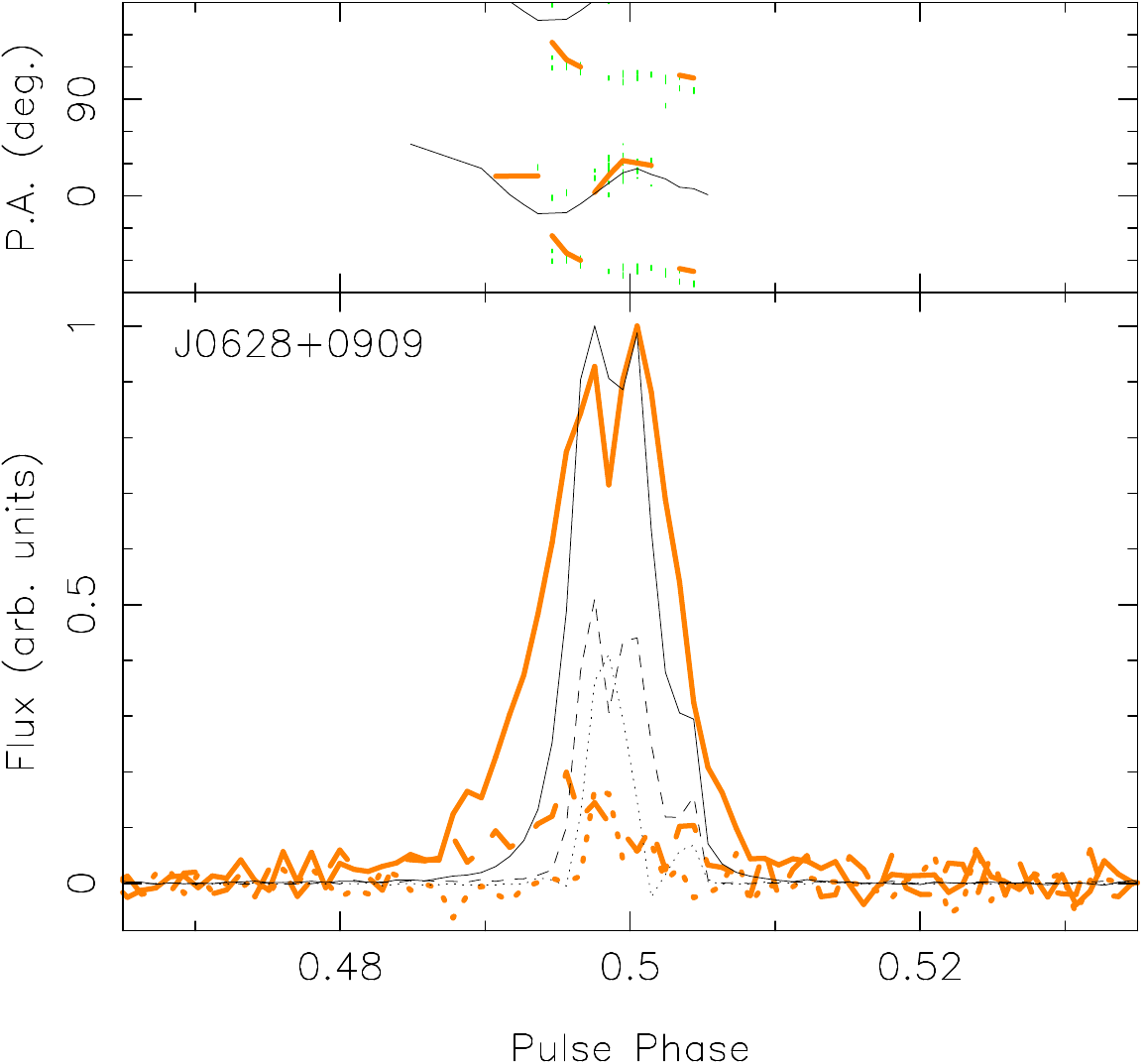}
  \includegraphics[width=0.45\columnwidth]{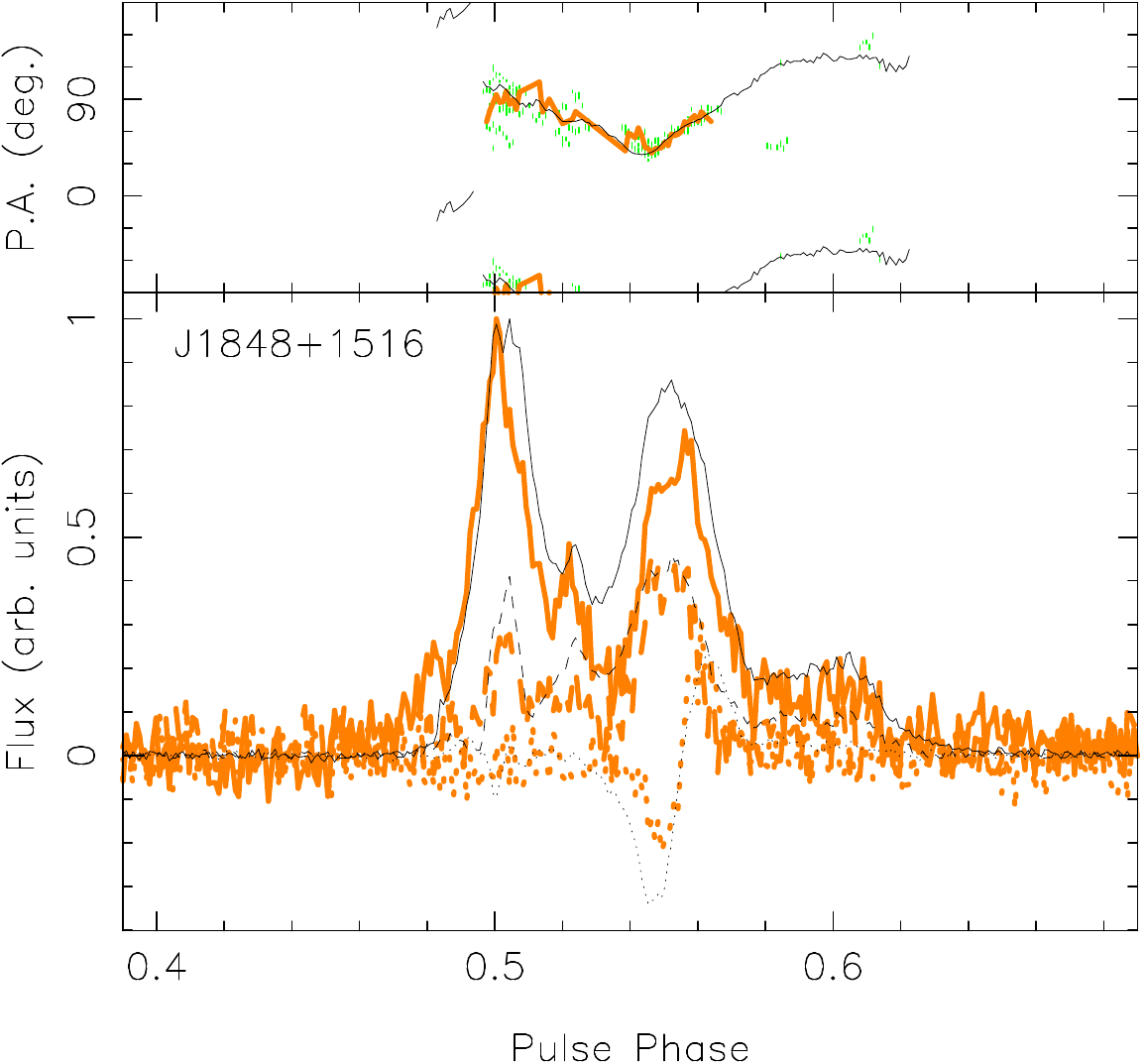}
  \includegraphics[width=0.45\columnwidth]{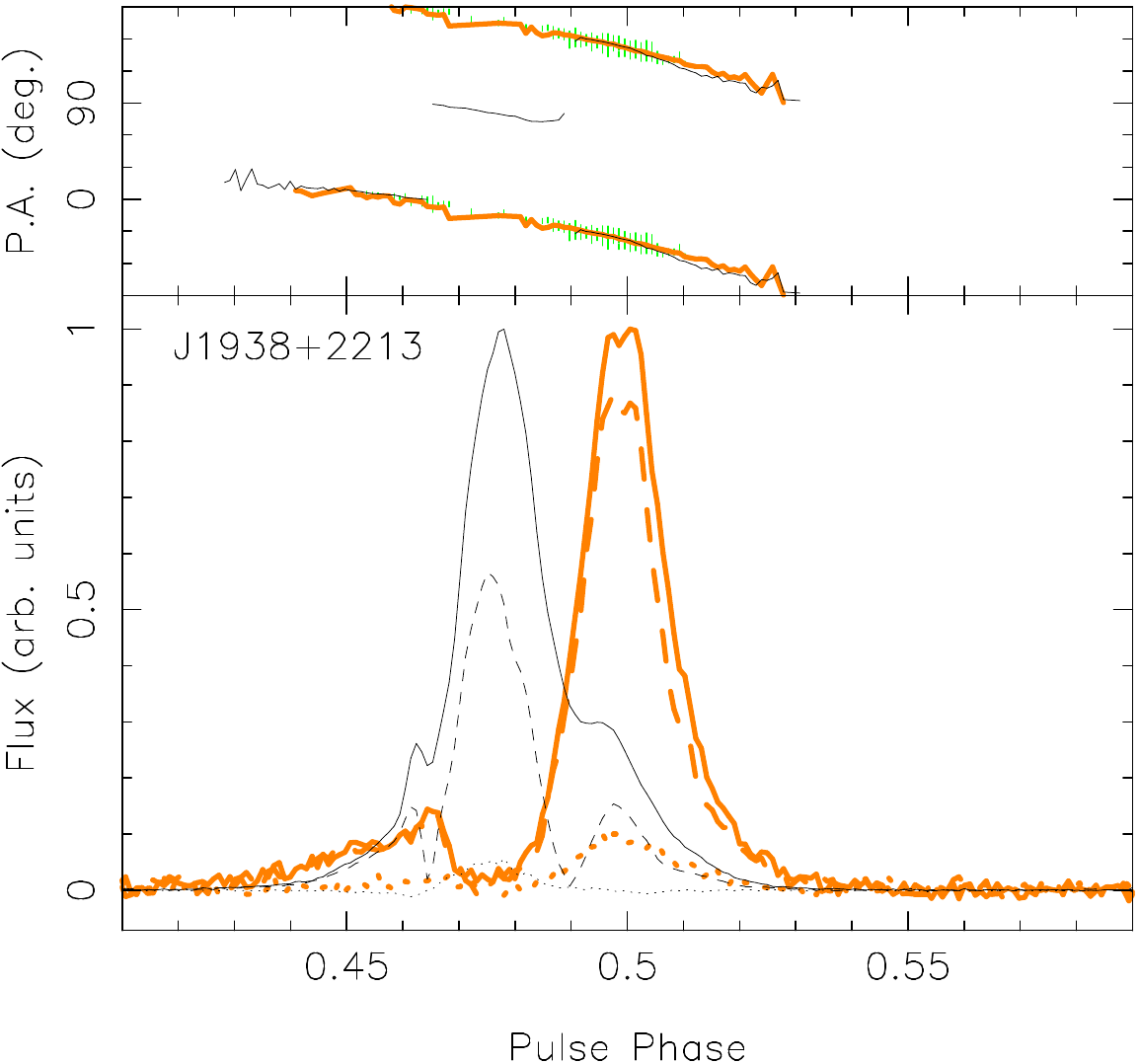}
  \caption{Polarization profiles of sparsely strong pulses (black thin lines) are compared with those of normal weaker pulses (thick lines), see the criteria marked in 
  the energy distribution in Figure~\ref{fig:knowEnergy3}, except for J1938+2213 for the bright pulses of $S/N>200$ in Figure~\ref{knownpulsarRRATlike}.
  In the top sub-panel the PA values in dark green come from bright pulses with linearly polarized intensities $>10\sigma$, black and orange lines are respectively for the averaged PA values of bright and weak pulses. The bottom sub-panel is for the total power (solid line), linear polarization (dashed line) and circular polarization (dotted line), with the profile peaks normalized to the unity.
  }
        \label{fig:know2pol}
\end{figure}

\subsection{Sparse Strong pulses and polarization}
%\subsection{The Energy distribution of Known RRATs}

For previously known RRATs, it is important to verify if these sparse strong pulses are very outstanding from the energy distribution of normal pulses, and if they have very different polarization from normal pulses. 

For known RRATs with a sufficient number ($>300$) of single pulses observed by FAST so that a well-defined average pulse profile has been obtained, we obtain the energy distribution of their single pulses. For each pulse, we sum the individual pulse energy in the on-pulse range, and then obtain the average energy of all single pulses. The energy of all individual pulses is then normalized by the averaged energy. 
As shown in Figure~\ref{fig:knowEnergy3}, the pulse energy distributions of the known RRATs rarely show any points with a scaled energy greater than 10 if they are believed as normal pulses, except for J0302+2252 and J1915+0639 which seems to be very unusual with two more peaks in the distribution in addition to the main component for nulling pulses around zero. 
For RRATs as extremely nulling pulsars, the distribution has a component for nulling together with a few periods with a high-energy tail for emission.
For RRATs with sparse strong pulses, J0628+0909, J1848+1515 and J1905+0414 and also J1913+1330, we do see these sparse pulses with unusual energy.

% \subsubsection{Polarization of bright and weak pulses of Known RRATs}

Figure~\ref{fig:know2pol} compares the polarization profiles for the normal weak pulses and these sparsely strong pulses, in the two sides of the normalized energy distributions Figure~\ref{fig:knowEnergy3}. In principle their number distributions should be fitted with a log-normal energy distribution. However, such fitting cannot be done well since only a small number of bright pulses are available. The sparse strong pulses can therefore be distinguished from the weak pulses in the distribution roughly by the lowest valley. We find that the polarization profiles from these small number of strong pulses in the high-energy tail have the same or the orthogonal polarized modes for their polarization angle distributions for weaker pulses, except for the case of PSR J1915+0639 which is hardly to understand. This indicates that mostly there is no difference between polarization profiles for sparse strong pulses and mean polarization profiles of normal pulses.

\section{Discussions and Conclusions} 
\label{sect5:Conclusions}

\begin{figure*}
  \centering
  \includegraphics[width=0.7\textwidth]{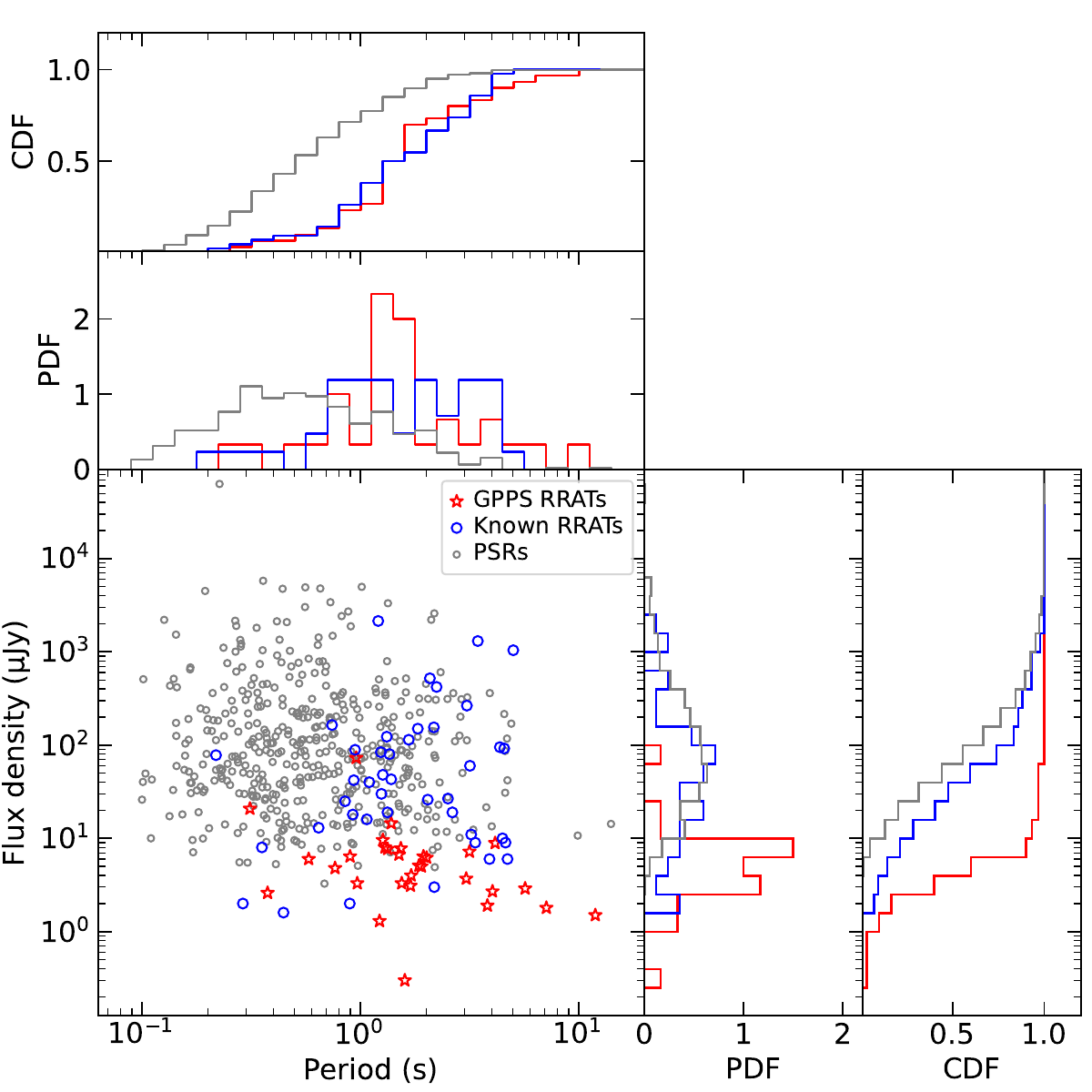}
    \caption{The distribution for the period and mean flux density of
     the newly discovered transient sources in this paper (red), compared with the known RRATs (blue) and the other pulsars in the GPPS survey Paper I~\citep[gray, ][]{Han2021RAA}. In the top and right panels are the cumulative distribution functions (CDFs) and the probability density functions (PDFs) for the period and mean flux density, respectively. 
    } % pentagram  frame  cross
  \label{fig:perioddist}
\end{figure*}

We developed an efficient single pulse searching module for generating  DM-time images and target detection with AI, so that it is computationally fast. Two processing strategies, the full bandwidth and the two halves of observation band of FAST are processed independently for single pulse search which can improve the detection of narrow or steep-spectrum pulses. We apply this new single pulse search module to the FAST GPPS survey, and discovered 76 new RRATs in our Galaxy. Among them, 26 sources have only several pulses detected by FAST, so that their rotation period cannot be derived from the limited duration of FAST tracking observations. We have got 16 sources identified as proto-RRATs which have sparse strong pulses with a recognized period. 10 sources of them are apparently extreme nulling pulsars, and the other 24 sources are weak pulsars with sparse strong pulses.

For many newly discovered transient sources, the occasional strong emission state can help them to be found if the pulses are strong enough over the detection threshold. This has some implications. One is there must be a large neutron star population in our Galaxy to uncover but mostly undetectable. Second, single pulse searching is a desired step to catch them. Third, the occasionally brightened emission state of these neutron stars must be related to some physical conditions which must be understood. 

Noticed that these 76 transient objects here, including 12 objects discovered by the single pulse search that reported in the paper I~\citep{Han2021RAA} are about 15\% of newly discovered pulsars. In the other words, at least 15\% of neutron stars would not be detected from the traditional periodicity signal search. Therefore, the single pulse search is important on revealing a large population of unknown RRATs, i.e. these pulsars with sporadic strong pulses.
 
The  RRAT detection percentage of the FAST GPPS survey is much larger than the percentage of the known RRATs compared to all known pulsars, which is only 5\% for all known RRATs, including extremely nulling pulsars and weak pulsars with sparse strong pulses. Detection of RRATs by FAST data leads us to find pulsars at the sub-$\mu$Jy level. For some of them, we cannot get enough pulses to figure out the period yet. Finding their period by multiple observations or by increasing the duration of FAST observation is very necessary. 

Figure~\ref{fig:perioddist} shows the distribution for the period and mean flux density of newly discovered transient sources in this paper compared to values for the known RRATs and the other pulsars with a period of larger than 0.1\,s in the first FAST GPPS paper \citep{Han2021RAA}. The mean flux densities of these newly discovered sources are much lower than that of the known RRATs and even the GPPS pulsars. Most of them have a mean flux density of few $\mu$Jy, and J1921+1227g even has the lowest mean flux density of about 0.28$\mu$Jy. This demonstrates the population of weak radiate neutron stars inside the Milky Way to be discovered. For the Probability density functions (PDFs) and cumulative distribution functions (CDFs), there is no significant difference in period distribution between these transient sources and the previously known RRATs. The difference between the FAST GPPS pulsars and these RRATs is prominent, which apparently caused by different selection effects from the search method. Any pulsars discovered by the normal periodical search are not classified as RRATs. On the other hand, the single pulse detection also has limitations. The identification of extremely narrow transient \citep[$\rm \textless 1~ms$, like the ultranarrow burst of FRB~20191107B,][]{Gupta2022MNRAS} or even narrower band radiation in FRB, has to be improved in the future. The detection efficiency of signals at the boundary of the DM-time images has to be improved.

%%############################################ RRAT-like pulsars

Extremely sensitive observations are fundamental to understand the transient sources, especially RRATs. Observations of the known RRATs by using FAST could detect more weak pulses, in addition to previously detected sparse strong pulses. In our FAST GPPS survey area, a large number of known RRATs have the FAST data recorded already. Together with other FAST projects, we have got 48 known RRATs detected by FAST.  Four of them have only one pulse detected in our FAST observations and another 1 has four pulses detected in a five minutes observation session but not confirmed in the verification observation. 25 known RRATs are simply normal pulsars in the FAST observations, 5 of them are extremely nulling pulsars, and 13 objects are weaker pulsars with sparse strong pulses. Polarization observations show no unusual difference of these strong pulses from weaker pulses.

We therefore conclude that most RRATs are simply weaker pulsars with sparse strong pulses or very nulling pulsars. The sensitivity of the telescope and long observations are the key to understand the enigma of RRATs.

\normalem
\begin{acknowledgements}
%
%We thank Prof. R.T. Gangadhara and the referees, Prof. R.N. Manchester
%and Prof. Jim Cordes, for helpful comments.
%
This project, as one of five key projects, is being carried out by using
FAST, a Chinese national mega-science facility built and operated by
the National Astronomical Observatories, Chinese Academy of Sciences.
J.~L. Han is supported by the National Natural Science Foundation of
China (NSFC, Nos. 11988101 and 11833009) and the Key Research
Program of the Chinese Academy of Sciences (Grant No. QYZDJ-SSW-SLH021);
D.~J. Zhou is supported by the Cultivation Project for the FAST
scientific Payoff and Research Achievement of CAMS-CAS.
H.~G. Wang, C. Wang, P.~F. Wang and X.~P. You are supported by NSFC
No. 12133004; P.~F. Wang, C. Wang and H.~G. Wang are also partially
supported by the National SKA program of China No. 2020SKA0120200.
In addition, 
C. Wang is also partially supported by NSFC No. U1731120;
%
%X.Y. Gao is partially supported by NSFC No. U1831103;
%
P.~F. Wang is also partially supported by the NSFC No. 11873058;
H.~G. Wang is also partially supported by the Guangzhou Science and Technology Project No. 202102010466;
Jun Xu is partially supported by NSFC No. U2031115.
% R. Yuen is partly supported by Xiaofeng Yang's Xinjiang Tianchi Bairen  project and CAS Pioneer Hundred Talents Program.
%
% L.G. Hou thanks the support from the Youth Innovation Promotion
% Association CAS.
%
% (more to add).
\end{acknowledgements}

%\small
\section*{Authors contributions}  
The FAST GPPS survey is a key science project of FAST led by J.~L. Han.
D.~J. Zhou realized the single pulse search module and related software under the supervision of J.~L. Han, and processed all data presented in this paper. 
J.~L. Han coordinated the teamwork and coordinated computational resources. and also in charge of writing this paper.
Jun Xu observed some RRATs in his project and contributed the data of some known RRATs.
Chen Wang designed the survey observation plan for the GPPS survey, and fed all targets for each observation session and verification observations. 
P.~F. Wang realized the polarization processing pipeline which is used in this paper.
%observations.
%
Tao Wang initialized and realized parts of the data preparing module.
%
%
% Wei-Cong Jing worked on the DM and the upper limit estimations from the Galactic electron density models.
%
P.~F. Wang  and Jun Xu made fundamental contributions to the construction
and maintenance on the computation platform.
W.~C. Jing, Xue Chen, Yi Yan and W.~Q. Su %, Yi Yan, Z.~L.Yang and N. N. Cai 
joined many group discussions and commented on the results of this paper.
H.~Q. Gan, P. Jiang and J.~H. Sun ensured the success of FAST observations.
Other people jointly propose or monitor the project.
All authors contributed to the finalization of the paper.

\section*{Data and software availability}

Original FAST observational data will be open sources according to the
FAST data 1-year protection policy. The folded and calibrated data for
sources in this paper can be obtained from authors by kind requests. Pulsar profile data presented and the source codes in this paper are available on the webpage of the GPPS survey \url{http://zmtt.bao.ac.cn/GPPS/}.

\bibliographystyle{mnras}
\bibliography{bibfile}{}

%\small
\section*{Appendix}  
The single pulse search module for the GPPS survey uncovers many radio transient sources in the sky. 

Table~\ref{tab:fewPulses} (published in the online version of this paper) list parameters for each pulse, including the TOA of the pulse peak, signal-to-noise ratio of dedispersed pulse $R$, pulse width in millisecond, and the fluence discussed in Sect.\ref{sect2.2.2}.

Figure~\ref{fig:AppfewPulses} (published in the online version of this paper) show all water-fall plots for 26 radio transients with only a few pulses detected for each source, including the 3 example plots in Fig.~\ref{fig:fewPulses}. 

Figure~\ref{fig:AppnewRRATs} (published in the online version of this paper) show plots for 16 RRATs discovered in the GPPS survey, while Figure~\ref{fig:newRRATs} is only one example.

Figure~\ref{fig:Appnewnulling} (published in the online version of this paper) show plots for 10 extremely nulling pulsars as discovered in the GPPS survey, while Figure~\ref{fig:newnullingpulsars} is only one example.

Figure~\ref{fig:AppnewweakPulsars} (published in the online version of this paper) show plots for 24 weak pulsars with sparse strong pulses discovered in the GPPS survey, while Figure~\ref{fig:newWeakPulsars} is only one example.

Figure~\ref{fig:APPknownRRAT1} (published in the online version of this paper) show plots for 25 previously known RRATs shown as normal pulsars in the FAST observations, while Figure~\ref{knownRRAT1} just two examples.

Figure~\ref{fig:APPknownRRAT2} (published in the online version of this paper) show plots for 5 previously known RRATs shown as extremely nulling pulsars in the FAST observations, while Figure~\ref{knownRRAT2} just two examples.

Figure~\ref{fig:APPknownRRAT3} (published in the online version of this paper) show plots for 13 previously known RRATs shown as % extremely nulling pulsars and 
pulsars with sparse strong pulses in the FAST observations, while Figure~\ref{knownRRAT3} just examples respectively.

\begin{table}
    \centering
    {\footnotesize
    \caption{Properties of single pulses for newly discovered transient sources with few pulses detected by FAST.}
    \label{tab:fewPulses}
    \setlength{\tabcolsep}{4.0pt}
    \begin{tabular}{crcrrr}
    \hline%\noalign{\smallskip}
    ObsDate & No. & TOA            & $R$  & $W$    & $F$      \\
            &     & (MJD)          &      & (ms)   & (mJy ms) \\
      (1)   &  (2) & (3)          & (4)   & (5)   & (6) \\
\hline
    \multicolumn{6}{c}{J0637+0332g = gpps0528}            \\
    20220909& 1   & 59831.03261773 & 9.9  & 7.6  &47.8    \\
            & 2   & 59831.03501918 & 10.6 & 12.2 & 45.8   \\
            & 3   & 59831.03555291 & 11.7 & 8.8  & 40.0   \\
\hline
    \multicolumn{6}{c}{J1828-0003g = gpps0501}            \\
    20200803& 1   & 59794.62034865 & 40.5 & 31.7 &600.5   \\
            & 1   & 59794.62673776 & 16.1 & 37.9 &143.3   \\
    20221114& 1   & 59897.31460359 & 95.6 & 42.4 &1830.0  \\
\hline
    \multicolumn{6}{c}{J1847-0046g = gpps0282}            \\
    20211031& 1   & 59518.38408605 &  7.6 & 12.1 & 60.9   \\
    20211113& 1   & 59531.34693849 & 11.4 & 43.5 &111.7   \\
            & 2   & 59531.35255450 & 31.8 & 17.0 &483.1   \\
    20220606& 1   & 59735.77706666 & 32.9 & 26.8 &572.8   \\
            & 2   & 59735.78296057 & 22.4 & 13.9 &367.9   \\
            & 3   & 59735.78746444 & 12.9 & 34.1 &178.5   \\
\hline
    \multicolumn{6}{c}{J1850-0004g = gpps0280}            \\
    20200415& 1   & 58953.93718024 & 22.6 & 5.8  & 147.0  \\
            & 2   & 58953.93992108 & 25.0 & 6.9  & 206.0  \\
    20200902& 1   & 59094.53330366 & 21.3 & 7.7  & 107.7  \\
    20220608& 1   & 59737.77205668 & 62.2 & 7.4  & 479.8  \\
            & 2   & 59737.77357873 & 23.8 & 7.1  & 176.5  \\
    20221006& 1   & 59889.35833273 & 35.5 & 7.4  & 122.0  \\
            & 3   & 59737.77825393 & 22.9 & 15.8 & 163.3  \\
            & 4   & 59737.78097067 & 16.9 & 10.1 & 120.4  \\
\hline
    \multicolumn{6}{c}{J1853+0209g = gpps0502}            \\
    20200812& 1   & 59073.58342402 & 30.9 & 120.6& 773.9  \\
    20220824& 1   & 59815.56281964 & 60.7 & 217.7& 1923.4 \\
\hline
    \multicolumn{6}{c}{J1853+0353g = gpps0281}            \\
    20210514& 1   & 59347.82755635 & 11.8 & 42.6 & 166.7  \\
            & 2   & 59347.83650747 & 22.0 & 64.6 & 415.1  \\
    20210624& 1   & 59388.76374658 & 26.1 & 57.4 & 476.0  \\
\hline
    \multicolumn{6}{c}{J1855-0211g = gpps0526}            \\
    20220810& 1   & 59801.62433801 & 9.6  & 74.7 & 85.9   \\
    20221205& 1   & 59918.27588921 & 57.1 & 25.5 & 660.7  \\
    20221205& 2   & 59918.27671860 & 33.0 & 41.0 & 466.9  \\
    20221205& 3   & 59918.28352914 & 77.9 & 48.0 &1695.7  \\
    20221205& 1   & 60008.03900500 & 111.3& 102.5&3474.8  \\
\hline
    \multicolumn{6}{c}{J1855-0154g = gpps0503}                    \\
    20210514& 1   & 59464.52884265 & 18.6 & 12.0 & 134.9  \\
            & 2   & 59464.52880855 & 21.4 & 10.2 & 170.4  \\
            & 3   & 59464.52884265 & 12.9 & 9.1  & 78.5   \\
\hline
    \multicolumn{6}{c}{J1855-0054g = gpps0504}                    \\
    20220210& 1   & 59678.94763145 & 26.1 & 61.9 & 529.6  \\
    20220602& 1   & 59731.78531778 & 9.9  & 55.9 & 93.2   \\
            & 2   & 59731.78540454 & 14.6 & 30.1 & 204.0  \\
            & 3   & 59731.78846381 & 19.3 & 108.0& 391.2  \\
            & 4   & 59731.78994988 & 22.4 & 79.1 & 434.5  \\
\hline
    \multicolumn{6}{c}{J1855+0033g = gpps0283}            \\
    20200328& 1   & 58935.98123561 & 15.9 & 53.2 & 112.5  \\
    20210328& 1   & 59300.95680323 & 11.9 & 11.6 & 72.7   \\
            & 2   & 59300.97261027 & 32.3 & 46.0 & 436.2  \\
            & 3   & 59300.98378786 & 33.8 & 19.3 & 338.2  \\
\hline
    \multicolumn{6}{c}{J1856+0528g = gpps0284}            \\
    20210608& 1   & 59372.79606129 & 12.7 & 86.2 & 99.5   \\
            & 2   & 59372.79608372 & 12.2 & 47.6 & 70.4   \\
            & 3   & 59372.79920230 & 10.5 & 50.3 & 69.9   \\
    20210826& 1   & 59452.57993757 & 17.0 & 87.8 & 169.8  \\
            & 2   & 59452.58749961 & 13.0 & 43.5 & 86.8   \\
            & 3   & 59452.58950899 & 10.4 & 84.9 & 65.6   \\
\hline
\end{tabular}}
	\begin{tablenotes}
	\item {\footnotesize
Notes: Column (1)-(3): Observation date, pulse number, TOA of \\
pulse peak in MJD; Column (4)-(6): signal-to-noise ratio, and \\
pulse width in ms and also the fluence.
}
  \end{tablenotes}
\end{table} 
\addtocounter{table}{-1}
\begin{table}[!ht]
    \centering
    {\footnotesize
   \caption{{\it -- continued }.}
    \setlength{\tabcolsep}{4.0pt}
    \begin{tabular}{crcrrr}
    \hline%\noalign{\smallskip}
    ObsDate & No. & TOA            & $R$  & $W$    & $F$      \\
%            &     & (MJD)          &      & (ms)   & (mJy ms) \\
      (1)   &  (2) & (3)          & (4)   & (5)   & (6) \\
\hline
    \multicolumn{6}{c}{J1859+0832g = gpps0505}                    \\
    20220526& 1   & 59724.77170952 & 50.6 & 7.7  & 425.2  \\
    20220526& 2   & 59724.77171976 & 18.5 & 78.8 & 239.0  \\
    20221107& 1   & 59890.34972011 & 33.3 & 8.5  & 173.6  \\
\hline
    \multicolumn{6}{c}{J1900+0908g = gpps0527}           \\
    20220522& 1   & 59720.79041037 & 41.5 & 21.5 & 391.5 \\
            & 2   & 59720.79043544 & 7.3  & 10.8 & 32.5  \\
\hline
    \multicolumn{6}{c}{J1902+0557g = gpps0525}           \\
    20221018& 1   & 59870.44448379 & 14.7 &  3.1 & 61.6  \\
            & 2   & 59870.44530378 & 12.1 &  21.0& 58.8  \\
    20230212& 1   & 59987.12293867 & 15.1 & 20.3 & 73.3  \\
    
\hline
    \multicolumn{6}{c}{J1916+1142Ag = gpps0287}          \\
    20200302& 1   & 58910.09452353 & 7.2  & 10.5 & 14.9  \\
    20210128& 2   & 59222.20075862 & 12.9 & 3.5  & 44.7  \\
\hline
    \multicolumn{6}{c}{J1918+0342g = gpps0506}                      \\ 
    20211202& 1   & 59550.31150165 & 33.9 & 12.0 & 75.7     \\  
    20220602& 1   & 59731.85140692 & 11.9 & 25.7 & 88.3     \\  
\hline
    \multicolumn{6}{c}{J1918+1514g = gpps0507}                      \\
    20200531& 1   & 58999.81141189 & 8.31 & 12.3 & 44.1     \\
            & 2   & 58999.81210714 & 10.6 & 15.2 & 79.8     \\
\hline
    \multicolumn{6}{c}{J1921+1629g = gpps0288}              \\
    20210822& 1   & 59448.63792384 & 14.3 & 16.7 & 80.7     \\
    20211004& 1   & 59491.54677614 & 38.6 & 5.5  & 237.6    \\
            & 2   & 59491.54955334 & 49.5 & 6.63 & 376.0    \\
\hline
    \multicolumn{6}{c}{J1924+1734g = gpps0289}              \\
    20210822& 1   & 59448.65213287 & 42.5 & 18.8 & 578.5    \\
            & 2   & 59448.65362293 & 32.4 & 33.8 & 536.2    \\
    20211005& 1   & 59492.41449285 & 36.0 & 31.7 & 457.0    \\
            & 2   & 59492.41467918 & 19.2 & 24.4 & 189.1    \\
            & 3   & 59492.41598276 & 39.3 & 20.7 & 568.4    \\
\hline
    \multicolumn{6}{c}{J1927+1940g = gpps0290}              \\
    20190327& 1   & 58569.02255643 & 10.2 & 8.7  & 49.2     \\
    20210624& 1   & 59388.71211556 & 9.0  & 21.1 & 71.7     \\
\hline
    \multicolumn{6}{c}{J1932+2126g = gpps0508}                       \\
    20220323& 1   & 59661.04248301 & 21.6 & 13.1 & 157.1     \\
    20220608& 1   & 59737.82493655 & 24.0 & 8.0  & 123.5     \\
    20220720& 1   & 59780.62376501 & 11.8 & 21.7 & 80.8      \\
\hline
    \multicolumn{6}{c}{J1933+2401g = gpps0291}              \\
    20210301& 1   & 59274.07468225 & 13.2 & 33.4 & 87.1     \\
    20210626& 1   & 59390.78088520 & 9.1  & 30.4 & 52.0     \\
\hline
    \multicolumn{6}{c}{J1934+2341g = gpps0292}              \\
    20210301& 1   & 59274.07585975 & 12.5 & 12.0 & 89.2     \\
            & 2   & 59274.07263818 & 10.5 & 8.1  & 89.2     \\
            & 3   & 59274.07382939 & 19.1 & 5.0  & 127.4    \\
    20210624& 1   & 59388.81110584 & 15.1 & 6.6  & 88.0     \\
            & 2   & 59388.81117695 & 9.7  & 16.3 & 48.7     \\
            & 3   & 59388.81247483 & 16.3 & 24.5 & 125.9    \\
            & 4   & 59388.81427651 & 8.5  & 17.0 & 34.8     \\
            & 5   & 59388.81443653 & 30.6 & 5.6  & 199.7    \\
    20221106& 1   & 59889.41136795 & 10.7 & 10.6 & 54.1     \\
            & 2   & 59889.43586809 & 38.2 & 5.9  & 229.7    \\
\hline
    \multicolumn{6}{c}{J2001+4209g = gpps0293}              \\
    20210802& 1   & 59427.71952664 & 12.9 & 11.6 & 63.3     \\
    20211004& 1   & 59491.56354889 & 17.9 & 9.0  & 107.8    \\
            & 2   & 59491.56567598 & 10.1 & 13.2 & 40.6     \\
\hline
    \multicolumn{6}{c}{J2005+3154g = gpps0294}              \\
    20210804& 1   & 59430.74319212 & 11.5 & 8.5  & 45.5     \\
            & 2   & 59430.74537058 & 14.5 & 11.8 & 62.7     \\
    20210805& 1   & 59431.65409682 & 16.8 & 7.5  & 98.2     \\
            & 2   & 59431.64156848 & 14.7 & 15.7 & 73.4     \\
            & 3   & 59431.64187222 & 8.3  & 9.8  & 41.4     \\
\hline
            & 1   & 59431.64377282 & 9.5  & 26.3 & 52.9     \\
    20211009& 1   & 59496.54708948 & 28.9 & 7.2  & 127.2    \\
            & 2   & 59496.56473674 & 11.7 & 10.1 & 71.4     \\
            & 3   & 59496.56707139 & 19.5 & 5.1  & 113.8    \\
\hline
\end{tabular}}
\end{table} 
\addtocounter{table}{-1}
\begin{table}[!ht]
    \centering
    {\footnotesize
   \caption{{\it -- continued and ended}.}
    \setlength{\tabcolsep}{4.0pt}
    \begin{tabular}{crcrrr}
    \hline%\noalign{\smallskip}
    ObsDate & No. & TOA            & $R$  & $W$    & $F$      \\
%            &     & (MJD)          &      & (ms)   & (mJy ms) \\
      (1)   &  (2) & (3)          & (4)   & (5)   & (6) \\
\hline
    \multicolumn{6}{c}{J2030+3833g = gpps0295}              \\
    20210220& 1   & 59265.16695103 & 9.0  & 28.7 & 102.8    \\
            & 2   & 59265.16802449 & 32.8 & 45.3 & 185.7    \\
    20210317& 1   & 59290.09440744 & 9.3  & 38.3 & 249.1    \\
            & 2   & 59290.09889063 & 10.9 & 41.5 & 715.3    \\
    20210624& 1   & 59388.82883282 & 9.5  & 32.2 & 32.4     \\
            & 2   & 59388.83097994 & 11.6 & 40.7 & 67.1     \\
\hline
\end{tabular}}
\end{table} 

\begin{figure*}[!t]
\centering
\includegraphics[width=0.33\textwidth]{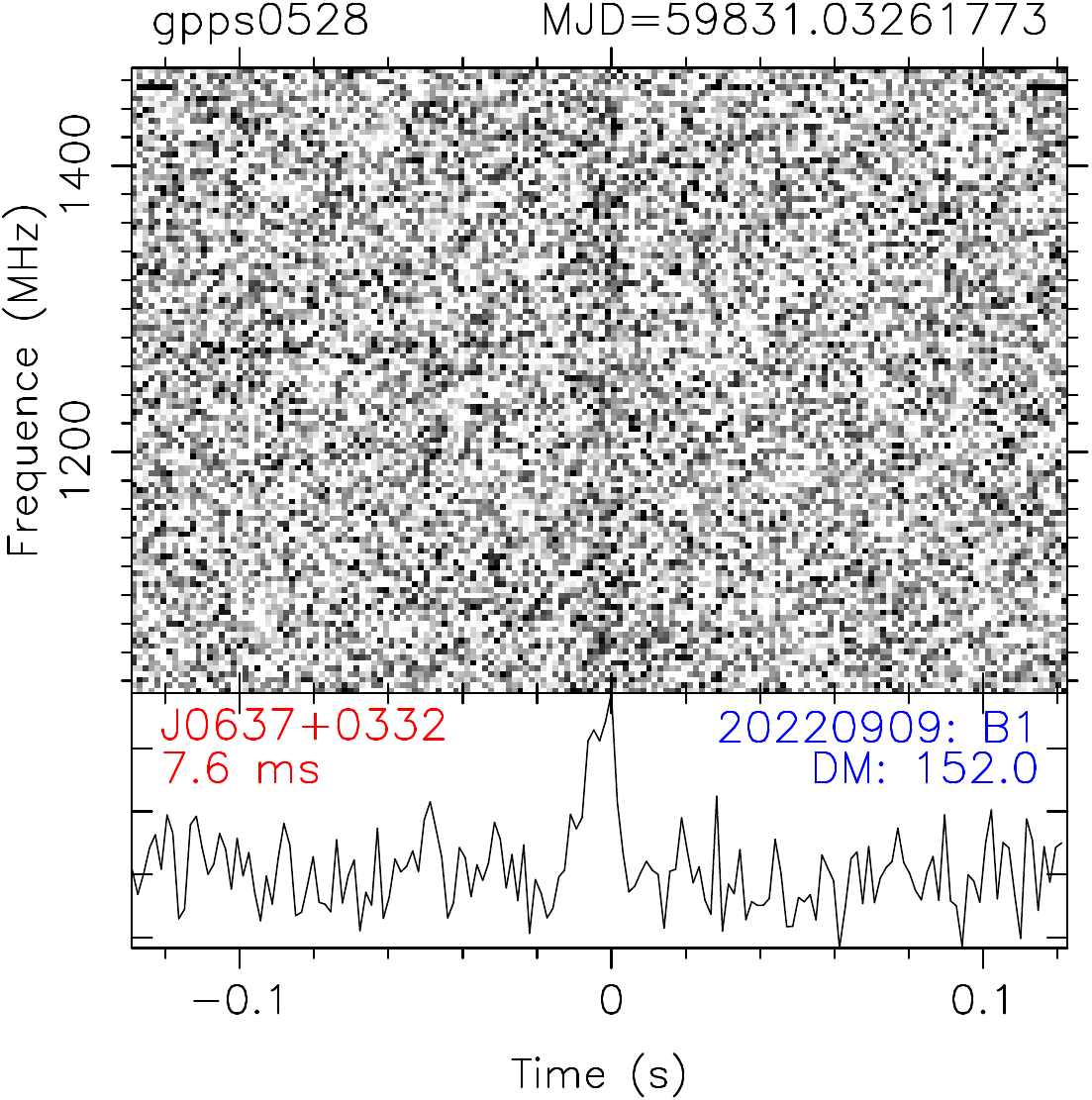}
\includegraphics[width=0.33\textwidth]{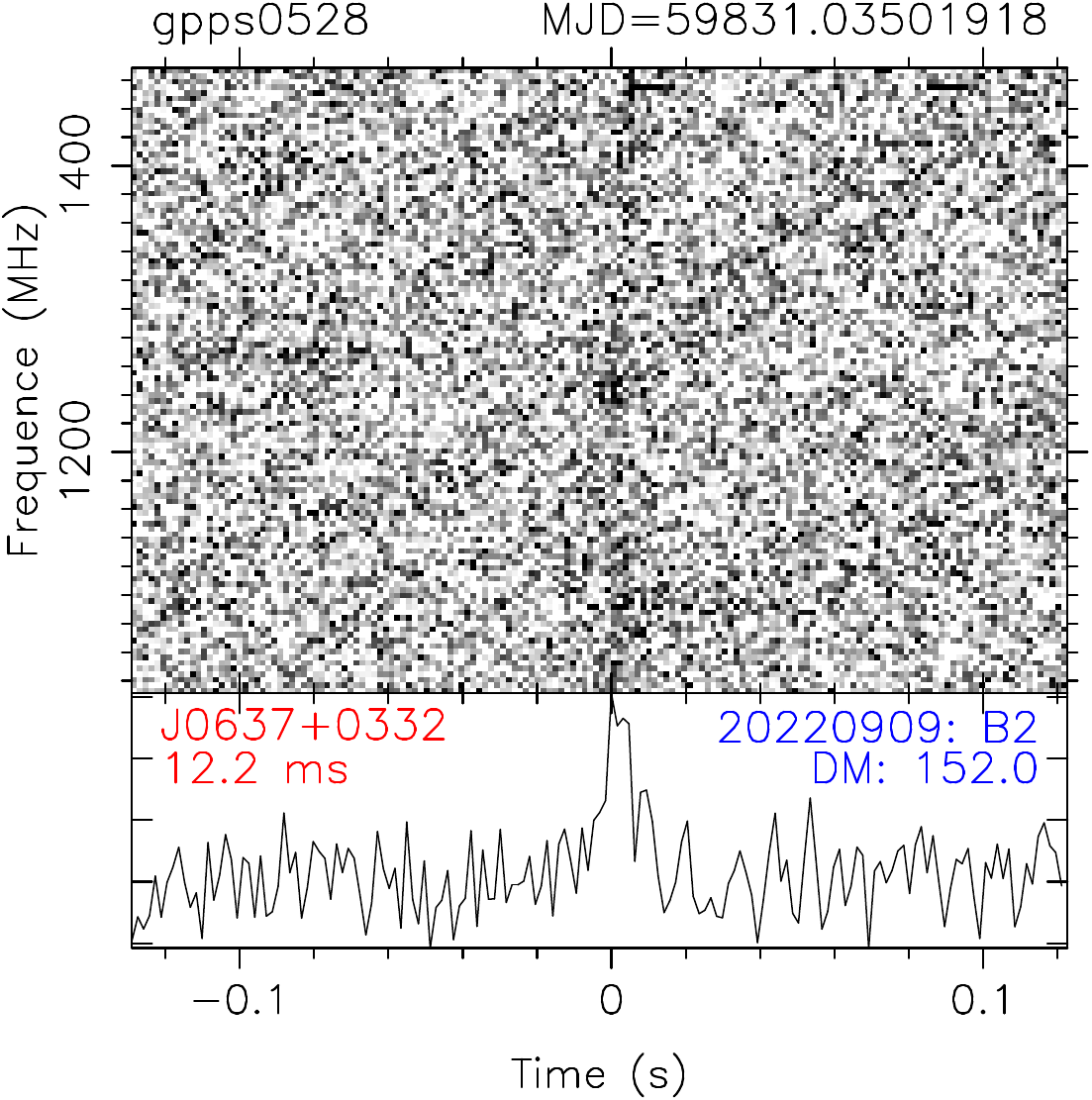}
\includegraphics[width=0.33\textwidth]{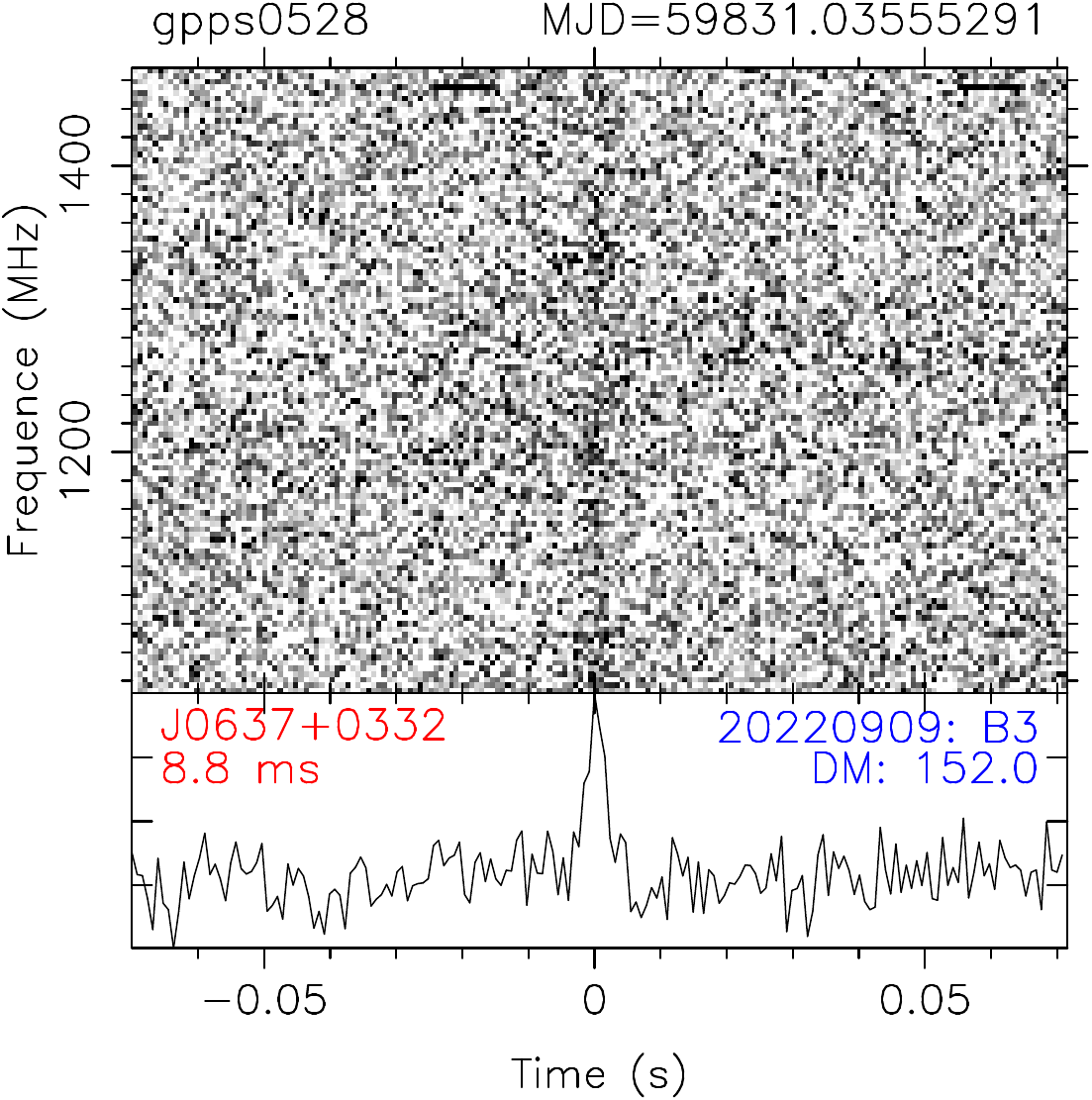}\\[0.2mm]
\includegraphics[width=0.33\textwidth]{New-figs/J1828-0003/G30.20+4.91_20220803_snapshot-M19-P2-c2048b1.fits-DM-192.6-T-202.1319-202.4842-1.pdf}
\includegraphics[width=0.33\textwidth]{New-figs/J1828-0003/G30.20+4.91_20220803_snapshot-M19-P4-c2048b1.fits-DM-192.6-T-114.1522-114.4542-1.pdf}
\includegraphics[width=0.33\textwidth]{New-figs/J1828-0003/J182853-000340sp_20221114_swiftcalibration-M01-P1-c2048b1.fits-DM-192.6-T-658.3883-658.8916-1.pdf}\\[0.2mm]
\includegraphics[width=0.33\textwidth]{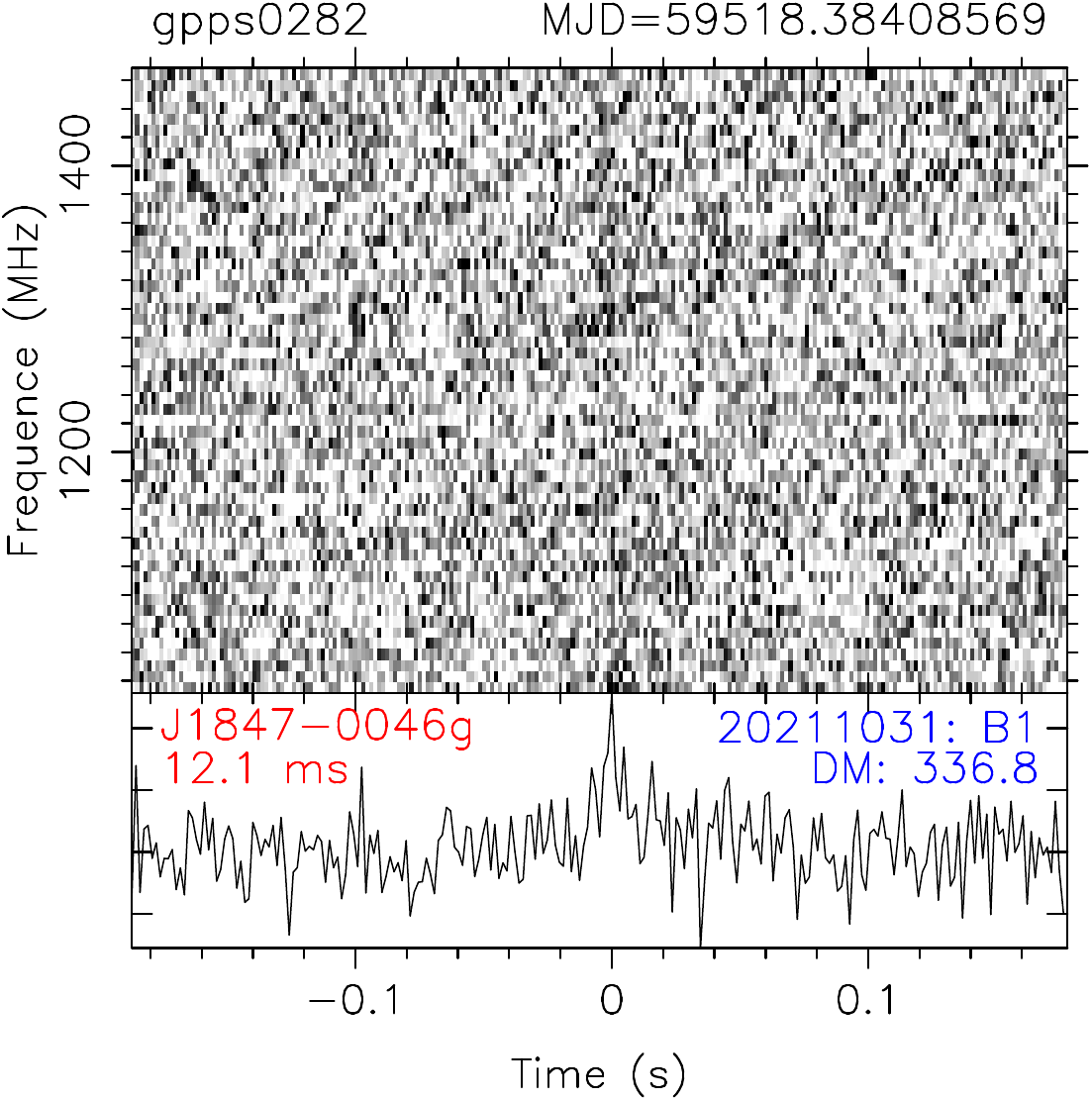}
\includegraphics[width=0.33\textwidth]{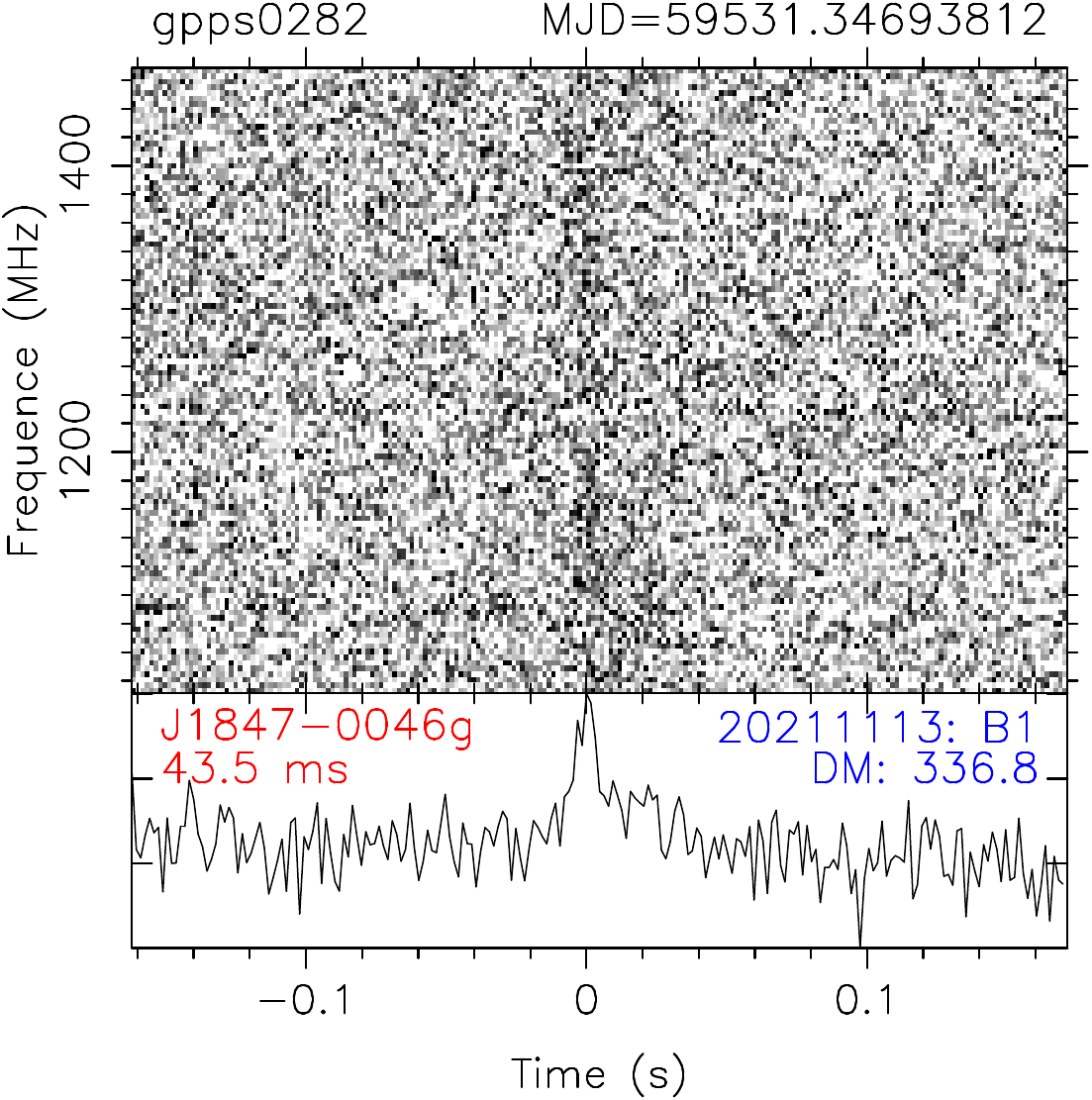}
\includegraphics[width=0.33\textwidth]{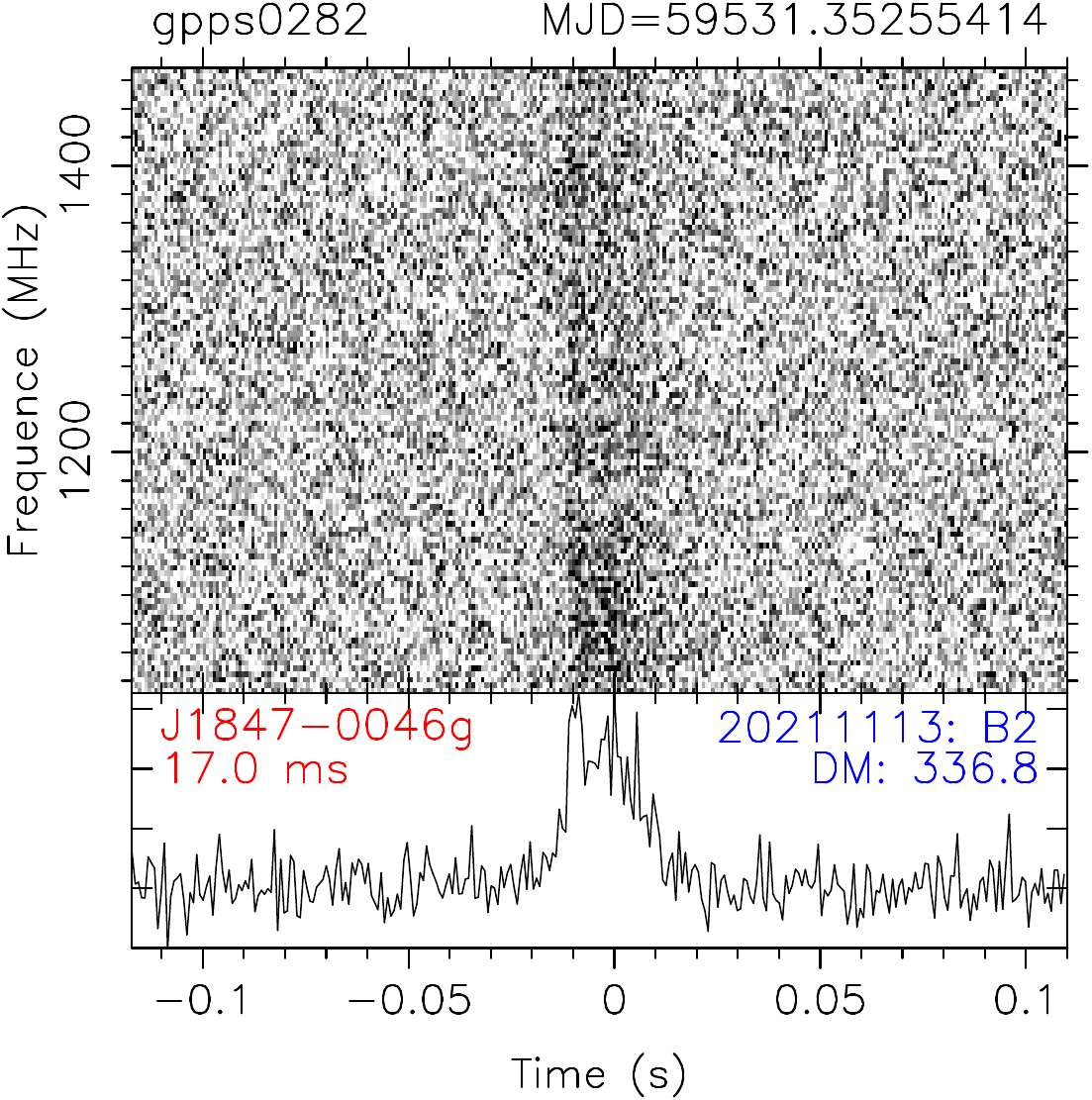}\\[0.2mm]
\includegraphics[width=0.33\textwidth]{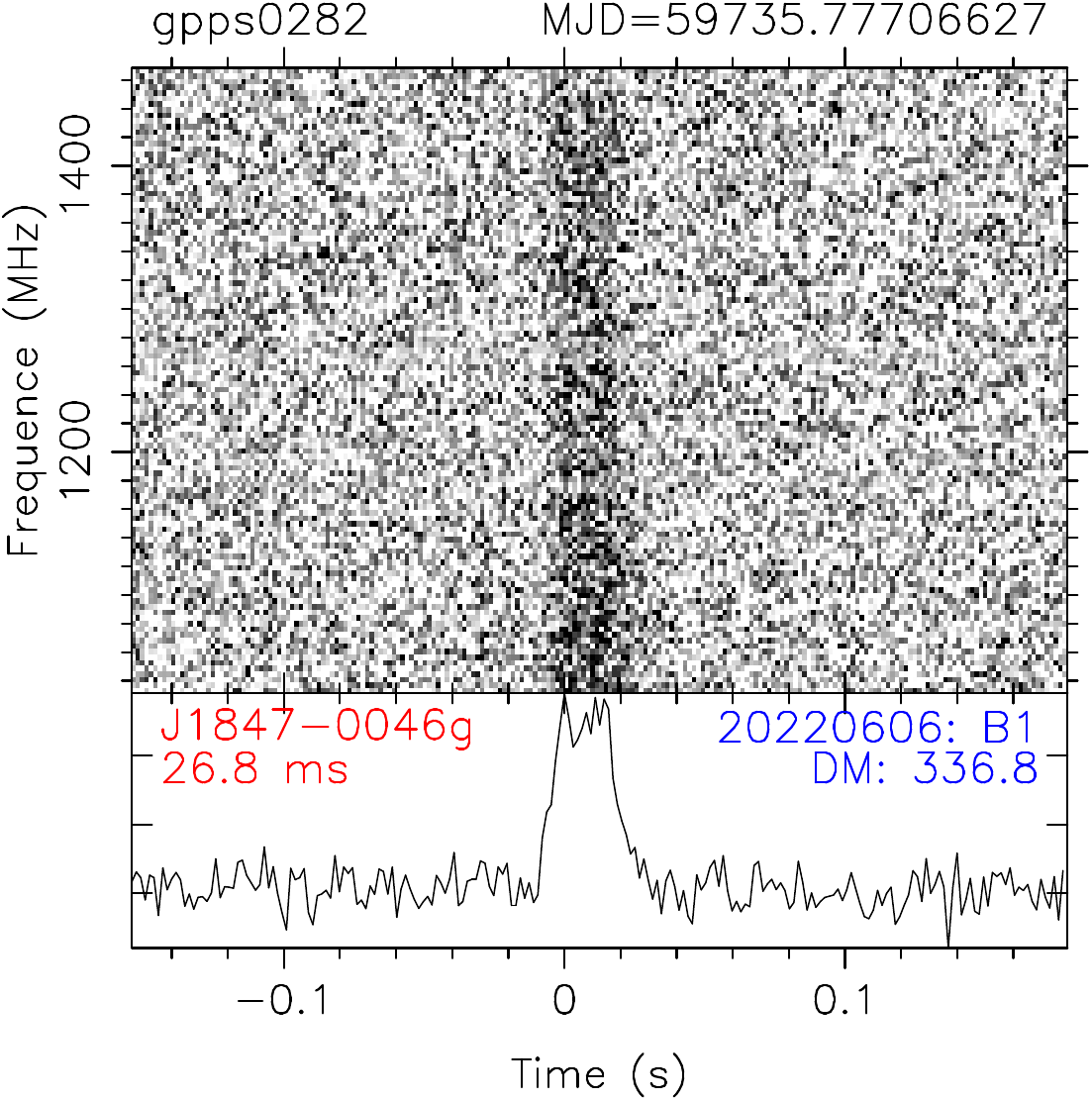}
\includegraphics[width=0.33\textwidth]{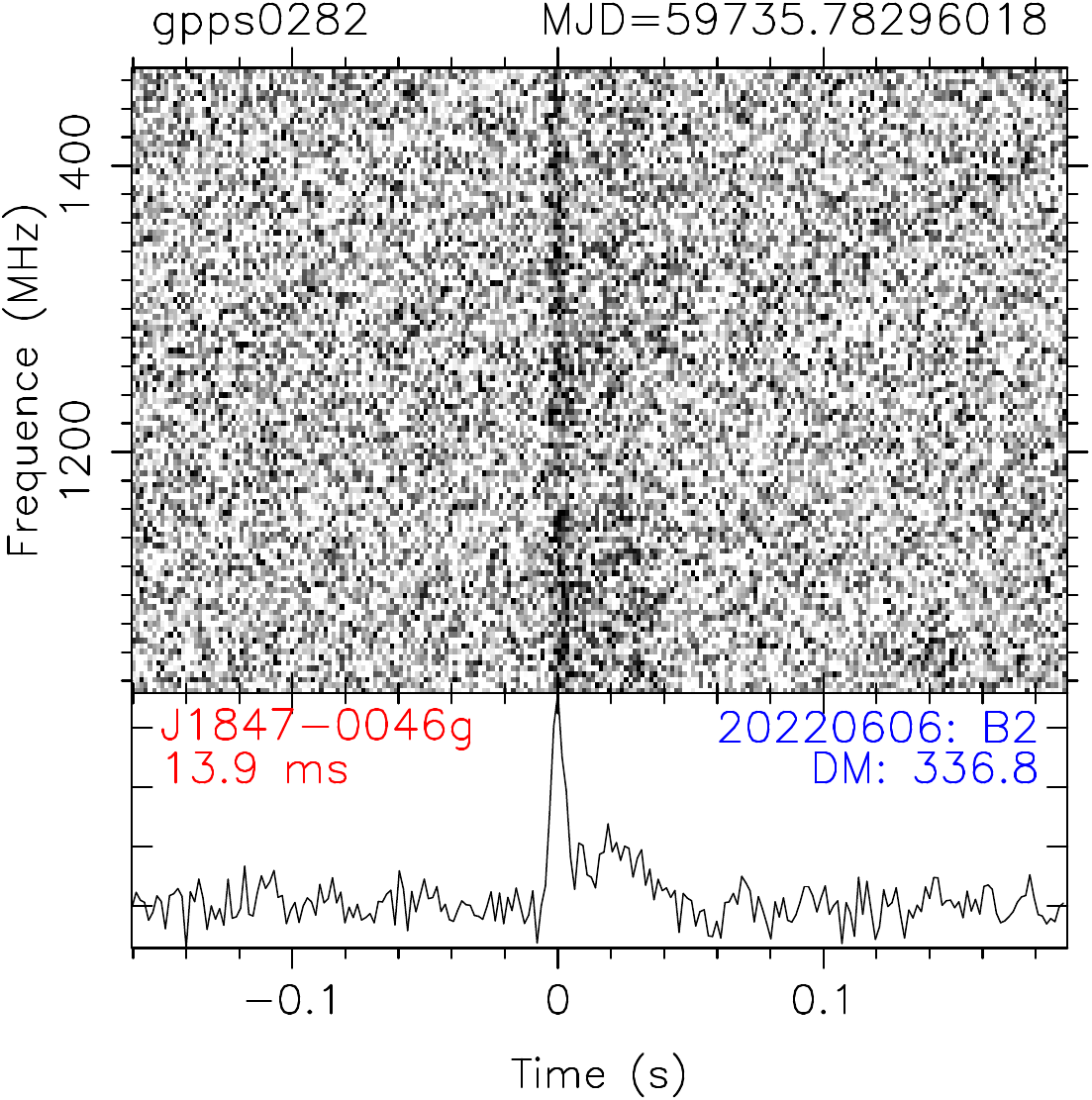}
\includegraphics[width=0.33\textwidth]{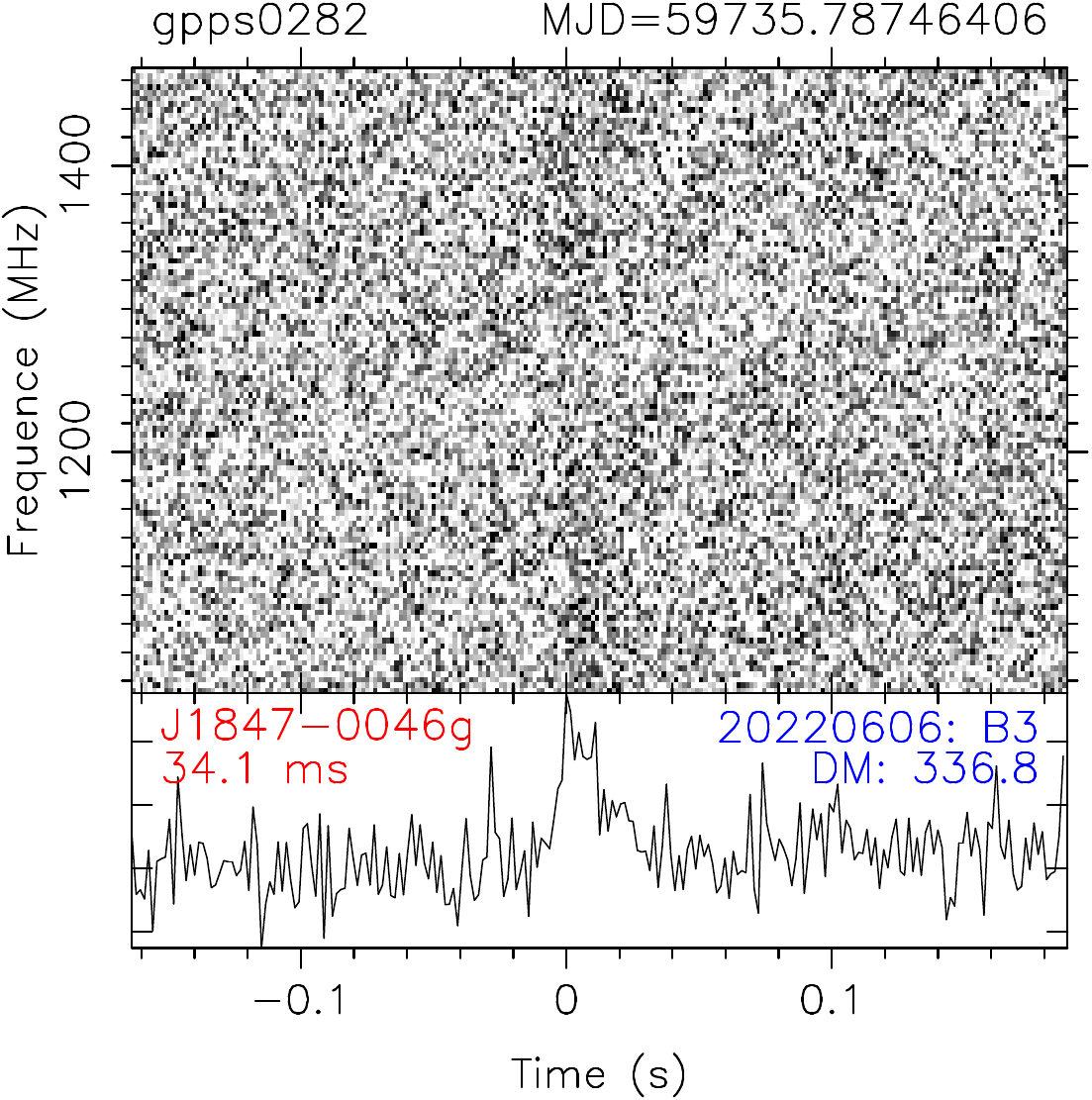}
\caption{The same as Figure~\ref{fig:fewPulses} but for 105 pulses detected by FAST from 26 radio transient sources.}
\label{fig:AppfewPulses}
\end{figure*}
\addtocounter{figure}{-1}
\begin{figure*}[!t]
\centering
\includegraphics[width=0.33\textwidth]{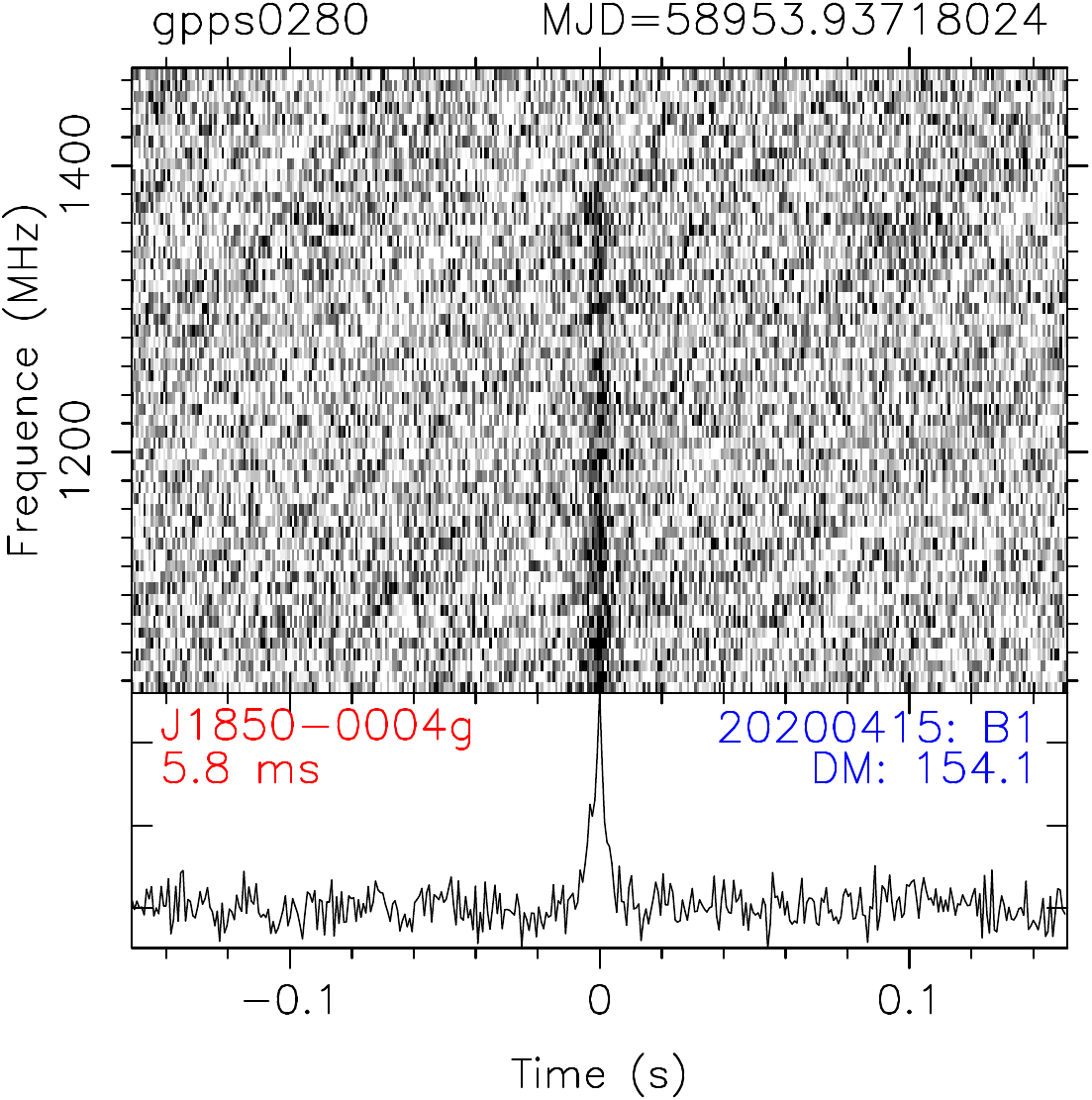}
\includegraphics[width=0.33\textwidth]{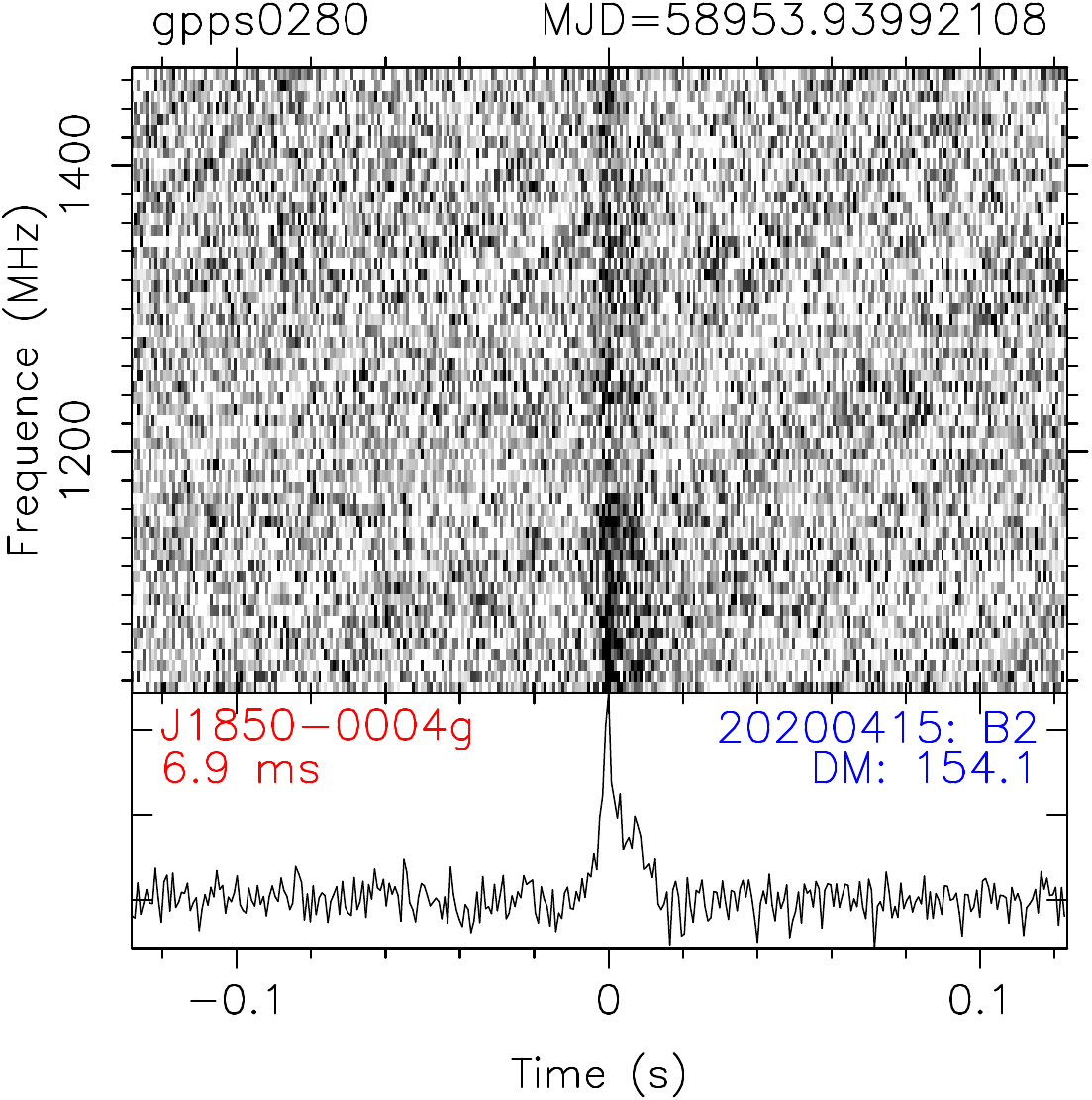}
\includegraphics[width=0.33\textwidth]{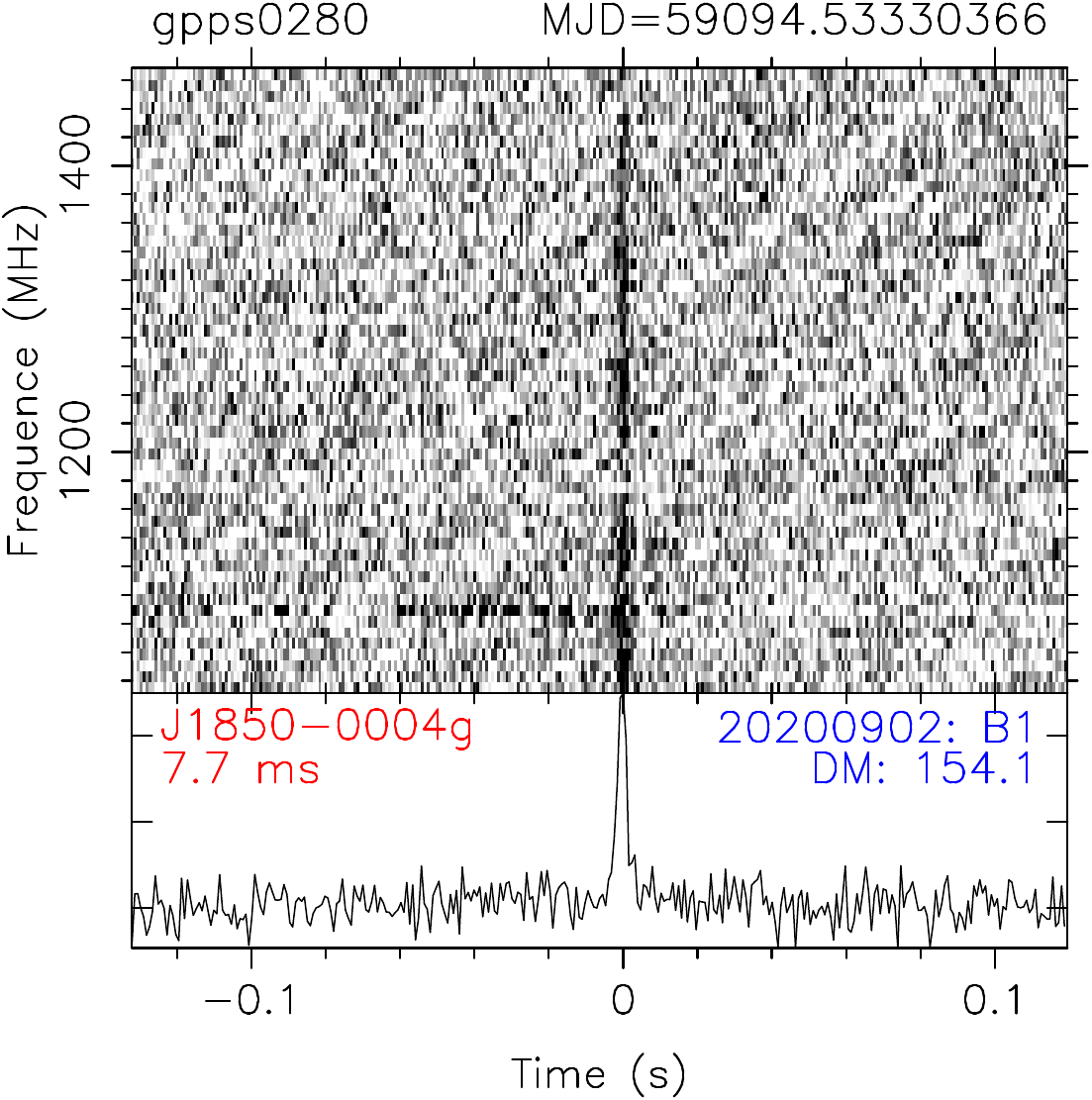}\\[0.2mm]
\includegraphics[width=0.33\textwidth]{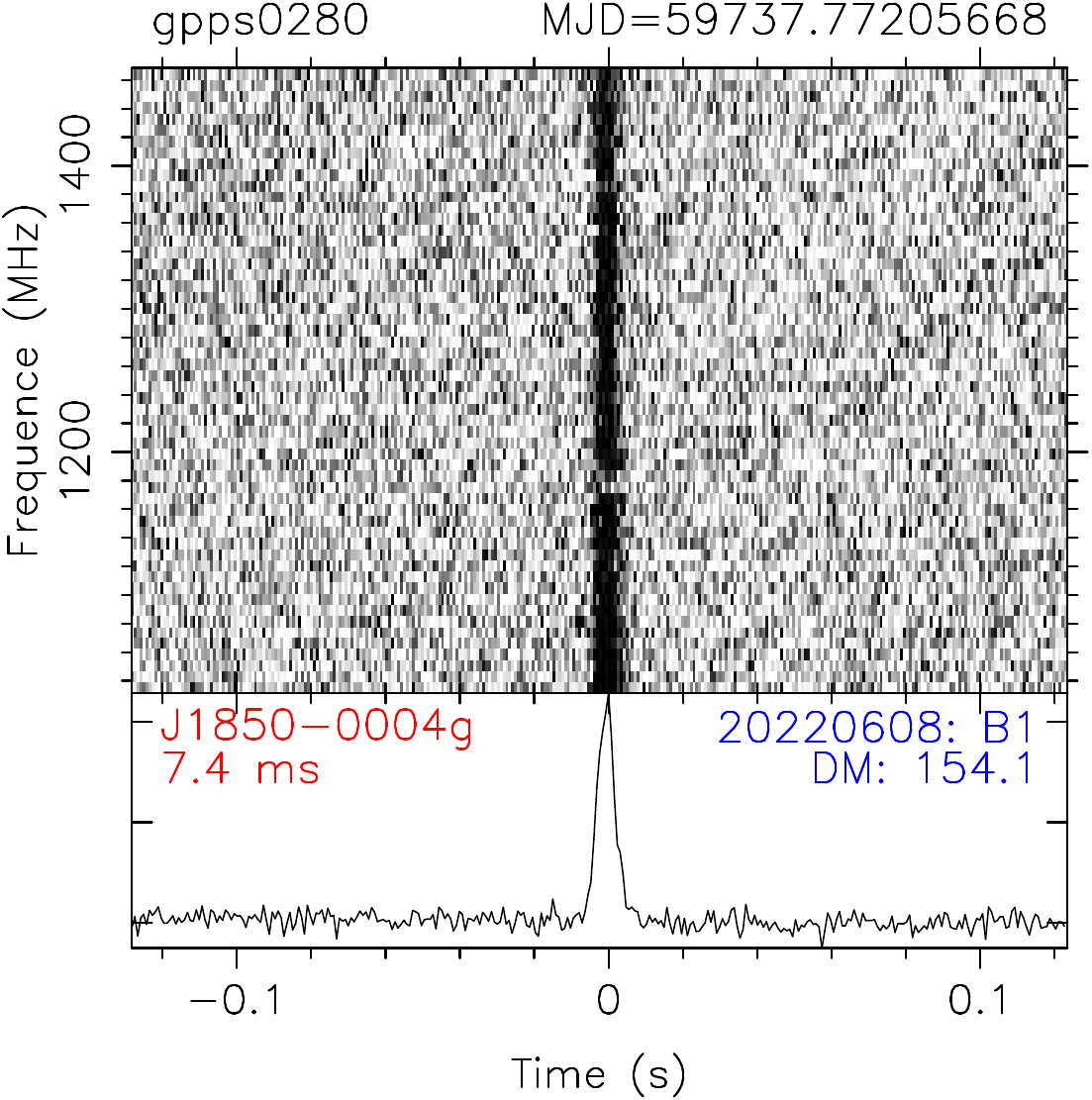}
\includegraphics[width=0.33\textwidth]{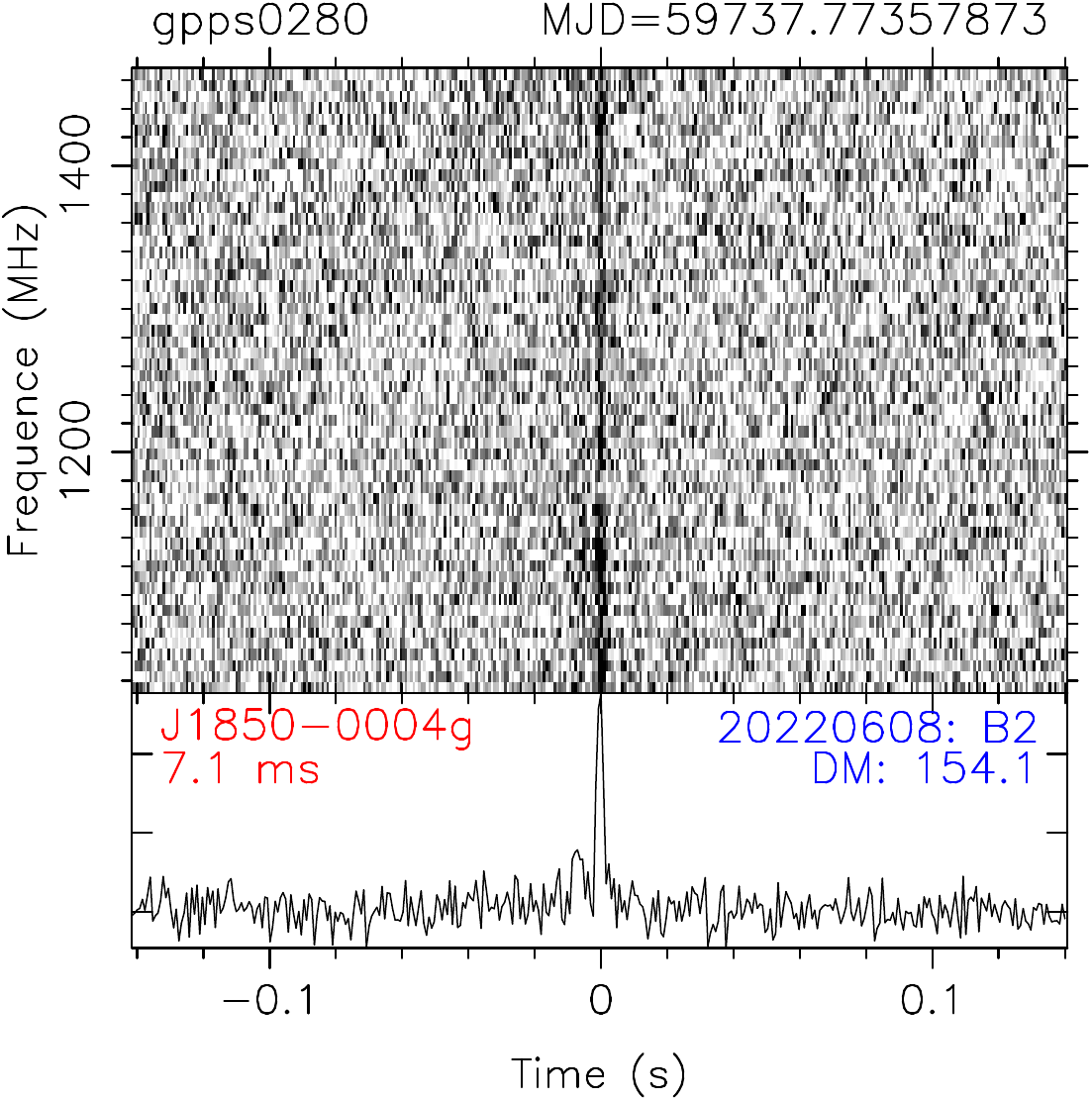}
\includegraphics[width=0.33\textwidth]{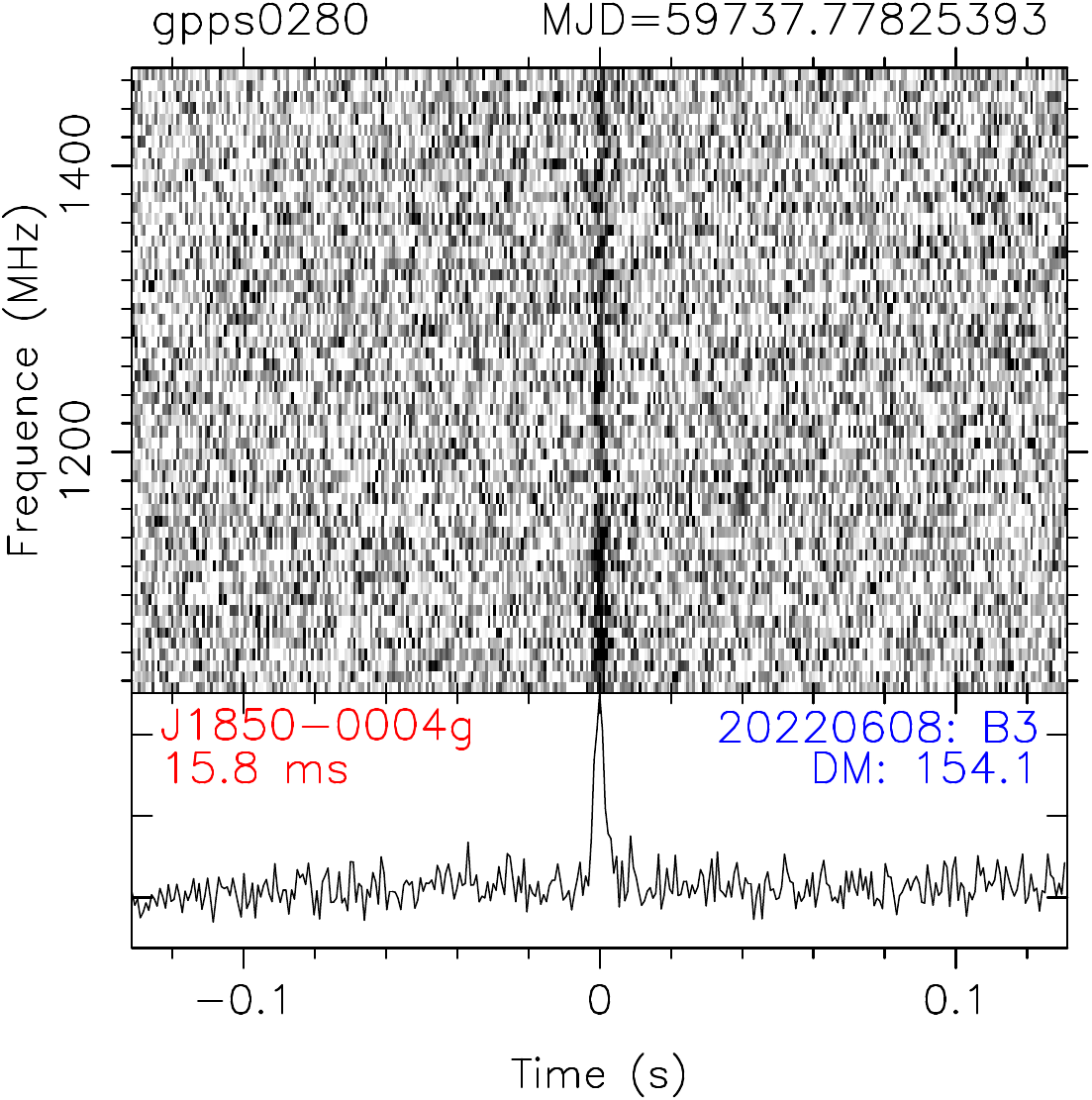}\\[0.2mm]
\includegraphics[width=0.33\textwidth]{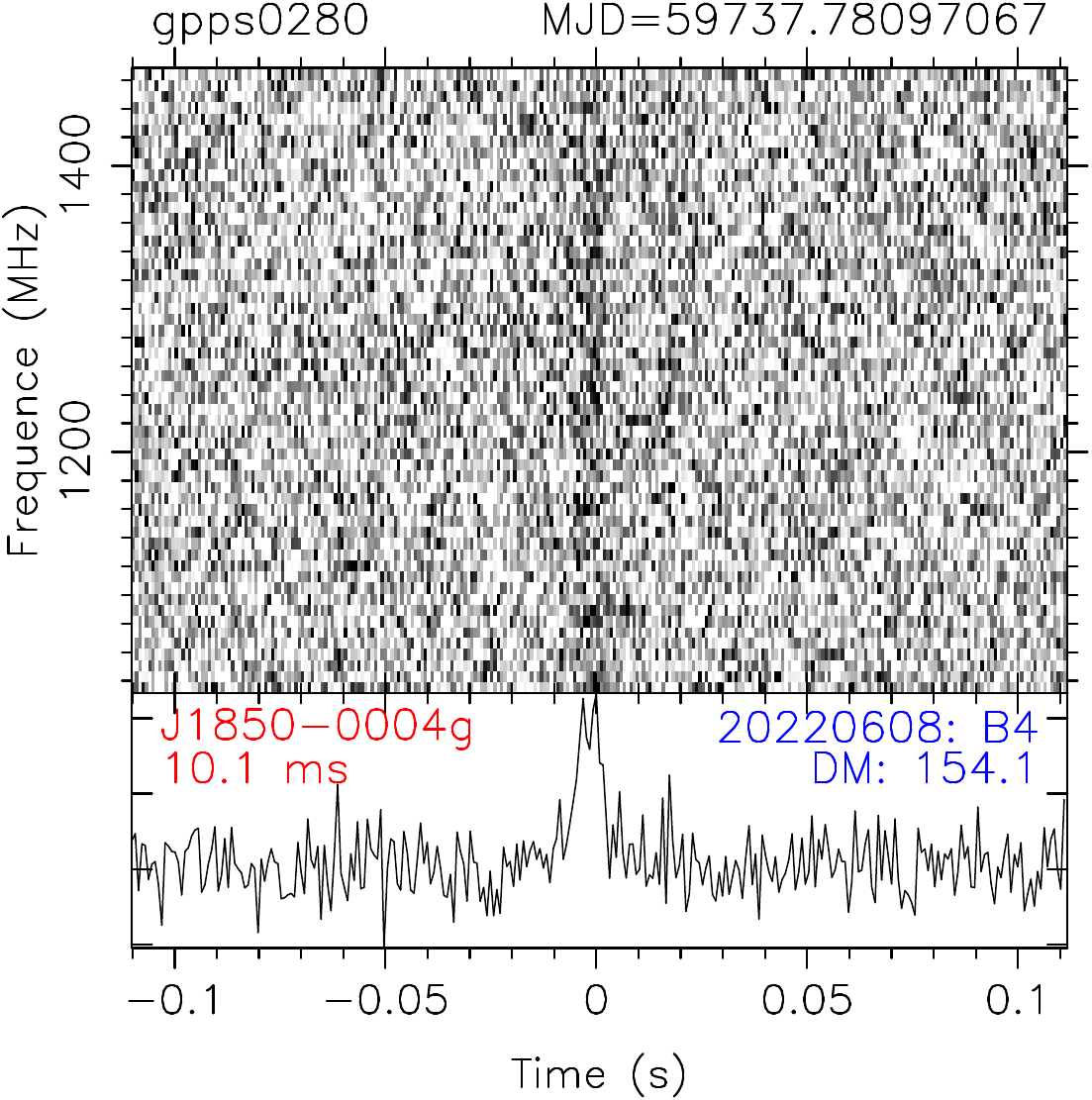}
\includegraphics[width=0.33\textwidth]{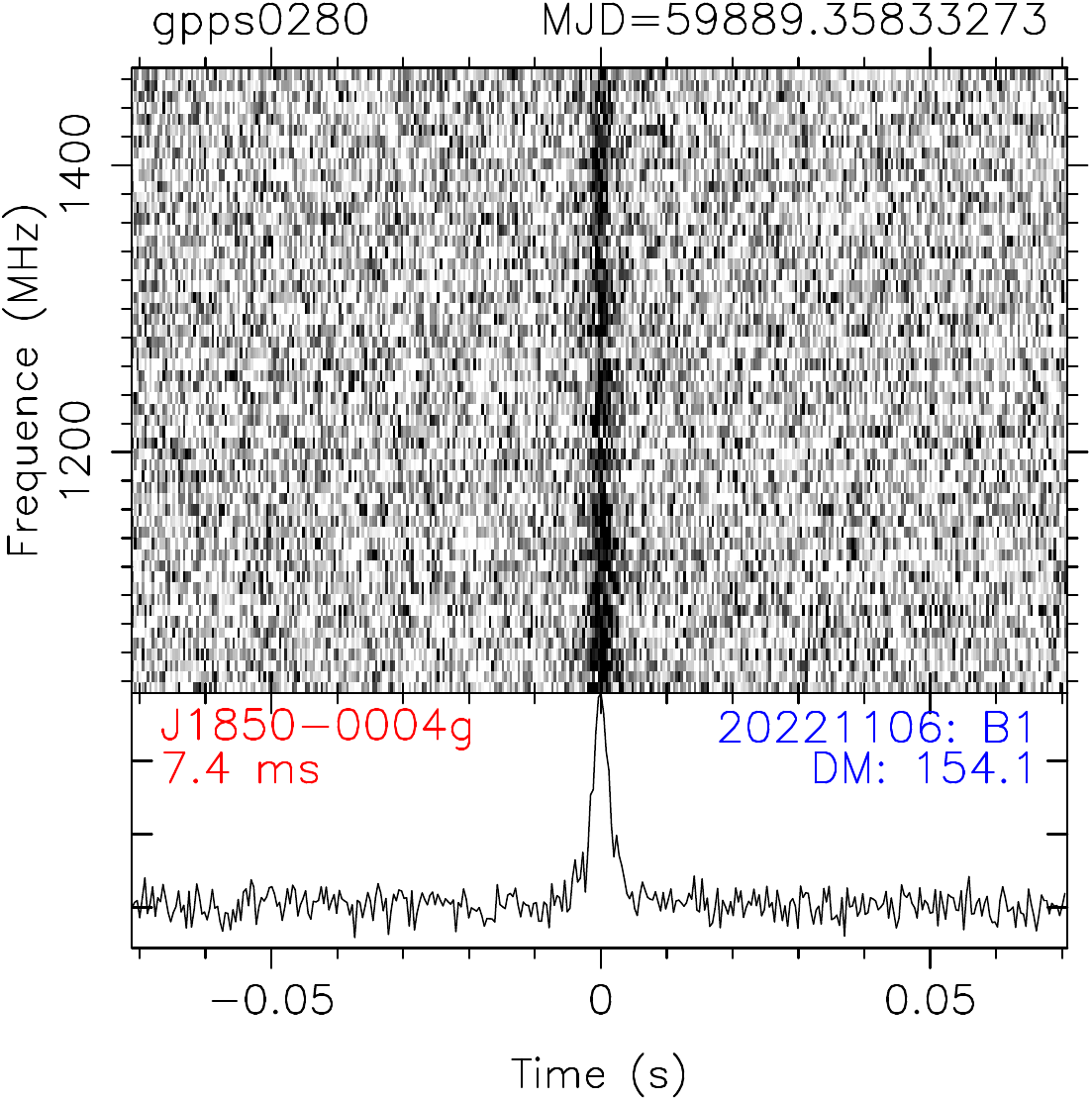}
\includegraphics[width=0.33\textwidth]{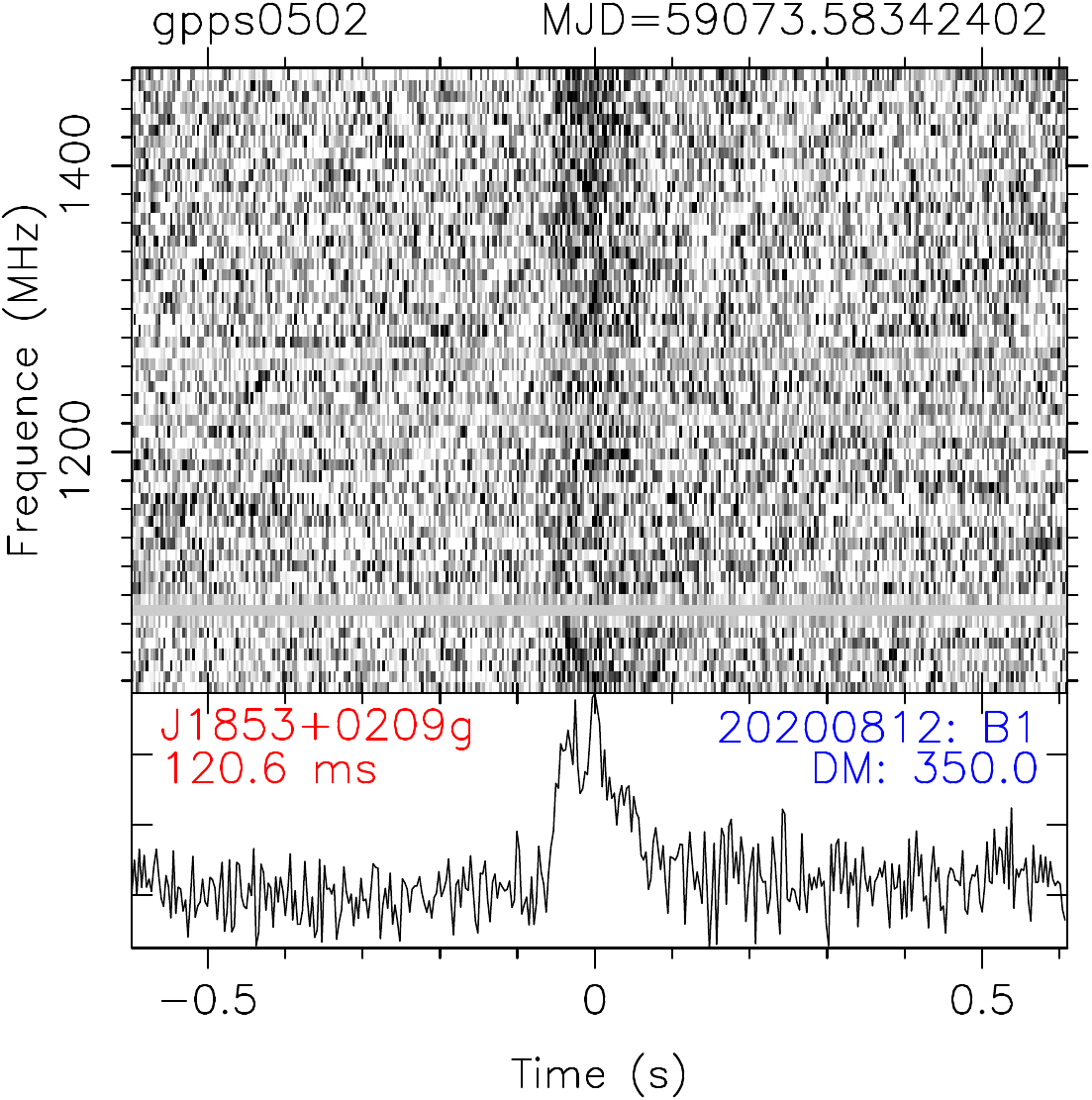}\\[0.2mm]
\includegraphics[width=0.33\textwidth]{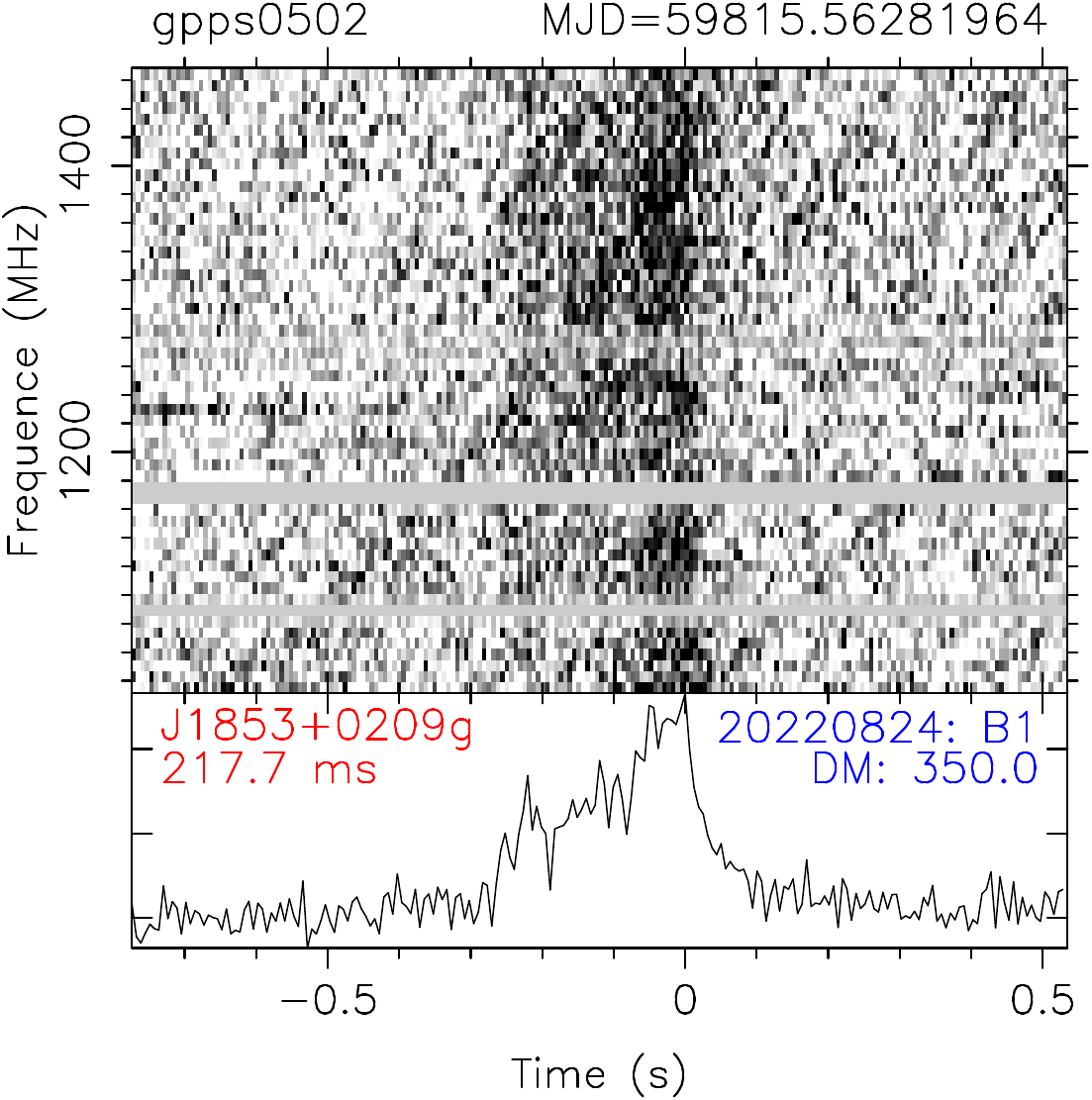}
\includegraphics[width=0.33\textwidth]{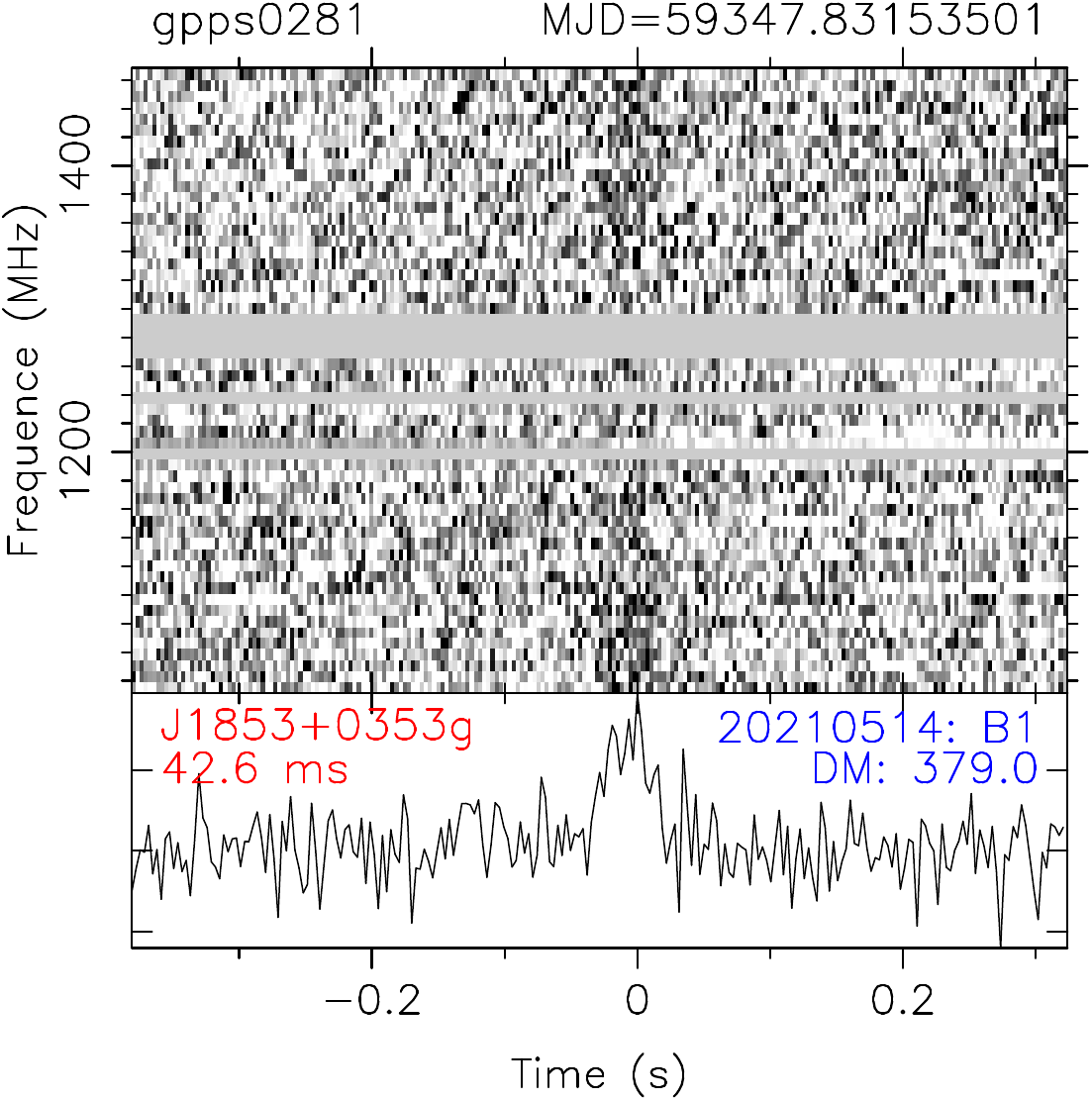}
\includegraphics[width=0.33\textwidth]{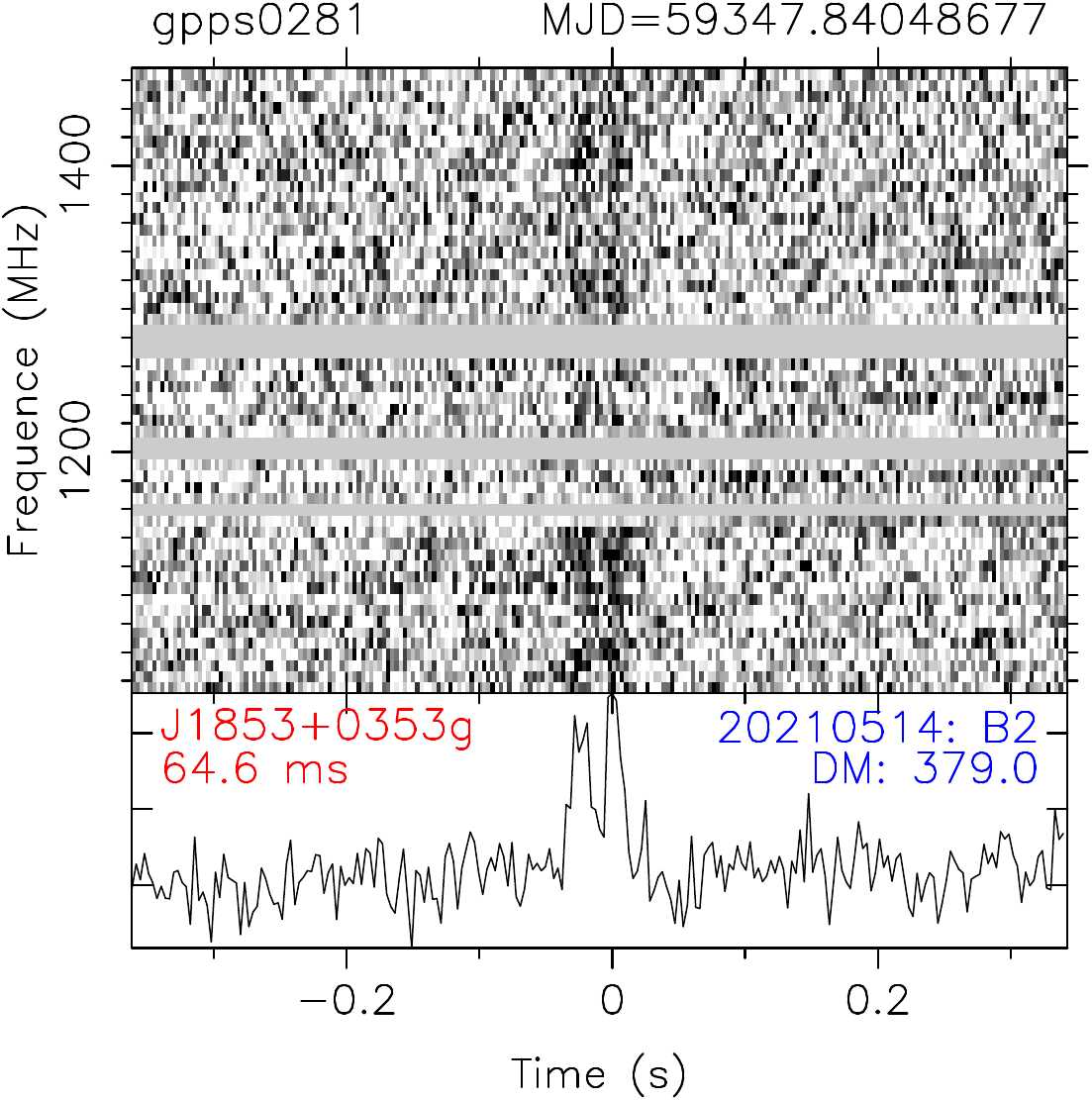}
\caption{(Continued.)}
\label{fig:AppfewPulses}
\end{figure*}
\addtocounter{figure}{-1}
\begin{figure*}[!t]
\centering
\includegraphics[width=0.33\textwidth]{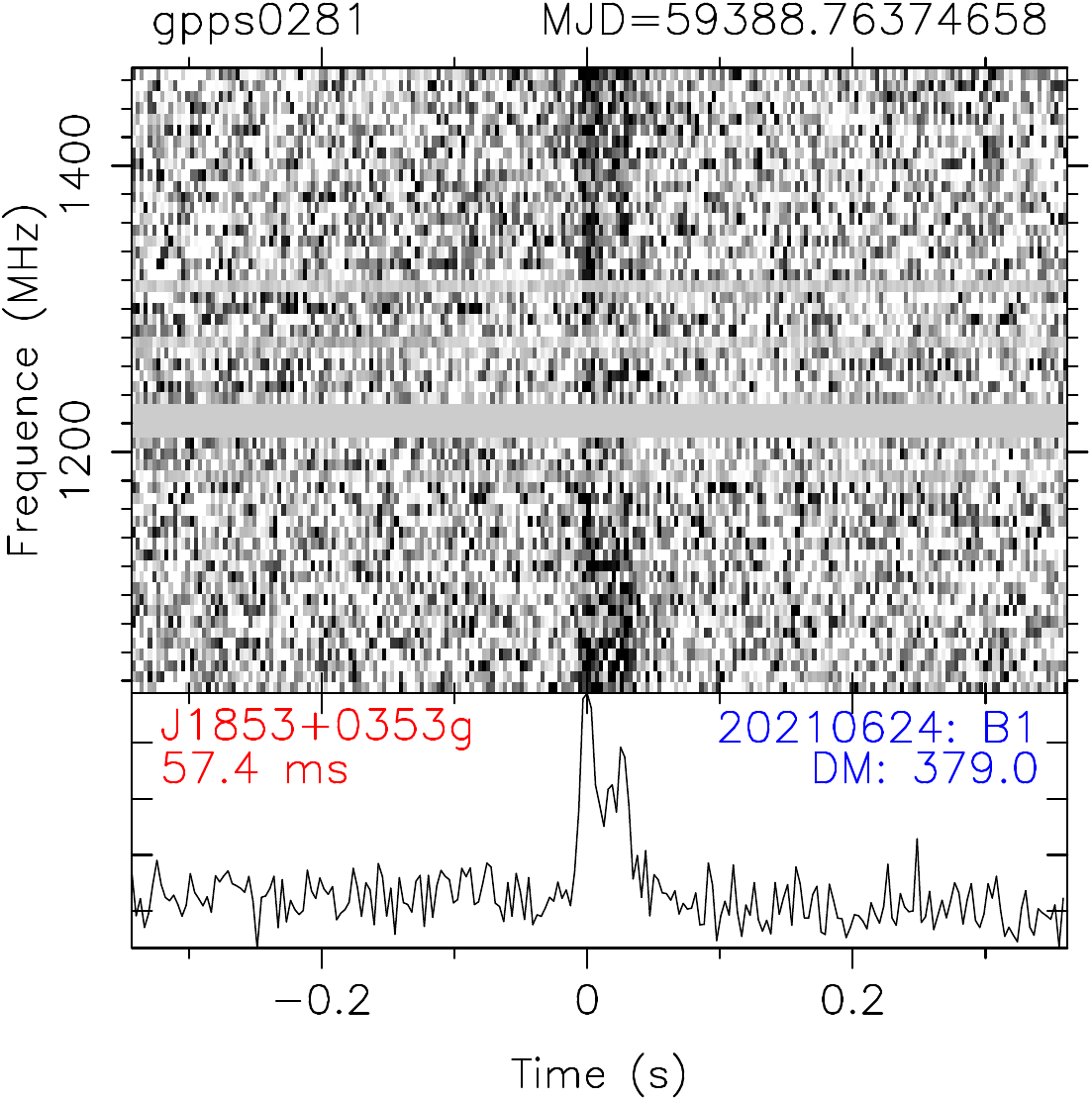}
\includegraphics[width=0.33\textwidth]{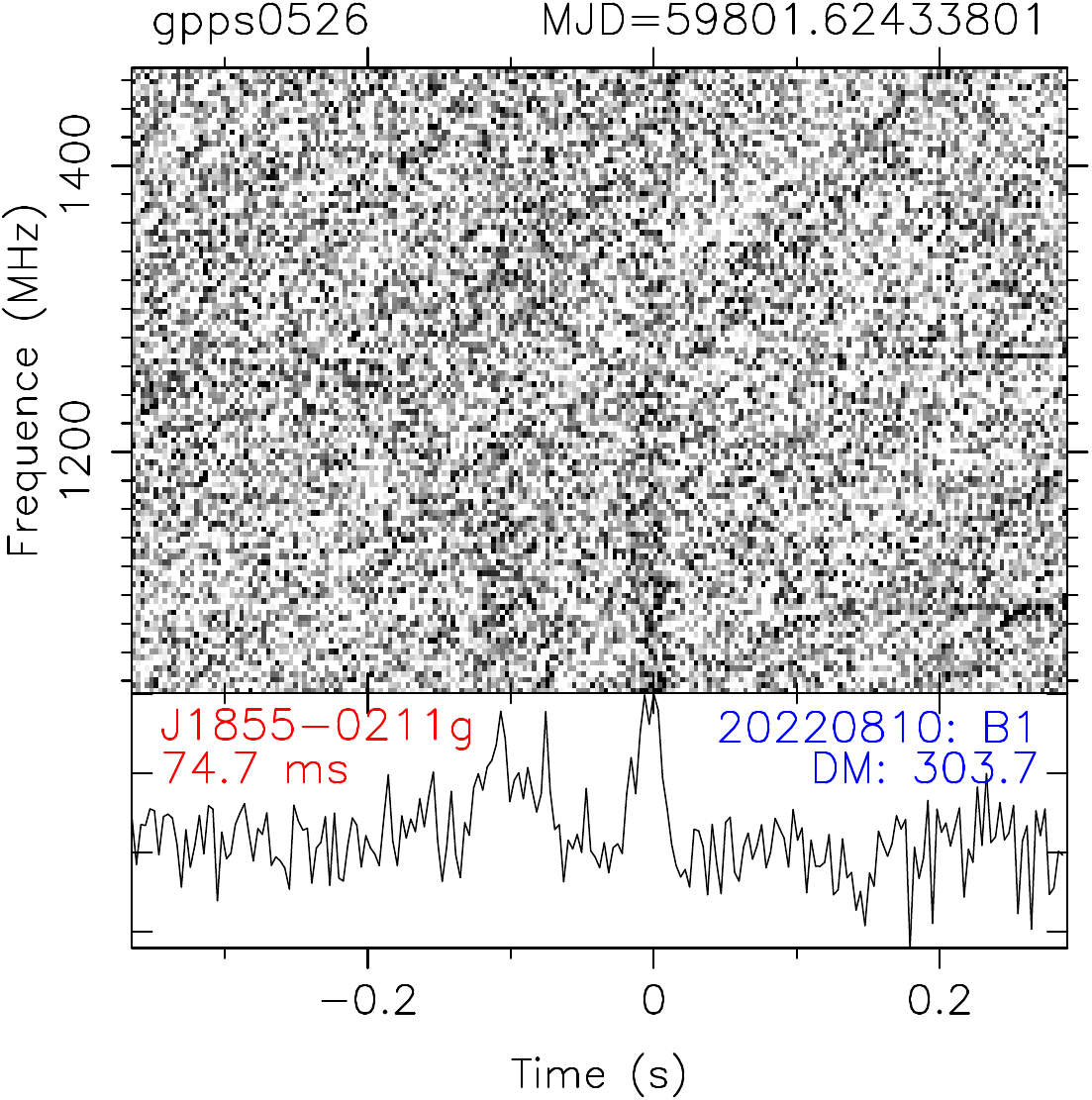}
\includegraphics[width=0.33\textwidth]{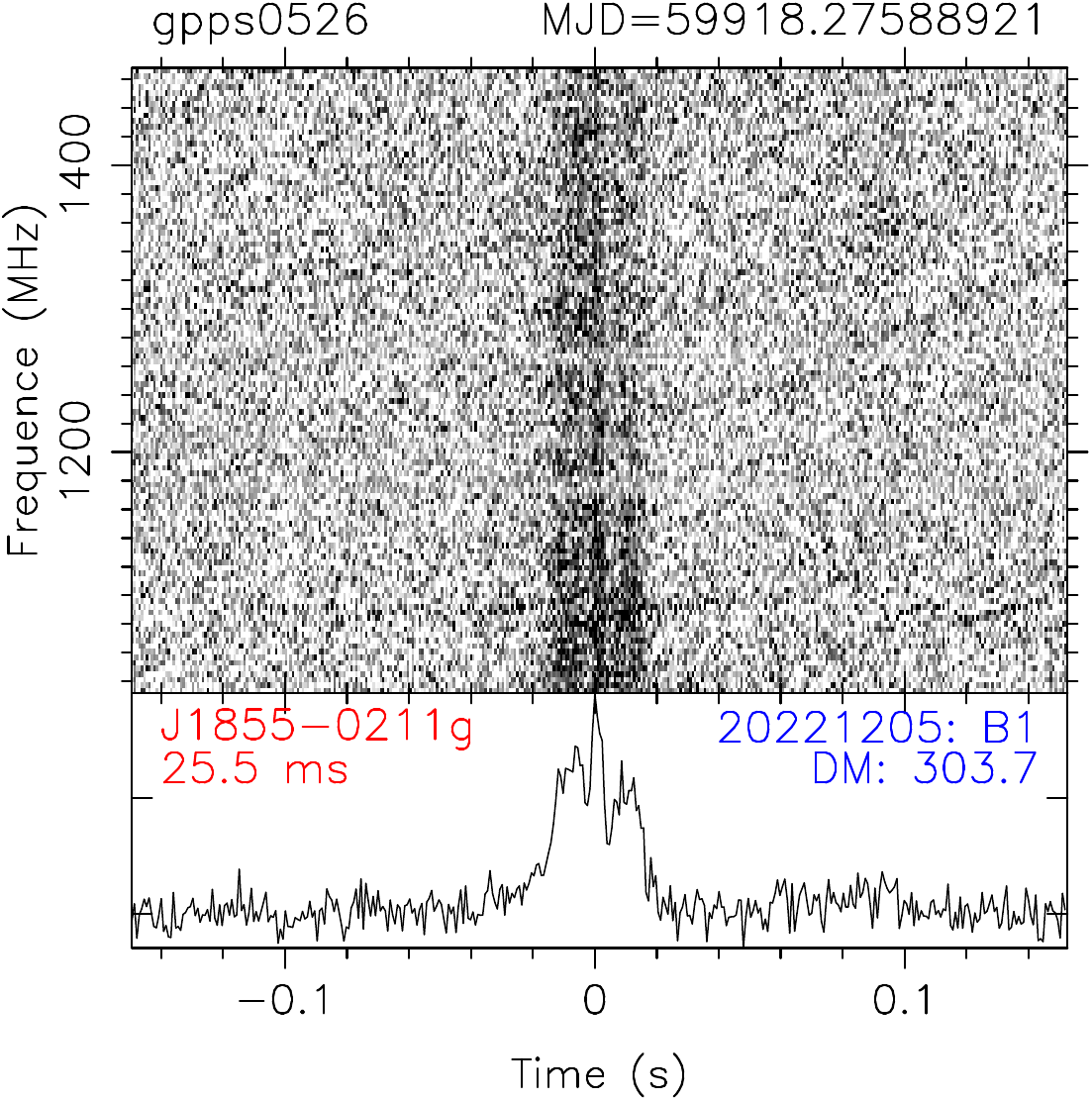}\\[0.2mm]
\includegraphics[width=0.33\textwidth]{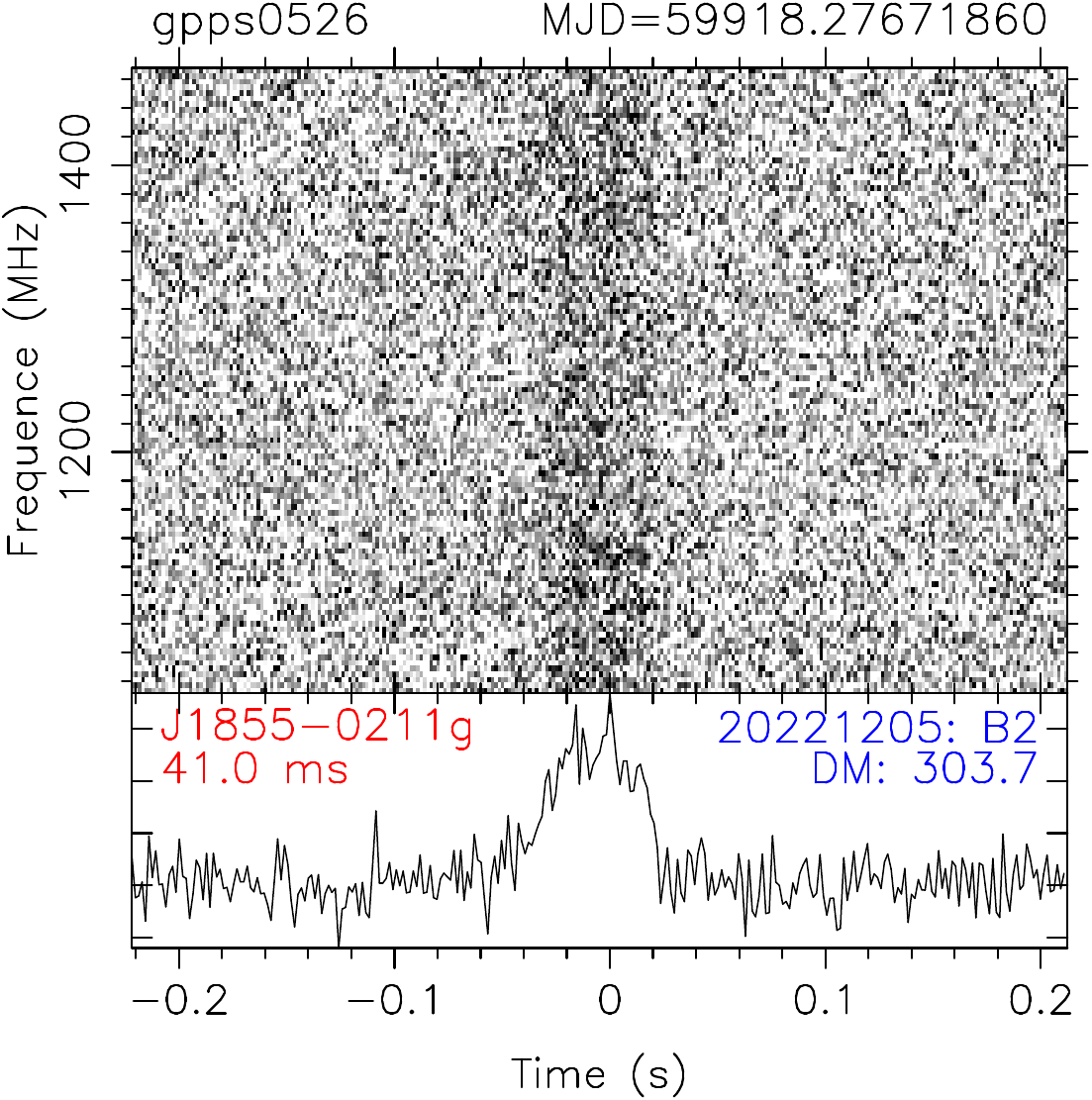}
\includegraphics[width=0.33\textwidth]{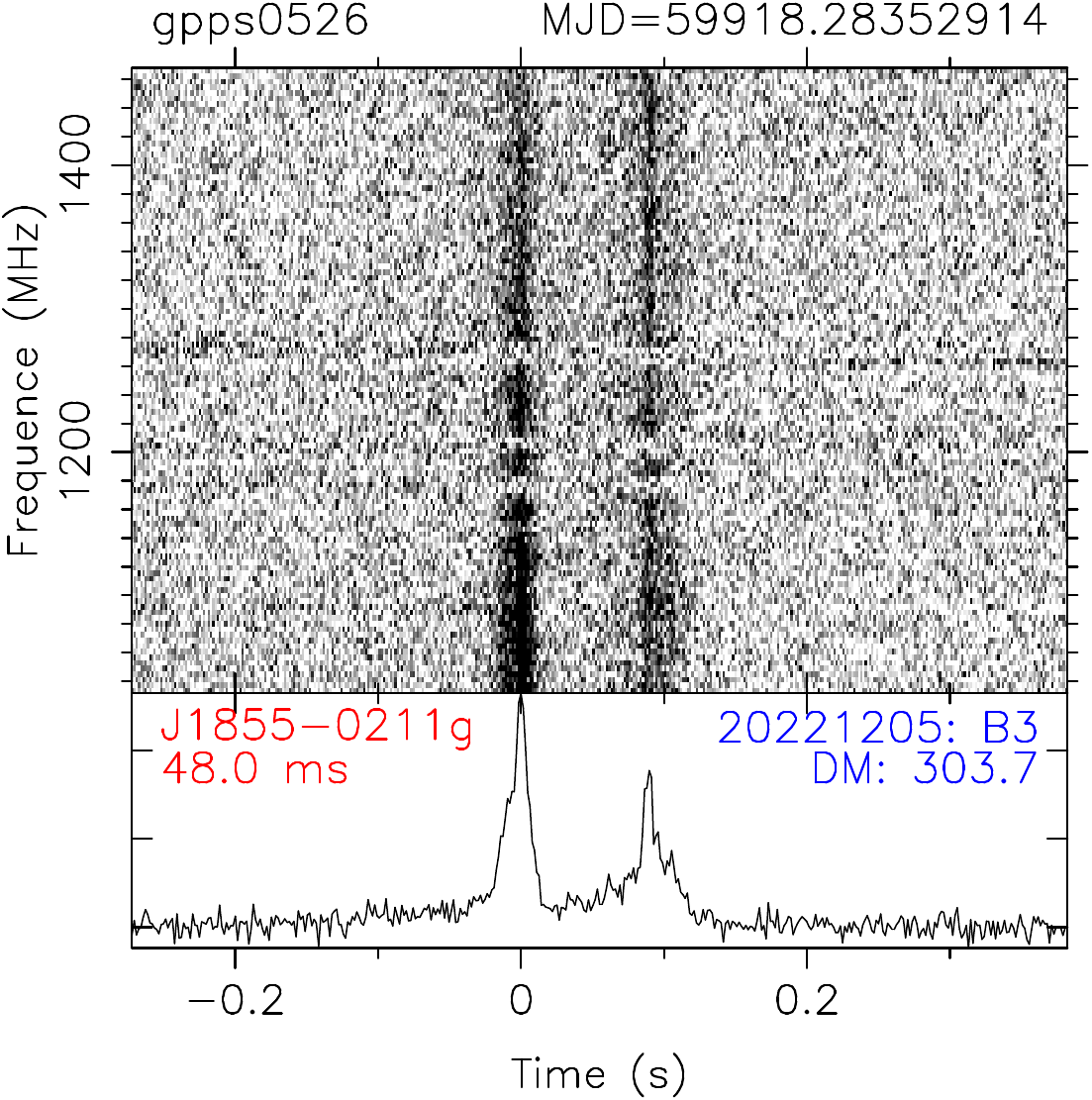}
\includegraphics[width=0.33\textwidth]{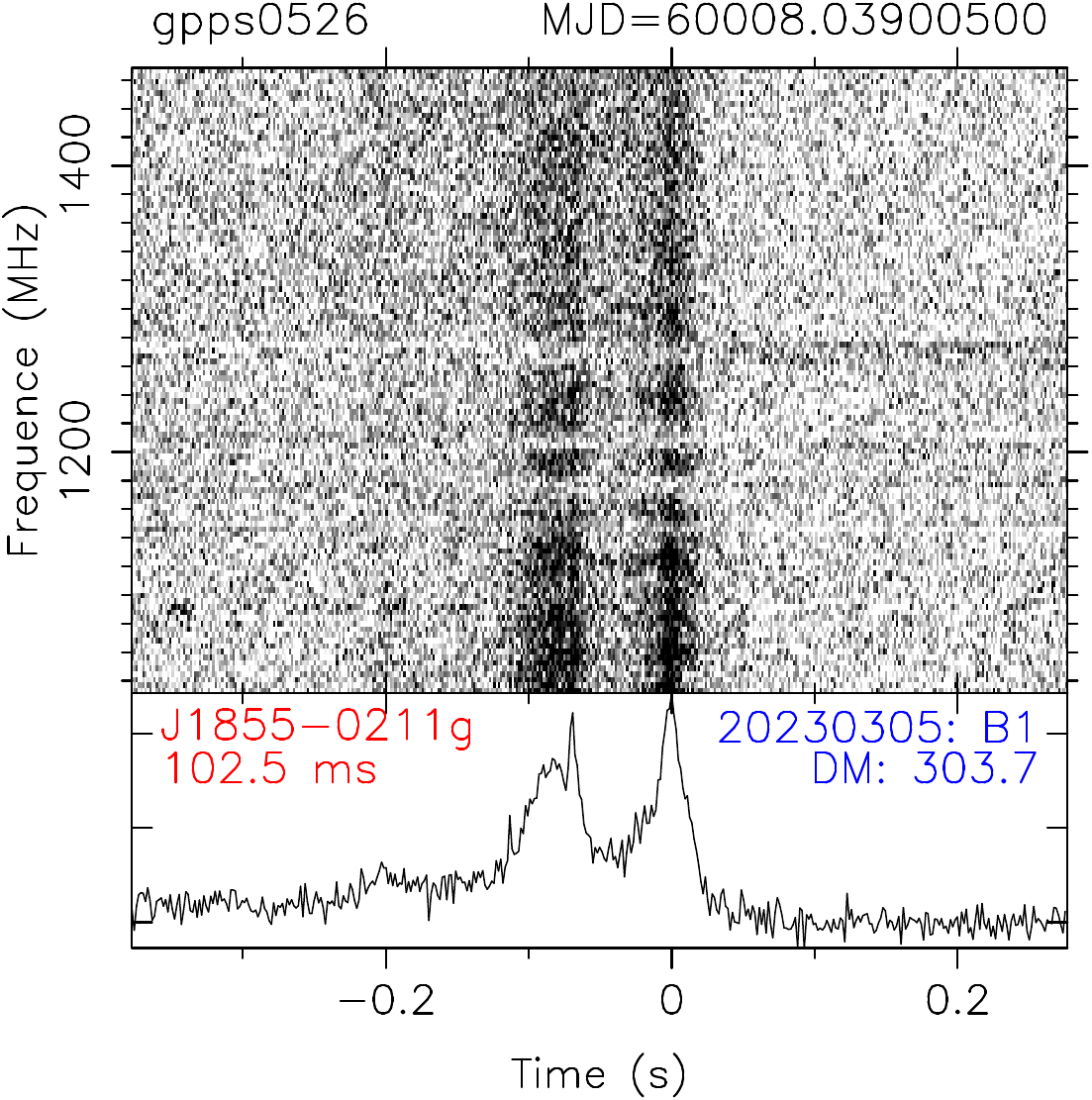}\\[0.2mm]
\includegraphics[width=0.33\textwidth]{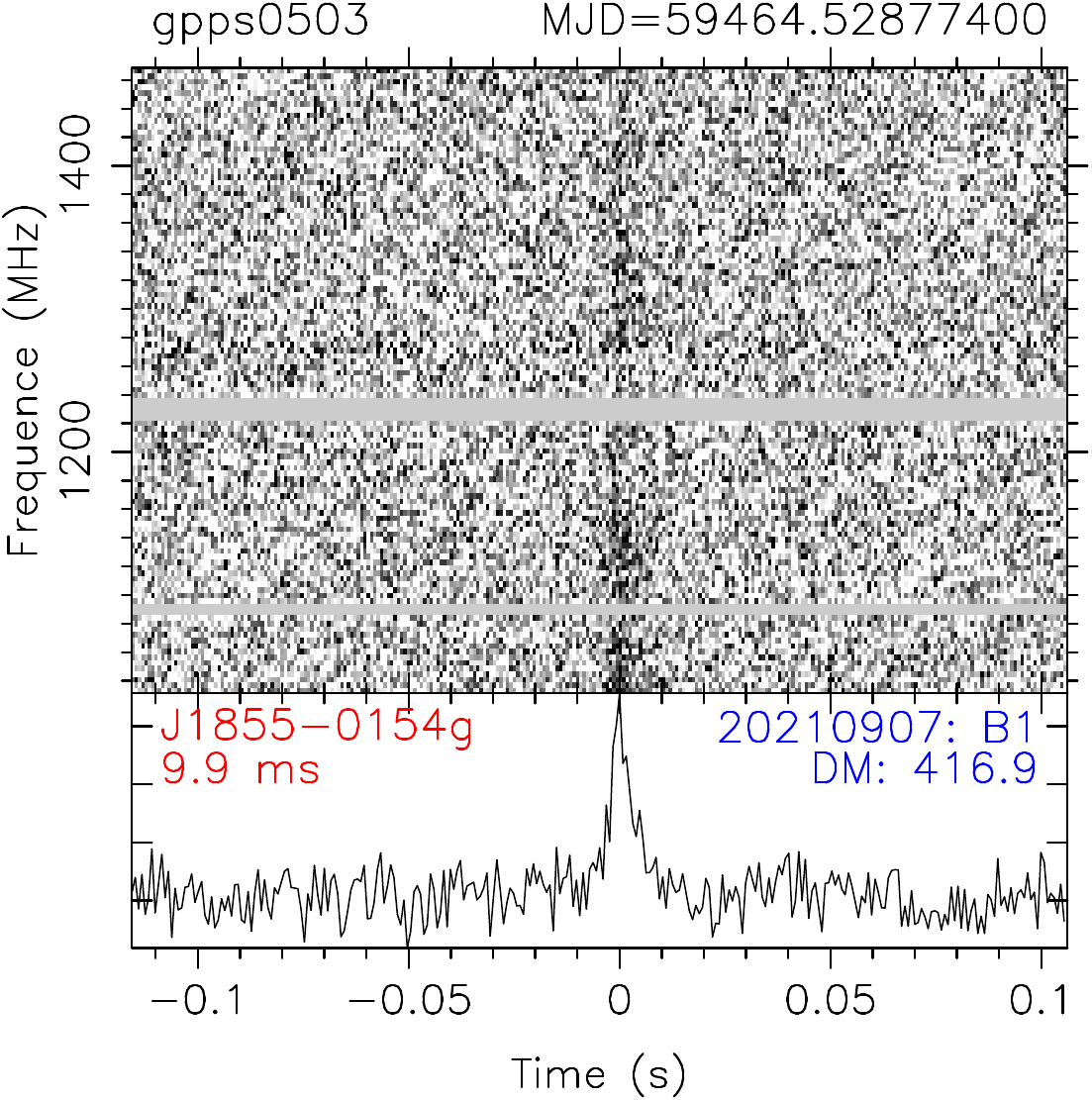}
\includegraphics[width=0.33\textwidth]{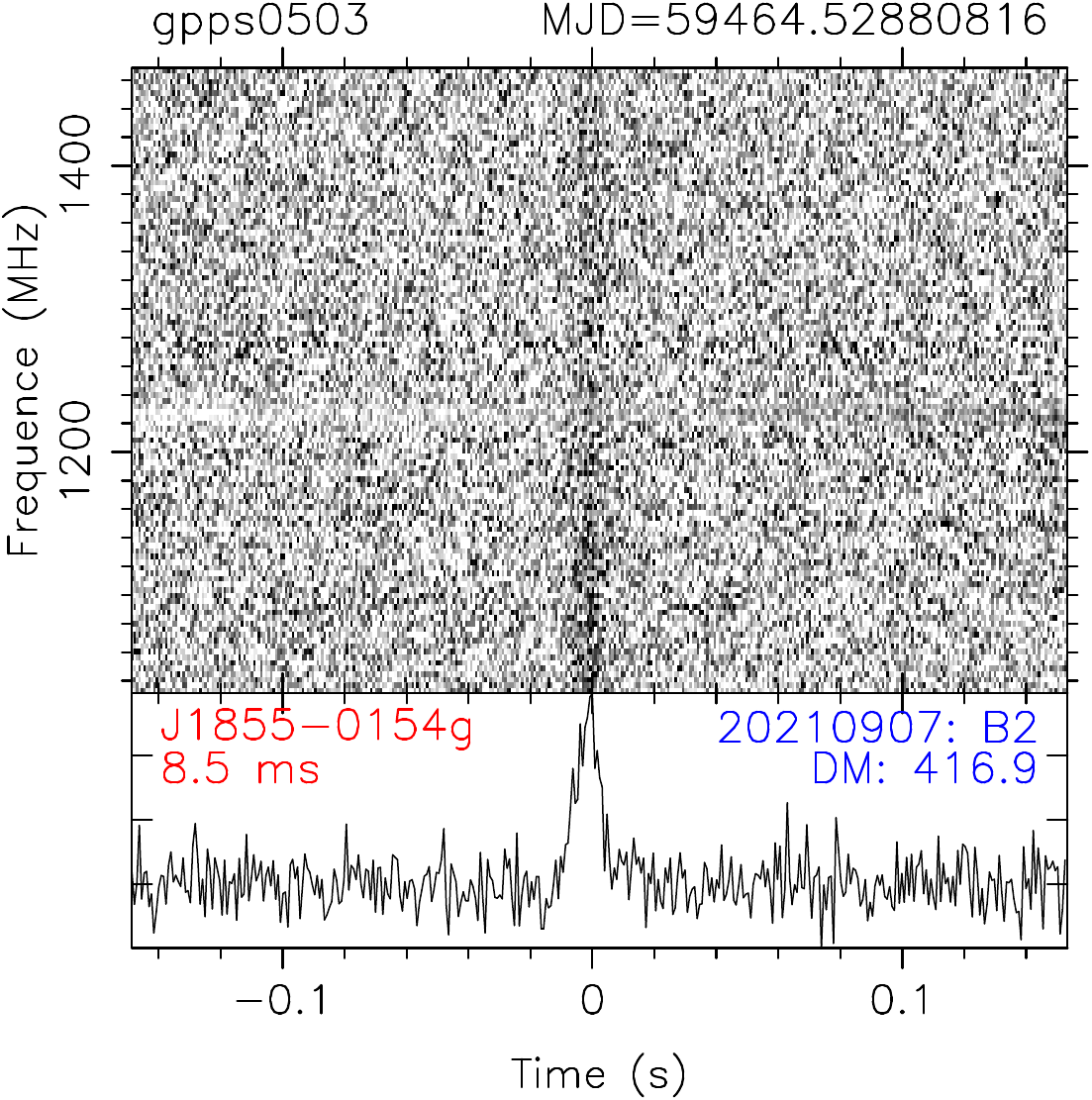}
\includegraphics[width=0.33\textwidth]{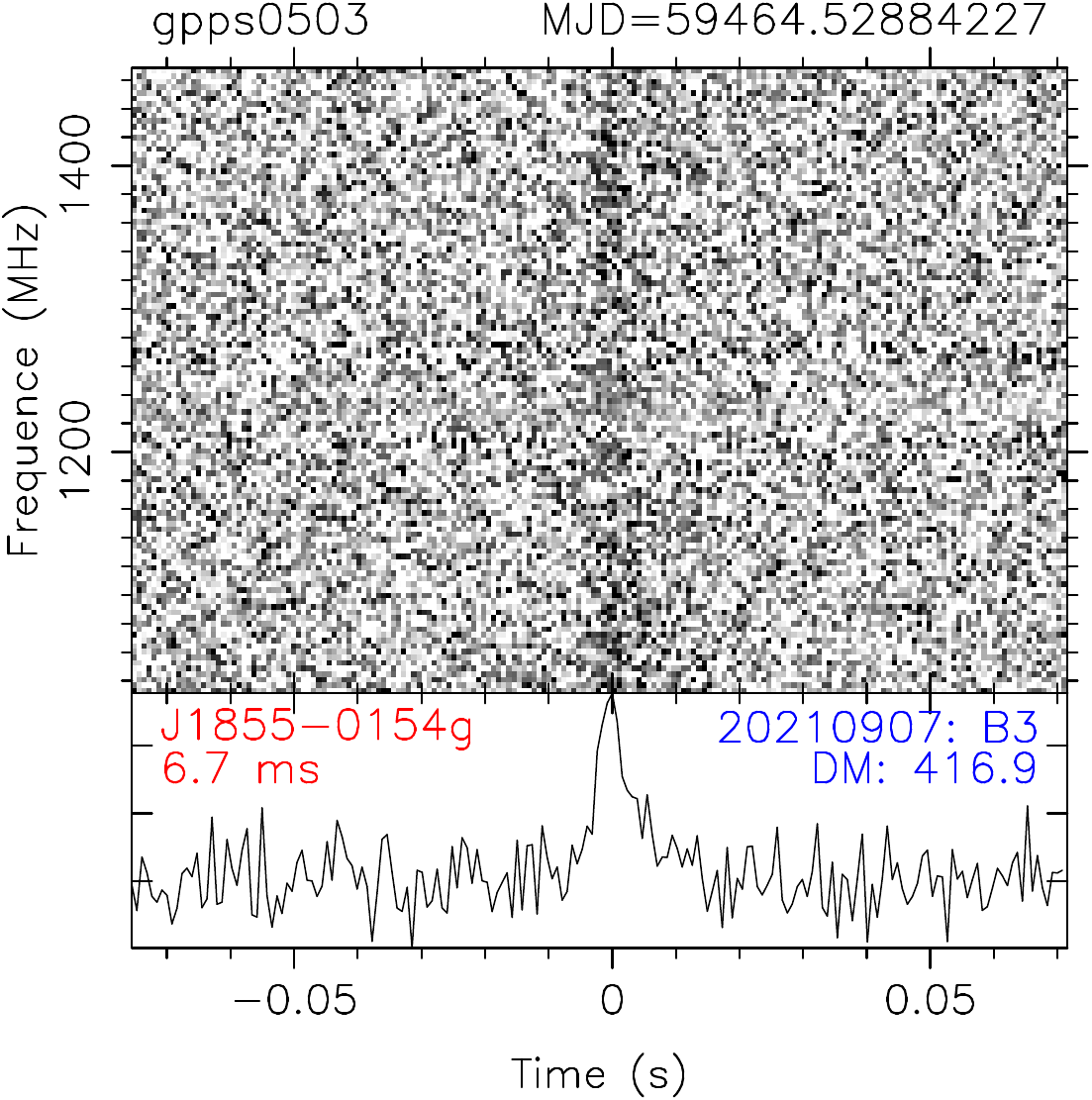}\\[0.2mm]
\includegraphics[width=0.33\textwidth]{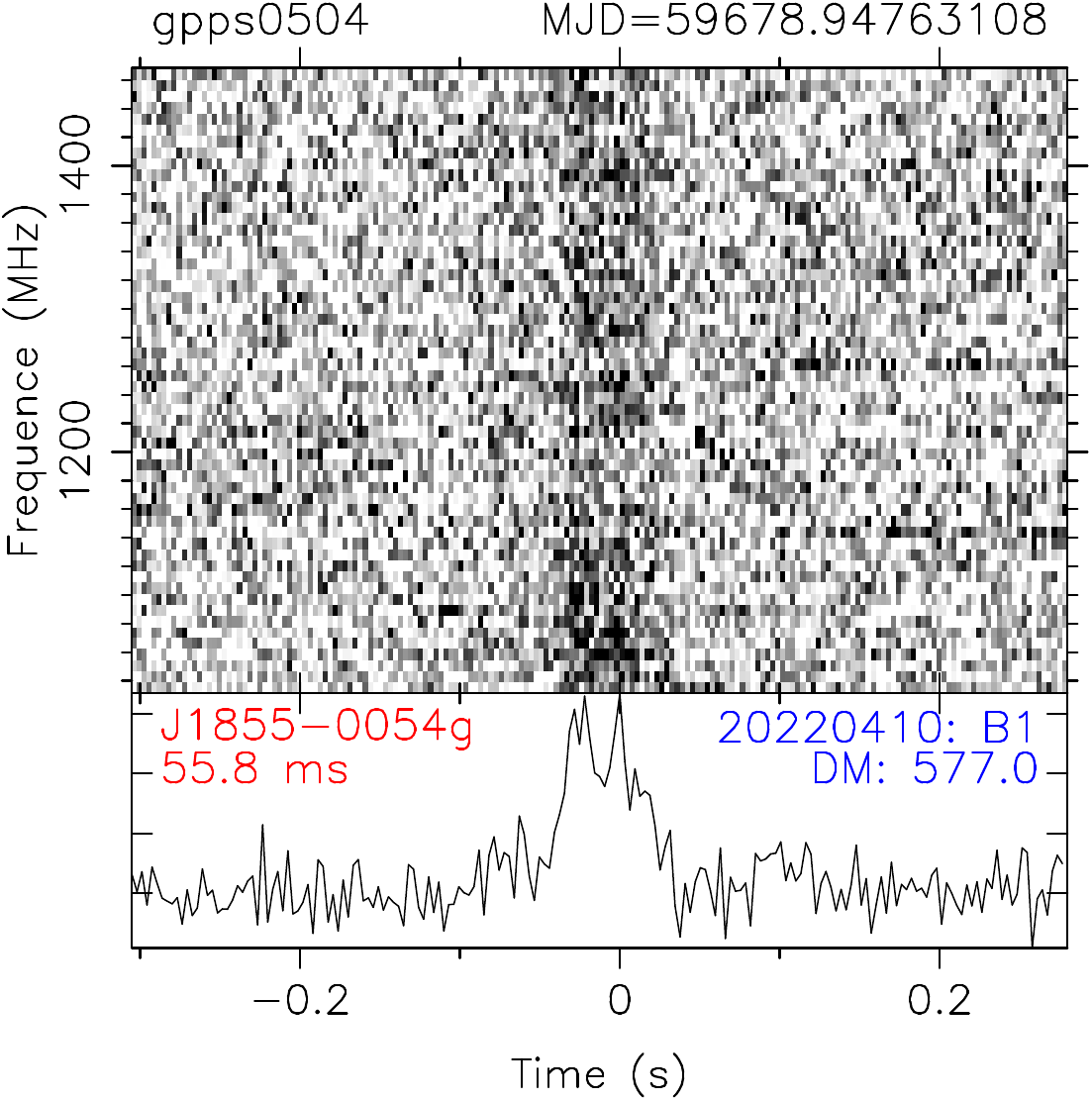}
\includegraphics[width=0.33\textwidth]{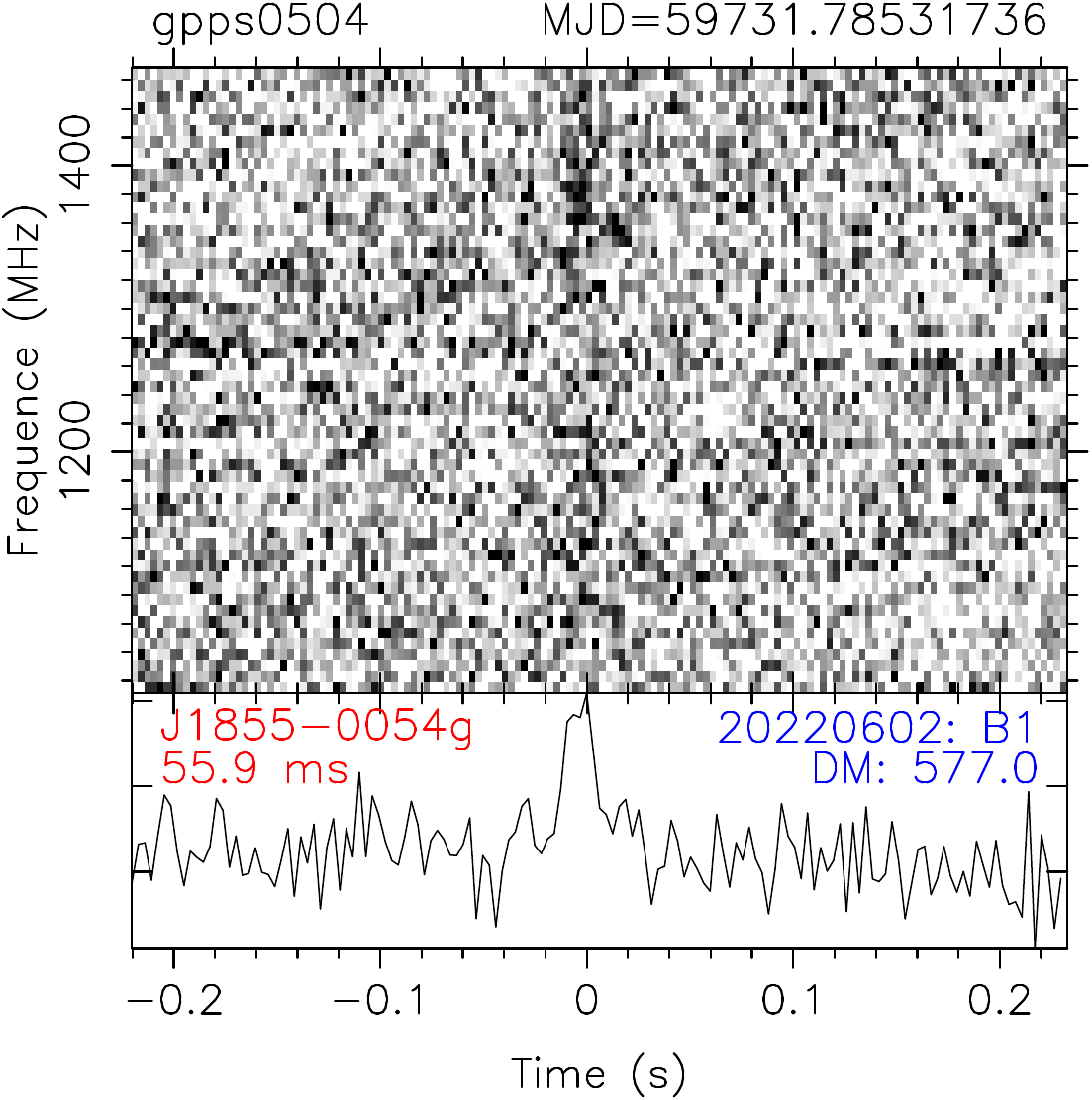}
\includegraphics[width=0.33\textwidth]{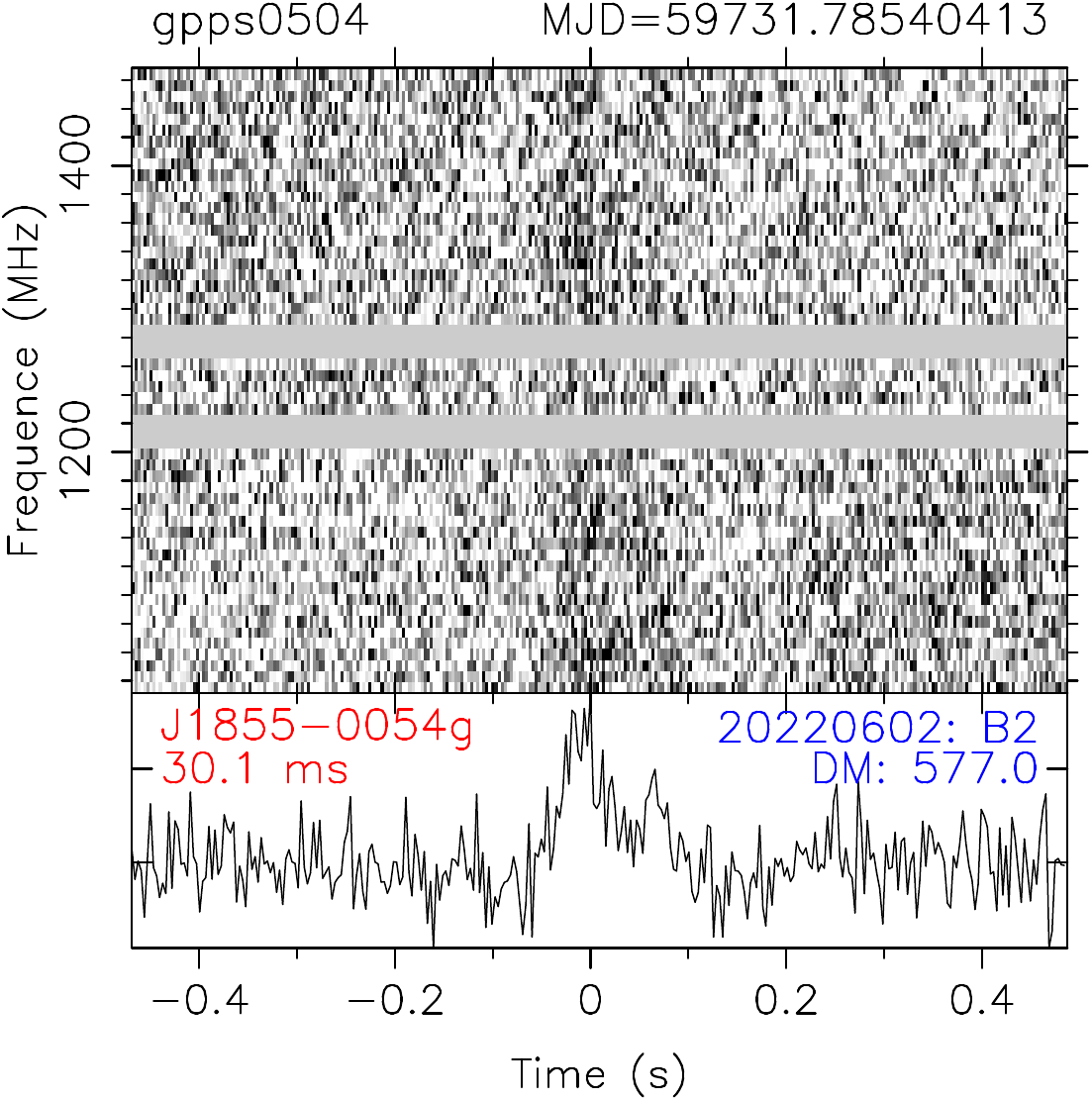}
\caption{(Continued.)}
\end{figure*}
\addtocounter{figure}{-1}
\begin{figure*}[!t]
\centering
\includegraphics[width=0.33\textwidth]{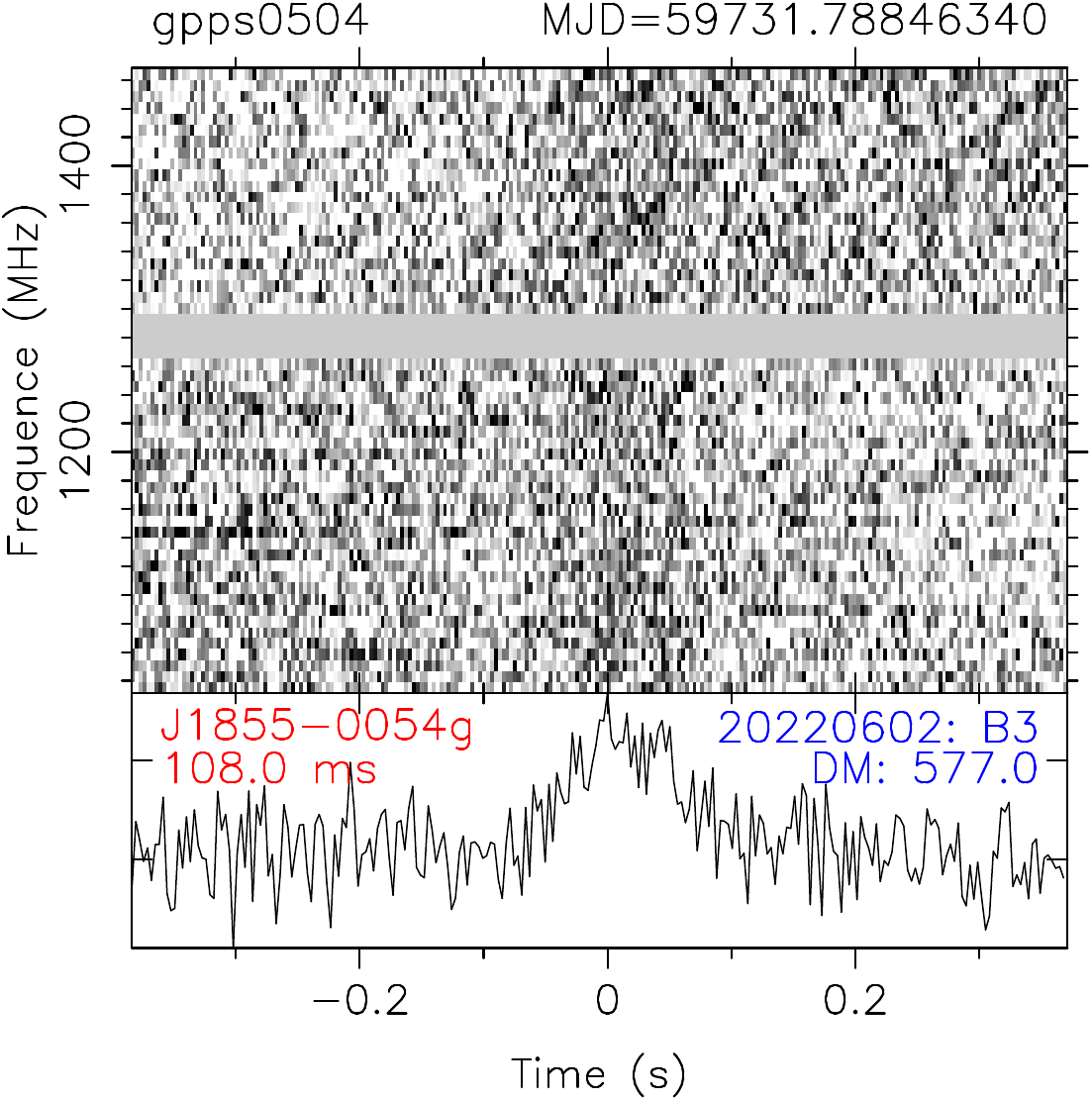}
\includegraphics[width=0.33\textwidth]{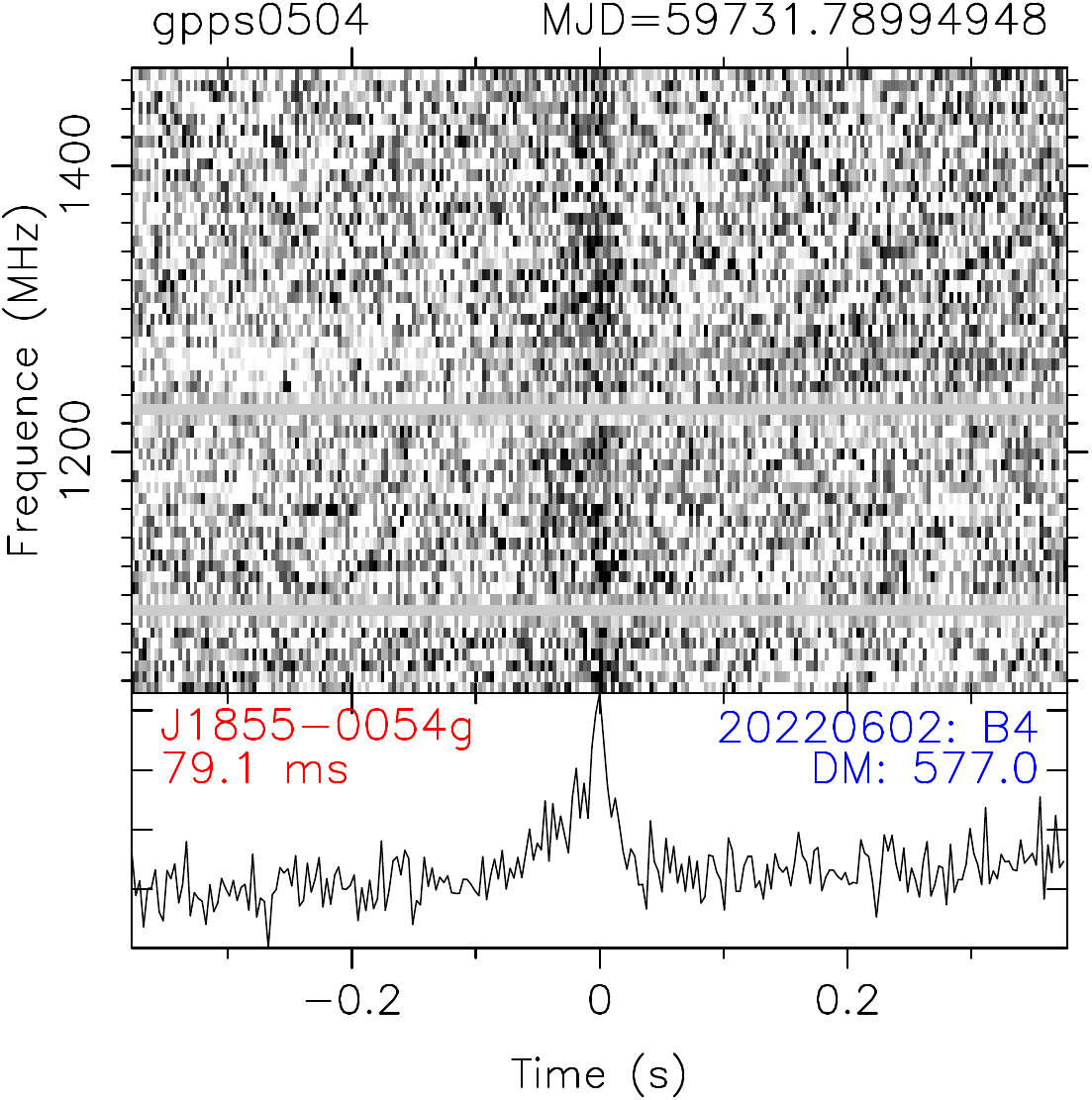}
\includegraphics[width=0.33\textwidth]{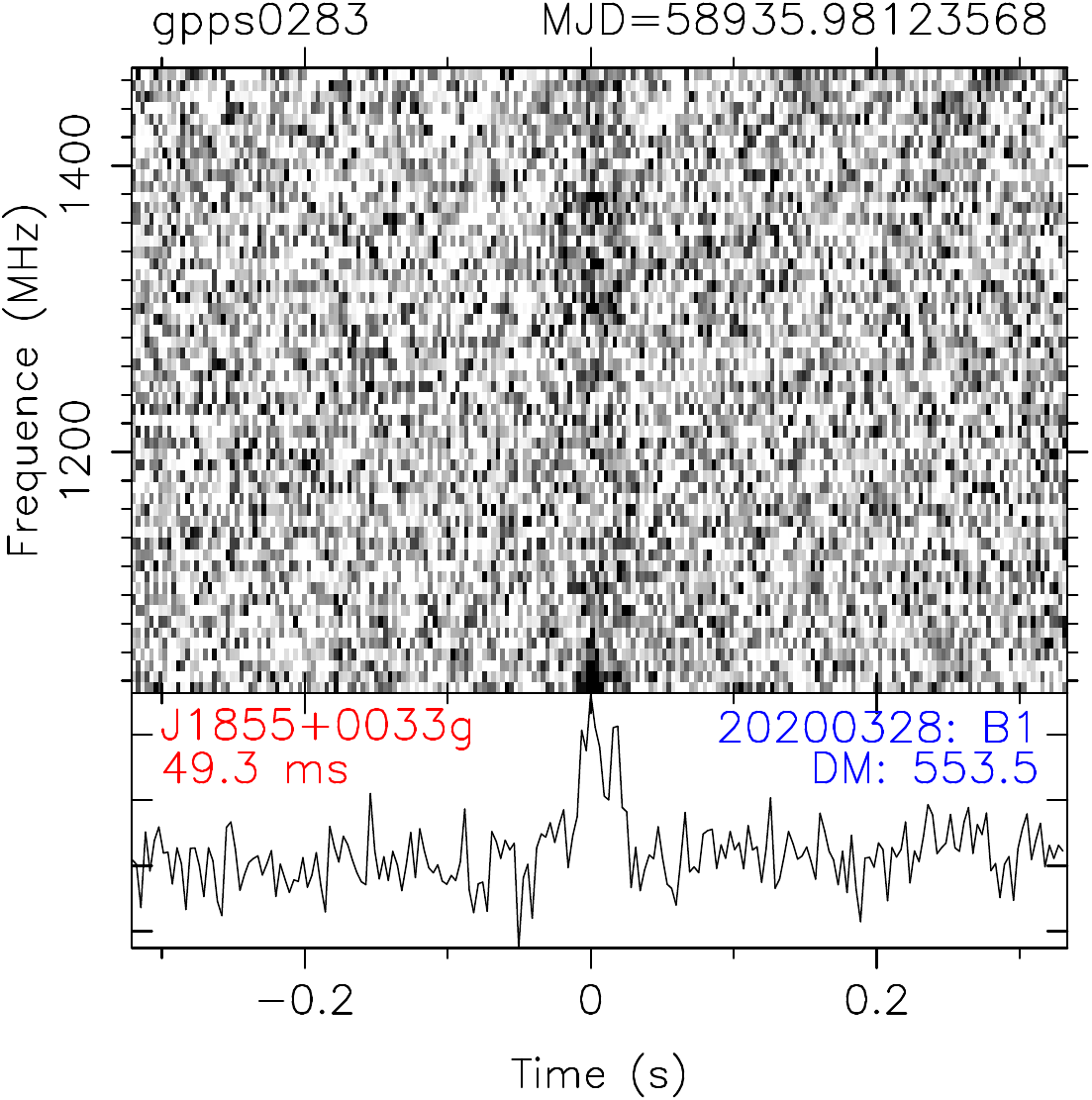}\\[0.2mm]
\includegraphics[width=0.33\textwidth]{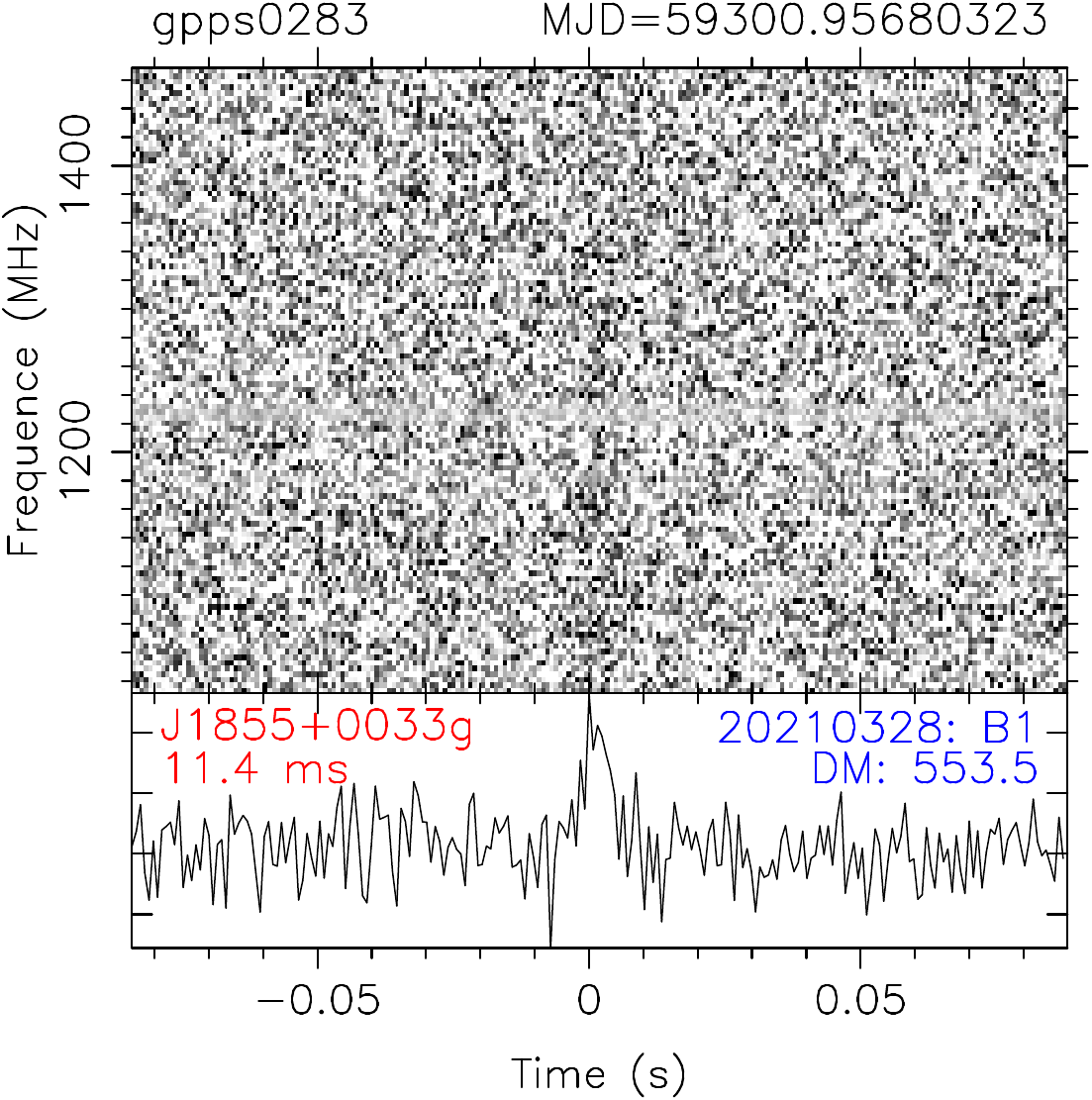}
\includegraphics[width=0.33\textwidth]{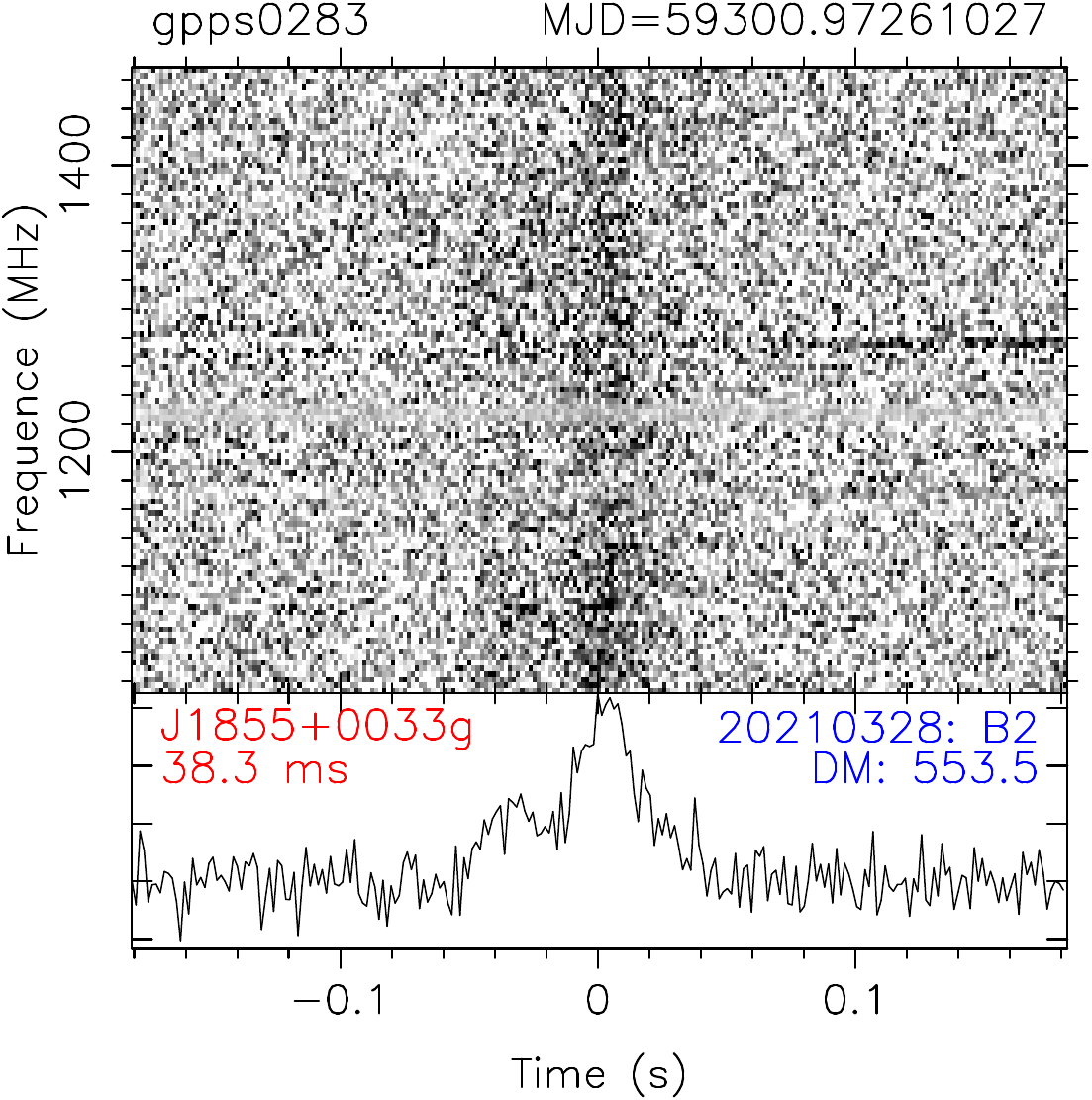}
\includegraphics[width=0.33\textwidth]{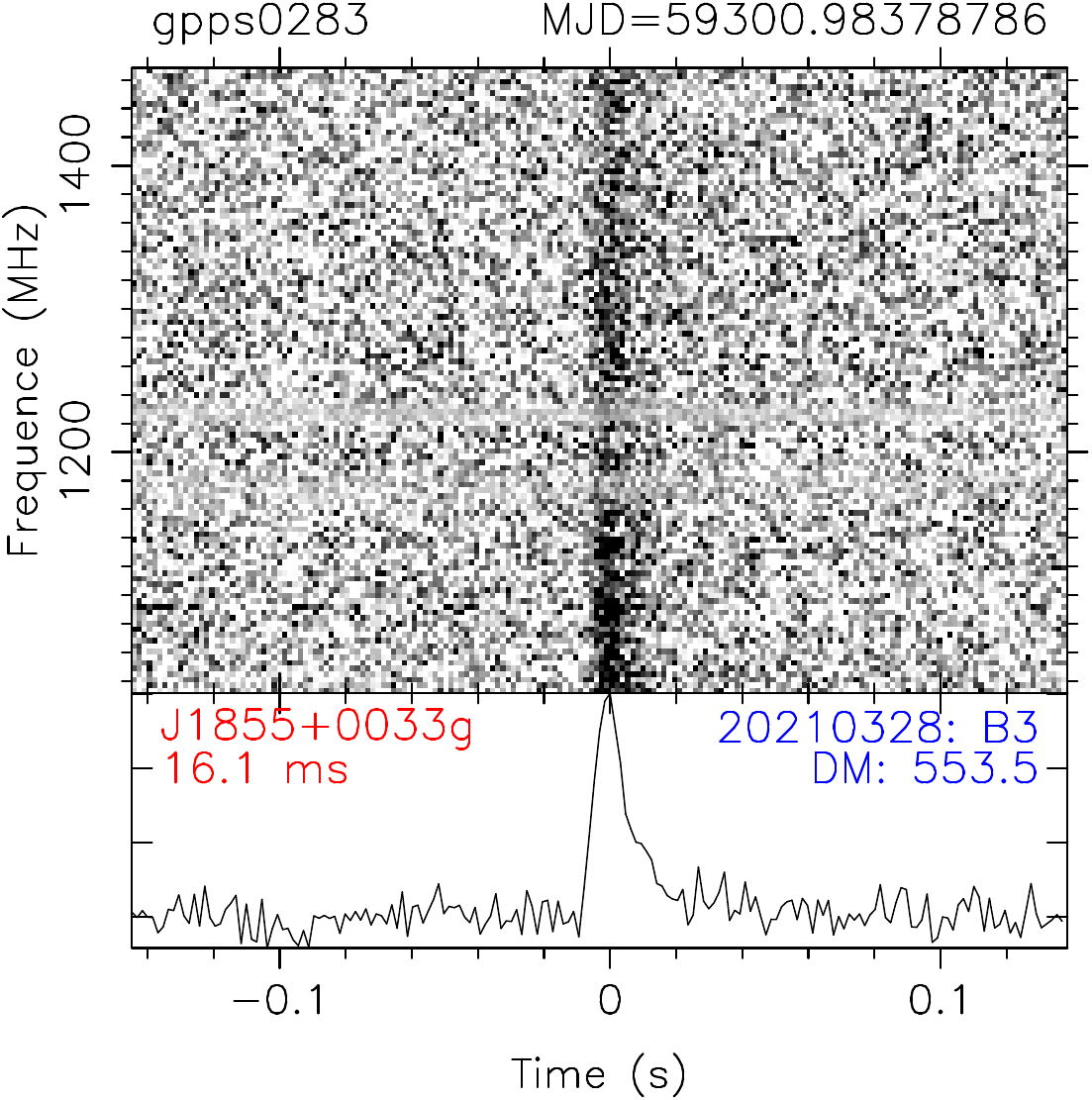}\\[0.2mm]
\includegraphics[width=0.33\textwidth]{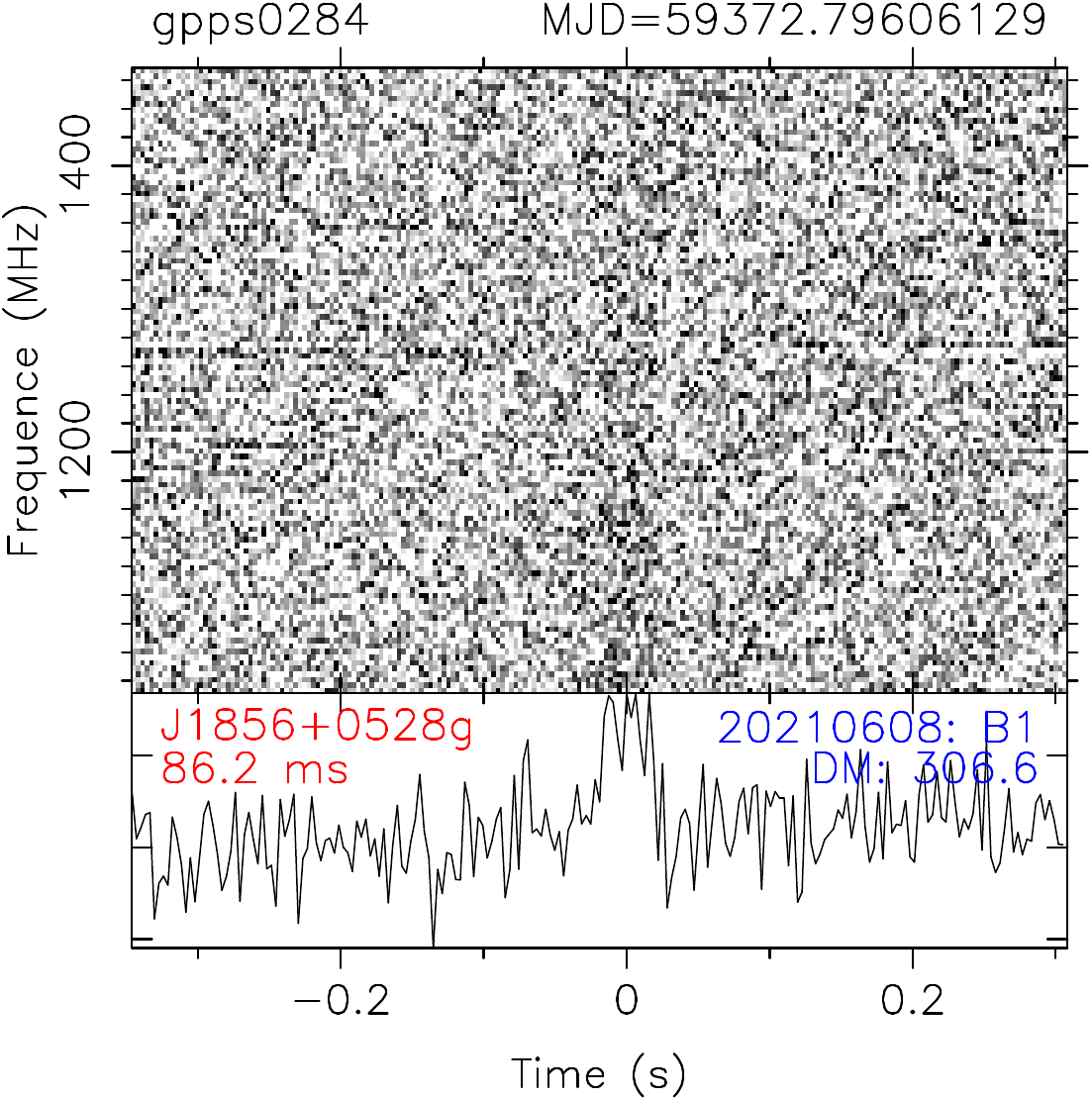}
\includegraphics[width=0.33\textwidth]{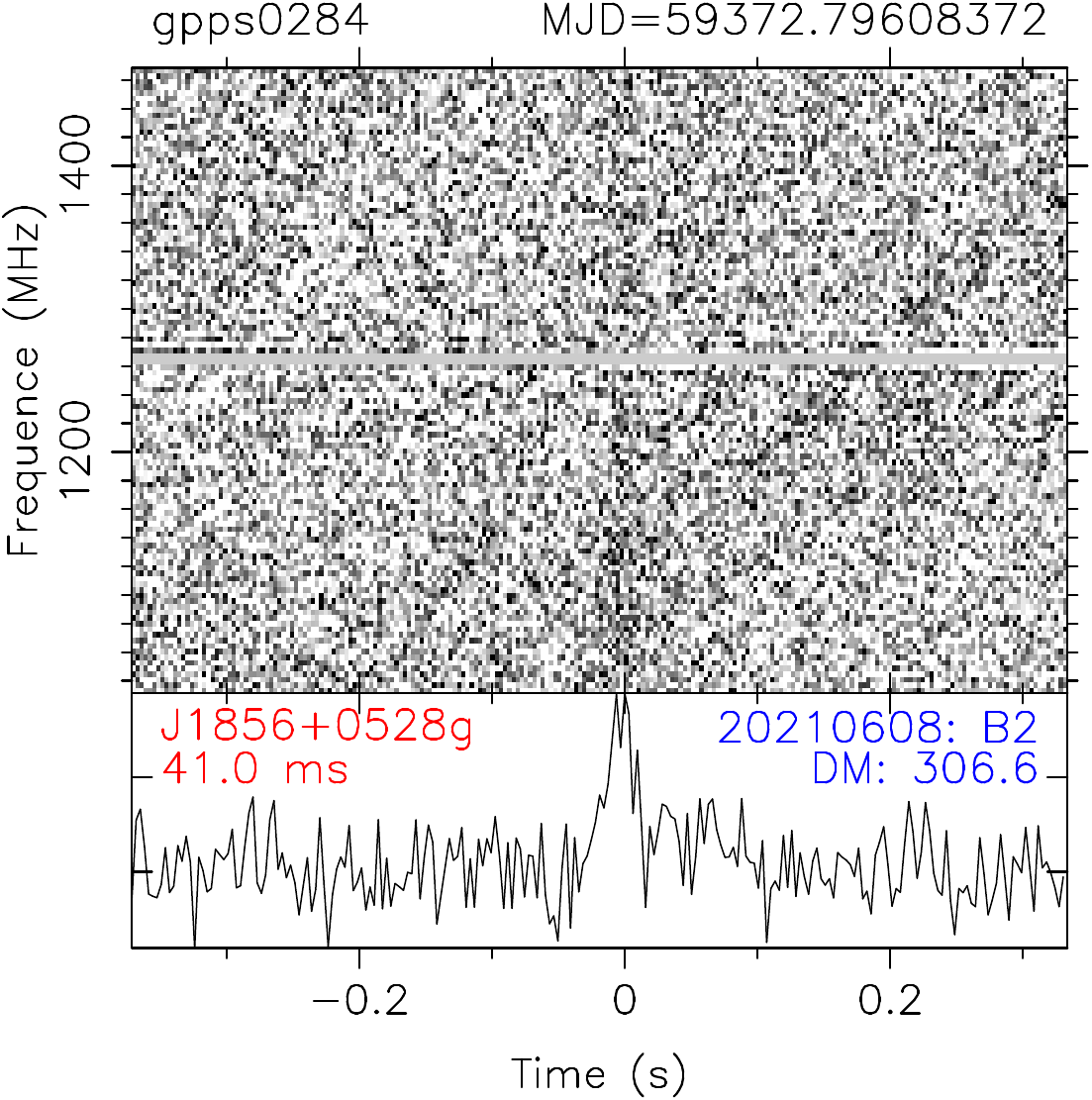}
\includegraphics[width=0.33\textwidth]{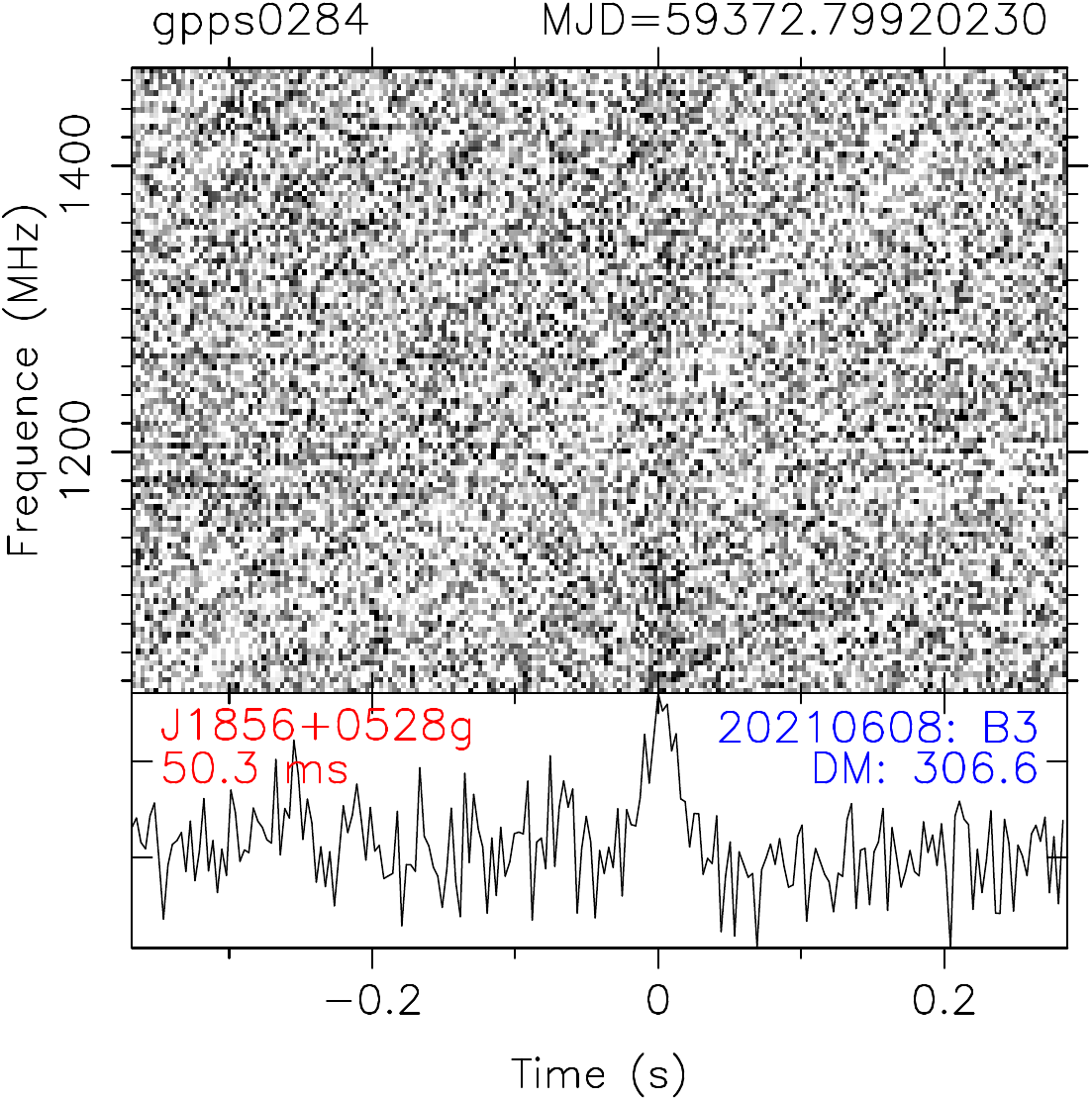}\\[0.2mm]
\includegraphics[width=0.33\textwidth]{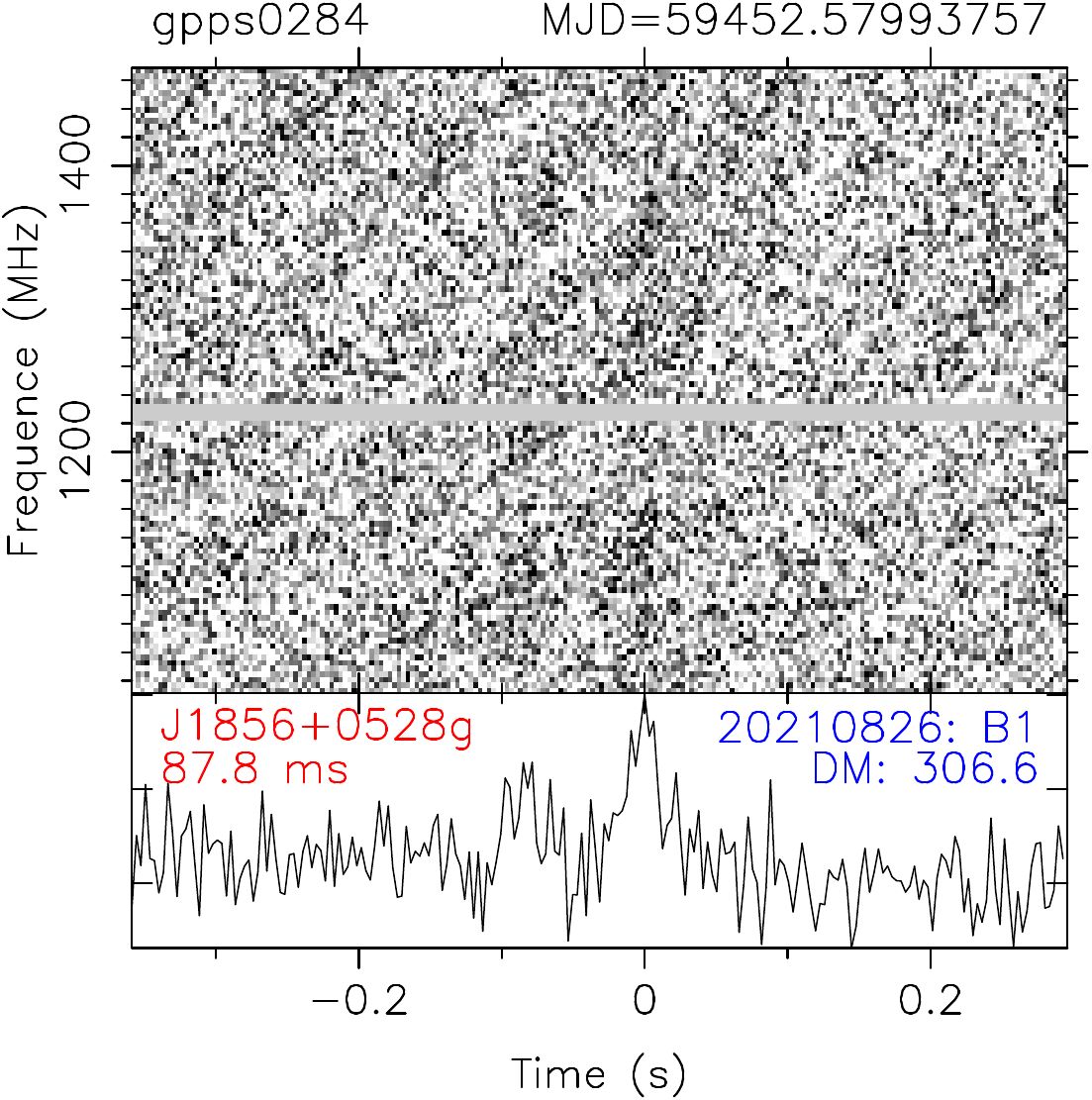}
\includegraphics[width=0.33\textwidth]{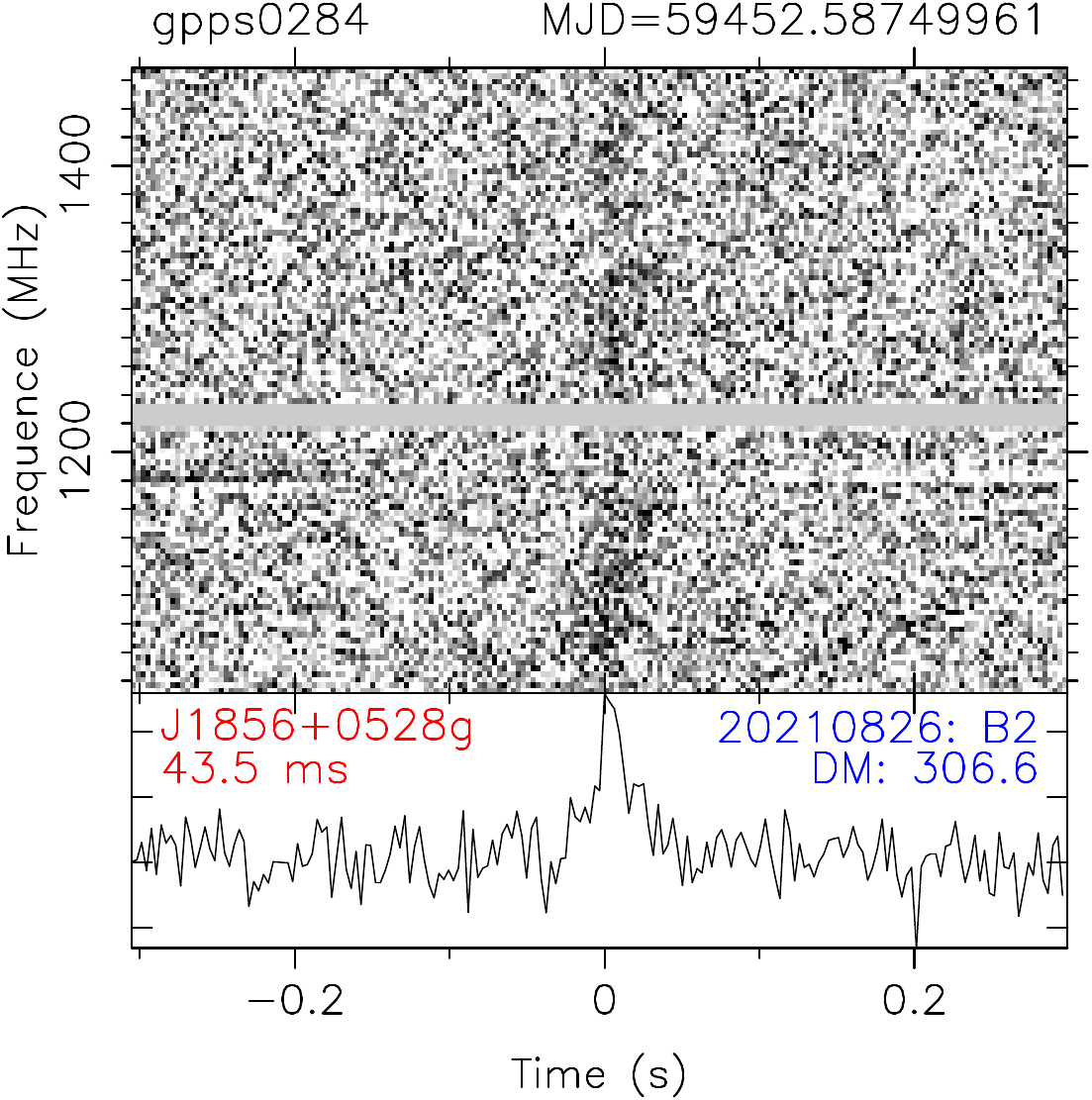}
\includegraphics[width=0.33\textwidth]{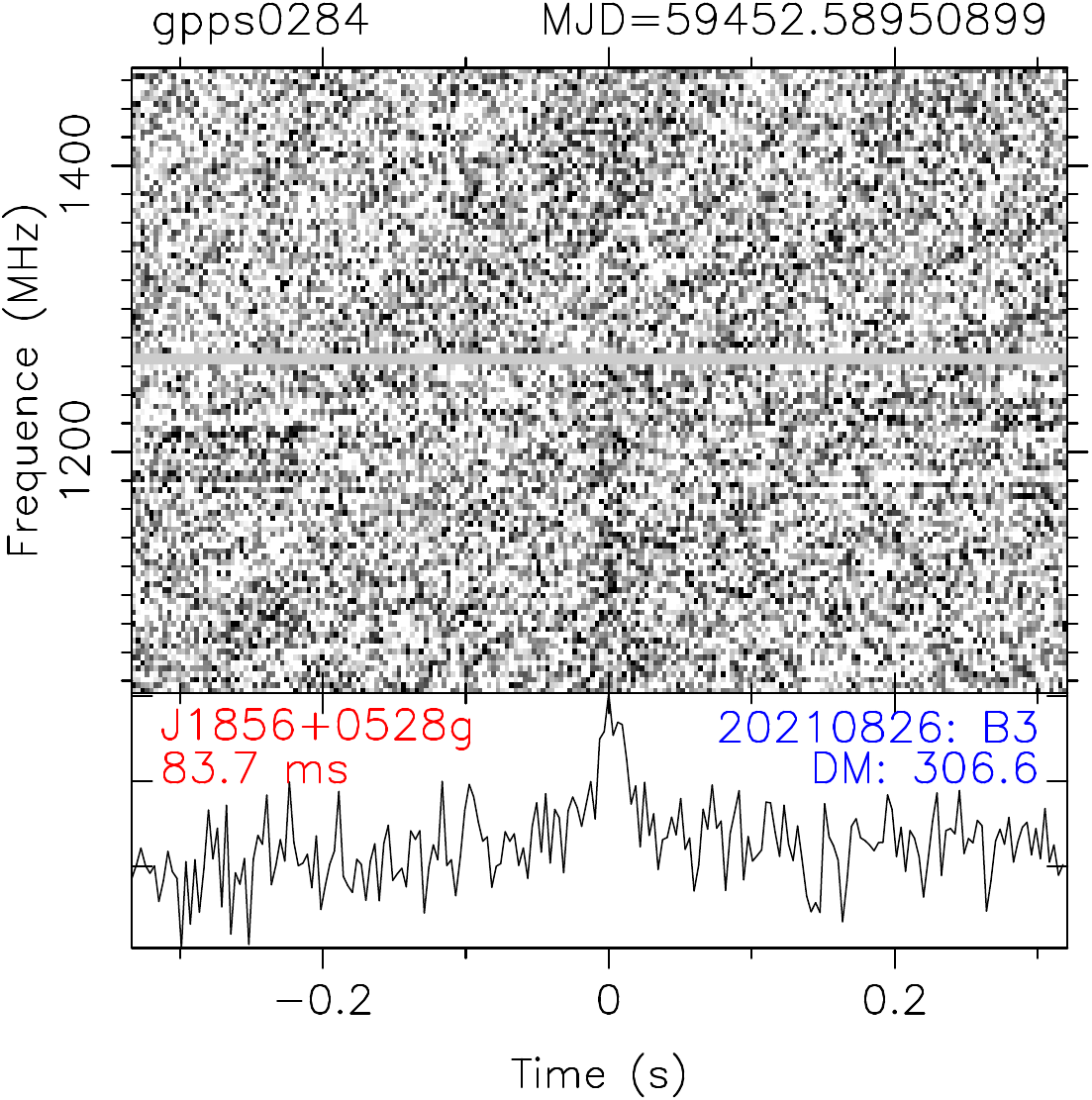}
\caption{(Continued.)}
\end{figure*}
\addtocounter{figure}{-1}
\begin{figure*}[!t]
\centering
\includegraphics[width=0.33\textwidth]{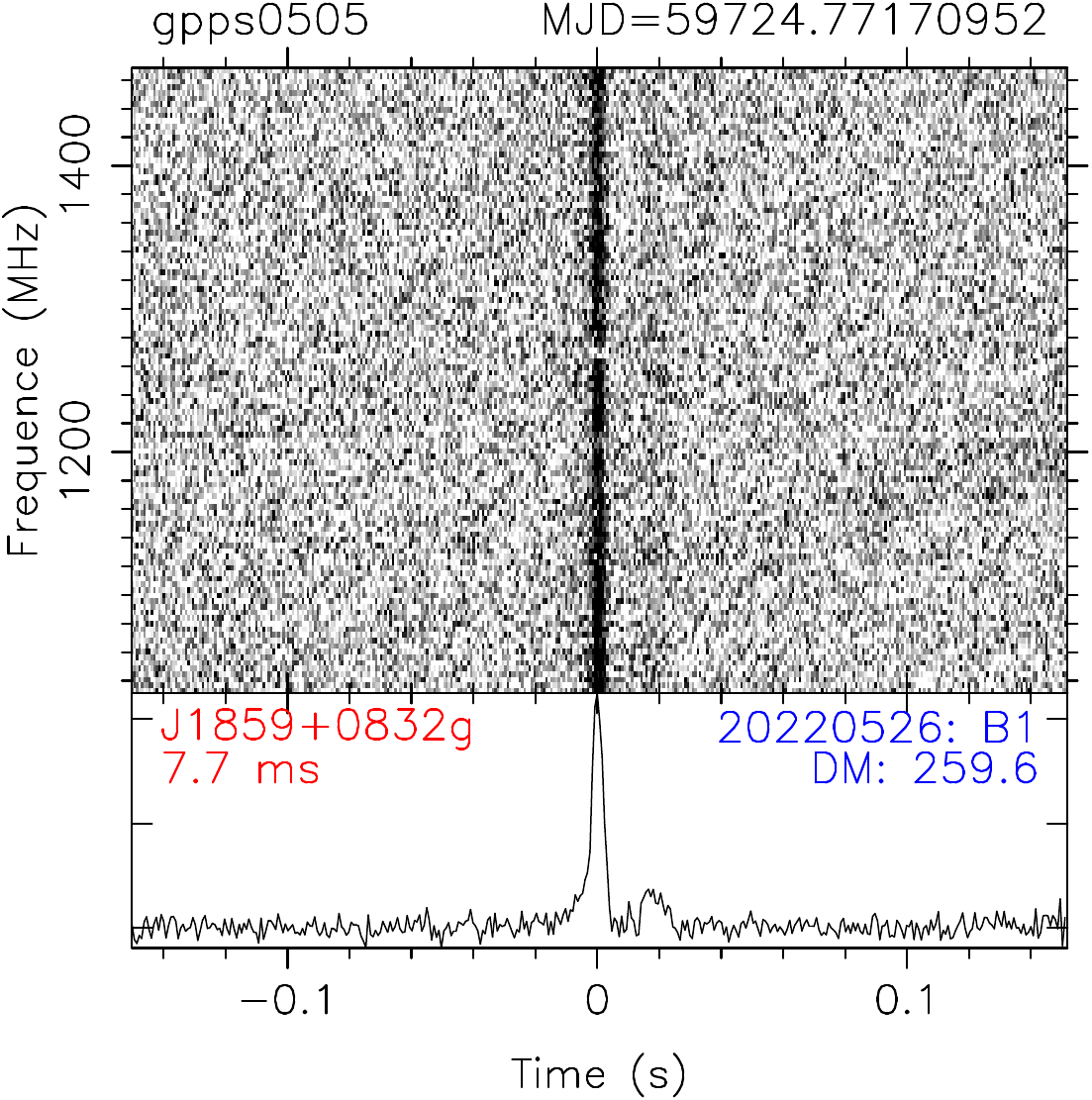}
\includegraphics[width=0.33\textwidth]{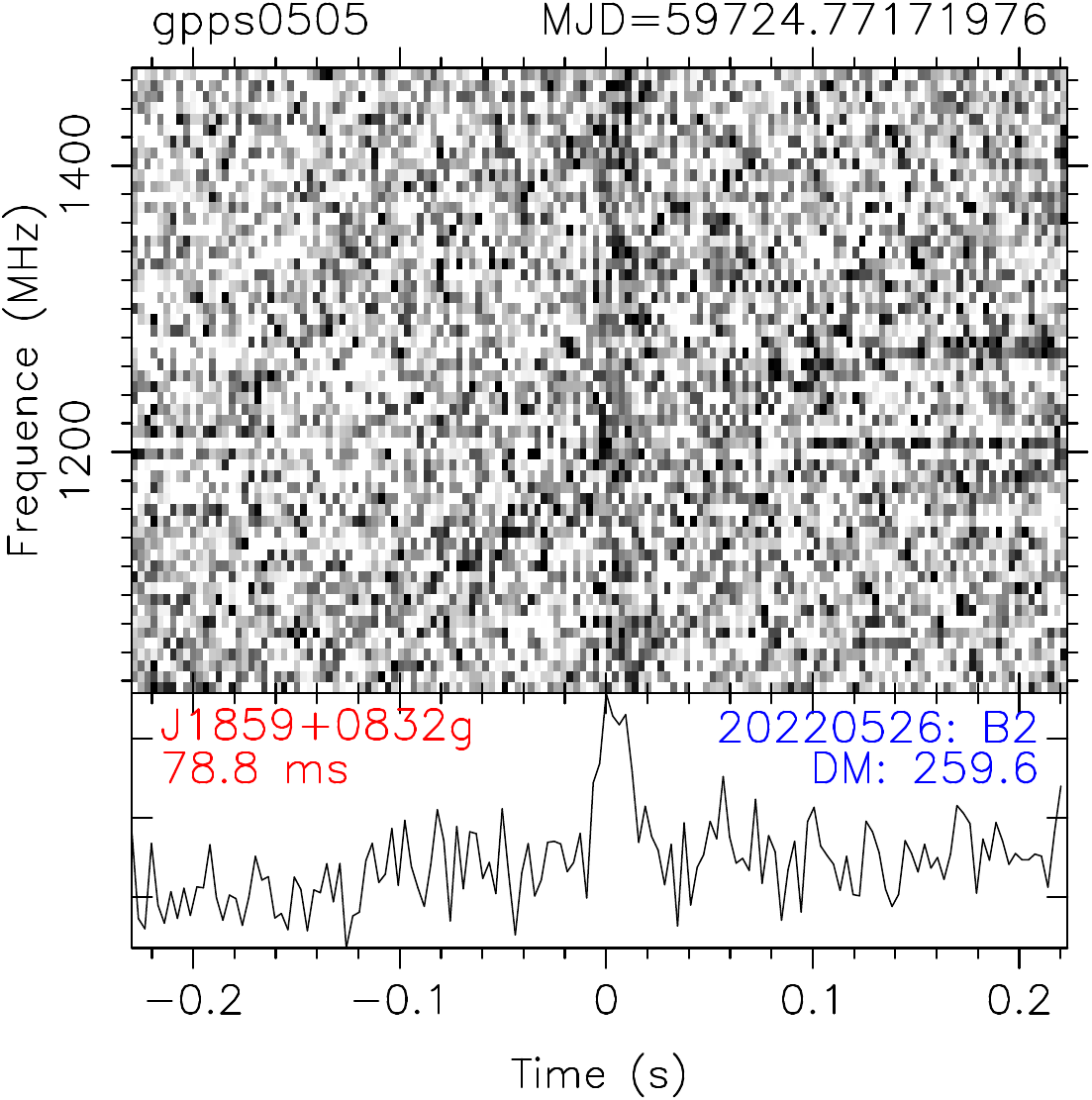}
\includegraphics[width=0.33\textwidth]{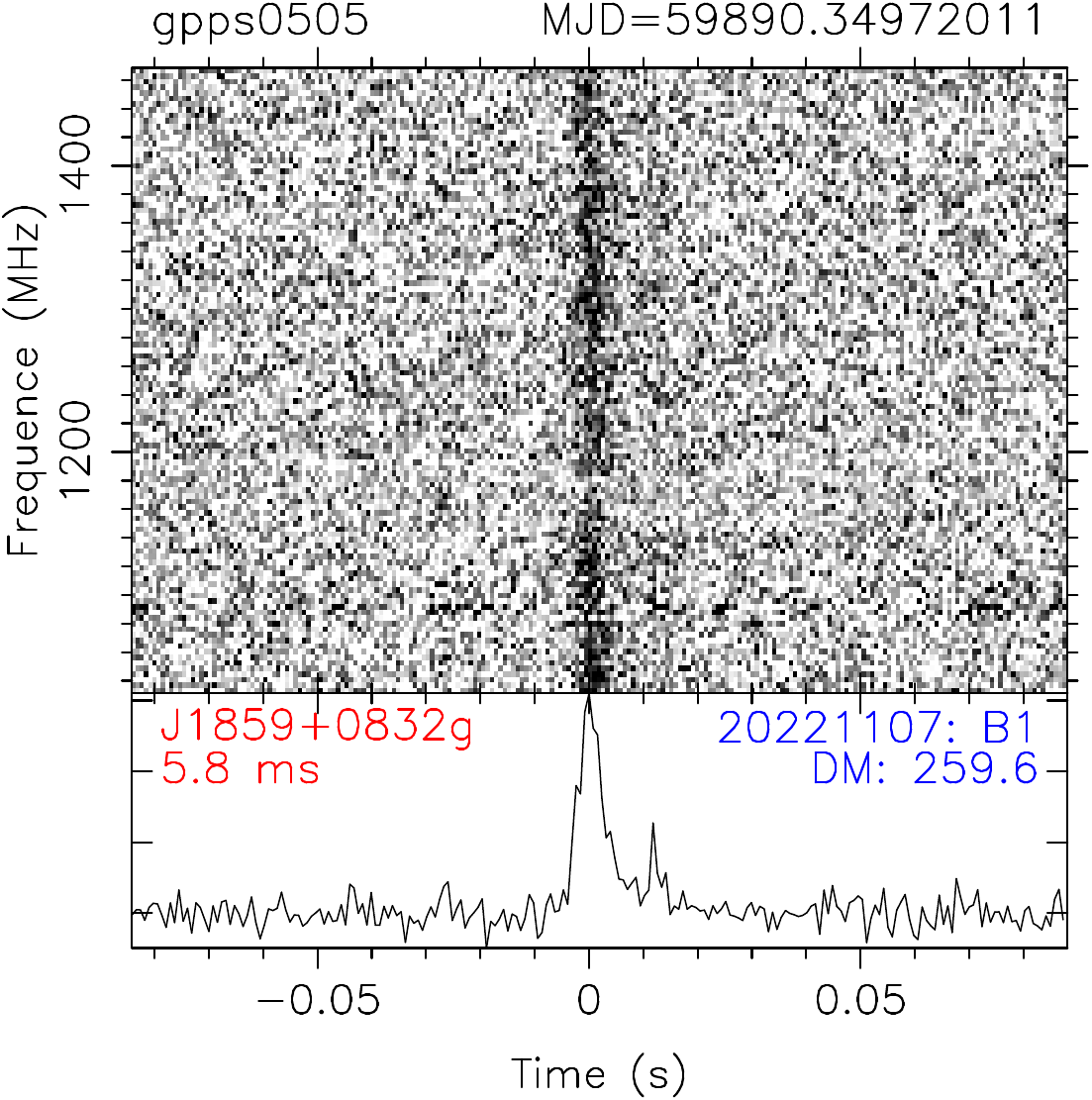}\\[0.2mm]
\includegraphics[width=0.33\textwidth]{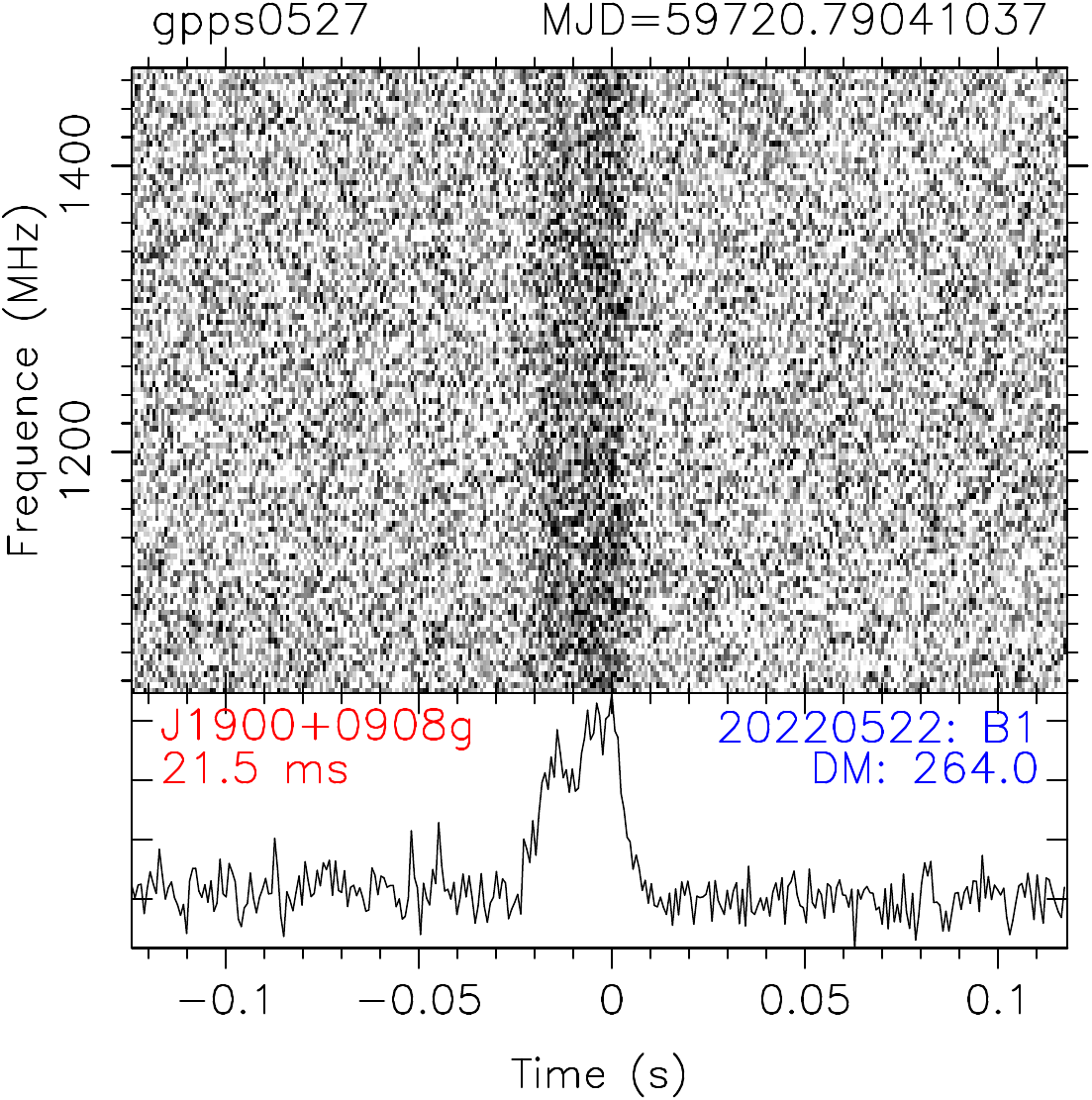}
\includegraphics[width=0.33\textwidth]{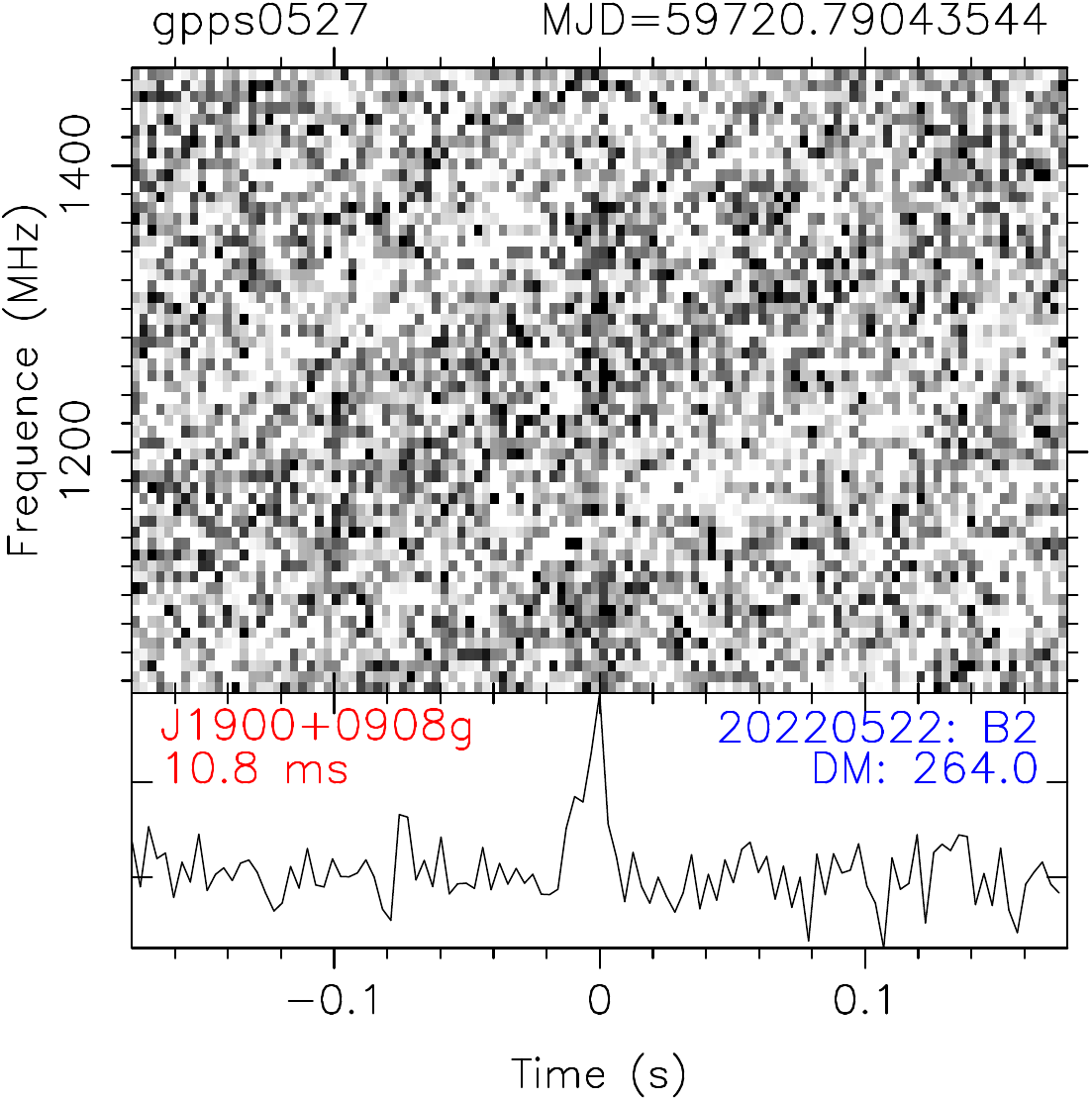}
\includegraphics[width=0.33\textwidth]{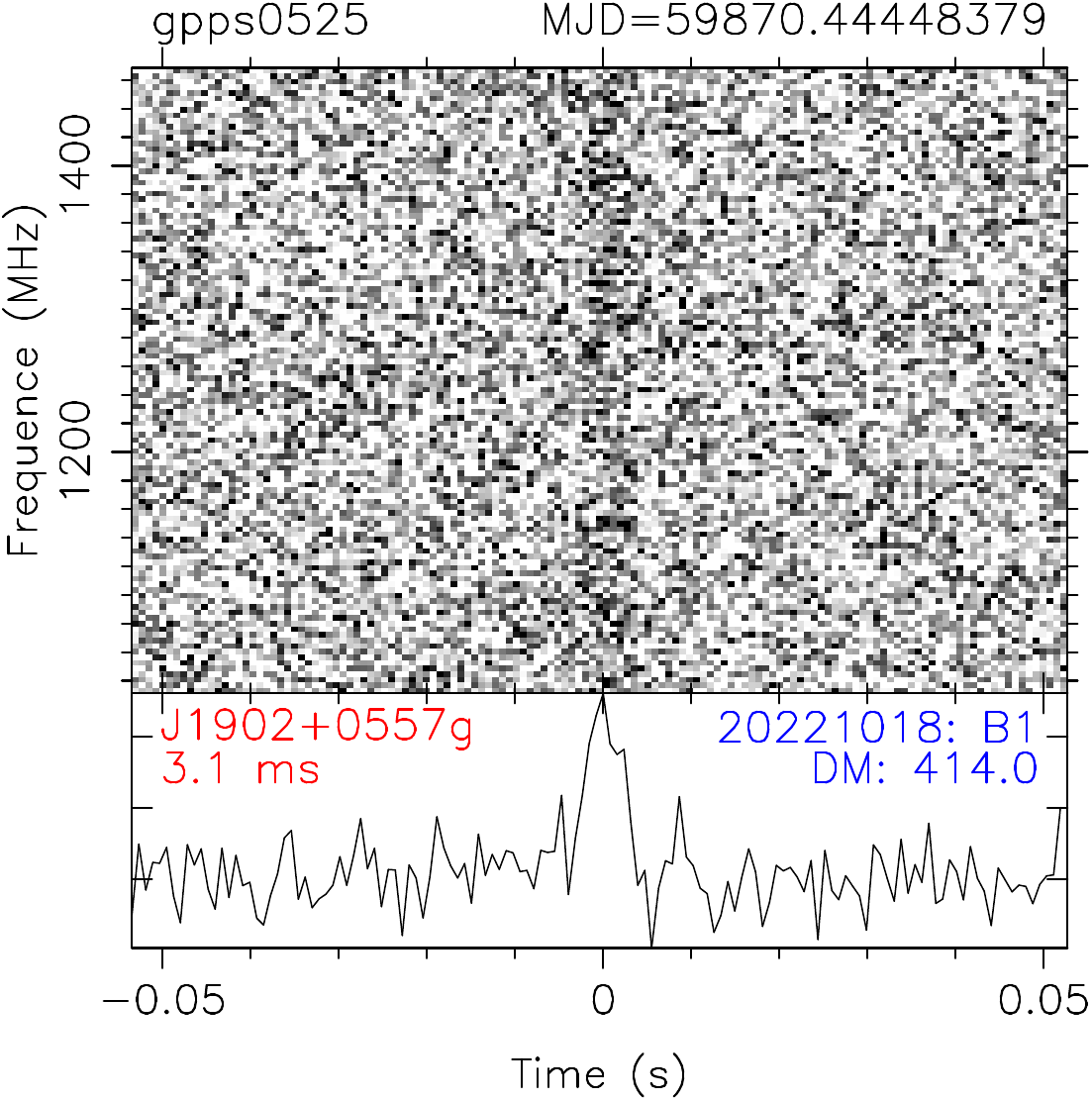}\\[0.2mm]
\includegraphics[width=0.33\textwidth]{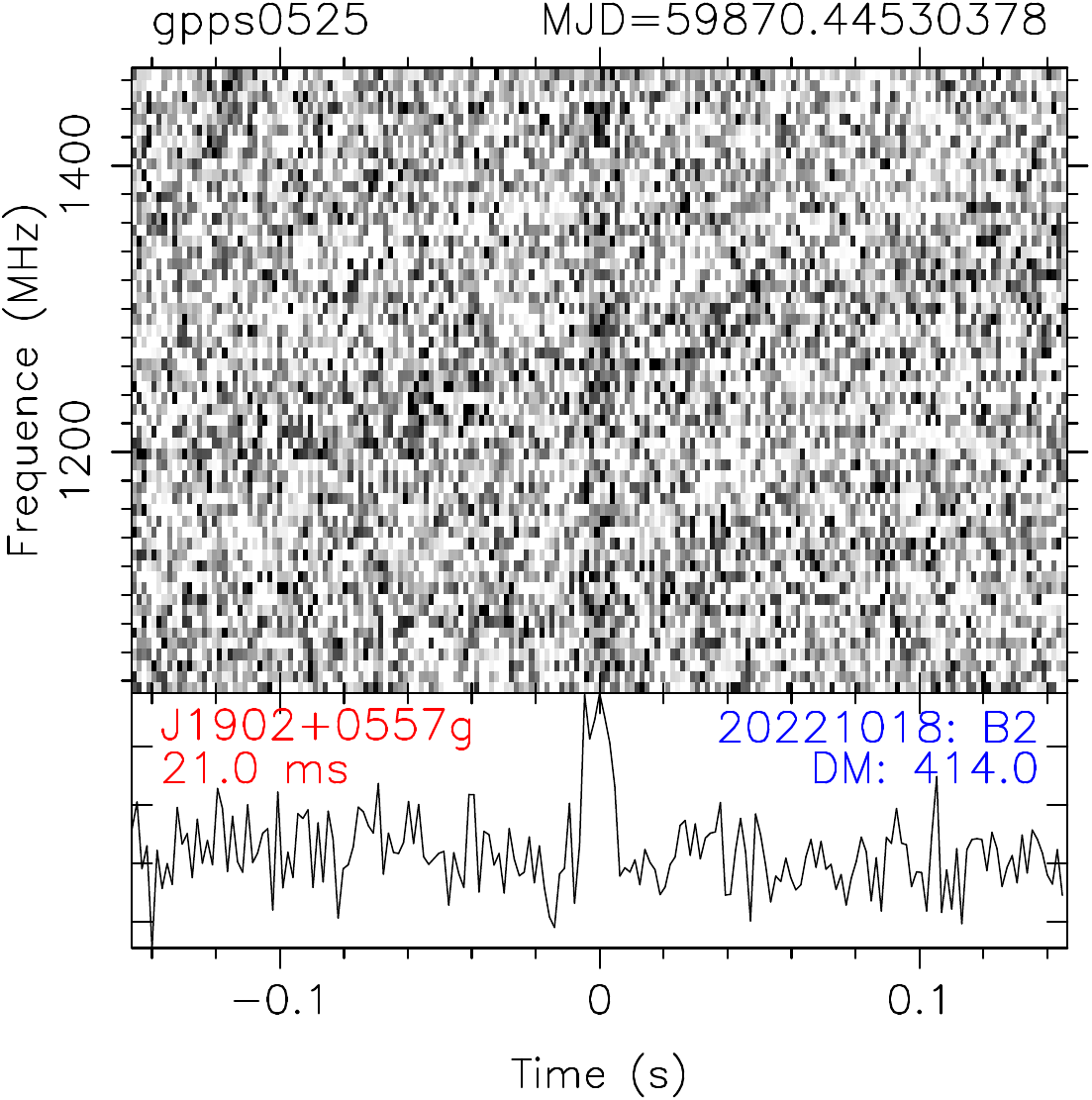}
\includegraphics[width=0.33\textwidth]{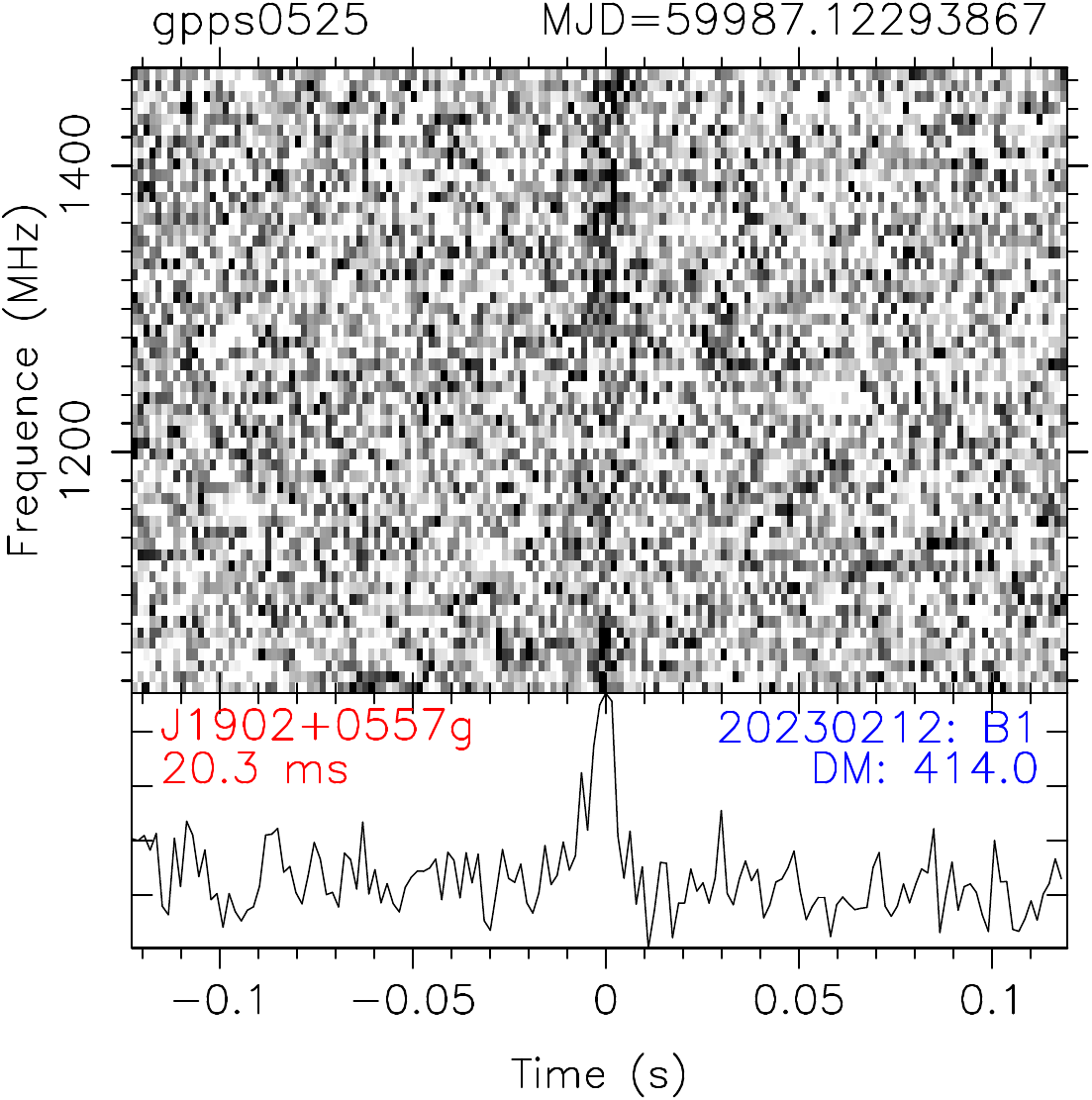}
\includegraphics[width=0.33\textwidth]{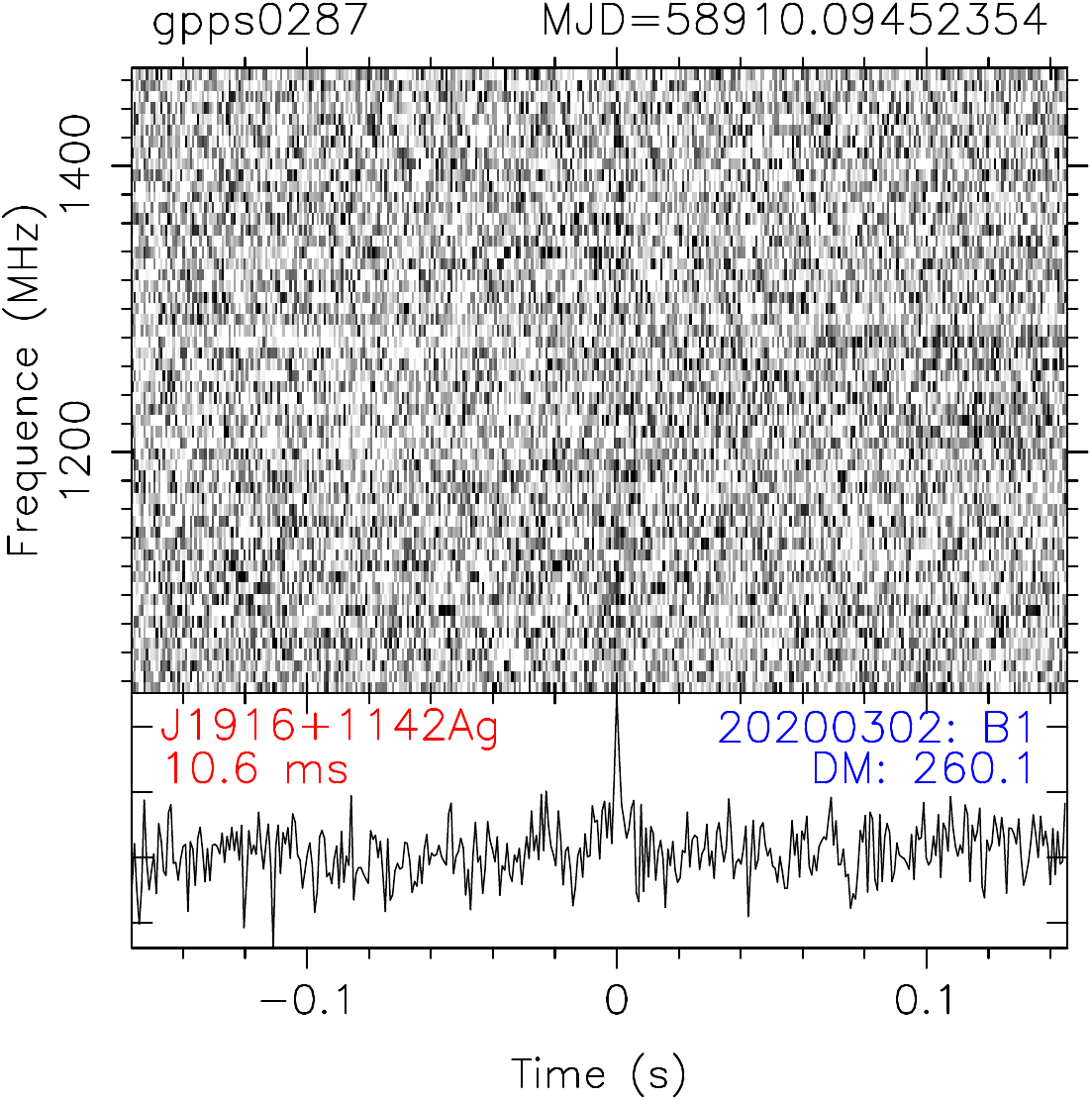}\\[0.2mm]
\includegraphics[width=0.33\textwidth]{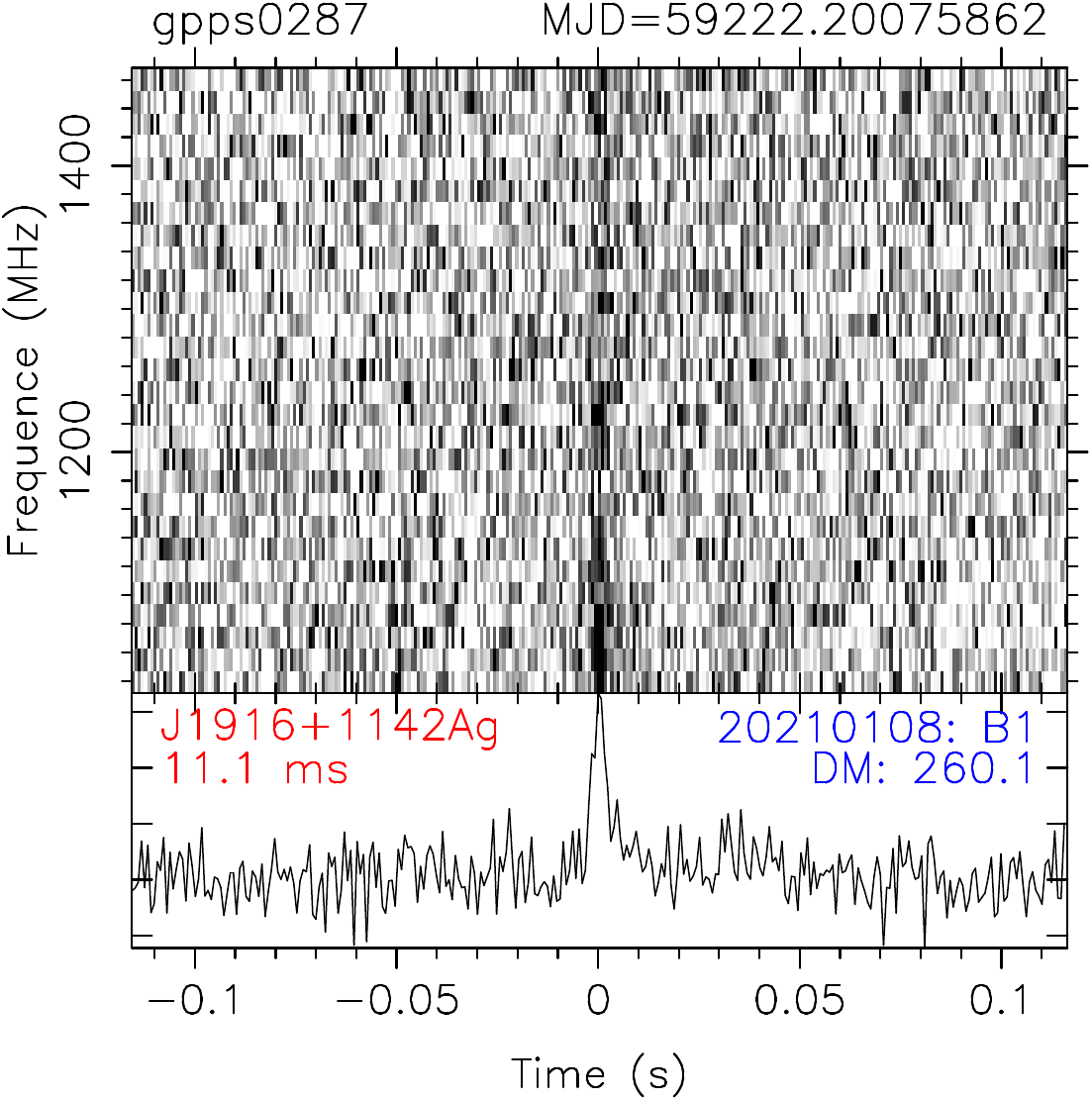}
\includegraphics[width=0.33\textwidth]{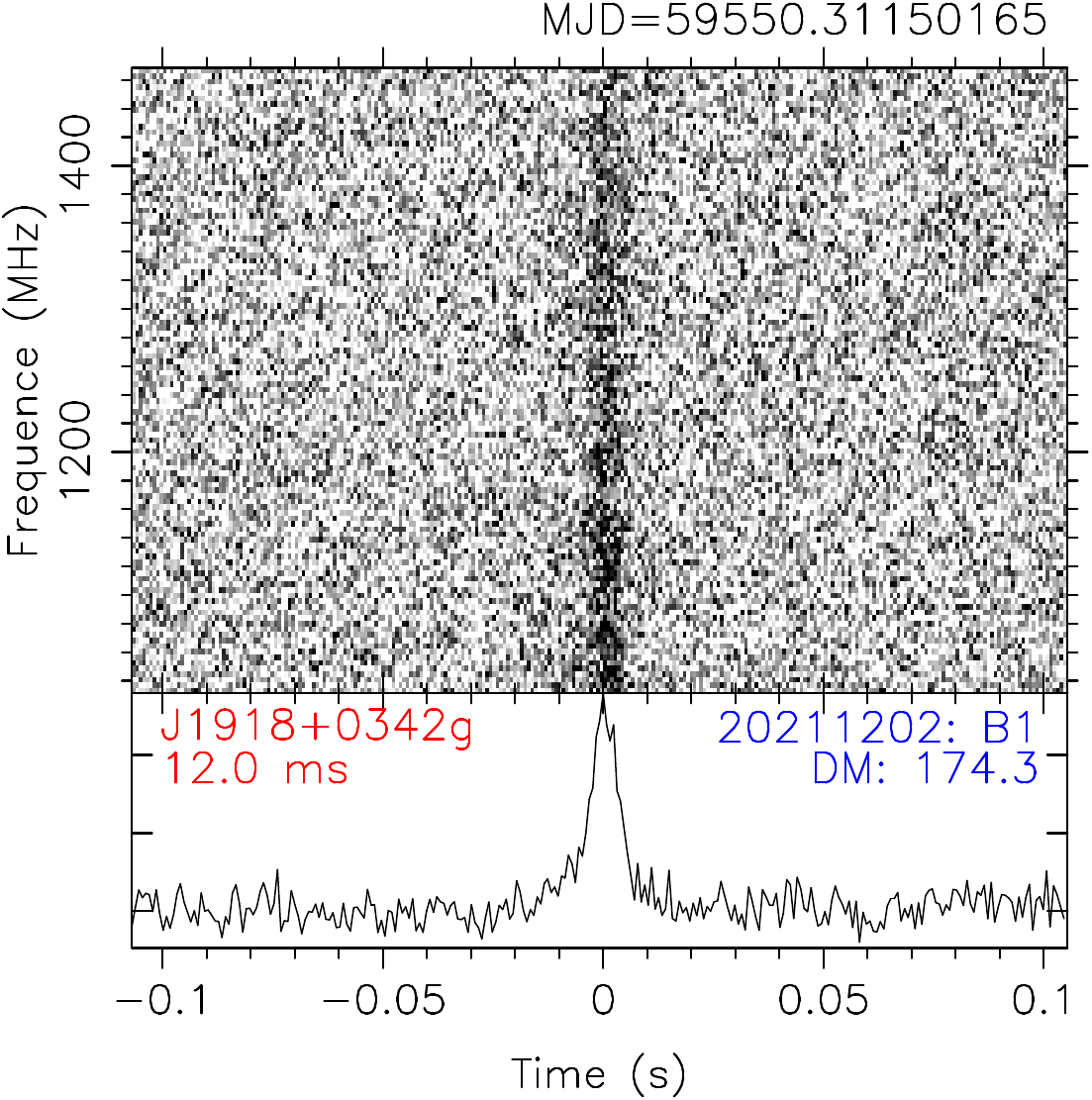}
\includegraphics[width=0.33\textwidth]{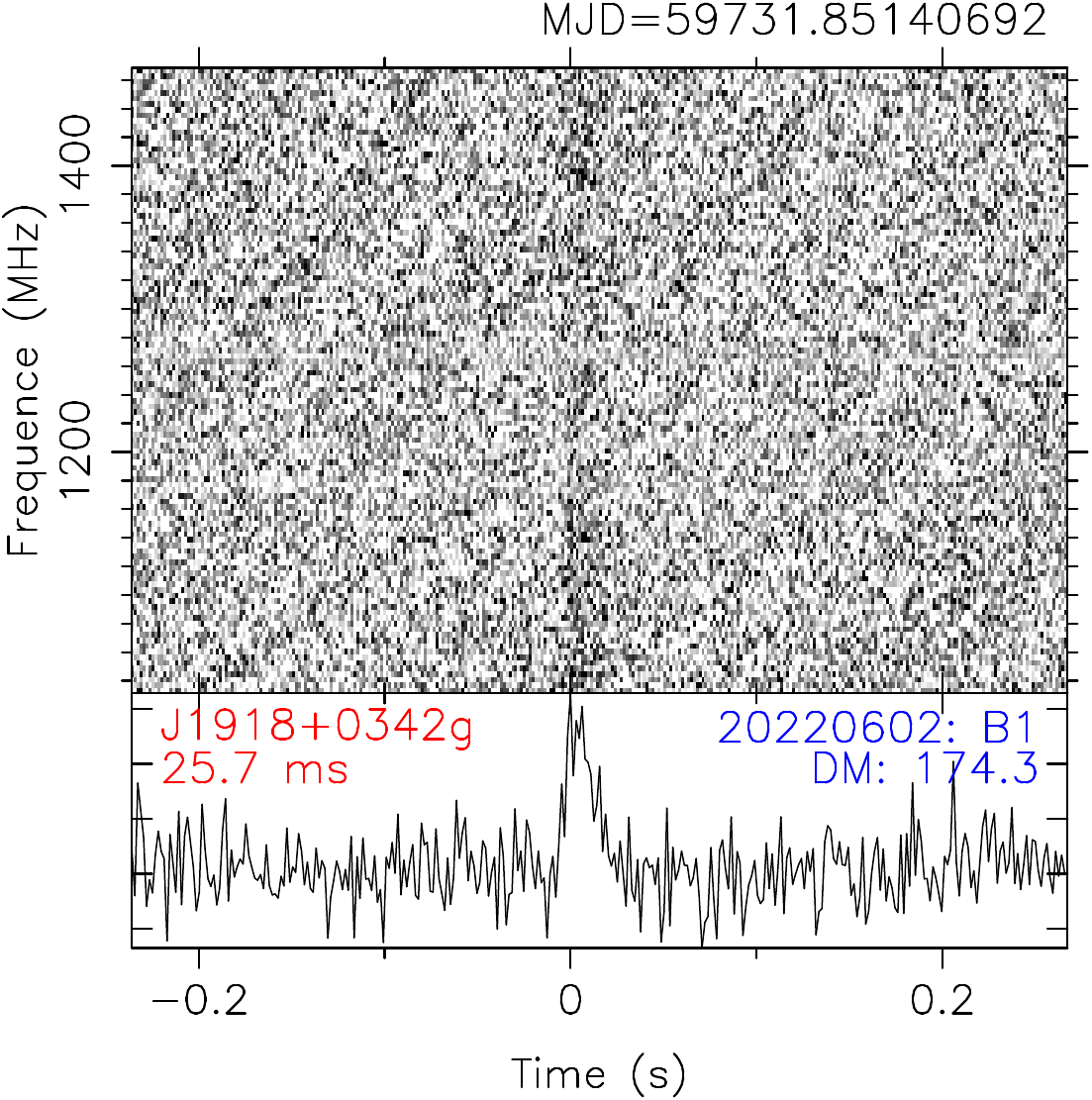}
\caption{(Continued.)}
\end{figure*}
\addtocounter{figure}{-1}
\begin{figure*}[!t]
\centering
\includegraphics[width=0.33\textwidth]{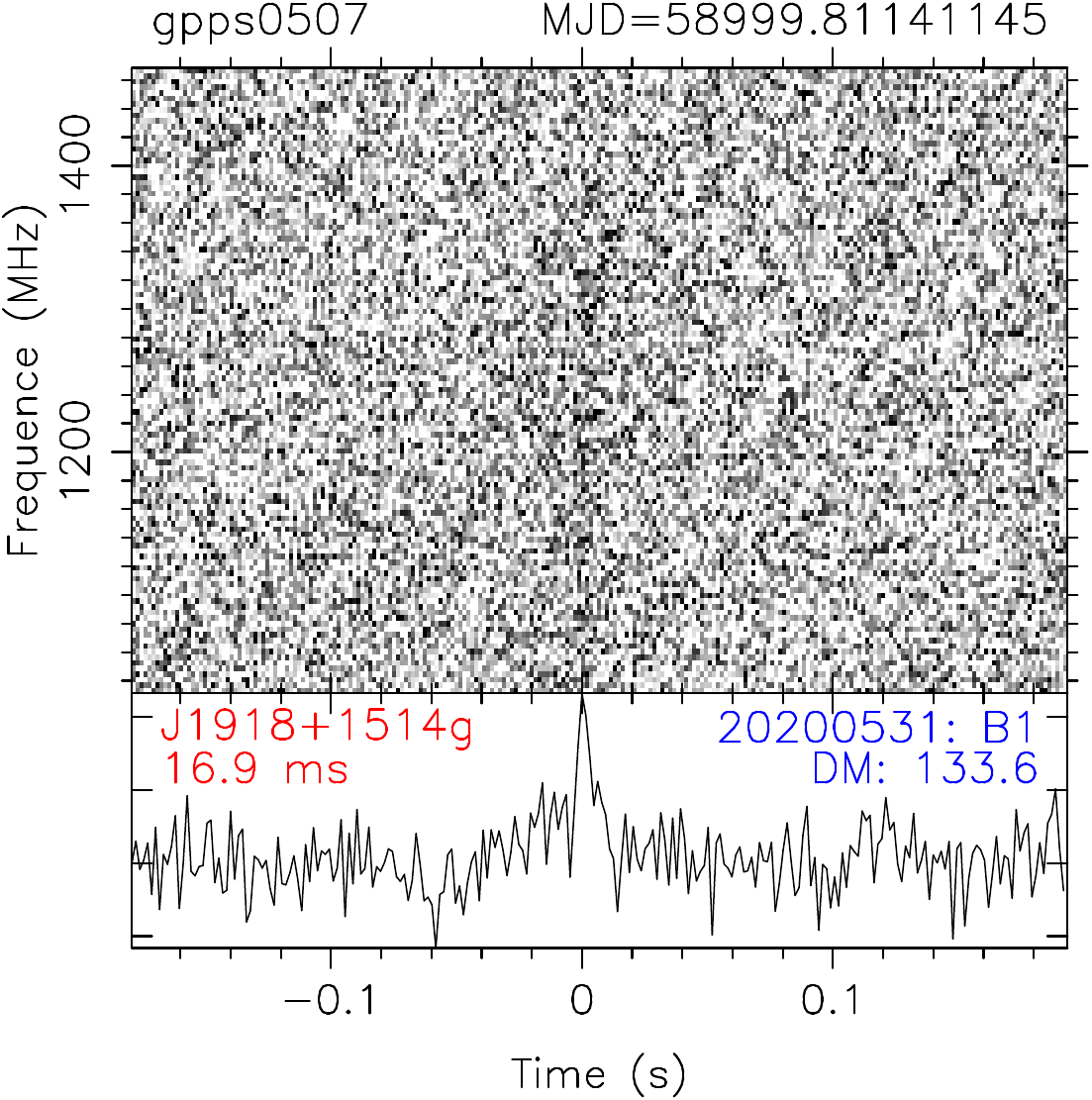}
\includegraphics[width=0.33\textwidth]{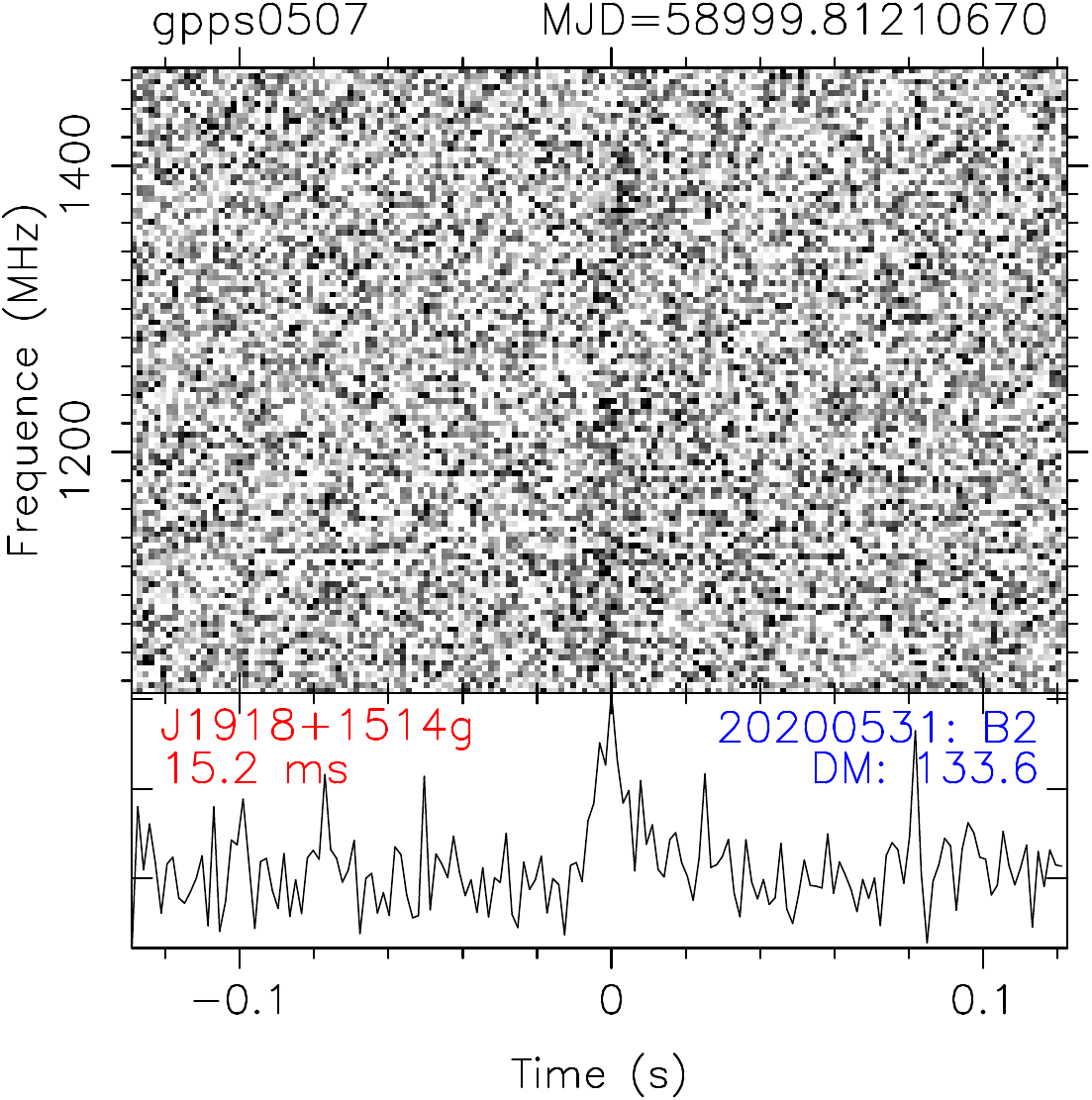}
\includegraphics[width=0.33\textwidth]{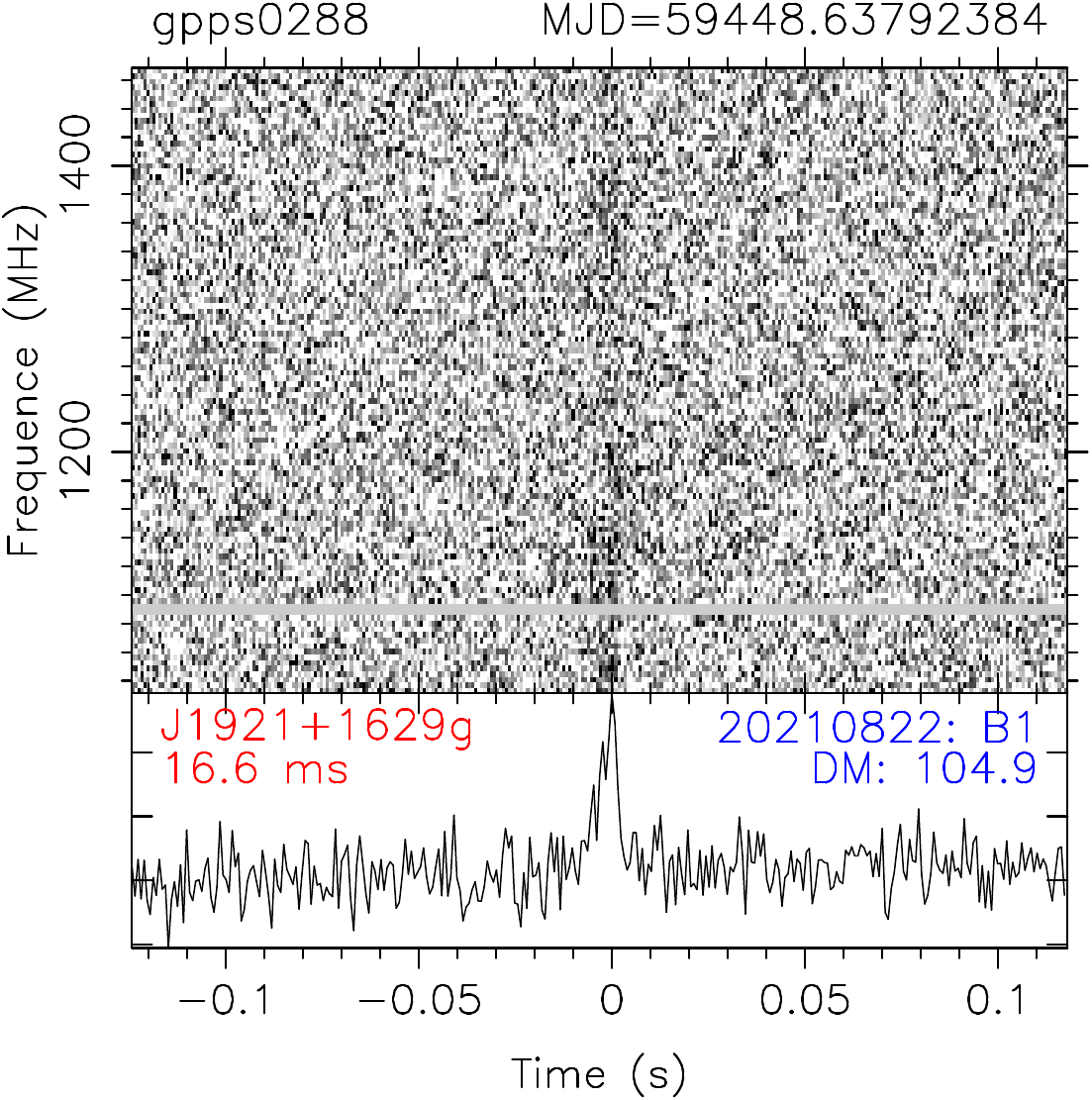}\\[0.2mm]
\includegraphics[width=0.33\textwidth]{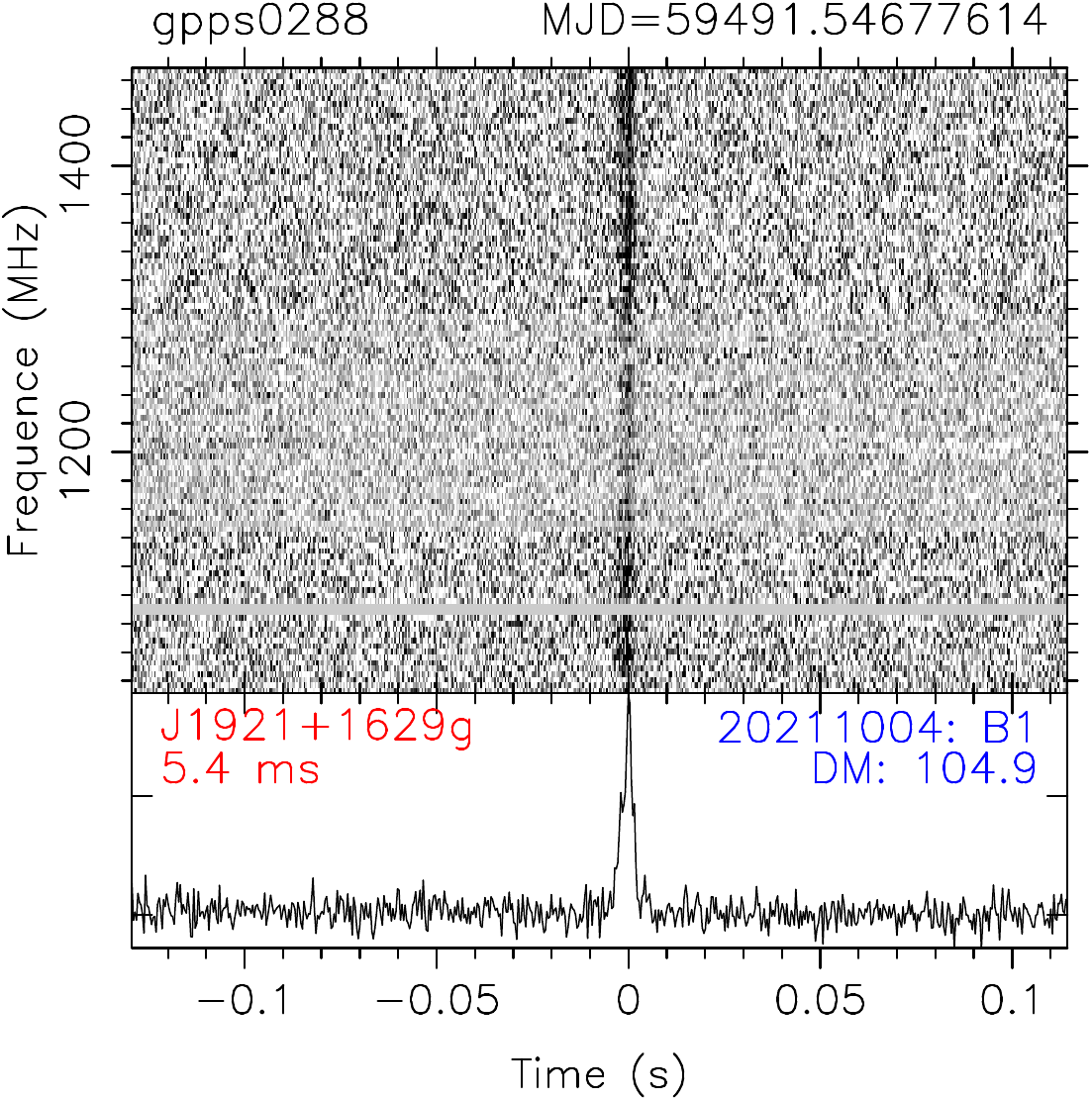}
\includegraphics[width=0.33\textwidth]{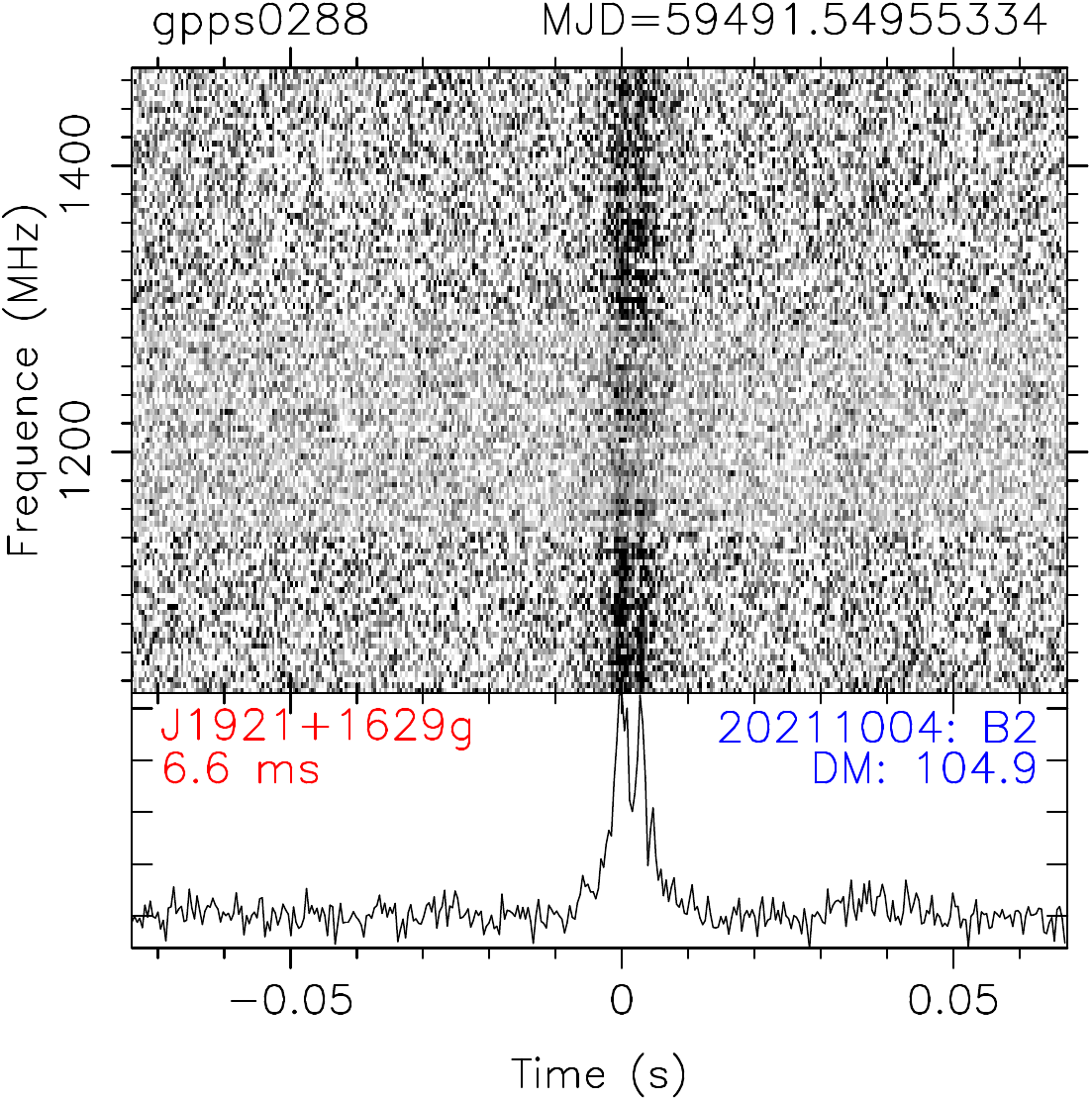}
\includegraphics[width=0.33\textwidth]{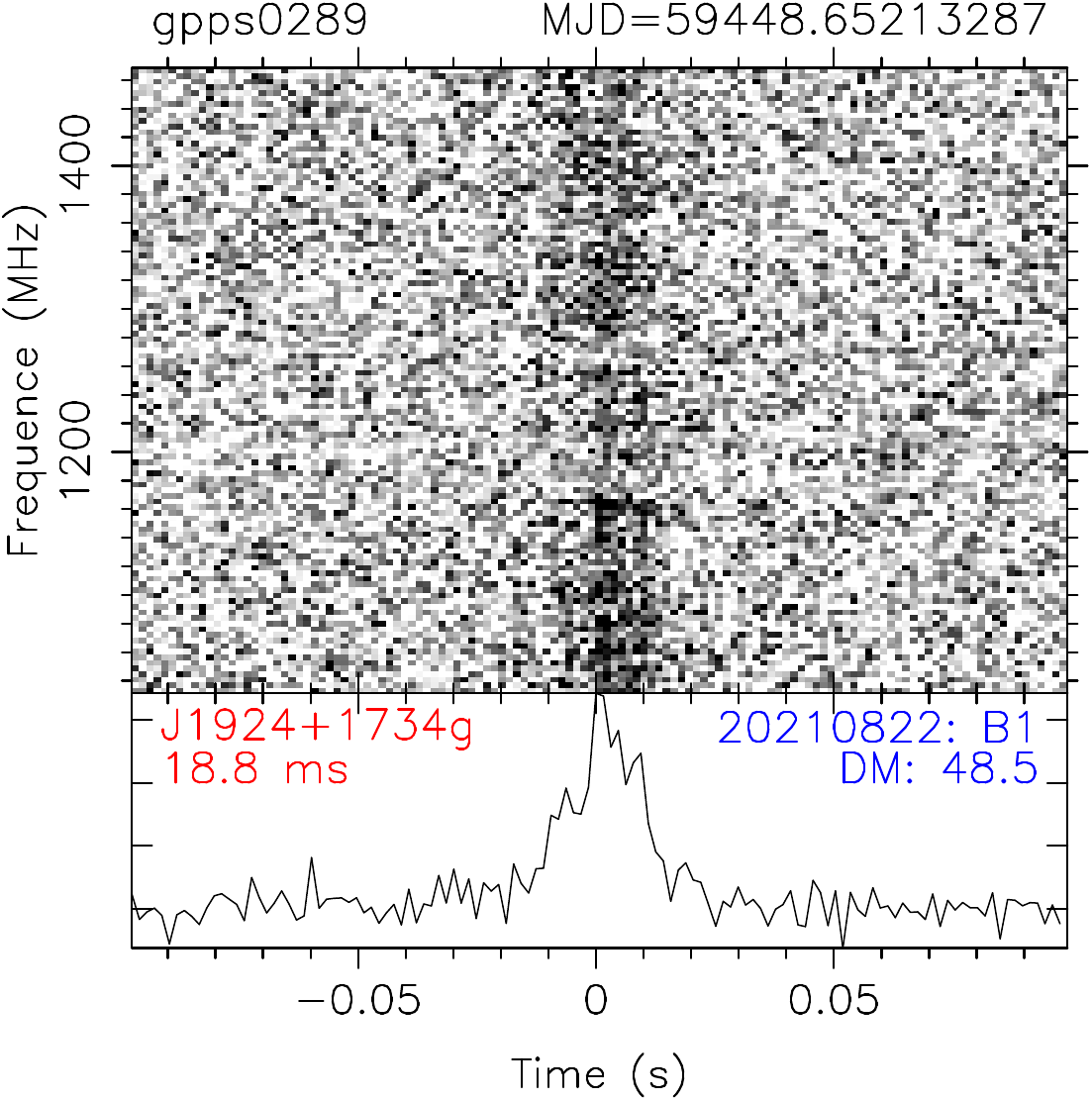}\\[0.2mm]
\includegraphics[width=0.33\textwidth]{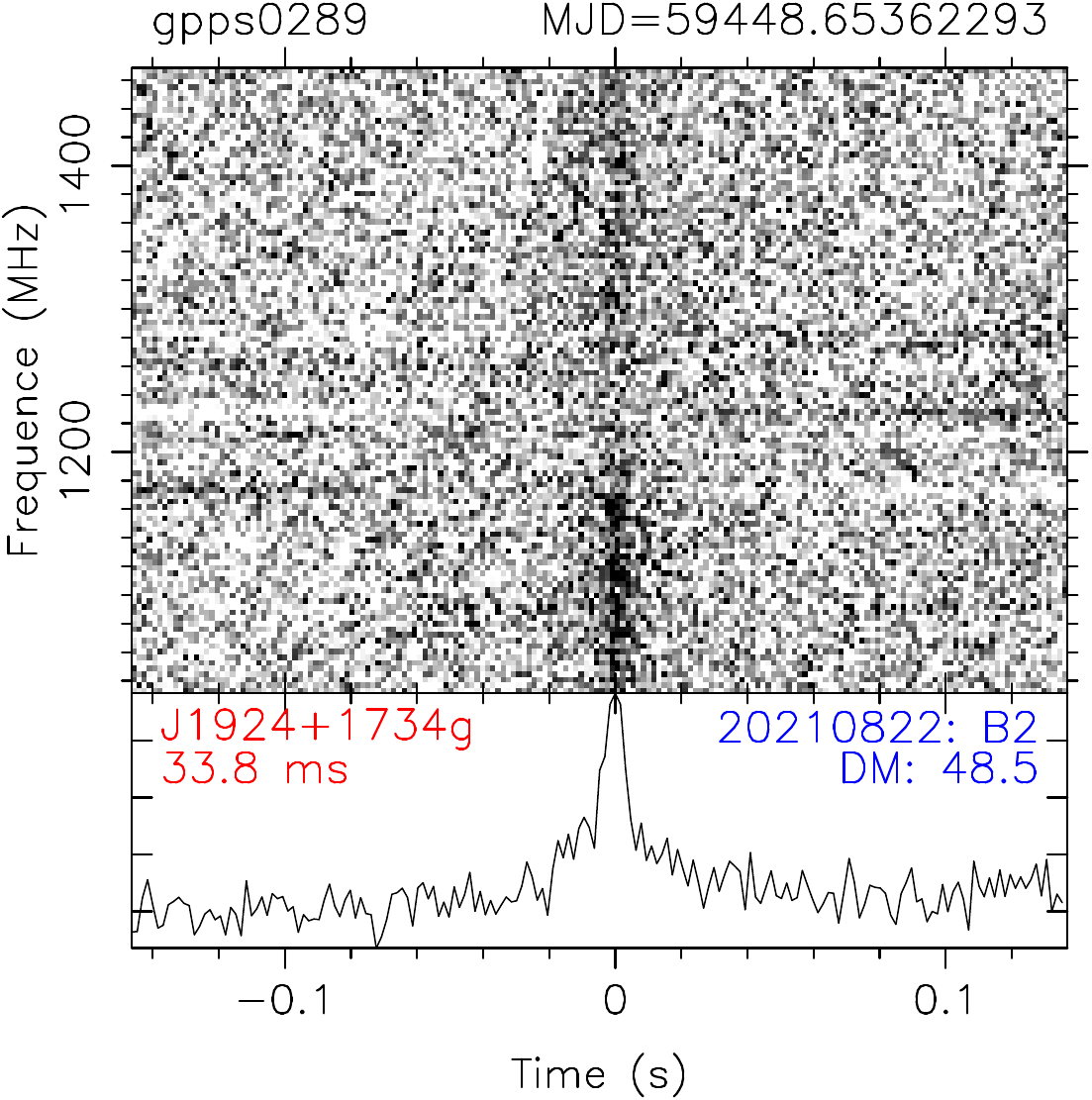}
\includegraphics[width=0.33\textwidth]{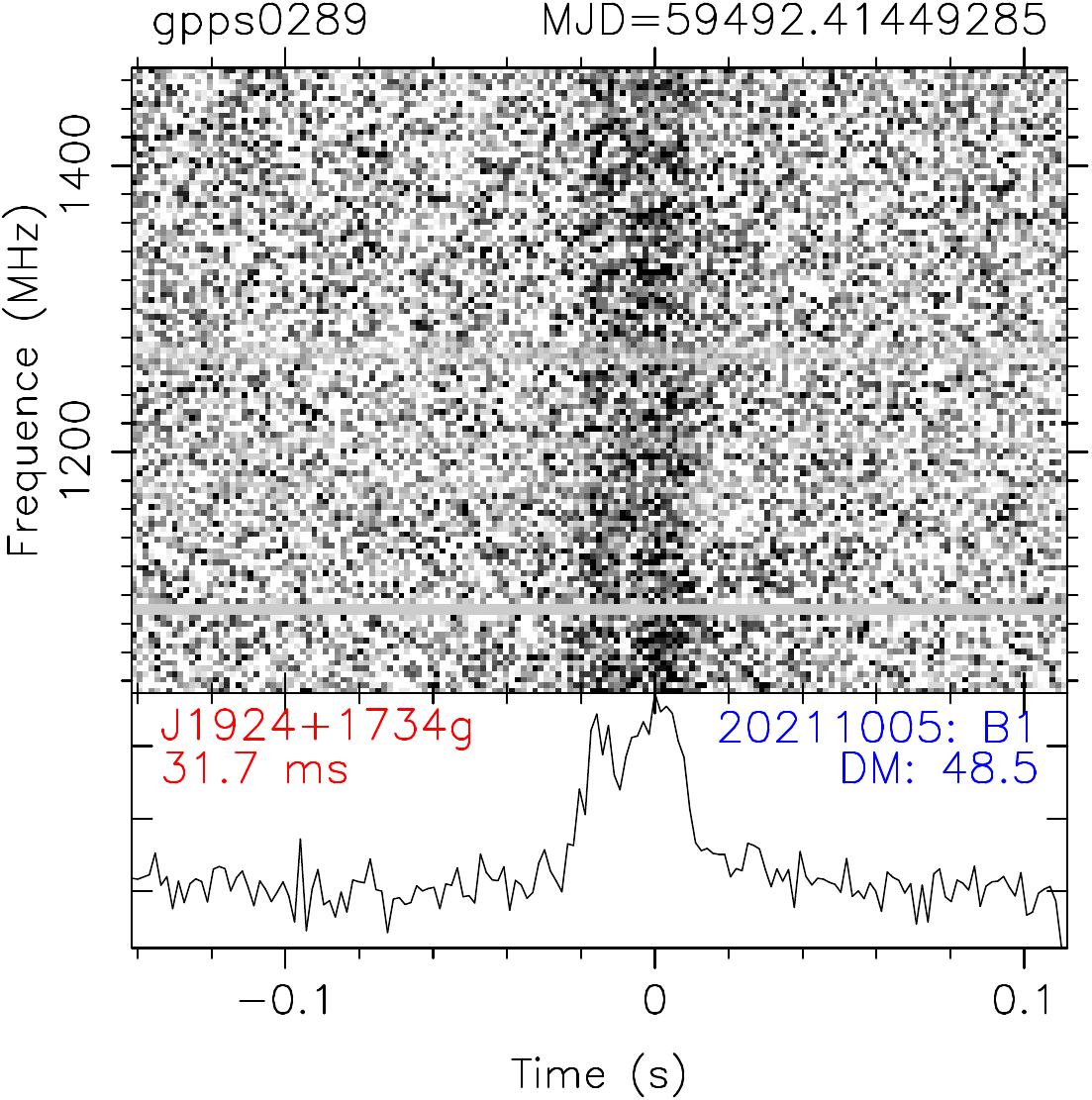}
\includegraphics[width=0.33\textwidth]{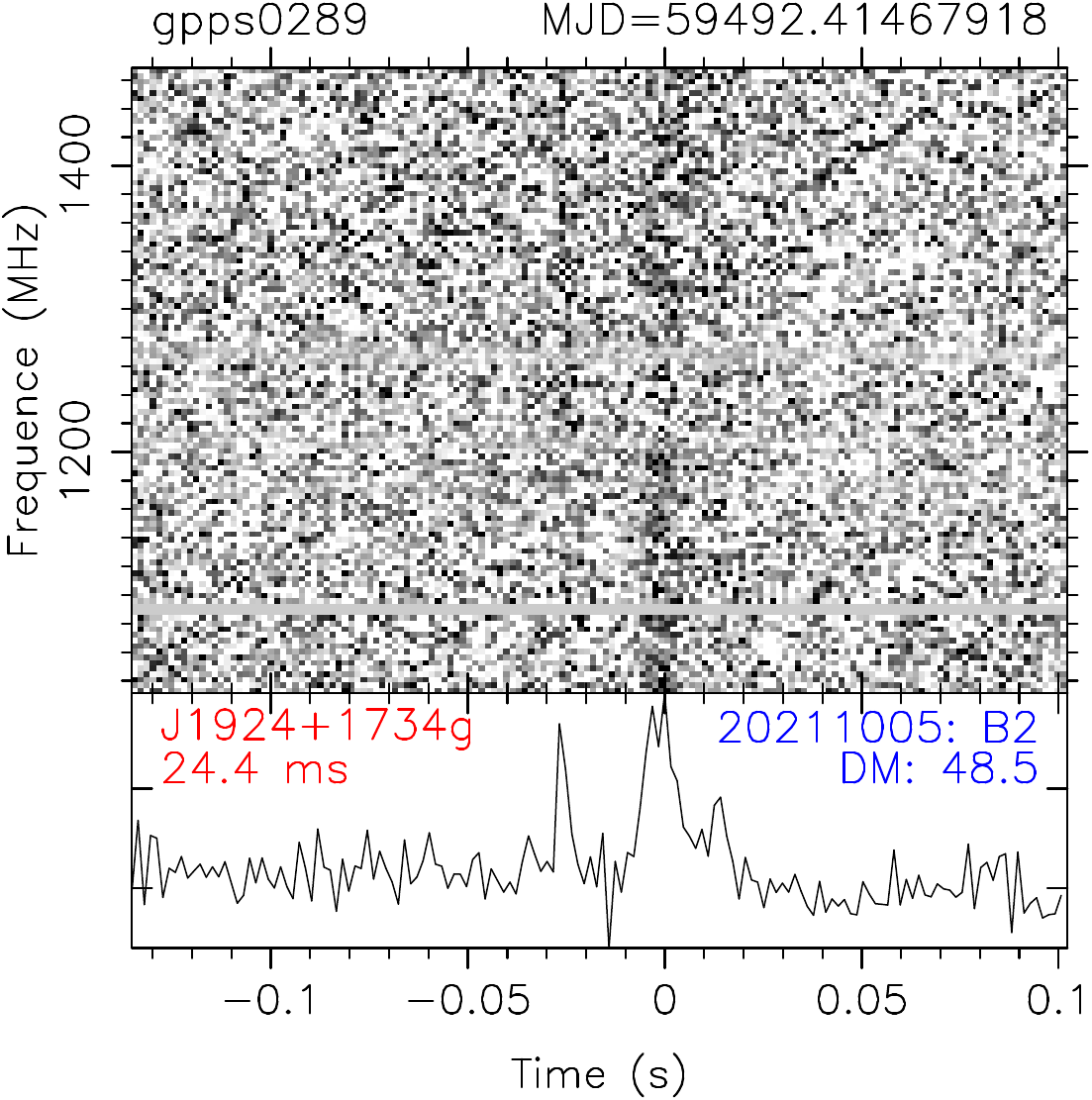}\\[0.2mm]
\includegraphics[width=0.33\textwidth]{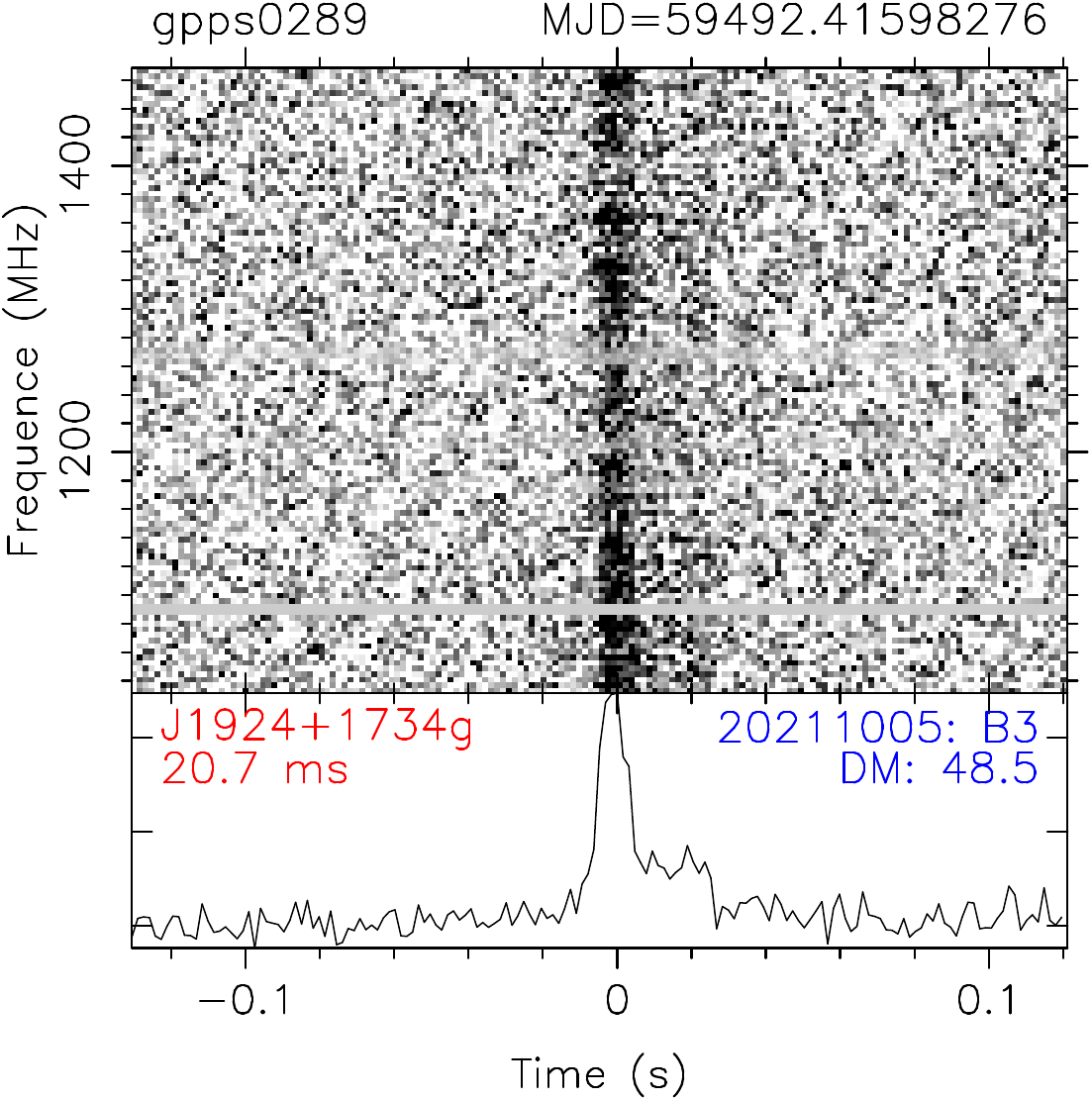}
\includegraphics[width=0.33\textwidth]{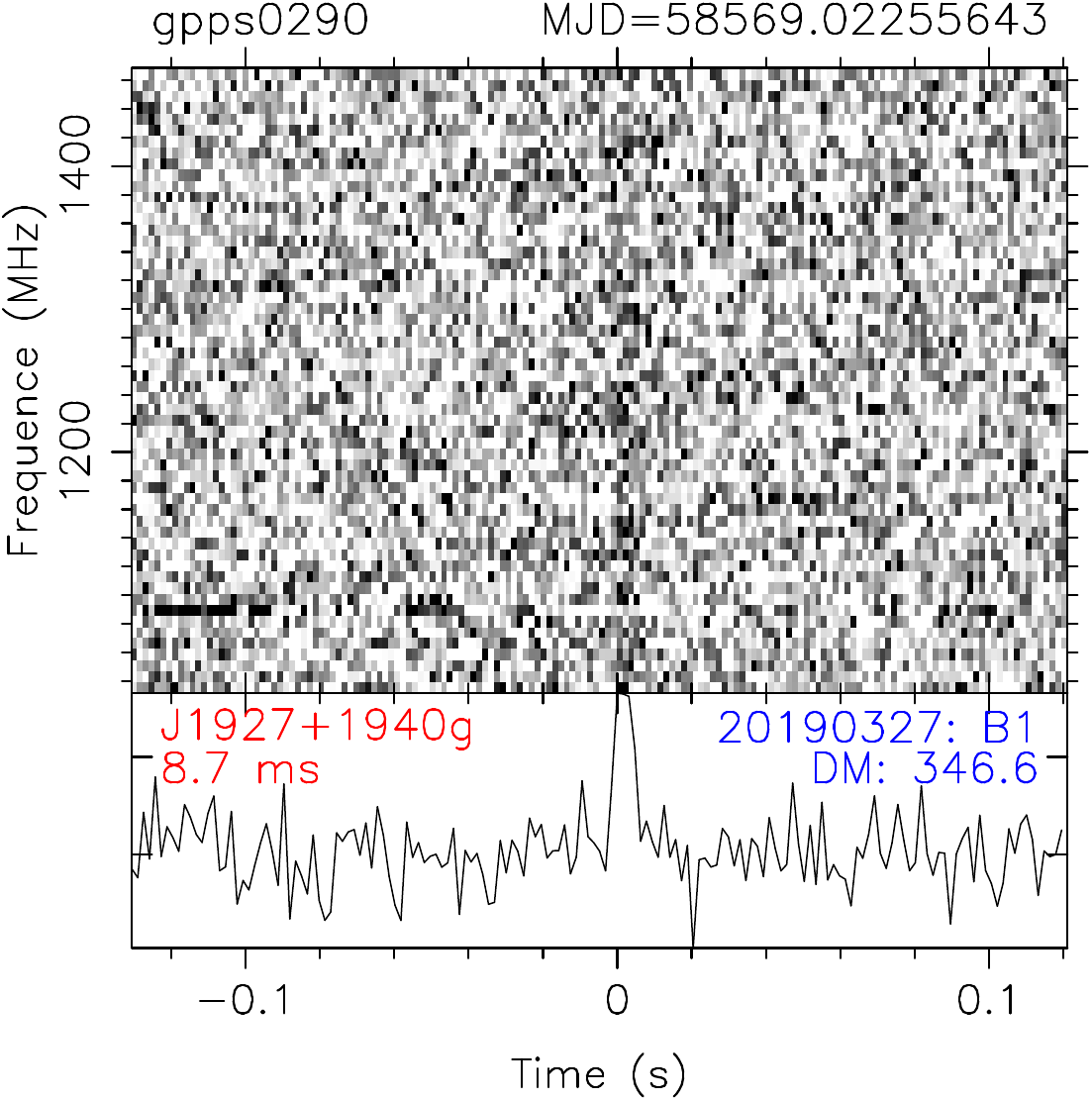}
\includegraphics[width=0.33\textwidth]{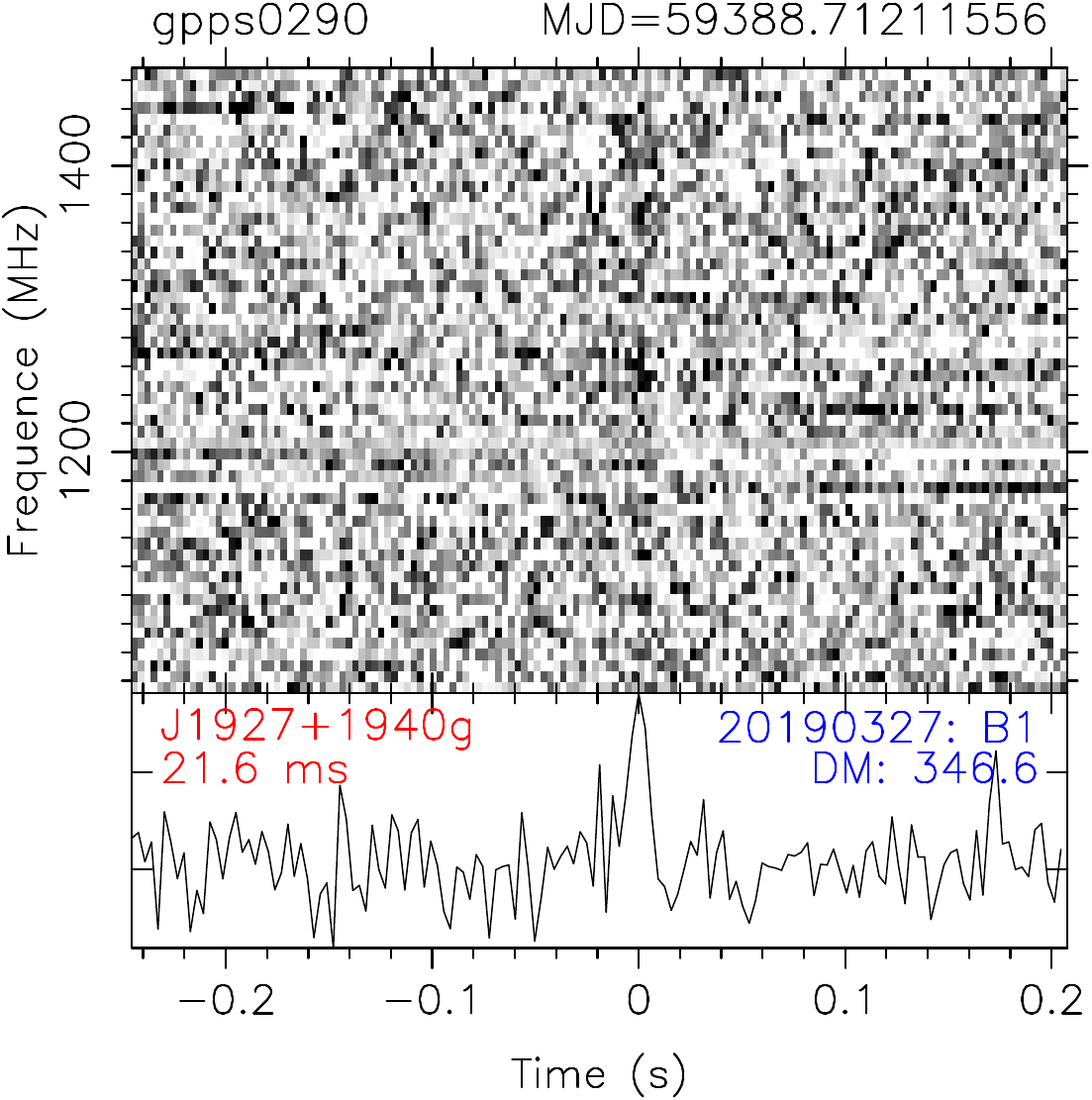}
\caption{(Continued.)}
\end{figure*}
\addtocounter{figure}{-1}
\begin{figure*}[!t]
\centering
\includegraphics[width=0.33\textwidth]{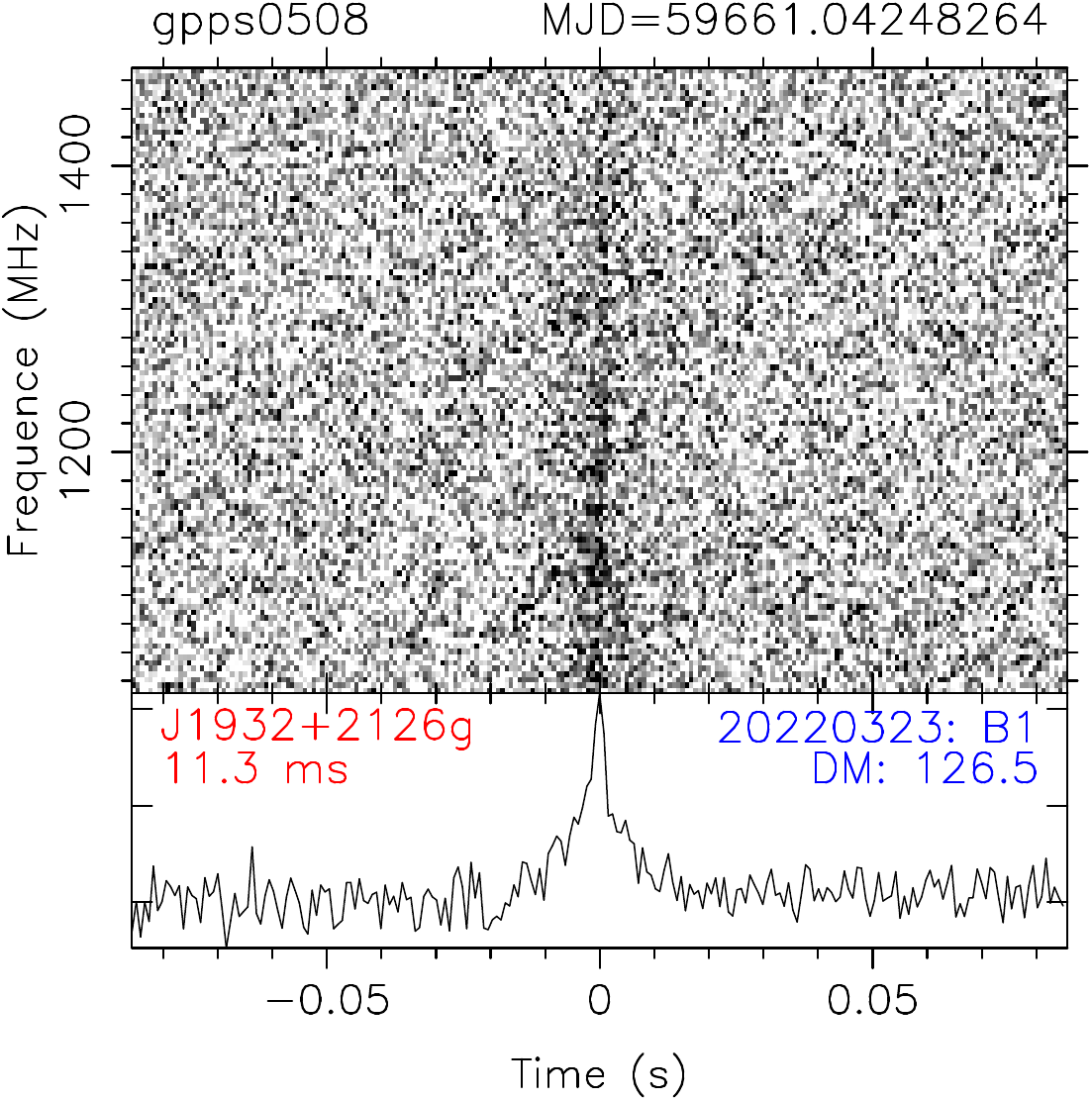}
\includegraphics[width=0.33\textwidth]{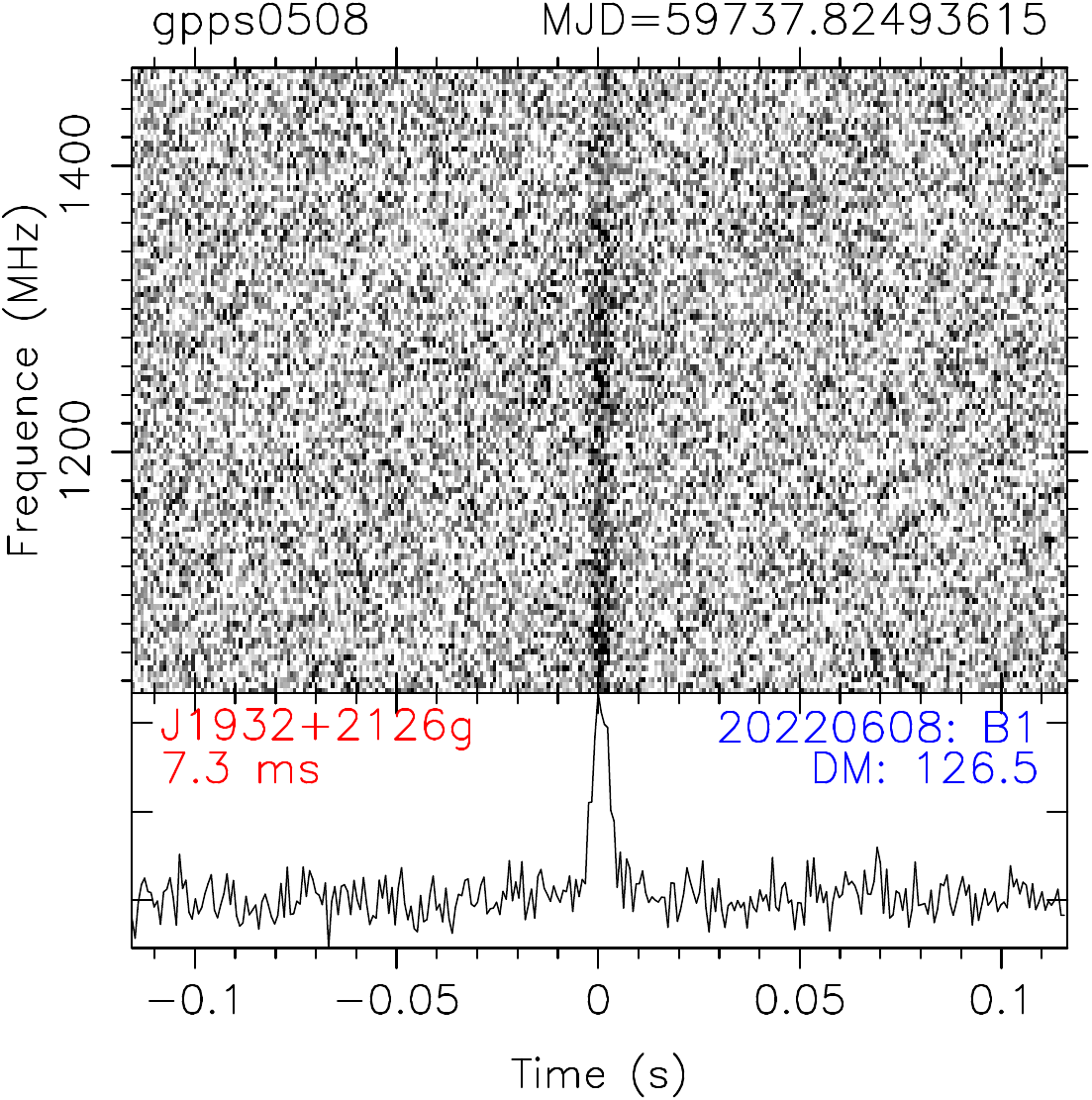}
\includegraphics[width=0.33\textwidth]{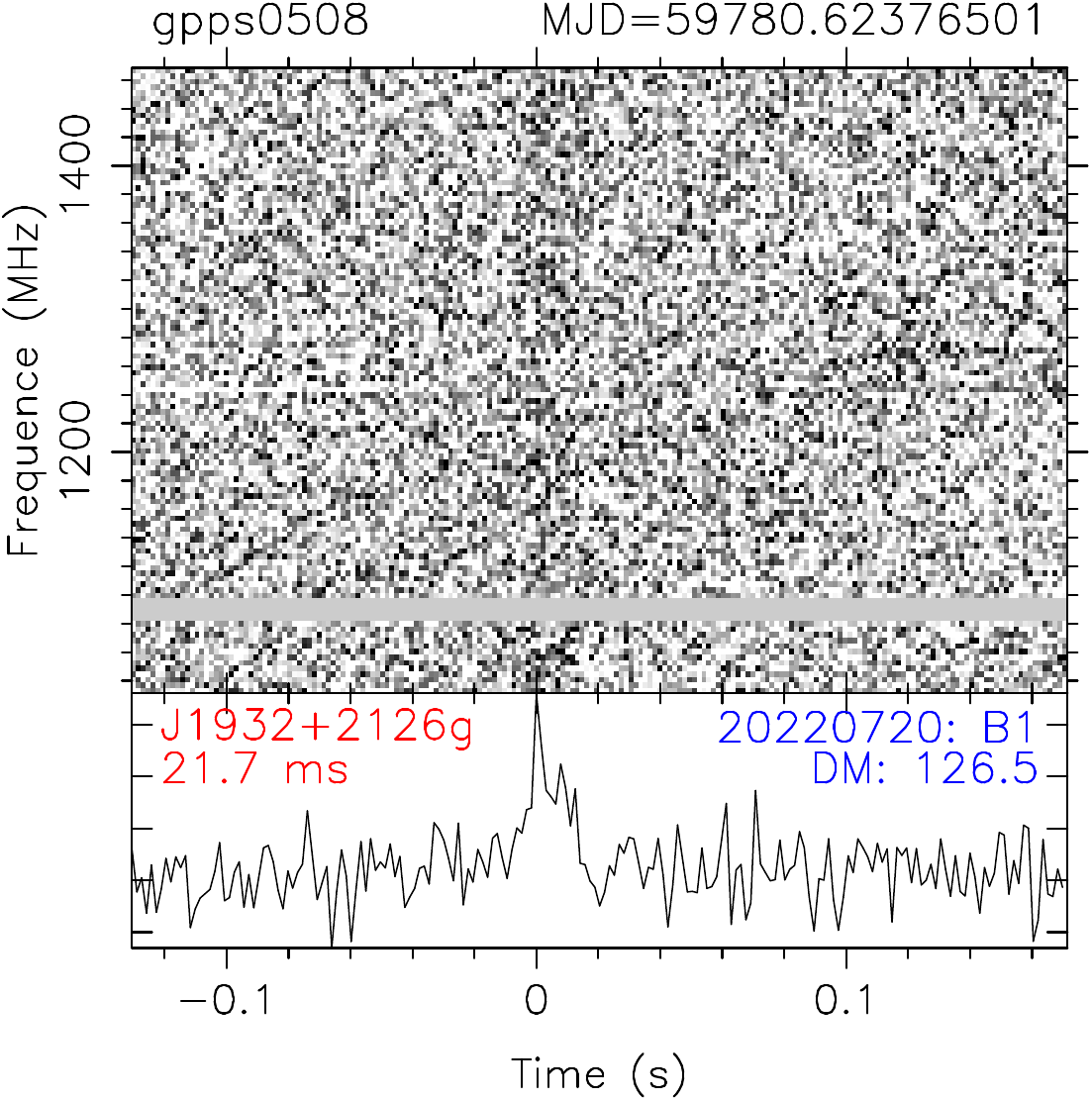}\\[0.2mm]
\includegraphics[width=0.33\textwidth]{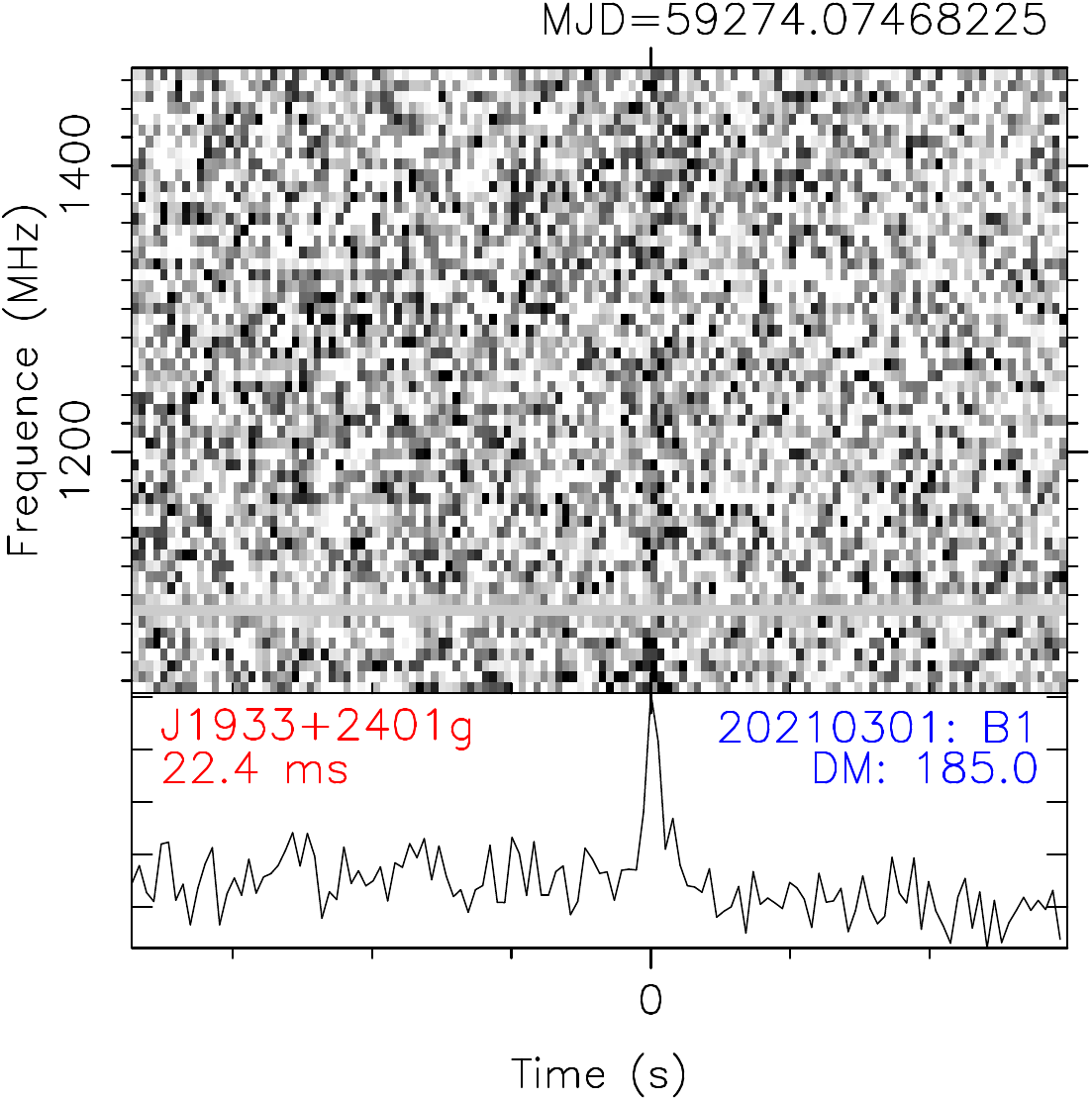}
\includegraphics[width=0.33\textwidth]{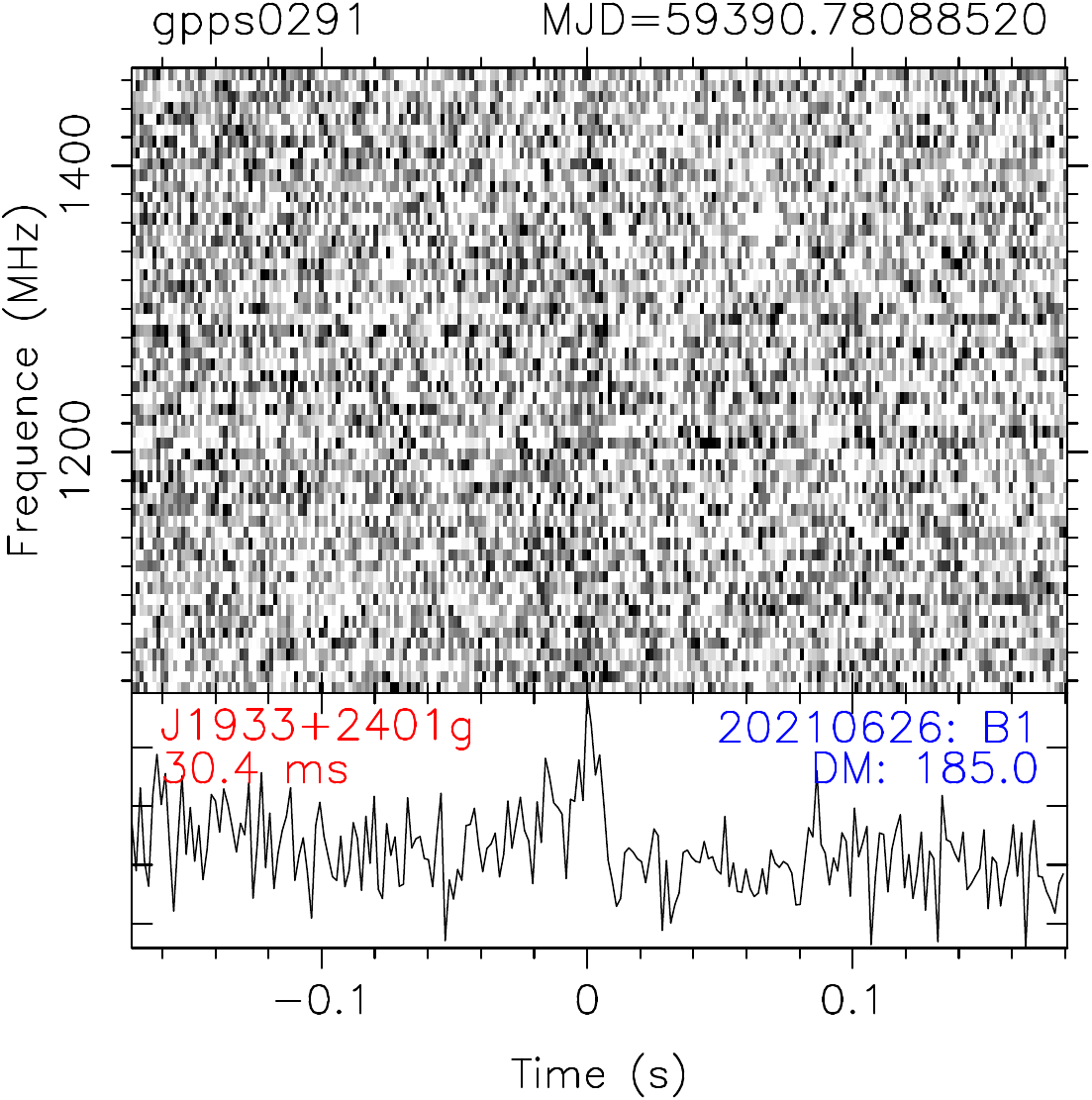}
\includegraphics[width=0.33\textwidth]{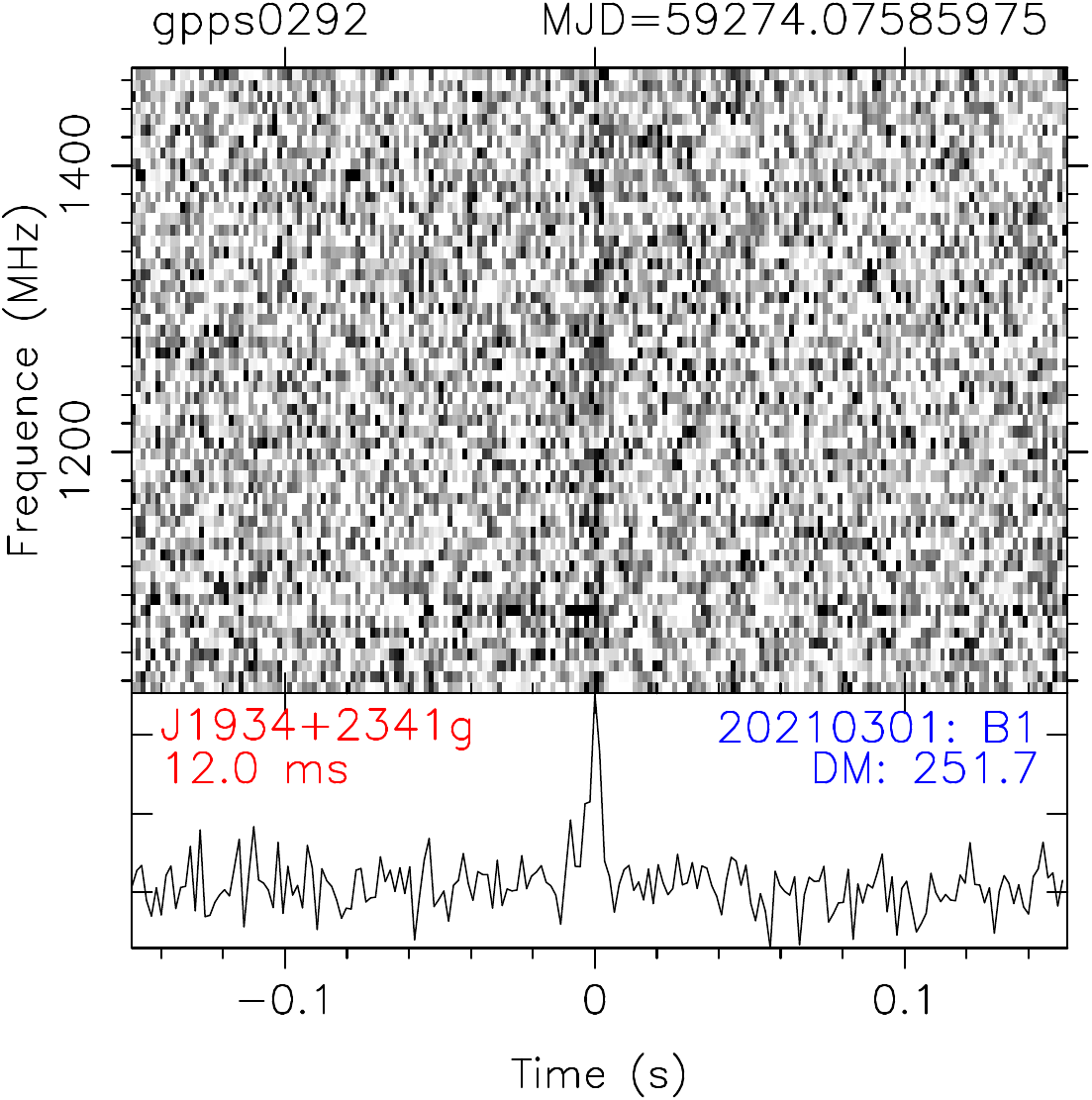}\\[0.2mm]
\includegraphics[width=0.33\textwidth]{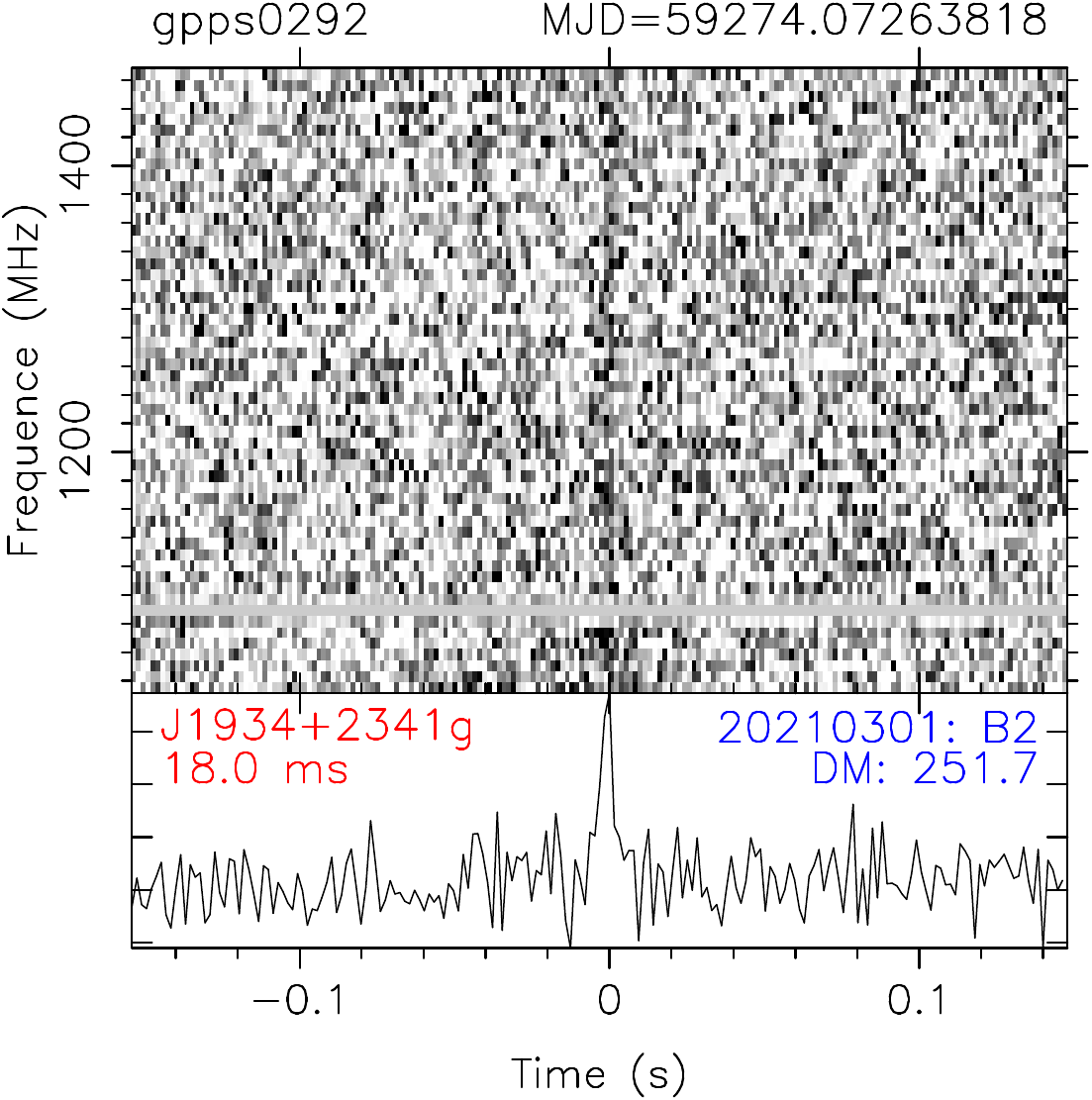}
\includegraphics[width=0.33\textwidth]{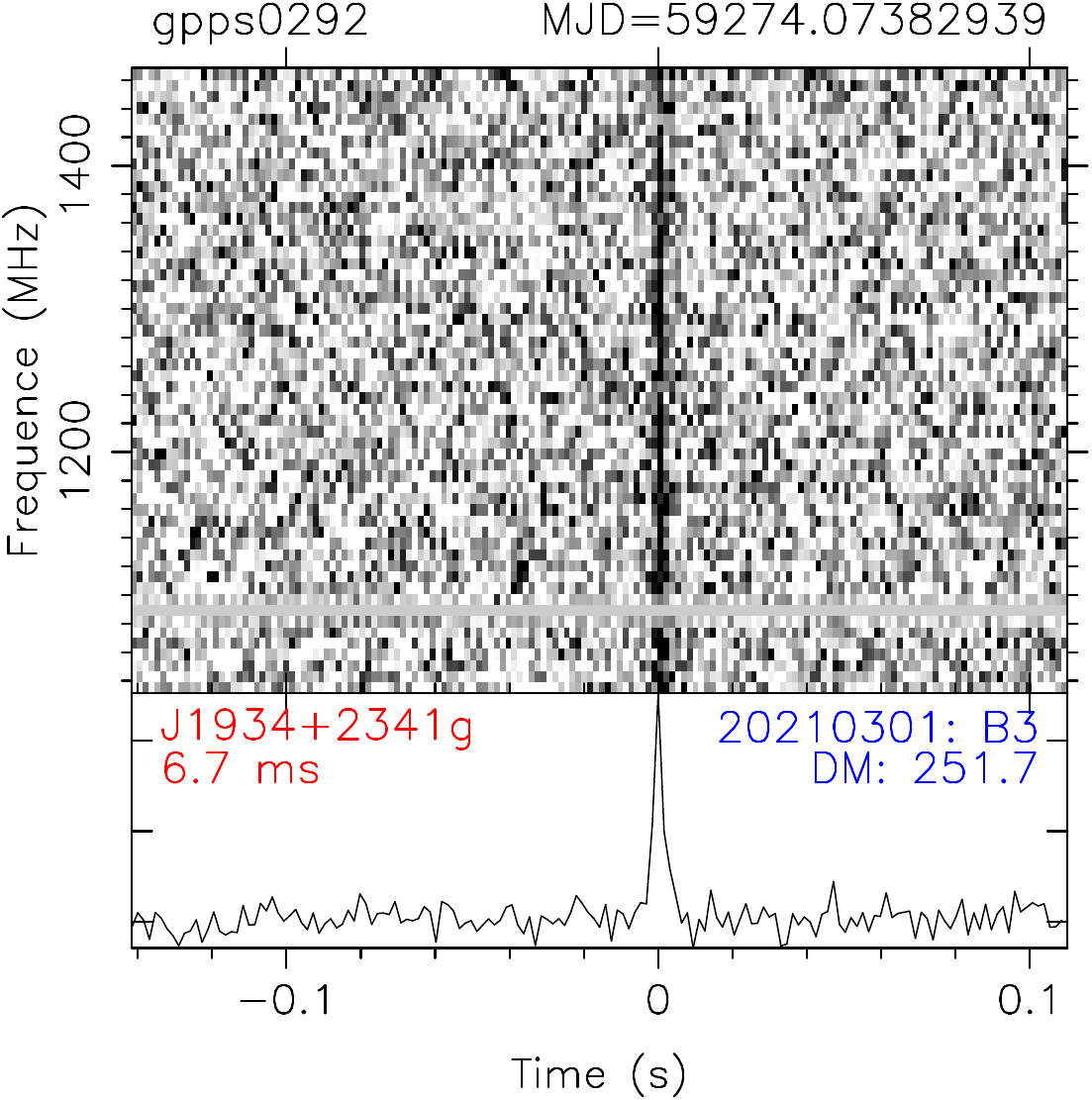}
\includegraphics[width=0.33\textwidth]{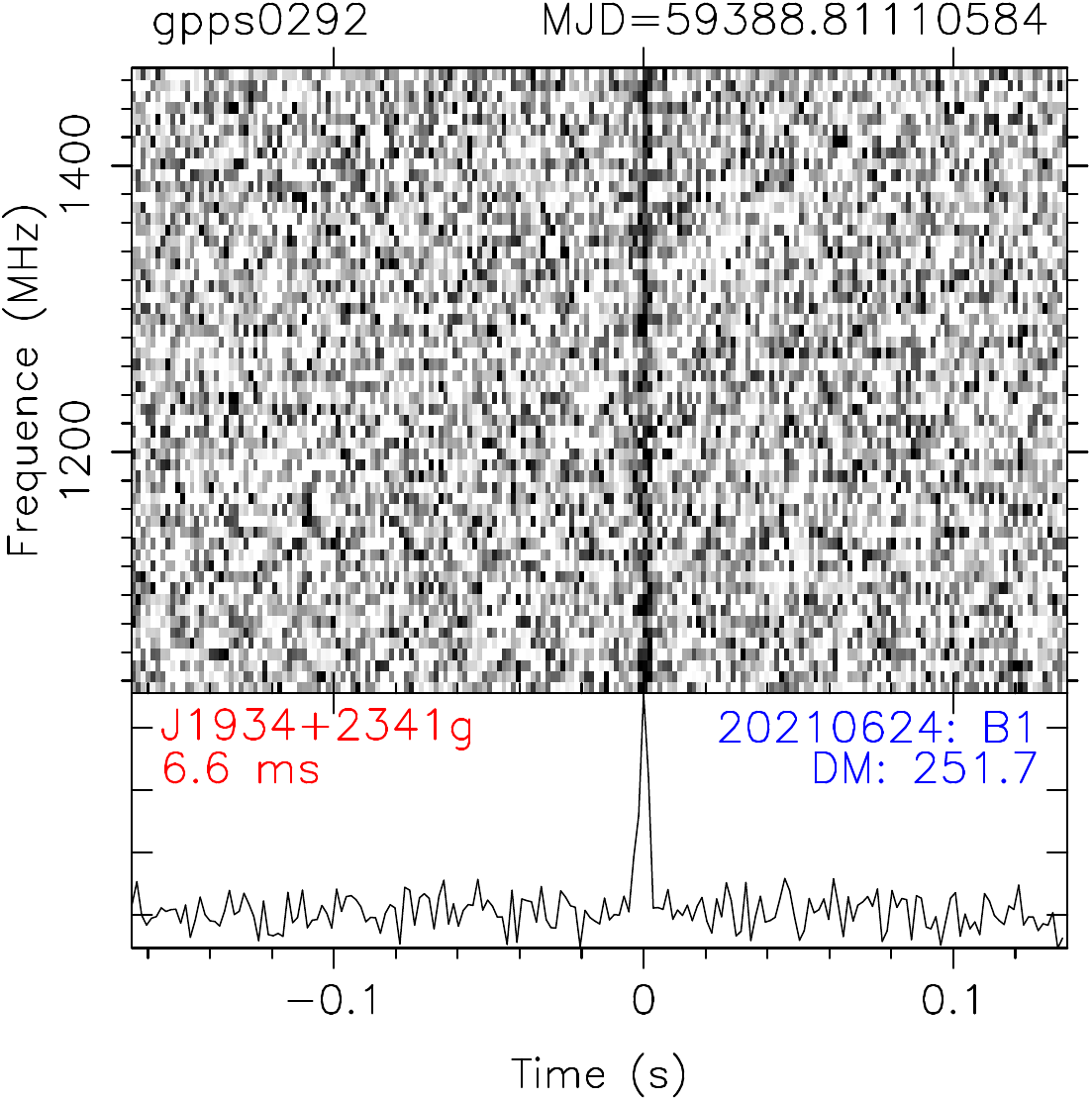}\\[0.2mm]
\includegraphics[width=0.33\textwidth]{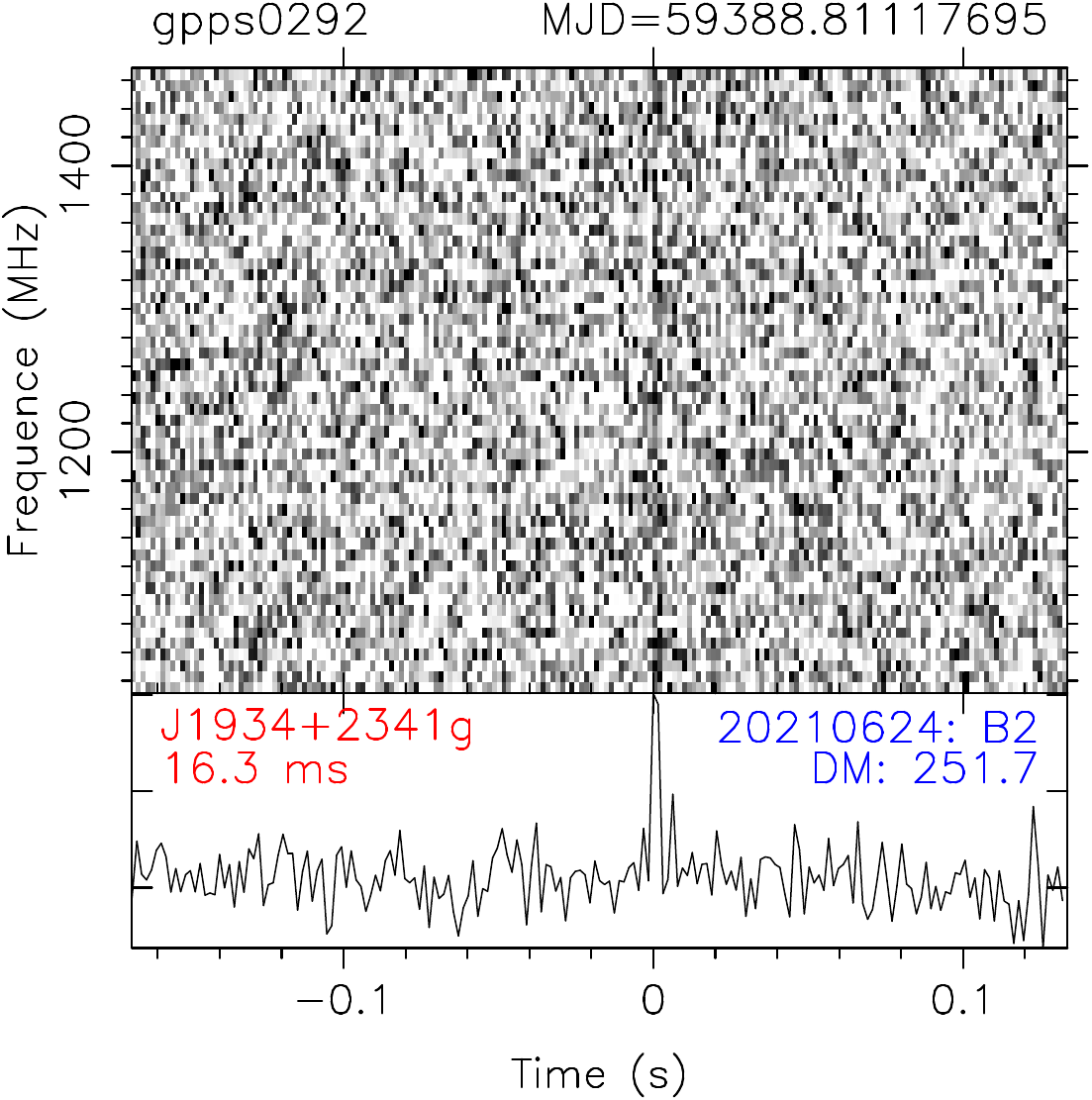}
\includegraphics[width=0.33\textwidth]{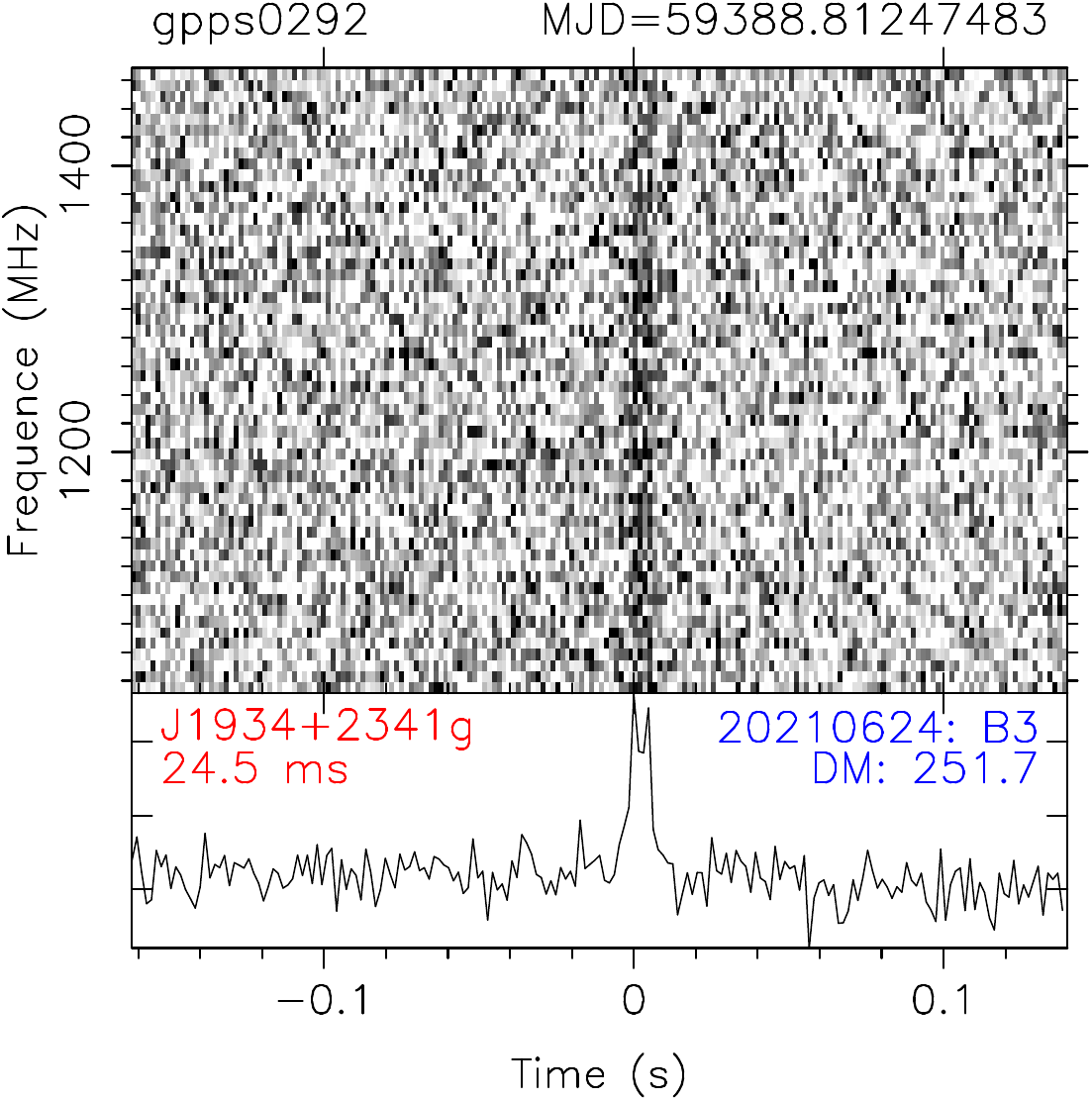}
\includegraphics[width=0.33\textwidth]{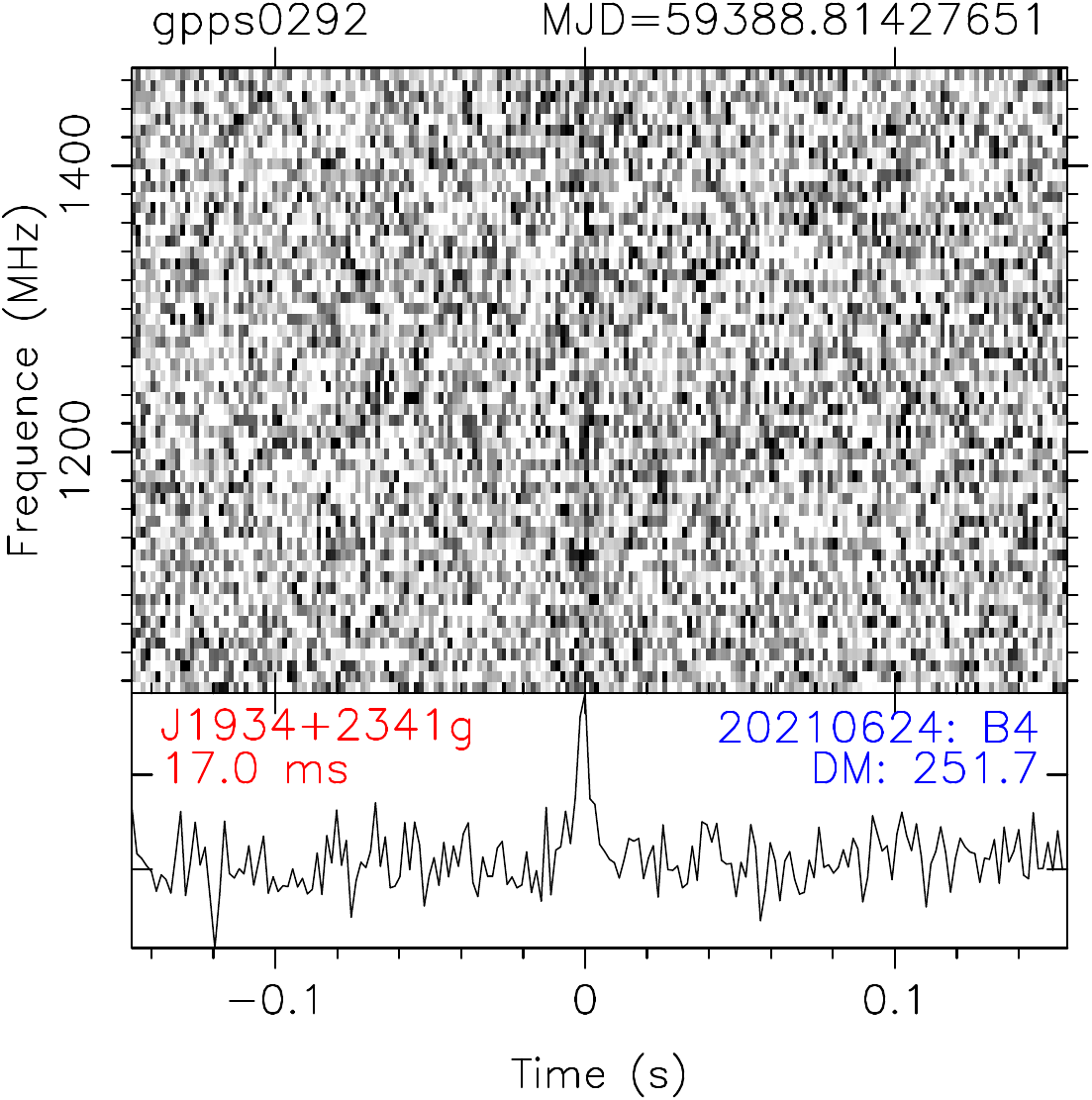}
\caption{(Continued.)}
\end{figure*}
\addtocounter{figure}{-1}
\begin{figure*}[!t]
\centering
\includegraphics[width=0.33\textwidth]{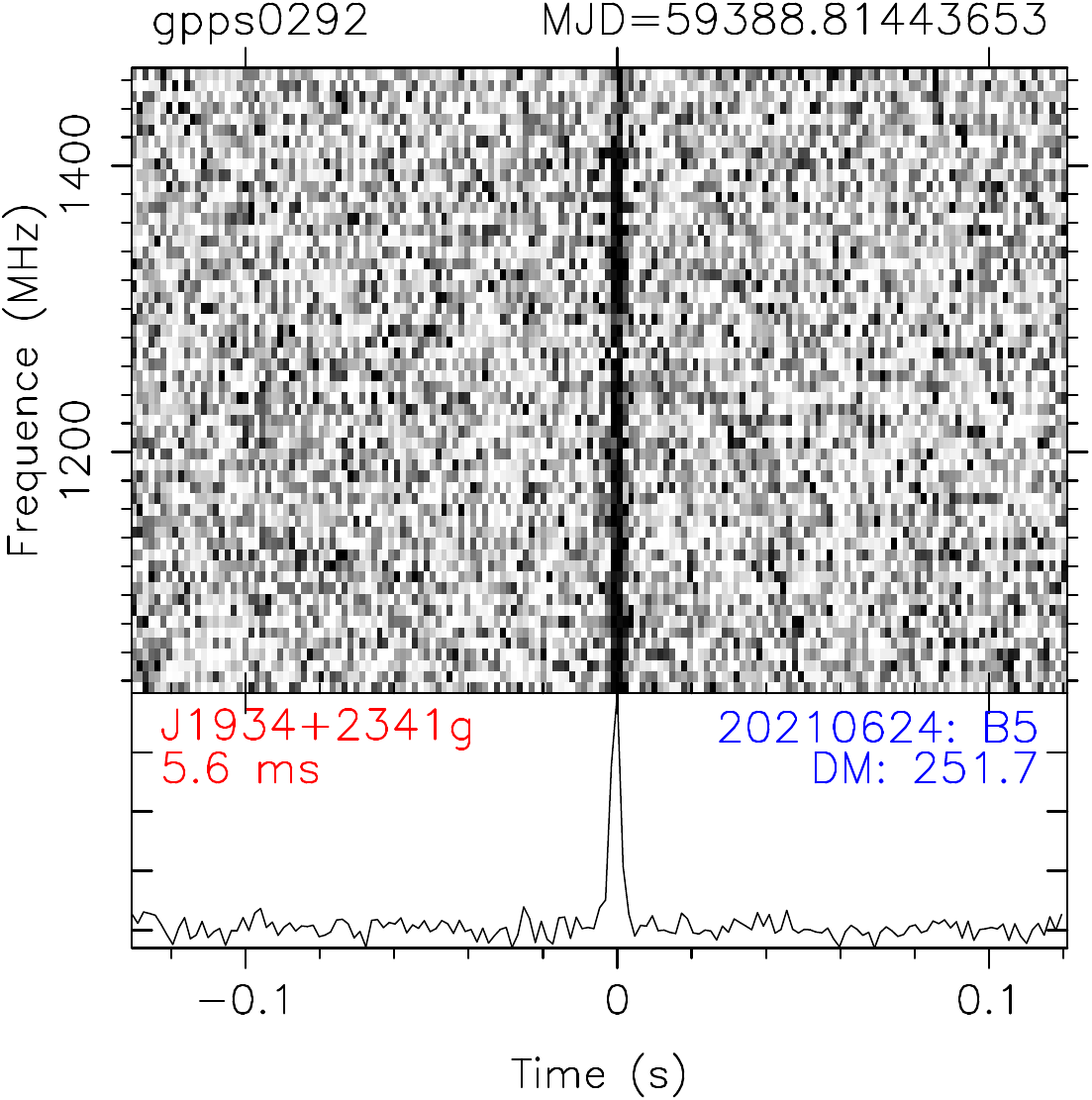}
\includegraphics[width=0.33\textwidth]{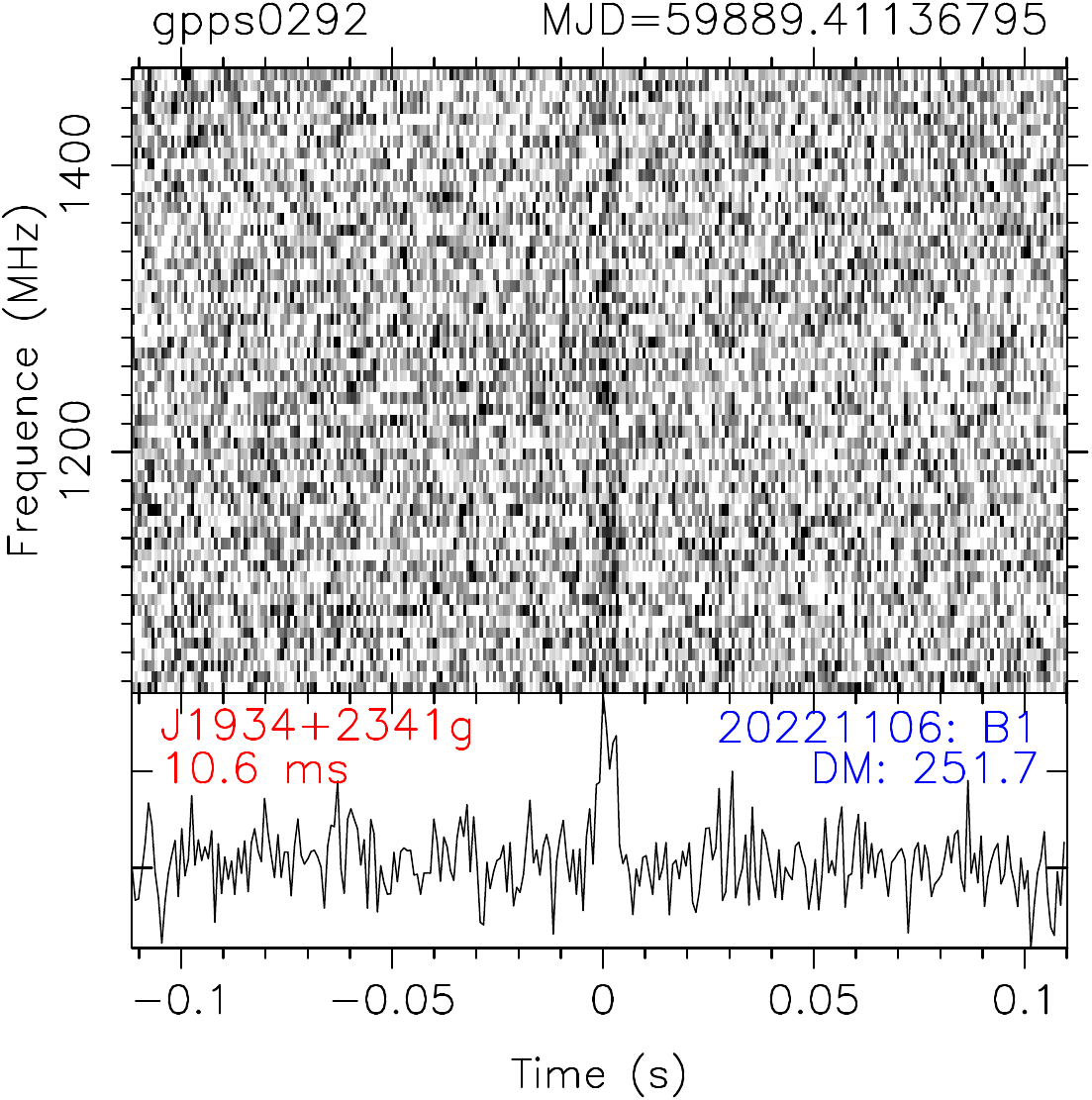}
\includegraphics[width=0.33\textwidth]{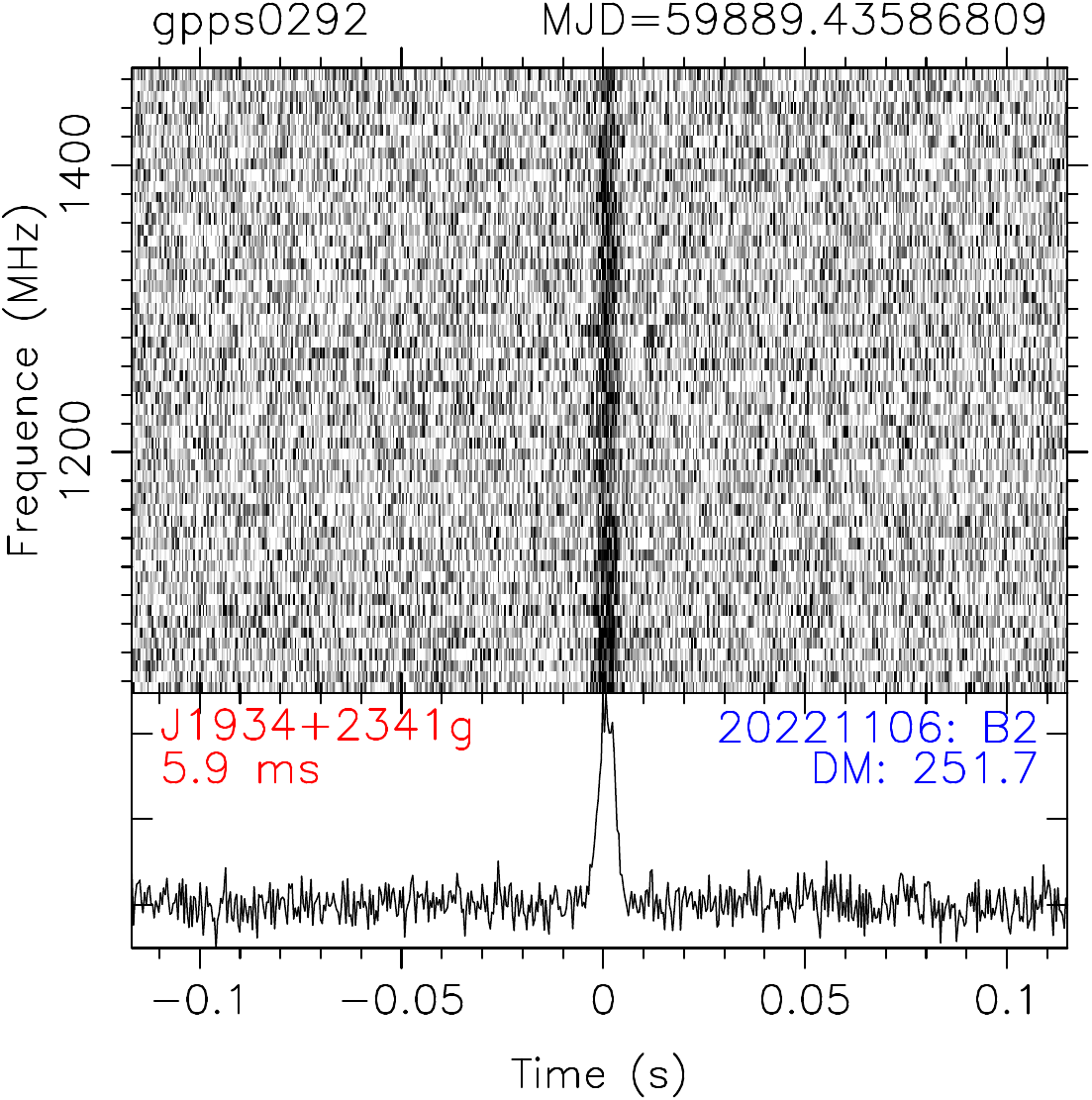}\\[0.2mm]
\includegraphics[width=0.33\textwidth]{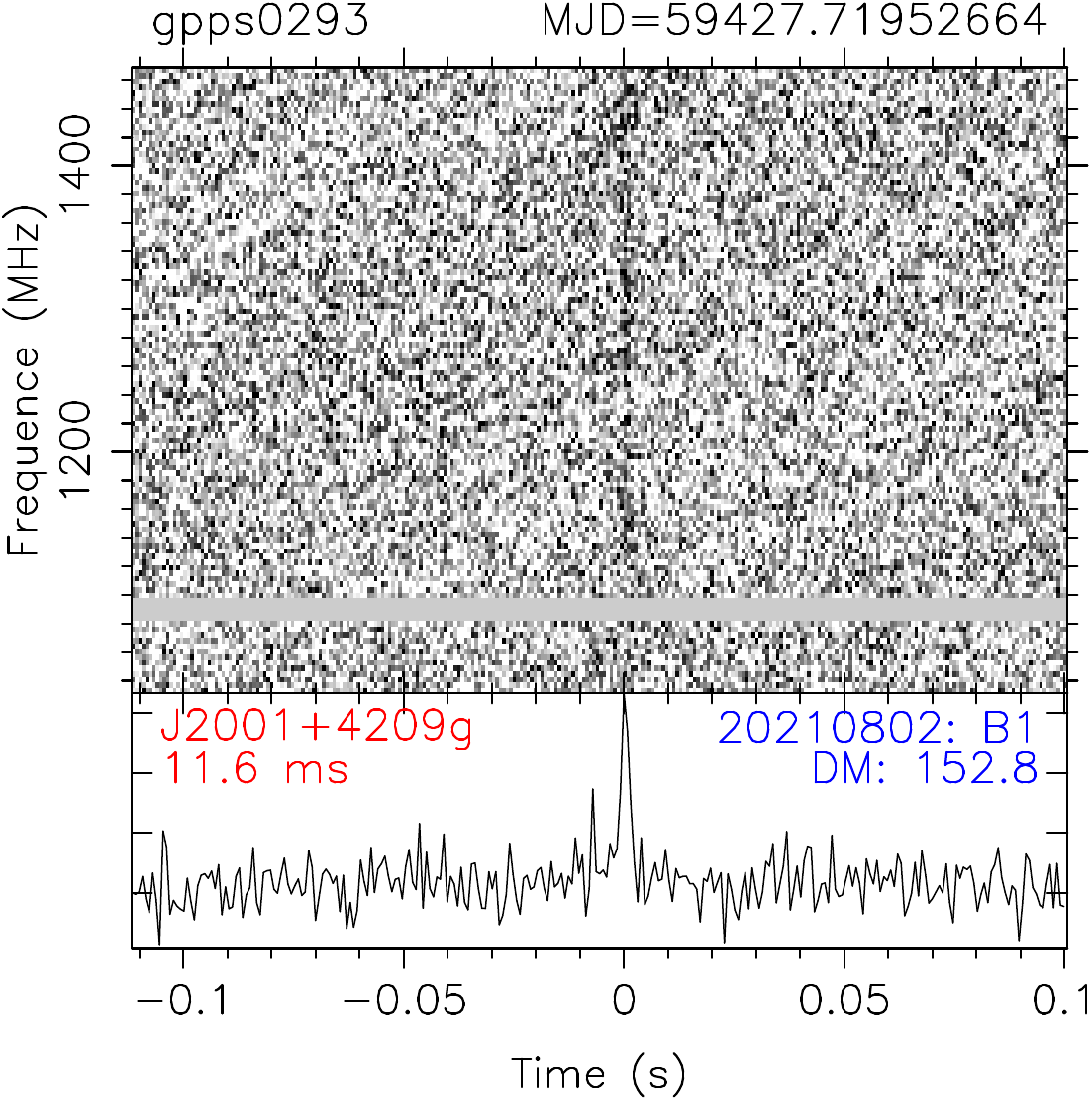}
\includegraphics[width=0.33\textwidth]{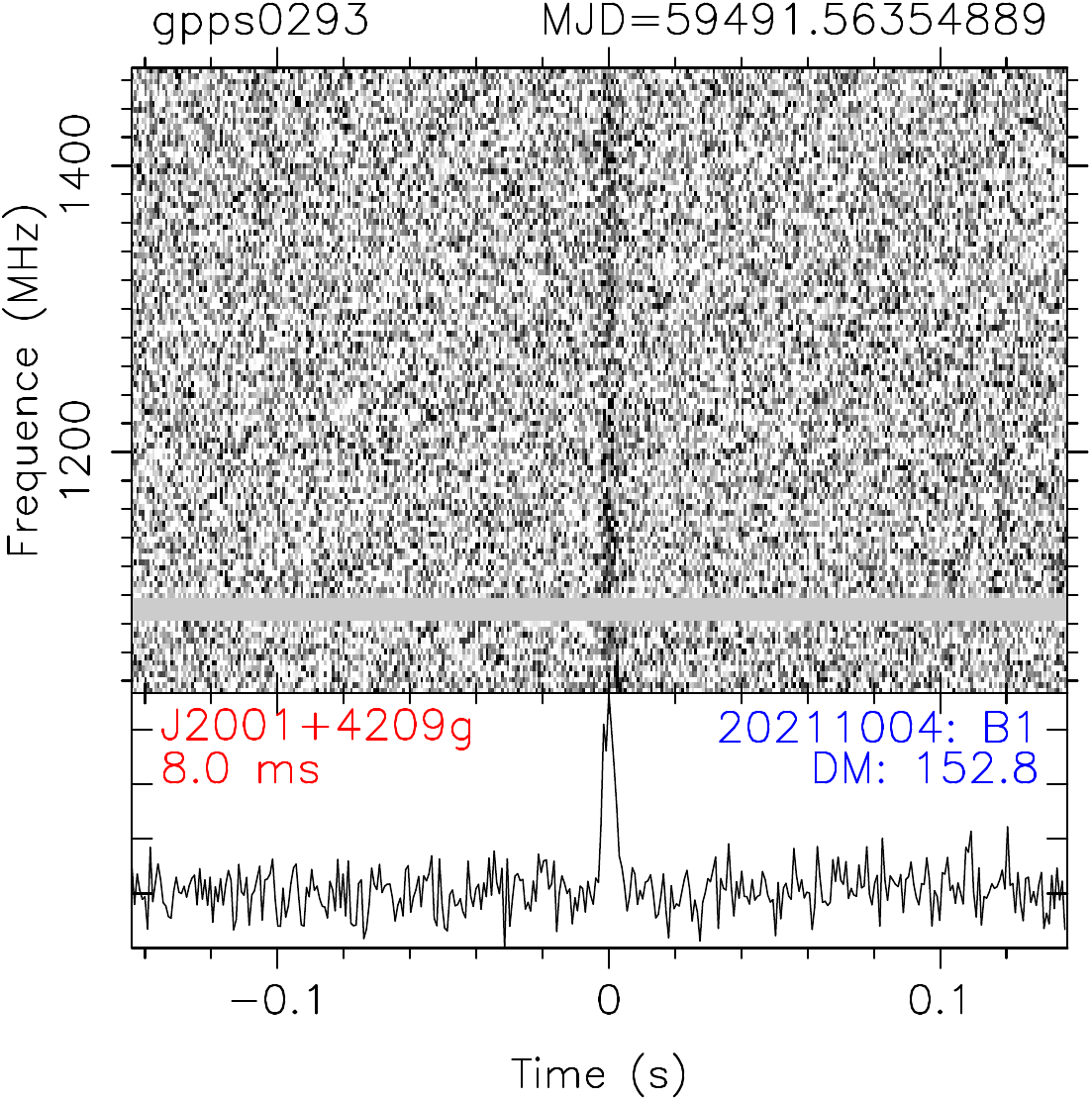}
\includegraphics[width=0.33\textwidth]{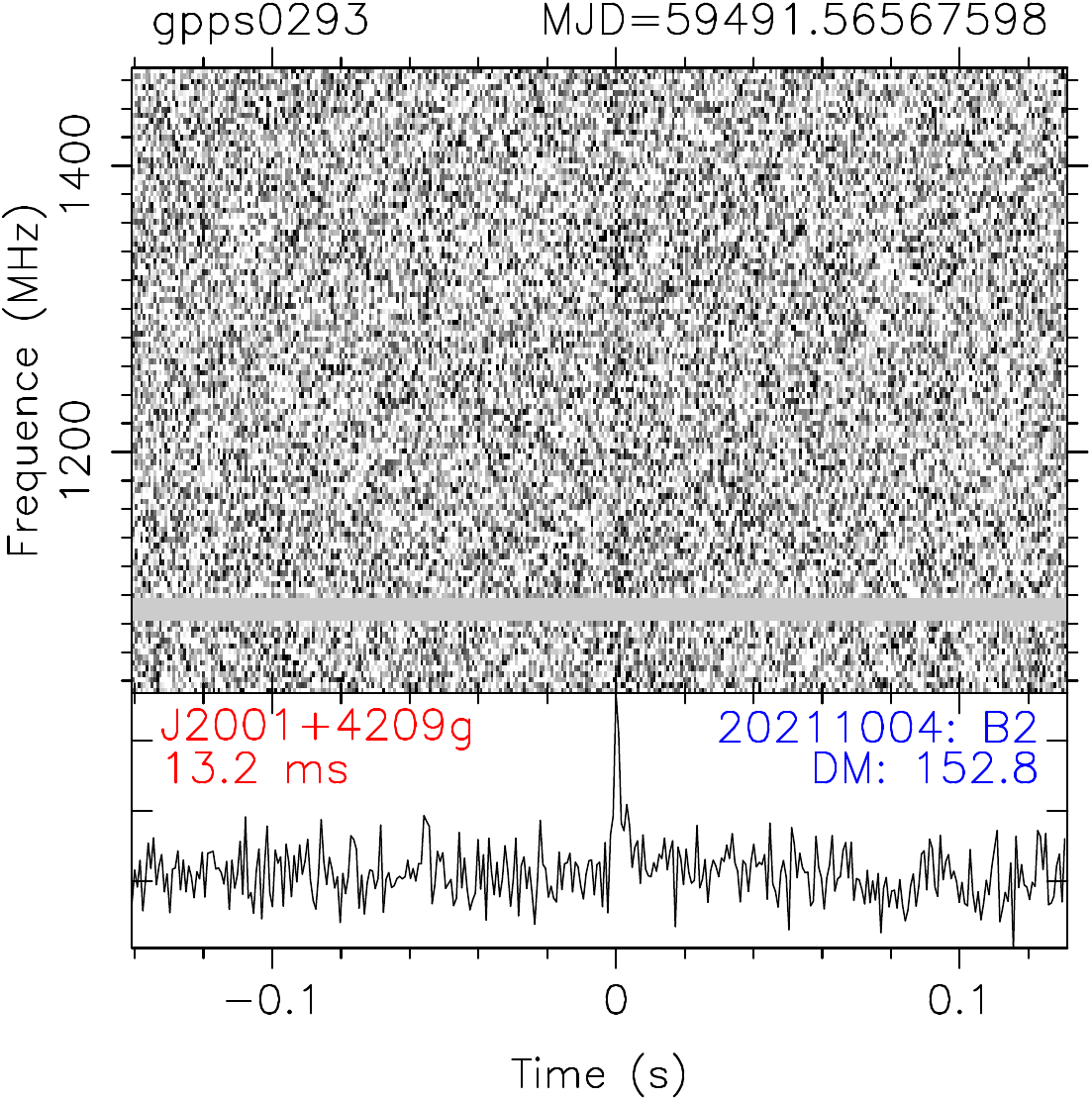}\\[0.2mm]
\includegraphics[width=0.33\textwidth]{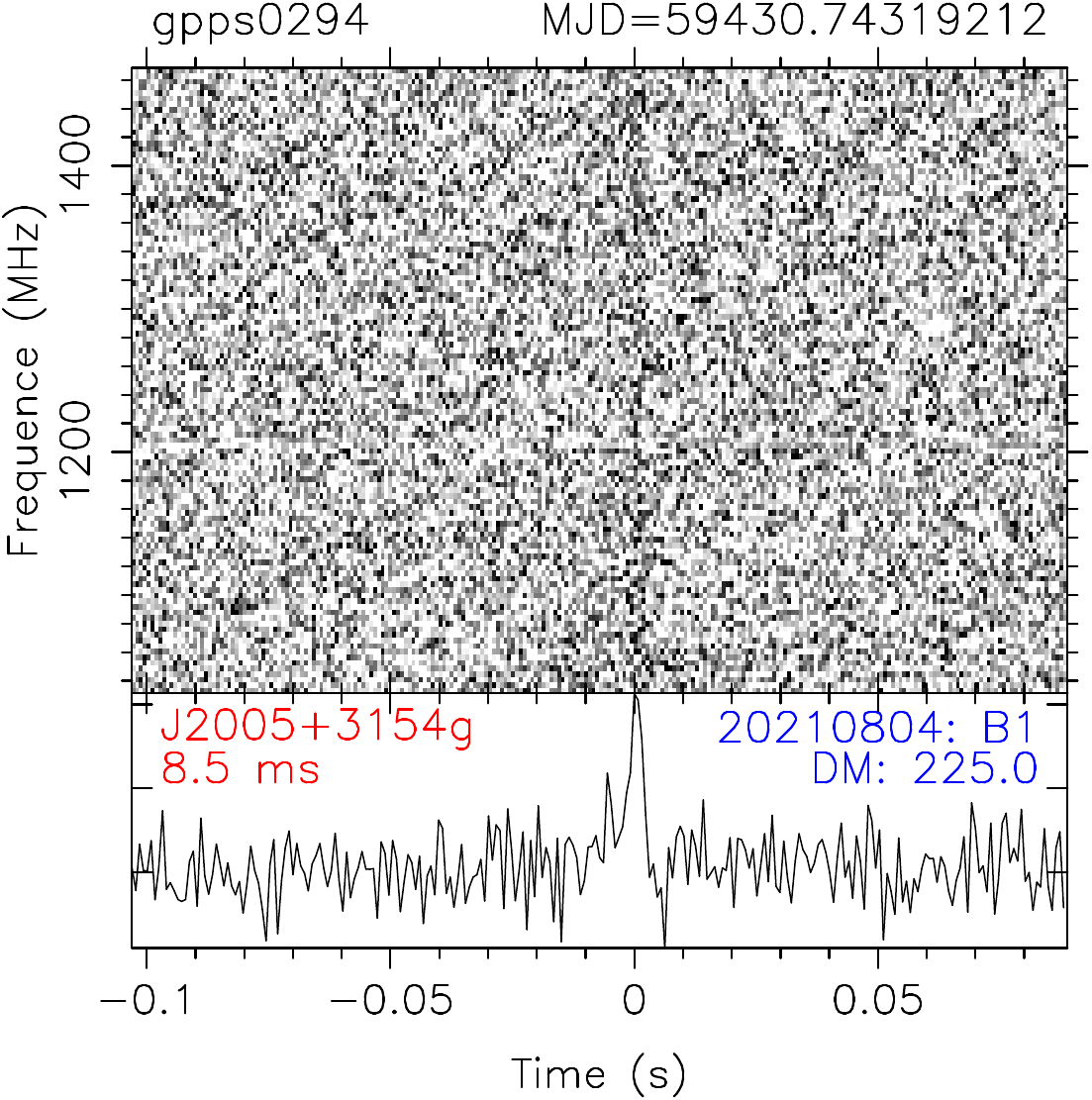}
\includegraphics[width=0.33\textwidth]{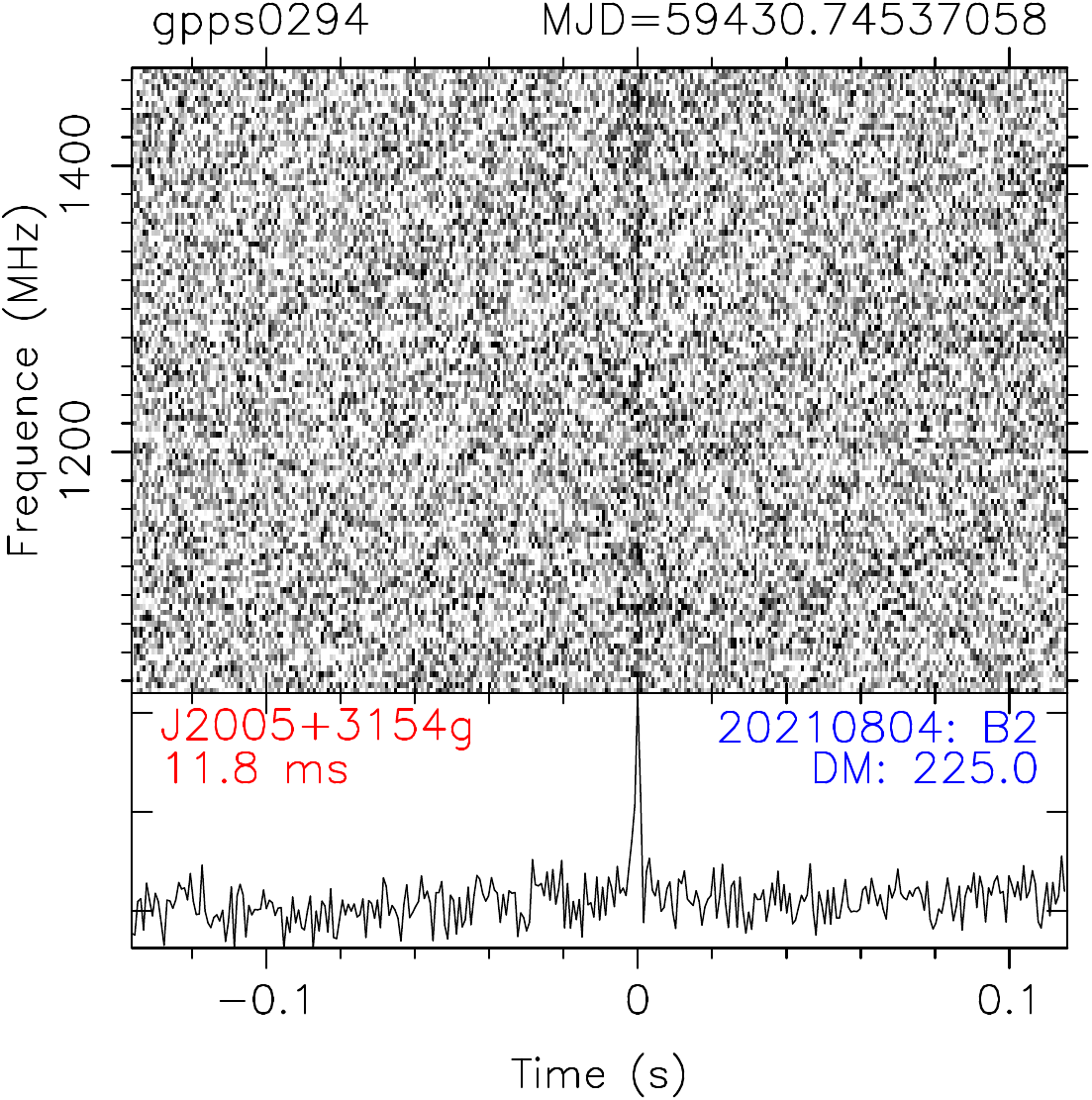}
\includegraphics[width=0.33\textwidth]{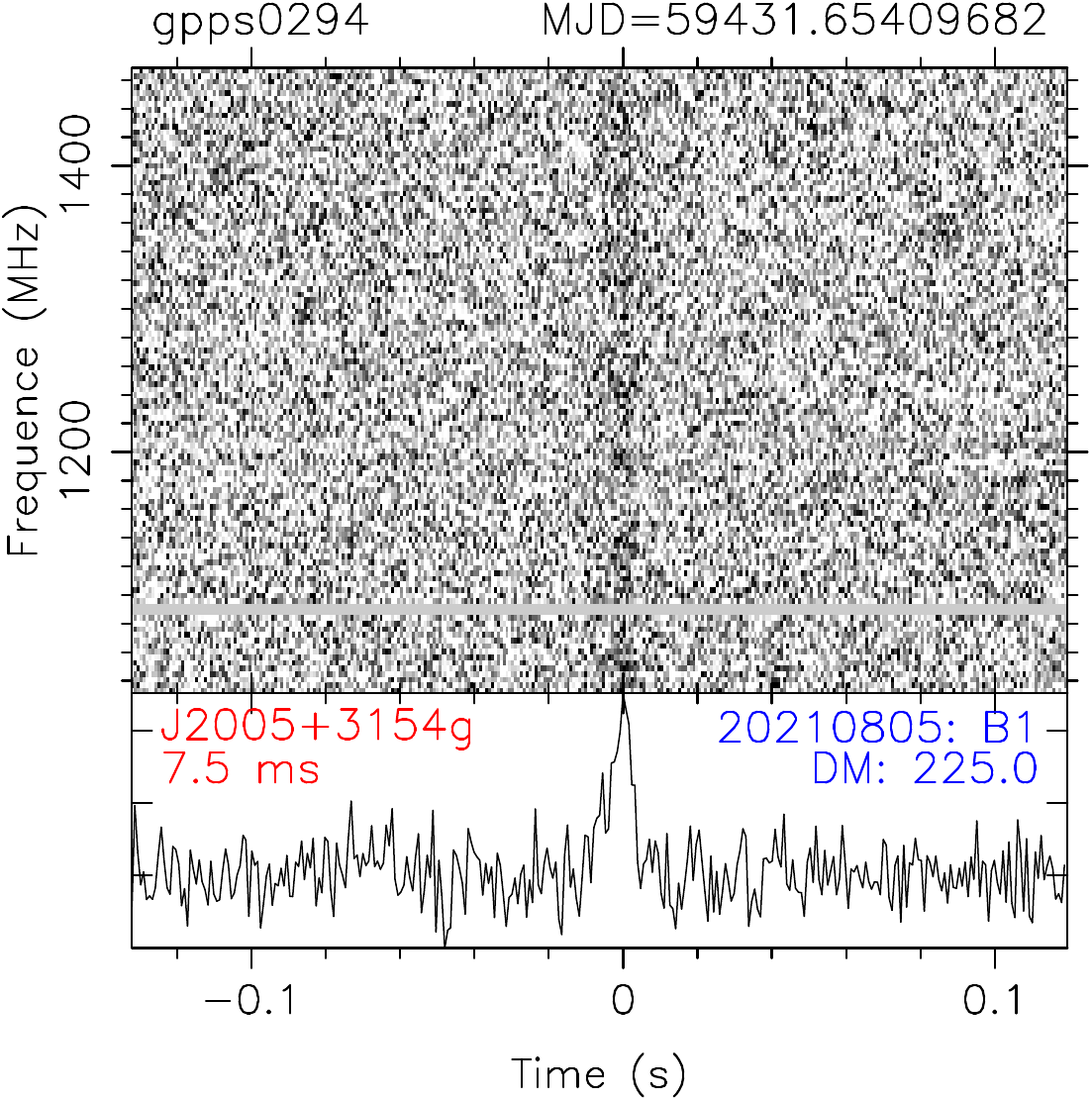}\\[0.2mm]
\includegraphics[width=0.33\textwidth]{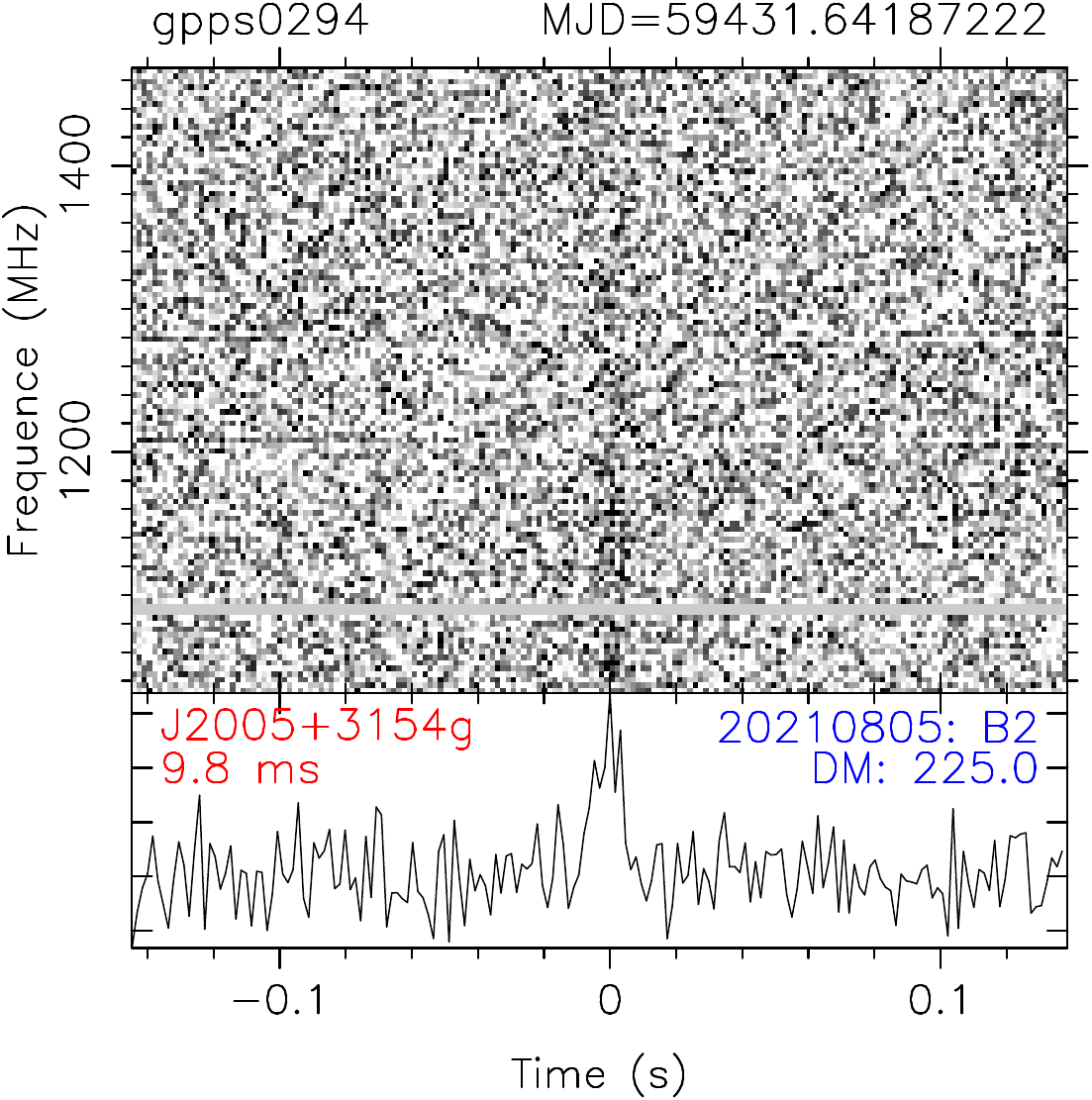}
\includegraphics[width=0.33\textwidth]{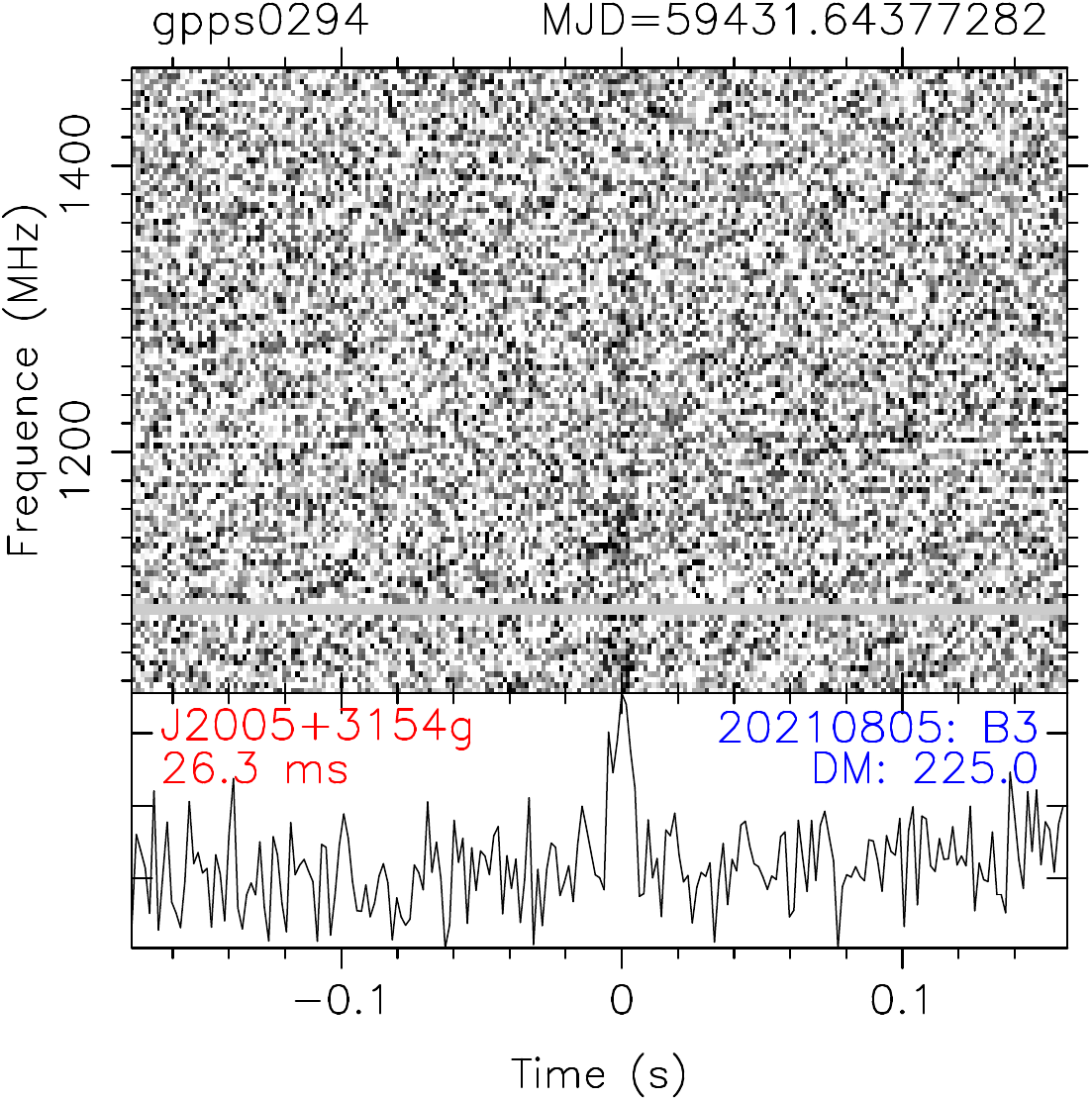}
\includegraphics[width=0.33\textwidth]{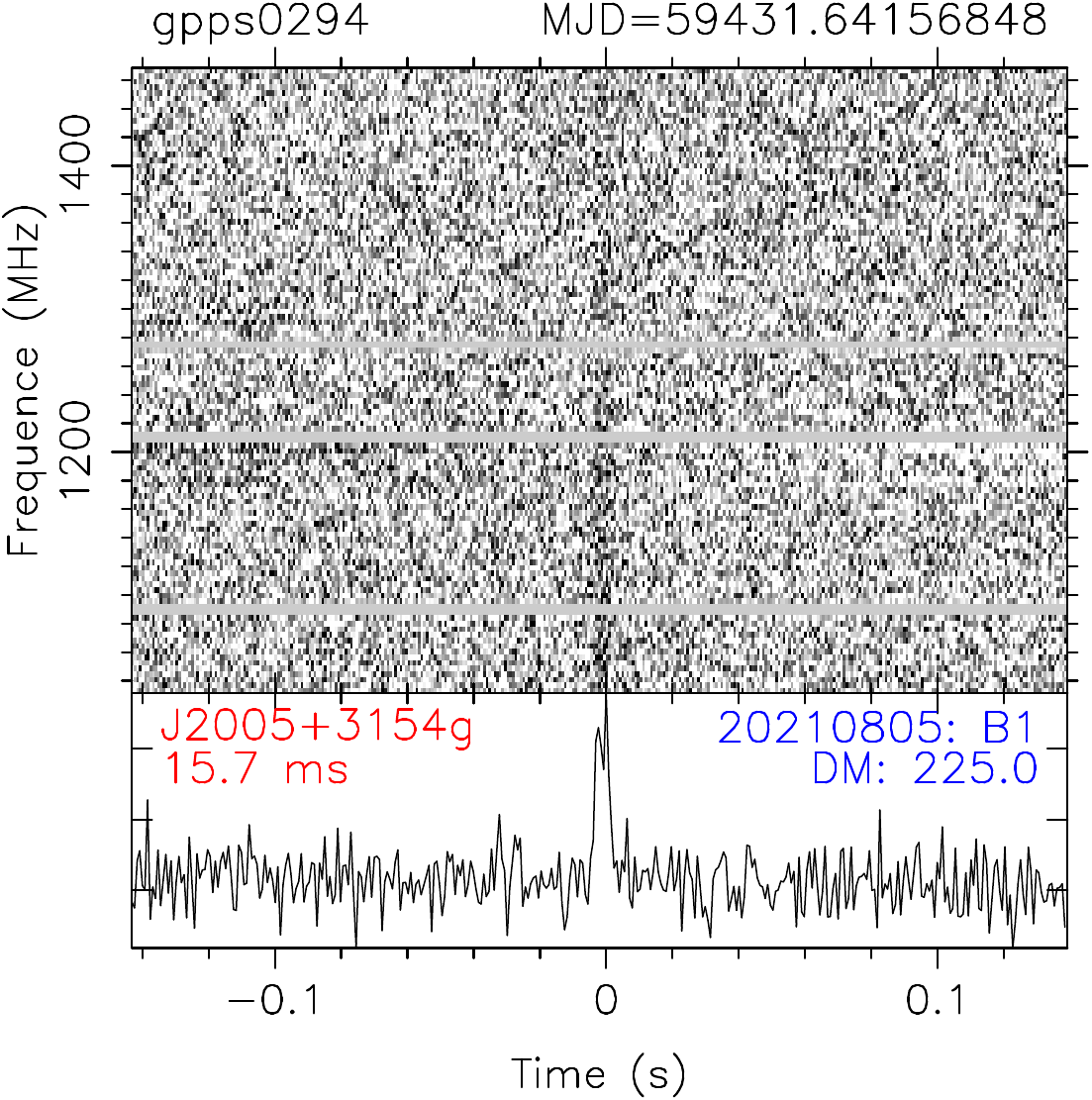}
\caption{(Continued.)}
\end{figure*}
\addtocounter{figure}{-1}
\begin{figure*}[!t]
\centering
\includegraphics[width=0.33\textwidth]{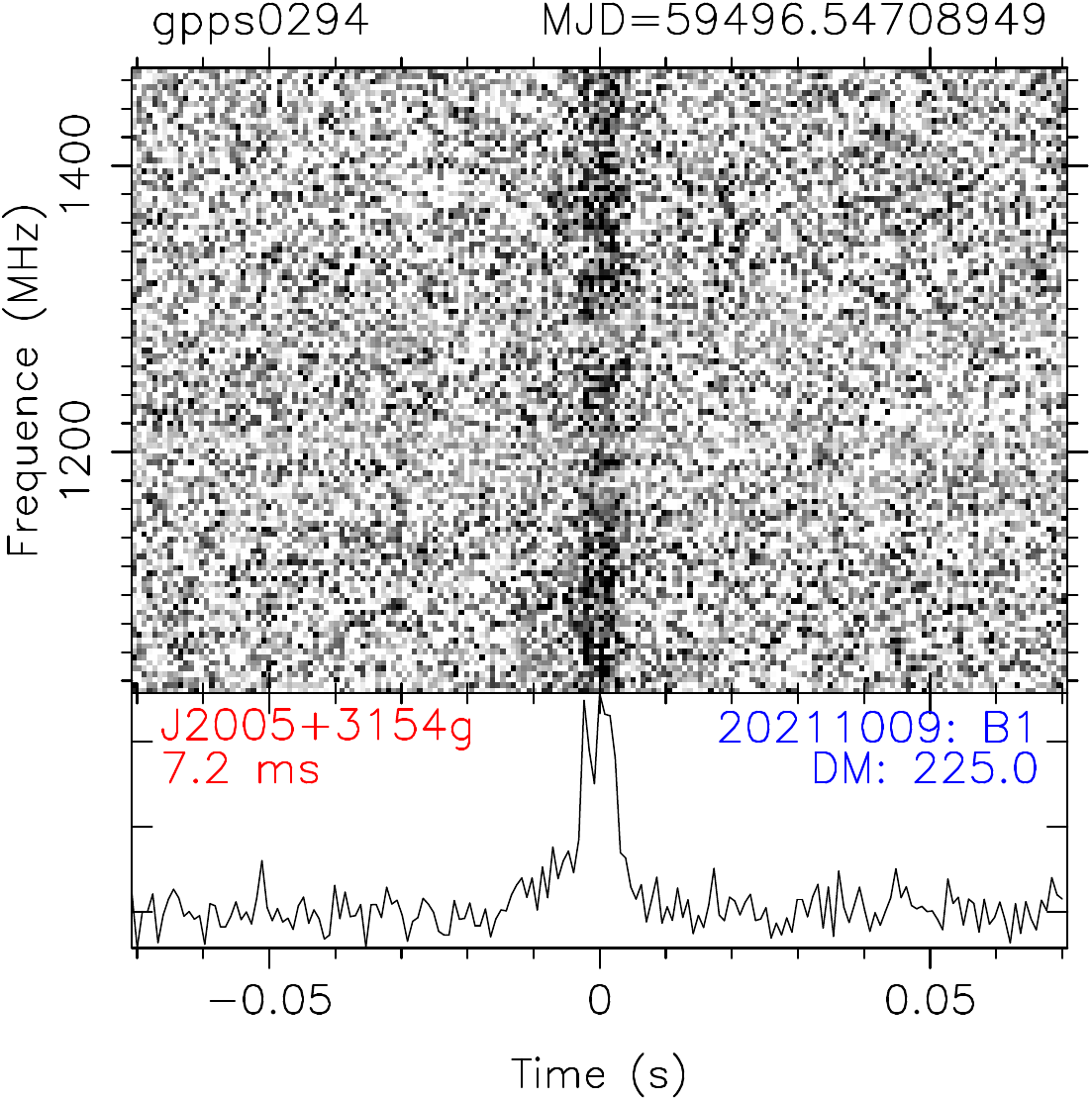}
\includegraphics[width=0.33\textwidth]{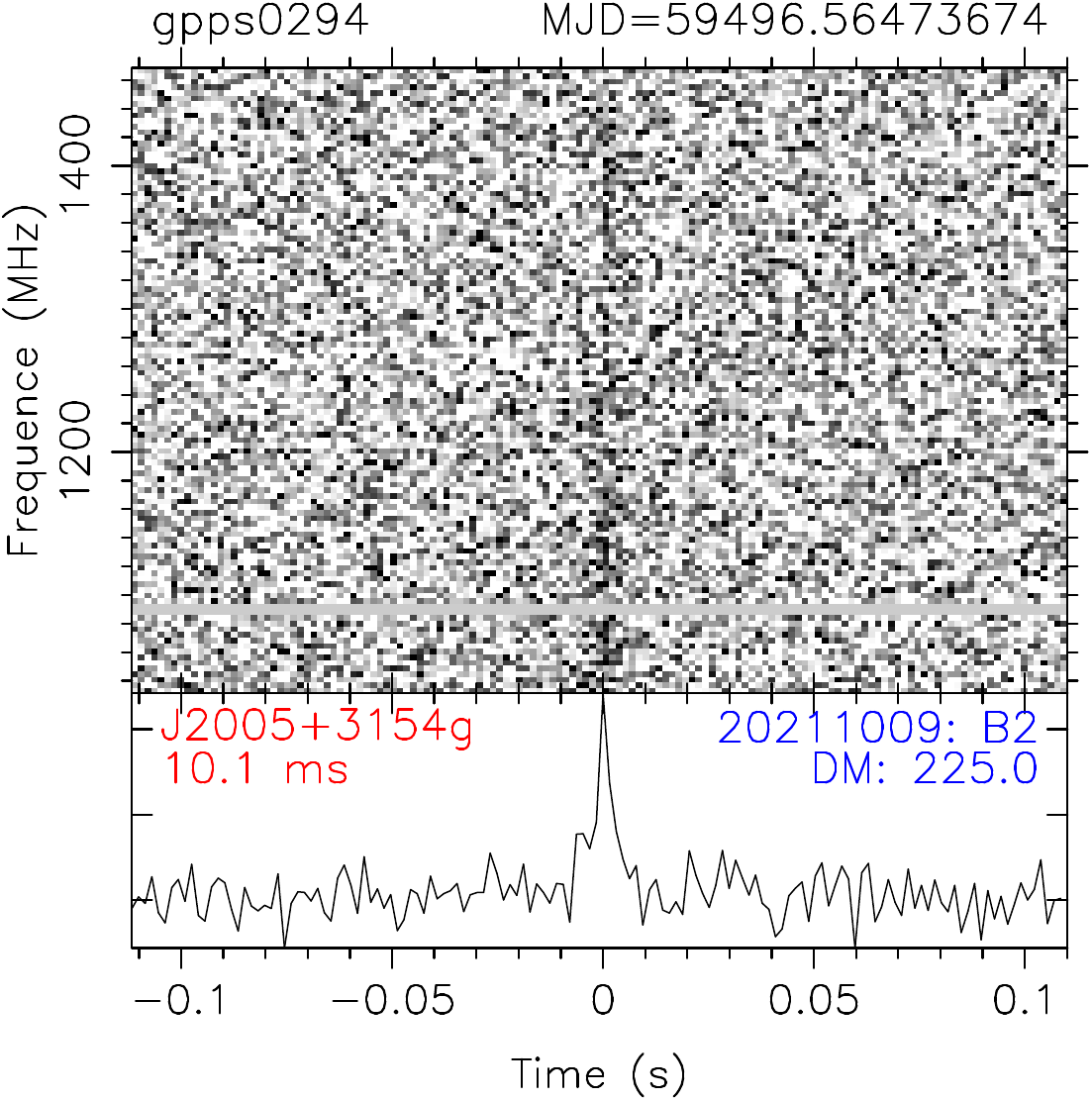}
\includegraphics[width=0.33\textwidth]{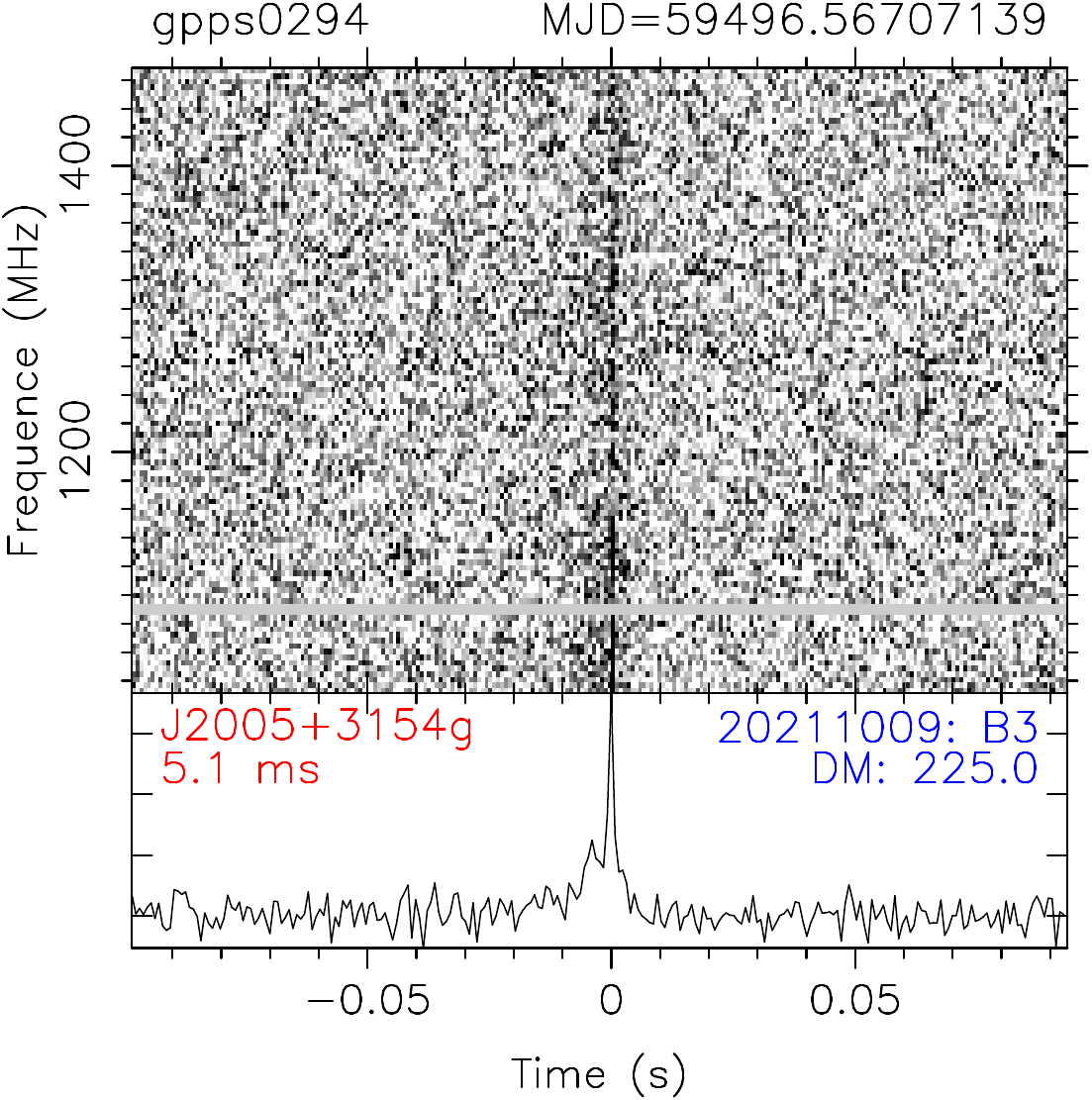}\\[0.2mm]
\includegraphics[width=0.33\textwidth]{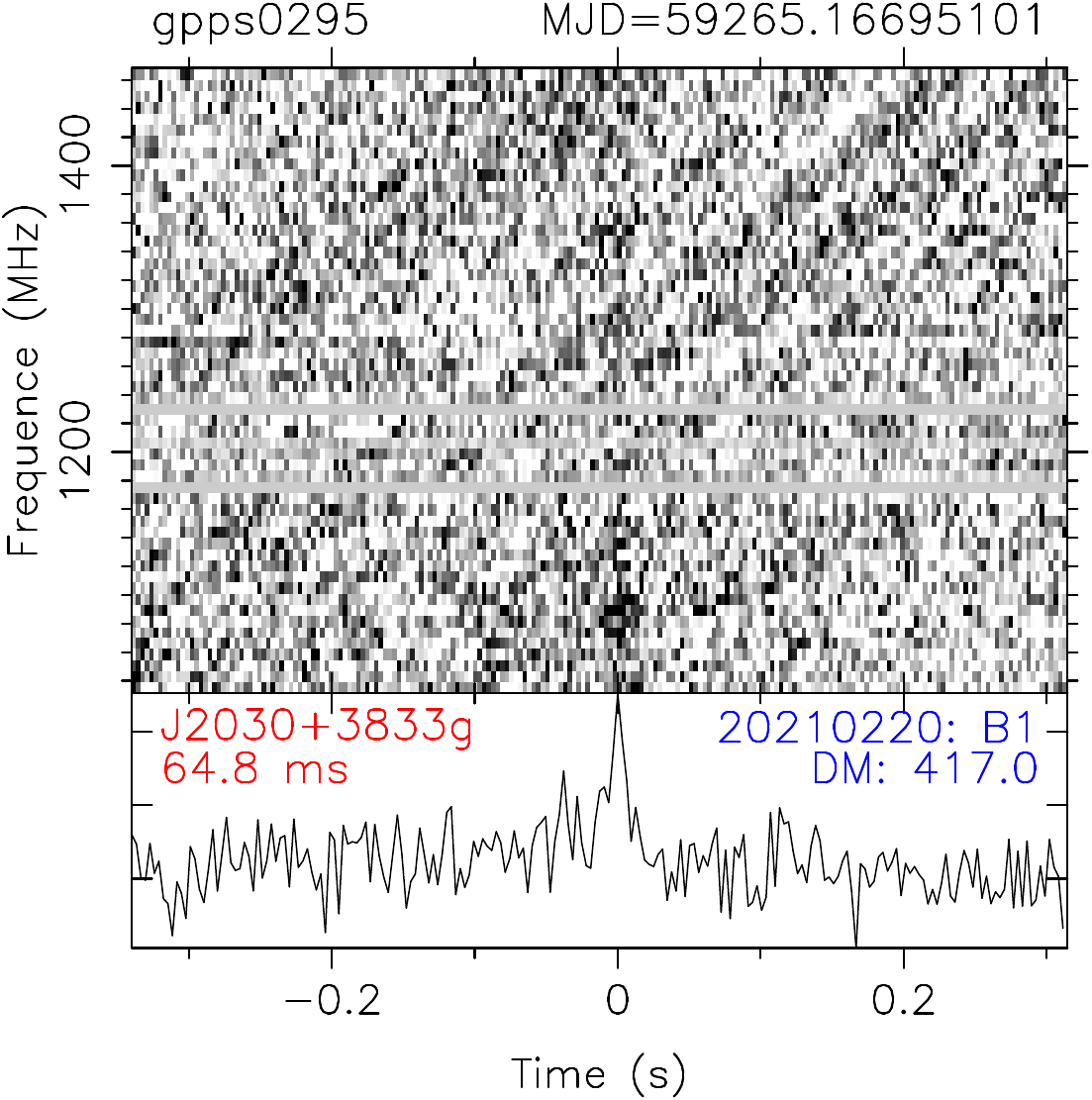}
\includegraphics[width=0.33\textwidth]{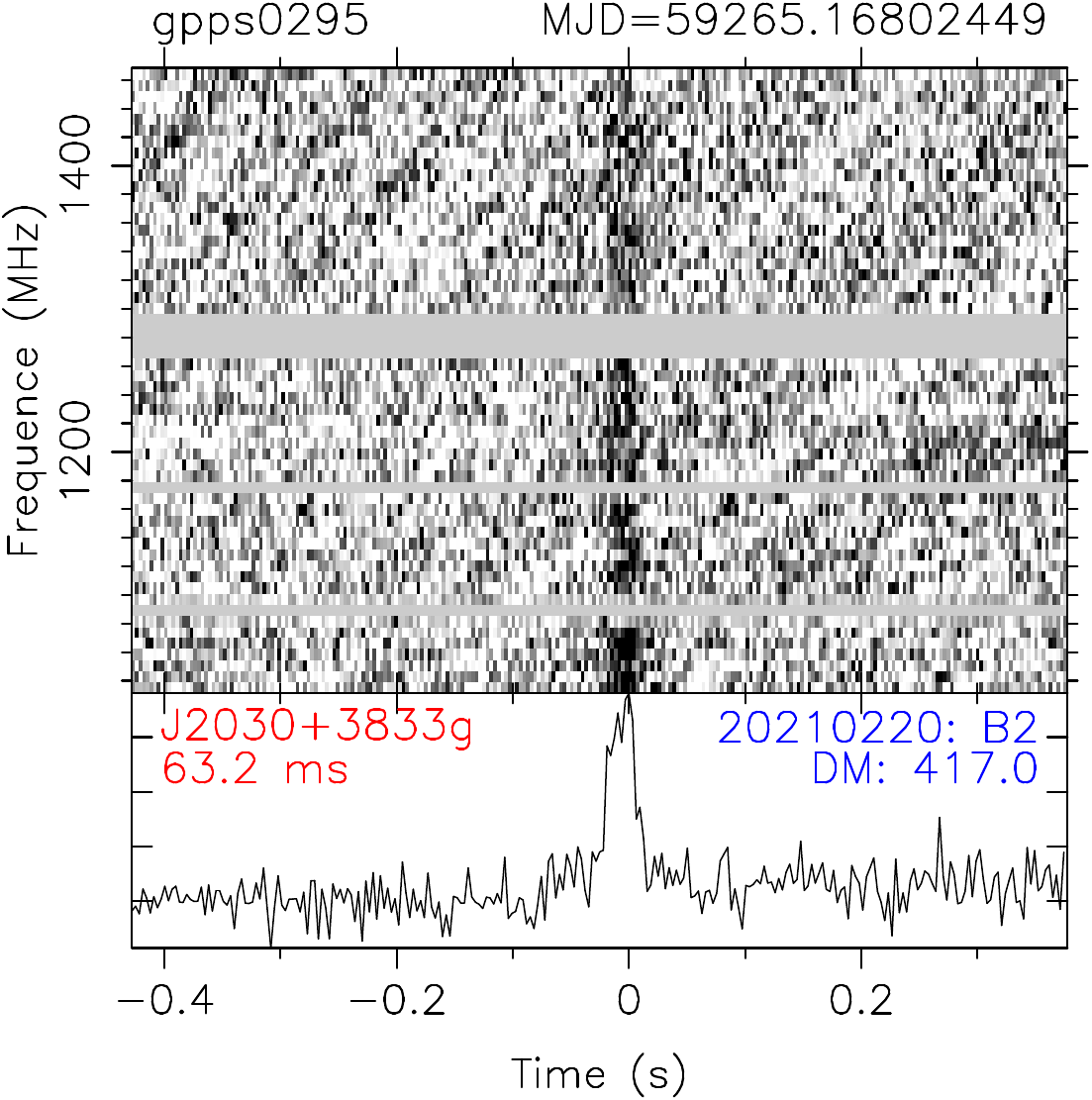}
\includegraphics[width=0.33\textwidth]{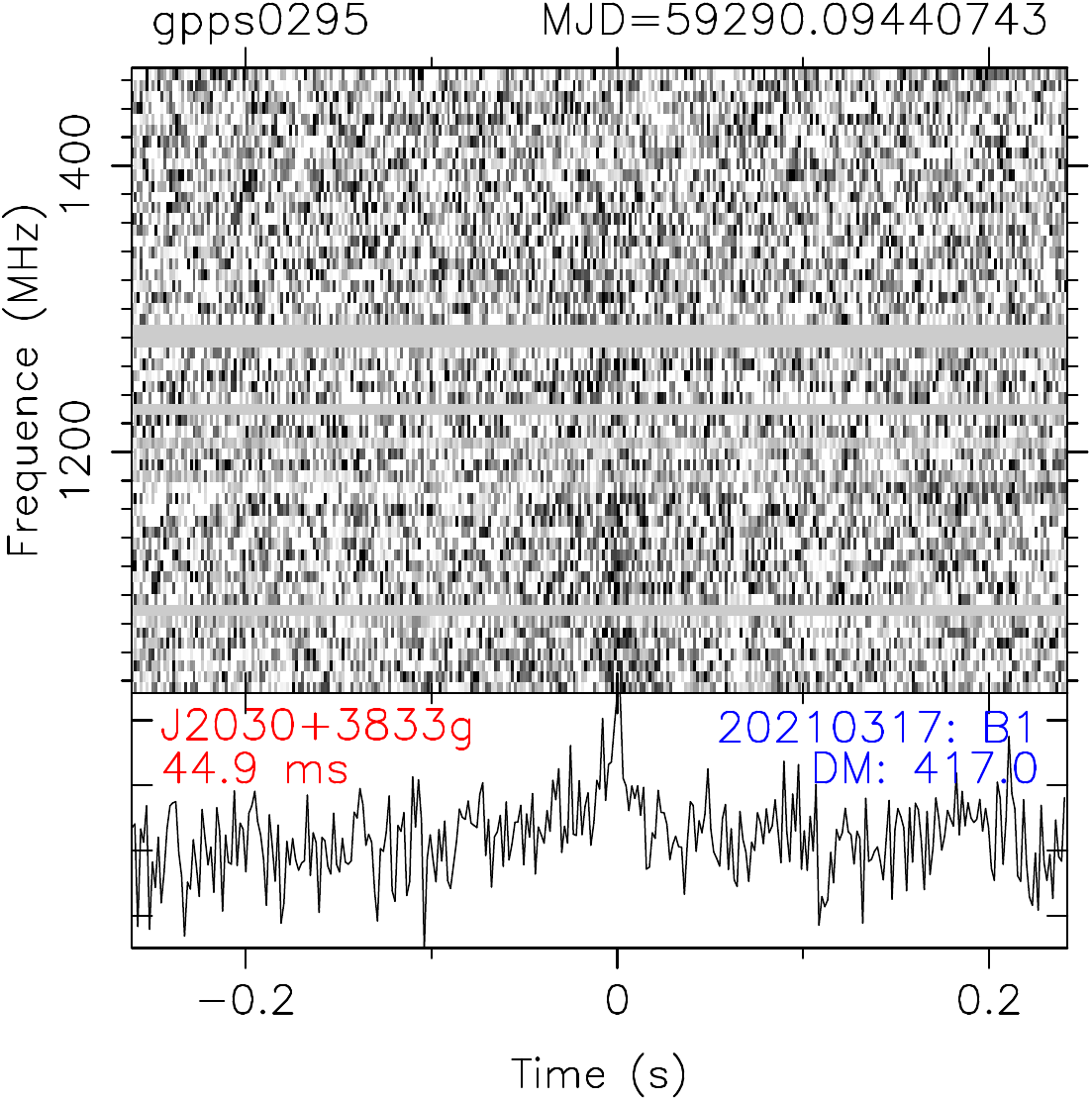}\\[0.2mm]
\includegraphics[width=0.33\textwidth]{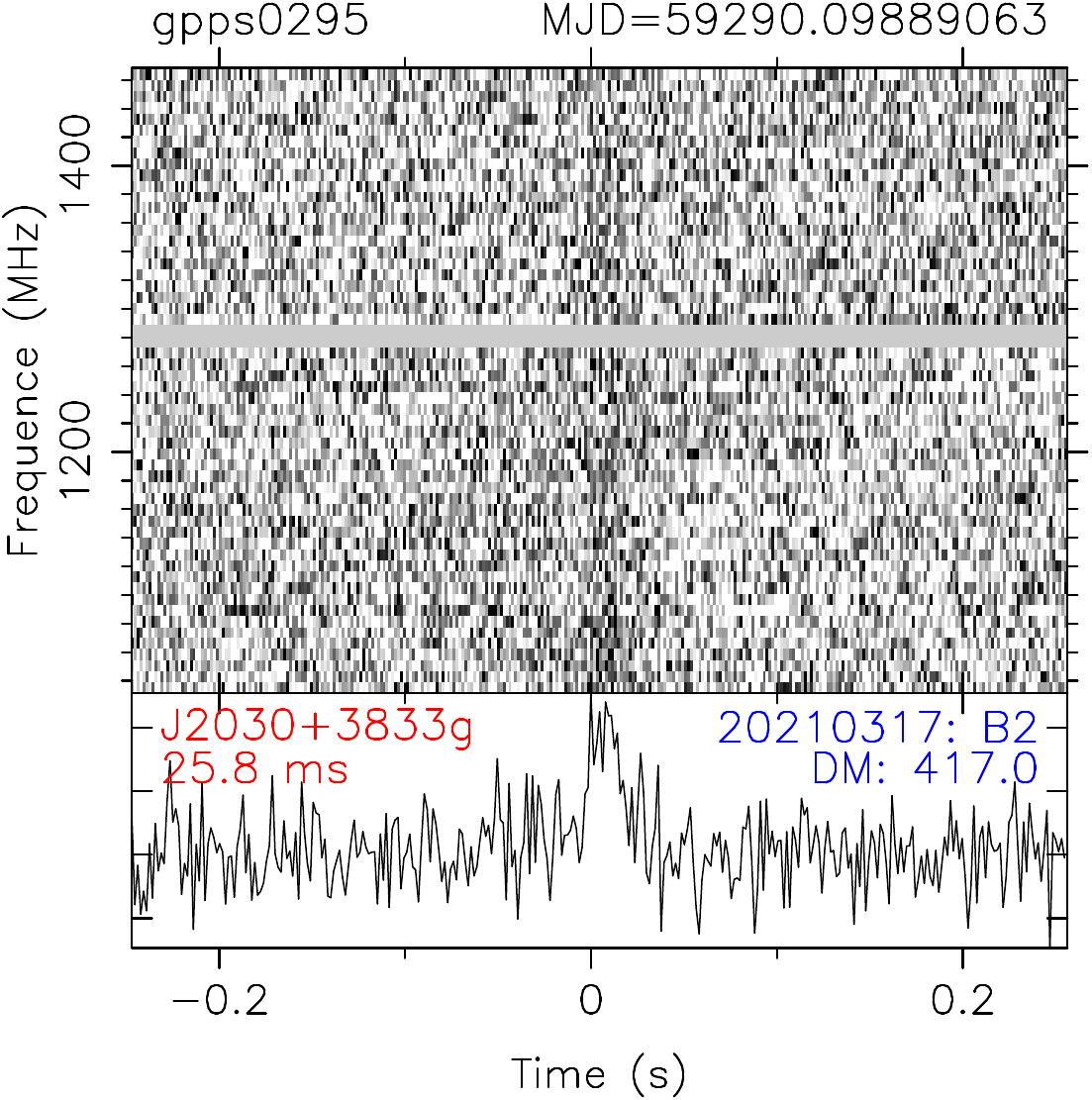}
\includegraphics[width=0.33\textwidth]{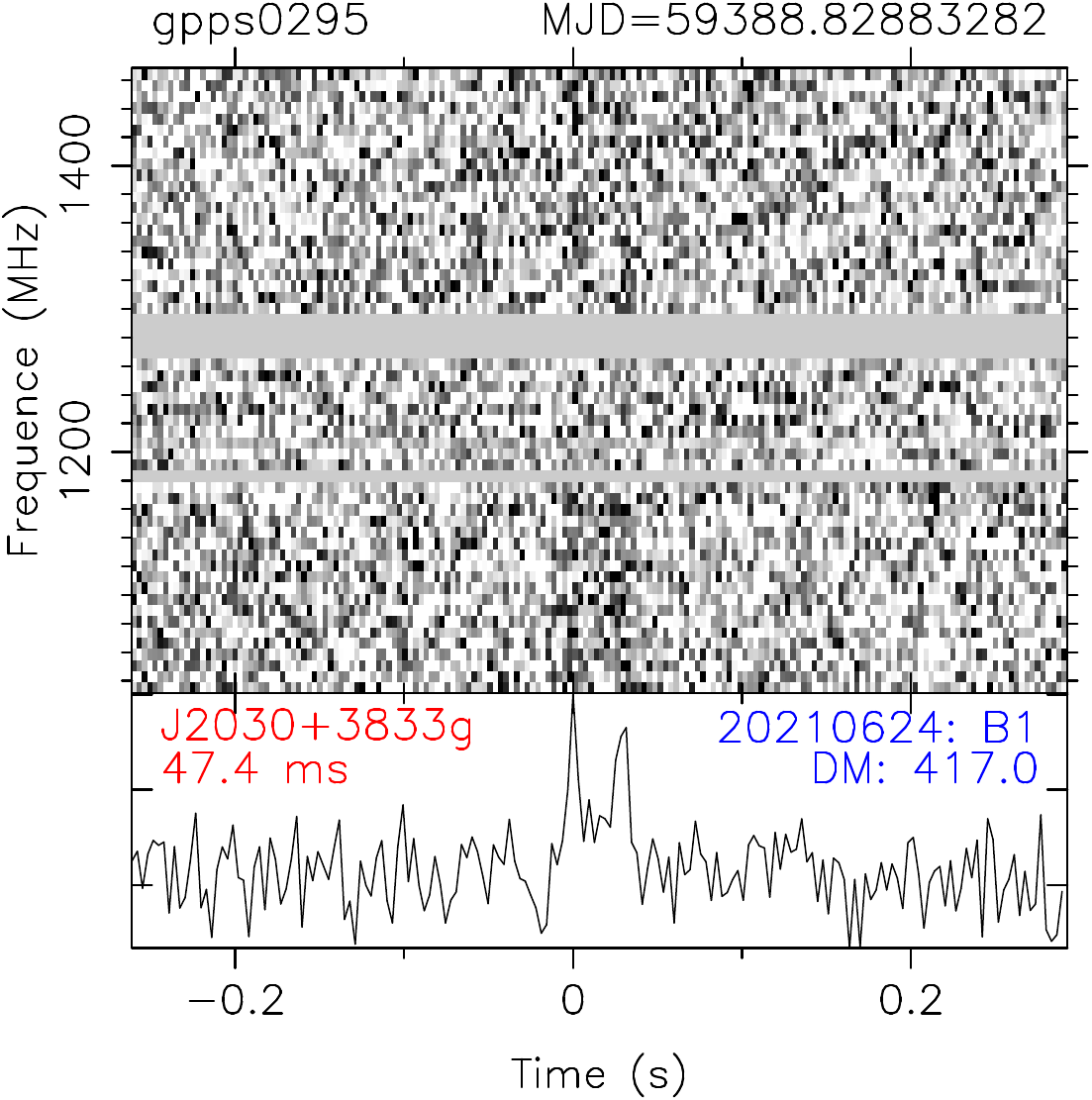}
\includegraphics[width=0.33\textwidth]{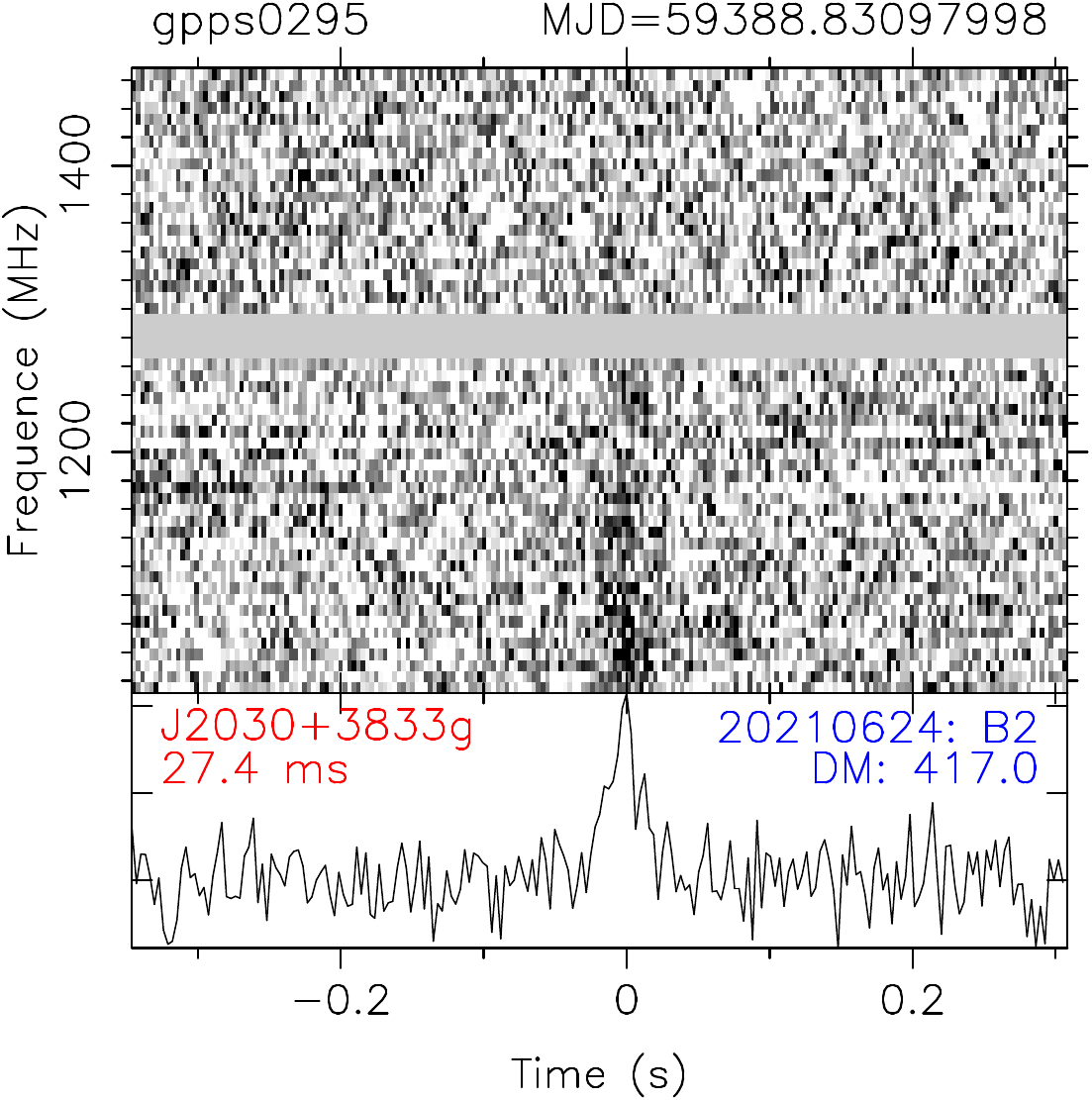}
\caption{(Continued.)}
\end{figure*}

\begin{figure*}
  \centering
    \includegraphics[width=0.33\textwidth]{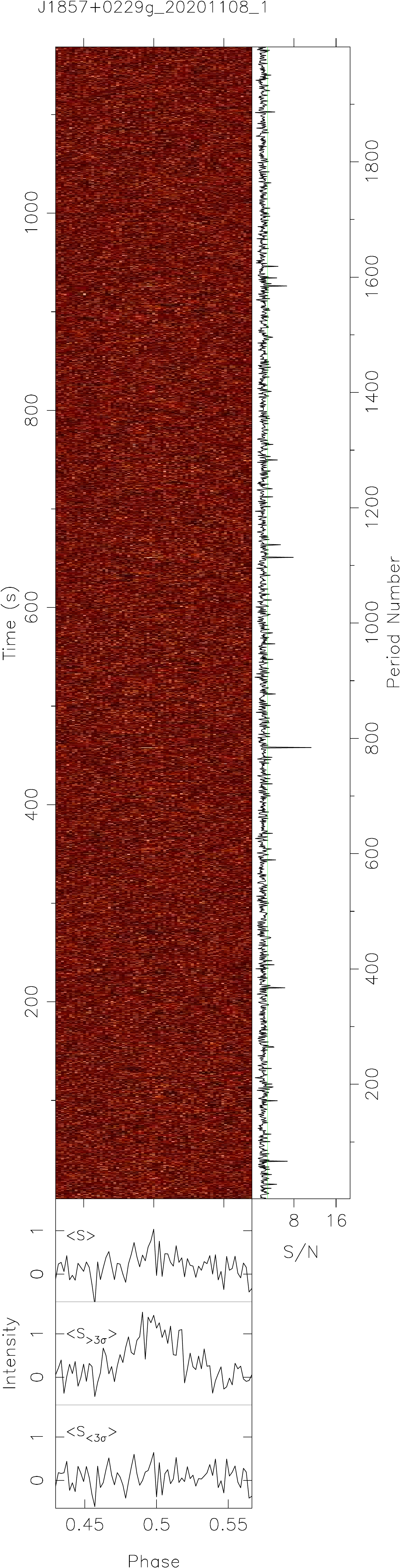}
    \includegraphics[width=0.33\textwidth]{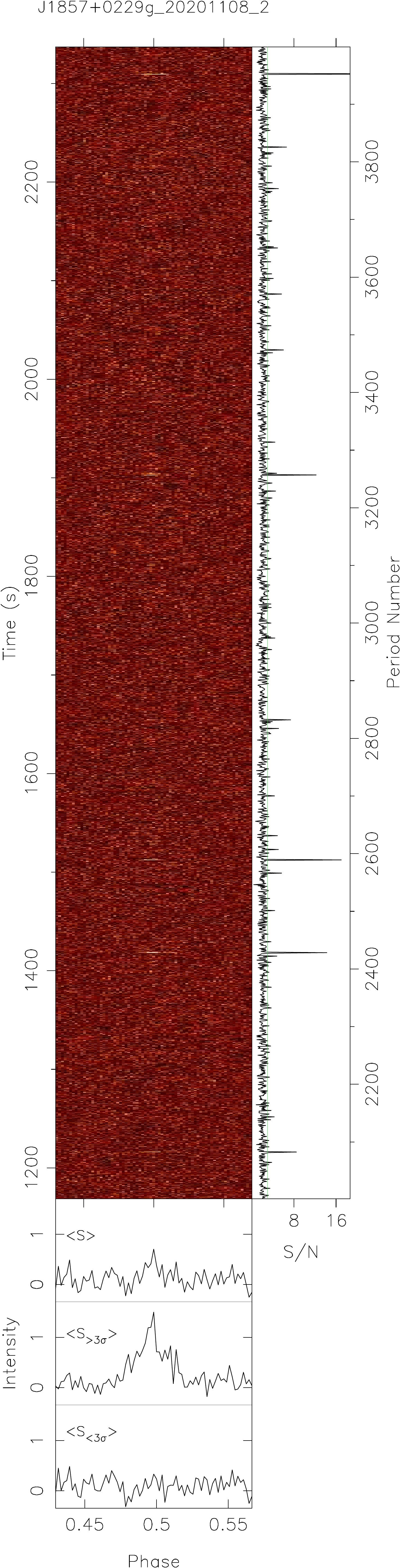}
    \includegraphics[width=0.33\textwidth]{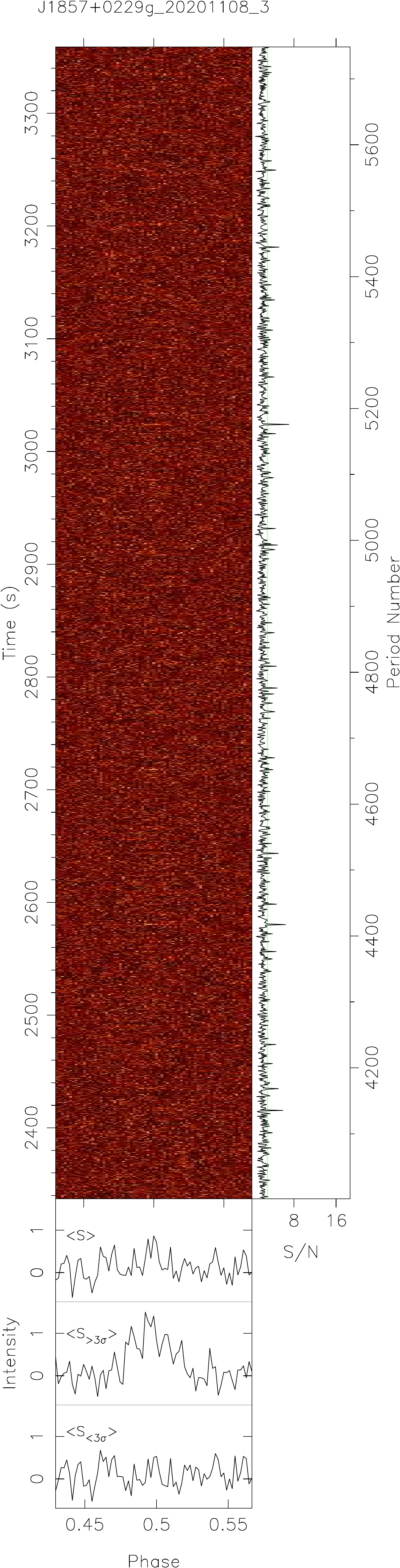} 
  \caption{The same as Figure~\ref{fig:newRRATs} but for 16 proto-RRATs discovered in the GPPS survey.}
  \label{fig:AppnewRRATs}
\end{figure*}
\addtocounter{figure}{-1}
\begin{figure*}
  \centering
    \includegraphics[width=0.33\textwidth]{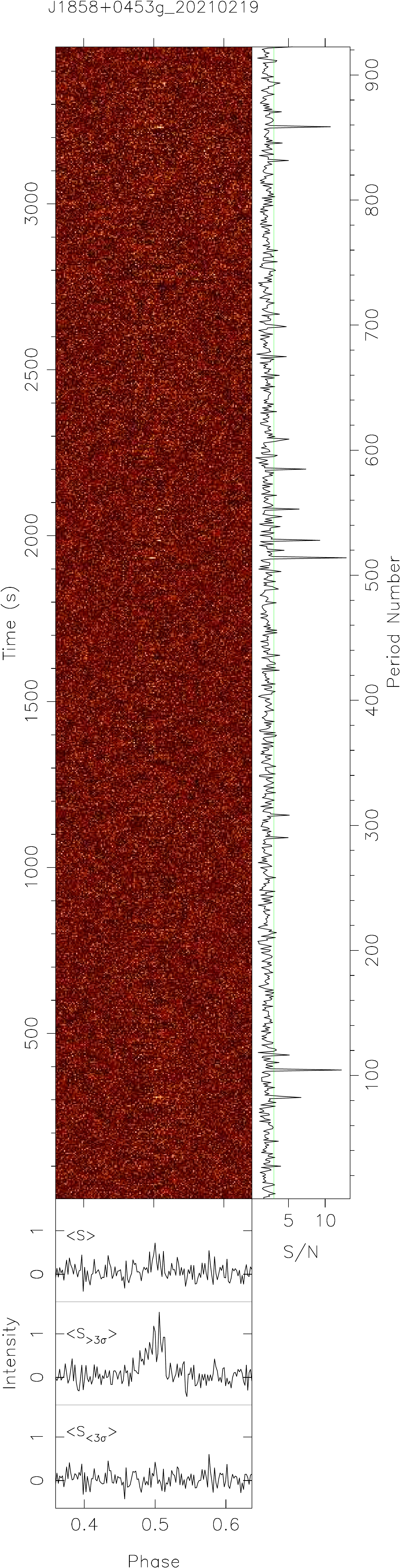} 
    \includegraphics[width=0.33\textwidth]{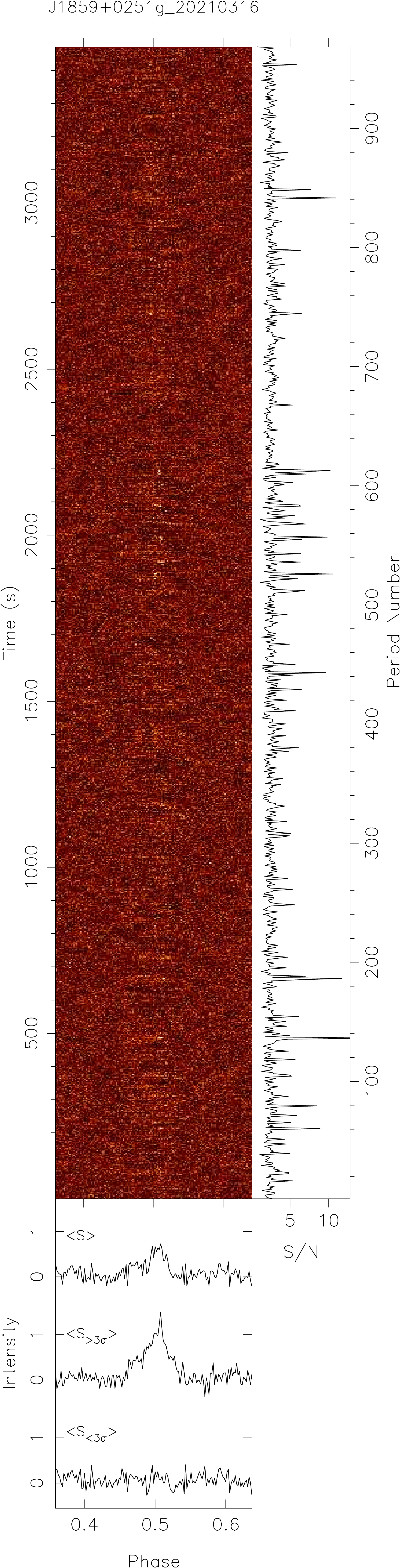}
    \includegraphics[width=0.33\textwidth]{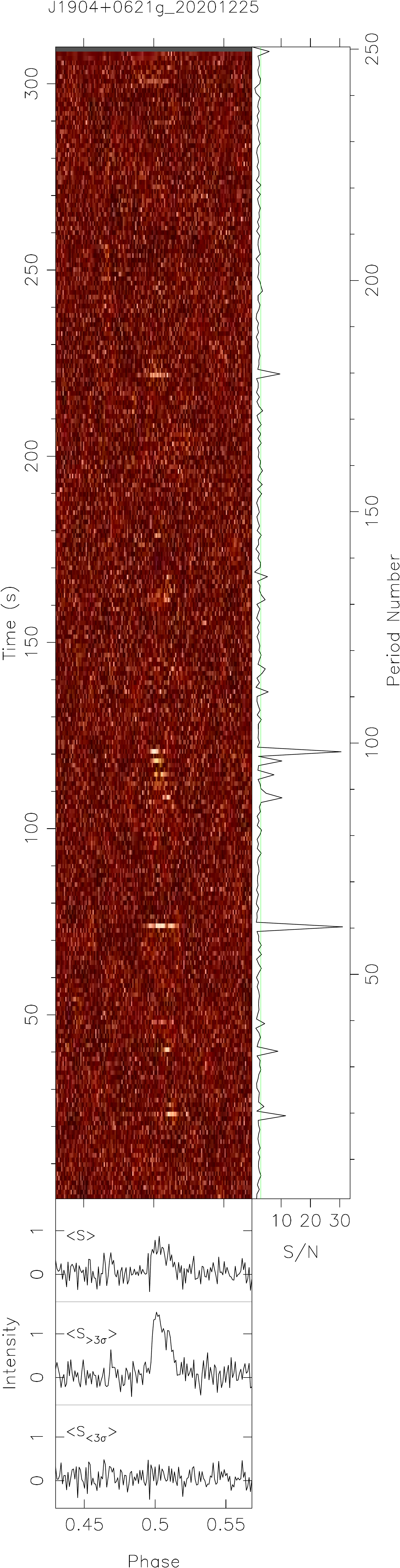}
  \caption{{\it -- continued}.}
\end{figure*}
\addtocounter{figure}{-1}
\begin{figure*}
  \centering 
    \includegraphics[width=0.33\textwidth]{New-figs/J190455+062136/J190455+062136sp_20210318_tracking-M01-P1-c2048b1.zapFp.debased.pdf} 
    \includegraphics[width=0.33\textwidth]{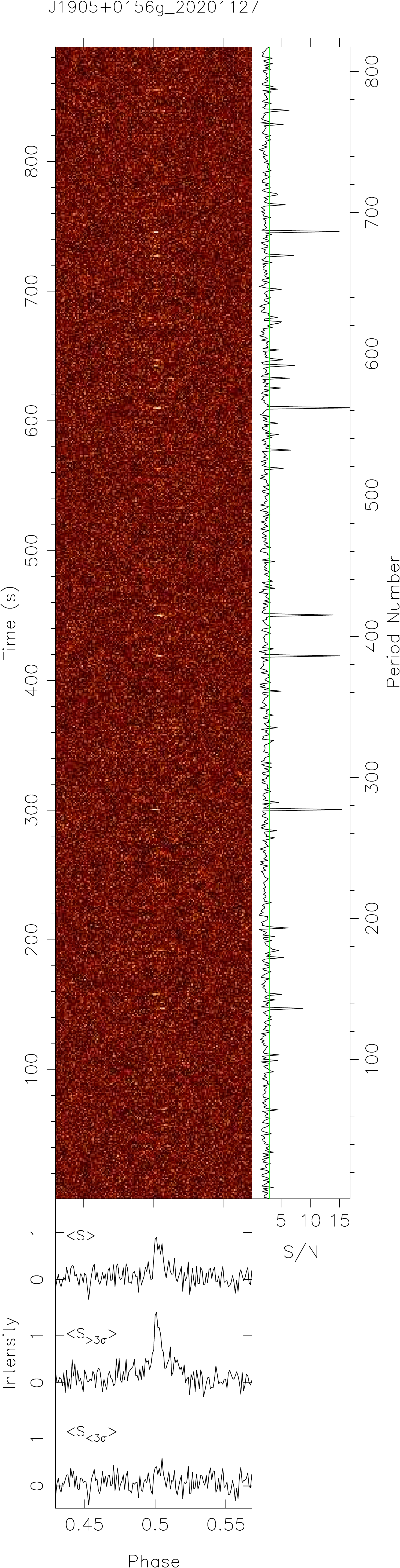} 
    \includegraphics[width=0.33\textwidth]{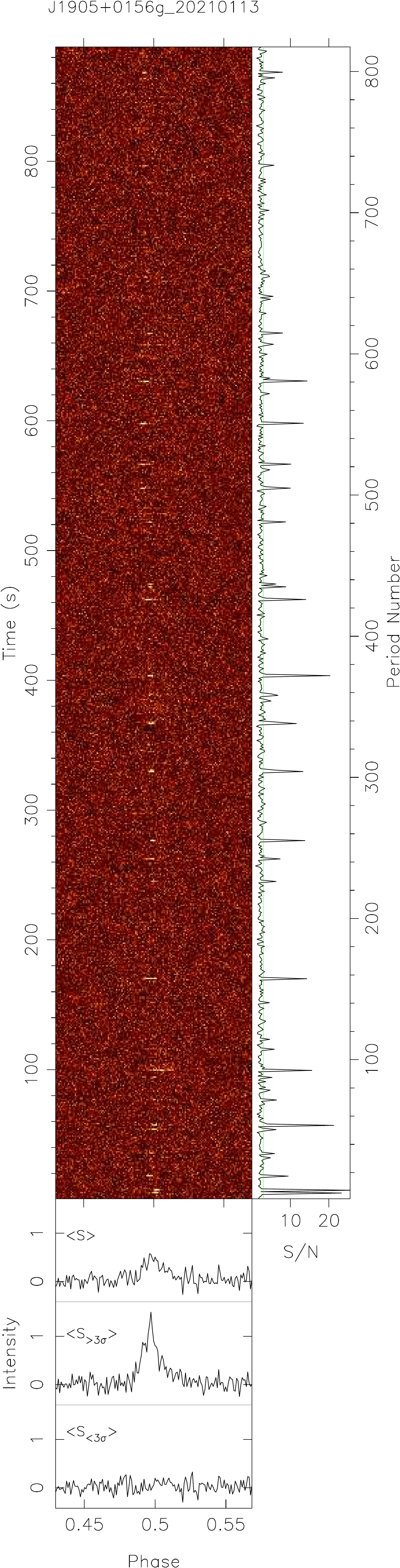} 
  \caption{{\it -- continued}.}
\end{figure*}
\addtocounter{figure}{-1}
\begin{figure*}
  \centering
    \includegraphics[width=0.33\textwidth]{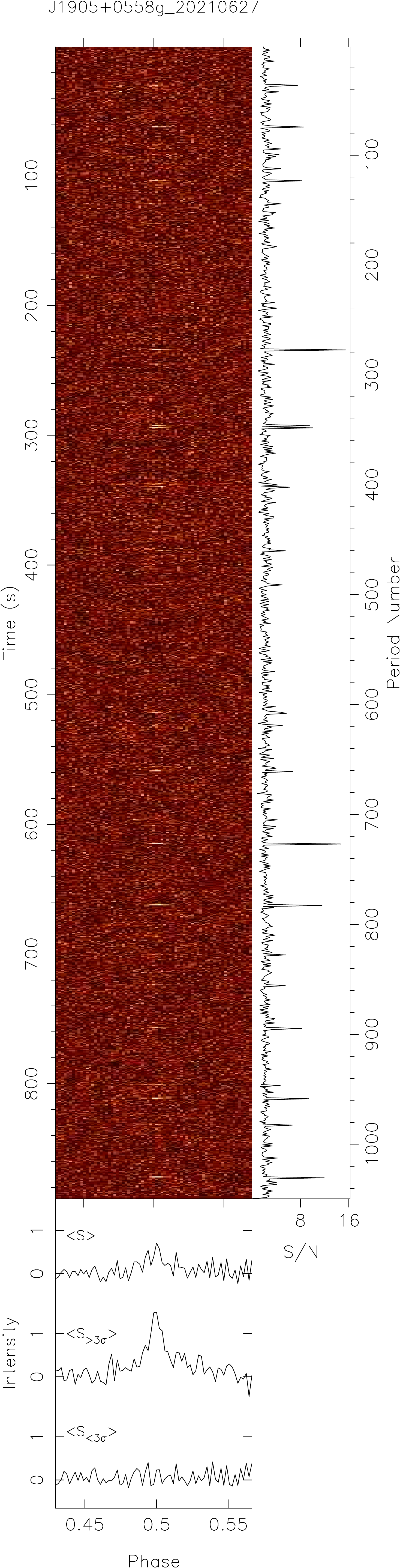}
    \includegraphics[width=0.33\textwidth]{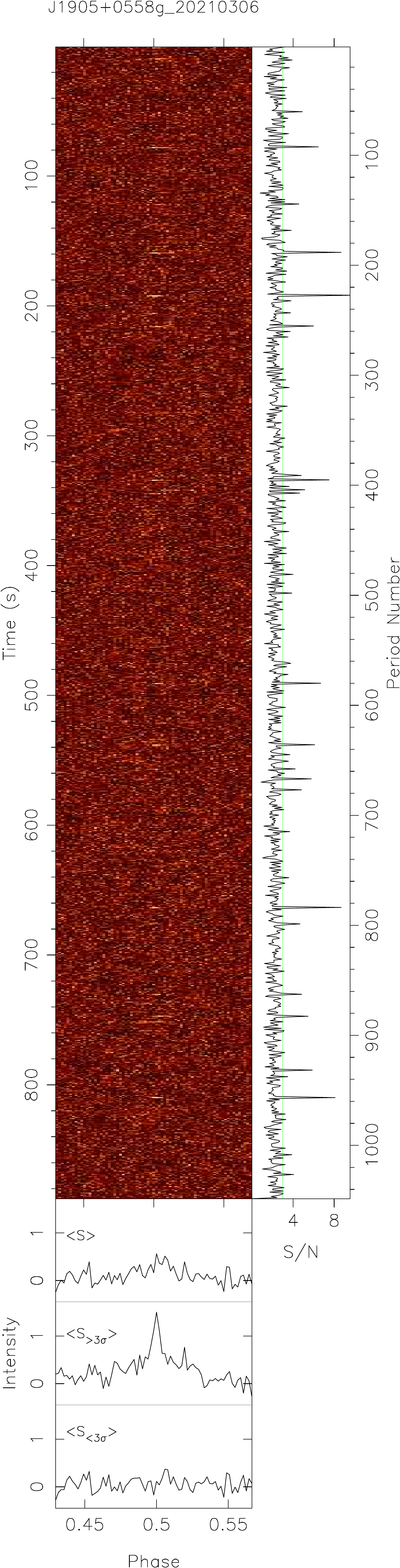}
    \includegraphics[width=0.33\textwidth]{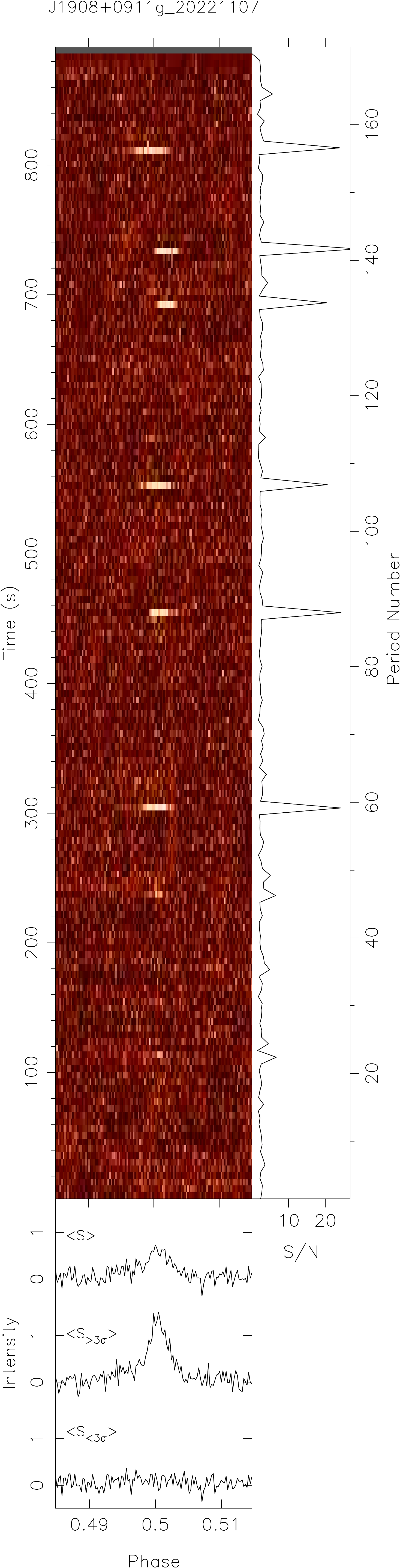} 
  \caption{{\it -- continued}.}
\end{figure*}
\addtocounter{figure}{-1}
\begin{figure*}
  \centering
    \includegraphics[width=0.33\textwidth]{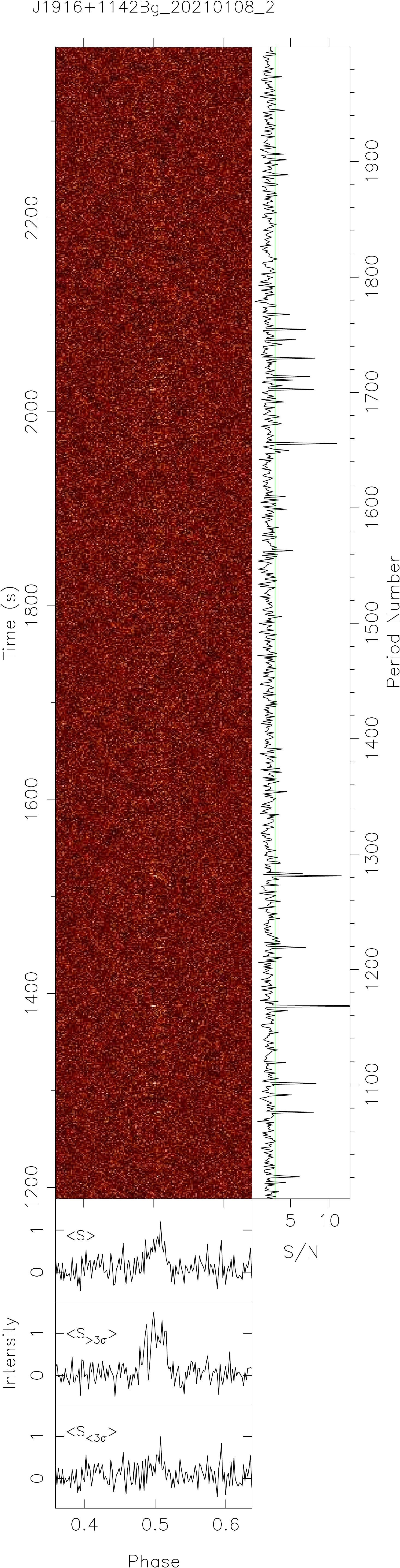}
    \includegraphics[width=0.33\textwidth]{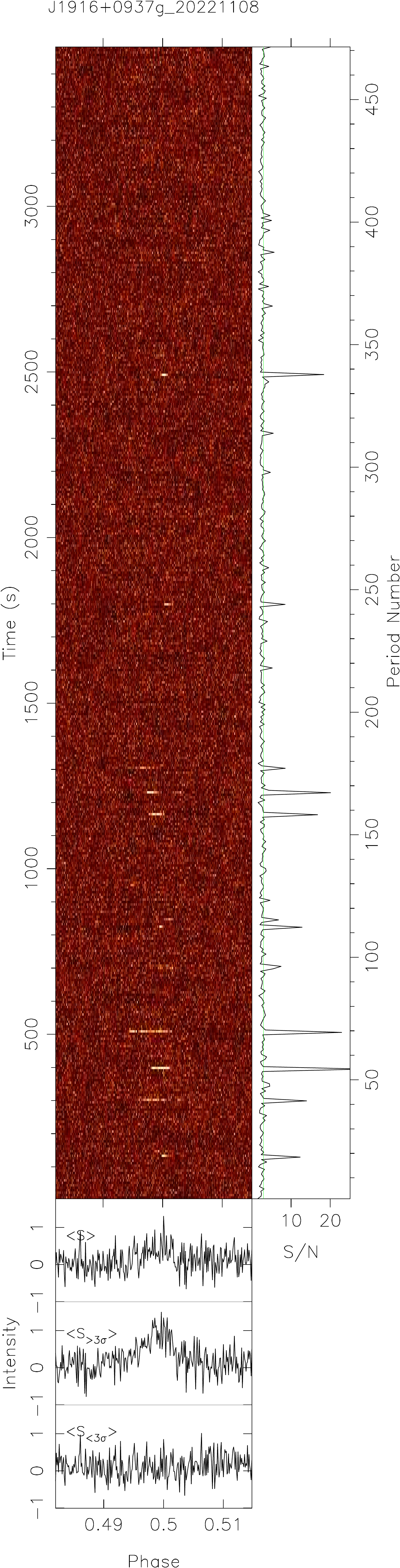} 
    \includegraphics[width=0.33\textwidth]{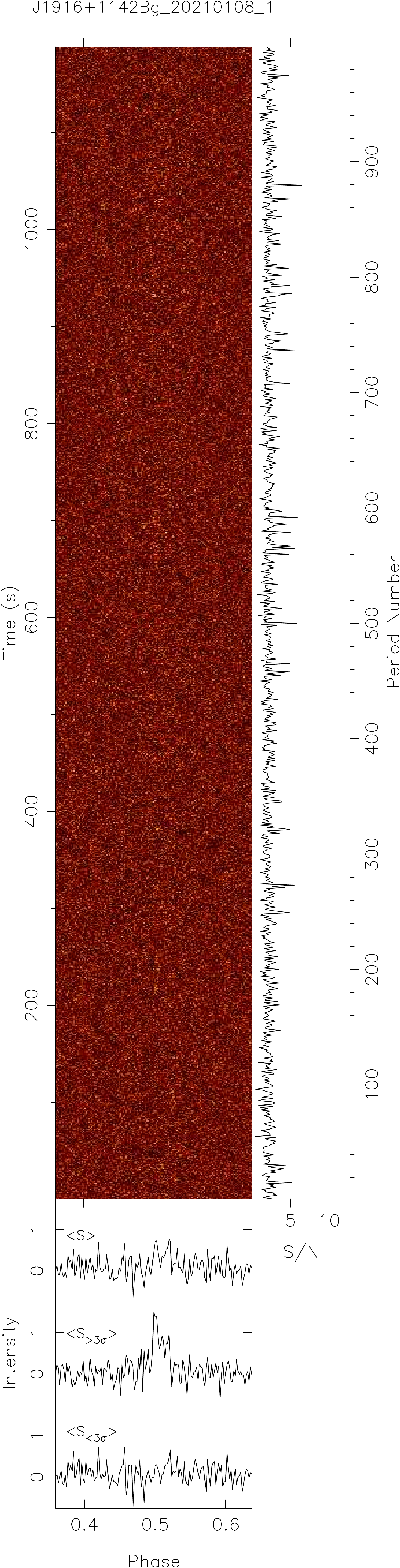}
  \caption{{\it -- continued}.}
\end{figure*}
\addtocounter{figure}{-1}
\begin{figure*}
\centering
    \includegraphics[width=0.33\textwidth]{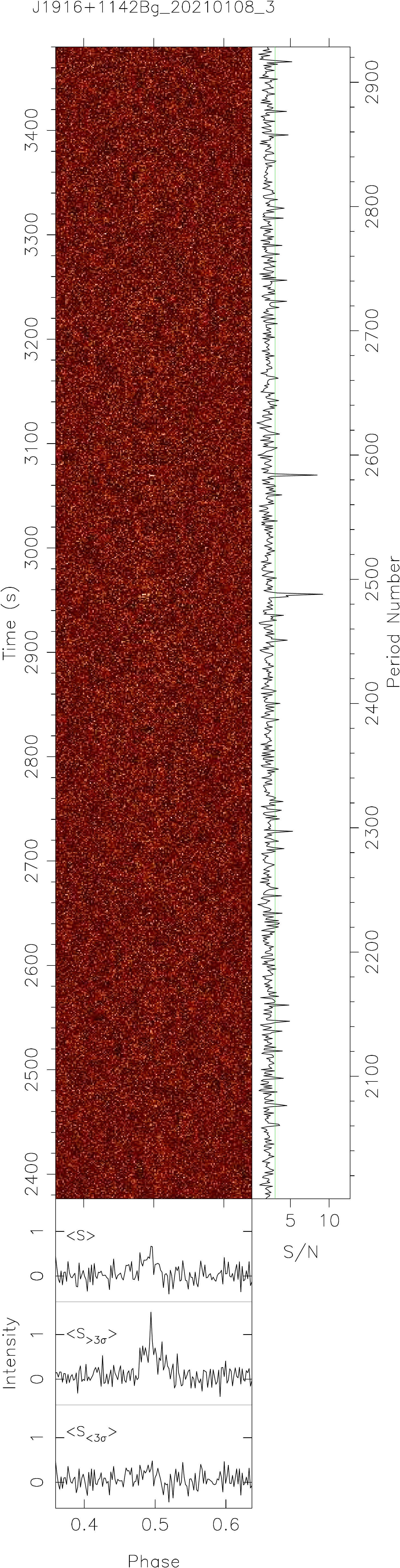} 
    \includegraphics[width=0.33\textwidth]{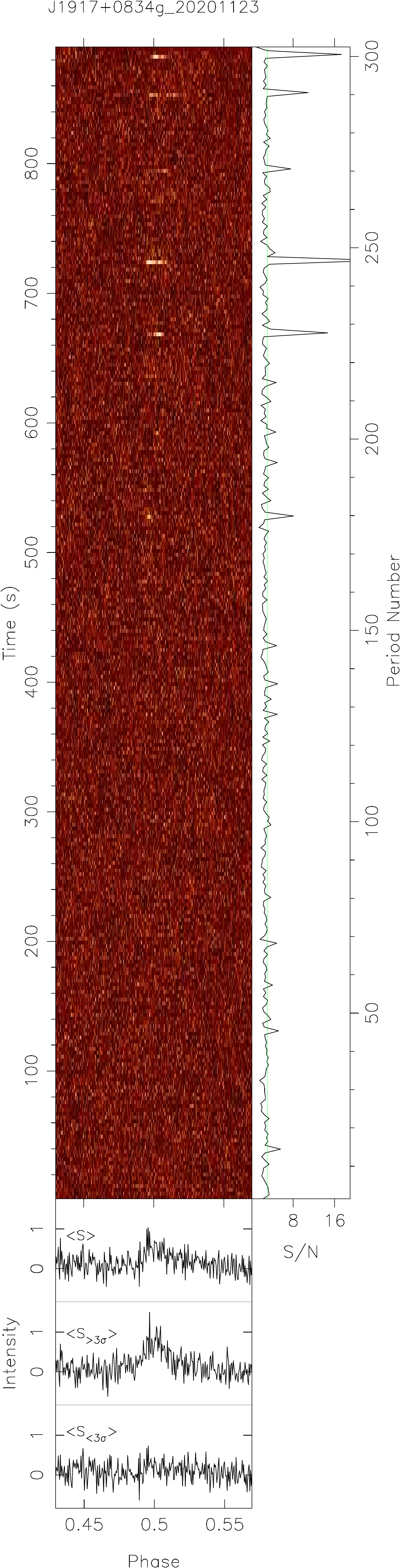} 
    \includegraphics[width=0.33\textwidth]{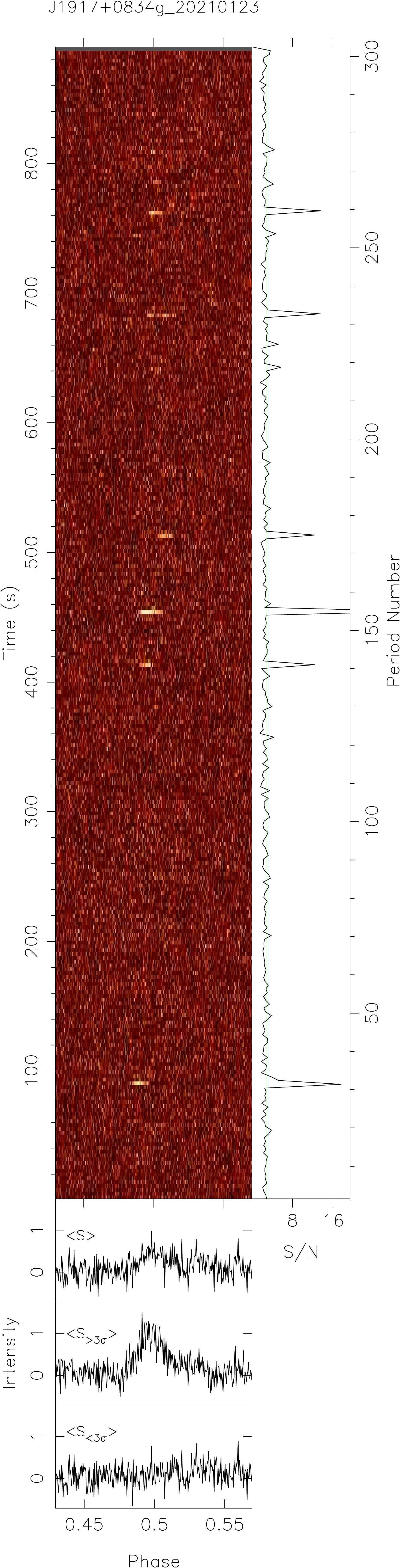}
  \caption{{\it -- continued}.}
\end{figure*}
\addtocounter{figure}{-1}
\begin{figure*}
\centering
    \includegraphics[width=0.33\textwidth]{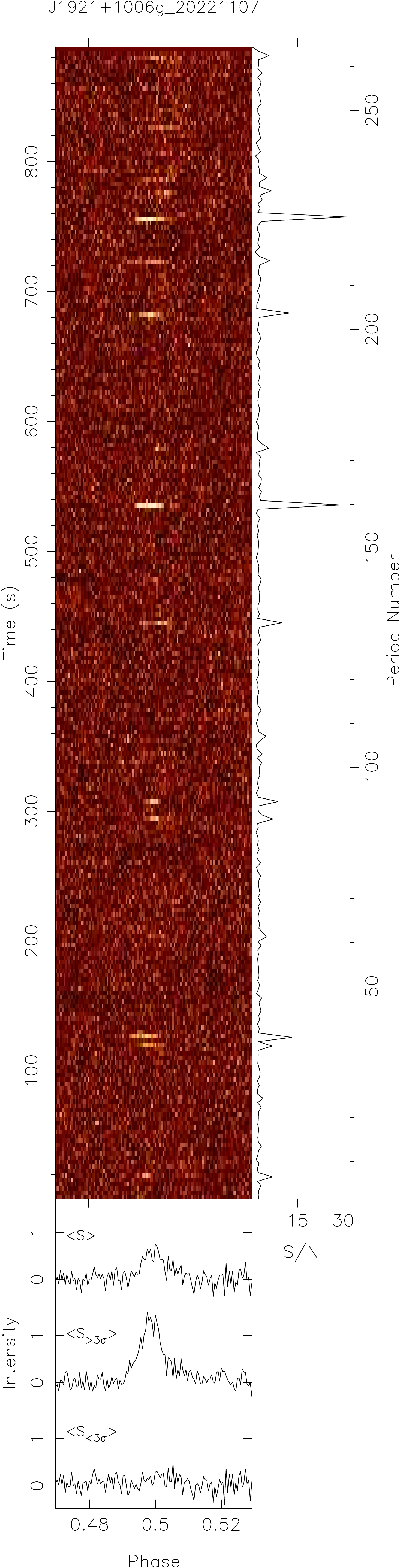} 
    \includegraphics[width=0.33\textwidth]{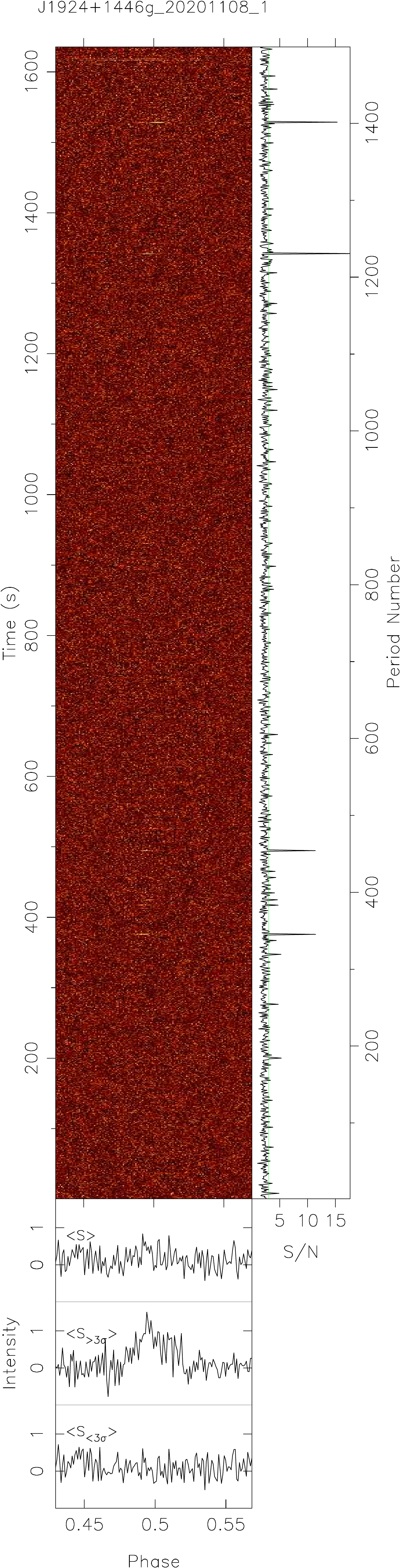} 
    \includegraphics[width=0.33\textwidth]{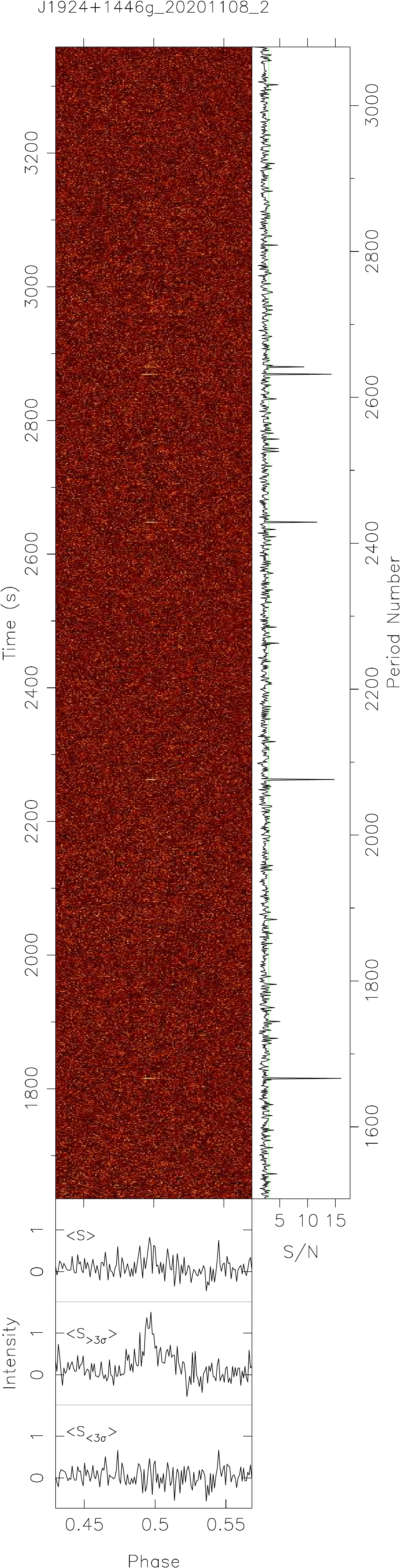}
  \caption{{\it -- continued}.}
\end{figure*}
\addtocounter{figure}{-1}
\begin{figure*}
\centering
    \includegraphics[width=0.33\textwidth]{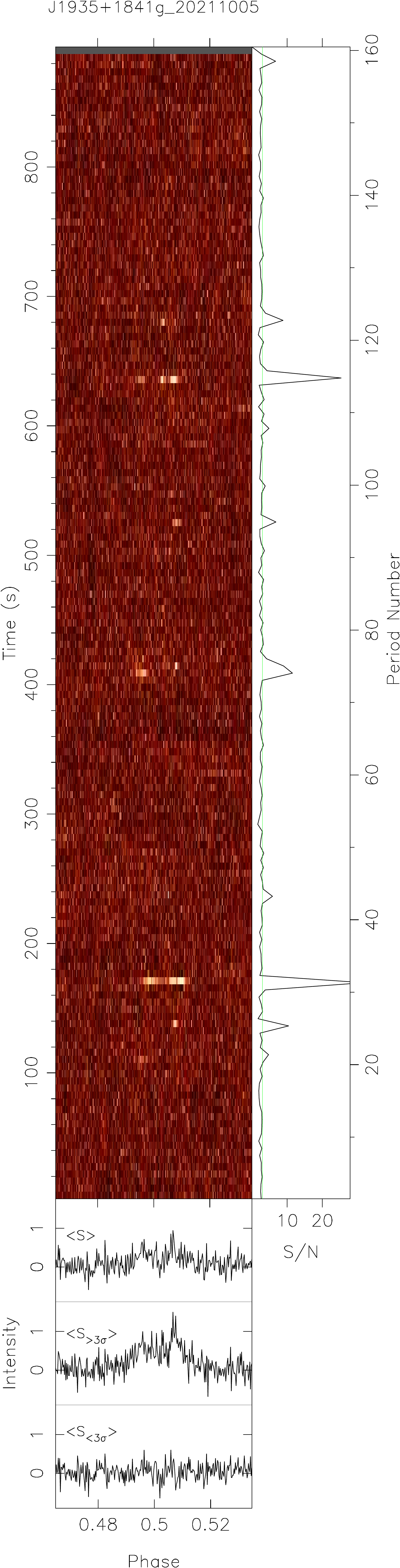} 
    \includegraphics[width=0.33\textwidth]{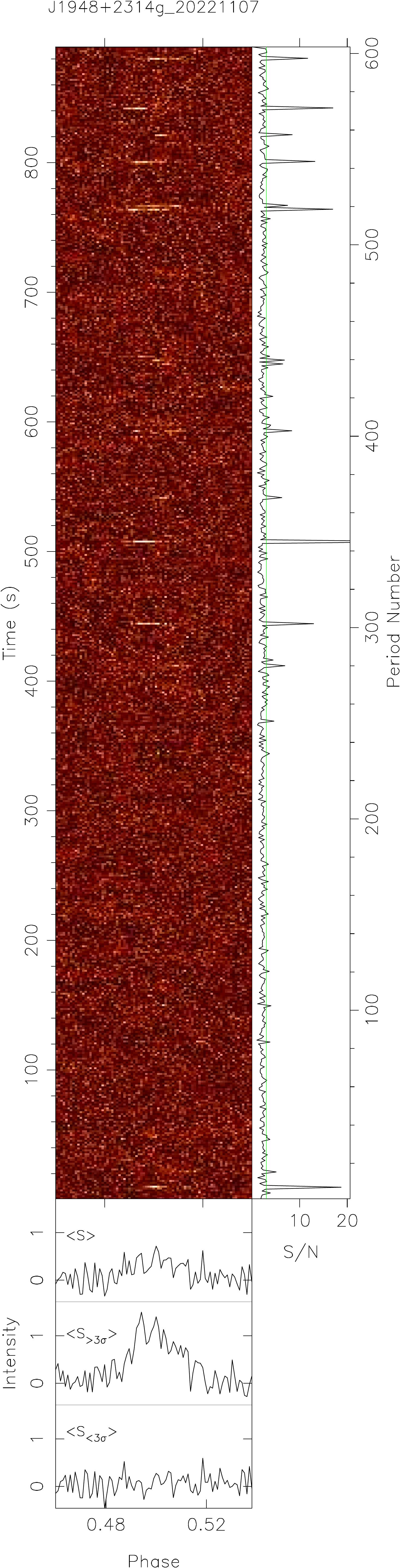} 
    \includegraphics[width=0.33\textwidth]{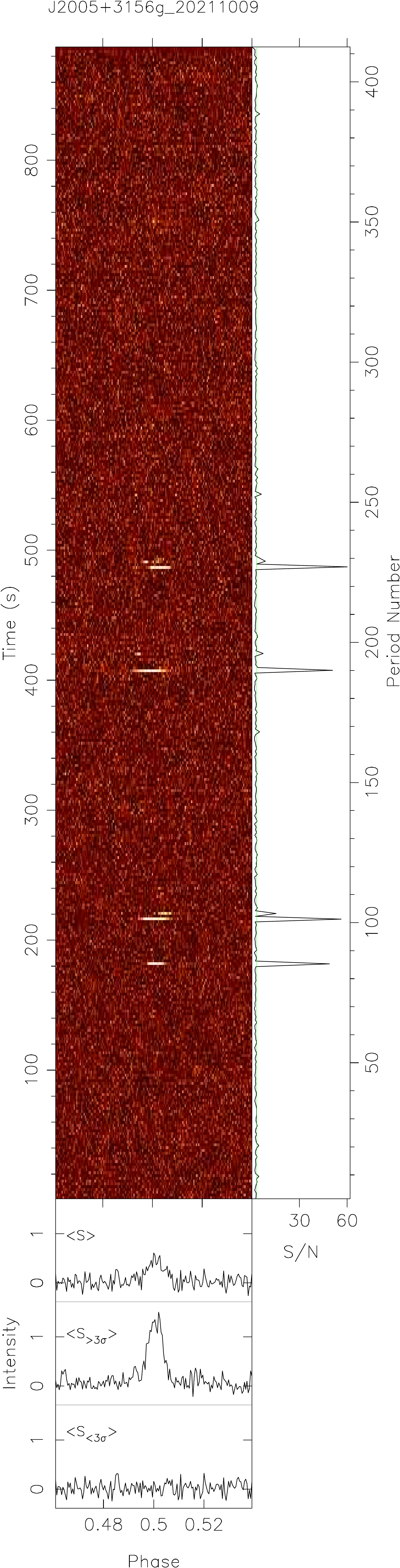}
  \caption{{\it -- continued}.}
\end{figure*}
\addtocounter{figure}{-1}
\begin{figure*}
\centering
    \includegraphics[width=0.33\textwidth]{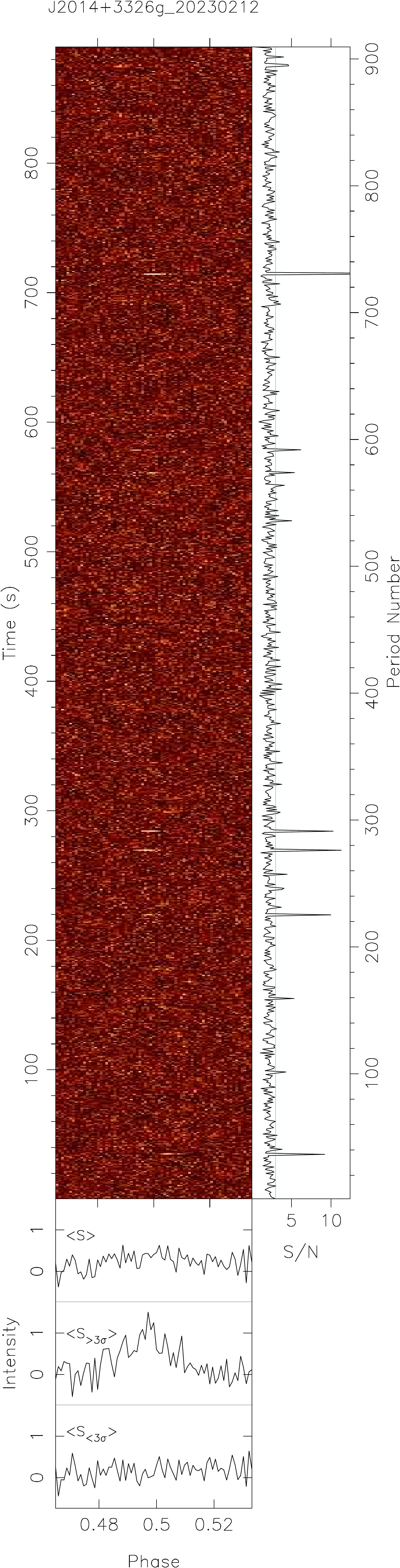}
  \caption{{\it -- continued}.}
\end{figure*}

\begin{figure*}
    \centering
    \includegraphics[width=0.33\textwidth]{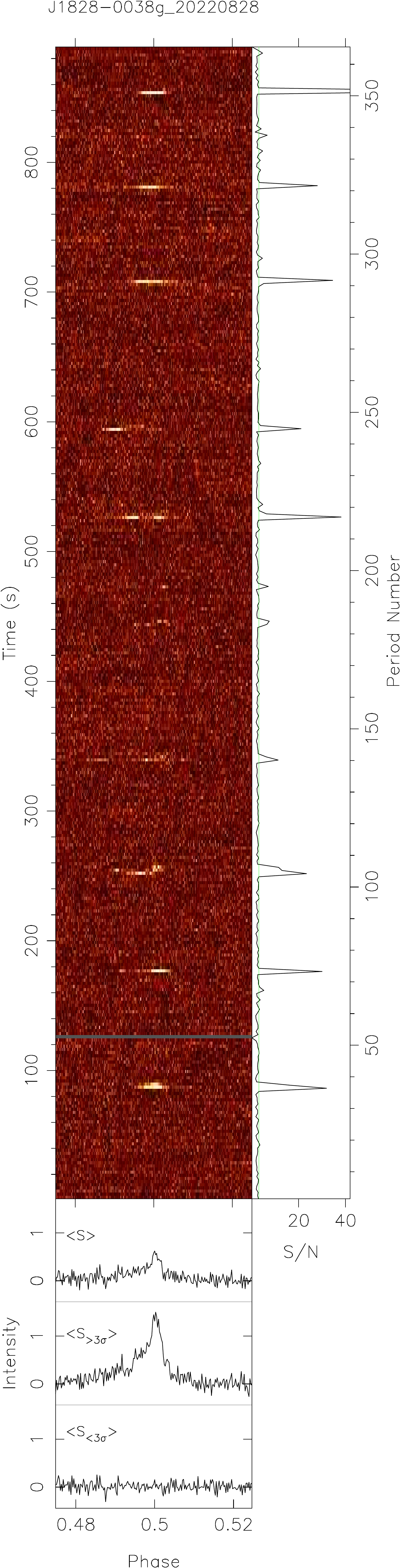} 
    \includegraphics[width=0.33\textwidth]{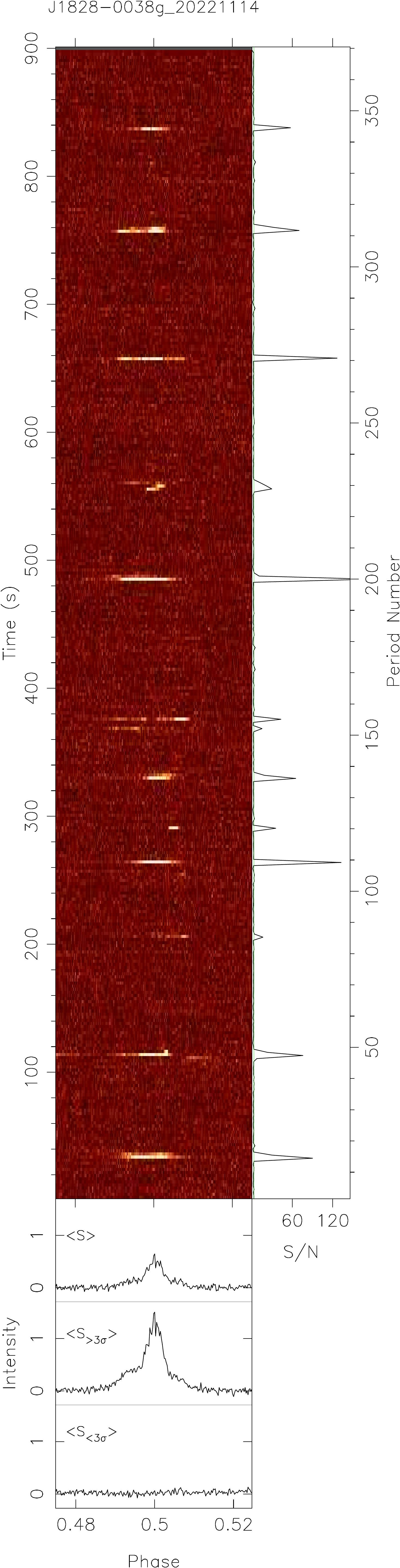}
    \includegraphics[width=0.33\textwidth]{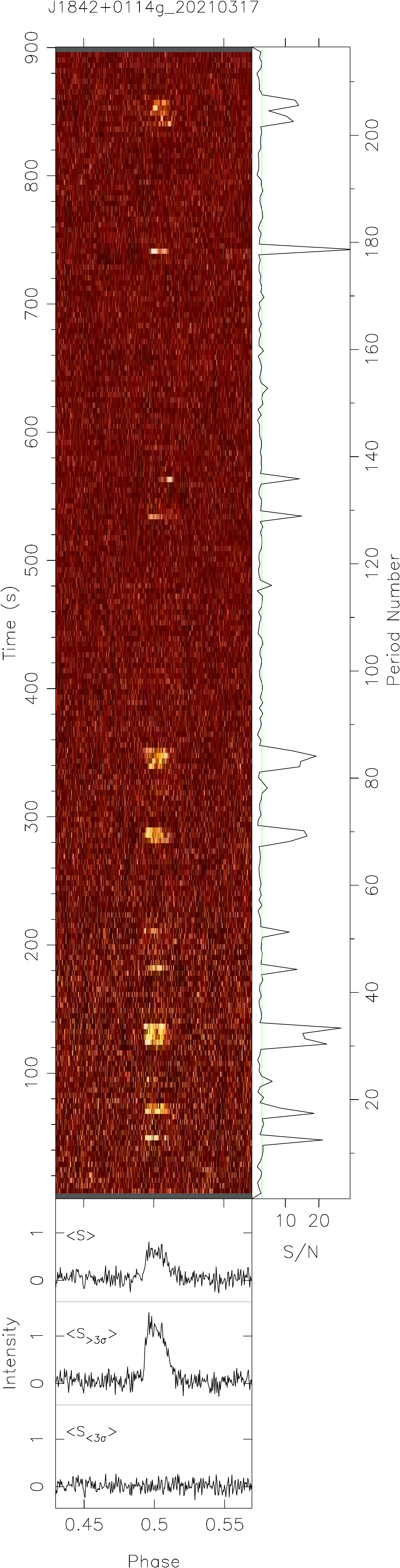}\\[1mm]
    \caption{Very nulling pulsars discovered by GPPS.}
    \label{fig:Appnewnulling}
\end{figure*}
\addtocounter{figure}{-1}
\begin{figure*}
    \centering
     
    \includegraphics[width=0.33\textwidth]{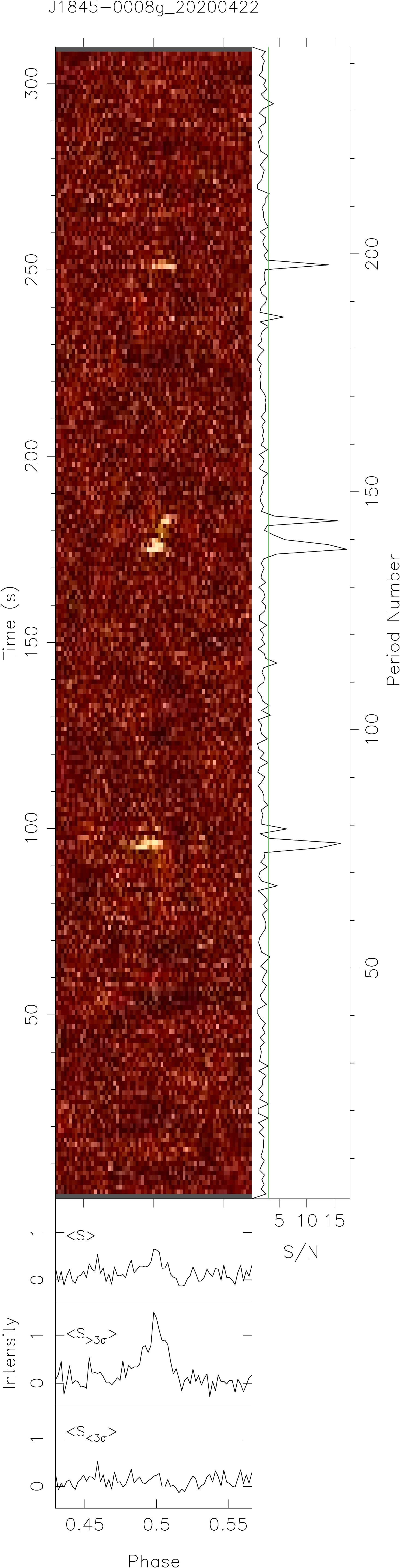} 
    \includegraphics[width=0.33\textwidth]{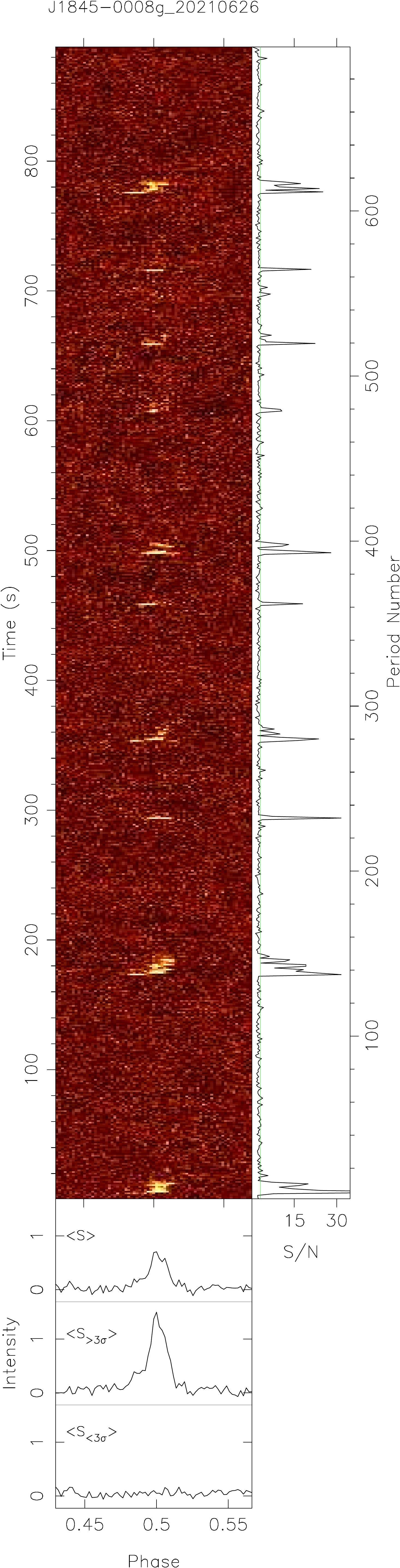}  
    \includegraphics[width=0.33\textwidth]{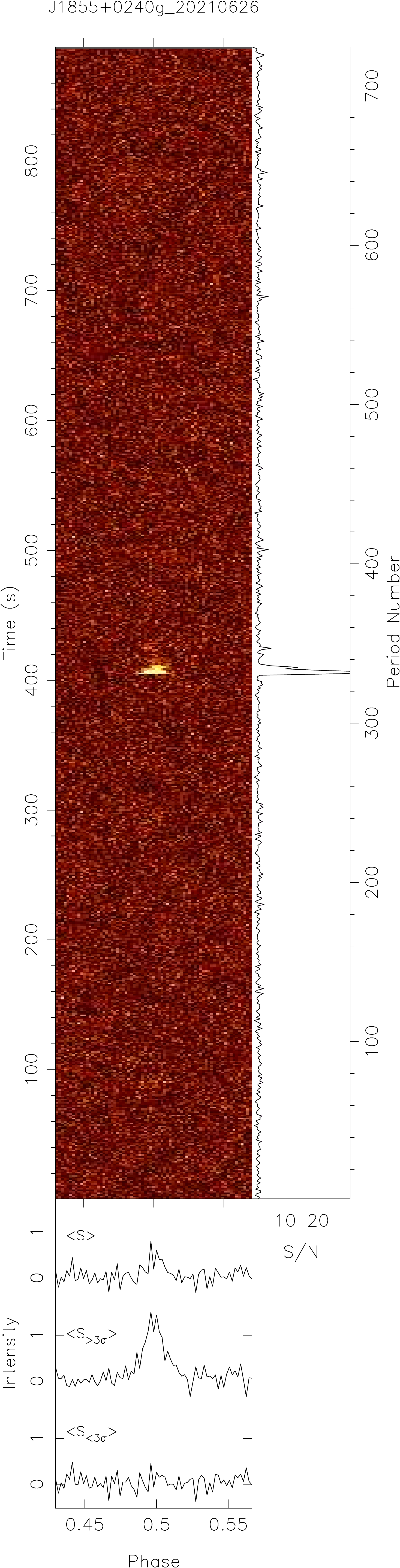}\\[1mm]
    \caption{{\it -- continued}.}
\end{figure*}
\addtocounter{figure}{-1}
\begin{figure*}
    \centering
    \includegraphics[width=0.33\textwidth]{New-figs/J1858-0113/J185852-011306_20220403_tracking-M01-P1-c2048b1.ar.zap.Fp.debased.pdf} %SP080
    \includegraphics[width=0.33\textwidth]{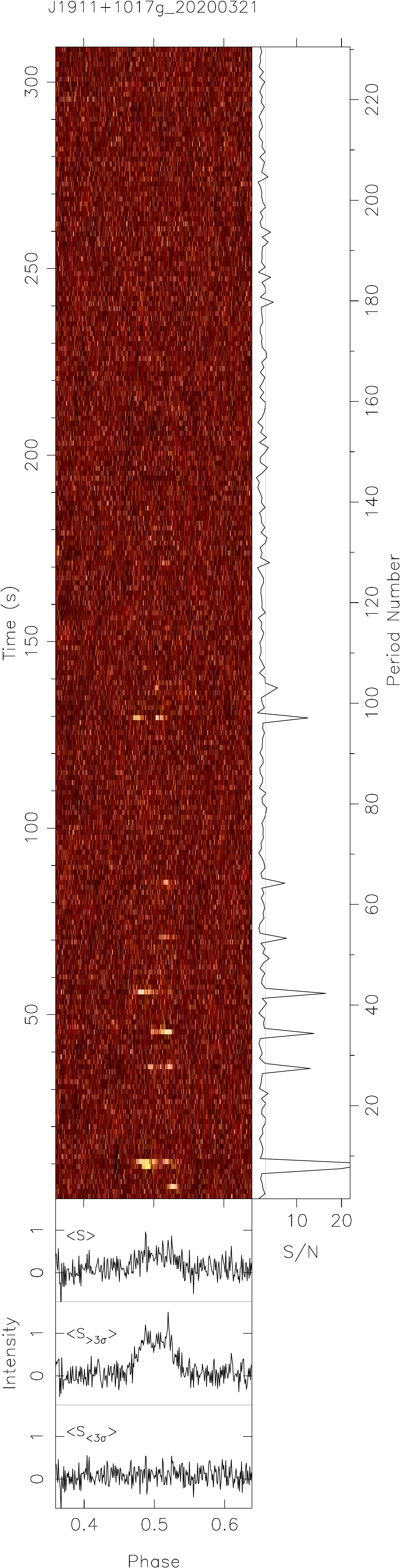} 
    \includegraphics[width=0.33\textwidth]{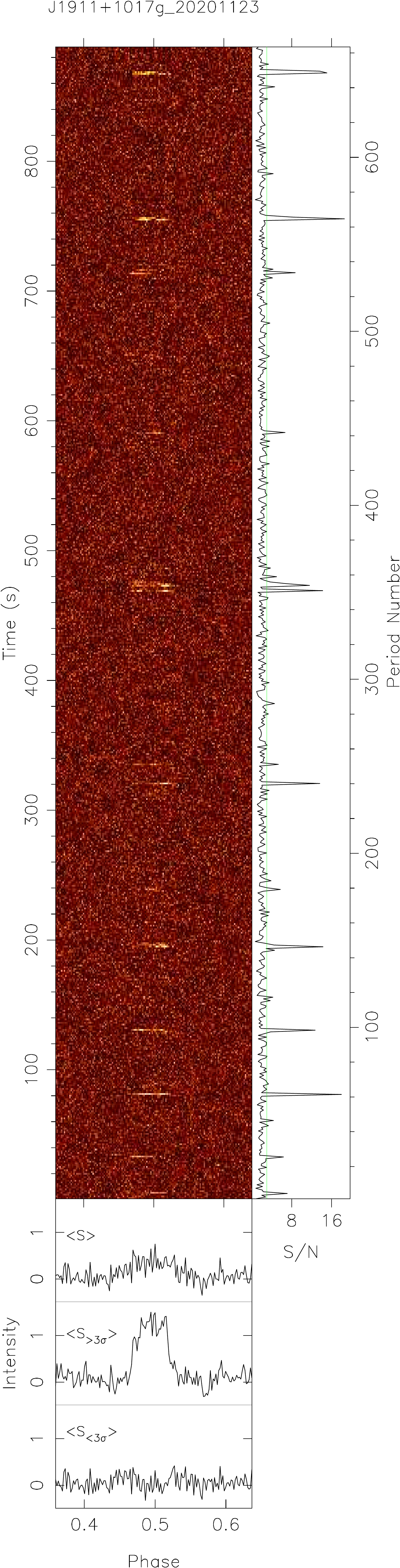} \\[1mm]
    \caption{{\it -- continued}.}
\end{figure*}
\addtocounter{figure}{-1}
\begin{figure*}
    \centering
    \includegraphics[width=0.33\textwidth]{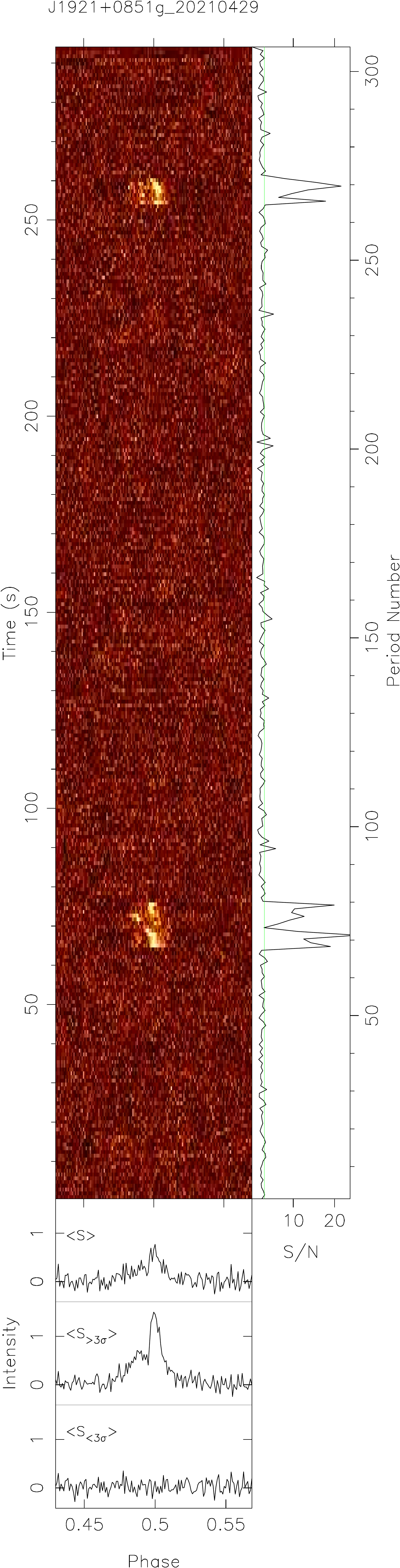}
    \includegraphics[width=0.33\textwidth]{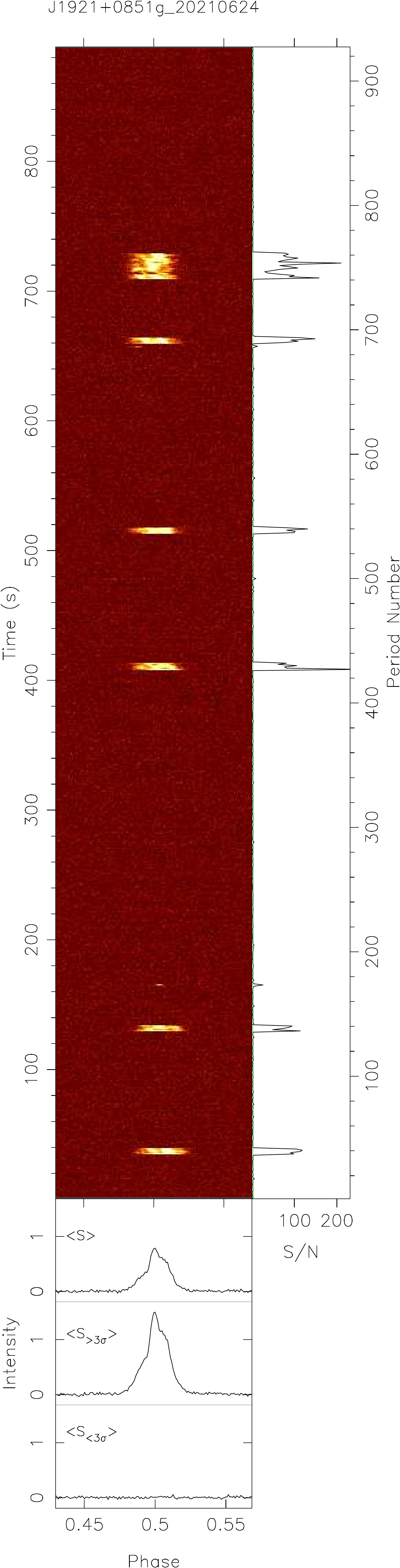}  
    \includegraphics[width=0.33\textwidth]{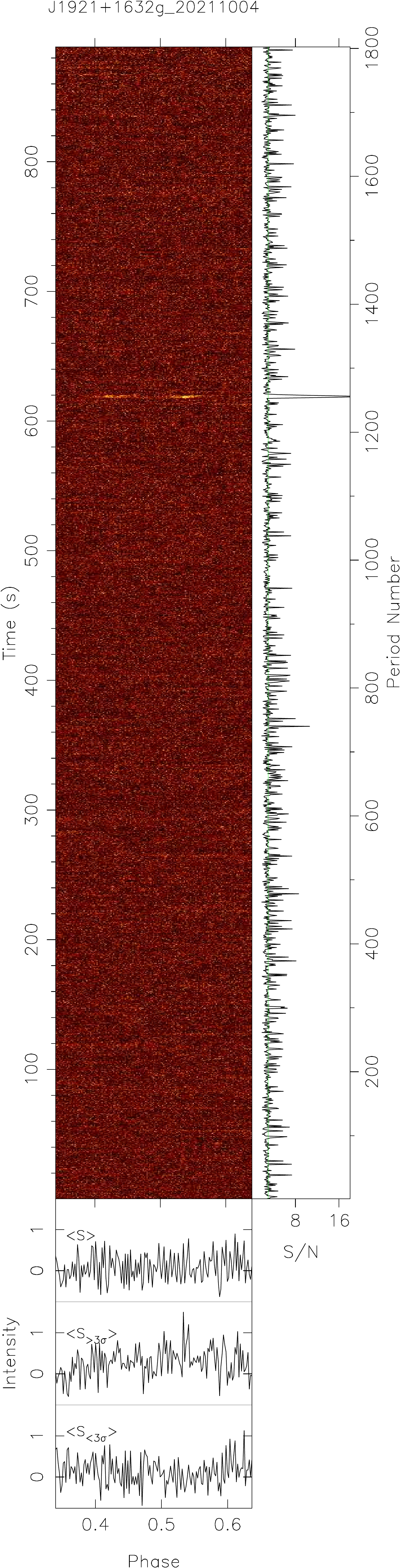}\\[1mm]
    \caption{{\it -- continued}.}
\end{figure*}
\addtocounter{figure}{-1}
\begin{figure*}
    \centering
    \includegraphics[width=0.33\textwidth]{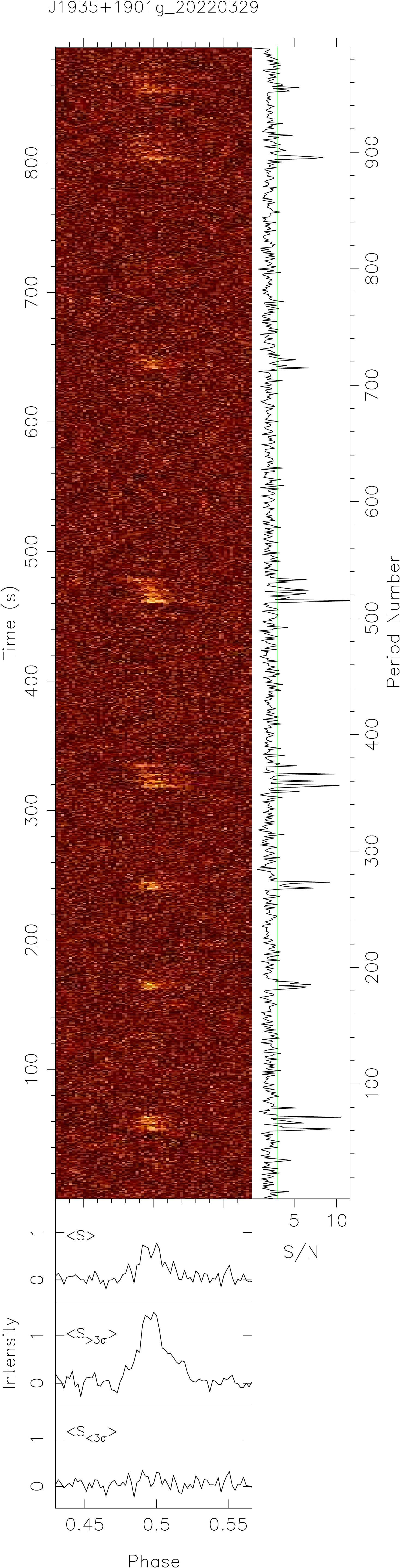} 
    \includegraphics[width=0.33\textwidth]{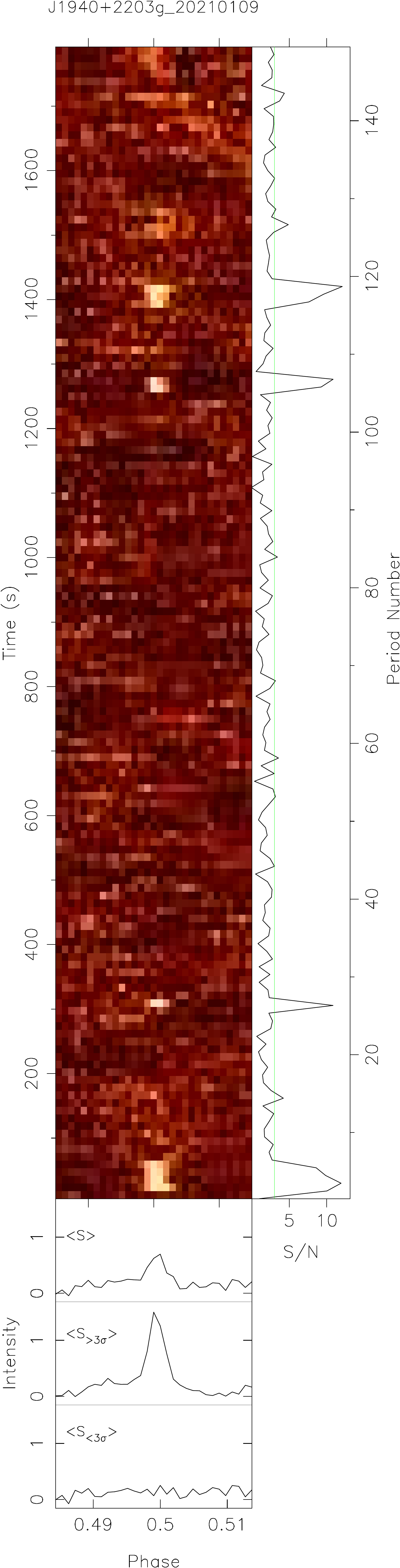} \\[1mm]
    \caption{{\it -- continued}.}
\end{figure*}

\begin{figure*}
  \centering
  \includegraphics[width=38mm]{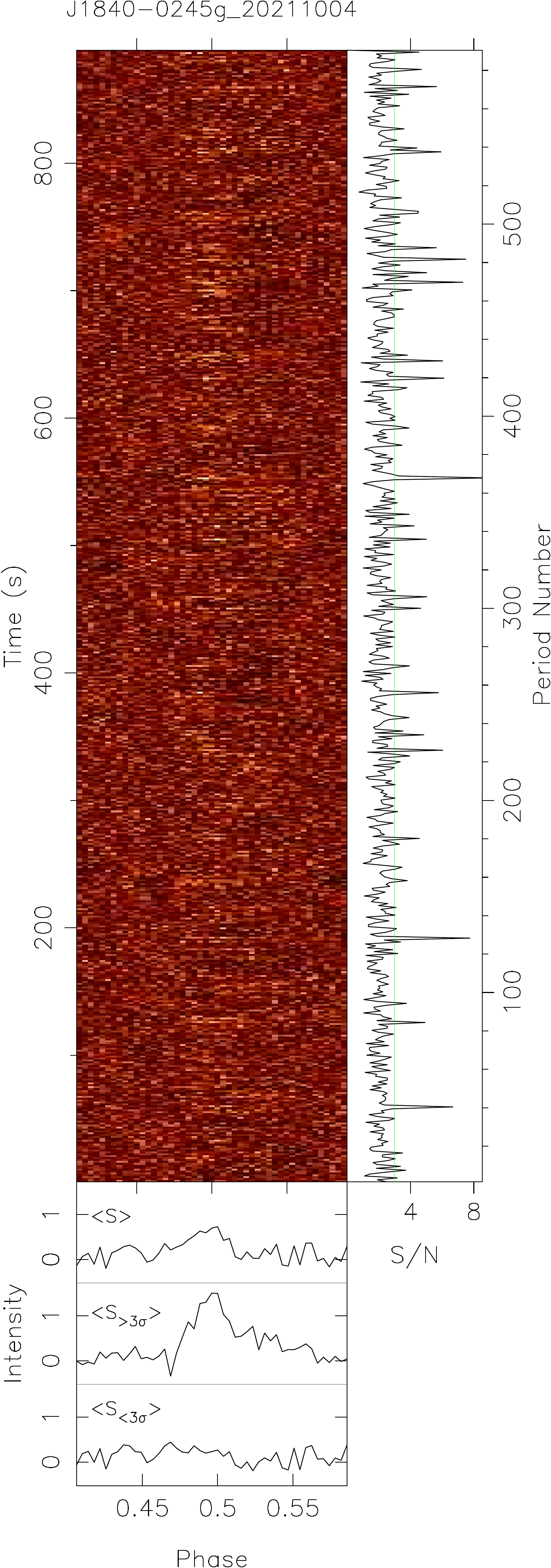} 
  \includegraphics[width=38mm]{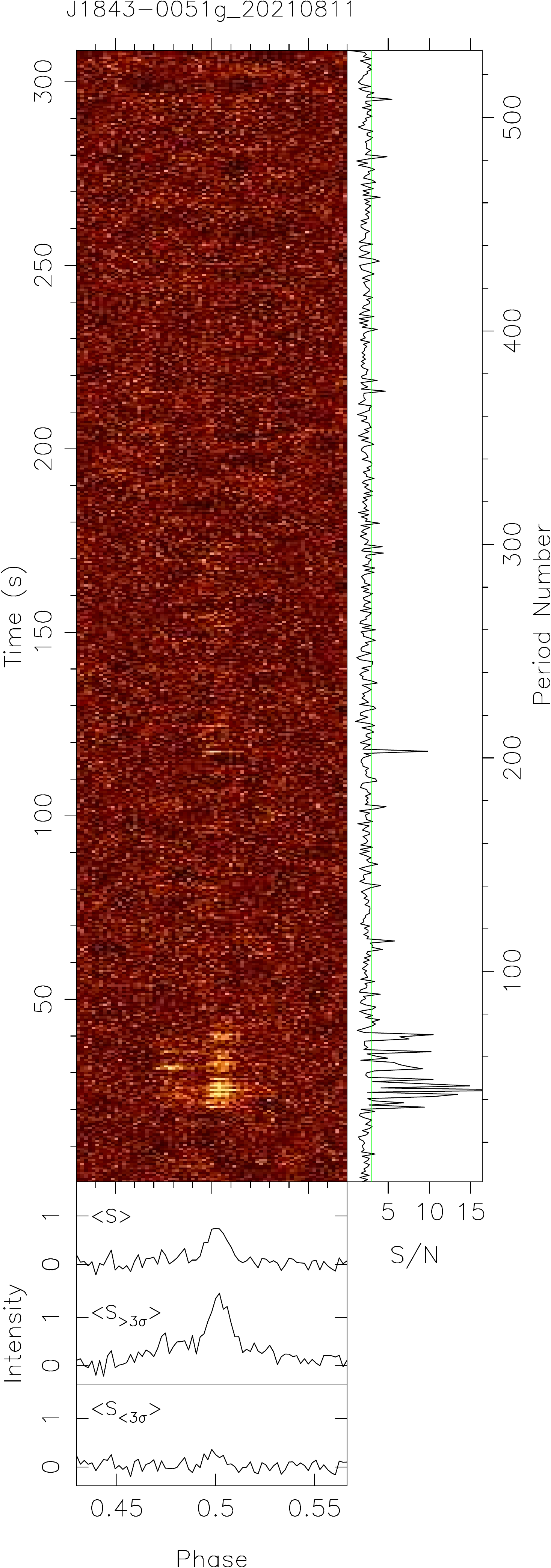} 
  \includegraphics[width=38mm]{New-figs/J184332-005100/J184332-005100sp_20211009_tracking-M01-P1-c2048b1.ar.E.zap.Fp.debased.pdf}
  \includegraphics[width=38mm]{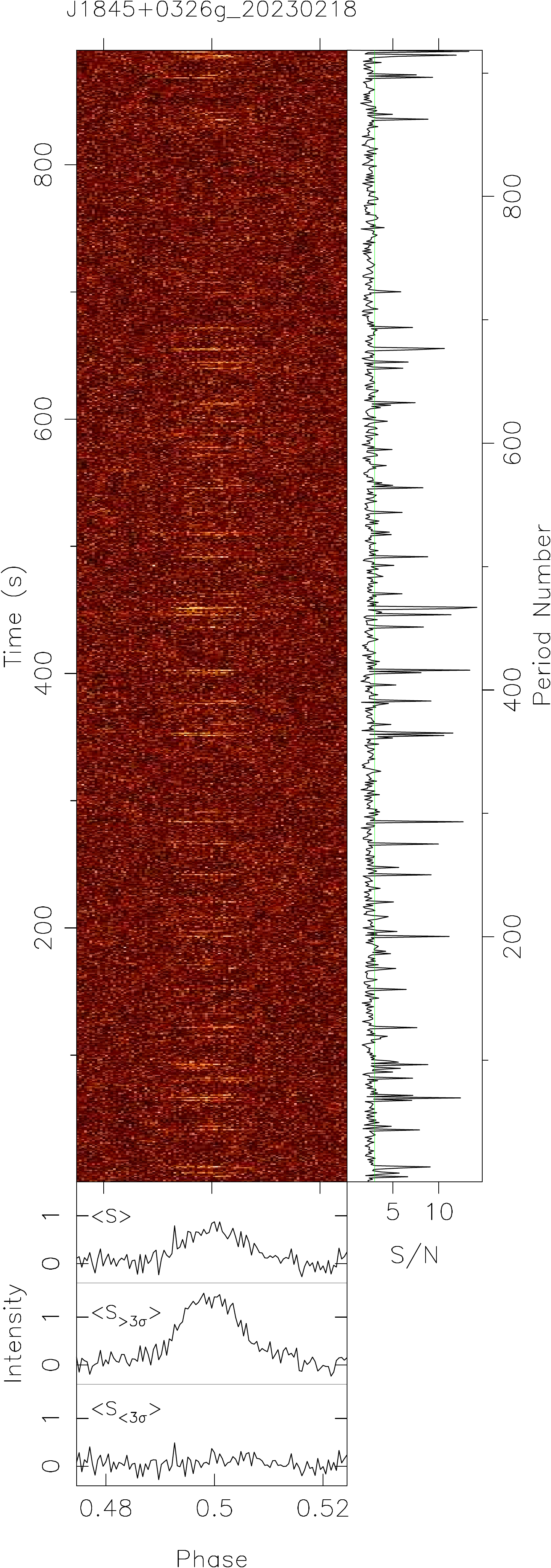} \\[1mm]
  \includegraphics[width=38mm]{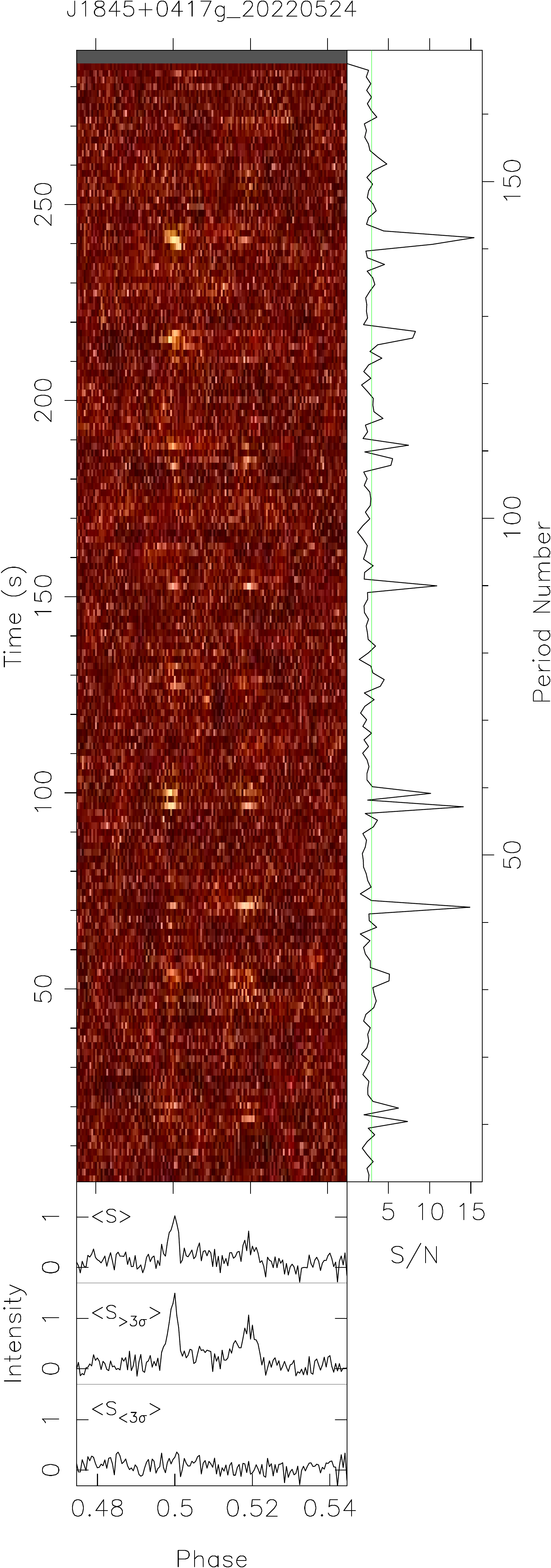}
  \includegraphics[width=38mm]{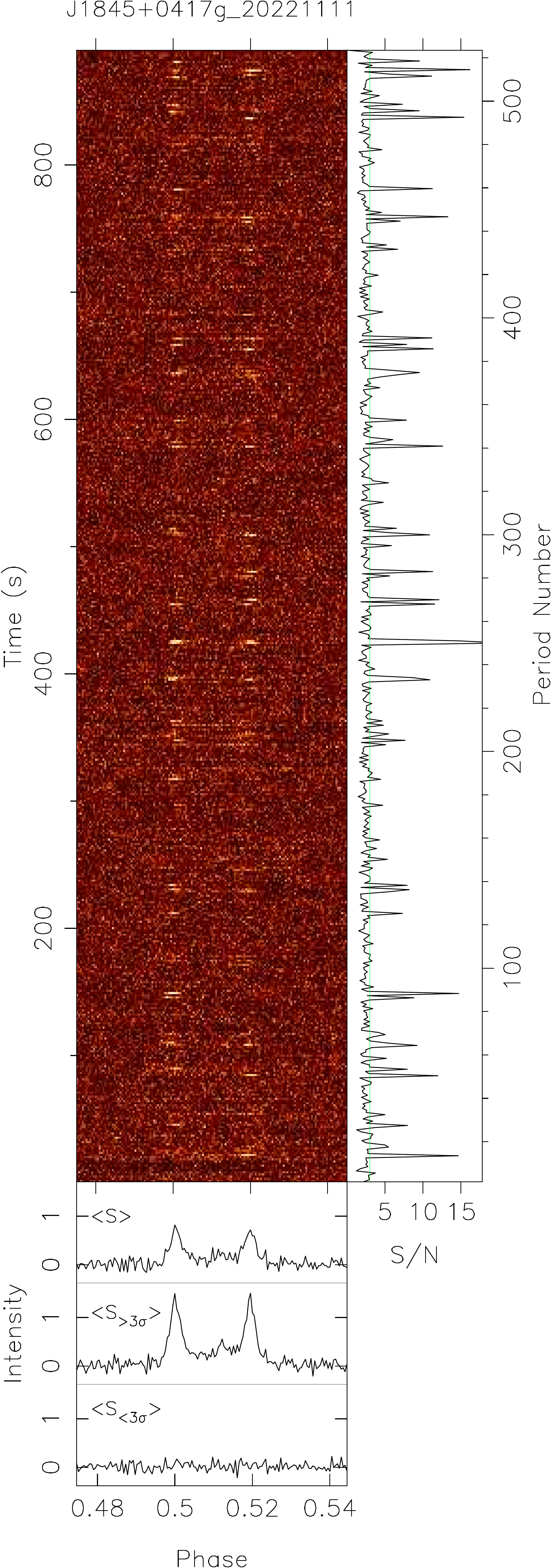} 
  \includegraphics[width=38mm]{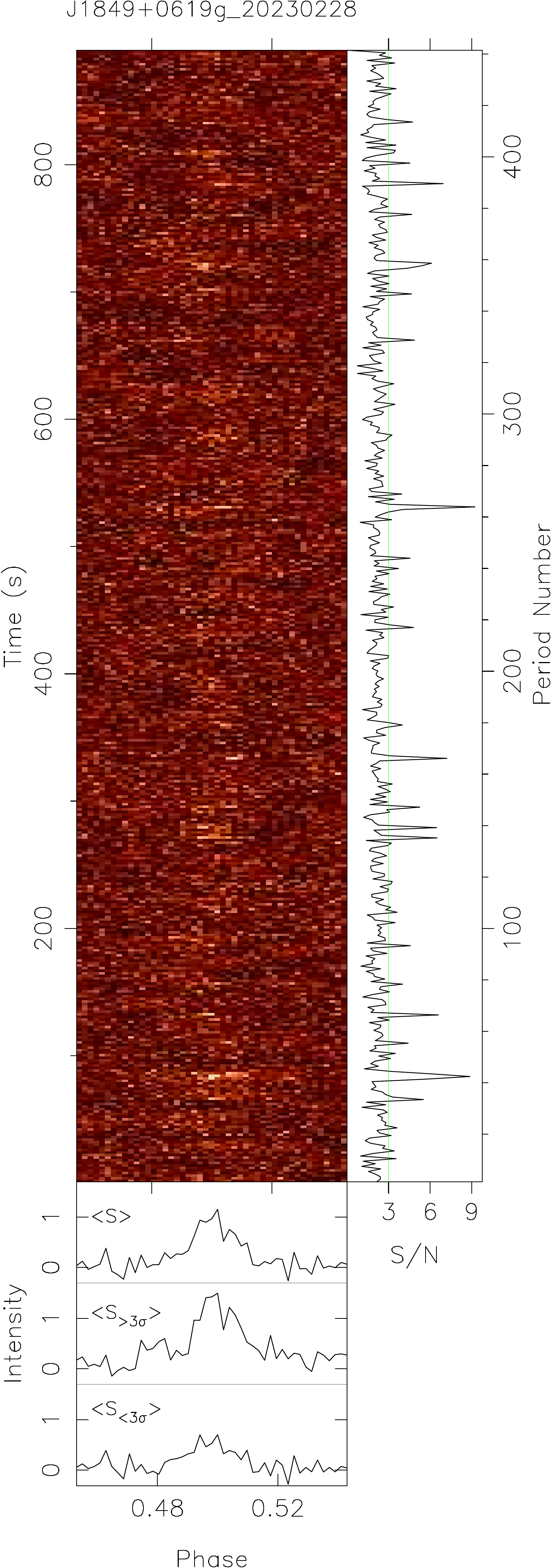}
  \includegraphics[width=38mm]{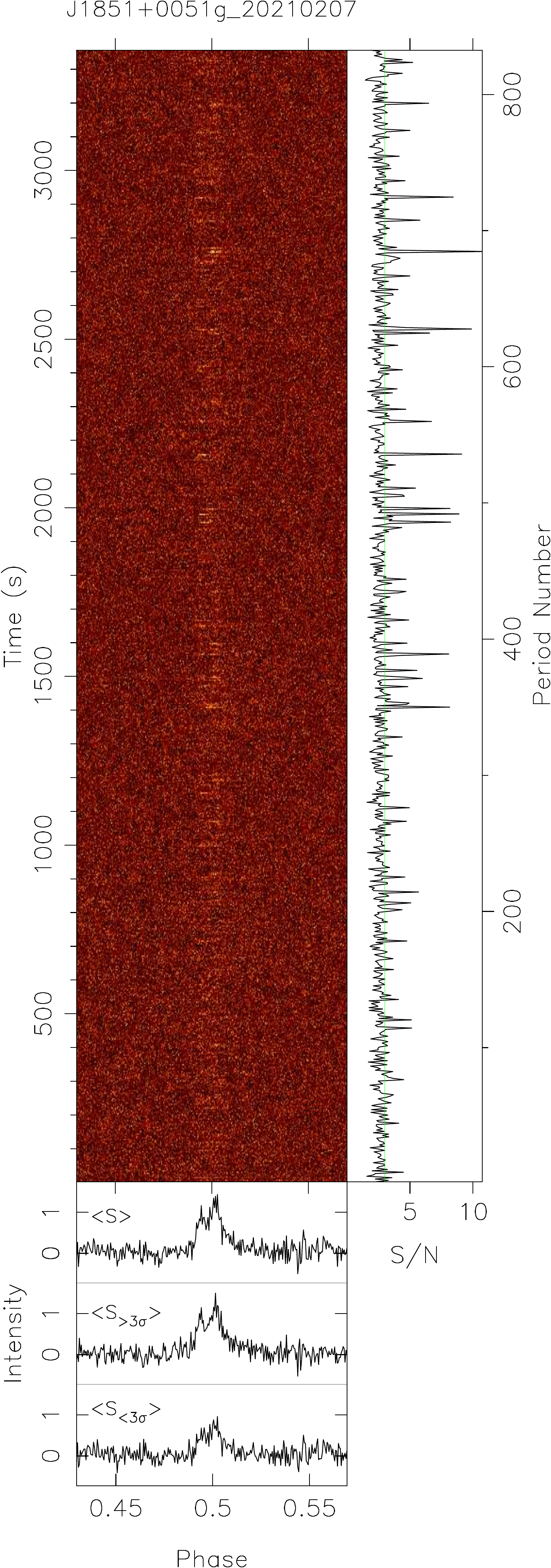} 
\caption{ Weak pulsars by GPPS.}
\label{fig:AppnewweakPulsars}
\end{figure*}
\addtocounter{figure}{-1}
\begin{figure*}
  \centering
  \includegraphics[width=38mm]{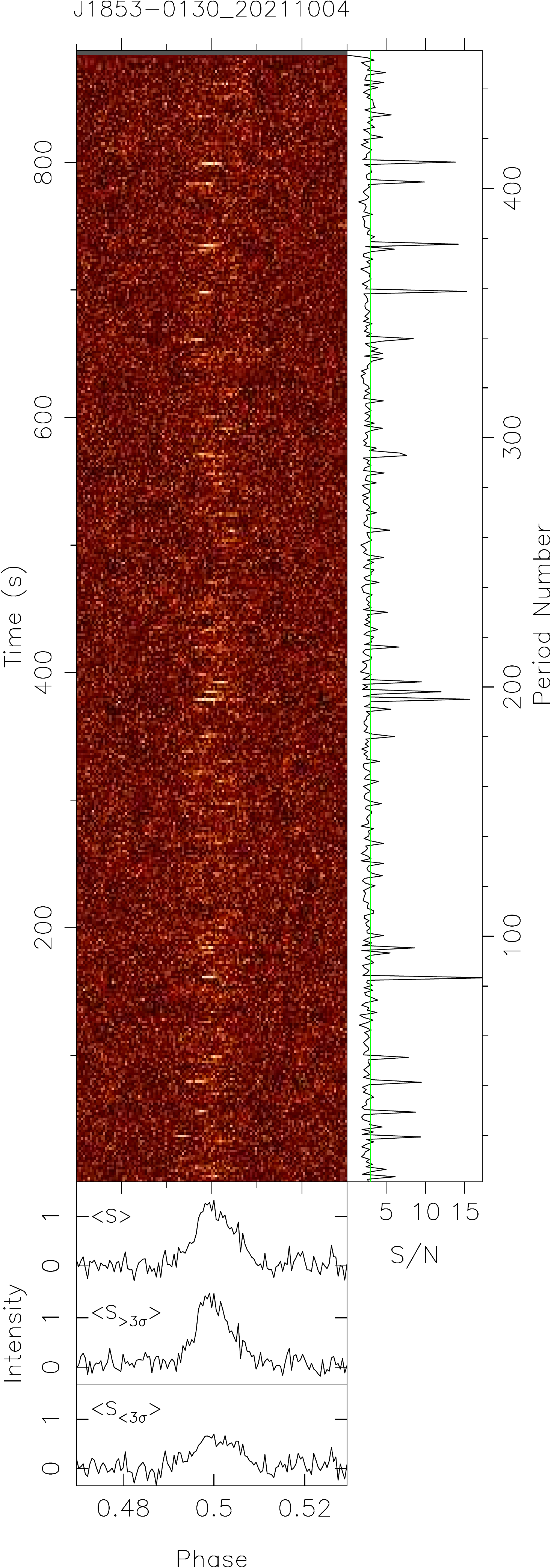} 
  \includegraphics[width=38mm]{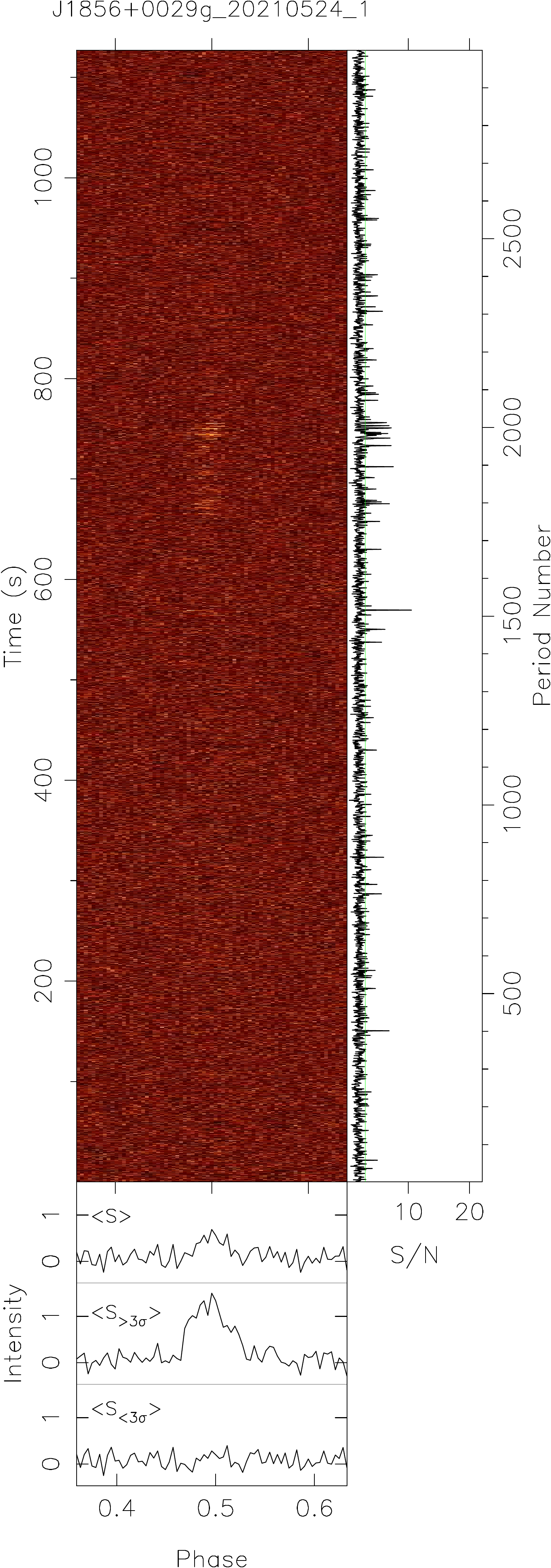} 
  \includegraphics[width=38mm]{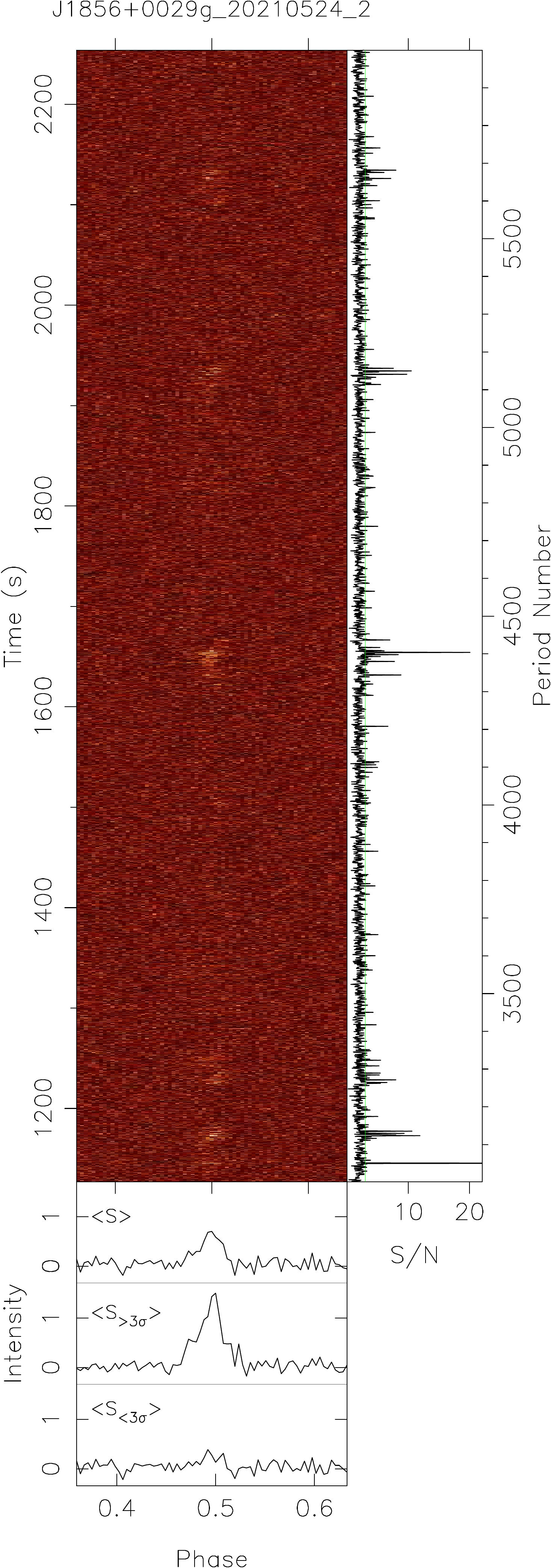} 
  \includegraphics[width=38mm]{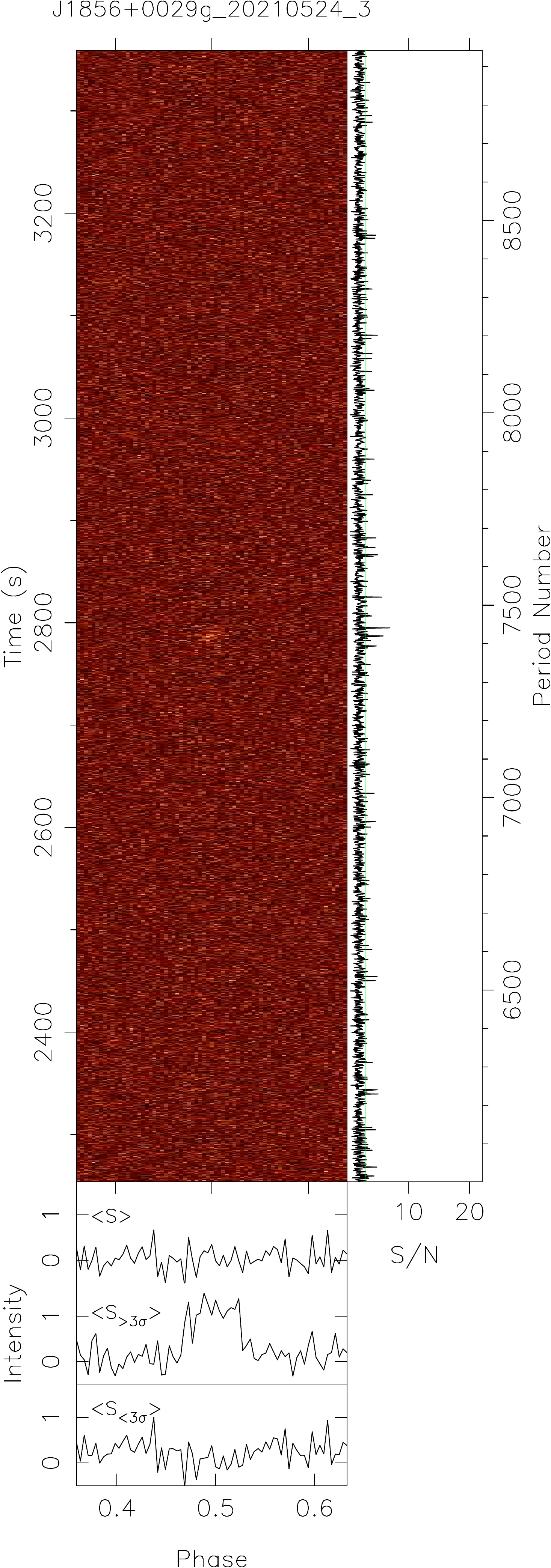}  \\[1mm]
  \includegraphics[width=38mm]{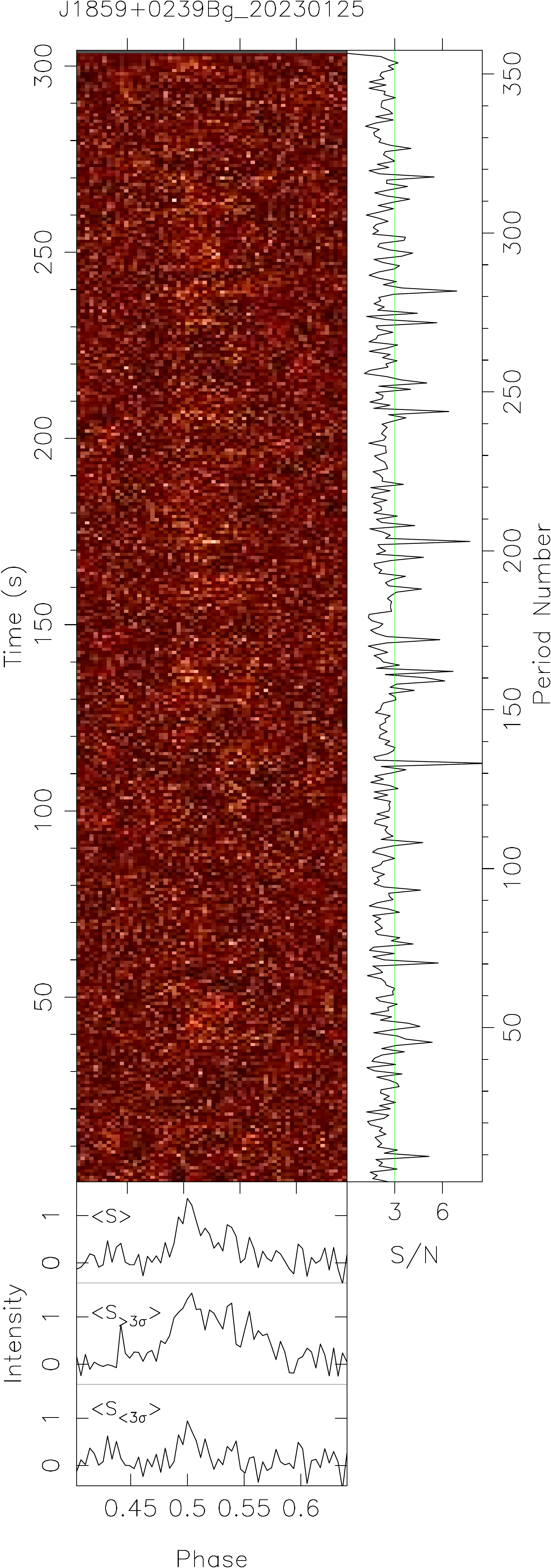} 
  \includegraphics[width=38mm]{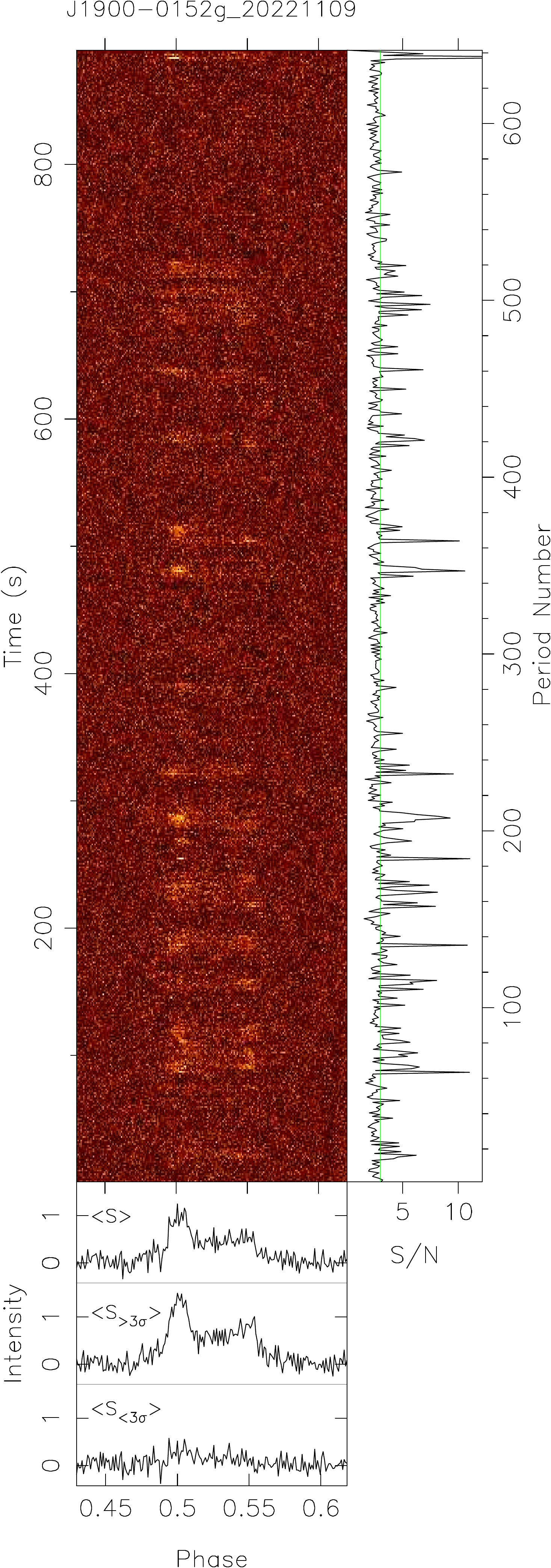} 
  \includegraphics[width=38mm]{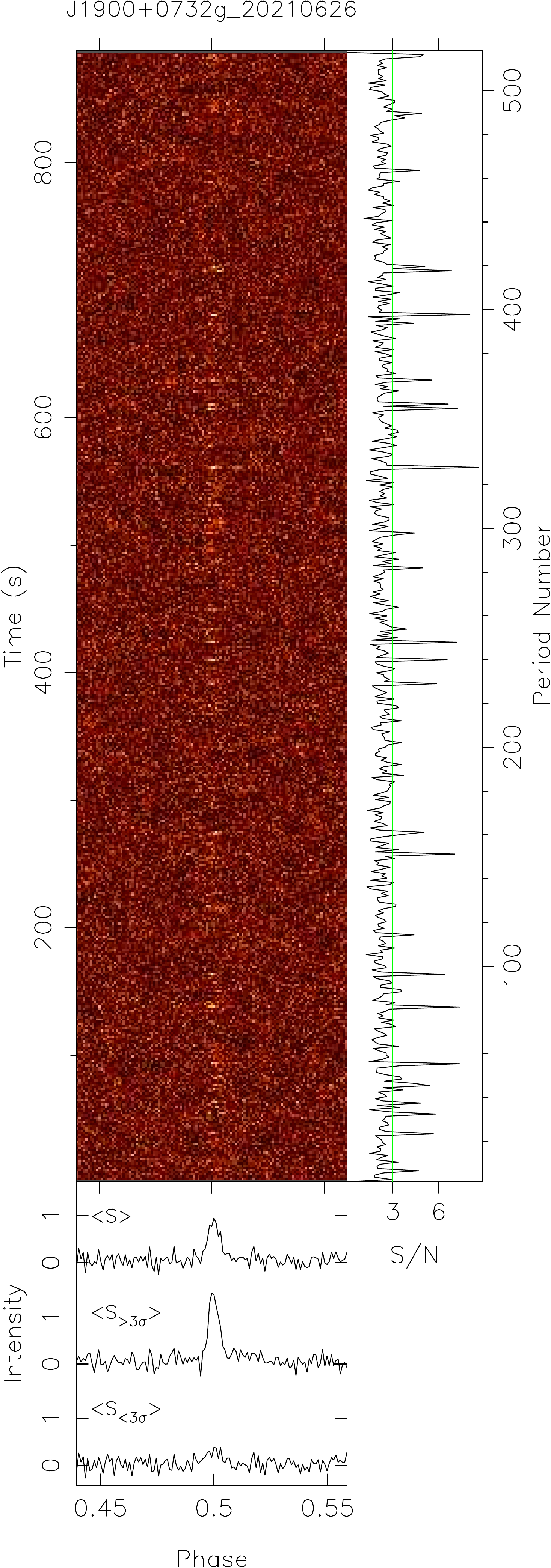}
  \includegraphics[width=38mm]{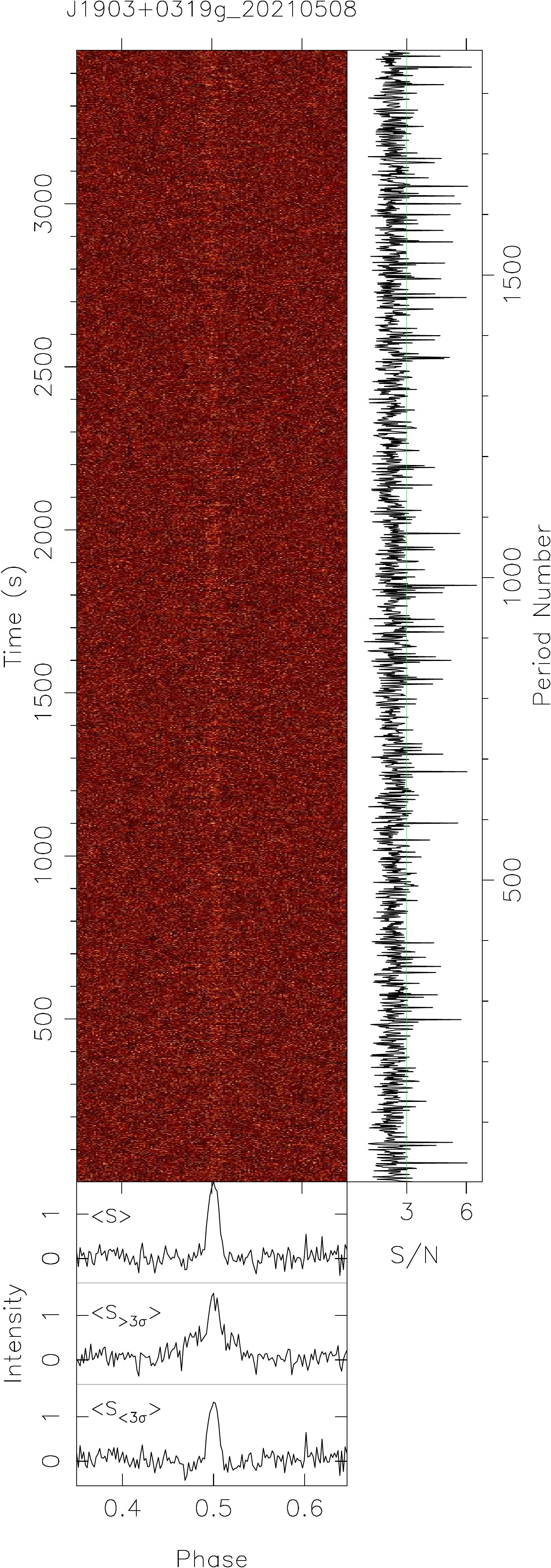} 
\caption{{\it -- continued}.}
\end{figure*}
\addtocounter{figure}{-1}
\begin{figure*}
  \centering  
  \includegraphics[width=38mm]{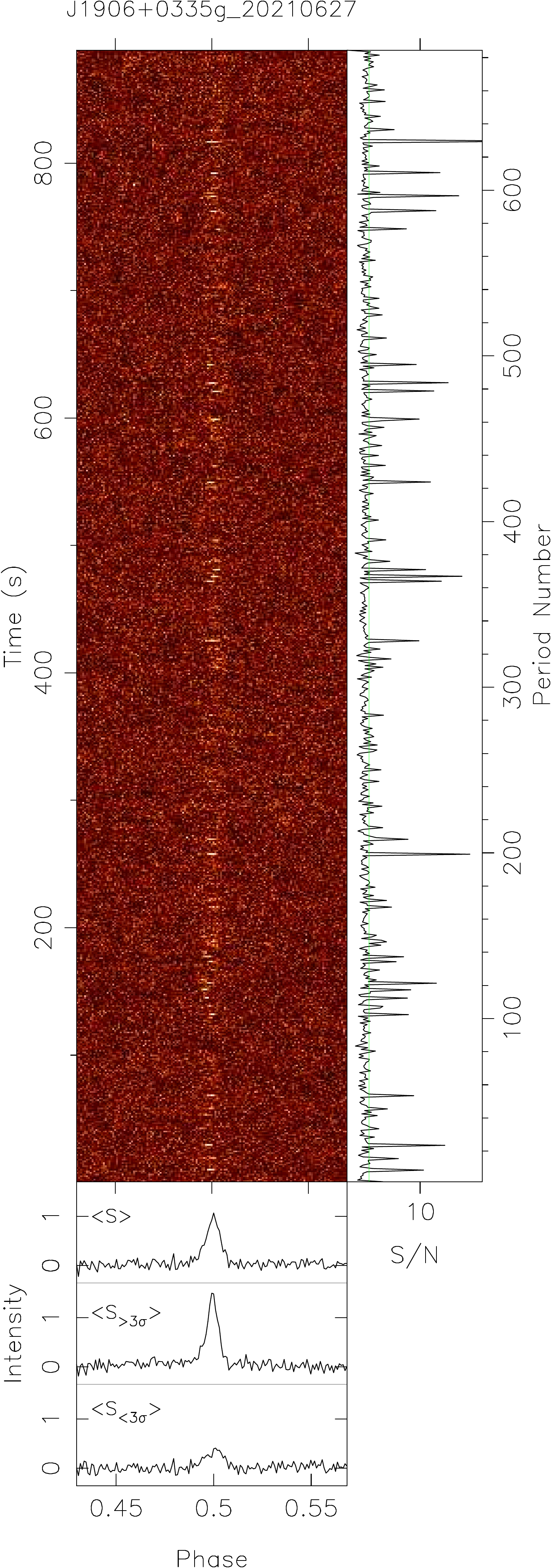} 
  \includegraphics[width=38mm]{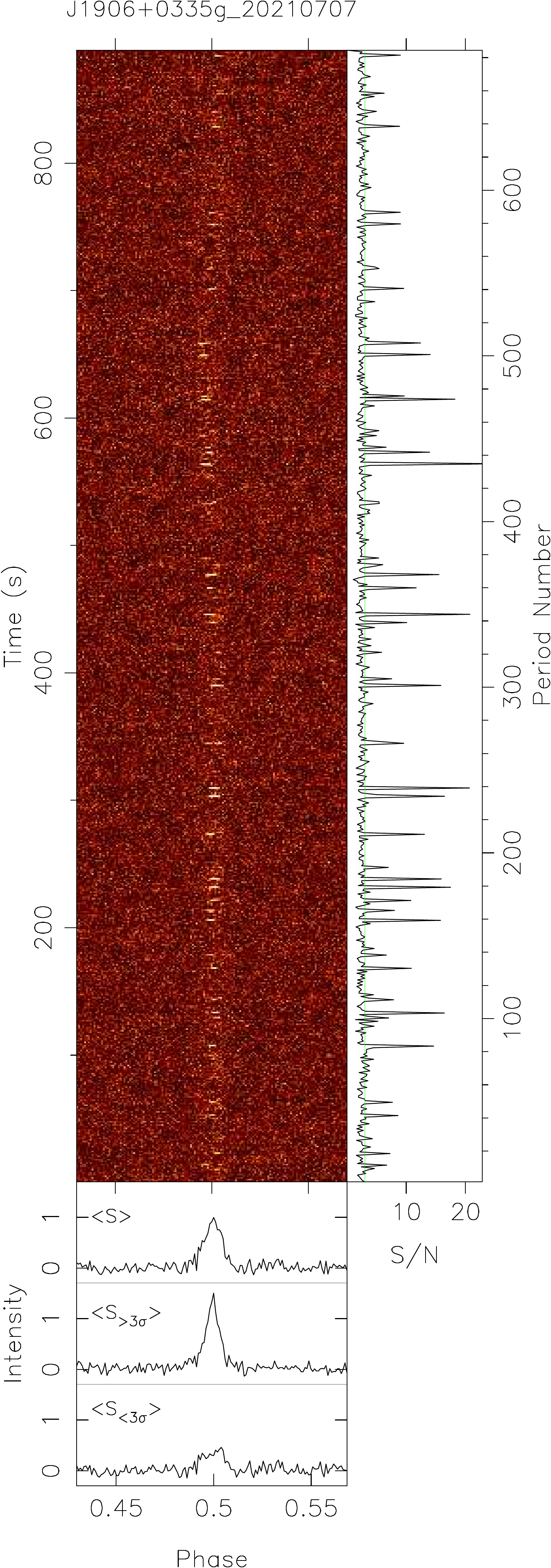} 
  \includegraphics[width=38mm]{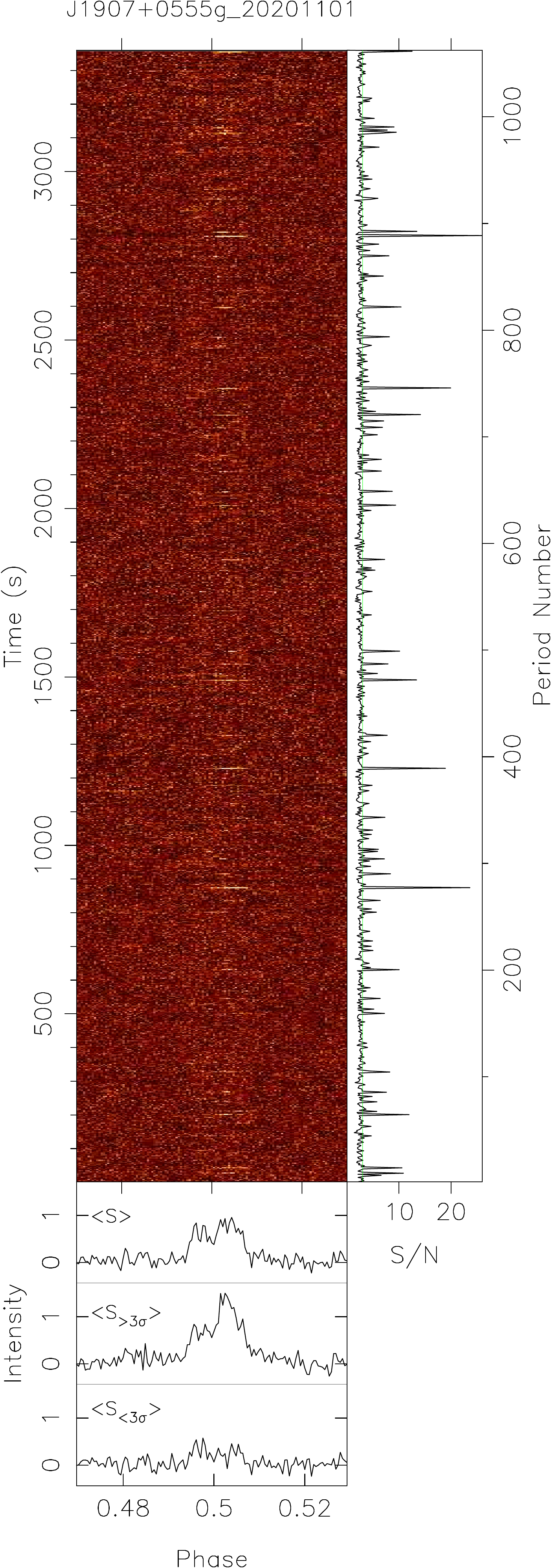} 
  \includegraphics[width=38mm]{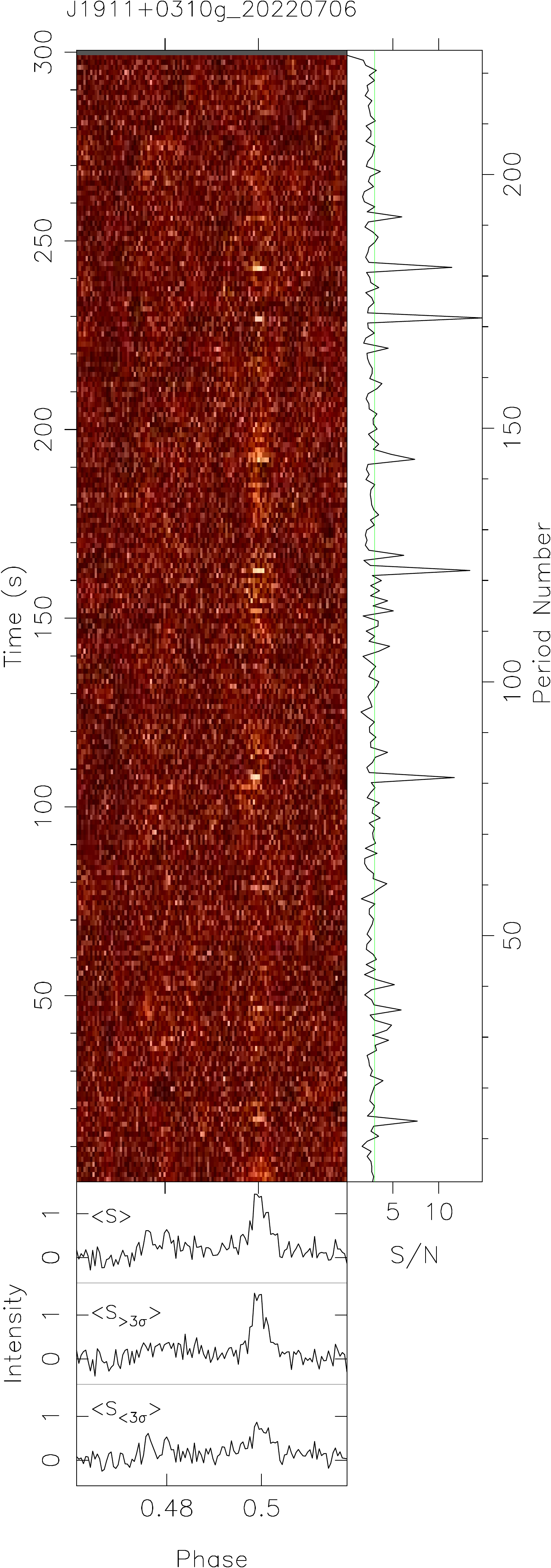} 
  \includegraphics[width=38mm]{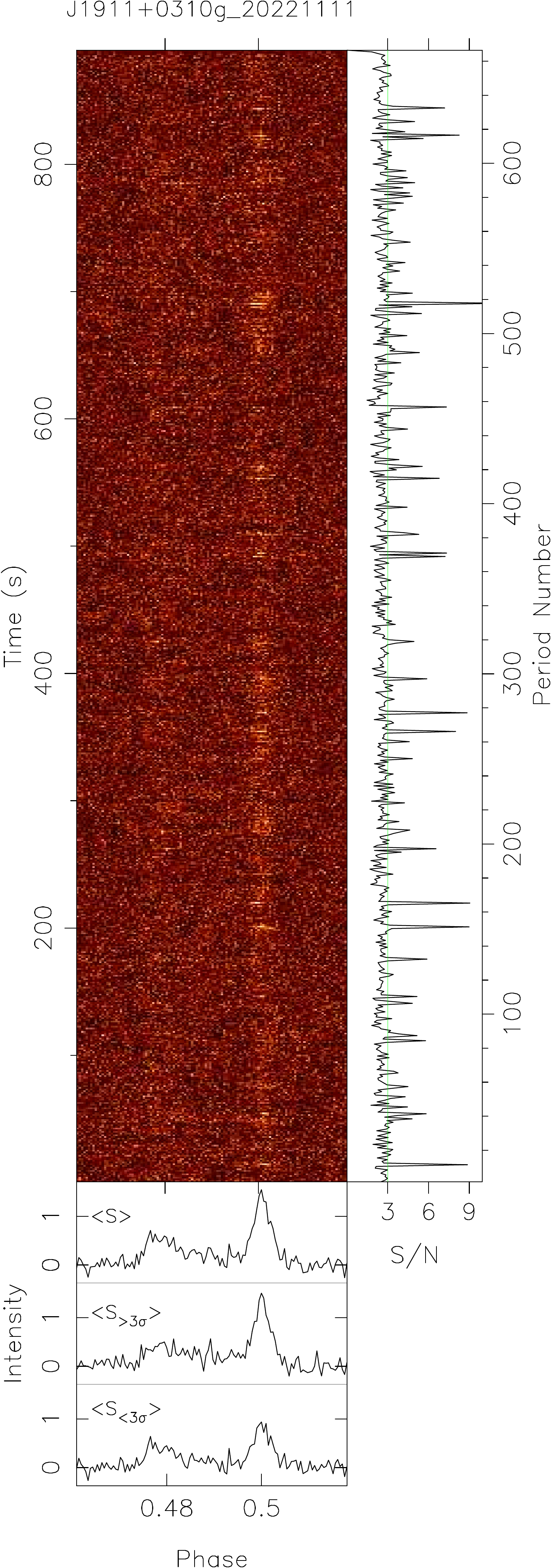} 
  \includegraphics[width=38mm]{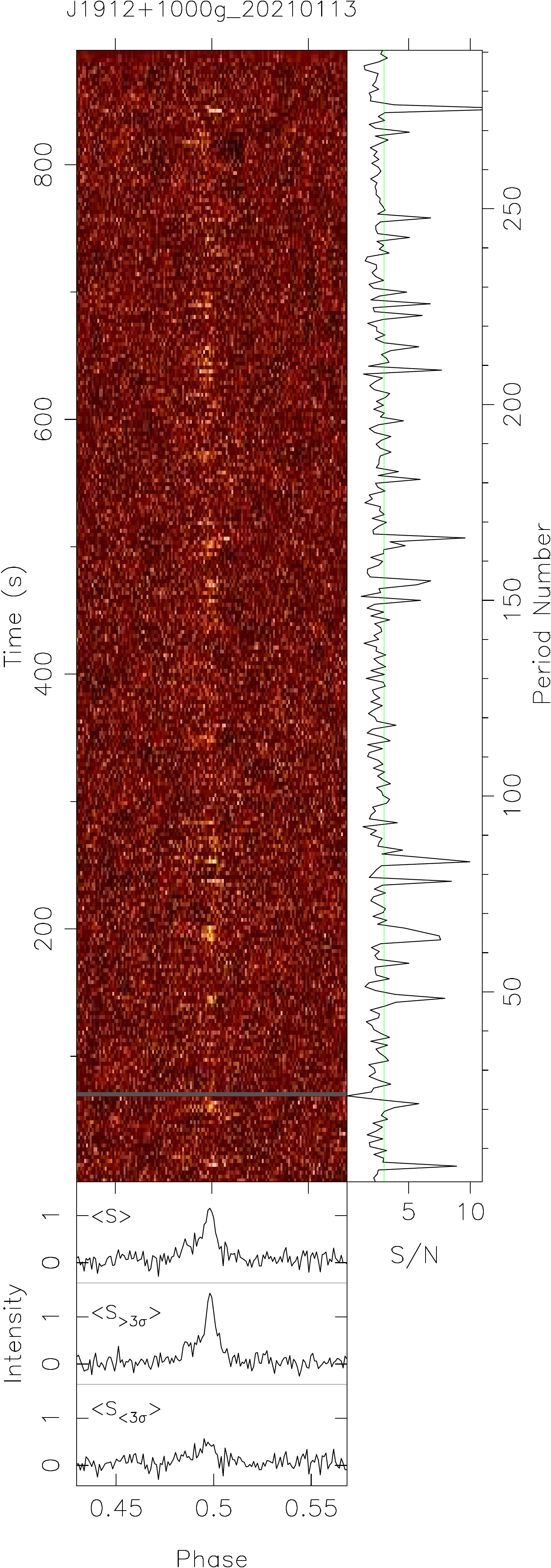} 
  \includegraphics[width=38mm]{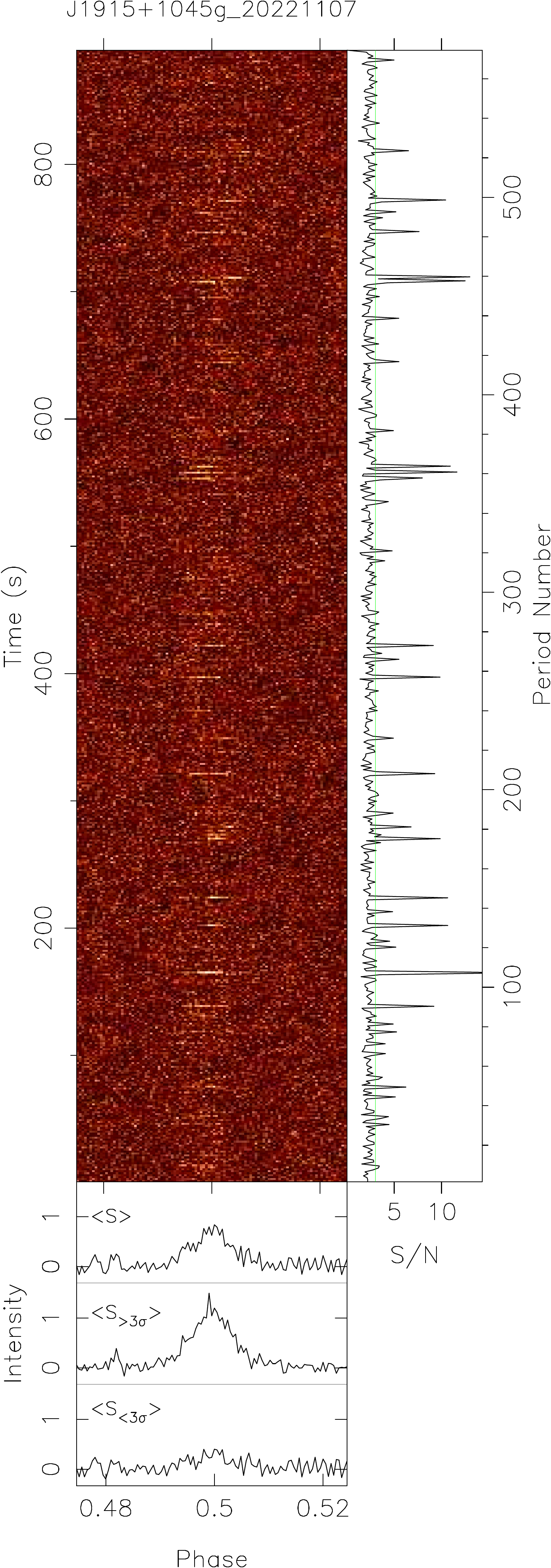}
  \includegraphics[width=38mm]{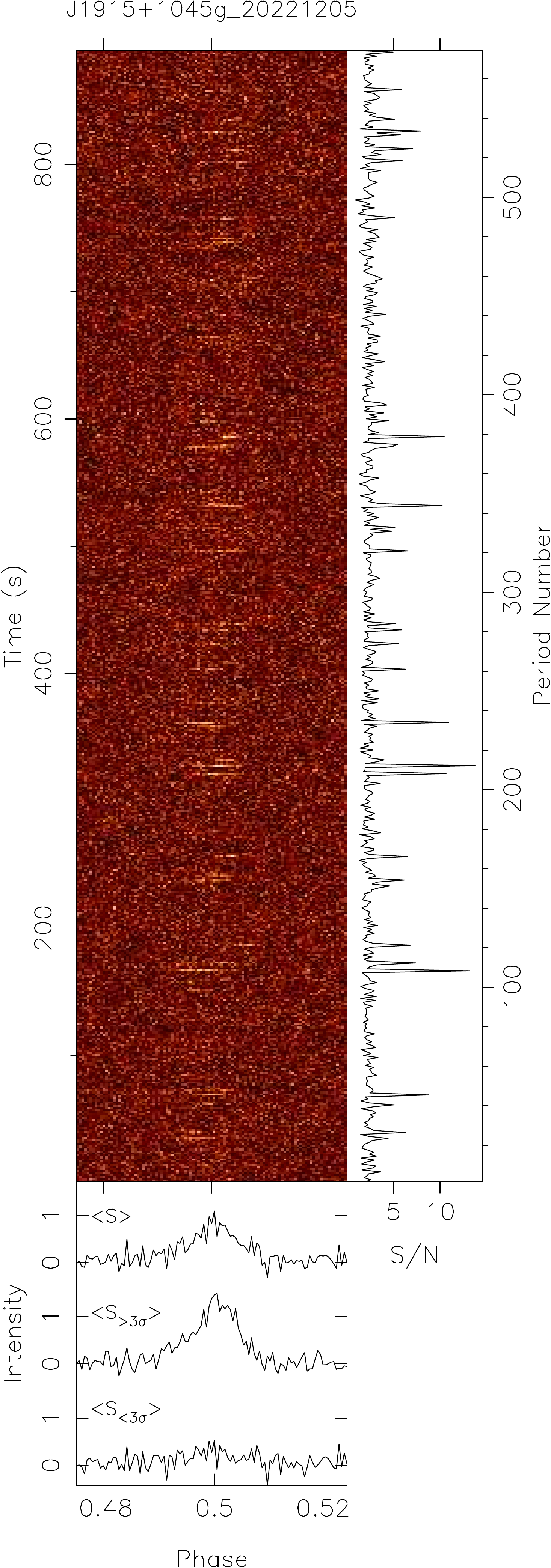}
\caption{{\it -- continued}.}
\end{figure*}
\addtocounter{figure}{-1}
\begin{figure*}
  \centering
  \includegraphics[width=38mm]{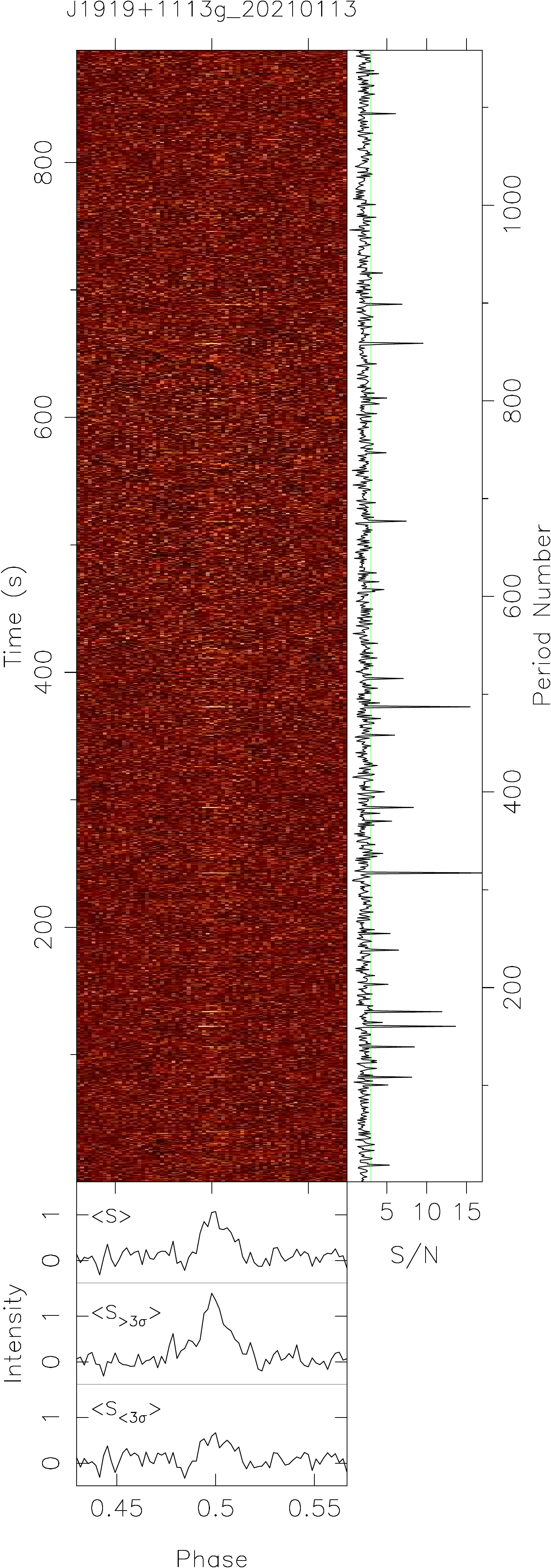} 
  \includegraphics[width=38mm]{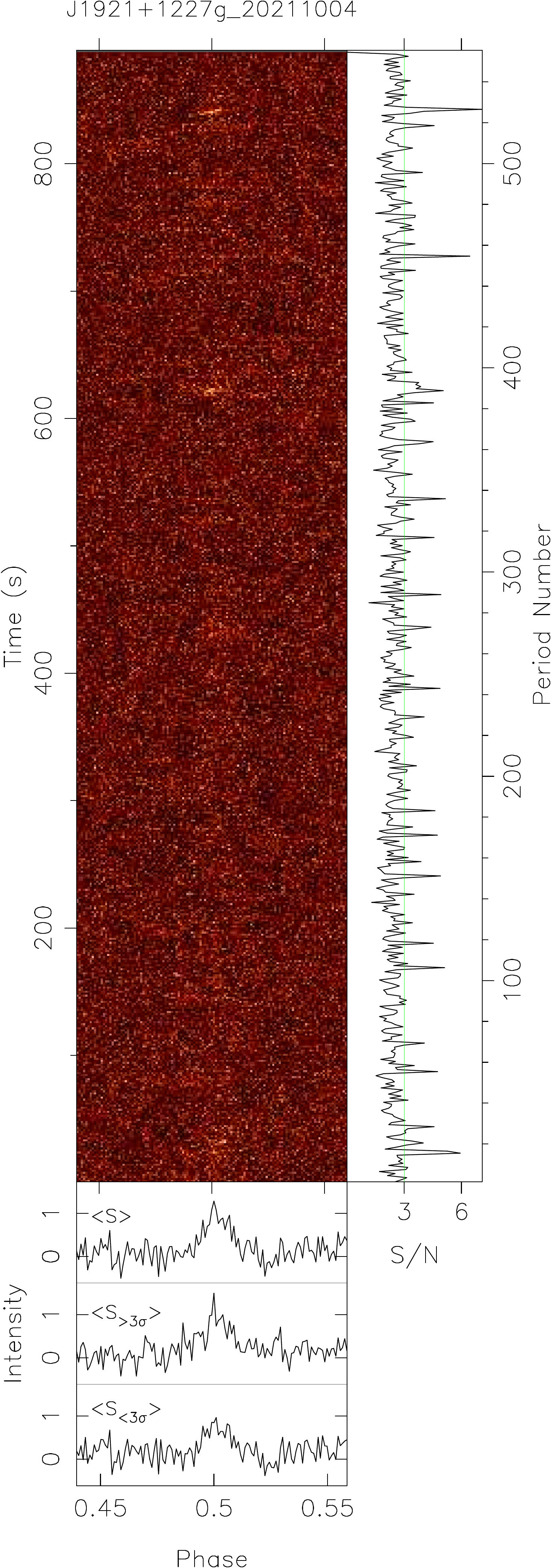} 
  \includegraphics[width=38mm]{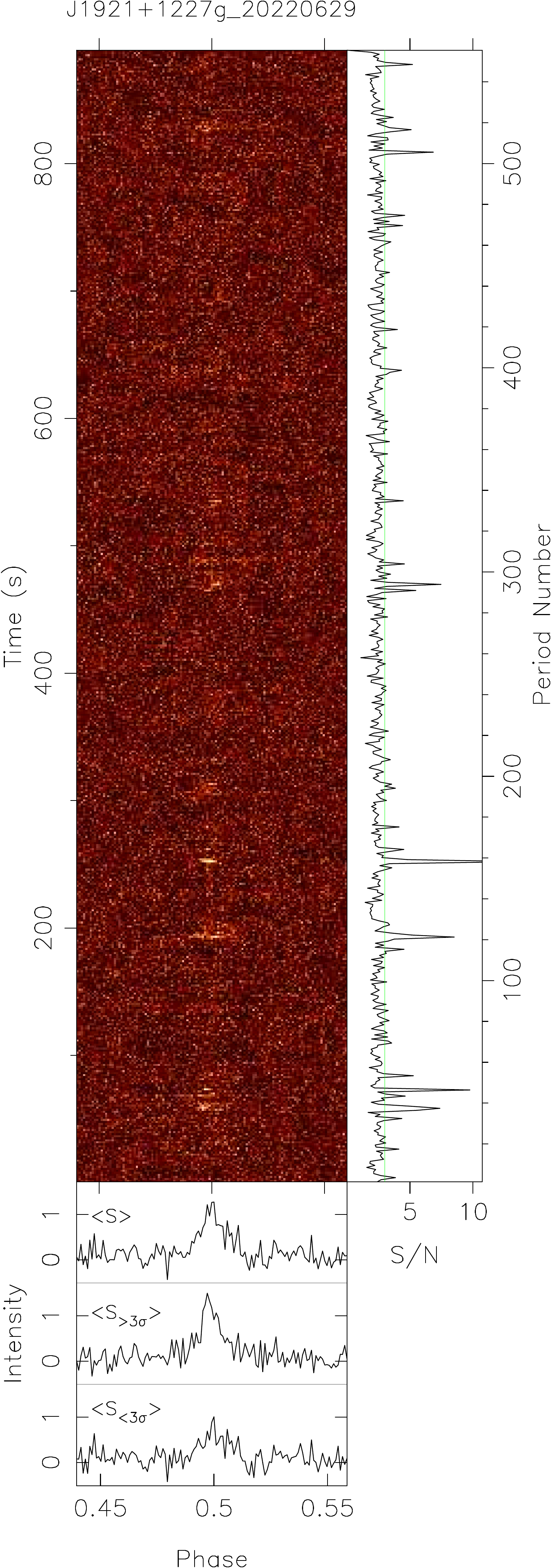} 
  \includegraphics[width=38mm]{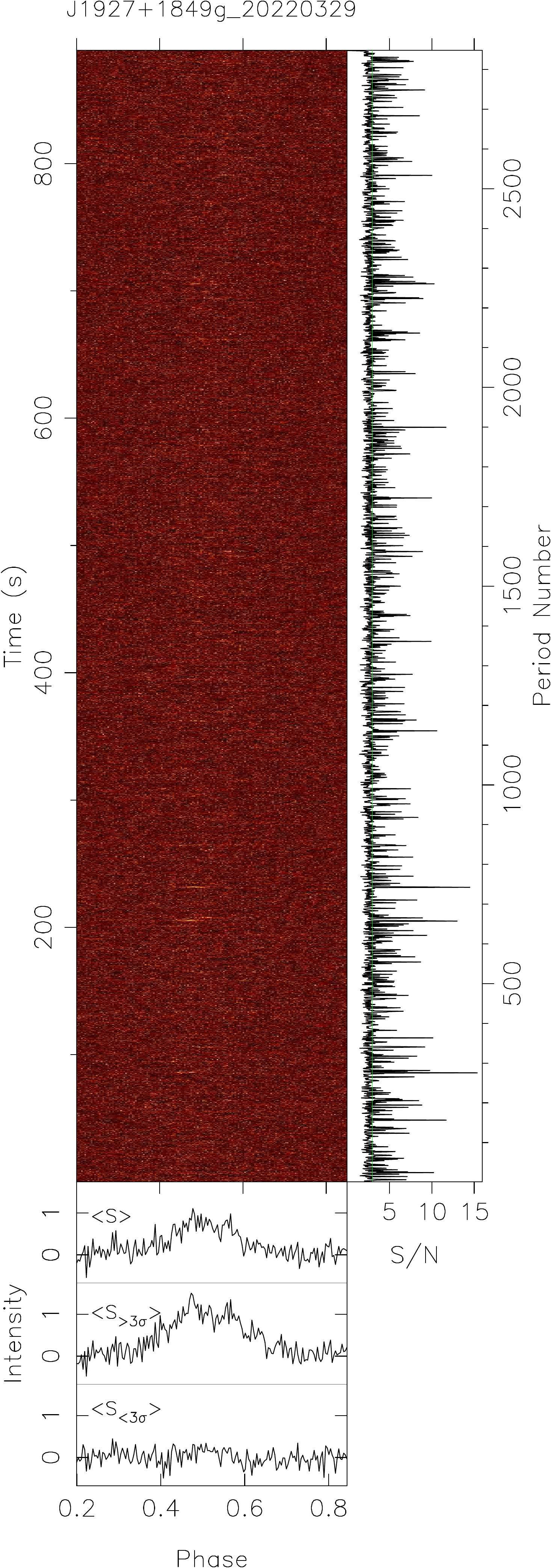} 
  \includegraphics[width=38mm]{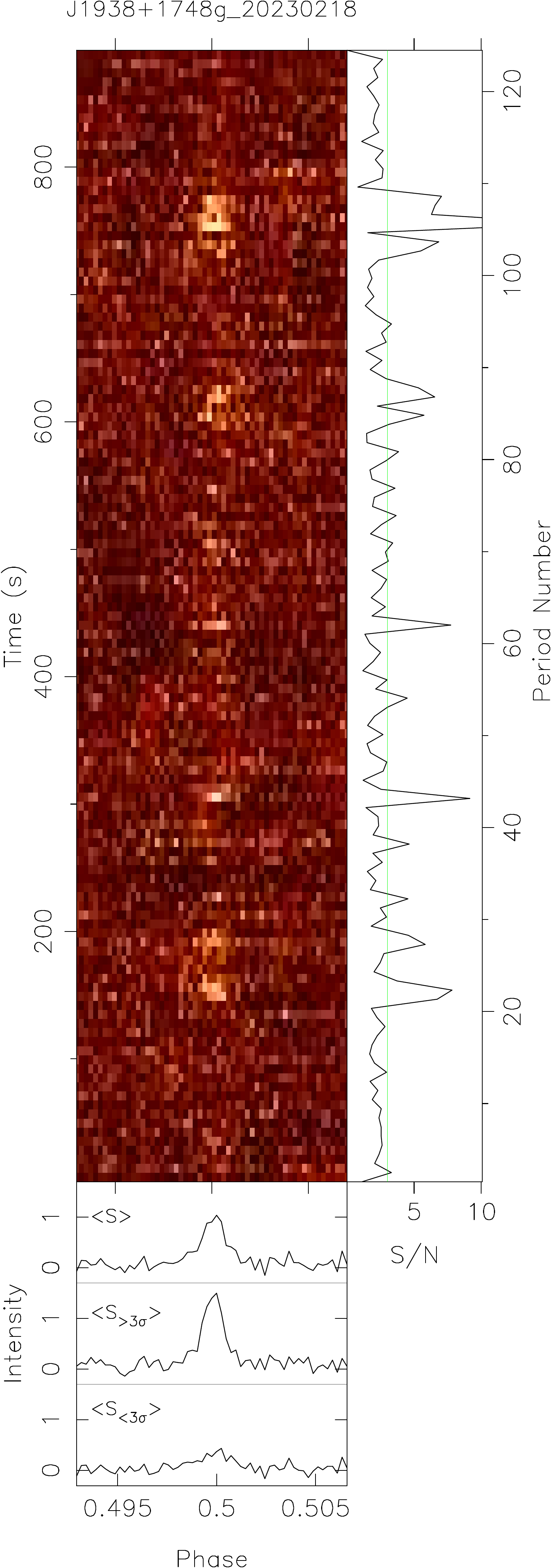}
  \includegraphics[width=38mm]{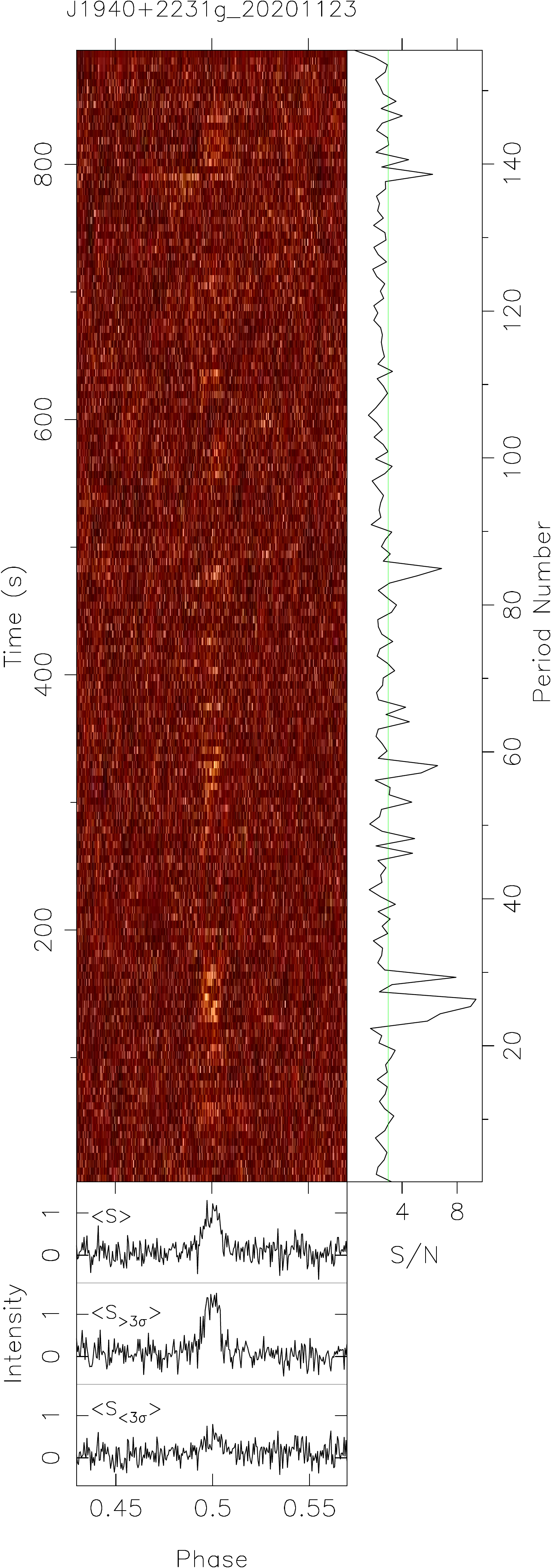} 
  \includegraphics[width=38mm]{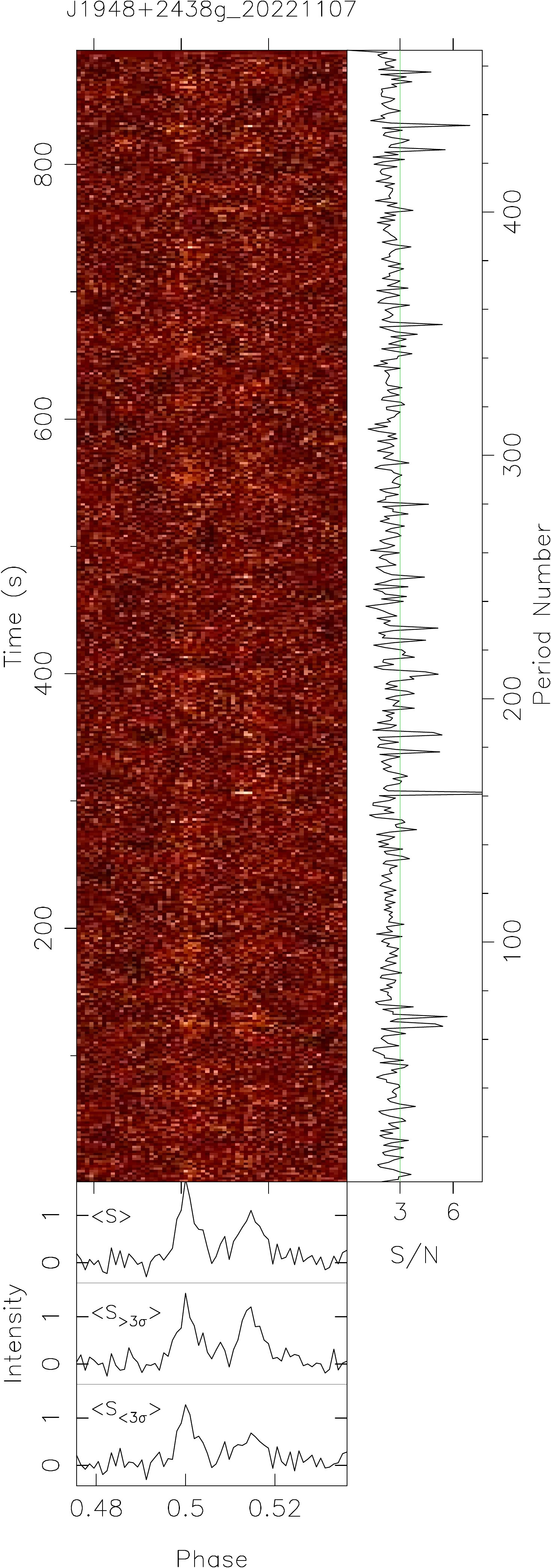}
  \includegraphics[width=38mm]{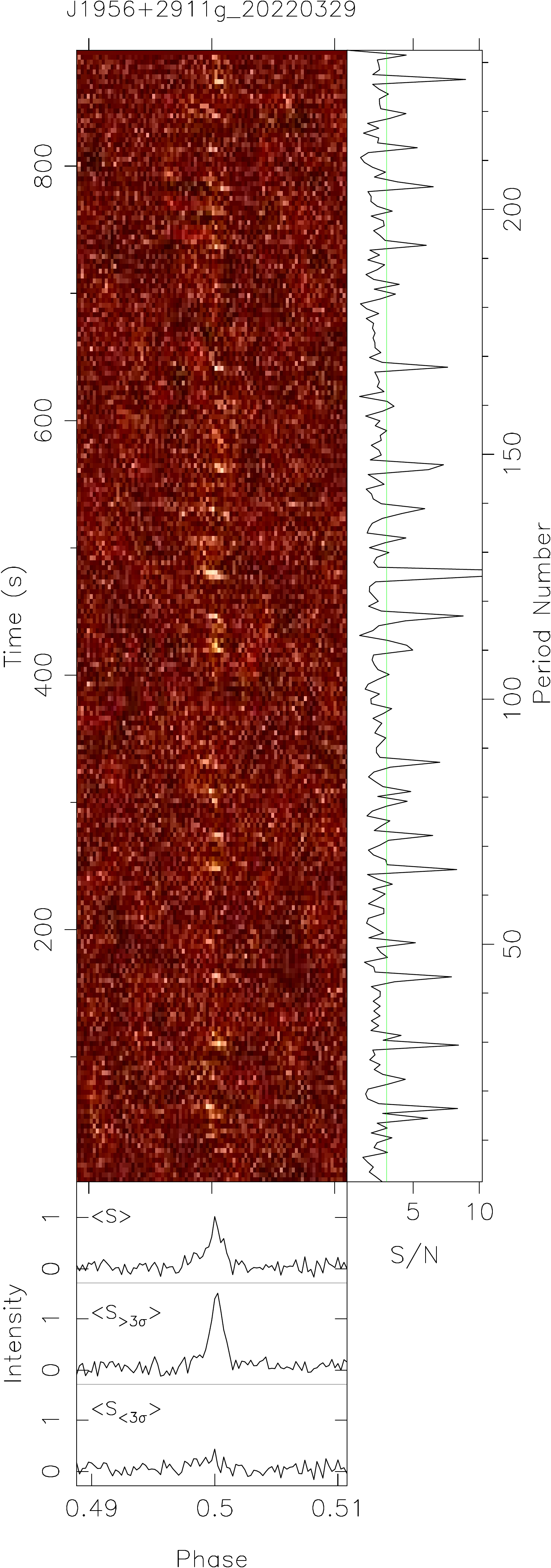} %SP077
\caption{{\it -- continued and ended}.}
\end{figure*}

%%%%%%%%%%%%%%%%%%%%%%%%%%%%%%%%%%%%%%%%%%%%% Normal pulsar or just few nulling fractions
\begin{figure*}[!htp]
  \centering
  \includegraphics[width=38mm]{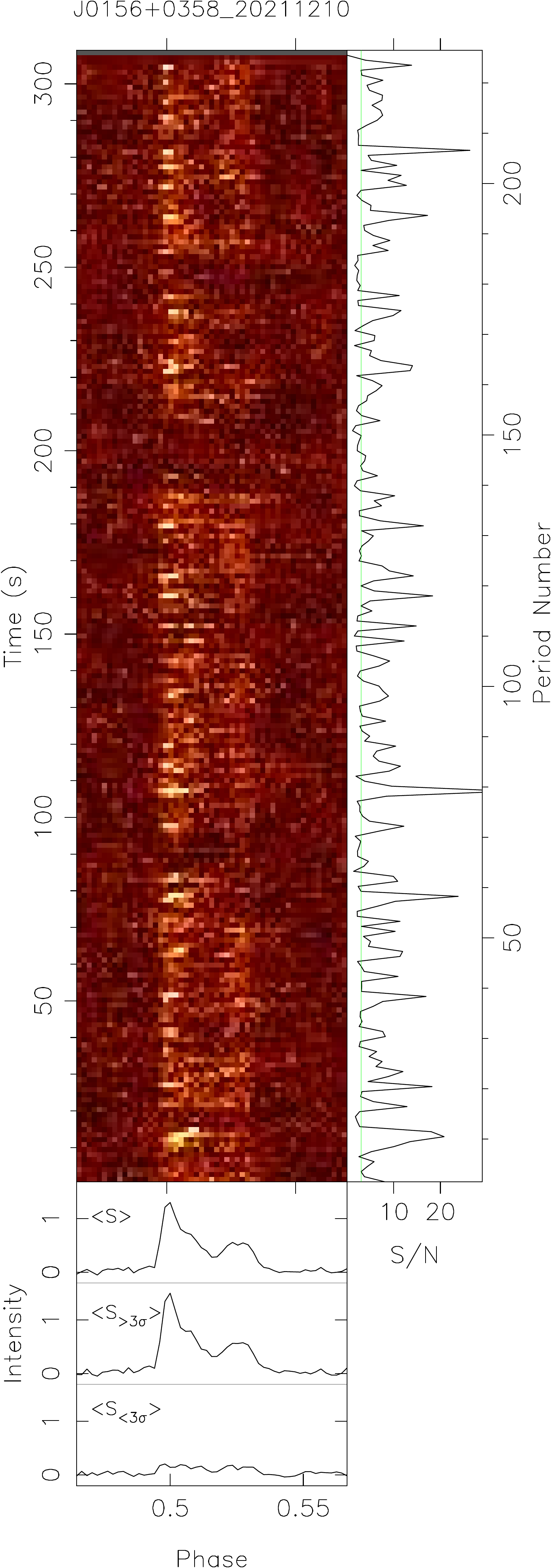} %J0156+0400
  \includegraphics[width=38mm]{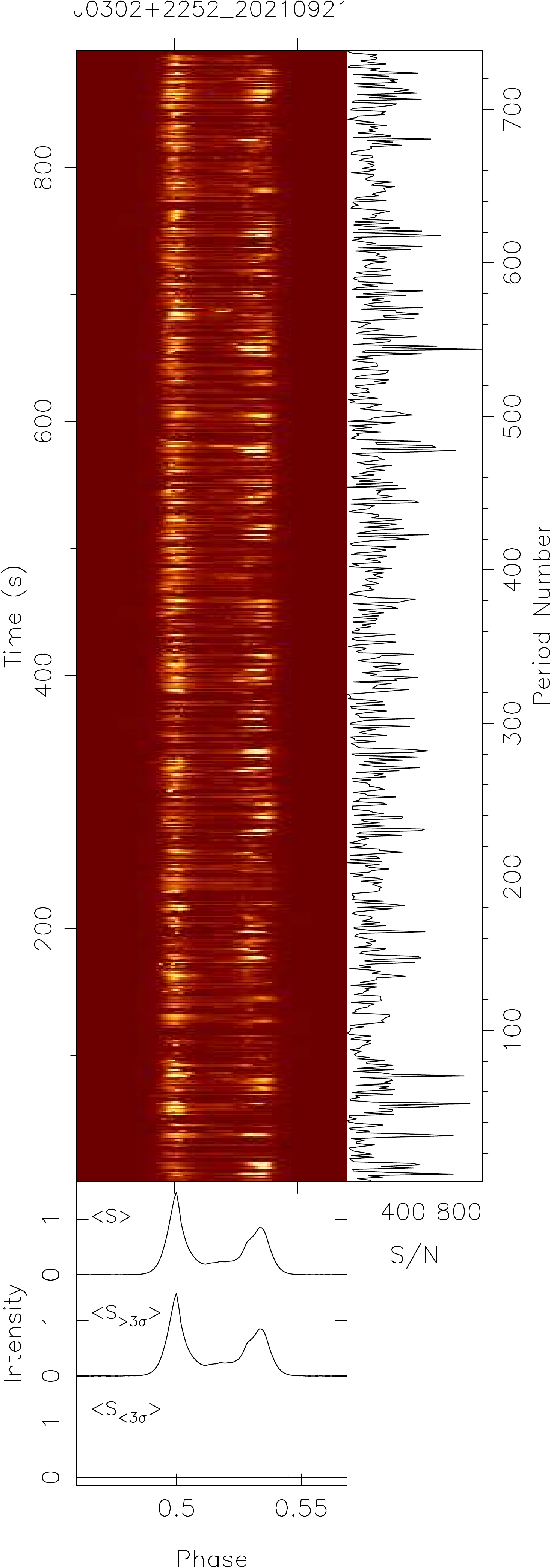} 
  \includegraphics[width=38mm]{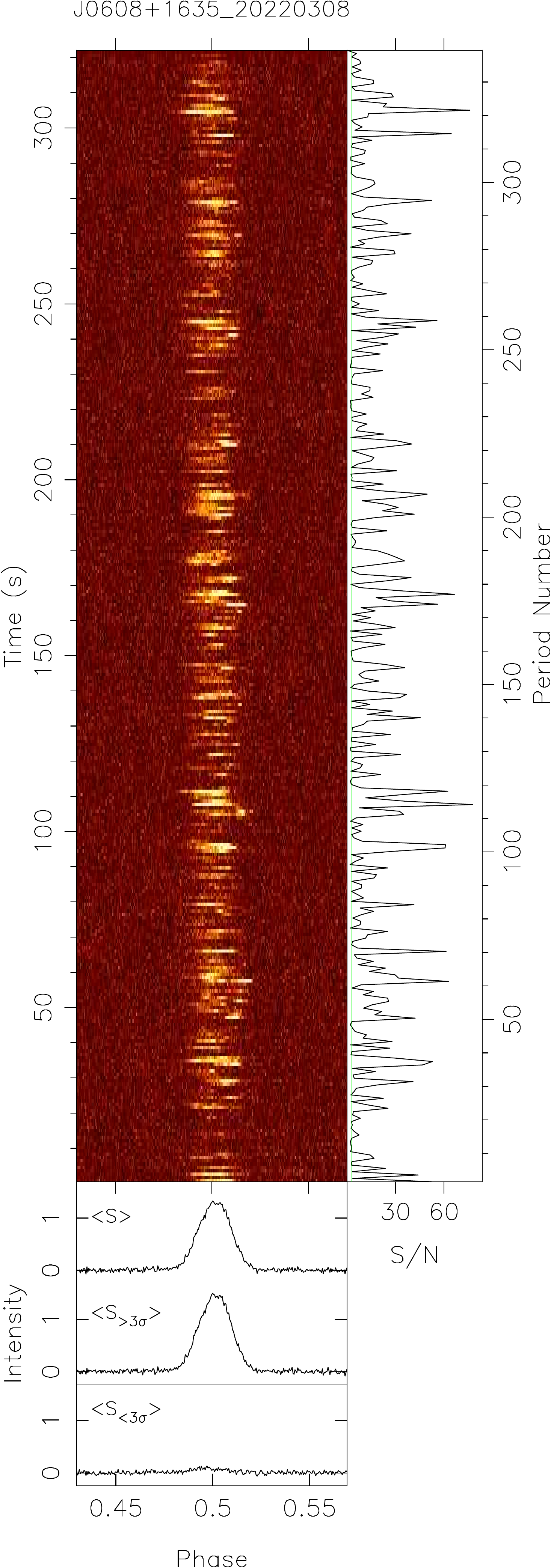} %J0608+1635
  \includegraphics[width=38mm]{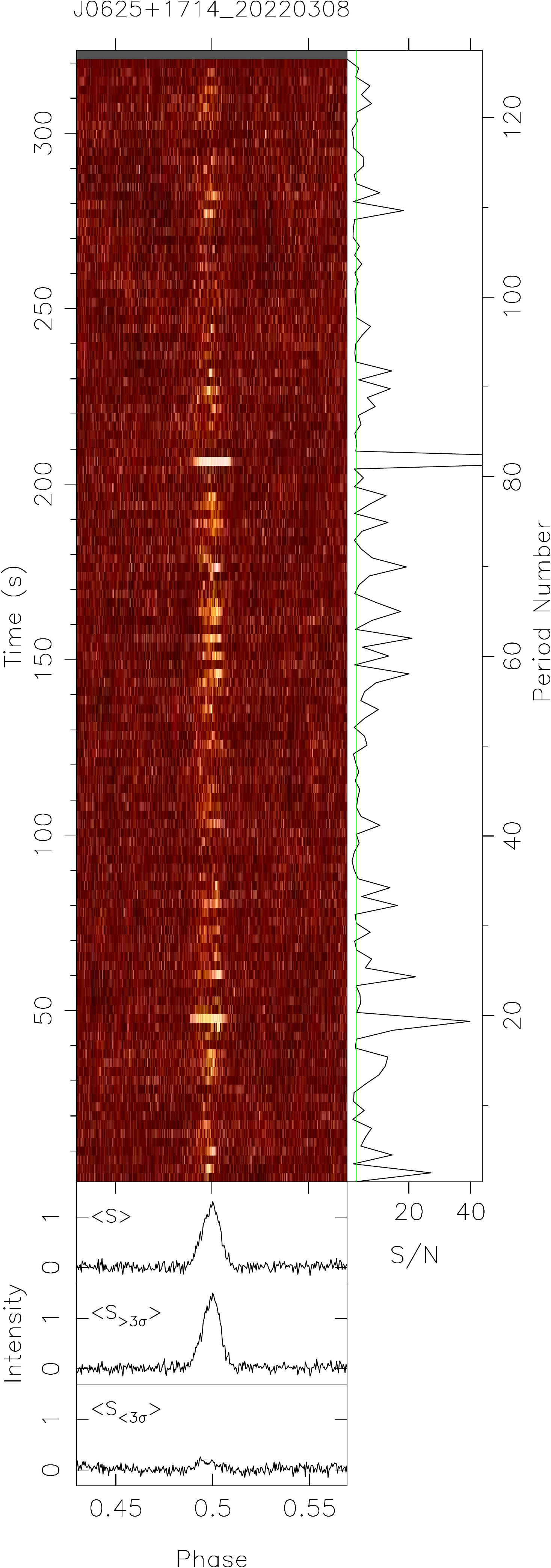} \\[1.0mm]%J0625+1714
  \includegraphics[width=38mm]{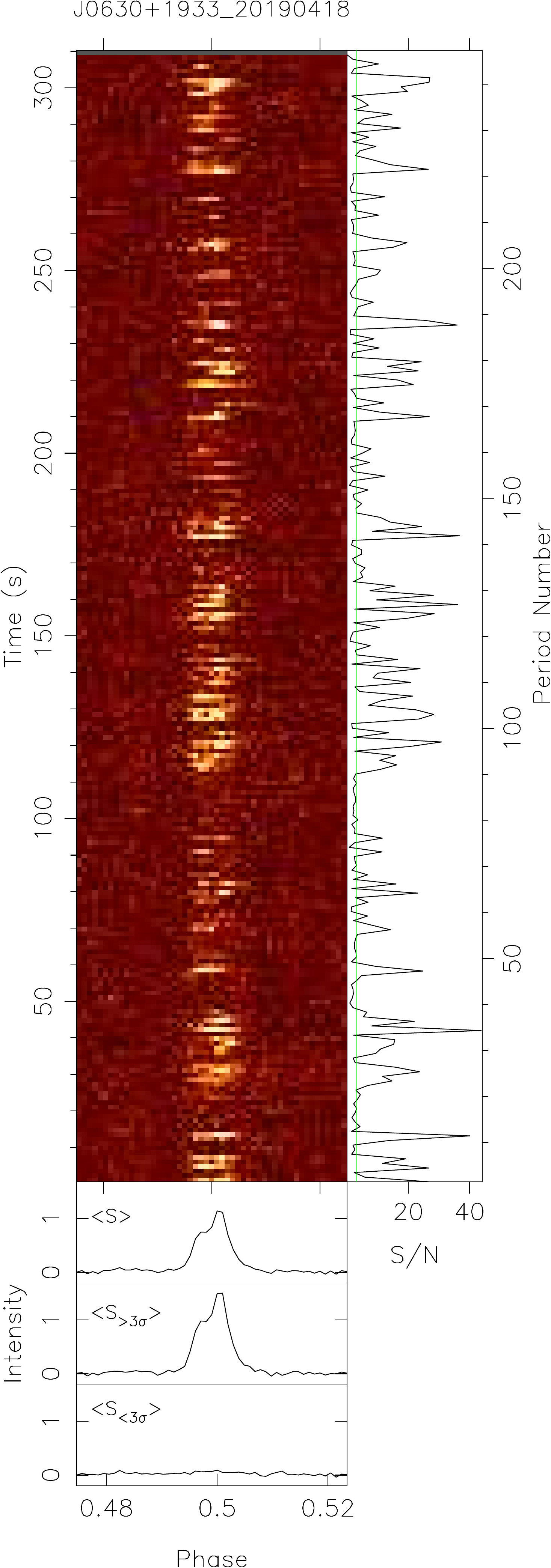} %J0630$+$1933
  \includegraphics[width=38mm]{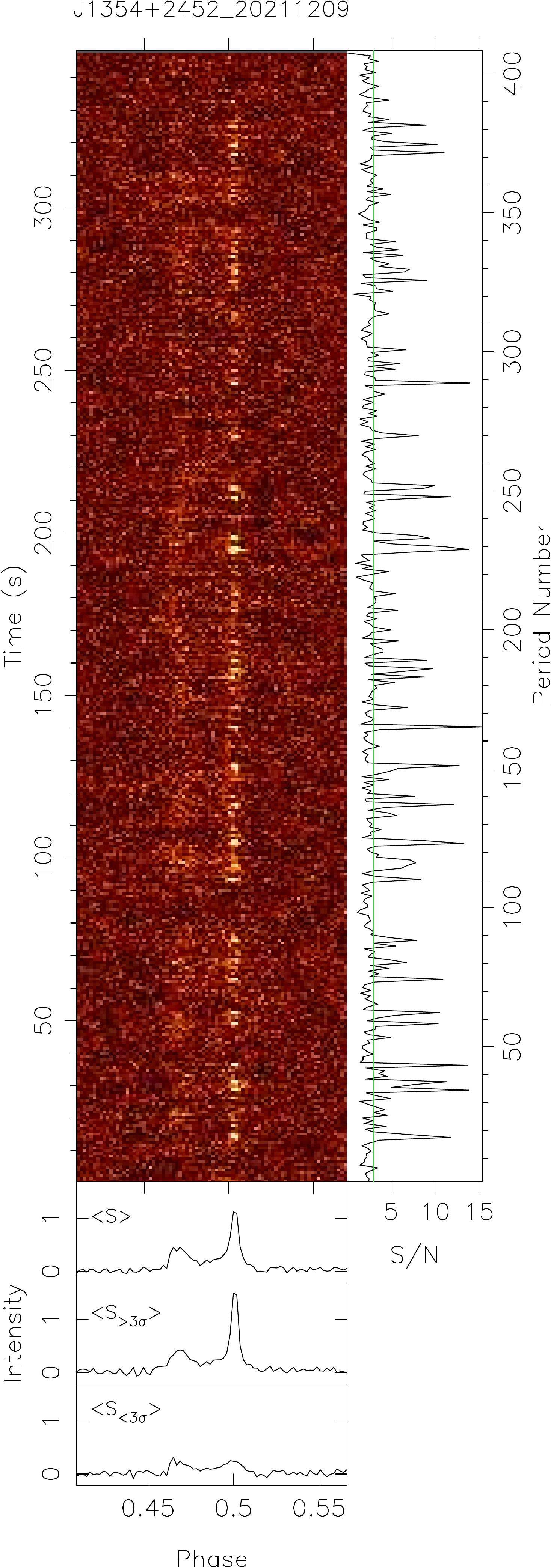} %J1354+2400
  \includegraphics[width=38mm]{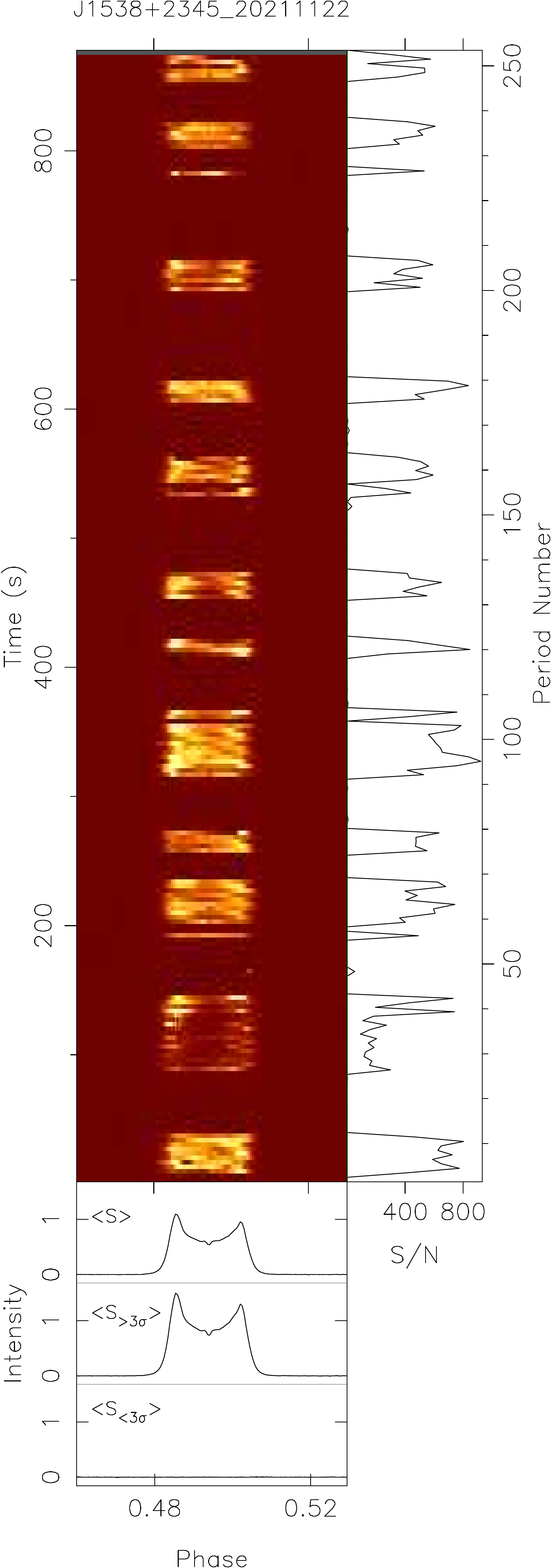}
  \includegraphics[width=38mm]{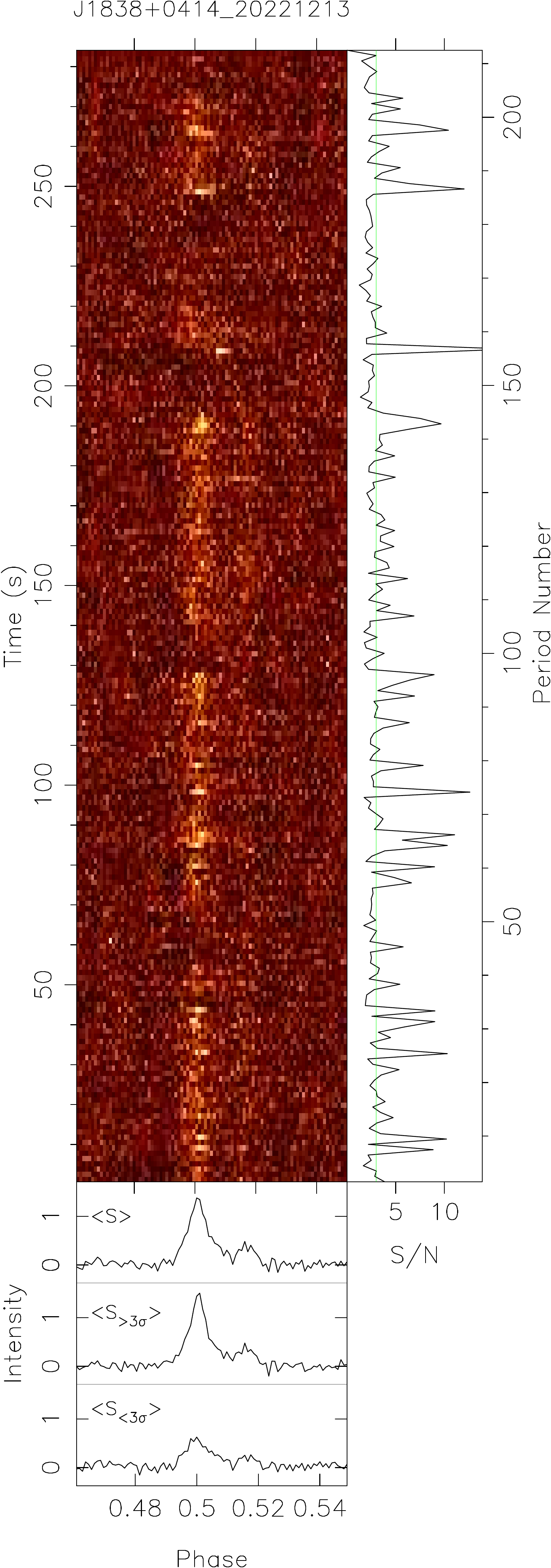}
  \caption{The known RRATs for just normal pulsars though nulling features
  occasionally emerges.}
  \label{fig:APPknownRRAT1}
\end{figure*}
\addtocounter{figure}{-1}
\begin{figure*}[!htp]
\centering
  \includegraphics[width=38mm]{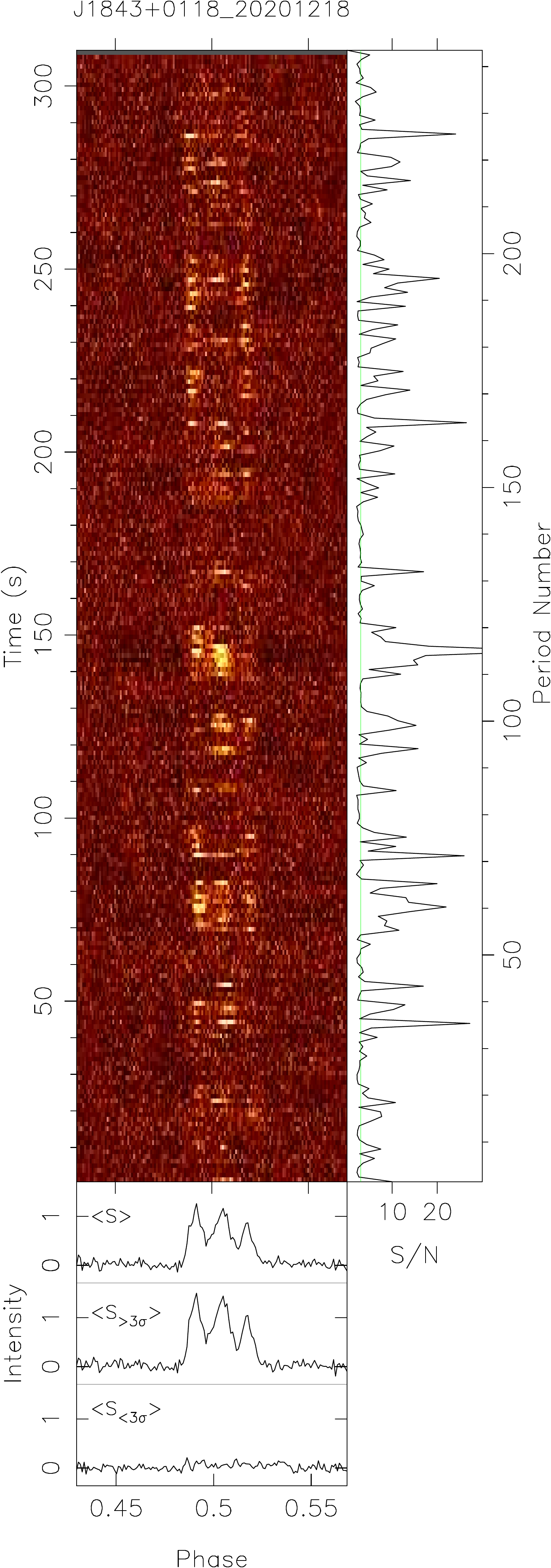} 
  \includegraphics[width=38mm]{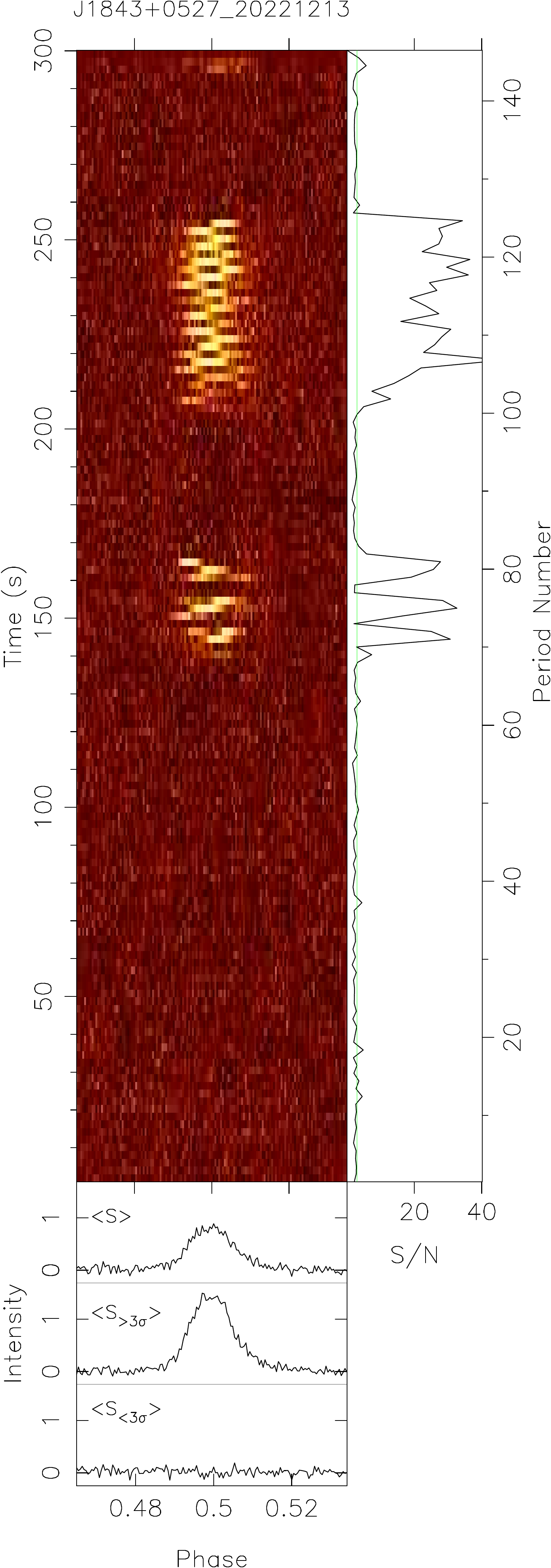} 
  \includegraphics[width=38mm]{KnowRRATs/J1849+0106sp_20211014_tracking-M01-P1-c2048b1.ar.r.psh.Fp.debased.pdf}
  \includegraphics[width=38mm]{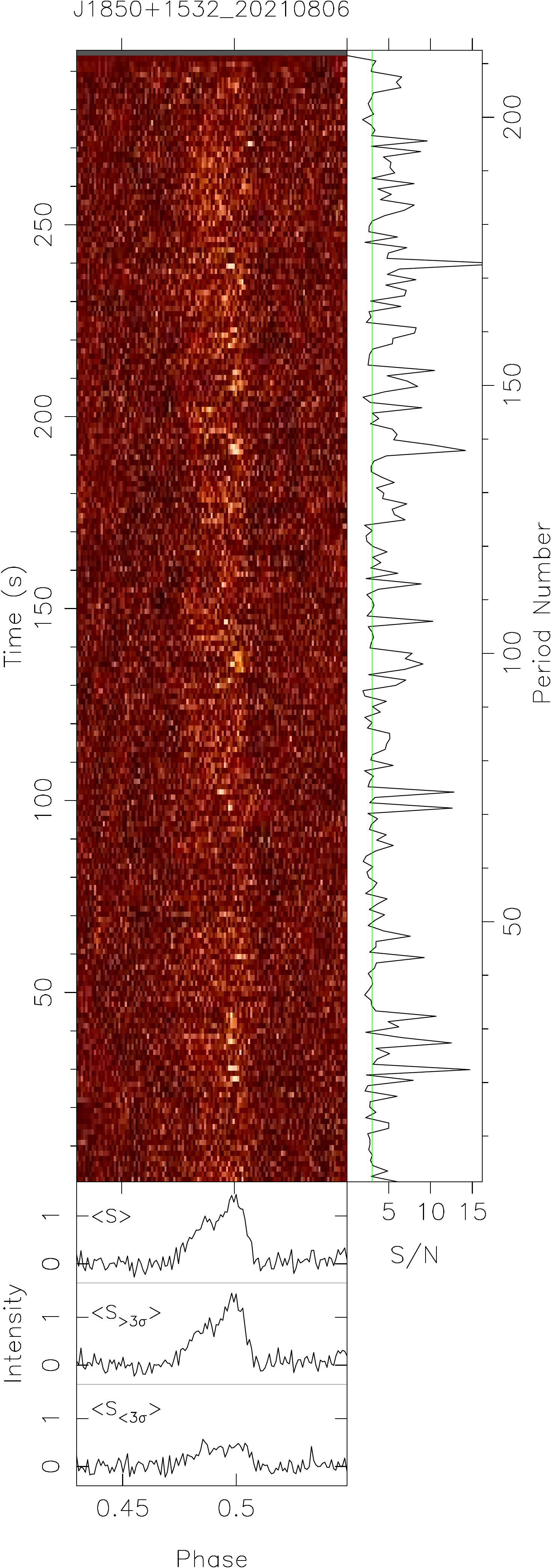} %1850+1532
  \includegraphics[width=38mm]{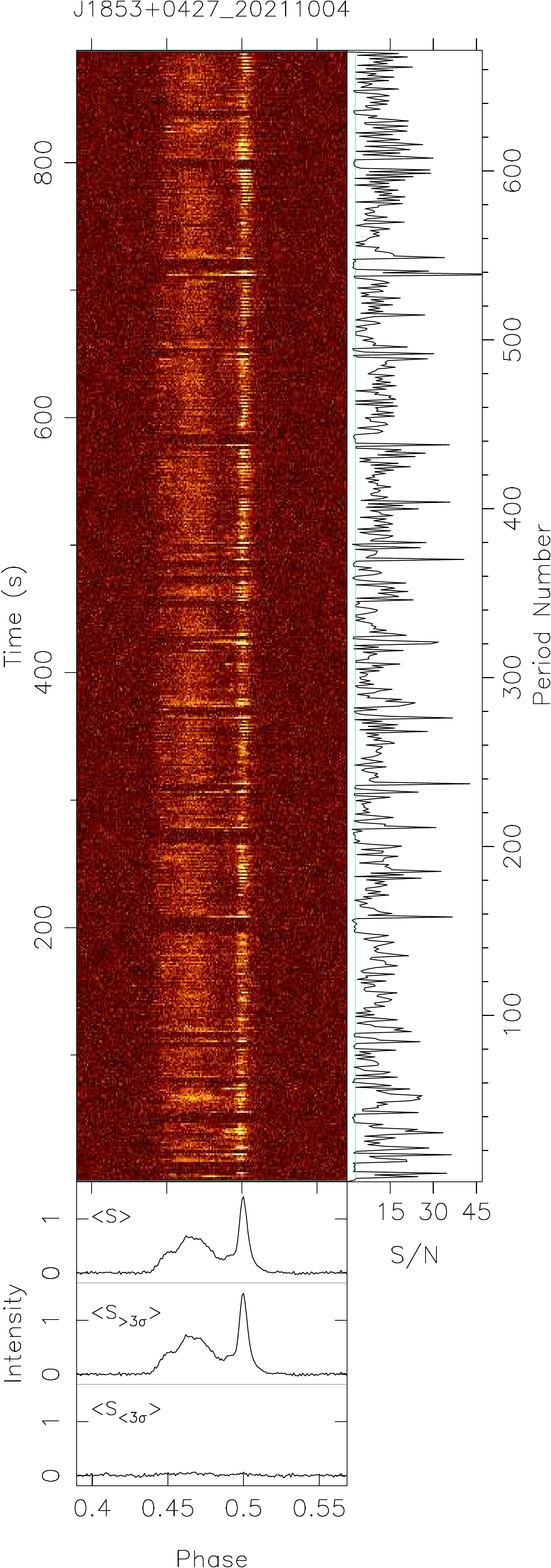} 
  \includegraphics[width=38mm]{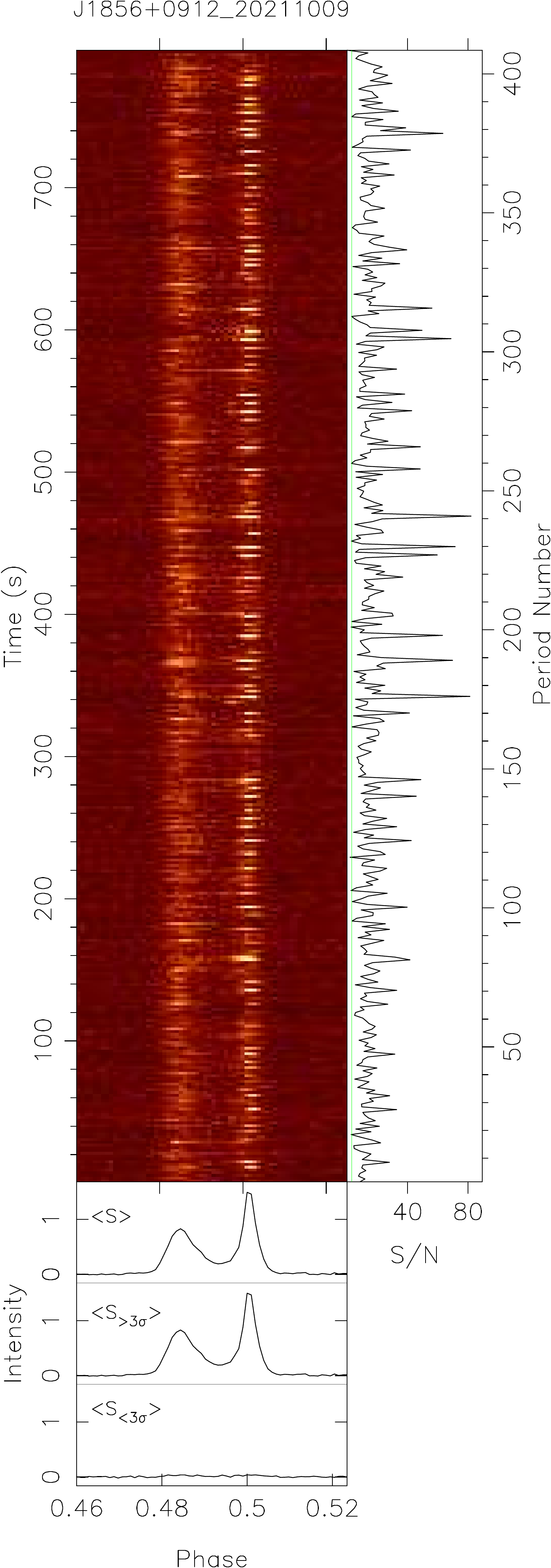} 
  \includegraphics[width=38mm]{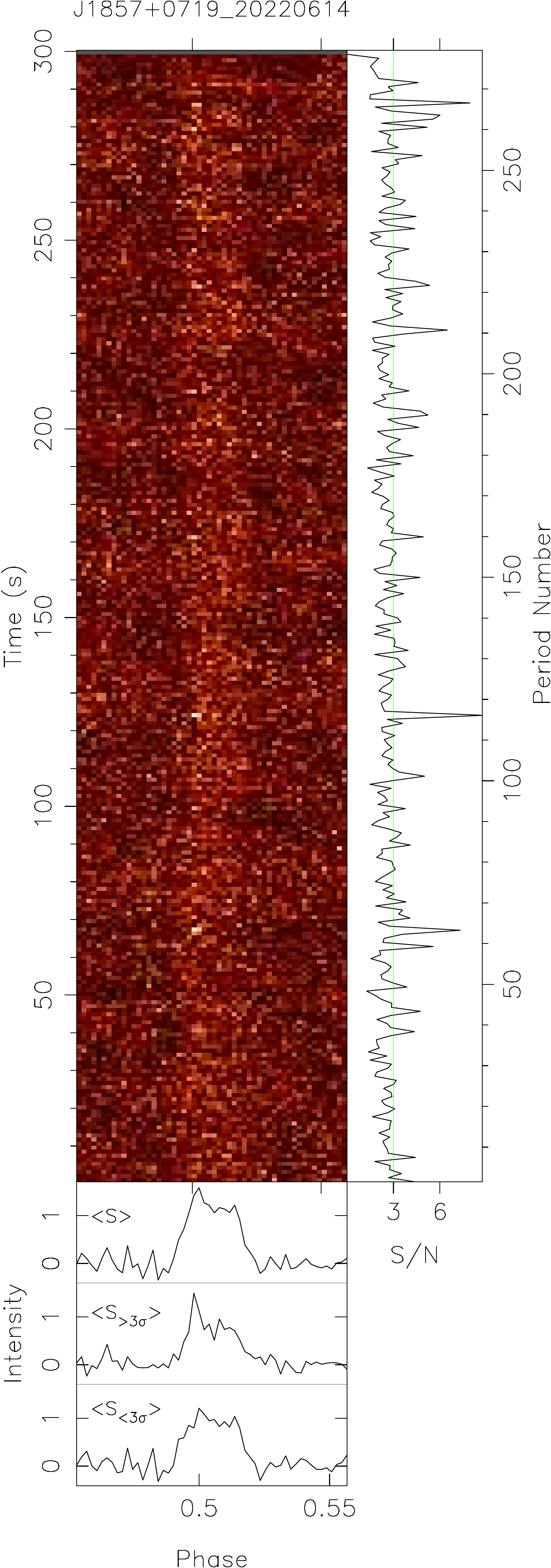} %\\[1.0mm]%J1857+0719
  \includegraphics[width=38mm]{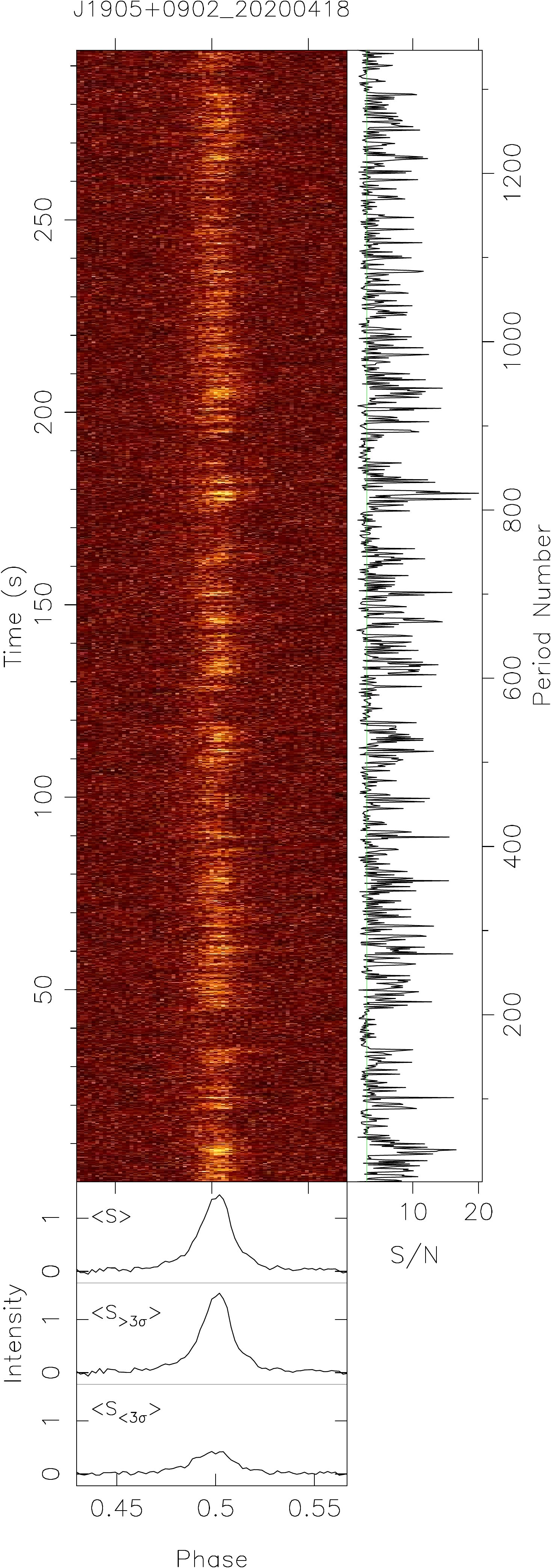} 
  \caption{{\it -- continued}.}
\end{figure*}
\addtocounter{figure}{-1}
\begin{figure*}
\centering
  \includegraphics[width=38mm]{KnowRRATs/J1908+1351sp_20211004_tracking-M01-P1-c2048b1.ar.zap.Fp.debased.pdf} 
  \includegraphics[width=38mm]{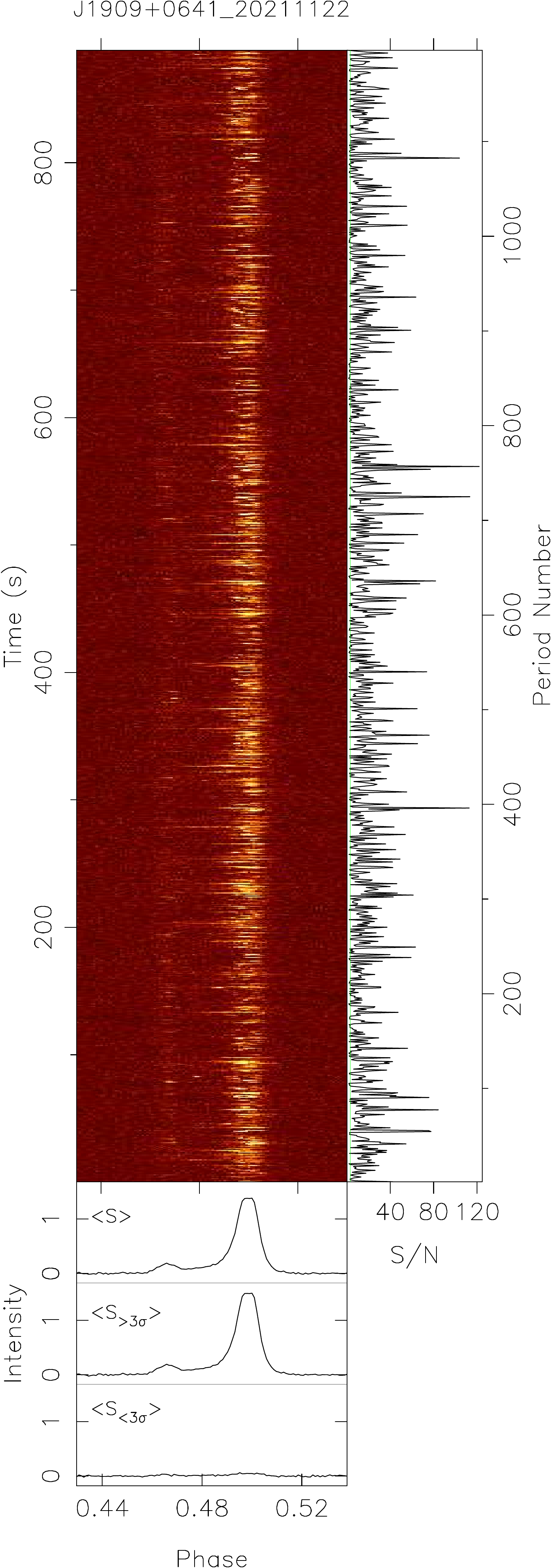} %J1909+0641
  \includegraphics[width=38mm]{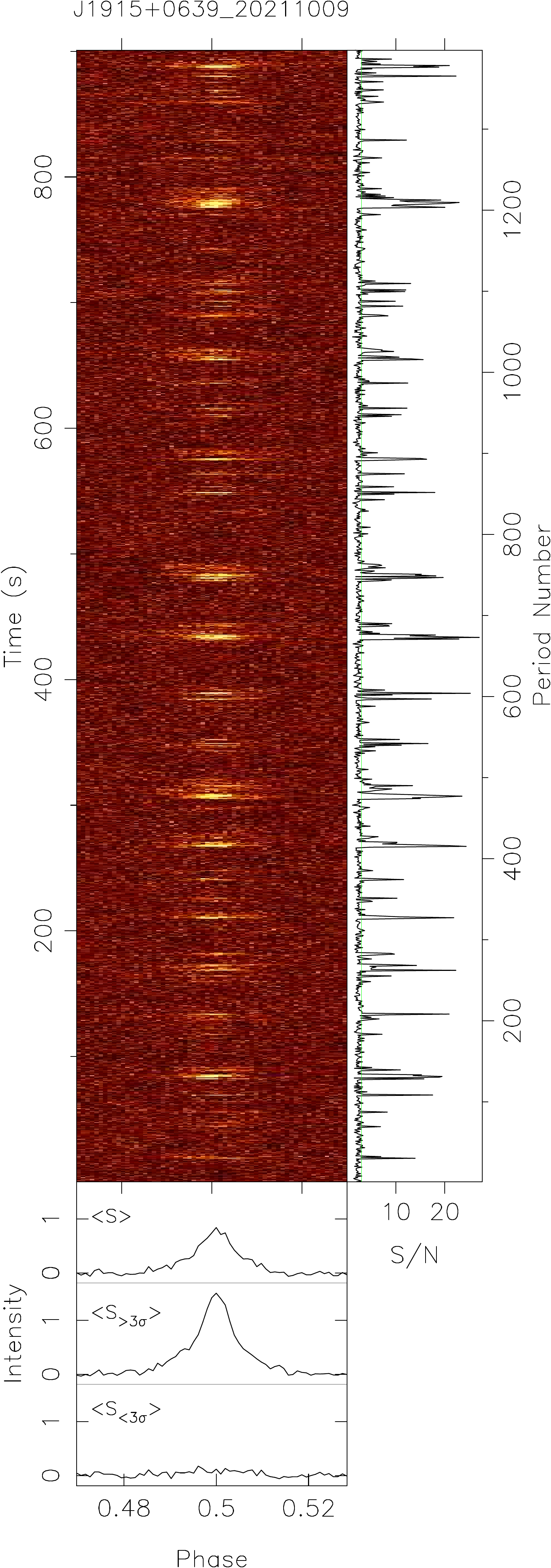}
  \includegraphics[width=38mm]{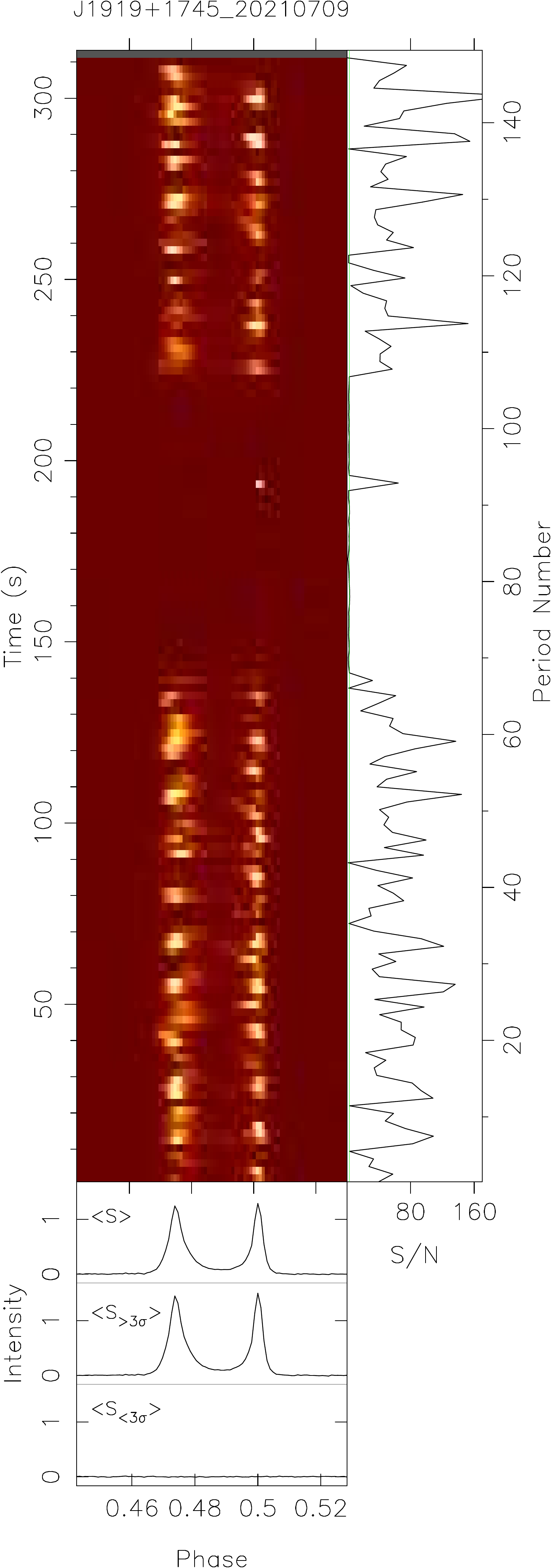} 
  \includegraphics[width=38mm]{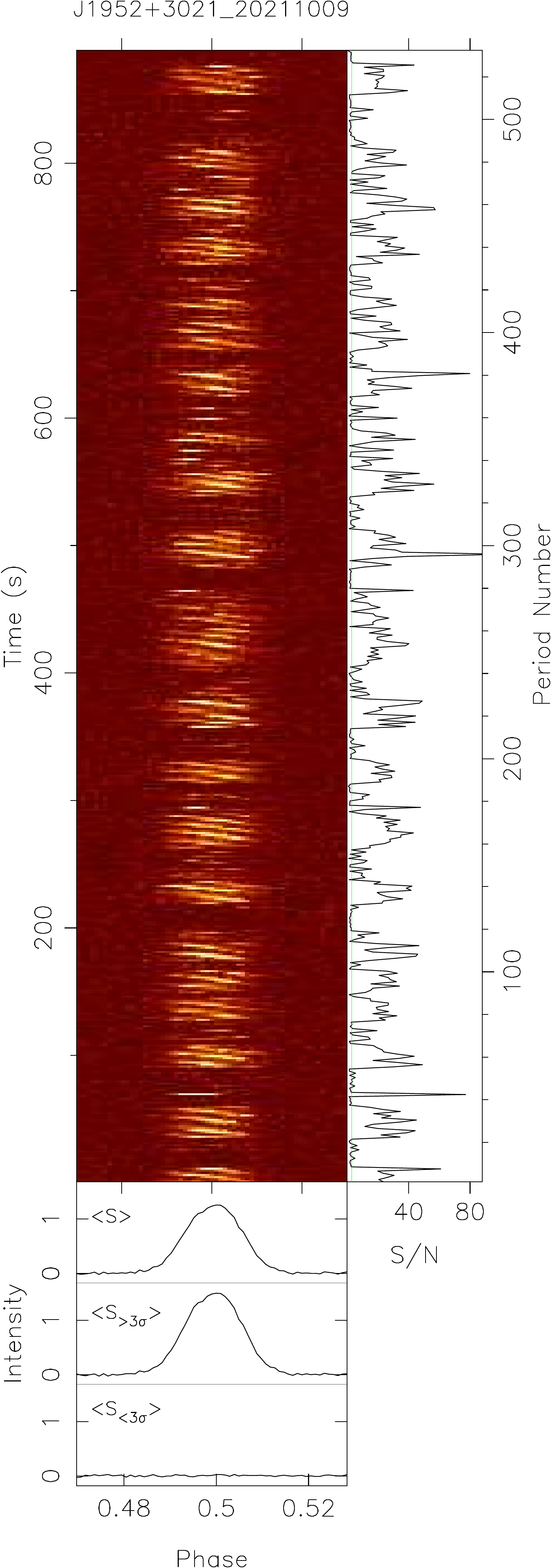} 
  \includegraphics[width=38mm]{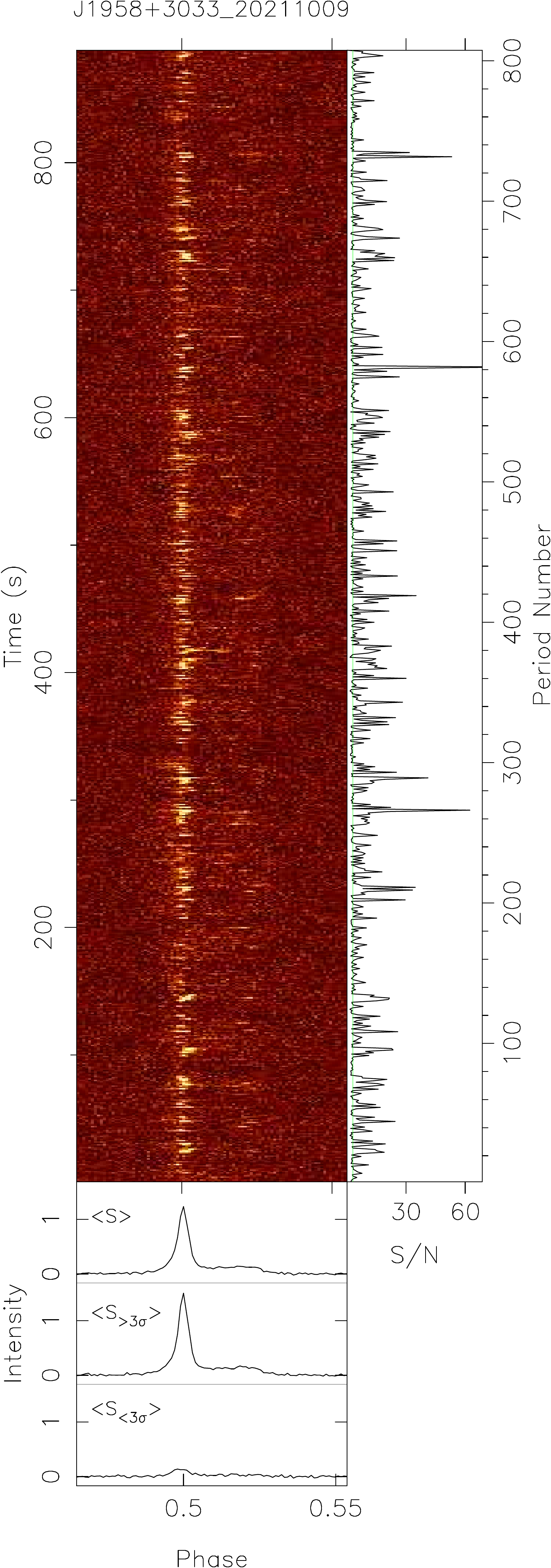} 
  \includegraphics[width=38mm]{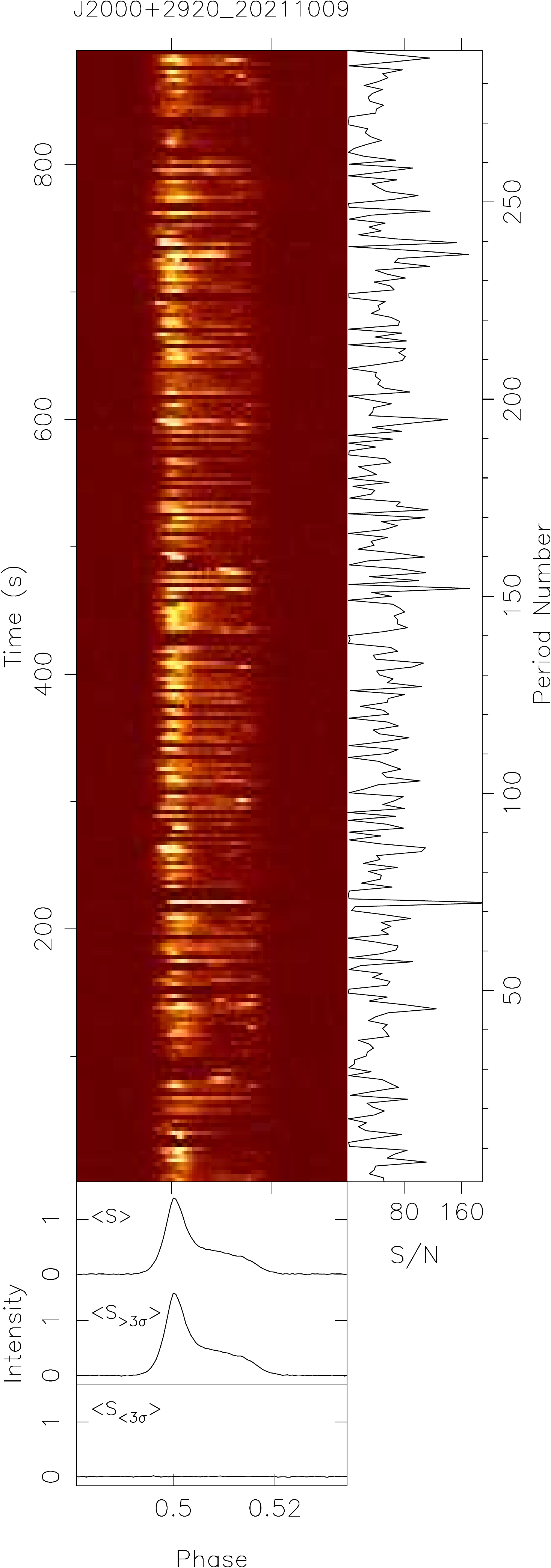}
  \includegraphics[width=38mm]{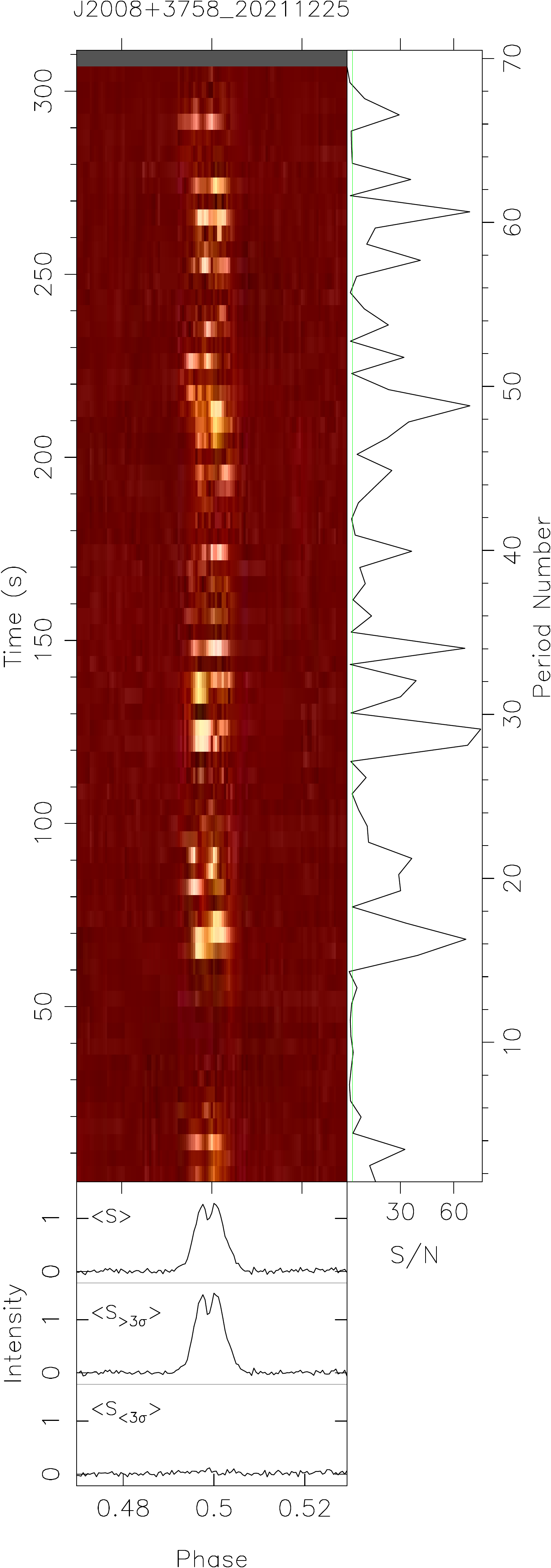} %J2008+3757, just RRAT-like pulsar, J2008+37 ,CHIME
  \caption{{\it -- continued}.}
\end{figure*}
\addtocounter{figure}{-1}
\begin{figure*}
\centering
  \includegraphics[width=38mm]{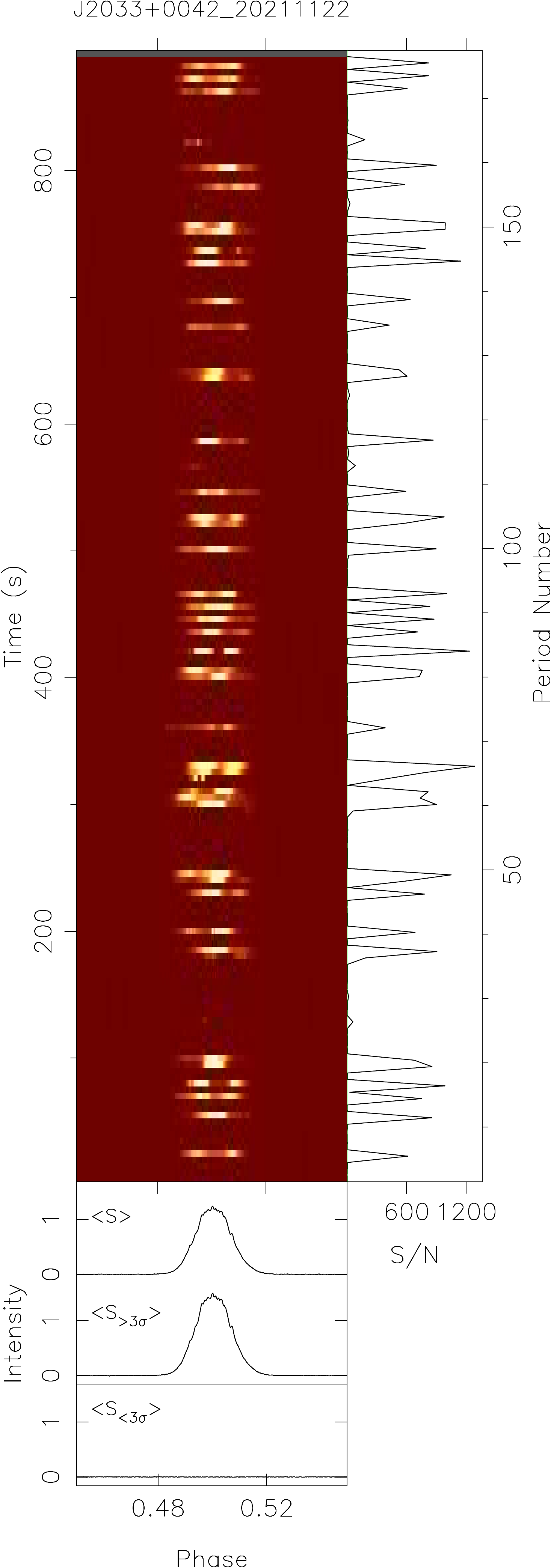} %J2033+0042
  \caption{{\it -- continued and ended}.}
\end{figure*}

%%%%%%%%%%%%%%%%%%%%%%%%%%%%%%%%%%%%%%%%%%%%% Nulling Known RRAT

\begin{figure*}%[!htp]
  \centering
\includegraphics[width=38mm]{KnowRRATs/J0103+5354_J0103+54_20210927_snapshotdec-M04-P4-c2048b1.ar.zap.Fp.debased.pdf}
  \includegraphics[width=38mm]{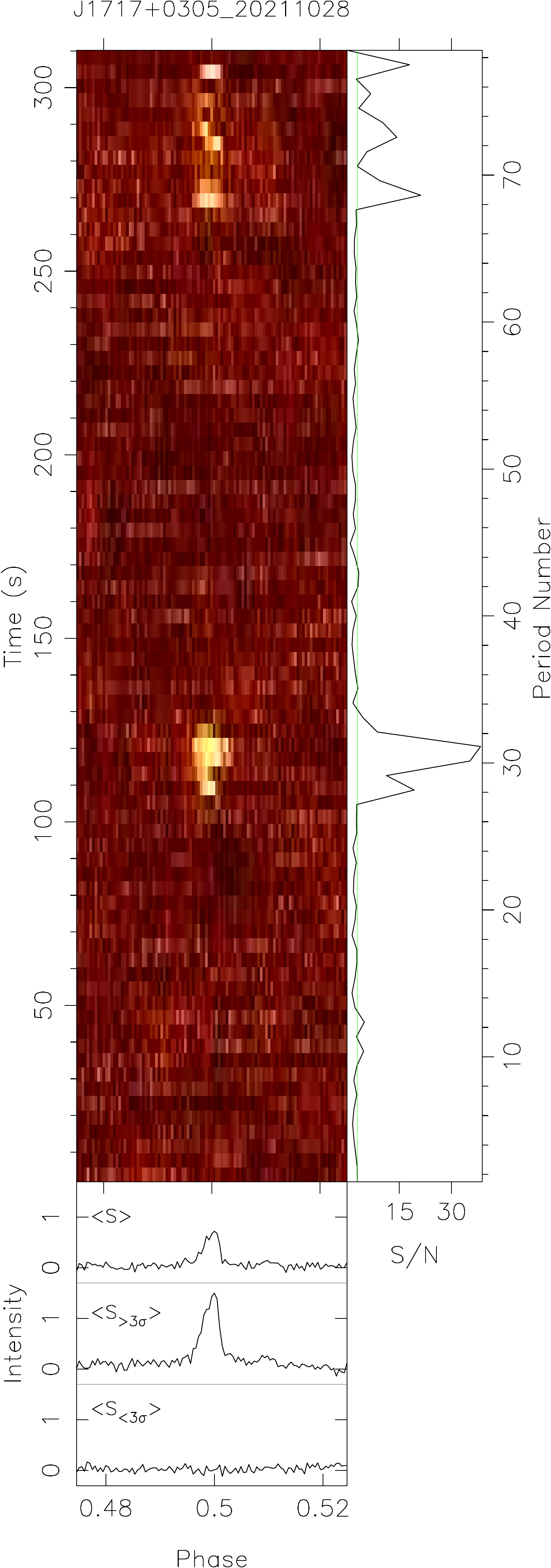} %J1717+03
  \includegraphics[width=38mm]{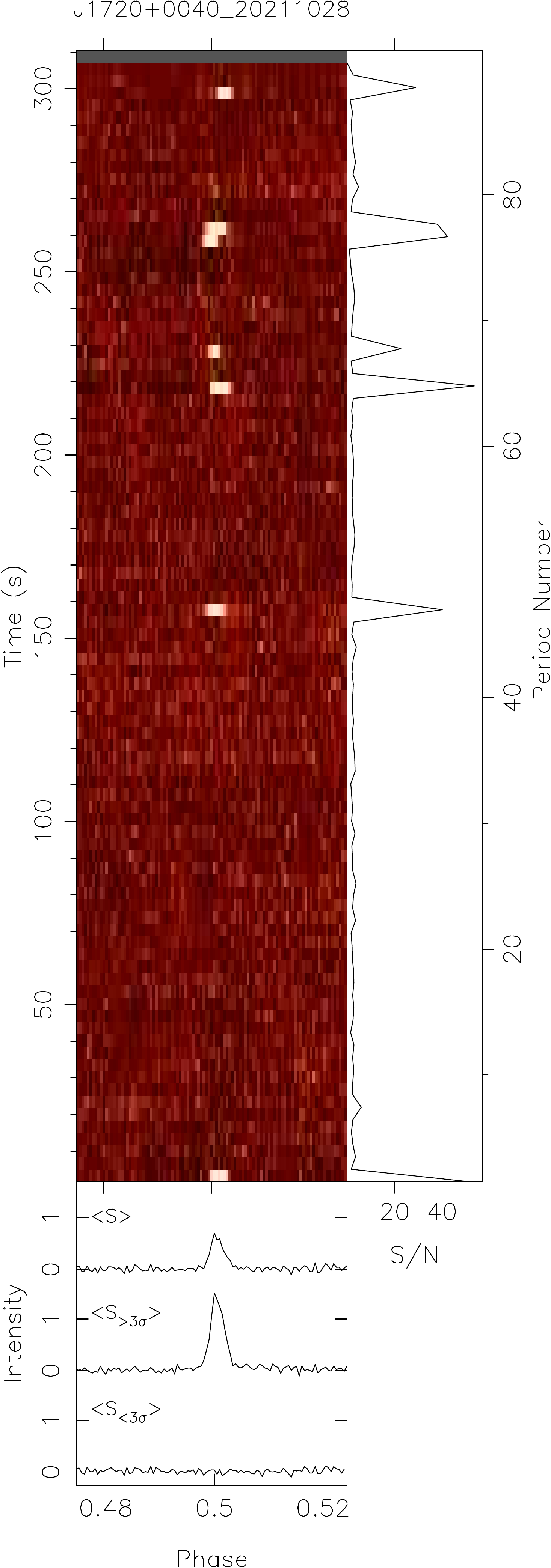}
  \includegraphics[width=38mm]{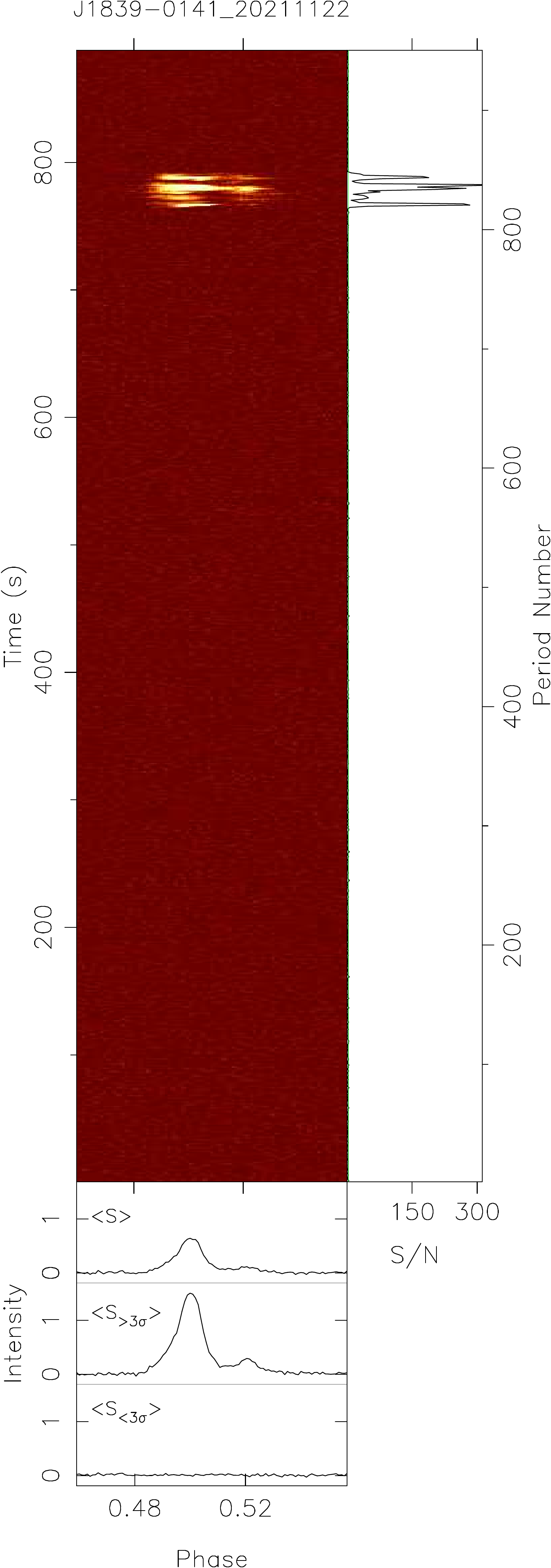} %J1839-0141
  \includegraphics[width=38mm]{KnowRRATs/J192904+173105_20210514_tracking-M04-P1-c2048b1.Fp.debased.pdf} %J1928+1725
  \caption{The known RRATs are shown as group of extremely nulling pulsars.}
    \label{fig:APPknownRRAT2}
\end{figure*}

%%%%%%%%%%%%%%%%%%%%%%%%%%%%%%%%%%%%%%%%%%%%%  Known RRAT as weak pulsars with sparse strong pulses
%\addtocounter{figure}{-1}
\begin{figure*}[!htp]
  \centering
  \includegraphics[width=38mm]{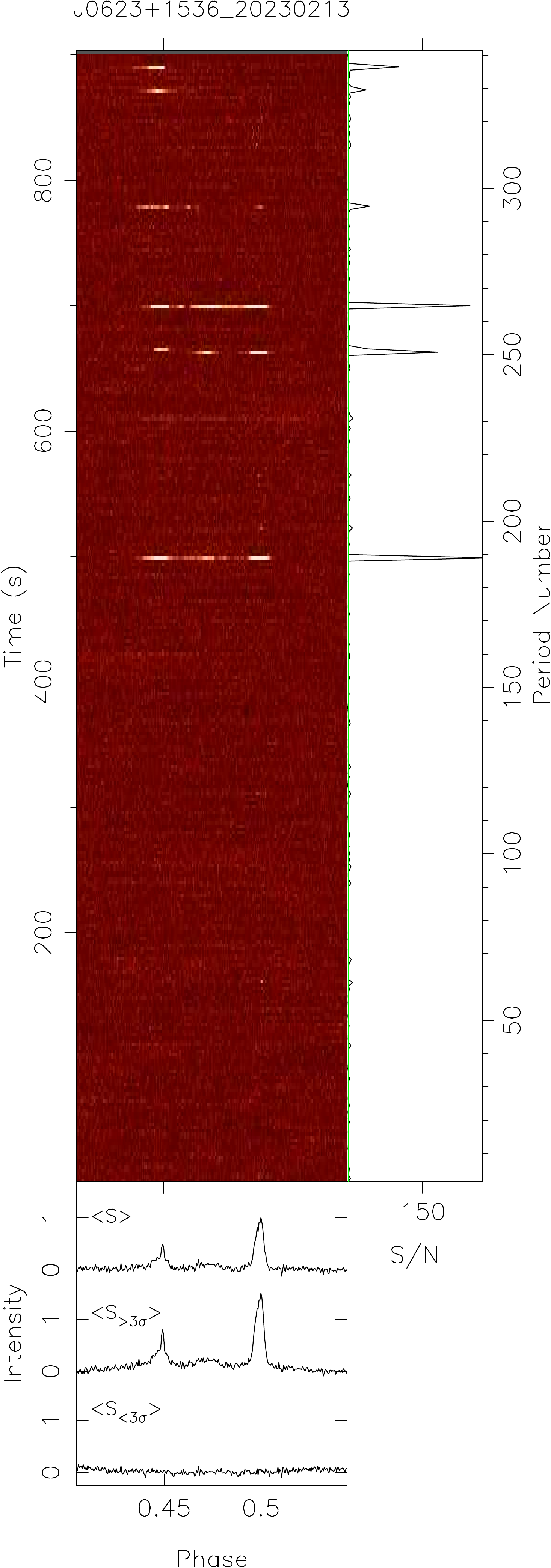}
  \includegraphics[width=38mm]{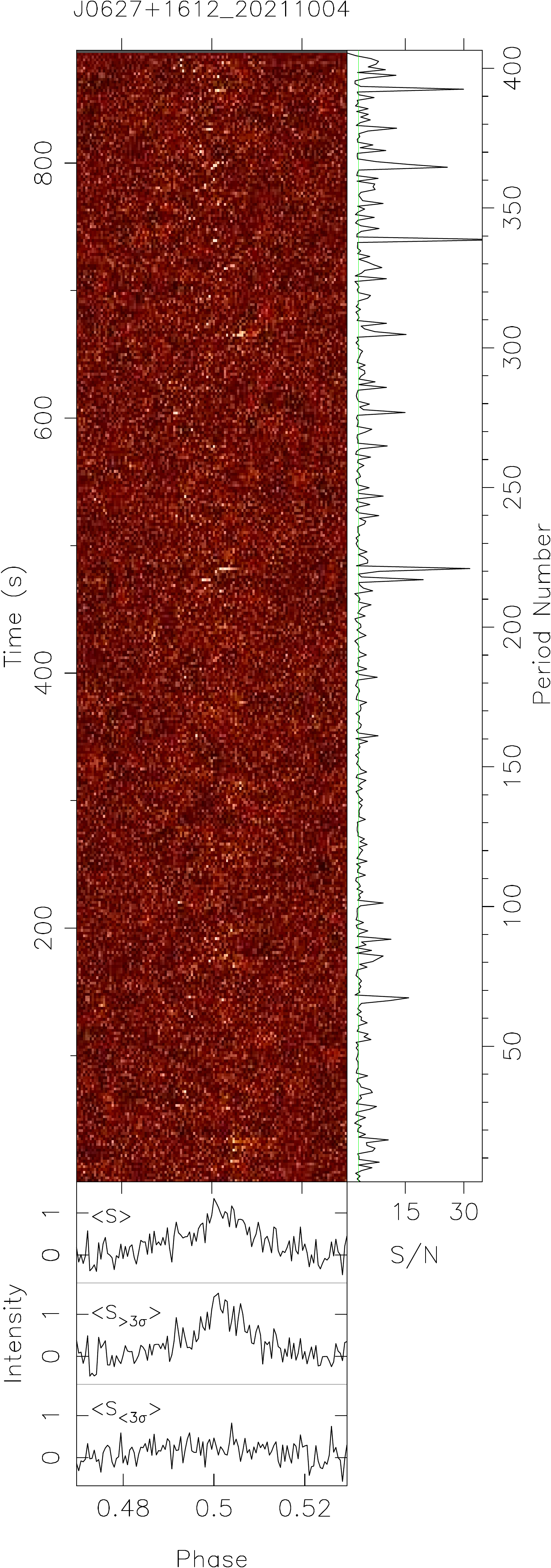} 
  \includegraphics[width=38mm]{KnowRRATs/J0628+0909_20210922_tracking-M01_0011_fn.Czap.R.CFS.pdf} 
  \includegraphics[width=38mm]{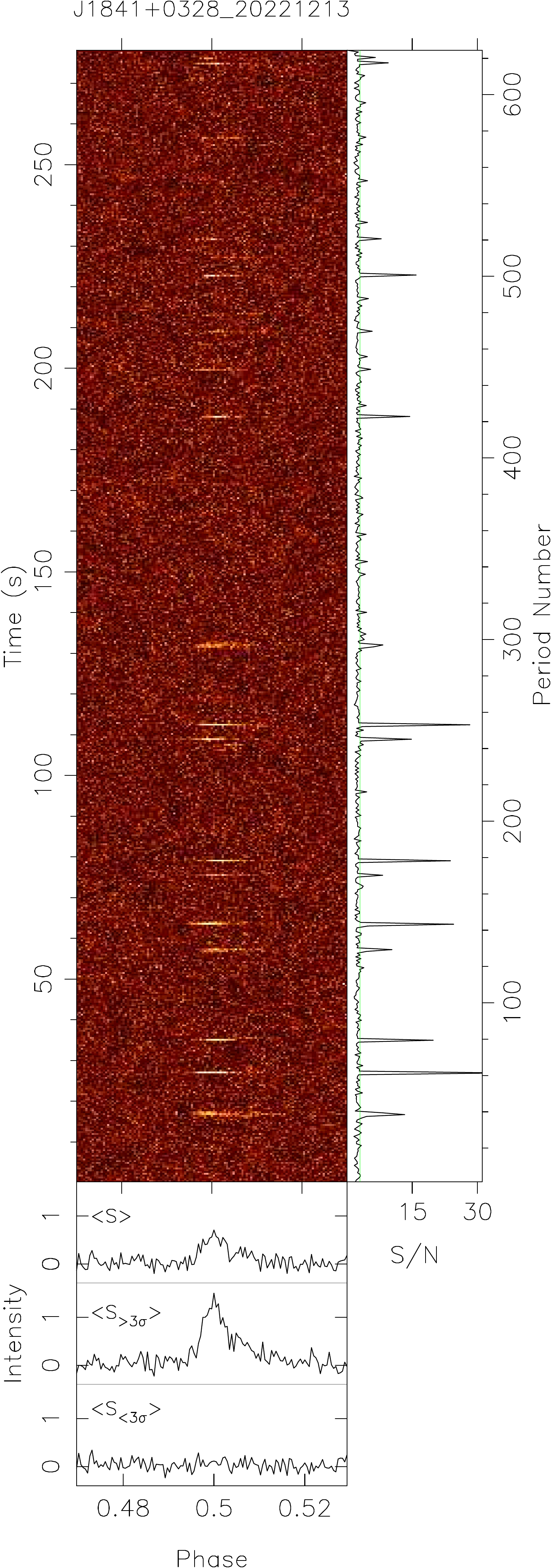} \\[1.0mm]%
  \includegraphics[width=38mm]{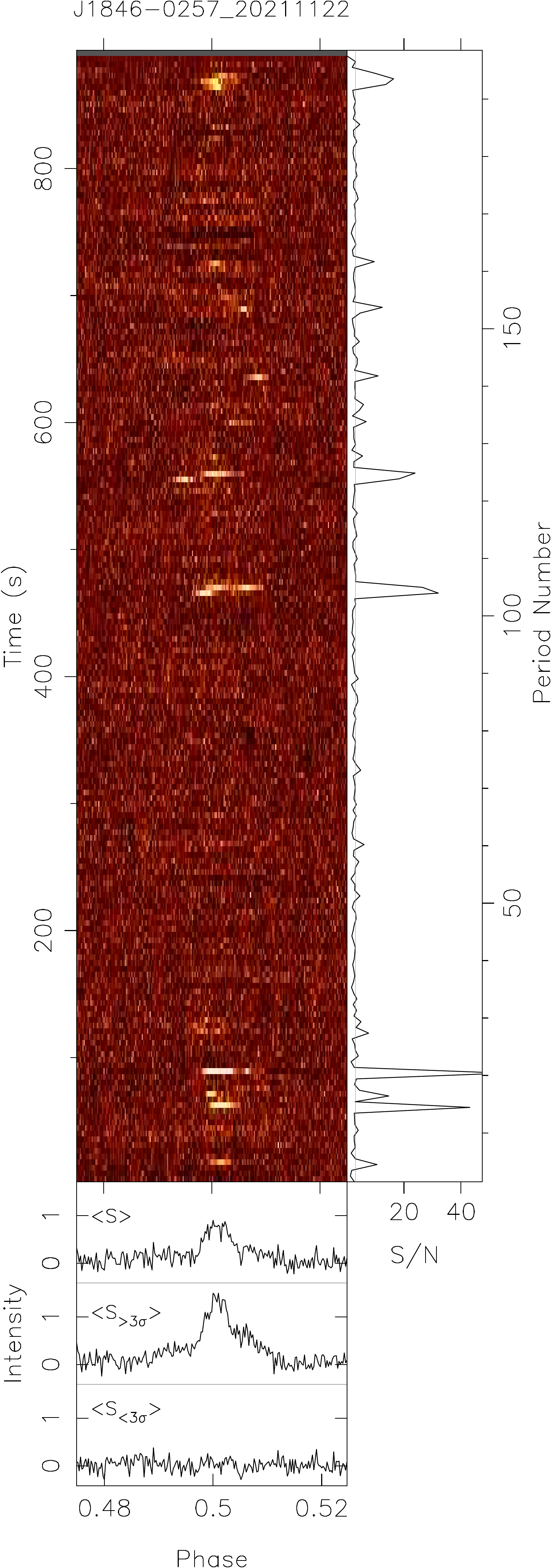} %J1846-0257
  \includegraphics[width=38mm]{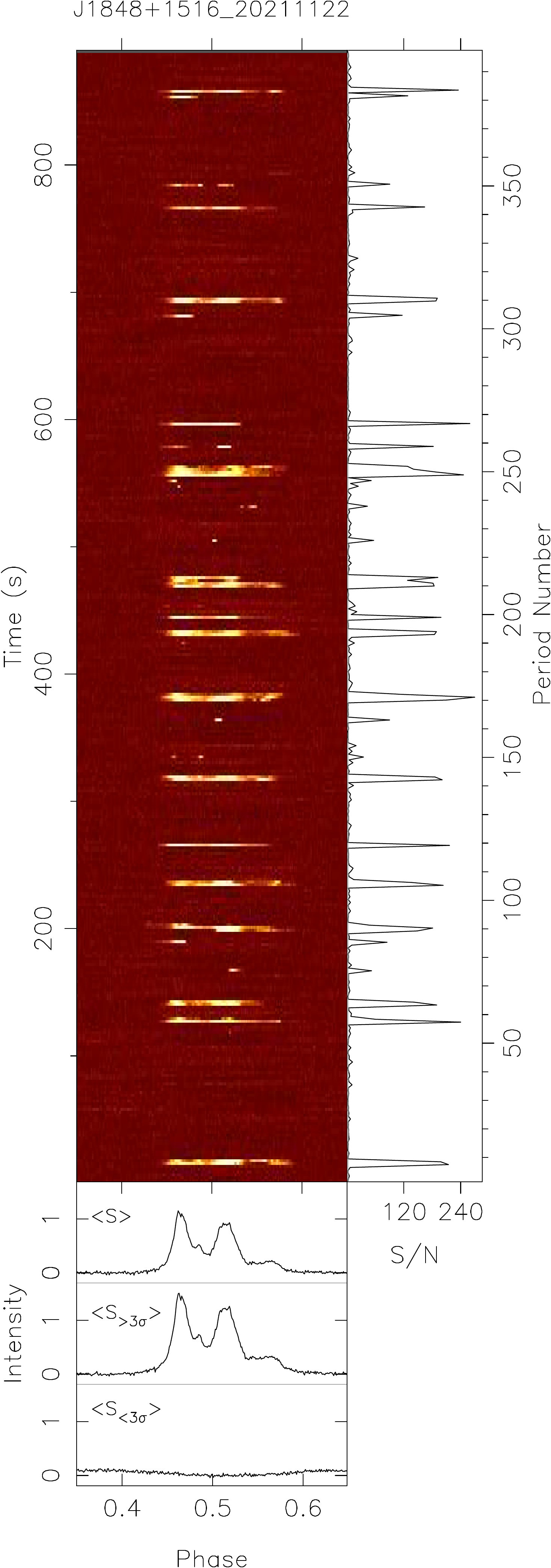} %J1848+1516
  \includegraphics[width=38mm]{KnowRRATs/J1854+0306_20211122_tracking-M01-P1-c2048b1.ar.z.psh.z.Fp.debased.pdf} %J1854+0306
  \includegraphics[width=38mm]{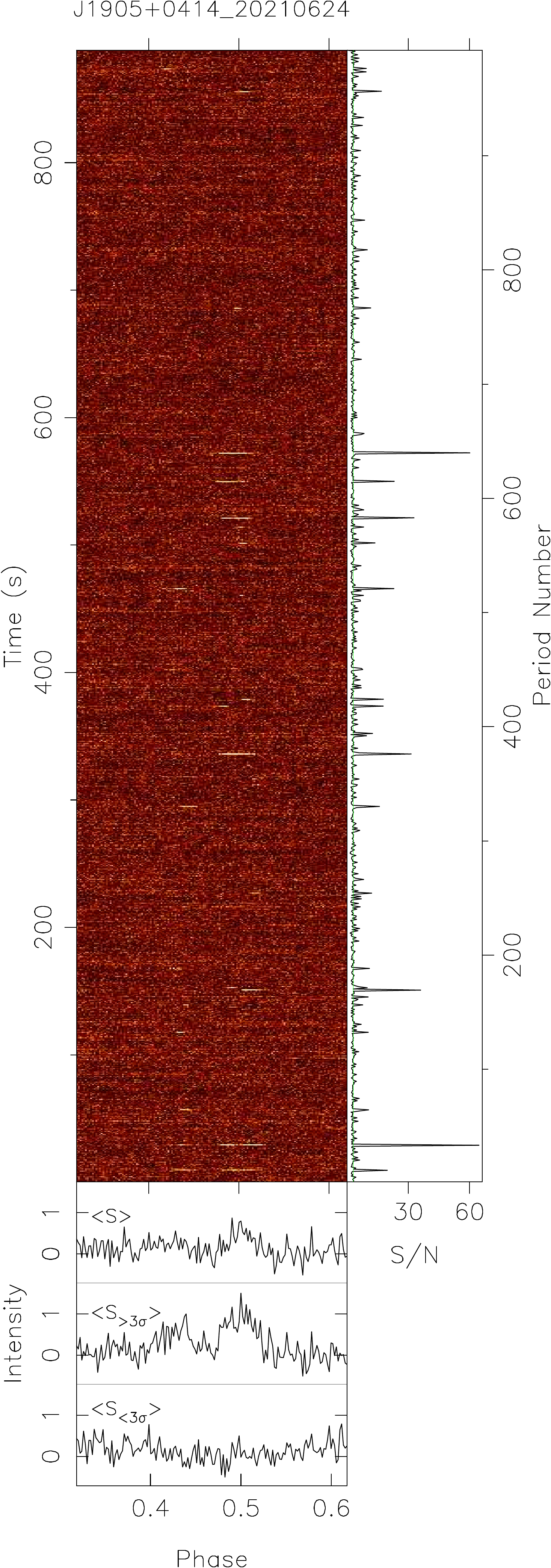} 
  \caption{Pulse-stacks for the Known RRATs of the group for weak pulsars with sparse strong pulses.}
    \label{fig:APPknownRRAT3}
\end{figure*}
\addtocounter{figure}{-1}
\begin{figure*}[!htp]
  \centering
  \includegraphics[width=38mm]{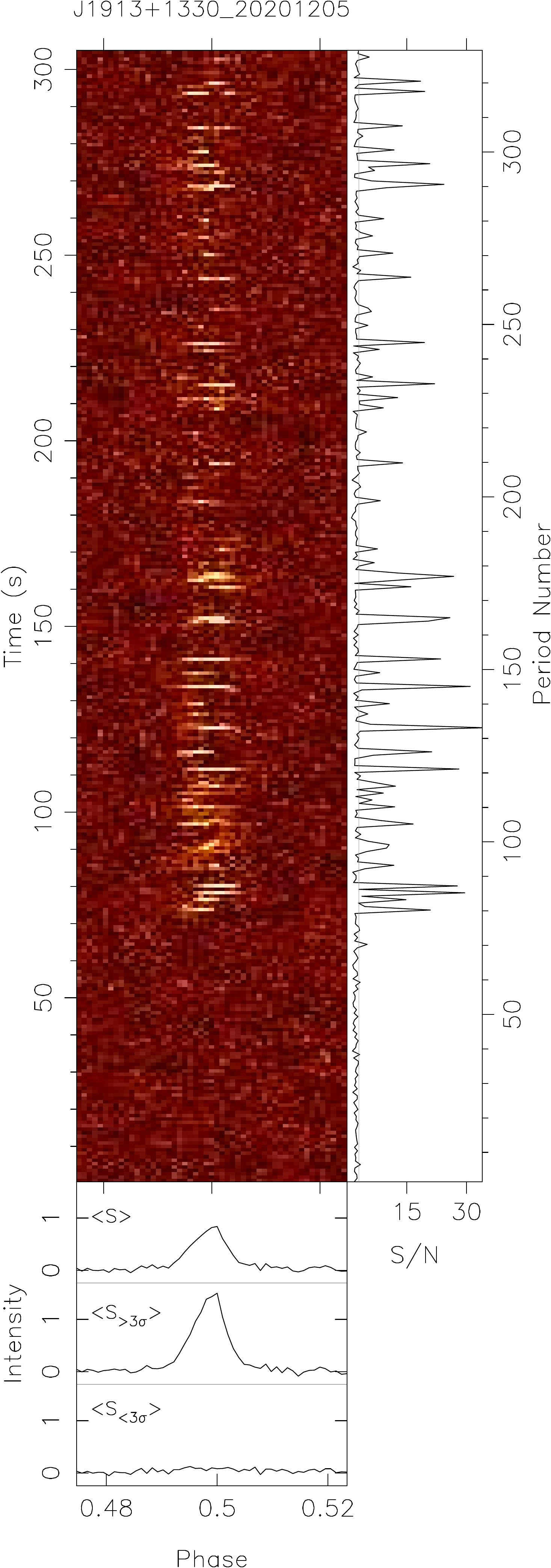} 
  \includegraphics[width=38mm]{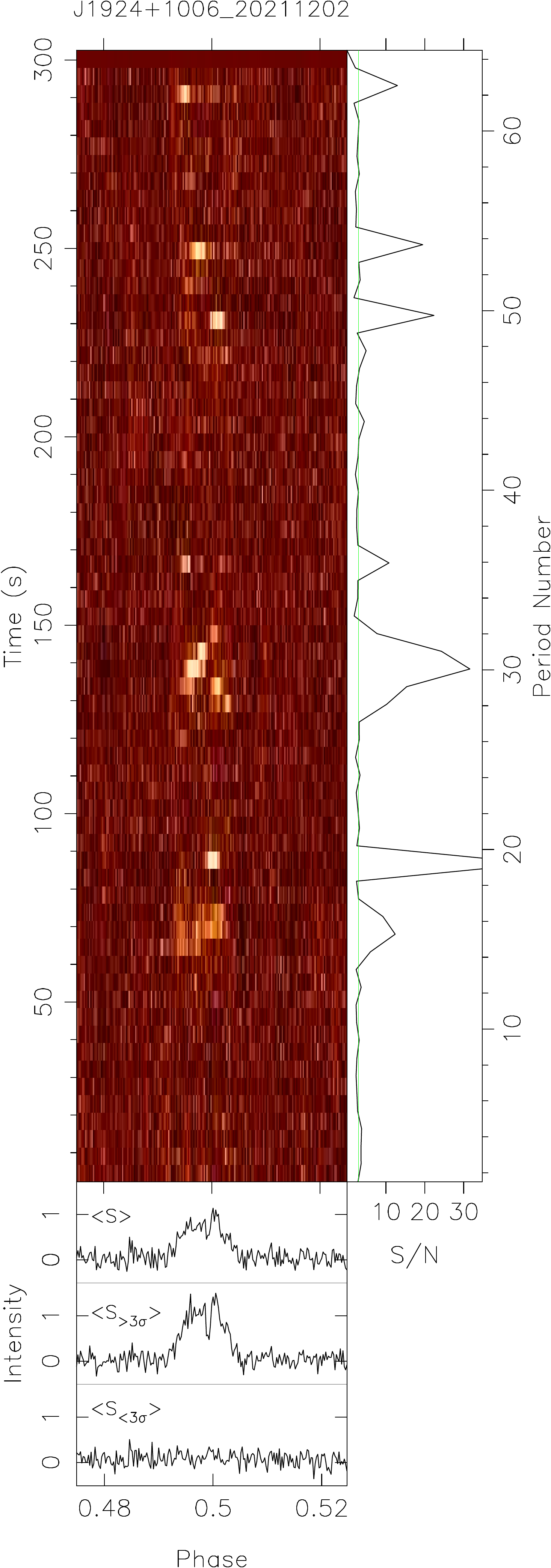} %J1924+1006
  \includegraphics[width=38mm]{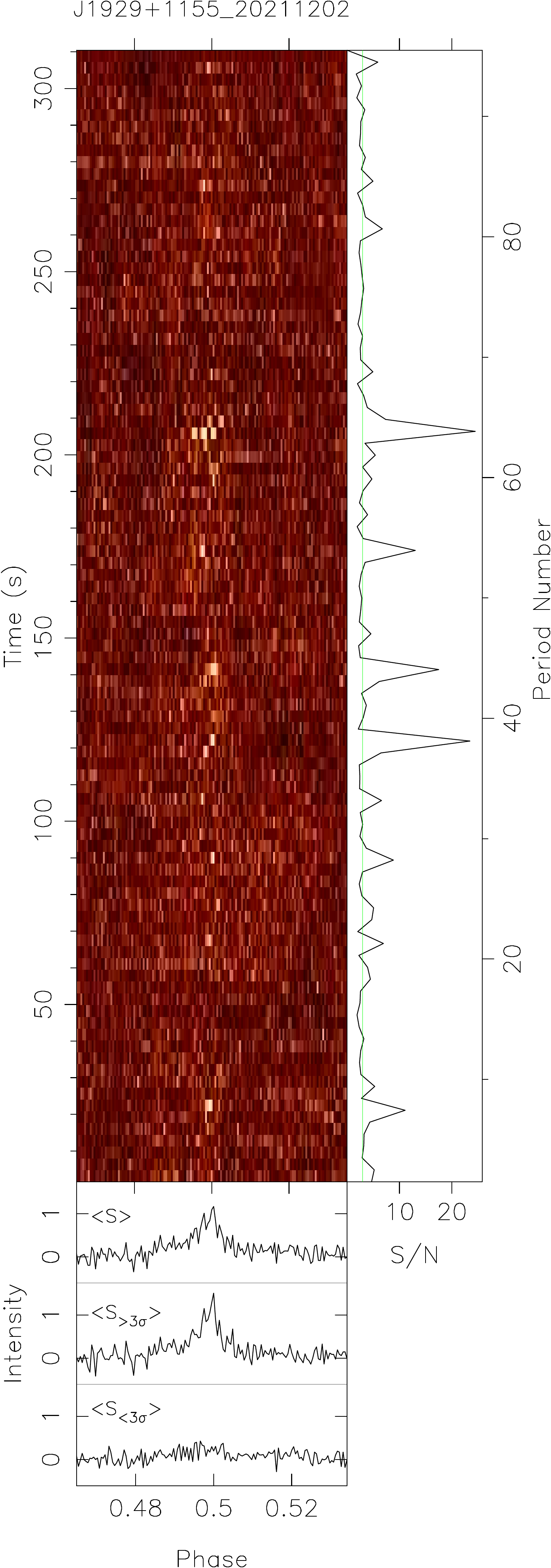} %J1929+1155
  \includegraphics[width=38mm]{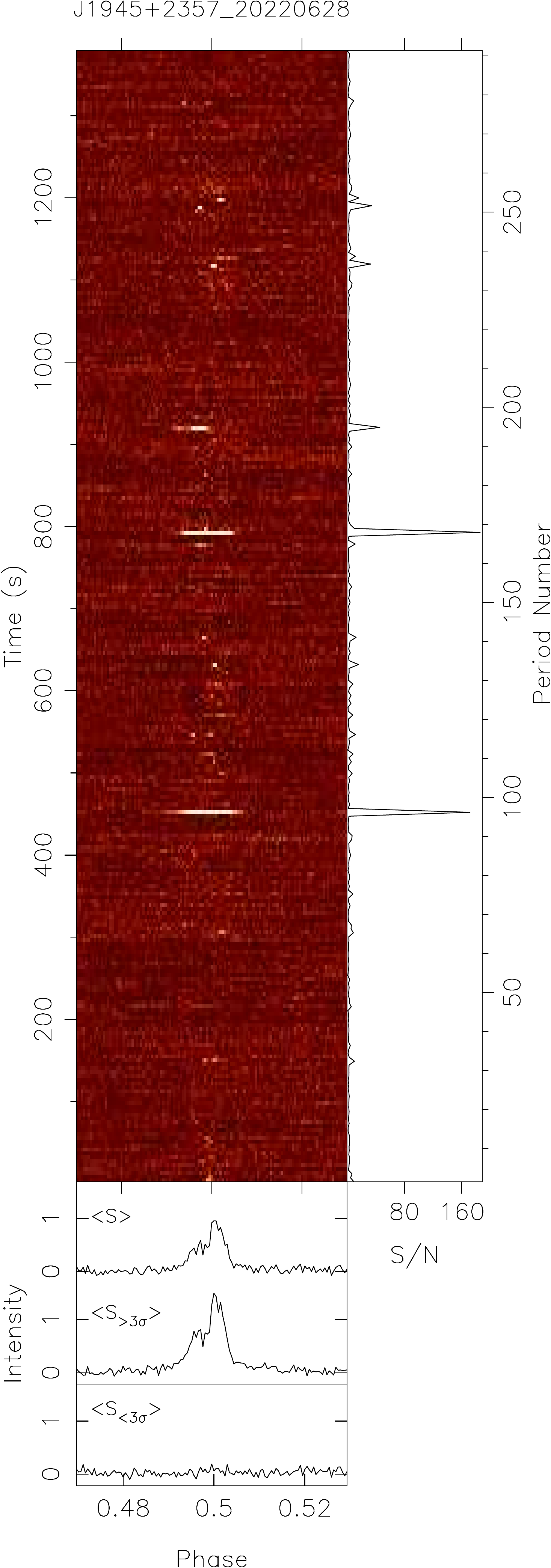} \\[1.0mm]%J1945+2357
  \includegraphics[width=38mm]{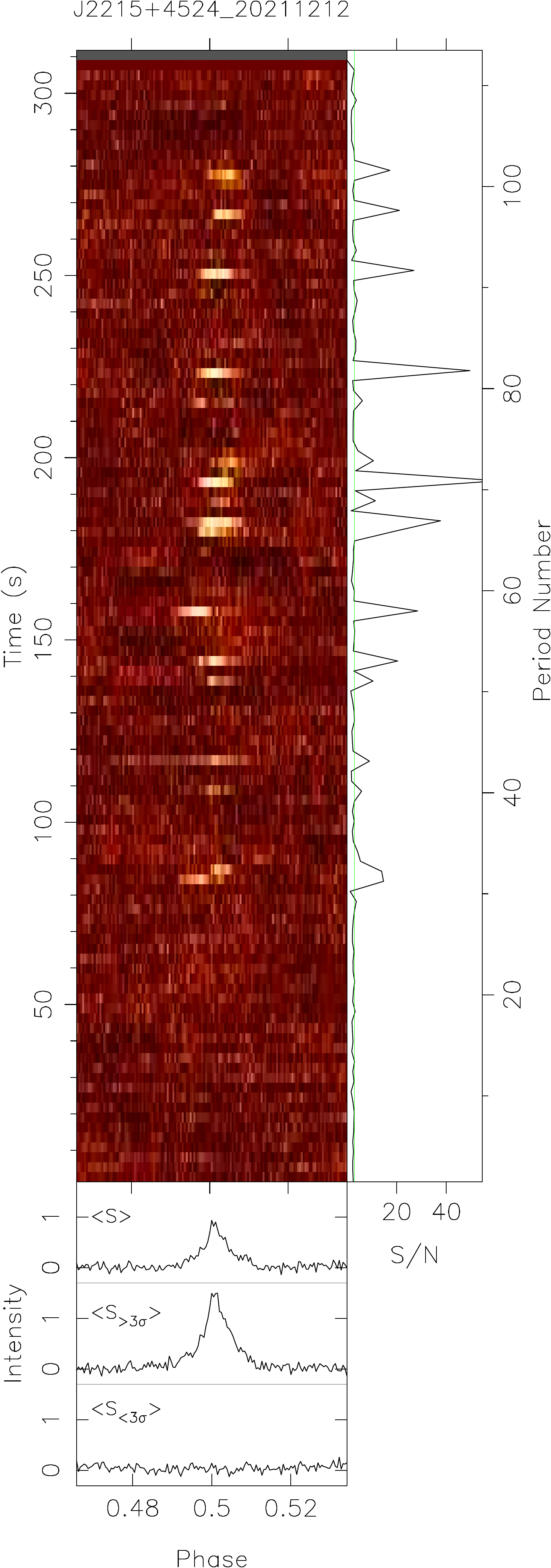} %J2215+4524
  \caption{{\it -- continued and ended}.}
\end{figure*}

\end{document}